\documentclass[a4paper,11pt,openright,twoside]{report}

\def\title{Provably Secure Networks: \\ Methodology and Toolset for Configuration Management}
\def\titlepdf{Provably Secure Networks: Methodology and Toolset for Configuration Management} 
\def\author{Cornelius Diekmann}
\def\date{\LaTeX\ build date: \today}

\usepackage[utf8]{inputenc}
\usepackage[T1]{fontenc}

\usepackage[dvipsnames]{xcolor}

\usepackage[english,ngerman]{babel}
\usepackage[babel]{csquotes}

\usepackage{IEEEtrantools}

\usepackage{vmargin}
\setpapersize{A4}
\setmarginsrb{3cm}{1.5cm}{3cm}{1.5cm}{6mm}{7mm}{5mm}{15mm}

\usepackage[pdftex]{graphicx}
\usepackage[labelfont=bf,font=small]{caption}
\usepackage[labelfont=bf,font=small]{subcaption}

\usepackage{amsmath}
\usepackage{amssymb}
\usepackage{amsthm}
\usepackage{nicefrac}

\usepackage[protrusion=true,expansion,kerning,spacing,tracking]{microtype}
\usepackage{setspace}

\usepackage{array}
\usepackage{booktabs}
\usepackage[emulate=units,binary-units=true]{siunitx}
\usepackage{url}

\usepackage{tikz}
\usetikzlibrary{shapes}
\usetikzlibrary{calc}
\usetikzlibrary{arrows}
\usetikzlibrary{arrows.meta} 

\tikzset{myptr/.style={-{Latex[scale=1.5]}}}%
\tikzset{myptrdouble/.style={{Latex[scale=1.5]}-{Latex[scale=1.5]}}}%
\tikzset{myptrdotted/.style={myptr,dashed}}%

\usepackage{enumitem}
\setlist[description]{leftmargin=2\parindent,labelindent=\parindent}

\usepackage{chngpage}
\usepackage{calc}

\usepackage{xspace}

\usepackage{ellipsis}

\newcommand{\hairspace}{\hspace{1pt}}
\newcommand{\eg}{\mbox{e.\hairspace{}g.,} }  
\newcommand{\ie}{\mbox{i.\hairspace{}e.,} }  
\newcommand{\cf}{\mbox{cf.}\ }
\newcommand{\etal}{\mbox{et~al.}\ }

\usepackage{fancyhdr}  	
\usepackage{listings} 
\definecolor{lightestyellow}{rgb}{1,1,0.9}

\colorlet{punct}{red!60!black}
\definecolor{background}{HTML}{EEEEEE}
\definecolor{delim}{RGB}{20,105,176}
\colorlet{numb}{magenta!60!black}

\lstdefinelanguage{json}{
    basicstyle=\ttfamily\scriptsize,
    numberstyle=\scriptsize,
    stepnumber=1,
    numbersep=8pt,
    showstringspaces=false,
    breaklines=true,
    breakatwhitespace=true,
    frame=lines,
    backgroundcolor=\color{background},
    literate=
     *{0}{{{\color{numb}0}}}{1}
      {1}{{{\color{numb}1}}}{1}
      {2}{{{\color{numb}2}}}{1}
      {3}{{{\color{numb}3}}}{1}
      {4}{{{\color{numb}4}}}{1}
      {5}{{{\color{numb}5}}}{1}
      {6}{{{\color{numb}6}}}{1}
      {7}{{{\color{numb}7}}}{1}
      {8}{{{\color{numb}8}}}{1}
      {9}{{{\color{numb}9}}}{1}
      {null}{{{\color{numb}{null}}}}{4}
      {:}{{{\color{punct}{:}}}}{1}
      {,}{{{\color{punct}{,}}}}{1}
      {\{}{{{\color{delim}{\{}}}}{1}
      {\}}{{{\color{delim}{\}}}}}{1}
      {[}{{{\color{delim}{[}}}}{1}
      {]}{{{\color{delim}{]}}}}{1},
}

\usepackage[pdftex,hyperfootnotes=false,pdfpagelabels]{hyperref}

\hypersetup{
    colorlinks=true,
    linktocpage=true,
    breaklinks=true,
    bookmarksnumbered,
    bookmarksopen=true,
    bookmarksopenlevel=1,
    pdfhighlight=/O,
    urlcolor=black,
    linkcolor=black,
    citecolor=black,
    pdftitle={\titlepdf},
    pdfauthor={\author},
    pdfsubject={Doktorarbeit in der Informatik},
}

\usepackage{fancyvrb}

\usepackage{mathpartir}

\mathchardef\mhyphen="2D 

\usepackage{bbding}

\usepackage{pifont}
\newcommand{\cmark}{\ding{51}}%
\newcommand{\xmark}{\ding{55}}%

\usepackage{components/moeptikz}

\newif\ifnetcover
\netcovertrue
\netcoverfalse

\ifnetcover
\usepackage{pdfpages}

\makeatletter
\newcommand*{\cleartoleftpage}{%
	\clearpage
	\if@twoside
	\ifodd\c@page
	\hbox{}\newpage
	\if@twocolumn
	\hbox{}\newpage
	\fi
	\fi
	\fi
}
\makeatother
\fi

\newcommand{\mvar}[1]{\ensuremath{\mathit{#1}}}
\newcommand{\mdef}[1]{\ensuremath{\mathsf{#1}}}
\newcommand{\mfun}[1]{\mdef{#1}}
\newcommand\mconstr[1]{\mdef{#1}}
\newcommand\mctrl[1]{\ensuremath{\mathbf{#1}}}

\newcommand{\topos}{\emph{topoS}}
\newcommand{\fffuu}{\emph{{f}{f}{f}uu}}

\newcommand\iptables{{iptables}}

\newcommand{\BigO}{\mathcal{O}}

\DeclareMathSymbol{\mlq}{\mathord}{operators}{``}
\DeclareMathSymbol{\mrq}{\mathord}{operators}{`'}

\newcommand\lstapp{\ensuremath{\mathbin{:\mkern-1mu:\mkern-1mu:}}} 
\newcommand\lstcons{\ensuremath{\mathbin{:\mkern-1mu:}}} 

\theoremstyle{plain}
\newtheorem{theorem}{Theorem}
\newtheorem{lemma}{Lemma}

\newtheorem{corollary}{Corollary}

\theoremstyle{definition}

\newtheorem{definition}{Definition}

\makeatletter
\def\examplefont#1{\def\ex@font{#1}}         \def\ex@font{}
\examplefont{}
\newlength{\exampleindent}    
\newenvironment{example}%
   {\begin{list}{}{%
    \setlength{\leftmargin}{\exampleindent}}
    \ex@font \item[]}
   {\end{list}}
\makeatother

\definecolor{LightYellow}{RGB}{255, 238, 170}
\definecolor{LightGrey}{RGB}{230, 230, 230}

\tikzset{MyDoubleArrow/.style={double arrow, draw=black, anchor=west, align=center, text width=1em}}
\tikzset{MySingleArrow/.style={single arrow, draw=black, anchor=west, align=center, text width=1em}}
\tikzset{MySingleLeftArrow/.style={single arrow, rotate=180, draw=black, anchor=east, align=center, text width=1em}}

\tikzset{MyRoundedBox/.style={
		draw,
		rounded corners=3pt,
		inner sep=5pt,
		anchor=west,
		text width=10em,
		align=center
	}
}

\newcommand*\circled[1]{\tikz[baseline=-3pt, scale=0.5, every node/.style={scale=0.5}]{\node[shape=circle,draw,inner sep=1pt,minimum size=16pt] (char) {#1};}}

\newcommand\allow{\ensuremath{\textnormal{\circled{\small \Checkmark}}}}
\newcommand\deny{\ensuremath{\textnormal{\circled{\small \XSolidBrush}}}}
\newcommand\undecided{\ensuremath{\textnormal{\circled{\textnormal{\large \textbf{?}}}}}}

\newcommand\matchop[1]{\ensuremath{\mfun{match}\ {#1}}}
\newcommand\matches[1]{\ensuremath{\matchop\gamma \ #1 \ p}}
\newcommand\nmatches[1]{\ensuremath{\neg\; \matchop\gamma \ #1 \ p}}
\newcommand\bigstep[3]{\ensuremath{\Gamma,\gamma,p \vdash\big\langle #1,\; #2 \big\rangle \Rightarrow #3}}

\newcommand\iptaction[1]{\ensuremath{\mathtt{#1}}}

\newcommand\bigstepapprox[4]{\ensuremath{\Gamma,\gamma,p \vdash\big\langle #1,\; #2 \big\rangle \Rightarrow_{#4} #3}}

\allowdisplaybreaks

\begin{document}
\bstctlcite{IEEEexample:BSTcontrol}

\pgfdeclarelayer{background}
\pgfdeclarelayer{foreground}
\pgfsetlayers{background,main,foreground}

\ifnetcover
\pagenumbering{Alph}
\phantomsection
\pdfbookmark[1]{Cover}{tableofcontents}
\thispagestyle{empty}
\includepdf[pages={1},offset=2.54cm -2.54cm]{cover_front.pdf}
\thispagestyle{empty}
\cleardoublepage
\pagenumbering{alph}
\fi


\begin{titlepage}
\setlength{\parindent}{0cm}
\onehalfspacing
\begin{center}
	\large
  TECHNISCHE UNIVERSITÄT MÜNCHEN\\
  Institut für Informatik\\
  Lehrstuhl für Netzarchitekturen und Netzdienste\\
\end{center}

\vfill
  
\begin{center}
	\Large
  \textbf{
  \title{}}\\[1cm]
  
  \vspace*{1ex}

 	\large
  Cornelius Hermann Diekmann
\end{center}

\vfill

Vollständiger Abdruck der von der Fakultät für Informatik 
der Technischen Universität München zur Erlangung des akademischen Grades eines
\begin{center}
Doktors der Naturwissenschaften (Dr. rer. nat.)
\end{center}
genehmigten Dissertation.

\vspace{0.5cm}

\begin{center}
\setlength{\tabcolsep}{0pt}
\begin{tabular}{p{4.7cm}p{0.7cm}p{9.3cm}}
Vorsitzender:               &    & Prof.\ Tobias Nipkow, Ph.D.\\
Prüfer der Dissertation:    & 1. & Prof.\ Dr.-Ing.\ Georg Carle\\
                            & 2. & Prof.\ Steven M.\ Bellovin (Columbia University)
                           
\end{tabular}
\end{center}

\vspace{0.5cm}

Die Dissertation wurde am 28.03.2017 bei der Technischen Universität München
eingereicht und durch die Fakultät für Informatik am 12.07.2017 angenommen.

\end{titlepage}

\thispagestyle{empty}

\vspace*{1cm}
\vfill

\begin{flushleft}
\noindent
Cataloging-in-Publication Data\\
\author{}\\
\emph{\titlepdf{}}\\
Dissertation, Juli 2017\\
Network Architectures and Services, Department of Computer Science\\
Technische Universität München\\[1cm]
ISBN 978-3-937201-57-3\\
ISSN 1868-2634 (print)\\
ISSN 1868-2642 (electronic)\\
DOI 10.2313/NET-2017-07-2\\
Network Architectures and Services NET-2017-07-2\\
Series Editor: Georg Carle, Technische Universität München, Germany\\
\copyright\ 2017, Technische Universität München, Germany
\end{flushleft}


  \selectlanguage{english}

  \pagenumbering{roman}
  \pagestyle{plain}

\pdfbookmark[1]{Abstract}{Abstract}
\begingroup
\let\clearpage\relax
\let\cleardoublepage\relax
\let\cleardoublepage\relax

%
%
%
%
%
%


\section*{Abstract}
\noindent
Network management and administration is an inherently complex task, in particular when it comes to security. 
Configuration complexity in this domain leads to human error, which is often only uncovered when it is too late: after a successful attack.  

This thesis focuses on the security of network configurations, \ie network-level access control and network-level information flow security. 
The objective is to employ formal methods to prevent, uncover, and prove lack of security-related configuration errors. 
We contribute methods and tools to translate between security components on various abstraction levels and to verify their conformance. 
We prove correctness of our tools with the Isabelle interactive proof assistant.

First, we propose a method to construct new networks from scratch. 
We present our tool \topos{} which enables automation of the design, requiring only a specification of the security requirements. 
Second, we present a method to understand and analyze existing network security device configurations, focusing on the iptables firewall. 
We present our fully automated tool \fffuu{} for this task. 
Finally, we show how both approaches can interact with each other. 

Our experience has shown that a solution to the presented problems must be usable, must not expose over-formalism to the administrator, must leave the administrator in full low-level control, must support legacy configurations, and must be non-invasive, \ie must not require that an administrator completely relinquishes control to a tool. 
We demonstrate that our proposed tool-supported methodology fulfills its goals as follows: 
By its very nature, access control lists scale quadratically in the number of networked entities or roles.
We propose a methodology to specify security requirements which can scale better than linear (depending on its usage). 
Our methodology works on well-defined intermediate results and gives the administrator full control over them. 
A policy computed by this approach can be deployed to a network directly. 
Or, our methodology can be used completely non-invasive: It can statically verify that an existing iptables ruleset conforms to the policy and requirements. 
In general, we provide a method to compute a clear overview of the policy enforced by an existing (legacy) iptables firewall. 
Both directions (synthesizing new policies vs.\ verifying existing policies) are compatible with each other and an administrator may freely choose to which extent she wants to migrate to our methodology and to which extent she wants to remain in full low-level control. 
Ultimately, it is possible to use our toolset in a full circle. 

We evaluated our tools, among others, on an aircraft cabin data network, Android measurement app, and on the largest collection of public, real-world iptables dumps (made available by us). 
We showed further applicability in the domain of microservice management, SDN configuration, cyber physical systems, software architectures, and privacy. 


\vfill
\newpage

\selectlanguage{ngerman}
\pdfbookmark[1]{Kurzfassung}{Kurzfassung}

\section*{Kurzfassung}
\noindent
Administrierung und Management eines Netzwerkes ist eine inhärent komplexe Aufgabe, insbesondere im Hinblick auf Security. Konfigurationskomplexität führt zu menschlichem Versagen, welches erst erkannt wird, wenn es zu spät ist: nach einem erfolgreichen Angriff.

Diese Dissertation beschäftigt sich mit der Sicherheit von Netzwerkkonfigurationen, d.h. Access Control und Information Flow Security auf Netzwerkebene. Das erklärte Ziel ist es formale Methoden einzusetzen, um sicherheitsrelevante Konfigurationsfehler zu verhindern, erkennen und deren Abwesenheit zu beweisen. Wir stellen Tools und Methoden bereit, um zwischen Sicherheitskomponenten auf verschiedenen Abstraktionsebenen zu übersetzen und deren Konformität zu verifizieren.
Wir beweisen die Korrektheit unserer Tools mit dem interaktiven Theorembeweiser Isabelle. 

Im ersten Teil der Arbeit schlagen wir eine Methode vor, um Netzwerke von Grund auf neu zu designen. Wir präsentieren unser Tool \topos{}, welches diesen Designprozess automatisiert und dafür nur eine Spezifikation der Sicherheitsanforderungen benötigt. Im zweiten Teil stellen wir eine Methode vor, um bestehende Netzwerksicherheitsgerätekonfigurationen zu verstehen und zu analysieren. Dabei fokussieren wir uns auf die iptables Firewall und stellen unser automatisiertes Tool \fffuu{} vor. Im finalen Teil der Arbeit zeigen wir, wie beide Ansätze ineinandergreifen.

Unsere Erfahrung hat gezeigt, dass Lösungen, die den genannten Problemen gerecht werden wollen, benutzbar sein müssen, dem Administrator keine Überformalisierung aussetzen dürfen, dem Administrator low-level Kontrolle zugestehen müssen, Legacy-Konfigurationen unterstützen müssen und nicht invasiv sein dürfen, d.h. dass sie nicht fordern dürfen, dass ein Administrator komplett die Kontrolle an ein Tool abgibt. Wir zeigen, dass unser vorgeschlagener, toolgestützter Ansatz diese Ziele wie folgt erfüllt: Es liegt in der Natur der Sache, dass Access Control Listen quadratisch mit der Anzahl der Geräte bzw. Rollen skalieren. Wie schlagen eine Methode vor, die es erlaubt Sicherheitsanforderungen zu spezifizieren, welche besser als linear skaliert (abhängig von der Nutzung). Unsere Methode arbeitet auf wohldefinierten Zwischenergebnissen und überlässt dem Administrator die volle Kontrolle über diese. Eine so berechnete Policy kann direkt in einem Netzwerk ausgerollt werden, oder unsere Methode kann komplett nicht-invasiv eingesetzt werden: Die Übereinstimmung existierender iptables Regelsätze mit der Policy oder den Anforderungen kann statisch überprüft werden. Im Allgemeinen stellen wir eine Methode vor, um eine Übersicht über die Policy zu berechnen, welche eine (legacy) iptables Firewall umsetzt. Beide Richtungen (neue Policies synthetisieren vs.\ existierende Policies verifizieren) sind untereinander kompatibel und es obliegt dem Administrator zu entscheiden, zu welchem Grad sie auf unsere Methode umstellen möchte und zu welchem Grad sie die komplette low-level Kontrolle behalten möchte. Es ist möglich unsere Tools iterativ einzusetzen. 

Wir haben unsere Tools unter Anderem in einem Flugzeugkabinennetzwerk, einer Android Messapp und der größten öffentlichen iptables Kollektion (die von uns bereitgestellt wurde), getestet. Wir zeigten weitere Anwendbarkeit im Bereich des Microservicemanagement, SDN Konfiguration, Cyberphysicalsystems, Softwarearchitekturen und Privacy.

\endgroup			

\selectlanguage{english}

\vfill

  \cleardoublepage


\phantomsection
\pdfbookmark[1]{\contentsname}{tableofcontents}

\sloppy
\tableofcontents
\fussy

\cleardoublepage










  \cleardoublepage


  \pagenumbering{arabic}
  \pagestyle{fancy}

  \chapter{Problem Statement \& Goals}  
  \label{chap:introduction}

  \begin{quote}
\textit{Simplicity is a great virtue but it requires hard work to achieve it and education to appreciate it. And to make matters worse: complexity sells better.}
\end{quote}
\hfill E.\ W.\ Dijkstra, On the nature of Computing Science (1984)~\cite{EWD:EWD896}.

\section{Introduction}
Network administration is a challenging task and requires competent network administrators. 
Handling user complaints, improving performance, reacting to hardware failures, account management, low-level troubleshooting, complying with high-level corporate policies, standard conformance, \dots, and security are among the daily tasks of a network administrator~\cite{burges2004sysadminbook,bellovin2009configuration}. 
Traditionally, network management is a low-level, manual, ad-hoc task.  
%
%
It is said that our networks are kept running by ``Masters of Complexity''~\cite{sck11futurenetworkspastprotocols,mck2012sdntame}. 
%
%
%
Yet, ``controlling complexity is a core problem in information security''~\cite{guttman05rigorous}. 
Unsurprisingly, security issues exist in many networks~\cite{firwallerr2004,wool2010firewall,netsecconflicts,sherry2012making,diekmann2014forte,fwviz2012,fireman2006,ZhangAlShaer2007flip,databreach2009src,nelson2010margrave,survery2012networktroubleshooting}.

%
Network segmentation, isolation, and controlled access are the fundamental building blocks for the baseline security of a computer network~\cite[\cf B 4.1, M 5.111]{bsigrundschutz}. 
Administrating these security-related aspects of a network is a highly complex task. 
Human error, in particular configuration errors, are a central cause for network problems~\cite{sherry2012making,networkdowntime2009,oppenheimer2003internet}. 
Configuration errors which lead to security problems are sometimes attributed to the (accidental) complexity of the low-level languages which are used to configure network security mechanisms \cite{Cuppens2005orbacxmlfirewall,pozo2009model}. 
For example, the default Linux firewall iptables~\cite{iptables} features more than 200 matching features~\cite{maniptablesextensions}. 
The firewall configuration language cannot be simplified by removing features because they are actively used~\cite{diekmanngithubnetnetwork,cloudflare2014blogbpf,serverfaultiptables}. 
In general, administrators need low-level control over their rulesets since, often, performance and other network-related issues apart from security must also be implemented in a firewall ruleset.

In addition, legacy configurations of enormous complexity have evolved over time. 
For example, the iptables firewall is over ten years old and there are also rulests of that age which are still deployed on core firewalls but are no longer understood by the administrator~\cite{diekmanngithubnetnetwork}. 
Even though simple, high-level languages for network configuration have been proposed~\cite{netcore12,nelson2014flowlog,zhao2011policyremanet,anderson2014netKATsemantics,mignis2014,bartal1999firmato,Cuppens2005orbacxmlfirewall,ZhangAlShaer2007flip,hinrichs2009practical,soule2014merlin,policy2010berapolicyformalenterprise}, the question of how to deal with legacy configuration often remains unanswered. 

\medskip
In this thesis, we address research questions about the security of network configurations. 
We focus on network-level access control and network-level information flow security. 
The declared goal of this thesis is to provide means to help administrators to increase the security of their network configuration. 
We begin by designing a high-level language for security requirements which can be translated in several steps to configurations for network security mechanisms, \eg iptables. 
This process is unique in that it still allows low-level control for the administrator as well as guaranteeing soundness. 
In the second part, we take the opposite direction and translate legacy iptables firewall configurations of enormous complexity to a simplified, high-level view. 

\begin{center}
$\bullet$ \emph{The ipta({\hskip-1pt}b{\hskip-2pt})les -- There and Back Again} $\bullet$
\end{center}

The complete formal theory, as well as executable tools, have been machine-verified with the interactive theorem prover Isabelle/HOL~\cite{isabelle2016}. 
Several contributions to the archive of formal proofs have been made~\cite{Network_Security_Policy_Verification-AFP,IP_Addresses-AFP,Simple_Firewall-AFP,Iptables_Semantics-AFP,Routing-AFP,LOFT-AFP}. 
During his research, the author advanced the state of the art, both in the world of formal methods~\cite{diekmann2015fm} as well as in the world of computer networks~\cite{diekmann2016networking}. 
While the theoretical work is ``substantial''~\cite{iptablesafpmaillarry} and ``shiny''~\cite{iptablesafpmailgerwin}, the practical applicability has also been demonstrated to hundreds of hackers~\cite{diekmann32c3firewall}\footnote{Around 500 people attended the talk on-site; as of November 2016, the video recording of the talk has over 4000 views.} \cite{diekmann1curry,diekmann2curry,diekmann3curry}.


\section{Research Objectives}
\label{sec:researchobjectives}
%
The declared goal of this thesis is to improve the situation in the field of network security administration. 
At the end of the day, in order to minimize attack surface, we want an answer to the question \textit{``Which machines should be allowed to speak to each other?''} and we want to know whether the answer to the question is also practically enforced. 
To put this overall goal statement in concrete terms, we first present a model of security components and afterwards split the overall goal into several research questions according to the model. 

\paragraph*{Security Components}
\begin{figure}[h!]
\centering
 \begin{center}
    \resizebox{0.99\textwidth}{!}{%
 	\begin{tikzpicture}
	 \node [MyRoundedBox, fill=LightYellow](sinvar) at (0,0) {Security Requirements};
	 \node [MyDoubleArrow](arr1) at (sinvar.east) {};
	 \node [MyRoundedBox, fill=LightYellow](policy) at (arr1.east) {Security Policies};
	 \node [MyDoubleArrow](arr2) at (policy.east) {};
	 \node [MyRoundedBox, fill=LightYellow](mechanism) at (arr2.east) {Security Mechanisms};
	 \end{tikzpicture}
	 }
 \end{center}
   \caption[Security Components]{Security Components\footnotemark}
  \label{fig:intro:securitycomponents}
\end{figure}
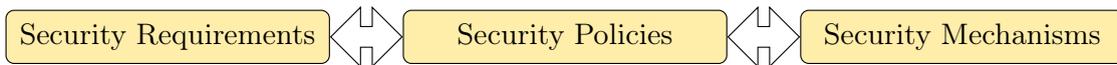
%
%
\footnotetext{The image was inspired by Bishop~\cite{bishop2003compsec} and first appeared in the author's master's thesis~\cite{cornythesis}. Since 2013, it is also used in the lecture ``Network Security'' at TUM.}%
\noindent
Security can be divided into three components~\cite{bishop2003compsec}, as illustrated by Figure~\ref{fig:intro:securitycomponents}: 
The \emph{security requirements} specify on a high level of abstraction the scenario-specific security goals. 
The \emph{security policy} specifies rules which implement the requirements. 
Finally, the \emph{security mechanisms} enforce the policy; requiring low-level configuration.

\begin{example}%
	\textbf{Example. }%
We can imagine the security requirements as a text document written in natural language. 
The security policy could be expressed by an access control matrix. 
A firewall, the security mechanism, can be configured with a ruleset to implement the desired policy. 
\end{example}

Security problems arise if, on the one hand, the components 
 are not consistent with each other, \eg a policy does not correctly reflect some security requirements or a security mechanism is misconfigured and does not implement the policy.  
On the other hand, security problems may also arise if the specification of the security requirements does not express the desired security properties.

\paragraph*{Research Questions}
The overall research question is 

\begin{center}
\textit{``How can we provide means to help the administrator to configure secure networks and verify the security of existing network configurations?''}
\end{center}

Given the model of security components and the scope of this thesis, a secure system must fulfill two properties: First, the security requirements must express the desired security properties and, second, the three components must be consistent with each other. 
We consider the notion of \textit{``secure''} in the overall question by these two aspects. 
The posed question contains further aspects as it asks about developing new configurations vs.\ analyzing existing configurations. 
We further divide the question into these aspects. 
This yields the following two questions: 
First, we ask the question (\textbf{Q1}) \textit{``How can we design secure networks from scratch?''}. 
Second, we ask the question (\textbf{Q2}) \textit{``How can we analyze and verify existing configurations?''}. 

Finally, we need to consider the last aspect of the overall question \textit{``How can we provide means to help the administrator?''}. 
This last aspect corresponds to additional, generic, non-functional quality requirements which restrain the possible outcome of \textbf{Q1} and \textbf{Q2}. 
Hence, we will state them first.
\medskip

\noindent
We divide the non-functional requirements (\textbf{NF}) into the following aspects: 
\begin{description}
	\item[NF1] \textit{Can we provide automated tools for the solutions to \textbf{Q1} and \textbf{Q2}?}\medskip\newline
		A theory or abstract process which answers \textbf{Q1} and \textbf{Q2} is helpful from a scientific point of view. 
		However, to actually help administrators, working tools are required~\cite{survery2012networktroubleshooting}. 
	
	\item[NF2] \textit{Can the correctness of the tools be justified?}\medskip\newline
		For a tool to be useful, it must be trustworthy. 
		In particular, if security-critical decisions and processes are offloaded to a tool, its correctness is crucial. 
		Therefore, we require a formal, machine-verifiable correctness proof of our tools and, consequently, the theory they are built upon. 
	
	\item[NF3] \textit{Is over-formalism exposed to the administrator?}\medskip\newline
		Possible tools must usable. 
		While the focus of this thesis is not on user studies and usability, by evaluating related work, we discovered anecdotally that tools which expose an excessive amount of formalism are easily rejected by our administrator. 
	
	\item[NF4] \textit{Are the solutions to \textbf{Q1} and \textbf{Q2} compatible?}\medskip\newline
		A framework which takes away low-level control from the administrators and takes control over config files is not desired. 
		In particular, it is generally inadvisable to touch an administrator's configuration~\cite{debadminhandbook2015}, 
		and administrators need the possibility to manually apply low-level modifications to configurations. 
		Therefore, it must be possible to go back and forth between the solutions to \textbf{Q1} and \textbf{Q2}. 
		For example, it must be possible that an administrator makes low-level changes to rules which are generated by high-level requirements and it must be verified again that the low-level changes do not violate high-level requirements. 
		In different scenarios, rules generated from high-level requirements must co-exist with legacy rules without negative security implications. 
		
\end{description}
\bigskip

\noindent
We now detail on the first question (\textbf{Q1}) \textit{``How can we design secure networks from scratch?''}. 
This corresponds to the left-to-right direction of Figure~\ref{fig:intro:securitycomponents}. 
We divide it into the following aspects: 
\begin{description}
	\item[Q1.1] \textit{How can the security requirements be specified?}\medskip\newline
	A language to specify security requirements is required. 
	For the definition of \textit{``secure''}, some means for the administrator to check that the specified requirements express the desired meaning is necessary. 
	A solution which also satisfies \textbf{NF3} must expose low manual configuration overhead and little formalism to the administrator. 
	
	\item[Q1.2] \textit{How can a security policy be derived from the requirements?}\medskip\newline
	To satisfy \textbf{NF1}, a process which is completely automatic is required. 
	In addition, to satisfy \textbf{NF4}, it should also be possible to verify a policy w.r.t.\ the requirements. 
	
	\item[Q1.3] \textit{How can a policy be deployed to real network security mechanisms?}\medskip\newline
	Also for this step, to satisfy \textbf{NF1}, a process which is completely automatic is required. 
	The model assumptions which need to be fulfilled by the real-world security mechanism to enforce the policy need to be explicitly stated. 
	Different possible choices for security mechanisms need to be evaluated, \eg firewalls, OpenFlow-enabled switches, and containers. 
\end{description}
\bigskip

\noindent
We now detail on the second question (\textbf{Q2}) \textit{``How can we analyze and verify existing configurations?''}. 
This corresponds to the right-to-left direction of Figure~\ref{fig:intro:securitycomponents}. 
We divide it into the following aspects: 
\begin{description}
	\item[Q2.1] \textit{What are the semantics of a security mechanism?}\medskip\newline
		The behavior of a network security mechanism needs to be described. 
		This behavior can be very complicated as the example of iptables shows. 
		Yet, to fulfill \textbf{NF2}, a precise and formal model about the low-level behavior is required. 

	\item[Q2.2] \textit{How does an entity in a security mechanism configuration correspond to an entity in a policy?}\medskip\newline
		A policy may use symbolic names for entities whereas an entity in a security mechanism is usually identified by a network address. 
		Network addresses, \eg IP addresses, can be easily spoofed whereas a symbolic name in a policy is assumed to genuinely name an entity. 
		Therefore, to lift raw network addresses as they occur in a mechanism's configuration to entities in a policy, in an additional step, it must be ensured that addresses cannot be spoofed. 

	\item[Q2.3] \textit{How can a high-level policy be derived from a low-level security mechanism configuration?}\medskip\newline
		Given a low-level configuration of a security mechanism, a high-level policy which abstracts over all unnecessary low-level details needs to be derived. 
		For example, given an iptables ruleset with its over 200 different matching features and complex chain semantics, it needs to be simplified to a simple access control matrix. 

	\item[Q2.4] \textit{Can a derived high-level policy be verified w.r.t.\ a given set of security requirements?}\medskip\newline
		This question does not ask about deriving the requirements from a policy, since this process is not possible without guessing the intention of a policy author. 
		Because of \textbf{NF2}, we refrain from guessing. 
		Note that \textbf{Q1.2} has been strengthened such that a successful answer to it must already entail an answer to this question. 
		We ask this question to ultimately ensure that the required answers to \textbf{Q1} and \textbf{Q2} do not exist in isolation, but must be compatible in both directions. 
\end{description}

\section{Structure of this Thesis}
This thesis is structured to follow the research questions. 
Question \textbf{Q1} is answered in Part~\ref{part:greenfield} and question \textbf{Q2} is answered in Part~\ref{part:existing-configs}. 
We conclude, demonstrate applicability, and combine the answers to both questions in Part~\ref{part:applicability}. 

\begin{description}
	\item[Part~\ref{part:greenfield}] We answer \textbf{Q1} by contributing a method to specify security requirements with low manual configuration effort and present the first fully verified translation of high-level security requirements to low-level security mechanism configurations.
 	\item[Part~\ref{part:existing-configs}] We answer \textbf{Q2} by contributing the first fully verified tool to analyze existing iptables filtering rules which understands all match conditions and can extract a high-level policy overview. 
 	\item[Part~\ref{part:applicability}] We demonstrate the interplay of our answers to \textbf{Q1} and \textbf{Q1}, summarize applicability, and conclude. 
\end{description}

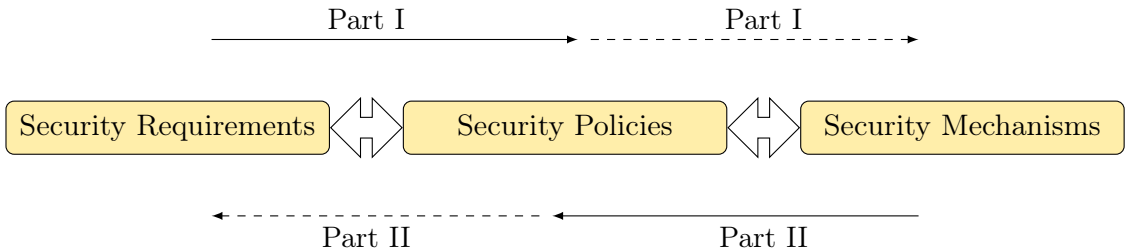
\begin{figure}[!htb]
	\centering
	\begin{center}
		\resizebox{0.99\textwidth}{!}{%
			\begin{tikzpicture}
			\node [MyRoundedBox, fill=LightYellow](sinvar) at (0,0) {Security Requirements};
			\node [MyDoubleArrow](arr1) at (sinvar.east) {};
			\node [MyRoundedBox, fill=LightYellow](policy) at (arr1.east) {Security Policies};
			\node [MyDoubleArrow](arr2) at (policy.east) {};
			\node [MyRoundedBox, fill=LightYellow](mechanism) at (arr2.east) {Security Mechanisms};
			\draw [shorten <=-2cm,shorten >=-1ex,-latex] ($(arr1) + (0,3em)$)--($(policy) + (0,3em)$);
			\node [anchor=south] at ($(arr1) + (0,3em)$) {Part~\ref{part:greenfield}};
			\draw [shorten <=2ex,shorten >=-2cm,dashed,-latex] ($(policy) + (0,3em)$)--($(arr2) + (0,3em)$);
			\node [anchor=south] at ($(arr2) + (0,3em)$) {Part~\ref{part:greenfield}};
			\draw [shorten <=-2cm,shorten >=-1ex,-latex] ($(arr2) + (0,-3em)$)--($(policy) + (0,-3em)$);
			\node [anchor=north] at ($(arr2) - (0,3em)$) {Part~\ref{part:existing-configs}};
			\draw [shorten <=2ex,shorten >=-2cm,dashed,-latex] ($(policy) + (0,-3em)$)--($(arr1) + (0,-3em)$);
			\node [anchor=north] at ($(arr1) - (0,3em)$) {Part~\ref{part:existing-configs}};
			\end{tikzpicture}
		}
	\end{center}
	\caption{Overview of the Parts of this Thesis.}
	\label{fig:intro:chapoverview}
\end{figure}

\bigskip

The overall structure, close to Figure~\ref{fig:intro:securitycomponents}, is illustrated in Figure~\ref{fig:intro:chapoverview}. 
The solid lines mean that these translation steps are fully verified in Isabelle/HOL. 
The dashed line below Part~\ref{part:greenfield} indicates that there is a final, small, syntactic rewriting step which is not formally verified. 
However, we will use Part~\ref{part:existing-configs} to verify the results of this step afterwards. 
The dashed line above Part~\ref{part:existing-configs} means that we cannot compute security requirements, given only a policy. 
Such an attempt would correspond to reverse engineering and ultimately lead to guessing a user's intent. 
We provide means to verify a policy given the security requirements, but we make no attempt of any reverse engineering. 
We describe the individual chapters in the following.

\bigskip

\noindent
Part~0: Introduction
\begin{description}
	\item[Chapter~\ref{chap:currentsituation}] provides an overview of the current situation and hints at the relevance of our research questions. 
	\item[Chapter~\ref{chap:introisablle}] gives a brief overview of Isabelle, the interactive proof assistant used to machine-verify the results of this thesis (\textbf{NF2}). 
\end{description}

\noindent
Part~\ref{part:greenfield}: \nameref{part:greenfield}. A detailed overview of this part can be found in Chapter~\ref{chap:greenfield:overview}. 
\begin{description}
	\item[Chapter~\ref{chap:forte14}] presents a method to formalize security requirements, answering \textbf{Q1.1}. 
		We show how a specification can be securely auto-completed, which increases usability and decreases exposed formalism (\textbf{NF3}). 
		To give an administrator feedback about the specified requirements, we show how they can be directly visualized as policy or how a policy can be verified, given a set of requirements and visualizing all possible violations. 
		This directly answers \textbf{Q1.2}. 
	\item[Chapter~\ref{sec:forte14:model-library}] presents a library of ready-to-use templates to prevent exposing any formalism. Only attributes need to be assigned to define security requirements (\textbf{NF3}). 
	\item[Chapter~\ref{sec:forte14:evalandcasestudy}] finally presents our tool, a case study, and further demonstrates applicability in an example. It also provides an outlook to provide a forward reference to introduce the problems which are not solved until Chapter~\ref{sec:forte14:evalandcasestudy}. 
	\item[Chapter~\ref{chap:hilbertsoffending}] discusses a weakness of the automated policy construction method and subsequently improves it with regard to completeness and performance (\textbf{Q1.2}, \textbf{NF1}).
	\item[Chapter~\ref{chap:esss14}] does one step towards automatic (\textbf{NF1}) translation to security mechanisms (\textbf{Q1.3}). It shows how to translate connection-level policies to stateful network-level policies. 
	\item[Chapter~\ref{chap:mansdnnfv}] finally presents deployment to a real network. Different security mechanisms are presented. This answers \textbf{Q1.3}. 
\end{description}

\noindent
Part~\ref{part:existing-configs}: \nameref{part:existing-configs}. A detailed overview of this part can be found in Chapter~\ref{chap:existing:overview}. 
\begin{description}
	\item[Chapter~\ref{chap:fm15}] presents a formal semantics of the filtering behavior of iptables, providing an answer to \textbf{Q2.1}.
	\item[Chapter~\ref{chap:nospoof}] presents a novel algorithm to certify spoofing protection of a firewall configuration, providing an important part of the answer to question \textbf{Q2.2}. 
	\item[Chapter~\ref{chap:networking16}] presents an algorithm to partition the complete IPv4 and IPv6 address space into classes with equal access rights. 
	This provides the missing piece to the answer for question \textbf{Q2.2}. 
	Building on this partitioning, we also present a method to translate a complex low-level iptables filtering ruleset with arbitrary match conditions to a simple firewall model and abstract it to an access control matrix which only considers IP addresses. 
	This answers question \textbf{Q2.3}. 
\end{description}

\noindent
Part~\ref{part:applicability}: \nameref{part:applicability}. 
A detailed overview of this part can be found in Chapter~\ref{chap:partthree:overview}. 
\begin{description}
	\item[Chapter~\ref{chap:dynamicdocker}] introduces the applicability and compatibility of our developed solutions~(\textbf{NF4}) by a simple example. 
		It shows how our tools help operating a Docker-based environment. 
	\item[Chapter~\ref{chap:puttingtogether}] presents the interplay of our tools~(\textbf{NF4}) in a real-world case study. 
		It shows a privacy audit of the MeasrDroid platform. 
	\item[Chapter~\ref{chap:answers}] summarizes our answers to the scientific questions and summarizes the achieved results of this thesis. 
	\item[Chapter~\ref{chap:relatedworktable}] defines a list of criteria for tools which help in managing network access control. Based on these criteria, it then compares this work to the state of the art. 
	\item[Chapter~\ref{chap:applicability}] summarizes applicability of our work with regard to generic policy management and reasoning, iptables firewall analysis, and software-defined networking.  
\end{description}

\bigskip

\paragraph*{Meta Structure}
All chapters start with a short abstract which summarizes the chapter's contributions in the big picture of this thesis. 
All parts which are based on joint work have an explicit statement on the author's contributions. 
If no such statement exists, the part is the single-handed contribution of Cornelius Diekmann.

\section{Publications in the Context of this Thesis}

\begin{description}
	\item[Chapter~\ref{chap:forte14}] \begin{sloppypar}Cornelius Diekmann, Stephan-A.\ Posselt, Heiko Niedermayer, Holger \mbox{Kinkelin}, Oliver Hanka, and Georg Carle. \emph{Verifying Security Policies using Host Attributes}. In FORTE -- 34th IFIP International Conference on Formal Techniques for Distributed Objects, Components and Systems, volume 8461, pages 133-148, Berlin, Germany, June 2014. Springer.\end{sloppypar}

	\item[Chapter~\ref{chap:esss14}] Cornelius Diekmann, Lars Hupel, and Georg Carle. \emph{Directed Security Policies: A Stateful Network Implementation}. In Engineering Safety and Security Systems, volume 150 of Electronic Proceedings in Theoretical Computer Science, pages 20-34, Singapore, May 2014. Open Publishing Association.

	\item[Chapter~\ref{chap:mansdnnfv}] Cornelius Diekmann, Andreas Korsten, and Georg Carle. \emph{Demonstrating topoS: Theorem-Prover-Based Synthesis of Secure Network Configurations}. In 2nd International Workshop on Management of SDN and NFV Systems, manSDN/NFV, Barcelona, Spain, November 2015. 	
	
	\item[Chapter~\ref{chap:fm15}] Cornelius Diekmann, Lars Hupel, and Georg Carle. \emph{Semantics-Preserving Simplification of Real-World Firewall Rule Sets}. In 20th International Symposium on Formal Methods, pages 195-212, Oslo, Norway, June 2015. Springer.
	
	\item[Chapter~\ref{chap:nospoof}] Cornelius Diekmann, Lukas Schwaighofer, and Georg Carle. \emph{Certifying Spoofing-protection of Firewalls}. In 11th International Conference on Network and Service Management, CNSM, Barcelona, Spain, November 2015.
	
	\item[Chapter~\ref{chap:networking16}] Cornelius Diekmann, Julius Michaelis, Maximilian Haslbeck, and Georg Carle, \emph{Verified iptables Firewall Analysis}. In IFIP Networking 2016, Vienna, Austria, May 2016.
	
	\item[Chapter~\ref{chap:puttingtogether}] Marcel von Maltitz, Cornelius Diekmann and Georg Carle, \emph{Taint Analysis for System-Wide Privacy Audits: A Framework and Real-World Case Studies}. In 1st Workshop for Formal Methods on Privacy, Limassol, Cyprus, November 2016. Note: no proceedings published. 

\end{description}

\noindent
Our formalization has been published in the Archive of Formal Proofs in the following entries: 
\begin{itemize}
	\item Cornelius Diekmann, \emph{Network Security Policy Verification}. 
	\item Cornelius Diekmann, Julius Michaelis and Lars Hupel, \emph{IP Addresses}. 
	\item Cornelius Diekmann, Julius Michaelis and Max Haslbeck, \emph{Simple Firewall}. 
	\item Cornelius Diekmann and Lars Hupel, \emph{Iptables Semantics}. 
	\item Julius Michaelis and Cornelius Diekmann, \emph{Routing}. 
	\item Julius Michaelis and Cornelius Diekmann, \emph{LOFT -- Verified Migration of Linux Firewalls to SDN}
\end{itemize}

  \chapter{Current Situation \& Problem Analysis}
\label{chap:currentsituation}
In this chapter, we describe the current state of network administration, configuration, and management with regard to security issues. 
Afterwards, we analyze the root cause of configuration complexity by comparing networks to software. 




\section{Evaluation of the Situation}
%
A 2009 whitepaper by Netcordia~\cite{networkdowntime2009} describes that networks ``often fail, at great expense, not because of underlying equipment problems, but because of human error in setting them up and running them.'' 
The document concludes that ``a primary (if not the primary) cause'' for network downtime is human error. 
%
%
%
%
%
A 2003 survey~\cite{oppenheimer2003internet} indicates that most Internet services fail because of human operator error, where configuration errors are the largest category of those human errors. 
A study~\cite{firwallerr2004}, featuring 37 enterprise firewall configurations from the years 2000 and 2001, reveals that many firewalls are misconfigured. 
The study also reveals that the firewalls' configuration quality improves with new releases of the firewall product, which is mainly attributed to better default rule sets. 
Hence, better default settings provide less surface for human error. 
However, the study concludes ``that there are no good high-complexity rule sets''~\cite{firwallerr2004}. 
Several years later, the situation has not actually improved~\cite{wool2010firewall}. 
Mansmann \etal\cite{fwviz2012} also hint that historically grown firewall rule sets are insufficiently understood. 
In 2007, a survey of 70 large ISPs revealed that management of access control lists were considered as the ``most critical missing or limited vendor security feature'' for infrastructure protection~\cite{infrastructure2007report,bellovin2009configuration}. 
In 2016, the same report series~\cite{infrastructure2016report} still lists access control lists as one of the most widely and most actively used technique.\footnote{The 2016 version of the report does no longer include an ``Infrastructure Shortcomings'' section.}
%
A 2012 survey~\cite{sherry2012making} of 57 enterprise network administrators confirms that a ``majority of administrators stated [estimated] misconfiguration as the most common cause of failure'' \cite{sherry2012making}. 
%
%
%
%
A large 2013 study~\cite{imc2013demystifying}, conducted over two years across more than 10 large datacenters, reveals that there exists a variety of misconfigurations in network management. 
%
%
%
%
Based on Wool's findings~\cite{firwallerr2004}, Casado\etal\cite{ethane07} also conclude that ``most networks today require substantial manual configuration by trained operators [\dots] to achieve even moderate security''. 
%
%
%
%
Burns \etal\cite{autosecpolicymgnt01} predict that ``the scope of management is rapidly exceeding human capabilities because of the acceleration of changes in technology and topology''~\cite{autosecpolicymgnt01}. 
They see the need to eliminate low-level technical device configuration and focus on the desired behavior of a network. 
``Policies should define the intent of the administrator independently of the mechanisms used to implement the policy.'' \cite{autosecpolicymgnt01}.

%
%
%
%
This implies that the manual configuration complexity in network security management is a key aspect for failure. 
Advanced tools to support the administrator with the configuration complexity are barely deployed. 
``Paradoxically, most mission critical IT networks are configured and managed with little help from automated systems, but by humans working with few tools''~\cite{networkdowntime2009}. 
%
But ``administrators desire for newer, more sophisticated tools.''~\cite{survery2012networktroubleshooting}. 
However, ``the security requirements of distributed systems are hard to specify and hard to formalize''~\cite{arxivnsfworkshopformalprivacy2016}. 
A recent Dagstuhl seminar on ``Formal Foundations for Networking'' concludes that ``[t]here is a growing need for tools and methodologies that provide rigorous guarantees about performance, reliability, and security''~\cite{bjorner_et_al:DR:2015:5044}. 

%
A study~\cite{databreach2009src} conducted by Verizon from 2004 to 2009 and the United States Secret Service during 2008 and 2009 reveals that data leaks are also often caused by configuration errors~\cite{databreach2009}. 
The authors estimate that this might be due to the fact that ``attackers know most users are over-privileged'' \cite{databreach2009src}. 
In 2016~\cite{verizon2016databreach}, privilege misuse (and misconfiguration) are still among the top causes of data breaches, which has also been demonstrated in a very concrete example~\cite{faultyaccess2016online}. 
This indicates that the complexity in network access policies which define who can communicate with whom cannot just be simplified, but on the contrary, should be expanded to reduce the attack surface by stricter, hence more complicated, access control policies.

%
Many vendor-specific devices with their own configuration interface exist~\cite{telekom2013kuvs}. 
A survey among enterprise administrators confirms that ``typical enterprise networks are a complex ecosystem of firewalls, IDSes, web proxies, and other devices.''~\cite{sherry2012making}.  
``Managing many heterogeneous devices requires broad expertise and consequently a large management team.''~\cite{sherry2012making} 
A 2006 study~\cite{netsecconflicts} with 38 network administrators reveals the configuration complexity of security devices that require setting up low-level security policies, such as IPsec  gateways, increases the probability of human error. 
The study finds that ``even the expert administrators created policy conflicts''~\cite{netsecconflicts}. 

\section{A Problem Classification}
\label{sec:intro:problemclassification-and-sdn}
In this section, we try to trace back the symptoms of management complexity in networks to their root causes. 

We know that even a set of simple switches which support round-robin load balancing is Turing-Complete~\cite{EPFL-REPORT-187131}. 
But computer networks are becoming increasingly more software-defined: 
SANE~\cite{casado2006sane} inspired Ethane~\cite{ethane07}, which itself inspired OpenFlow~\cite[§3.2 Example 1]{mckeown2008openflow}, which is now the de facto standard for Software-Defined Networking (SDN), which is used in the industry~\cite{openflowatgoogle2012,b42013googlesdn}. 
With networks resembling more and more to software, we compare network management complexity to a field where complexity is well-studied since many decades: software engineering.

\subsection{Types of Complexity in Software Engineering}
In 1987, Brooks published his thousandfold-cited paper ``No Silver Bullet: Essence and Accidents of Software Engineering''~\cite{brooks1987no} in the IEEE Computer magazine. 
In software engineering nowadays, a good understanding about the complexity challenges exists~\cite[\S 1.2. The Inherent Complexity of Software]{safari2007softwareeng}. 
In this section, we discuss the complexity challenges in software engineering as identified by Brooks~\cite{brooks1987no} and later convey the results to the challenges in network management. 

In software engineering, one distinguishes between \emph{accidental} difficulties and \emph{essential} difficulties\footnote{sometimes also referred to inherent difficulties}. 
Essential difficulties are the difficulties which are inherent in the nature of software, whereas accidental difficulties are those that are not inherent. 
For example, designing, conceptualizing, and defining the requirements and interfaces of a business application is an inherently complex task. 
The domain-specific challenges, which are inherently complex, must be imaged by the application and a huge amount of relationships between data items, business processes, and algorithms must be specified. 
In contrast, implementing the application in the \texttt{C} programming language and dealing with memory errors is an accidental complexity that could have been avoided by selecting a memory-safe programming language.

Essential difficulties in software engineering refer to~\cite{brooks1987no}: 
\begin{itemize}
\item The \emph{complexity} of software itself and in particular the complexity challenges in the problem domain that are mirrored by the software. 
\item The \emph{conformity} that software must comply with existing or legacy system interfaces. 
\item The \emph{changeability} of software and that software is often used beyond its original purpose. 
\item The \emph{invisibility} of software and the fact that it cannot be adequately visualized. 
\end{itemize}

While accidental difficulties can be tackled, essential difficulties are inherently hard to overcome and Brooks projects that ``there is no single development, in either technology or in management technique, that by itself promises even one order-of-magnitude improvement in productivity, in reliability, in simplicity''. 

\subsection{Types of Complexity in Network Security Management}
In computer networks, administrators are ``touching low-level configurations all the time''~\cite{techtraget2013feamstersdn}. 
But problems with low-level configuration languages are comparable to the problems which arise from the use of the \texttt{C} programming language. 
Thus, low-level languages can be classified as accidental complexity of network management.

We now interpret the essential difficulties of software engineering in the context of network security management. 
We classify the essential difficulties as 
\begin{itemize}
\item The \emph{complexity} of security requirements themselves. 
\item The \emph{conformity} with legacy systems in a network and the understanding of legacy configuration. 
\item The \emph{changeability} of network traffic and that attackers may exploit weaknesses in unforeseeable ways. 
\item The \emph{invisibility} of network configurations and the fact that it cannot be adequately visualized. 
\end{itemize}

In addition, compared to software engineering, our tooling for networks is ``pathetic''~\cite{mck2012sdntame}.

\paragraph*{Related Work}
In a closely related work, Benson \etal\cite{benson2009unraveling} present metrics to measure ``inherent complexity'' of network (router) configurations, ``abstracting away all the details of the underlying configuration language''~\cite{benson2009unraveling}. 
In other words, they measure essential difficulties, explicitly abstracting over accidental difficulties. 

In a study and interview with several system administrators, they empirically uncover the following essential difficulties, which are very close to the difficulties we have identified: 
The inherent \emph{complexity} itself, expressed as a network's reachability policy. 
A network's evolution over time and legacy configuration parts, which is related to \emph{conformity}. 
Finally, the interviews with the administrators reveal that complexity metrics are ``helping operators visualize and understand networks''~\cite[\S\hairspace7]{benson2009unraveling}, which relates to the \emph{invisibility} of network configuration. 
They do not identify the \emph{changeability} of network traffic but identify that some networks are more complex than necessary because they are optimized for monetary cost.

  \chapter{Brief Introduction to Isabelle and Notation}
\label{chap:introisablle}
We implemented our theory and the formal proofs in the Isabelle/HOL theorem prover~\cite{isabelle2016}. 
Isabelle is a generic and interactive proof assistant. 
We use its standard Higher-Order Logic (HOL).

Internally, Isabelle is an LCF-style theorem prover.
This means, a fact can only be proven if it is accepted by a mathematical inference kernel. 
Proof steps can be done by either the user or by (embedded or external) automated proof tactics and solvers. 
All proof steps must pass this kernel, hence, a faulty prover does not introduce unsoundness because the kernel would reject unsound steps which it cannot reproduce. 
The correctness of a proof only depends on the correctness of the kernel. 
This architecture makes the system highly trustworthy, because the proof kernel consists only of little code, is widely used (and has been for over a decade) and is rigorously manually checked. 
This makes errors very unlikely, which has been demonstrated by Isabelle's success over the past 20 years. 
In fact, there has not been a known bug in the Isabelle kernel in the past 20 years which affected a user's proof.\footnote{But there have been bugs (which were all fixed) for artificially constructed corner cases.}

Standards such as Common Criteria~\cite{cc2012p3} require formal verification for their highest \emph{Evaluation Assurance Level} (EAL7) and the Isabelle/HOL theorem prover is suitable for this purpose~\cite[\S{}A.5]{cc2012p3}. 
Therefore, our approach is not only suitable for verification, but also a first step towards certification.

To stay focused, we usually only present the intuition behind proofs or even omit a proof completely. 
Whenever we omit a proof for a claim which is not obvious, we add a footnote that points to our formalization. 
In addition, for better readability and brevity, we will not present all proven statements as theorem but present some facts in natural language within a sentence. 
We point the interested reader to the proof or definition by a footnote. 
For example, when the text states within a sentence that foo\footnotemark[42] holds, the machine-verified proof for the claim `foo' can be found by following the corresponding footnote.

\section{Notation}
We will now explain the notational conventions we apply throughout this thesis. 
In general, we use pseudo code close to SML, Haskell, and Isabelle.

\paragraph*{Functions}
A total function from type $\mathcal{A}$ to type $\mathcal{B}$ is denoted by \mbox{$\mathcal{A} \Rightarrow \mathcal{B}$}. 
In contrast, the logical implication is written with a long arrow ``$\Longrightarrow$''. 
Function application is written without parentheses: $\mfun{f} \ x \ y$ denotes function ``$\mfun{f}$ applied to parameter $x$ and parameter $y$''. 

\paragraph*{Lists}
We write $\lstcons$ for prepending a single element to a list, \eg $a \lstcons b \lstcons [c,\,d] = [a,\,b,\,c,\,d]$, and $\lstapp$ for appending lists, \eg $[a,\,b] \lstapp [c,\,d] = [a,\,b,\,c,\,d]$.
The empty list is written as~$[]$.
We write list comprehension as $[\mfun{f}\ a.\ a \leftarrow l]$, which denotes applying $\mfun{f}$ to every element $a$ of list $l$. 
Also, $[\mfun{f}\ x\ y.\ \ x \leftarrow l_1,\; y \leftarrow l_2]$ denotes the list comprehension where $\mfun{f}$ is applied to each combination of elements of the lists $l_1$ and $l_2$.
For $\mfun{f}\ x \ y = (x,\,y)$, this returns the cartesian product of $l_1$ and $l_2$.


\paragraph*{Types}
The set of Boolean values is denoted by the symbol $\mathbb{B} = \left\lbrace \mconstr{True},\ \mconstr{False}\right\rbrace$. 
To explicitly write down the type of an object, we annotate it with `$::$'.  
The two colons for type annotations have more spacing that the list operations; they can usually be distinguished by the context. 
For example, $\mconstr{True} :: \mathbb{B}$ or $\mfun{f} :: \mathcal{A} \Rightarrow \mathcal{B}$ are type annotations. 
We use polymorphic types, \eg $\mfun{f}$ could be applied to integers and return a Boolean but it could also be applied to graphs and return a string.

\paragraph*{Definitions}
Whenever applicable, we write definitions with `$\cdot \equiv \cdot$' to distinguish the operator from the mathematical equality operator `$\cdot = \cdot$'. 
This increases readability since the equality operator may also occur in definitions. 
We only use `$\cdot \equiv \cdot$' for formulas, not for types. 

\paragraph*{Control Statements}
Control statements, for example $\mctrl{if} \cdot \mctrl{then} \cdot \mctrl{else}$, are set in bold font. 

\paragraph*{Typesetting} 
To further increase readability, we stick to the following typesetting. 
Polymorphic types, whenever applicable, are set calligraphic, \eg $\mathcal{A}$, $\mathcal{B}$, $\mathcal{C}$. 
Usually, $\mathcal{G}$ denotes a graph where the nodes may be of arbitrary type. 
Specific types are set in italic or in normal text, depending on the context, \eg $\mathit{firewall\mhyphen{}rule}\ \mathit{list}$ or $\mathcal{A}\ \mathit{list}$ which is a list over arbitrary types. 
Functions and constants, in general everything that is not a free variable, are set in sans serif font, \eg $\mfun{f}$, $\mconstr{True}$. 
Variables and locally-bound objects are set italic, \eg $\mvar{a}$, $\mvar{v}_1$. 
Within an example, we also set entity names which are only valid for the example in italic. 
Linux shell commands are set in \texttt{typewriter} font. 

\begin{example}
\textbf{Example. }%
Let graph $\mvar{G}$ of type $\mathcal{G}$ be $(\lbrace v_1, v_2 \rbrace,\ \emptyset)$. 
$\mvar{G}$ only contains two vertices and no edges. 
The vertices could be of arbitrary type, \eg represented by strings or integers. 
We will say the vertices are of arbitrary type $\mathcal{V}$, then $\mathcal{G} = (\mathcal{V}\;set) \times ((\mathcal{V}{\times}\mathcal{V})\;set)$. 
For all examples, we assume that the entities referenced in the example are distinct, here, $v_1 \neq v_2$. 
We can have a function $\mfun{f}$ which maps any graph to $\mconstr{True}$. 
Then, $\mfun{f} :: \mathcal{G} \Rightarrow \mathbb{B}$. 
We could implement $\mconstr{f}$ by the lambda expression which ignores its first argument and always returns $\mconstr{True}$ as follows: 
$(\lambda \_.\ \mconstr{True})$. 
Then, $\mfun{f}\ (\lbrace v_1, v_2 \rbrace,\ \emptyset)$ holds. 
\end{example}

\section{Availability of our Formalization}
The Archive of Formal Proofs (AFP)~\cite{afpall} is the de-facto place to find Isabelle theories. 
It is organized similar to a scientific journal and all submissions are peer reviewed. 
The peer review assures that the submitted theories conform to the Isabelle style rules and that the proofs are properly accepted by Isabelle. 
Once a submission is accepted in the AFP, it is maintained and updated by the community for future Isabelle releases.

Technically, an entry which is accepted in the AFP is only guaranteed to contain sound proofs, it does not guarantee that actually something useful has been proven. 
The meaning and applicability of our theory are demonstrated in this thesis and in the non-AFP publications. 
However, the AFP entries (which are both manually-reviewed and machine-verified) created during this thesis provide a very strong guarantee about the correctness of our proofs and ensure that the theoretical results are easily reproducible and accessible, even with future versions of Isabelle.

The following entries have been created during this thesis with major contributions by Cornelius Diekmann \cite{Network_Security_Policy_Verification-AFP,IP_Addresses-AFP,Simple_Firewall-AFP,Iptables_Semantics-AFP}: 

\begin{itemize}
	\item Cornelius Diekmann, \emph{Network Security Policy Verification}. 
	\item Cornelius Diekmann, Julius Michaelis and Lars Hupel, \emph{IP Addresses}. 
	\item Cornelius Diekmann, Julius Michaelis and Max Haslbeck, \emph{Simple Firewall}. 
	\item Cornelius Diekmann and Lars Hupel, \emph{Iptables Semantics}. 
\end{itemize}

The following entries have been created during this thesis with contributions by Cornelius Diekmann~\cite{Routing-AFP,LOFT-AFP}: 
\begin{itemize}
	\item Julius Michaelis and Cornelius Diekmann, \emph{Routing}. 
	\item Julius Michaelis and Cornelius Diekmann, \emph{LOFT -- Verified Migration of Linux Firewalls to SDN}
\end{itemize}

 \part{Green-Field Approach}
 \label{part:greenfield}
 
 \chapter{Overview}
 \label{chap:greenfield:overview}
As shown in Figure~\ref{fig:intro:securitycomponents}, this thesis focuses on the consistency between {security requirements}, a {security policy}, and {security mechanisms}. 
This part presents the left-to-right direction: 
Given the security requirements, in a greenfield approach, we provide a method to construct the security policy and the configuration of security mechanisms. 
Our tool to support the process is called \topos{}. 

 \begin{center}
    \resizebox{0.99\textwidth}{!}{%
 	\begin{tikzpicture}
	 \node [MyRoundedBox, fill=LightYellow](sinvar) at (0,0) {Security Requirements};
	 \node [MySingleArrow](arr1) at (sinvar.east) {};
	 \node [MyRoundedBox, fill=LightYellow](policy) at (arr1.east) {Security Policies};
	 \node [MySingleArrow](arr2) at (policy.east) {};
	 \node [MyRoundedBox, fill=LightYellow](mechanism) at (arr2.east) {Security Mechanisms};
	 \draw [shorten <=-2cm,shorten >=-19ex,-latex] ($(arr1) + (0,2.5em)$)--($(policy) + (0,2.5em)$);
	 \node [anchor=south] at ($(policy) + (0,2.5em)$) {\topos};
	 \draw [shorten <=20ex,shorten >=-2cm,dashed,-latex] ($(policy) + (0,2.5em)$)--($(arr2) + (0,2.5em)$);
	 \end{tikzpicture}
	 }
 \end{center}

For this task, we have divided the three security components into four components. 

 \begin{center}
    \resizebox{0.99\textwidth}{!}{%
 	\begin{tikzpicture}
	 \node [MyRoundedBox, fill=LightYellow](sinvar) at (0,0) {Security Invariants};
	 \node [MySingleArrow](arr1) at (sinvar.east) {};
	 \node [MyRoundedBox, fill=LightYellow](policy) at (arr1.east) {Security Policy};
	 \node [MySingleArrow](arr2) at (policy.east) {};
	 \node [MyRoundedBox, fill=LightYellow](statepolicy) at (arr2.east) {Stateful Policy};
	 \node [MySingleArrow](arr3) at (statepolicy.east) {};
	 \node [MyRoundedBox, fill=LightYellow](mechanism) at (arr3.east) {Security Mechanisms};
	 \end{tikzpicture}
	 }
 \end{center}
 
A set of \emph{security invariants} formalizes the security requirements. 
The security policy has been split into the actual \emph{security policy} and the \emph{stateful policy}. 
This is motivated by the world of computer networks: 
The security policy expresses who may set up new connections and the stateful policy answers the question whether packets which belong to such an established connection are allowed bidirectionally. 
Finally, the \emph{security mechanisms} are network components; we will demonstrate our tool for the Linux iptables firewall and an OpenFlow-enabled switch. 

This part is structured as follows. 
In Chapter~\ref{chap:forte14}, we focus on the consistency between security invariants and a security policy. 
Given a specification of the security invariants, we show how to verify an existing policy or how to construct a new policy from scratch. 
Chapter~\ref{sec:forte14:model-library} presents a library of ready-to-use security invariant templates. 
Chapter~\ref{sec:forte14:evalandcasestudy} presents a case study.
Chapter~\ref{chap:hilbertsoffending} will improve the policy construction method. 
 \begin{center}
    \resizebox{0.99\textwidth}{!}{%
 	\begin{tikzpicture}
	 \node [MyRoundedBox, fill=LightYellow](sinvar) at (0,0) {Security Invariants};
	 \node [MyDoubleArrow](arr1) at (sinvar.east) {};
	 \node [MyRoundedBox, fill=LightYellow](policy) at (arr1.east) {Security Policy};
	 \node [MyDoubleArrow](arr2) at (policy.east) {};
	 \node [MyRoundedBox, fill=LightGrey](statepolicy) at (arr2.east) {Stateful Policy};
	 \node [MySingleArrow](arr3) at (statepolicy.east) {};
	 \node [MyRoundedBox, fill=LightGrey](mechanism) at (arr3.east) {Security Mechanisms};
	 \end{tikzpicture}
	 }
 \end{center}

In Chapter~\ref{chap:esss14}, we will cover the conformity of a security policy with a stateful policy, given the security invariants. 
We also show how to automatically compute a stateful policy from the security policy and the security invariants. 
 \begin{center}
    \resizebox{0.99\textwidth}{!}{%
 	\begin{tikzpicture}
	 \node [MyRoundedBox, fill=LightGrey](sinvar) at (0,0) {Security Invariants};
	 \node [MyDoubleArrow](arr1) at (sinvar.east) {};
	 \node [MyRoundedBox, fill=LightYellow](policy) at (arr1.east) {Security Policy};
	 \node [MyDoubleArrow](arr2) at (policy.east) {};
	 \node [MyRoundedBox, fill=LightYellow](statepolicy) at (arr2.east) {Stateful Policy};
	 \node [MySingleArrow](arr3) at (statepolicy.east) {};
	 \node [MyRoundedBox, fill=LightGrey](mechanism) at (arr3.east) {Security Mechanisms};
	 \end{tikzpicture}
	 }
 \end{center}
 
In Chapter~\ref{chap:mansdnnfv}, we put everything together and additionally demonstrate the translation to real network security devices. 
 \begin{center}
    \resizebox{0.99\textwidth}{!}{%
 	\begin{tikzpicture}
	 \node [MyRoundedBox, fill=LightGrey](sinvar) at (0,0) {Security Invariants};
	 \node [MyDoubleArrow](arr1) at (sinvar.east) {};
	 \node [MyRoundedBox, fill=LightGrey](policy) at (arr1.east) {Security Policy};
	 \node [MyDoubleArrow](arr2) at (policy.east) {};
	 \node [MyRoundedBox, fill=LightYellow](statepolicy) at (arr2.east) {Stateful Policy};
	 \node [MySingleArrow](arr3) at (statepolicy.east) {};
	 \node [MyRoundedBox, fill=LightYellow](mechanism) at (arr3.east) {Security Mechanisms};
	 \end{tikzpicture}
	 }
 \end{center}

\section*{Availability}
Our Isabelle/HOL theory files with the formalization and the referenced correctness proofs and our tool \topos{} are available at 
\begin{center}%
\url{https://github.com/diekmann/topoS} \ \ and \ \ the AFP~\cite{Network_Security_Policy_Verification-AFP}
\end{center}%

\newcommand{\studyTotalEntries}{15}
\newcommand{\studyTotalUtilityTool}{3.0&/&3.3&/&0.8}
\newcommand{\studyTotalUtilityToolNotbl}{3.0/3.3/0.8}
\newcommand{\studyTotalUtilityIdea}{4.0&/&3.5&/&0.9}
\newcommand{\studyTotalUtilityIdeaNotbl}{4.0/3.5/0.9}
\newcommand{\studyTotalHelps}{93\%}
\newcommand{\studyTotalWoulduse}{100\%}
\newcommand{\studyTotalTopoValid}{15.0&/&13.5&/&4.1}
\newcommand{\studyTotalTopoInvalid}{2.0&/&3.8&/&4.5}
\newcommand{\studyTotalTopoMissing}{7.0&/&8.1&/&4.5}
\newcommand{\studyTotalTopoError}{9.0&/&11.9&/&7.2}
\newcommand{\studyNovicEntries}{5}
\newcommand{\studyNovicUtilityTool}{4.0&/&3.2&/&1.2}
\newcommand{\studyNovicUtilityIdea}{4.0&/&3.2&/&1.2}
\newcommand{\studyNovicHelps}{80\%}
\newcommand{\studyNovicWoulduse}{100\%}
\newcommand{\studyNovicTopoValid}{12.0&/&10.6&/&5.7}
\newcommand{\studyNovicTopoInvalid}{4.0&/&6.6&/&4.9}
\newcommand{\studyNovicTopoMissing}{11.0&/&11.2&/&6.2}
\newcommand{\studyNovicTopoError}{17.0&/&17.8&/&9.1}
\newcommand{\studyInterEntries}{5}
\newcommand{\studyInterUtilityTool}{3.0&/&3.2&/&0.4}
\newcommand{\studyInterUtilityIdea}{4.0&/&4.0&/&0.0}
\newcommand{\studyInterHelps}{100\%}
\newcommand{\studyInterWoulduse}{100\%}
\newcommand{\studyInterTopoValid}{14.0&/&14.0&/&1.4}
\newcommand{\studyInterTopoInvalid}{1.0&/&1.6&/&1.9}
\newcommand{\studyInterTopoMissing}{7.0&/&7.4&/&1.0}
\newcommand{\studyInterTopoError}{8.0&/&9.0&/&2.6}
\newcommand{\studyExperEntries}{5}
\newcommand{\studyExperUtilityTool}{3.0&/&3.4&/&0.5}
\newcommand{\studyExperUtilityIdea}{3.0&/&3.2&/&0.7}
\newcommand{\studyExperHelps}{100\%}
\newcommand{\studyExperWoulduse}{100\%}
\newcommand{\studyExperTopoValid}{16.0&/&15.8&/&1.7}
\newcommand{\studyExperTopoInvalid}{1.0&/&3.2&/&4.4}
\newcommand{\studyExperTopoMissing}{5.0&/&5.6&/&2.1}
\newcommand{\studyExperTopoError}{9.0&/&8.8&/&3.8}

\newcommand{\sinvar}[0]{\mfun{m}}
\newcommand{\nP}[0]{P}

\newcommand{\eads}[0]{Airbus Group}

\chapter{Verifying Security Policies using Host Attributes}
\label{chap:forte14}


This chapter, Chapter~\ref{sec:forte14:model-library}, and Section~\ref{sec:forte14:case-study} in Chapter~\ref{sec:forte14:evalandcasestudy} are an extended version of the following paper~\cite{diekmann2014forte}:
\begin{itemize}
	\item Cornelius Diekmann, Stephan-A.\ Posselt, Heiko Niedermayer, Holger Kinkelin, Oliver Hanka, and Georg Carle. \emph{Verifying Security Policies using Host Attributes}. In FORTE -- 34th IFIP International Conference on Formal Techniques for Distributed Objects, Components and Systems, volume 8461, pages 133-148, Berlin, Germany, June 2014. Springer.
\end{itemize}

\noindent
The following major improvements and new contributions were added:
\begin{itemize}
	\item A full documentation of the security invariant library: Chapter~\ref{sec:forte14:model-library}. 
	\item Several improvements and more configuration options for several invariant templates.
	\item New invariant templates. 
	\item Generalized the $\Phi$-structure, requiring reworking of all corresponding proofs. 
	\item Improved the algorithm for policy construction: Chapter~\ref{chap:hilbertsoffending}. 
	\item Re-Implementation of the \topos{} tool in Isabelle: Section~\ref{sec:forte14:impl}. 
	\item New full-stack example: Section~\ref{sec:example:imaginary-factory-network}. 
	\item Updated and extended related work, added analogy to software engineering. 
\end{itemize}

\paragraph*{Statement on author's contributions}
All improvements with regard to the paper are the work of the author of this thesis. 
For the original paper, the author of this thesis provided major contributions for the ideas, realization, formalization, analysis, and proof of the overall model and the invariant templates. 
He implemented the prototypical tool, researched related work, and conducted the user feedback session. 
The case study (A Cabin Data Network) was designed with the help of Oliver Hanka. 
It was evaluated and formalized by the author of this thesis.

The following ideas have been previously presented in the author's master's thesis~\cite{cornythesis}: 
Security requirements modeled as Boolean predicates are composable, the idea of offending flows, the secure default parameter, and the observation that the security strategy may be linked to a Boolean value related to the offending host. 
While the core ideas of the master's thesis remain, the complete formal foundations have been reworked for this Ph.D.\ thesis.  
The author's master's thesis relied on many inconvenient assumptions. 
For this Ph.D.\ thesis, the author completely re-implemented his master's thesis to improve and rework the formalization, and discovered many new insights which allowed getting rid of all unpleasant assumptions. 
For many parts of the author's master's thesis (\eg composition), only the idea or prototypical, unverified code was presented. 
For this Ph.D.\ thesis, everything has been completely formalized and proven with Isabelle/HOL, requiring also a large rework or rebuild of existing theory. 
A preliminary prototype of \topos{} in Scala was developed during the author's master's thesis. 
The final \topos{} tool presented in this Ph.D.\ thesis is a completely new implementation (based on the new results of this thesis) in Isabelle/HOL. 

\medskip

\paragraph*{Abstract}
In this chapter, we focus on the relationship between \emph{Security Invariants} and a connection-level \emph{Security Policy}, as illustrated in Chapter~\ref{chap:greenfield:overview}.  
We present a formalization of security invariants and show how they can be used to verify a policy and how a policy can be constructed from scratch, given only the security invariants. 

\medskip

\section{Introduction}
A distributed system, from a networking point of view, is essentially a set of interconnected hosts. 
Its connectivity structure comprises an important aspect of its overall attack surface, which can be dramatically decreased by giving each host only the necessary access rights. 
Hence, it is common to protect networks using firewalls and other forms of enforcing network-level access policies.
Such access policies can be seen as means to describe which flows between hosts are allowed, and which are not.
However, raw sets of such policy rules \eg firewall rules, ACLs, or access control matrices, scale quadratically with the number of hosts and ``controlling complexity is a core problem in information security''~\cite{guttman05rigorous}. 
A case study, conducted in Chapter~\ref{sec:forte14:evalandcasestudy}, reveals that even a policy with only 10 entities may cause difficulties for experienced administrators. 
Expressive policy languages can help to reduce the complexity. 
However, the question whether a policy fulfills certain security invariants and how to express these often remains. 

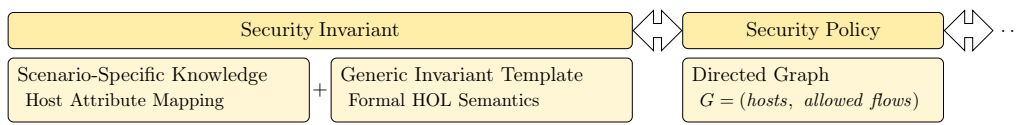
\begin{figure}[h!tb]
  \centering
   	\resizebox{0.9\textwidth}{!}{%
   		\begin{tikzpicture}
   		\node [MyRoundedBox, fill=LightYellow, text width=30em](sinvar) at (0,0) {Security Invariant};
   		\node [MyDoubleArrow](arr1) at (sinvar.east) {};
   		\node [MyRoundedBox, fill=LightYellow, text width=12em](policy) at (arr1.east) {Security Policy};
   		\node [MyDoubleArrow](arr2) at (policy.east) {};
   		\node [anchor=west](dots) at (arr2.east) {\dots};

   		\node [MyRoundedBox, fill=LightYellow!50, text width=14em, anchor=north west, align=left](sinvleft) at ($(sinvar.south west) + (0,-1ex)$) {Scenario-Specific Knowledge\newline{}\small{\strut{}\hspace*{1ex}Host Attribute Mapping}};
   		\node [MyRoundedBox, fill=LightYellow!50, text width=14em, anchor=north east, align=left](sinvright) at ($(sinvar.south east) + (0,-1ex)$) {Generic Invariant Template\newline{}\small{\strut{}\hspace*{1ex}Formal HOL Semantics}};
   		\node at ($(sinvleft)!0.5!(sinvright)$) {$+$};

   		\node [MyRoundedBox, fill=LightYellow!50, text width=12em, anchor=north west, align=left](dirgraph) at ($(policy.south west) + (0,-1ex)$) {Directed Graph\newline{}\small{\strut{}\hspace*{1ex}$G = \left(\mathit{hosts},\ \mathit{allowed\ flows}\right)$}};
   		\end{tikzpicture}%
   	}\caption{Formal Objects: Security Invariant and Security Policy.}%
   \label{fig:intro:formalobjects}%
\end{figure}

Using an attribute-based~\cite{abac2005} approach, we model simple, static, positive security policies with expressive, Mandatory Access Control (MAC) security invariants. 
The formal objects, illustrated in Figure~\ref{fig:intro:formalobjects}, are carefully constructed for their use-case. 
The policy is simply a graph, which can for example be extracted from or translated to firewall rules (cf.\ Chapter~\ref{chap:esss14}, Chapter~\ref{chap:mansdnnfv}, and Part~\ref{part:existing-configs}). 
The security invariants are split into the formal semantics, accessible to formal analysis, and scenario-specific knowledge, easily configurable by the end user.  
This model landscape enables verification of security policies. 
Primarily, we contribute the following universal insights for constructing security invariants. 
\noindent\begin{enumerate}
	\item Both provably \emph{secure} and \emph{permissive} default values for host attributes can be found. 
		This enables auto completion when specifying security invariants and hence decreases the user's configuration effort. 
	\item The security strategy, information flow or access control, determines whether a security violation occurs either at the sender's or at the receiver's side.
	\item A violated invariant can always be repaired by tightening the policy \emph{if and only if} the invariant holds for the deny-all policy. 
\end{enumerate}
In this Chapter~\ref{chap:forte14}, we focus on theoretical aspects: 
We formally introduce the underlying model in Section~\ref{sec:forte14:formalmodel} and conduct a formal analysis in Section~\ref{sec:forte14:analysis}. 
Ultimately, we derive an algorithm to construct a security policy automatically in Section~\ref{subsec:policyconstruction}. 
In the subsequent chapters, we focus on application: 
We present our security invariant template library in Chapter~\ref{sec:forte14:model-library}. 
Our implementation, a case study, and an example are presented in Section~\ref{sec:forte14:impl}, Section~\ref{sec:forte14:case-study}, and Section~\ref{sec:example:imaginary-factory-network}. 
Related work is described in Section~\ref{sec:forte14:relatedwork}.
We conclude in Section~\ref{sec:forte14:conclusion}. 

\section{Formal Model}
\label{sec:forte14:formalmodel}
In this section, we formally introduce the underlying model. 

\subsection{Terminology}
This research intersects with the field of study of policies, which may lead to a clash of terminology. 
Whenever there is an ambiguity, we will use network terminology. 
In particular, we will use the term \emph{host} for any entity which may appear in a policy, \eg a host may be a collection of IP addresses, a name, or even a role. 
In contrast to common policy terminology~\cite{samarati2001accessdickerbroken,jajodia2001flexibledatabaseaccesscontrol}, we do not differentiate between subjects and objects (sometimes called targets) as they are usually indistinguishable on the network level and a host may act as both. 
For example, data may be written to an entity $A$. 
Then $A$ could be interpreted as object. 
However, $A$ may not be a traditional file, but $A$ could also be a process which is analyzing the data and probably writing it to a different location. 
Then $A$ could be interpreted as subject. 
Since our entities are usually networked hosts, it is natural that they may act as both, subject and object. 
This model assumption is in line with McIlroy and Reed's data flow model~\cite{mcilroy1992multilevel}. 
They notice that interprocess communication makes it necessary ``to identify some subjects also as objects''~\cite{mcilroy1992multilevel}. 

We justify this choice with the another example, inspired by the goal that we ultimately want to enforce the policy on a network.  
A common terminology in the field of policies for filesystems is that users are subjects, files are targets, and access rights are either \emph{read}, \emph{write}, or \emph{execute}. 
For example, a subject performs a read access to a target. 
However, when considering access rights from a network administrator's point of view who is setting up a simple router ACL, there is only the choice between allowing or disallowing the communication. 
It is possible to distinguish between sending and receiving hosts. 
On the application layer, network communication may cause read, write, or execute actions. 
However, a packet from one host to another as seen by a router could be a request to read a file, the contents of a file, the instruction to write to a file, or even be executable code itself. 
This is application layer information, which is not available on the network layer. 
Hence, a distinction between those three actions is not possible, which also means that a distinction between subjects and targets is not possible. 
Consequently, there is only one kind of generic entity: a host. 

Having only the generic notion of a host and packets exchanged between hosts, it is still possible to distinguish between sender and receiver.
Likewise, considering connections between hosts, it is possible to distinguish between initiator and accepting host. 
Hence, the distinction between client and server is not lost. 

\subsection{A Model of Security Policies and Security Invariants}
According to Bishop, a \emph{security policy} is ``a specific statement of what is and is not allowed''~\cite{bishop2003compsec}. 
Narrowing its scope to network access control, a security policy is a set of rules which state the allowed communication relationships between hosts. 
It can be represented as a directed graph. 
This view is consistent with the model used in seL4: ``An access control policy is essentially a directed graph [\dots]''~\cite{sel42013IFS}. 
We will write $G = (V,\, E)$, with the hosts $V$ and the allowed flows $E$. 

\begin{definition}[Security Policy]
\label{def:securitypolicy}
	A security policy is a directed graph $G = (V,\, E)$, where the hosts $V$ are a set of type \mbox{$\mathcal{V}$} and the allowed flows $E$ are a set of type \mbox{$\mathcal{V}{\times}\mathcal{V}$}. 
	The type of $G$ is abbreviated by $\mathcal{G} = (\mathcal{V}\;set) \times ((\mathcal{V}{\times}\mathcal{V})\;set)$. 
\end{definition}

The policy we consider is on the abstraction level of connections or possibly application-level message flows. 
This abstraction level is important for unidirectional flows. 
For example, $v_1$ may write messages to a socket and $v_2$ reads them from a socket. 
Our policy may express that $v_1$ can send messages to $v_2$, but not the other way round. 
The unidirectional nature of this communication is adequate for reasoning on the connection level. 
However, when this policy is translated in a later chapter to a configuration for a network security mechanism, we need to consider network-level packet exchange. 
From the view of the network level, $v_1$ may transmit the message to $v_2$ using several packets over a TCP connection. 
While our connection-level policy models unidirectional message flow, a network-level TCP implementation requires bidirectional flow of packets between $v_1$ and $v_2$ for connection setup and acknowledgements. 
In an unusual manner, it is also possible to use only unidirectional UDP. 
We discuss the network-level implementation, the different cases, and how they are solved in Chapter~\ref{chap:esss14}. 
For both levels of abstraction---connection level and network level---a \emph{directed} graph has proven to be the appropriate model. 
In this chapter, the policy is just on the connection level. 

A policy defines rules (\emph{``how?''}). 
It does not justify the intention behind these rules (\emph{``why?''}). 
To reflect the \emph{why?}-question, we note that depending on a concrete scenario, hosts may have varying security-relevant attributes. 
We model a host attribute of arbitrary type $\Psi$ and establish a total mapping from the hosts $V$ to their scenario-specific attribute. 
Security invariants can be constructed by combining a \emph{host mapping} with a \emph{security invariant template}. 
Latter two are defined together because the same $\Psi$ is needed for a related {host mapping} and {security invariant template}; they are polymorphic over type $\Psi$. 
Different $\Psi$ may appear across several security invariants.

\begin{definition}[Host Mapping]
\label{def:securityinvarianttemplate1}
For scenario-specific attributes of type $\Psi$, a host \mbox{mapping $\nP$} is a total function which maps a host  to an attribute. 
$\nP$ is of type $\mathcal{V} \Rightarrow \Psi$. 
\end{definition}

\begin{definition}[Security Invariant Template]
\label{def:securityinvarianttemplate2}
A security invariant template $\sinvar$ is a predicate\footnote{A predicate is a total, Boolean-valued function.} of type $\mathcal{G} \Rightarrow (\mathcal{V} \Rightarrow \Psi) \Rightarrow \mathbb{B}$, defining the formal semantics of a security invariant. 
Its first argument is a security policy, its second argument a host attribute mapping. 
The predicate \mbox{${\sinvar}\ G\ \nP$} holds iff the security policy $G$ fulfills the security invariant specified by $\sinvar$ and $\nP$. 
\end{definition}

\begin{example}%
\label{example:forte14:blp}
\textbf{Example (BLP). }%
We have the goal to formalize a very simple invariant template to serve as example throughout this chapter. 
While the model is very simple, it has certain uses as Denning exemplifies for a ``government or military system''~\cite{denning1976lattice} and we also show an example in Section~\ref{sec:example:imaginary-factory-network}. 
We model label-based information flow security inspired by the Bell-LaPadula model~\cite{bell1973secure1,bell1973secure2,bell1973securerefinedmodel,bell1975padula3,blphistory}, but with simplifications outlined by Bishop's introductory informal description~\cite{bishop2003computer}. 
We exclude need-to-know vectors~\cite{bell1973secure1,bell1973secure2} (also called categories or compartments~\cite{bell1973secure1,bell1973secure2,bishop2003computer,bell1975padula3,eckert2013}). 
%
%
Hence, labels consist only of ``clearance levels'' for subjects and ``classifications'' of objects. 
Since we do not distinguish between subjects and objects in our theory, in our simplified model the label of an entity directly maps to its current \emph{security level}. 
The security levels model the host attributes for our security invariant template $\Psi = \left\lbrace\mdef{unclassified},\, \mdef{confidential},\, \mdef{secret},\, \mdef{top\-secret}\right\rbrace$. 
The Bell-LaPadula's no read-up and no write-down rules can be summarized by requiring that the security level of a \mbox{receiver $r$} should be greater-equal than the security level of the \mbox{sender $s$}, for all $(s,r) \in E$. 
With a total order `$\leq$' on $\Psi$, the security invariant template can be defined as $\sinvar\ (V,\,E)\ \nP \equiv {\forall (s,r) \in E.}\ \nP\ s \leq \nP\ r$. 

For simplicity, we decided for a total order, as opposed to the partial order of a lattice~\cite{denning1976lattice}. 
While a lattice structure is more expressive, we argue that using multiple simple invariants achieves the same expressiveness with better modularity. 
Theorem~\ref{thm:taint-iff-blp} will later underline this claim. 
Another lattice-based invariant template will be shown in Section~\ref{sinvar:domhierarchy}. 


Let the scenario-specific knowledge be that database $\mathit{db}_1 \in V$ is $\mdef{confidential}$ and all other hosts are $\mdef{unclassified}$. 
Using lambda calculus, the total function $\nP$ can be defined as $(\lambda h. \ \mctrl{if}\ h = \mathit{db}_1 \ \mctrl{then} \ \mdef{confidential} \ \mctrl{else} \ \mdef{unclassified})$. 
Hence $\nP\ \mathit{db}_1 = \mdef{confidential}$. 
If a host $v_1$ is $\mdef{unclassified}$, it may send data to $\mathit{db}_1$, but not the other way round. 
This is appropriate for the abstraction level of our policy.%
\footnote{
	However, on the network level, this implies that it is impossible that $v_1$ and $\mathit{db}_1$ establish a TCP connection. 
	Two hosts need exactly the same security label to establish a TCP connection~\cite{bellovin2004lookbackseciritytcp}. 
	Other invariants (in particular ACS, see next section) do not have this limitation. 
	Chapter~\ref{chap:esss14} details on how we derive a network-level implementation for a policy. 
	Examples and evaluation will show that this simplified Bell-LaPadula template---which primarily serves as example---is barely useful once considering the network level and TCP. 
	While there are some use cases where purely unidirectional packet flow is suitable (for example, we will use UDP in Section~\ref{sec:sdnnfv:casestudy}), trusted entities (Section~\ref{sinvar:blptrust}) will permit bidirectional data flow between different security levels. 
}

For any policy $G$, the predicate $\sinvar\ G\ \nP$ holds if $\mathit{db}_1$ does not leak confidential information (\ie there is no non-reflexive outgoing edge from $\mathit{db}_1$). 
Independent of any policy, $\sinvar\ \_\ \nP$ is a security invariant enriched with scenario-specific knowledge. 
\end{example}

\noindent
Security invariants formalize security goals. 
A template contributes the formal semantics. 
A host mapping contains the scenario-specific knowledge. 
This makes the scenario-independent semantics available for formal reasoning by treating $\nP$ and $G$ as unknowns. 
Even reasoning with arbitrary security invariants is possible by additionally treating $\sinvar$ (of type $\mathcal{G} \Rightarrow (\mathcal{V} \Rightarrow \Psi) \Rightarrow \mathbb{B}$) as unknown. 

With this modeling approach, the end user needs not to be bothered with the formalization of $\sinvar$, but only needs to specify $G$ and $\nP$. 
In the course of this chapter, we present a convenient method for specifying $\nP$. 
%

\section{Properties and Semantic Analysis of Security Invariants}
\label{sec:forte14:analysis}

\subsection{Security Strategies and Monotonicity}
\label{sec:security-strategies}
In IT security, one distinguishes between two main classes of security strategies: \emph{Access Control Strategies} (ACS) and \emph{Information Flow Strategies} (IFS) \cite[\S\hairspace6.1.4]{eckert2013}. 
An IFS focuses on confidentiality and an ACS on integrity or controlled access. 
We require that $\sinvar$ is in one of these classes.

Conventionally, IT security ``rests on confidentiality, integrity, and availability''~\cite[\S1.1]{bishop2003computer}. 
By limiting $\sinvar$ to IFS or ACS, we emphasize that availability is not in the scope of this work. 
Availability requires reasoning on a lower abstraction level, for example, to incorporate network hardware failure. 
Availability invariants could be expressed similarly, but would require inverse monotonicity (see below).

The two security strategies IFS and ACS have one thing in common: 
they prohibit illegal actions. 
From an integrity and confidentiality point of view, prohibiting more never has a negative side effect.  
Removing edges from the policy cannot create new accesses and hence cannot introduce new access control violations. 
Similarly, for an IFS, by statically prohibiting flows in the network, no new direct information leaks nor new side channels can be created. 
In brief, prohibiting more does not harm security. 
From this, it follows that if a policy $(V,\, E)$ fulfills its security invariant, for a stricter policy rule set $E' \subseteq E$, the policy $(V,\, E')$ must also fulfill the security invariant. 
We call this property \emph{monotonicity}. 

\begin{definition}[Monotonicity of Security Invariant Templates]
\label{def:securityinvariantmonotonicity}
A security invariant template $\sinvar$ is \emph{monotonic} if and only if for all $V$, $E$, $\nP$, and $E' \subseteq E$,  if $\sinvar\ (V,\, E)\ \nP$ holds, it must also hold that $\sinvar\ (V,\, E')\ \nP$. Formally: %
\begin{IEEEeqnarray*}{c}
{\forall V\; E\; \nP.}\ \sinvar\ (V,\, E)\ \nP \longrightarrow ({\forall E' \subseteq E.}\ \sinvar\ (V,\, E')\ \nP)
\end{IEEEeqnarray*}%
We require that all security invariant templates are monotonic.
\end{definition}

\begin{example}%
\textbf{Example. }%
We refer to the BLP example on page~\pageref{example:forte14:blp} and will prove that it is monotonic.  
The invariant template was defined as ${\forall (s,r) \in E.}\ \nP\ s \leq \nP\ r$ and we assume the invariant holds. 
Consequently, for any $E' \subseteq E$, it holds that ${\forall (s,r) \in E'.}\ \nP\ s \leq \nP\ r$. \qed

Therefore, if no information leak exists in the first place, disallowing additional flows cannot create a new information leak.
\end{example}

\subsection{Offending Flows}
Since $\sinvar$ is monotonic, if an IFS or ACS security invariant is violated, there must be some flows in $G$ that are responsible for the violation. 
By removing them, the security invariant should be fulfilled (if possible). 
We call a minimal set of such flows the \emph{offending flows}. 
Minimality is expressed by requiring that every single flow in the offending flows bears responsibility for the security invariant's violation.

\begin{definition}[Set of Offending Flows]
\label{def:set_offending_flows_def}
\begin{IEEEeqnarray*}{lcl}
\mdef{set\mhyphen{}offending\mhyphen{}flows}\ G\ \nP & \ \equiv \ &
\bigl\lbrace F \subseteq E  \mid  \neg\, \sinvar\ G\ \nP\ \ \wedge \ \sinvar\ (V,\, E \setminus F)\ \nP\ \ \wedge \ \\ 
 & & \qquad \forall (s,r) \in F.\ \neg\, \sinvar\ \left(V,\, \left(E \setminus F\right) \cup \left\lbrace \left( s,r \right) \right\rbrace \right)\  \nP \bigr\rbrace
\end{IEEEeqnarray*}
\end{definition}

Hence, the \emph{set of offending flows} for a policy $G = (V,\, E)$ with host attributes $\nP$ consists of all subsets of the graph's flows $F \subseteq E$, such that if the security requirement is violated, removing the flows $F$ renders the security invariant valid. 
In addition, $F$ must be minimal, \ie every single flow in $F$ can violate the security invariant when added to $E \setminus F$ again.

\begin{example}
\textbf{Example. }%
The definition does not require that the offending flows are uniquely defined. 
This is reflected in its type since it is a set of sets. 
For example, for $G = (\lbrace v_1, v_2, v_3 \rbrace, \lbrace (v_1, v_2), (v_2, v_3) \rbrace)$ and a security invariant that $v_1$ must not transitively access $v_3$, the invariant is violated: $v_2$ could forward requests. 
The set of offending flows is $\left\lbrace  \lbrace (v_1, v_2) \rbrace, \lbrace (v_2, v_3) \rbrace \right\rbrace$.
This ambiguity tells the end user that there are multiple options to fix a violated security invariant. 
The policy can be tightened by prohibiting one of the offending flows, $\eg \lbrace (v_1, v_2) \rbrace$. 
\end{example}

If $\sinvar\ G\ \nP$ holds, the set of offending flows is always empty\footnote{sinvar-no-offending}. 
Also, for every element in the set of offending flows, it is guaranteed that prohibiting these flows leads to a fulfilled security invariant\footnote{removing-offending-flows-makes-invariant-hold}. 
Without the minimality requirement of the offending flows, they are monotonic. 
\begin{lemma}[Monotonicity of non-minimal Offending Flows]
Assume $\neg\;\sinvar\ (V,\,E)\ \nP$ and $F$ is an offending flow \ie $\sinvar\ (V,\, E \setminus F)\ \nP$, then any $F'$ can be added to $F$ and $F \cup F'$ still repairs the policy violation: $\sinvar\ (V,\, E \setminus (F \cup F'))\ \nP$.
\end{lemma}

It is not guaranteed that the set of offending flows is always non-empty for a violated security invariant. 
Depending on $\sinvar$, it may be possible that no set of flows satisfies Definition~\ref{def:set_offending_flows_def}. 
For our security invariant templates, we require that whenever there is a violation, the set of offending flows is non-empty. 
This means, a violated invariant can always be repaired by tightening the policy. 
Theorem~\ref{thm:no-edges-validity} proves\footnote{valid-empty-edges-iff-exists-offending-flows} an important insight: this is possible if and only if the invariant holds for the deny-all policy.
%
%
%
\begin{theorem}[No Edges Validity]
\label{thm:no-edges-validity}
For $\sinvar$ monotonic, arbitrary \mbox{$V,\, E$, and $\nP,$} let $G = (V,\, E)$ and $G_\mathrm{deny\mhyphen{}all} \equiv (V, \emptyset)$. If $\;\neg\, \sinvar\ G\  \nP$ then%
\vskip-14pt
\begin{IEEEeqnarray*}{l}
\sinvar\ G_\mathrm{deny\mhyphen{}all}\ \nP \ \longleftrightarrow \ \mfun{set\mhyphen{}offending\mhyphen{}flows}\ G\ \nP \neq \emptyset
\end{IEEEeqnarray*}
\end{theorem}

We demand that all security invariants fulfill $\sinvar\ G_\mathit{deny\mhyphen{}all}\ \nP$. 
This means that violations are always fixable.

We call a host responsible for a security violation the \emph{offending host}. 
Given one offending flow, the violation either happens at the sender's or the receiver's side. 
The following difference between ACS and IFS invariant can be observed. 
If $\sinvar$ is an ACS, the host that initiated the request provokes the violation by violating an access control restriction. 
If $\sinvar$ is an IFS, the information leak only occurs when the information reaches the unintended receiver. 
This distinction is essential as it renders the upcoming Definition~\ref{def:secure_default} and Definition~\ref{def:unique_default} provable. 

\begin{definition}[Offending Hosts]
\label{def:offenders} For \mbox{$F \in \mfun{set\mhyphen{}offending\mhyphen{}flows}\ G\ \nP$}
\vskip-14pt
	\begin{IEEEeqnarray*}{lcll}
	\mdef{offenders}\ F & \ \equiv \ & \begin{cases} \left\lbrace \;s  \mid  (s, r) \in F \right\rbrace & \mctrl{if} \ \text{ACS}  \\
	\left\lbrace \;r  \mid  (s, r) \in F \right\rbrace  & \mctrl{if} \ \text{IFS}\end{cases}
	\end{IEEEeqnarray*}
\end{definition}

\subsection{Secure Auto Completion of Host Mappings}
\label{sec:forte14:securedefault} Since $\nP$ is a \emph{total} function $\mathcal{V} \Rightarrow \Psi$, a host mapping for \emph{every} element of $\mathcal{V}$ must be provided. 
However, an end user might only specify the \emph{security-relevant} host attributes. 
Let $\nP_C \subseteq \mathcal{V} \times \Psi$ be a finite, possibly incomplete host attribute mapping specified by the end user. 
For some $\bot \in \Psi$, the total function $\nP$ can be constructed by $\nP\ v \equiv (\mctrl{if}\ {(v, \psi) \in \nP_C} \ \mctrl{then} \ \psi \ \mctrl{else} \ \bot)$.
Intuitively, if no host attribute is specified by the user, $\bot$ acts as a default attribute. 

Given the user specified all security-relevant attributes, we observe that the default attribute can never solve an existing security violation. 
Therefore, we conclude that for a given security invariant $\sinvar$, a value $\bot$ can securely be used as a default attribute if it cannot mask potential security risks. 

We denote an update to $\nP$ which returns $\bot$ for $v$ by $\nP_{v \mapsto \bot}$, \ie $\nP_{v \mapsto \bot} \equiv (\lambda x. \ \mctrl{if}\ x = v \ \mctrl{then} \ \bot \ \mctrl{else} \ \nP\ x)$. 
In other words, a default attribute $\bot$ is secure w.r.t.\ the given information $\nP$ if for all \mbox{offenders $v$}, replacing $v$'s attribute by $\bot$ has the same amount of security-relevant information as the original $\nP$.


\begin{definition}[Secure Default Attribute]
\label{def:secure_default}
We call $\bot$ a secure default attribute if and only if for a fixed $\sinvar$ and for arbitrary $G$ and $\nP$ that cause a security violation, replacing the host attribute of any offenders by $\bot$ must guarantee that no security-relevant information is masked. 
Formally, 
\vskip-14pt
\begin{IEEEeqnarray*}{l}
\forall \; G \; \nP{}. \ \forall \; F \in \mfun{set\mhyphen{}offending\mhyphen{}flows}\ G\ \nP{}. 
\ \ \forall v \in \mfun{offenders}\ F.\ \neg\, \sinvar\ G\ \nP_{v \mapsto \bot}
\end{IEEEeqnarray*}
\end{definition}
\bigskip

\begin{example}
\textbf{Example. }%
In the simple Bell-LaPadula model, an IFS, let us assume information is leaked. 
The predicate `information leaks' holds, no matter to which lower security level the information is leaked.
In general, if there is an illegal flow, it is from a higher security level at the sender to a lower security level at the receiver. 
Replacing the security level of the receiver with the lowest security level, the information about the security violation is always preserved. 
Thus, $\mdef{unclassified}$ is the secure default attribute\footnote{interpretation BLPbasic: SecurityInvariant\_IFS}. 
In summary, if all classified hosts are labeled correctly, treating the rest as unclassified prevents information leakage. 
\end{example}

To elaborate on Definition~\ref{def:secure_default}, it can be restated as follows. 
It focuses on the available security-relevant information in the case of a security violation. 
The attribute of an offending host $v$ bears no information, except for the fact that there is a violation. 
A secure default attribute $\bot$ cannot solve security violations. 
Hence $\nP\ v$ and $\bot$ are equal w.r.t.\ the security violation. 
Thus, $\nP$ and $\nP_{v \mapsto \bot}$ must be equal w.r.t.\ the information about the security violation. 
Requiring this property for all policies, all possible security violations, all possible choices of offending flows, and all candidates of offending hosts, this definition justifies that $\bot$ never hides a security problem.

\begin{example}
\textbf{Example. }%
Definition~\ref{def:secure_default} can be specialized to the exemplary case in which a new host $x$ is added to a policy $G$ without updating the host mapping. 
Consulting an oracle, $x$'s real host attribute is $\nP\ x = \psi$. 
In reality, the oracle is not available and $x$ is mapped to $\bot$ because it is new and unknown. 
Let $x$ be an attacker. 
With the oracle's $\psi$-attribute, $x$  causes a security violation. 
We demand that the security violation is exposed even without the knowledge from the oracle. 
Definition~\ref{def:secure_default} satisfies this demand: if $x$ mapped by the oracle to $\psi$ causes a security violation, $x$ mapped to $\bot$ does not mask the security violation. 
\end{example}

Note that our theory assumes that a user specifies all security-relevant attributes. 
Hence, our default attribute is not designed to be secure if a user does not provide this information. 
This is in line with our goal that security invariants provide a language to describe scenario-specific security requirements: If a requirement is not listed in the requirement specification, our formalization provides no means to uncover this. 
However, our algorithm for automated policy construction (Section~\ref{subsec:policyconstruction}) may provide a feedback by computing an overly permissive policy which may hint at the missing requirement.

A `deny-all' default attribute is easily proven secure. 
Definition~\ref{def:secure_default} reads the following for this case: 
if an offender $v$ does something that violates $\sinvar\ G\ \nP$, then removing all of $v$'s rights ($\nP_{v \mapsto \mathrm{deny\mhyphen{}all}}$), a violation must persist. 
Hence, designing whitelisting security invariant templates with a restrictive default attribute is simple. 
However, to add to the ease-of-use, more permissive default attributes are often desirable since they reduce the manual configuration effort. 
In particular, if a security invariant only concerns a subset of a policy's hosts, no restrictions should be imposed on the rest of the policy. 
This is also possible with Definition~\ref{def:secure_default}, but may require a comparably difficult proof.

\begin{example}
\textbf{Example. }%
In the BLP example on page~\pageref{example:forte14:blp}, no matter how many (unconfigured) hosts are added to the policy, it is sufficient to only specify that $\mvar{db}_1$ is $\mdef{confidential}$. 
This confidentiality is guaranteed while no restrictions are put on hosts that do not interact with $\mvar{db}_1$. 
\end{example}

\begin{definition}[Default Attribute Uniqueness]
\label{def:unique_default}
A default attribute $\bot$ is called unique iff it is secure (Definition~\ref{def:secure_default}) and there is no $\bot' \neq \bot$ s.t. $\bot'$ is secure. 
\end{definition}%

We demand that all security invariants fulfill Definition~\ref{def:secure_default} and Definition~\ref{def:unique_default}. 
This means that there is only one unique secure default attribute $\bot$. 

\begin{example}%
\textbf{Example. }%
In the simple Bell-LaPadula model, since the security levels form a total order, the lowest security level is uniquely defined. 
\end{example}

With the experience of proving that default attributes of about 20 invariant templates fulfill Definition~\ref{def:secure_default} and Definition~\ref{def:unique_default}, the connection between offending host and security strategy was discovered. 
During our early research, we realized that a Boolean variable, fixed for $\sinvar$, indicating the offending host was necessary to make Definition~\ref{def:secure_default} and Definition~\ref{def:unique_default} provable. 
A classification of the different invariants revealed the important connection. 


\subsection[Phi-Structured Security Invariant Templates]{$\Phi$-Structured Security Invariant Templates}
\label{sec:phi-structured}
The Bell-LaPadula security invariant template was defined as ${\forall (s,r) \in E.}\ \nP\ s \leq \nP\ r$. 
That means, for all flows, a predicate over the sender's and receiver's security level is evaluated. 
Here, the predicate is $(\lambda \mvar{sl}_s\ \mvar{sl}_r.\ \mvar{sl}_s \leq \mvar{sl}_r)$. 
We found that a similar structure is also found in most other security invariant templates. 
Many security invariant templates presented in Section~\ref{sec:forte14:model-library} have a simple, common structure: a predicate is evaluated for all flows over the sender's and receiver's scenario-specific attributes. 
Let $\Phi$ be this predicate. 
$\Phi$ is of type $\Psi \Rightarrow \Psi \Rightarrow \mathbb{B}$. 
Then the security invariant template corresponds to ${\forall (s,r) \in E.}\ \Phi\ (\nP\ s)\ (\nP\ r)$. 
Consequently, we call all security invariant templates that follow this structure $\Phi$-structured.

\begin{definition}[$\Phi$-Structured Security Invariant Templates]
\label{def:phi-structured}
A security invariant template is called \emph{$\Phi$-structured} iff for some predicate $\Phi$, the invariant template can be expressed as ${\forall (s,r) \in E.}\ \Phi\ (\nP\ s)\ (\nP\ r)$.
\end{definition}%

It is possible to express lemmas and theorems generically for all $\Phi$-structured invariant templates. 
This simplifies proofs about specific invariant templates since the generic lemmas can be reused. 
In addition, this avoids proof duplication. 

\begin{example}%
\textbf{Example. }%
All $\Phi$-structured invariant templates immediately fulfill monotonicity\footnote{monotonicity-sinvar-mono}. 
\end{example}

Some security invariant templates slightly deviate from the $\Phi$-structured definition. 
However, the important results which hold for $\Phi$-structured invariant templates also hold for templates with the following structure. 
All have in common that some predicate is evaluated for all flows (\ie policy rules): 
\begin{itemize}
	\item The predicate also takes the name of the entity into account: \\
		${\forall (s,r) \in E.}\ \Phi\ (\nP\ s)\ s\ (\nP\ r)\ r$
	\item Reflexive flows are excluded: \\
		${\forall (s,r) \in E.}\ s \neq r \longrightarrow \Phi\ (\nP\ s)\ (\nP\ r)$
	\item A combination of both: \\
		${\forall (s,r) \in E.}\ s \neq r \longrightarrow \Phi\ (\nP\ s)\ s\ (\nP\ r)\ r$
\end{itemize}

For the sake of brevity, we only consider $\Phi$-structured invariant templates according to Definition~\ref{def:phi-structured}. 
We refer the reader to the proof document for the slightly deviated versions.


\subsection{Unique and Efficient Offending Flows}
\label{sec:forte14:maxpolicmethodandoffending}
We assume that $\sinvar$ is a $\Phi$-structured invariant template.
Definition~\ref{def:set_offending_flows_def} is defined over all subsets. 
Consequently, the naive computational complexity to compute the set of offending flows of is in \textbf{\textit{NP}}. 
This section shows that -- with knowledge about a concrete security invariant template $\sinvar$ -- it can be computed in linear time. 
For $\Phi$-structured invariants, the offending flows are always uniquely defined and can be described intuitively\footnote{ENF-offending-set}. 
The same holds for templates with a similar structure\footnote{ENFsr-offending-set, ENFnr-offending-set, ENFnrSR-offending-set}. 

\begin{theorem}[$\Phi$ Set of Offending Flows]
\label{thm:phi_set_offending_flows_def}
\begin{sloppypar}If $\sinvar$ is $\Phi$-structured, \ie \mbox{$\sinvar\ G\ \nP \equiv \forall (s,r) \in E.\ \Phi\ (\nP\ s)\ (\nP\ r)$}, then\end{sloppypar}
\vskip-12pt
	\begin{IEEEeqnarray*}{lcll}
	\mfun{set\mhyphen{}offending\mhyphen{}flows}\ G\ \nP & \ \equiv \ & \begin{cases} \lbrace\lbrace (s,r) \in E  \mid  \neg\;\Phi\ (\nP\ s)\ (\nP\ r) \rbrace\rbrace & \mctrl{if} \ \neg\;\sinvar\ G\ \nP  \\
	\emptyset  & \mctrl{if} \ \sinvar\ G\ \nP\end{cases}
	\end{IEEEeqnarray*}
\end{theorem}
\begin{example}
\textbf{Example. }%
For the Bell-LaPadula model, if no security violation exists the set of offending flows is $\emptyset$, else $\lbrace\lbrace (s,r) \in E  \mid  \nP(s) > \nP(r) \rbrace\rbrace$,\footnote{BLP-offending-set} \ie all flows from a higher to a lower security level. 
\end{example}
%

\paragraph*{Offending flows for templates similar to $\Phi$-structured invariants} 
In general, if no violation exists, the set of offending flows is always $\emptyset$\footnote{sinvar-no-offending}. 
If there exists a violation and the structure of the template corresponds to one of the structures similar to a $\Phi$-structured invariant, the following holds. 
If the structure of the template is ${\forall (s,r) \in E.}\ \Phi\ (\nP\ s)\ s\ (\nP\ r)\ r$, then $\mfun{set\mhyphen{}offending\mhyphen{}flows}\ G\ \nP$ is 
$\lbrace\lbrace (s,r) \in E  \mid  \neg\;\Phi\ (\nP\ s)\ s\ (\nP\ r)\ r) \rbrace\rbrace$\footnote{ENFsr-offending-set}.
If the structure is ${\forall (s,r) \in E.}\ s \neq r \longrightarrow \Phi\ (\nP\ s)\ (\nP\ r)$, then $\mfun{set\mhyphen{}offending\mhyphen{}flows}\ G\ \nP$ is 
$\lbrace\lbrace (s,r) \in E  \mid  s \neq r \ \wedge \neg\;\Phi\ (\nP\ s)\ (\nP\ r) \rbrace\rbrace$\footnote{ENFnr-offending-set}.
If the structure is ${\forall (s,r) \in E.}\ s \neq r \longrightarrow \Phi\ (\nP\ s)\ s\ (\nP\ r)\ r$, then $\mfun{set\mhyphen{}offending\mhyphen{}flows}\ G\ \nP$ is 
$\lbrace\lbrace (s,r) \in E  \mid  s \neq r \wedge \neg\;\Phi\ (\nP\ s)\ s\ (\nP\ r)\ r \rbrace\rbrace$\footnote{ENFnrSR-offending-set}.

\subsection{Composition of Security Invariants}
\label{subsec:forte:composability}
Usually, there is more than one security invariant for a given scenario. 
However, composition and modularity is often a non-trivial problem. 
For example, access control lists that are individually secure can introduce security breaches under composition~\cite{composable94}. 
Also, information flow security of individually secure processes, systems, and networks may be subverted by composition~\cite{restrictiveness}. 
This is known as the \emph{composition problem}~\cite{blphistory}. 

With our formalization, composability and modularity are enabled by design. 
For a fixed policy $G$ with $k$ security invariants, let $\sinvar{}_i$ be the security invariant template and $\nP{}_i$ the host mapping, for $i \in \left\lbrace 1 \dots k \right\rbrace$. 
The predicate $\sinvar_i\ G\ \nP_i$ holds if and only if the security invariant $i$ holds for the policy $G$. 
With this modularity, composition of all security invariants is straightforward\footnote{all-security-requirements-fulfilled}: all security invariants must be fulfilled. 
The monotonicity guarantees that having more security invariants provides greater or equal security. 
\begin{IEEEeqnarray*}{c}
\sinvar_1\ G\ \nP_1 \ \wedge \dots \wedge \ \sinvar_k\ G\ \nP_k
\end{IEEEeqnarray*}

This composition works due to two careful design decisions. 
First, a security policy can only have positive rules. 
This implies that there cannot be contradictory allow and deny rules; the policy is represented only by allow rules. 
Second, monotonicity ensures that the algorithms and definitions of the following sections behave as expected, also under composition.

Care must be taken with regard to the type when composing multiple security invariants. 
Each invariant $\sinvar_i$ may use its own type $\Psi_i$, hence, each invariant template is of type $\mathcal{G} \Rightarrow (\mathcal{V} \Rightarrow \Psi_i) \Rightarrow \mathbb{B}$. 
Some $\sinvar_j$ may be of type $\mathcal{G} \Rightarrow (\mathcal{V} \Rightarrow \Psi_j) \Rightarrow \mathbb{B}$, where $\Psi_i \neq \Psi_j$. 
For example, $\Psi_i$ could be security levels while $\Psi_j$ could be access control lists. 
Therefore, it is not possible to store several different security invariant templates in the same list. 
The list $[\sinvar_i,\ \sinvar_j]$ is not well-typed. 
However, a security invariant template applied to a policy and a host mapping, \eg $\sinvar_i\ G\ \nP_i$, is of type $\mathbb{B}$. 
Therefore, $[\sinvar_1\ G\ \nP_1,\ \dots,\ \sinvar_i\ G\ \nP_i,\ \sinvar_j\ G\ \nP_j,\ \dots,\ \sinvar_k\ G\ \nP_k]$ is a well-typed list of Booleans. 
It is also possible to partially apply the scenario-specific knowledge $\nP_i$ to $\sinvar_i$, \ie $(\lambda G.\ \sinvar_i\ G\ \nP_i)$. 
This leaves a function of type $\mathcal{G} \Rightarrow \mathbb{B}$. 
We call this a \emph{configured security invariant}. 
This means, for a specific scenario, all security invariants with their scenario-specific knowledge can be specified independently of the policy $G$. 
The list $[(\lambda G.\ \sinvar_1\ G\ \nP_1),\ \dots,\ (\lambda G.\ \sinvar_i\ G\ \nP_i),\ (\lambda G.\ m_j\ G\ \nP_j),\ \dots,\ (\lambda G.\ \sinvar_k\ G\ \nP_k)]$ is a well-typed list of predicates over a security policy.

The $\mfun{set\mhyphen{}offending\mhyphen{}flows}$ was defined for a policy and a host attribute mapping. 
Its type is $\mathcal{G} \Rightarrow (\mathcal{V} \Rightarrow \Psi) \Rightarrow (\mathcal{V} \times \mathcal{V})\ \mathit{set}\ \mathit{set}$. 
It can also be defined for a configured security invariant $(\lambda G.\ \mfun{set\mhyphen{}offending\mhyphen{}flows}\ G\ \nP)$. 
This leaves a function of type $\mathcal{G} \Rightarrow (\mathcal{V} \times \mathcal{V})\ \mathit{set}\ \mathit{set}$. 
To avoid notational overhead, we will apply $\mfun{set\mhyphen{}offending\mhyphen{}flows}$ to both invariant templates and configured invariants interchangeably. 

Notationally, we will write $\sinvar$ for a security invariant template and $m_c$ for a configured security invariant template.


\section{Policy Construction}
\label{subsec:policyconstruction}
In this section, we combine the results of the previous sections to present a simple algorithm to construct a security policy, given the list of configured security invariants. 

A policy that fulfills all security invariants can be constructed by removing all offending flows from an arbitrary starting policy. 
This approach is sound\footnote{generate-valid-topology-sound} for arbitrary $\sinvar$. 
In general, $G_{\mathrm{all}} \equiv (V, V{\times}V)$ is a good starting point. 

To simplify the following algorithm, we specify a helper function. 
\begin{IEEEeqnarray*}{lCl}
\mfun{delete\mhyphen{}edges}\ (V,\, E)\ E' & \equiv & (V,\, E \setminus E')
\end{IEEEeqnarray*}

Then, the following function constructs a security policy for a given list of configured invariants. 
%
\begin{IEEEeqnarray*}{lCl}%
  \IEEEeqnarraymulticol{3}{l}{\mfun{generate\mhyphen{}valid\mhyphen{}topology}\ :: (\mathcal{G} \Rightarrow \mathbb{B})\ \textnormal{list} \Rightarrow \mathcal{G} \Rightarrow \mathcal{G}}\\
  \mfun{generate\mhyphen{}valid\mhyphen{}topology}\ []\ G & = & G\\
  \mfun{generate\mhyphen{}valid\mhyphen{}topology}\ (m_c \lstcons \mvar{Ms})\ G & = & \\
  \IEEEeqnarraymulticol{3}{r}{\qquad\qquad \mfun{delete\mhyphen{}edges}\ \left(\mfun{generate\mhyphen{}valid\mhyphen{}topology}\ \mvar{Ms}\ G\right) \left(\bigcup \mfun{set\mhyphen{}offending\mhyphen{}flows}\ m_c\ G\right)}
\end{IEEEeqnarray*}

In each iteration, the algorithms removes the offending flows of one configured invariant and calls itself recursively with the updated policy. 
A different approach would be to compute the offending flows for all configured invariants for the starting policy and remove them at once. 
Both approaches yield the same result. 

\begin{lemma}
\label{lem:generate_valid_topology_def_alt}
\begin{IEEEeqnarray*}{c}
  \mfun{generate\mhyphen{}valid\mhyphen{}topology}\ \mvar{Ms}\ G = \\
  \mfun{delete\mhyphen{}edges}\ G \left(\bigcup \left(\bigcup m_c \in \mfun{set}\ \mvar{Ms}.\ \mfun{set\mhyphen{}offending\mhyphen{}flows}\ m_c\ G\right)\right)
\end{IEEEeqnarray*}
\end{lemma}

\begin{theorem}[Soundness of Policy Construction]
Let $M$ be a set of configured security invariants. This means, each element of $M$ is of type $\mathcal{G} \Rightarrow \mathbb{B}$, is monotonic, and always has defined offending flows. Then
\begin{IEEEeqnarray*}{c}
{\forall m_c \in M.} \ \ m_c\ \left(\mfun{generate\mhyphen{}valid\mhyphen{}topology} \ M\ G\right)
\end{IEEEeqnarray*}
\end{theorem}

For $\Phi$-structured security invariant templates and when starting with the allow-all policy $G_{\mathrm{all}} = (V, V{\times}V)$, the algorithm is also complete\footnote{generate-valid-topology-max-topo}. 
Completeness tells us that the constructed policy is most permissive, \ie it is not possible to add additional allowed flows to the policy. 
%
\begin{theorem}[Completeness of Policy Construction for $\Phi$-Structured Invariants]
\label{thm:generate_valid_topology_phi_complete}
Let $M$ be a set of configured, $\Phi$-structured security invariants. 
Let 
\begin{IEEEeqnarray*}{c}
(V,\, E_{\mathrm{max}}) = \mfun{generate\mhyphen{}valid\mhyphen{}topology} \ M\ (V,\, V \times V)
\end{IEEEeqnarray*}
Then all configured security invariants hold for the constructed policy $(V,\, E_{\mathrm{max}})$ but for all $e \in (V \times V) \setminus E_{\mathrm{max}}$, 
if $e$ is added to the constructed policy, at least one configured security invariant is violated.
\end{theorem}

For the rest of this thesis, we will always assume that we call our automated policy construction algorithm with $G_{\mathrm{all}}$.

\begin{example}%
\textbf{Example. }%
We refer to the BLP example on page~\pageref{example:forte14:blp} where only $\mvar{db}_1$ was labeled $\mdef{confidential}$. 
Constructing a valid policy for this setting means to remove all flows where $\mvar{db}_1$ is sending out information (except for the reflexive flow where $\mvar{db}_1$ sends to itself). 
This means that any other hosts $\neq \mvar{db}_1$ can freely interact and even send data to $\mvar{db}_1$. 
\end{example}

\begin{example}
\textbf{Example. }%
For $\Phi$-structured invariants, Theorem~\ref{thm:phi_set_offending_flows_def} tells that the offending flows are uniquely defined. 
Lemma~\ref{lem:generate_valid_topology_def_alt} tells that a valid policy is constructed by removing all those uniquely defined offending flows. 
Consequently, for $\Phi$-structured invariants, a maximum policy is uniquely defined.
\end{example}

\begin{example}
\textbf{Example. }%
If completely contradictory security invariants are given, the resulting (maximum) policy is the deny-all policy $G_\mathrm{deny\mhyphen{}all} = (V, \emptyset)$. 
\end{example}

\begin{example}
	\textbf{Example. }%
	We assume both a valid, manually-specified policy $G_{\mathrm{manual}} = (V,\, E)$ and a specification of the security invariants $M$ are available. 
	We can compute $G_{\mathrm{computed}} = \mfun{generate\mhyphen{}valid\mhyphen{}topology} \ M\ (V,\, V \times V)$.
	
	With this, one can test whether the specified invariants `mean' the right thing by comparing $G_{\mathrm{computed}}$ with $G_{\mathrm{manual}}$.
	
	If all invariants are $\Phi$-structured, any valid $G_{\mathrm{manual}}$ is a sub-policy of $G_{\mathrm{computed}}$.\footnote{enf-all-valid-policy-subset-of-max} 
	This means, if the specified invariants encode the `right' requirements and encode all requirements, $G_{\mathrm{computed}}$ should be equal to $G_{\mathrm{manually}}$. 
	
	Consequently, $G_{\mathrm{computed}}$ gives feedback about whether the specified security invariants carry the right meaning. 
\end{example}

\bigskip

\noindent
For $\Phi$-structured invariants, the algorithm has all properties on could wish for: 
It is simple, sound, complete, and fast. 
In Chapter~\ref{chap:hilbertsoffending}, we will improve this algorithm for non-$\Phi$-structured invariants.

\chapter{Security Invariant Template Library}
\label{sec:forte14:model-library}

The invariants of Section~\ref{sinvar:blptsimple}, \ref{sinvar:blptrust}, \ref{sinvar:domhierarchy}, and a simplified version of Section~\ref{sinvar:secgateway} have been presented in the following paper~\cite{diekmann2014forte}: 
\begin{itemize}
	\item Cornelius Diekmann, Stephan-A.\ Posselt, Heiko Niedermayer, Holger Kinkelin, Oliver Hanka, and Georg Carle. \emph{Verifying Security Policies using Host Attributes}. In FORTE -- 34th IFIP International Conference on Formal Techniques for Distributed Objects, Components and Systems, volume 8461, pages 133-148, Berlin, Germany, June 2014. Springer.
\end{itemize}

\noindent
The invariants of Section~\ref{sinvar:taint} and Section~\ref{sinvar:taintuntaint} have been presented in the following paper~\cite{maltitz2016fmpriv}:
\begin{itemize}
	\item Marcel von Maltitz, Cornelius Diekmann and Georg Carle, \emph{Taint Analysis for System-Wide Privacy Audits: A Framework and Real-World Case Studies}. In 1st Workshop for Formal Methods on Privacy, Limassol, Cyprus, November 2016. Note: no proceedings published. 
\end{itemize}

\paragraph*{Statement on author's contributions}
The formalization of all the presented invariant templates are the work of the author of this thesis. 
The following security invariants are based on the author's master's thesis~\cite{cornythesis}: 
 Section~\ref{sinvar:blptsimple}, \ref{sinvar:blptrust}, \ref{sinvar:dependability}, \ref{sinvar:domhierarchy}, \ref{sinvar:subnets}, \ref{sinvar:subnetsingw}, \ref{sinvar:secgateway}, \ref{sinvar:sink}, and \ref{sinvar:noninterference}. 
They have been re-implemented for the new underlying theory, extended, and improved. 

The following security invariants are based on joint work with Marcel von Maltitz: Section~\ref{sinvar:taint} and Section~\ref{sinvar:taintuntaint}. 
Von Maltitz contributed to the literature research and survey of the conceptualization of privacy. 
From this, von Maltitz derived a working (informal) definition of privacy. 
Both, the author of this thesis and von Maltitz, contributed to the research of related work with regard to taint analysis. 
Those sections are only outlined very briefly in the paragraph `\nameref{par:backgroundtainting}' of Section~\ref{sinvar:taint}. 
The author of this thesis provided major contributions for deriving the formulas, the implementation, realization, and proofs of the formalizations.

\medskip

\paragraph*{Abstract}
The previous chapter discussed the theoretic foundations of security invariant templates. 
In this chapter, we present our library of pre-defined security invariant templates and show some examples. 

\medskip

\section{Introduction}
In this chapter, we present all security invariant templates which are defined at the time of this writing. 
Our implementation currently features fifteen templates and grows. 
Common networking scenarios such as subnets, non-interference invariants, or access control lists are available. 
All can be inspected in the published theory files. 
They are summarized in \mbox{Table~\ref{tab:invariantemplates}} and will be illustrated in this chapter. 
The first column of the tables gives the template's name, the second column the section in which the template will be defined, the third column lists whether the template is $\Phi$-structured (cf.\ Section~\ref{sec:phi-structured}), the fourth column refers to the security strategy (Section~\ref{sec:security-strategies}), finally, column five provides a summarizing description of the template's intended use cases. 

\begin{table}[h!bt]
\centering
\caption{Security Invariant Templates}
\label{tab:invariantemplates}
\begin{tabular}[0.99\linewidth]{ @{} l @{\hspace*{0.8em}} l @{\hspace*{0.8em}} l @{\hspace*{0.8em}} c @{\hspace*{0.8em}} p{21em} @{}}%
	\toprule
	Name            & \S                        & $\Phi$ & Stgy & Description \\
	\midrule
	Simple BLP      & \ref{sinvar:blptsimple}   & \cmark & IFS & Simplified Bell-LaPadula \\
	Bell-LaPadula   & \ref{sinvar:blptrust}     & \cmark & IFS & Label-based Information Flow Security with trusted entities \\
	Comm.\ Partners & \ref{sinvar:commpartners} & \cmark & ACS & Simple ACLs (Access Control Lists) \\
	Comm.\ With     & \ref{sinvar:commwith}     & \xmark & ACS & White-listing transitive ACLs \\
	Not Comm.\ With & \ref{sinvar:notcommwith}  & \xmark & ACS & Black-listing transitive ACLs \\
	Dependability   & \ref{sinvar:dependability} & \xmark & ACS & Limit dependence on certain hosts \\
	Domain Hierarchy & \ref{sinvar:domhierarchy} & \cmark & ACS & Hierarchical control structures \\
	NoRefl          & \ref{sinvar:norefl}       & \cmark & ACS* & Allow/deny reflexive flows. Can lift symbolic policy identifiers to role names (\eg symbolic host name corresponds to an IP range.) \\
	NonInterference & \ref{sinvar:noninterference} & \xmark & IFS & Transitive non-interference properties \\
	PolEnforcePoint & \ref{sinvar:secgateway} & \cmark & ACS & Central application-level policy enforcement point. Master/Slave relationships. \\
	Sink            & \ref{sinvar:sink}        & \cmark & IFS & Information sink. Hosts (or host groups) must not publish any information \\
	Subnets         & \ref{sinvar:subnets}     & \cmark & ACS & Collaborating, protected host groups\\
	SubnetsInGW     & \ref{sinvar:subnetsingw} & \cmark & ACS & Simple, collaborating, protected or accessible host groups\\
	Simple Tainting & \ref{sinvar:taint}       & \cmark & IFS & Simplified label-based Privacy\\
	Tainting        & \ref{sinvar:taintuntaint} & \cmark & IFS & Label-based Privacy with untainting\\
 \bottomrule%
 \multicolumn{5}{c}{\S = Section. Stgy = Strategy: Access Control (ACS) or Information Flow (IFS)}
\end{tabular}
\end{table}

\paragraph*{Default Attributes}
The default attributes of all the templates were designed (whenever possible) to be as permissive as possible. 
This means, the default attributes should at least allow flows between all hosts which are set to the default attribute. 
This greatly adds to the ease-of-use, since the scope of an invariant is limited only to the explicitly configured hosts. 
The unconcerned parts of a security policy are not negatively affected. 
For example, for a policy with 100 hosts, one can easily express an invariant for only three hosts; the connectivity of the other 97 hosts among each other should not be negatively influenced since those 97 machines are set to $\bot$. 
For certain strict invariants, this is not always possible due to the proof obligations imposed by Definition~\ref{def:unique_default}. 

\paragraph*{Meta Invariant}
We present a meta security invariant template to model system boundaries in Section~\ref{sec:meta:systemboundaries}. 
We call it a \emph{meta} invariant because internally, it is implemented by two of the invariants of Table~\ref{tab:invariantemplates}.

\section{Simple Bell-LaPadula}
\label{sinvar:blptsimple}
A simplified version of the Bell-LaPadula model served as guiding example throughout the last chapter. 
In this section, it will be completely formalized. 

The following two paragraphs about the Bell-LaPadula Model in the literature and its history are based on two previously published paragraphs in the author's master's thesis~\cite{cornythesis}. 

\paragraph*{The Bell-LaPadula Model in the Literature}
The \emph{Bell-LaPadula} Model~\cite{bell1973secure1,bell1973secure2,bell1975padula3} (BLP) defines 5 universal access rights: read-only, append, execute, read-write, and control~\cite{bell1973secure2,eckert2013}. It resembles to military-style classifications~\cite{bishop2003computer}. Each subject is assigned a \emph{clearance level} and objects are assigned a \emph{classification}~\cite{bell1973secure1,bell1973secure2}. In this thesis, we call them both security levels. The classifications form an ordering with higher values representing more classified information. In simplified terms, the model introduces the famous no-read-up and no-write down rules.\footnote{The BLP model also features an additional access matrix.} With these system-global rules, BLP is a MAC model. No-read-up means that subjects are not allowed to access objects with a higher classifications than their own. No-write-down means that a subject with a high clearance is not allowed to write to objects with a lower classification. Note that the Bell-LaPadula model can also be classified as information flow security model.  
By enforcing only these two rules, objects are successively assigned ever increasing classification levels. Thus, the model introduces so-called \emph{trusted processes} which correspond to trusted subjects, allowed to decrease the classification of objects~\cite{eckert2013}.
\paragraph*{History}
Elliot Bell and Leonard J.\ La Padula produced the first volumes of the BLP Model~\cite{bell1973secure1,bell1973secure2} ``during two fiscal years''~\cite{blphistory}. After one year of modeling work, a gap between practical implementation and the model still existed. Engineers had problems utilizing the model to build secure prototypes, which resulted in simplifications of the model. The concept of a subject's current security level and trusted subjects were introduced~\cite{blphistory}. After two more years of work, the BLP model was published in its final version~\cite{bell1975padula3}. After 11 years, an interpretation of the BLP model for networks was developed~\cite{blphistory}. It featured ``hosts'' as active entities and ``connections'' as resources~\cite{blphistory}.

\bigskip

\paragraph*{Our Security Invariant Template}
As noticed, entities may act as subjects as well as objects. 
Since we do not distinguish the two, we only have sending entities and receiving entities. 
In addition, by looking at a single (possibly encrypted) connection, one cannot distinguish different access rights (such as read-only, append, execute, read-write, and control). 
For this interpretation, referring to the BLP example on page~\pageref{example:forte14:blp}, the no-read-up and no-write-down rules are fulfilled for packets with arbitrary requests if flows from higher to lower security levels are prevented. 
Therefore, the simplified Bell-LaPadula invariant template was introduced as $\sinvar\ (V,\,E)\ \nP \equiv {\forall (s,r) \in E.}\ \nP\ s \leq \nP\ r$. 

For the sake of presentiveness, the security levels were defined as $\lbrace\mdef{unclassified},\allowbreak{} \mdef{confidential},\allowbreak{} \mdef{secret},\allowbreak{} \mdef{top\-secret}\rbrace$. 
Technically, there is no reason that $\mdef{top\-secret}$ is the highest security level. 
We lift this restriction of a maximum level by modeling security levels as natural numbers: $\Psi = \mathbb{N}$. 
Hence, the template does not impose an upper limit on the ``confidentialness''. 

The lowest security level is $0 = \bot$, which can be understood as $\mdef{unclassified}$. 
Consequently, $1 = \mdef{confidential}$, $2 = \mdef{secret}$, $3 = \mdef{top\-secret}$, \dots. 
The total order of the security levels now corresponds to the total order of the natural numbers `$\leq$'. 
It is important that there is a lowest security level (\ie $0$), otherwise, a unique and secure default parameter could not exist. 
Hence, it is not possible to extend the security levels to $\mathbb{Z}$ to model unlimited ``un-confidentialness''. 

The formula of $\sinvar$, $\bot$, and the efficient version of the offending flows have already been presented. 
It is an information flow security strategy.

\medskip
\paragraph*{Alternative Definition}
For the following sections, we first introduce a helper function to simplify definitions. 
We call the function $\mfun{succ\mhyphen{}tran}$ and it represents the nodes that are transitively reachable in a graph by a starting node. 
Roughly speaking, $\mfun{succ\mhyphen{}tran}\ G\ v$ corresponds to all nodes reachable from $v$. 
It is defined with the help of the transitive closure of the edges of $G$. 

\begin{definition}[reachable hosts]
\begin{IEEEeqnarray*}{lcl}
\mfun{succ\mhyphen{}tran}\ (V,\, E)\ v & \ \equiv \ & \lbrace v'.\ (v, v') \in E^+ \rbrace
\end{IEEEeqnarray*}
\end{definition}

Executable code for $\mfun{succ\mhyphen{}tran}$ over lists is available in an AFP entry~\cite{Transitive-Closure-AFP}. 
Note that this executable code is not used for the simplified Bell-LaPadula model since the first definition without the transitive closure is way more efficient. 
However, it will be used for later definitions. 

With this helper, the formula of the simple Bell-LaPadula invariant template can now be expressed differently. 
It requires that for any host $v \in V$, for any reachable host $v'$, the security level of $v'$ is greater or equal to the one of $v$. 
\begin{IEEEeqnarray*}{l}
\sinvar\ (V,\, E)\ \nP{} \ \equiv \ \left(\forall v \in V.\ \ \forall v' \in \mfun{succ\mhyphen{}tran}\ (V,\, E)\ v.\ \ (\nP\ v) \leq (\nP\ v')\right)
\end{IEEEeqnarray*}
Both definitions are equal\footnote{sinvar-BLPbasic-tancl}. 
The original definition only requires a condition for the edges in the graph while the new definition imposes a global condition. 
Since both are equal, it can be said that the second definition is also $\Phi$-structured (although syntactically, it clearly is not) and that the offending flows are efficiently computable. 

In general, this is not the case for all invariant templates and the simple Bell-LaPadula template is a notable exception. 
When an invariant template is defined with the help of $\mfun{succ\mhyphen{}tran}$, all paths between two hosts may need to be considered. 
In this case, since the set of offending flows is defined over all subsets of the edges in the graph, it may not be possible to simplify this definition. 
Consequently, the set of offending flows may be of exponential size. 
Hence, it becomes infeasible to efficiently execute algorithms built on top of it, most notably, the policy construction. 
On the other hand, all templates which are marked as $\Phi$-structured in Table~\ref{tab:invariantemplates} are safe for the use in any algorithm.

\section{Simplified Bell-LaPadula with Trust}
\label{sinvar:blptrust}
A simplified version of the Bell-LaPadula model was outlined the previous section. 
In this section, we extend this template with a notion of trust by adding a Boolean flag \emph{trust} to the host attributes. 
This is a refinement to represent real-world scenarios more accurately and analogously happened to the original Bell-LaPadula model~\cite{blphistory}. 
For a host $v$, let $\mdef{level}\ (\nP\ v)$ denote $v$'s security level and $\mdef{trust}\ (\nP\ v)$ whether $v$ is trusted. 
A trusted host can receive information of any security level and may declassify it, \ie distribute the information with its own security level. 
For example, a trusted host is allowed to receive any information and with the $\mdef{unclassified}$ level, it is allowed to reveal it to anyone. 
The template is thus formalized as follows. 
\begin{IEEEeqnarray*}{ll}
  \sinvar\ (V,\, E)\ \nP \ \equiv \ \forall (s, r) \in E. \ \begin{cases} 
  	\mdef{True} & \mctrl{if} \ \ \mdef{trust}\ (\nP\ r) \\ 
  	\mdef{level}\ (\nP\ s) \; \leq \; \mdef{level}\ (\nP\ r) & \mctrl{otherwise}
  	\end{cases}
\end{IEEEeqnarray*}

The default attribute is $\bot = (\mdef{unclassified}, \mdef{untrusted})$,\footnote{interpretation BLPtrusted: SecurityInvariant-IFS } where $\mdef{untrusted}$ simply means $\mdef{False}$. 
It is $\Phi$-structured, where $\Phi = (\lambda c_1\ c_2.\ \ \mctrl{if}\ \mdef{trust}\ c_2 \ \mctrl{then}\ \mdef{True}\ \mctrl{else}\ \mdef{level}\ c_1 \leq \mdef{level}\ c_2 )$. 
As its simplified version, it is an information flow security strategy.

\section{Communication Partners}
\label{sinvar:commpartners}
This security invariant template can be understood as an implementation of Access Control Lists (ACLs). 
Traditional ACLs, for example represented as full access control matrix, scale quadratically in the number of hosts. 
This is hard to manage manually. 
To contain this quadratic configuration effort, this security invariant template was carefully designed to be contained to a subset of all policy entities. 
Three host attributes are defined:
\begin{IEEEeqnarray*}{l C l}
  \Psi & = & \mconstr{DontCare} \quad | \quad \mconstr{Care} \quad | \quad \mconstr{Master}\ \ \mathcal{V}\; \mathit{list}
\end{IEEEeqnarray*}

The host attributes can be understood as follows. 
For all hosts labeled as $\mconstr{DontCare}$, as long as they do not interact with $\mconstr{Master}$ hosts, no restriction is imposed on them. 
They are not part of the security requirement which is formalized by this invariant and correspond to the invariant's default attribute $\bot$. 
Hosts with a $\mconstr{Master}$ attribute are hosts which have an ACL enabled. 
The parameter of the $\mconstr{Master}$ attribute is the ACL, \ie a list of hosts which are allowed to access this host. 
Finally, the $\mconstr{Care}$ attribute tells that the host itself participates in the security requirement which is formalized, but does not have ACL restrictions by itself. 

\begin{example}
\textbf{Example. }%
Hosts labeled with $\mconstr{Master}\ []$ cannot be accessed at all. 

Usually, if a host $u$ is labeled with $\mconstr{Master}\ [ v ]$, then $v$ should be labeled as either $\mconstr{Care}$ or $\mconstr{Master}$. 
This means that $v$ is able to access $u$. 

If $v$ is labeled as $\mconstr{DontCare}$, it cannot access $u$, though it is in $u$'s access list. 
This covers the scenario that $u$ is temporarily disabled but one does not want to update all access lists which may mention $u$. 
For this scenario, $\mconstr{DontCare}$ can be used as a simple Boolean flag which temporarily disables $u$. 
\end{example}

A security requirement which is formalized with the Communication Partners invariant template focuses around $\mconstr{Master}$ hosts. 
Hosts with the $\mconstr{DontCare}$ attribute are completely out of focus for the formalized requirement. 
Hosts with $\mconstr{Care}$ attribute sit in-between: $\mconstr{Master}$ hosts may grant special rights to access them but they are also completely unconstrained on how they interact with the rest of the world, \ie $\mconstr{DontCare}$ hosts.

\begin{example}
\textbf{Example. }%
Let $\mvar{db}_1$ have the attribute $\mconstr{Master}\ [h_1,\, h_2]$ and $\mvar{db}_2$ the attribute $\mconstr{Master}\ [\mvar{db}_1]$.
Let both $h_1$ and $h_2$ have the $\mconstr{Care}$ attribute. 
Let all other hosts have the default attribute of $\mconstr{DontCare}$. 
Then, $h_1$ and $h_2$ can access $\mvar{db}_1$ and can also access all other hosts, excluding $\mvar{db}_2$. 
$\mvar{db}_1$ can access all hosts. 
All hosts which have not been mentioned can freely access each other and $h_1$ and $h_2$. 

\begin{center}
\begin{tikzpicture}
\node (db1) at (-1,2) {$\mvar{db}_1$};
\node (db2) at (1,2) {$\mvar{db}_2$};
\node (h1) at (-2,0) {$\mvar{h}_1$};
\node (h2) at (0,0) {$\mvar{h}_2$};
\node (h3) at (2,0) {$\mvar{h}_3$};

\draw[myptr] (db1) to (db2);
\draw[myptr] (db1) to (h3);
\draw[myptr] (db2) to (h1);
\draw[myptr] (db2) to (h2);
\draw[myptr] (db2) to (h3);

\draw[myptr] (db1) to[loop above] (db1);
\draw[myptr] (db2) to[loop above] (db2);
\draw[myptrdouble] (h1) to (db1);
\draw[myptr] (h1) to[loop below] (h1);
\draw[myptrdouble] (h2) to (db1);
\draw[myptrdouble] (h2) to (h1);
\draw[myptr] (h2) to[loop below] (h2);
\draw[myptrdouble] (h3) to [bend left=45] (h1);
\draw[myptrdouble] (h3) to (h2);
\draw[myptr] (h3) to[loop below] (h3);

\end{tikzpicture}
\end{center}

\end{example}

We formalize the invariant. 
The invariant template does not restrict reflexive accesses, \ie accesses from a host to itself. 
It can be expressed as one of the $\Phi$-structured invariants. 
It is necessary that the predicate (here $\mfun{allowed\mhyphen{}flow} :: \Psi \Rightarrow \mathcal{V} \Rightarrow \Psi \Rightarrow \mathcal{V} \Rightarrow \mathbb{B}$) has access to both the host attribute and the name of the host in order to look up the name of a host in the access control list.

\begin{IEEEeqnarray*}{l C l}
  \sinvar\ (V,\,E)\ \nP & \equiv & {\forall (s,r) \in E.}\ s \neq r \longrightarrow \mfun{allowed\mhyphen{}flow}\ (\nP\ s)\ s\ (\nP\ r)\ r
\end{IEEEeqnarray*}

The following table implements the access control restrictions $\mfun{allowed\mhyphen{}flow}$. 
The first column $\Psi_s$ is the host attribute of the sender, the second column the name of the sender $s$. 
Likewise, the third column is the host attribute of the receiver $\Psi_r$, followed by the name of the receiver $r$. 
Next, the result (rslt), and an explanation are given. 
Whenever the name of the sender, the receiver, or the ACL itself is not important for a rule, it is left out with an underscore `$\_$'. 

\begin{table}[h!tbp]
\centering
\begin{tabular}[0.99\linewidth]{ @{}l@{ \ }l@{ }|l@{ \ }l@{ }|c@{ \ }p{21em} }
	\toprule
	$\Psi_s$ & s & $\Psi_r$  &r & rslt & explanation \\
	\midrule
    $\mconstr{DontCare}$ & \_ & $\mconstr{DontCare}$ & \_ & \cmark & No restrictions \\
    $\mconstr{DontCare}$ & \_ & $\mconstr{Care}$ & \_ & \cmark & No restrictions \\
    $\mconstr{DontCare}$ & \_ & $\mconstr{Master}$ \_ & \_ & \xmark & Hosts which are not part of the formalized security requirement must not access hosts with ACL restrictions \\
    $\mconstr{Care}$ & \_ & $\mconstr{Care}$ & \_ & \cmark & No restrictions \\
    $\mconstr{Care}$ & \_ & $\mconstr{DontCare}$ & \_ & \cmark & No restrictions \\
    $\mconstr{Care}$ & s & $\mconstr{Master}\ M$ & \_ & $s \in M$ & $s$ must be in the ACL $M$ \\
    $\mconstr{Master}$ \_ & s & $\mconstr{Master}\ M$ & \_ & $s \in M$ & -- `` -- \\
    $\mconstr{Master}$ \_ & \_ & $\mconstr{Care}$ & \_ & \cmark & No restrictions \\
    $\mconstr{Master}$ \_ & \_ & $\mconstr{DontCare}$ & \_ & \cmark & No restrictions \\
 \bottomrule%
\end{tabular}%
\end{table}

The use of the $\mconstr{DontCare}$ attribute has an additional positive effect. 
It prevents that stale access list entries have an undesired effect. 

\begin{example}
\textbf{Example. }%
Let $u$ be labeled with $\mconstr{Master}\ [ v ]$. 
Now, host $v$ is removed but the access list at $u$ is not updated because it is expected that $v$ will be re-added again soon. 
However, it was forgotten about $v$ and the stale access list entry at $v$ remains. 
Of course, it is easy to warn about stale access list entries, yet, they may be desirable for a transition period. 
Somewhere in the future, a new host with accidentally the same name $v$ is added. 
Because $v$ is not part of the security requirement formalized  by this invariant, it is set to $\bot = \mconstr{DontCare}$. 
Therefore, though the stale access list entry still exists, $v$ is not granted access to $u$ and the stale access list can thus cause no immediate harm. 
\end{example}

\section{Comm.\ With}
\label{sinvar:commwith}
The ``Communicate With'' security invariant template is an experimental, white-listing, transitive access control list model. 
It is not $\Phi$-structured and we are not aware of an efficient implementation for the offending flows. 
Therefore, it serves primarily for demonstration purposes. 
With the improvements demonstrated in Chapter~\ref{chap:hilbertsoffending}, the invariant can be practically applied but there is no guarantee that a computed policy may be maximum. 
The same holds for all following non-$\Phi$-structured invariant templates. 

The type for the host attributes is a list of hosts, \ie $\Psi = \mathcal{V}\ \mathit{list}$ and represents to the only hosts that may be accessed, even transitively. 
The default attribute is the empty list $\bot = []$. 
The formula for $\sinvar{}$ states that for any host $v$, all nodes $a$ which are reachable from $v$ must be in $v$'s access list. 
\begin{IEEEeqnarray*}{l C l}
\sinvar\ (V,\, E)\ \nP & \ \equiv \ & \forall v \in V. \ \ \left(\forall a \in \mfun{succ\mhyphen{}tran}\ \left(V,\, E\right)\ v.\ \ a \in \nP\ v \right)
\end{IEEEeqnarray*}

Alternatively, the invariant can be expressed as for all edges in the transitive closure over $E$, \ie all transitive accesses, the accessed host $v_2$ must be in the access list of the accessing host $v_1$. 
\begin{IEEEeqnarray*}{l C l}
\sinvar\ (V,\, E)\ \nP & \ \equiv \ & \forall (v_1,v_2) \in E^+.\ \ v_2 \in (\nP\ v_1)
\end{IEEEeqnarray*}
Both definitions are equal\footnote{ACLcommunicateWith-sinvar-alternative}.

\begin{example}
\textbf{Example. }%
For example, assume $V = \lbrace v_1, v_2, v_3 \rbrace$ and let $\nP{}\ v_1 = [v_2,\, v_3]$. 
This means $v_1$ must only access $v_2$ and $v_3$ transitively. 
Let $E = \lbrace (v_1, v_2),\ (v_2, v_3)\rbrace$, then $v_1$ accesses $v_3$ transitively. 
With $v_1$'s host attributes, a direct access $(v_1, v_3)$ would be allowed. 
In order for the invariant to be fulfilled for $E$, additionally, $v_2$ must be allowed to access $v_3$.
\noindent
\begin{center}
\begin{minipage}{0.48\linewidth}\centering
\begin{tikzpicture}
\node (v1) at (0,0) {$\mvar{v}_1$};
\node (v2) at (1,0) {$\mvar{v}_2$};
\node (v3) at (2,0) {$\mvar{v}_3$};

\draw[myptr] (v1) to(v2);
\draw[myptr] (v2) to (v3);
\end{tikzpicture}
\end{minipage}
\end{center}
\medskip

Assume there is another node $v_4$ and there are additional edges from $v_1$ to $v_3$ and from $v_3$ to $v_4$. 
\noindent
\begin{center}
\begin{minipage}{0.48\linewidth}\centering
\begin{tikzpicture}
\node (v1) at (0,0) {$\mvar{v}_1$};
\node (v2) at (1,0) {$\mvar{v}_2$};
\node (v3) at (2,0) {$\mvar{v}_3$};
\node (v4) at (3,0) {$\mvar{v}_4$};

\draw[myptr] (v1) to (v2);
\draw[myptr] (v1) to[bend right=30] (v3);
\draw[myptr] (v2) to (v3);
\draw[myptr] (v3) to (v4);
\end{tikzpicture}
\end{minipage}
\end{center}
Assume $v_1$ must not access $v_4$. 
Then any combination of edges on any path from $v_1$ to $v_4$ which disconnects $v_1$ and $v_4$ may be an offending flow. 
Hence, the offending flows are not uniquely defined. 
The choice that the invariant incorporates transitive accesses results in the fact that the set of offending flows may be even exponential. 
For example, let $E = \lbrace (v_1,v_2),\ (v_1,v_3),\ (v_2,v_3),\ (v_3,v_4) \rbrace$ and let $\nP$ be such that every host can access all other hosts, 
except that $v_1$ must not access $v_4$. 
Then the set of offending flows is $\lbrace\lbrace(v_1, v_2), (v_1, v_3)\rbrace, \lbrace(v_1, v_3), (v_2, v_3)\rbrace, \lbrace(v_3, v_4)\rbrace\rbrace$. 

\begin{minipage}{0.29\linewidth}\centering
\begin{tikzpicture}
\node (v1) at (0,0) {$\mvar{v}_1$};
\node (v2) at (1.5,0) {$\mvar{v}_2$};
\node (v3) at (2.5,0) {$\mvar{v}_3$};
\node (v4) at (3.5,0) {$\mvar{v}_4$};

\draw[myptrdotted] (v1) to (v2);
\draw[myptrdotted] (v1) to[bend right=30] (v3);
\draw[myptr] (v2) to (v3);
\draw[myptr] (v3) to (v4);
\end{tikzpicture}
\end{minipage}
\hspace*{\fill}
\begin{minipage}{0.29\linewidth}\centering
\begin{tikzpicture}
\node (v1) at (0,0) {$\mvar{v}_1$};
\node (v2) at (1,0) {$\mvar{v}_2$};
\node (v3) at (2.5,0) {$\mvar{v}_3$};
\node (v4) at (3.5,0) {$\mvar{v}_4$};

\draw[myptr] (v1) to (v2);
\draw[myptrdotted] (v1) to[bend right=30] (v3);
\draw[myptrdotted] (v2) to (v3);
\draw[myptr] (v3) to (v4);
\end{tikzpicture}
\end{minipage}
\hspace*{\fill}
\begin{minipage}{0.29\linewidth}\centering
\begin{tikzpicture}
\node (v1) at (0,0) {$\mvar{v}_1$};
\node (v2) at (1,0) {$\mvar{v}_2$};
\node (v3) at (2,0) {$\mvar{v}_3$};
\node (v4) at (3.5,0) {$\mvar{v}_4$};

\draw[myptr] (v1) to (v2);
\draw[myptr] (v1) to[bend right=30] (v3);
\draw[myptr] (v2) to (v3);
\draw[myptrdotted] (v3) to (v4);
\end{tikzpicture}
\end{minipage}

\end{example}

\section{Not Comm.\ With}
\label{sinvar:notcommwith}
The ``Not Communicate With'' security invariant template is an experimental, black-listing, transitive access control list model. 
It is not $\Phi$-structured and we are not aware of an efficient implementation for the offending flows. 
Therefore, it serves primarily for demonstration purposes. 
It complements the previously presented ``Communicate With'' template. 

The type for the host attributes is a set of hosts, \ie $\Psi = \mathcal{V}\ \mathit{set}$ and corresponds to the hosts that a host must not transitively access. 
For example, let $\nP{}\ v = \lbrace v_1,\, v_2, \rbrace$, then $v$ must not access $v_1$ and $v_2$ transitively. 
The default attribute is the universe, \ie the set of all elements, $\bot = \mdef{UNIV}$. 
The universe is represented symbolically and executable code can be generated for any (finite or infinite) type $\mathcal{V}$ of nodes. 
Analogously to the previous invariant, the formula states that any hosts which are accessible from some starting host $v$ must not be in $v$'s ``not-access-list''. 
\begin{IEEEeqnarray*}{l C l}
\sinvar\ (V,\, E)\ \nP & \ \equiv \ & \forall v \in V. \ \ \left(\forall a \in \mfun{succ\mhyphen{}tran}\ \left(V,\, E\right)\ v.\ \ a \notin \nP\ v \right)
\end{IEEEeqnarray*}

Technically, this invariant is exactly the inverse of the previous invariant.\footnote{ACLcommunicateNotWith-inverse-ACLcommunicateWith} 
If all host attributes are inversed, \ie $\forall v.\ \nP_{\mathrm{new}}\ v = \mdef{UNIV} \setminus (\nP\ v)$, then the two invariants are equal. 
All insights about the ``Communicate With'' template also apply to this template. 
The main difference it that in this invariant, $\Psi$ is a set to represent $\mdef{UNIV}$, which is not possible in the previous invariant.


\section{Dependability}
\label{sinvar:dependability}
Hosts provide a service on which other hosts might depend. 
The ``Dependability'' invariant template was designed to limit a network's dependence on a service. 
Note that we do not model availability; this access control model describes a limit of how many hosts may at most access one specific host. 
Every host is assigned a dependability level, represented as natural number: $\Psi = \mathbb{N}$. 
This number encodes the maximal number of hosts which may transitively depend on a host. 

Let $\mdef{card}$ be the cardinality of a set. 
Then, the invariant can be expressed as follows: 
\begin{IEEEeqnarray*}{l C l}
\sinvar\ (V,\, E)\ \nP & \ \equiv \ & (\forall v \in V.\ \ \mdef{card}\ \left(\mfun{succ\mhyphen{}tran}\ (V,\, E)\ v\right) \leq (\nP\ v))
\end{IEEEeqnarray*}

The template can also be expressed as a predicate over all edges. 
Note that it is not $\Phi$-structured since the predicate depends on the graph. 
\begin{IEEEeqnarray*}{l C l}
\sinvar\ (V,\, E)\ \nP & \ \equiv \ & (\forall (v_1,v_2) \in E.\ \ \mdef{card}\ \left(\mfun{succ\mhyphen{}tran}\ (V,\, E)\ v_1\right) \leq (\nP\ v_1))
\end{IEEEeqnarray*}

The dependability level corresponds to the number of hosts a host may transitively access. 
Hence, it is an access control strategy. 
By default, a host does not have any access rights: $\bot = 0$. 

A high dependability level of a host may represent that many other hosts depend on the service provided by this hosts. 
Hence, high numbers are indicators for important hosts which may require special care.

The invariant is hard to use for the following two reasons. 
First, the default dependability level means that hosts may not communicate at all. 
Consequently, every host needs to be assigned a dependability level manually. 
Second, due to the use of $\mfun{succ\mhyphen{}tran}$, in case of a violation, the number of offending flows may be exponential. 
Therefore, we propose the following use of the invariant. 
The dependability levels can be set automatically to a valid host attribute configuration and the invariant should be excluded from automated policy construction. 
Setting the level for each host $v$ to $\mdef{card}\ \left(\mfun{succ\mhyphen{}tran}\ (V,\, E)\ v\right)$ provides a valid configuration\footnote{dependability-fix-nP-impl-correct}. 
This avoids manual configuration and makes sure that the offending flows are always empty. 
Whenever there is an update to the policy, the dependability levels can be recomputed. 
If a significant change in the levels due to the update has occurred, this raises a pointer for further manual inspection. 
This may uncover unintended side-effects of a policy change.

\begin{example}
\textbf{Example. }%
Assume $V = \lbrace v_1, v_2, v_3, v_4, v_5, v_6, v_7 \rbrace$ and 
$E = \lbrace (v_1,v_2),\allowbreak (v_2,v_1),\allowbreak (v_2,v_3),\allowbreak (v_4,v_5),\allowbreak (v_5,v_6),\allowbreak (v_6,v_7) \rbrace$. 
The nodes $v_1$, $v_2$, $v_3$ are disconnected from the other part of the graph. 
Below, the dependability levels are visualized for every node beneath its name. 
Noteworthy, $v_1$ needs a dependability level of at least $3$ because it can access $v_2$, $v_3$, and can also transitively access itself. 

\begin{minipage}{0.98\linewidth}\centering
\begin{tikzpicture}
\tikzset{every node/.style = {draw, circle split, align=center, inner sep=1, outer sep=0}}

\node (v1) at (0,0) {$\mvar{v}_1$\nodepart{lower}$3$};
\node (v2) at (1.5,0) {$\mvar{v}_2$\nodepart{lower}$3$};
\node (v3) at (3,0) {$\mvar{v}_3$\nodepart{lower}$0$};
\node (v4) at (5,0) {$\mvar{v}_4$\nodepart{lower}$3$};
\node (v5) at (6.5,0) {$\mvar{v}_5$\nodepart{lower}$2$};
\node (v6) at (8,0) {$\mvar{v}_6$\nodepart{lower}$1$};
\node (v7) at (9.5,0) {$\mvar{v}_7$\nodepart{lower}$0$};

\draw[myptrdouble] (v1) to (v2);
\draw[myptr] (v2) to (v3);

\draw[myptr] (v4) to (v5);
\draw[myptr] (v5) to (v6);
\draw[myptr] (v6) to (v7);
\end{tikzpicture}
\end{minipage}

Now assume we connect $v_3$ and $v_4$. 
Due to $v_3$'s dependability level, this only causes one offending flow, namely exactly the flow we added. 

\begin{minipage}{0.98\linewidth}\centering
\begin{tikzpicture}
\tikzset{every node/.style = {draw, circle split, align=center, inner sep=1, outer sep=0}}

\node (v1) at (0,0) {$\mvar{v}_1$\nodepart{lower}$3$};
\node (v2) at (1.5,0) {$\mvar{v}_2$\nodepart{lower}$3$};
\node (v3) at (3,0) {$\mvar{v}_3$\nodepart{lower}$0$};
\node (v4) at (5,0) {$\mvar{v}_4$\nodepart{lower}$3$};
\node (v5) at (6.5,0) {$\mvar{v}_5$\nodepart{lower}$2$};
\node (v6) at (8,0) {$\mvar{v}_6$\nodepart{lower}$1$};
\node (v7) at (9.5,0) {$\mvar{v}_7$\nodepart{lower}$0$};

\draw[myptrdouble] (v1) to (v2);
\draw[myptr] (v2) to (v3);

\draw[myptrdotted] (v3) to (v4);

\draw[myptr] (v4) to (v5);
\draw[myptr] (v5) to (v6);
\draw[myptr] (v6) to (v7);
\end{tikzpicture}
\end{minipage}

An approach to fix the violation by increasing $v_3$'s dependability level fails. 
Due to the use of $\mfun{succ\mhyphen{}tran}$, there are multiple possibilities for the offending flows. 
The following offending flows are created by this: 

\begin{minipage}{0.98\linewidth}\centering
\begin{tikzpicture}
\tikzset{every node/.style = {draw, circle split, align=center, inner sep=1, outer sep=0}}

\node (v1) at (0,0) {$\mvar{v}_1$\nodepart{lower}$3$};
\node (v2) at (1.5,0) {$\mvar{v}_2$\nodepart{lower}$3$};
\node (v3) at (3,0) {$\mvar{v}_3$\nodepart{lower}$2$};
\node (v4) at (5,0) {$\mvar{v}_4$\nodepart{lower}$3$};
\node (v5) at (6.5,0) {$\mvar{v}_5$\nodepart{lower}$2$};
\node (v6) at (8,0) {$\mvar{v}_6$\nodepart{lower}$1$};
\node (v7) at (9.5,0) {$\mvar{v}_7$\nodepart{lower}$0$};

\draw[myptrdouble] (v1) to (v2);
\draw[myptr] (v2) to (v3);
\draw[myptrdotted] (v3) to (v4);
\draw[myptr] (v4) to (v5);
\draw[myptr] (v5) to (v6);
\draw[myptr] (v6) to (v7);
\end{tikzpicture}
\end{minipage}

\begin{minipage}{0.98\linewidth}\centering
\begin{tikzpicture}
\tikzset{every node/.style = {draw, circle split, align=center, inner sep=1, outer sep=0}}

\node (v1) at (0,0) {$\mvar{v}_1$\nodepart{lower}$3$};
\node (v2) at (1.5,0) {$\mvar{v}_2$\nodepart{lower}$3$};
\node (v3) at (3,0) {$\mvar{v}_3$\nodepart{lower}$2$};
\node (v4) at (5,0) {$\mvar{v}_4$\nodepart{lower}$3$};
\node (v5) at (6.5,0) {$\mvar{v}_5$\nodepart{lower}$2$};
\node (v6) at (8.5,0) {$\mvar{v}_6$\nodepart{lower}$1$};
\node (v7) at (10,0) {$\mvar{v}_7$\nodepart{lower}$0$};

\draw[myptrdotted] (v1) to[bend left=30] (v2);
\draw[myptrdotted] (v2) to[bend left=30] (v1);
\draw[myptr] (v2) to (v3);
\draw[myptr] (v3) to (v4);
\draw[myptr] (v4) to (v5);
\draw[myptrdotted] (v5) to (v6);
\draw[myptr] (v6) to (v7);
\end{tikzpicture}
\end{minipage}

\begin{minipage}{0.98\linewidth}\centering
\begin{tikzpicture}
\tikzset{every node/.style = {draw, circle split, align=center, inner sep=1, outer sep=0}}

\node (v1) at (0,0) {$\mvar{v}_1$\nodepart{lower}$3$};
\node (v2) at (1.5,0) {$\mvar{v}_2$\nodepart{lower}$3$};
\node (v3) at (3,0) {$\mvar{v}_3$\nodepart{lower}$2$};
\node (v4) at (5,0) {$\mvar{v}_4$\nodepart{lower}$3$};
\node (v5) at (7,0) {$\mvar{v}_5$\nodepart{lower}$2$};
\node (v6) at (8.5,0) {$\mvar{v}_6$\nodepart{lower}$1$};
\node (v7) at (10,0) {$\mvar{v}_7$\nodepart{lower}$0$};

\draw[myptrdotted] (v1) to[bend left=30] (v2);
\draw[myptr] (v2) to[bend left=30] (v1);
\draw[myptr] (v2) to (v3);
\draw[myptr] (v3) to (v4);
\draw[myptrdotted] (v4) to (v5);
\draw[myptr] (v5) to (v6);
\draw[myptr] (v6) to (v7);
\end{tikzpicture}
\end{minipage}

\begin{minipage}{0.98\linewidth}\centering
\begin{tikzpicture}
\tikzset{every node/.style = {draw, circle split, align=center, inner sep=1, outer sep=0}}

\node (v1) at (0,0) {$\mvar{v}_1$\nodepart{lower}$3$};
\node (v2) at (1.5,0) {$\mvar{v}_2$\nodepart{lower}$3$};
\node (v3) at (3,0) {$\mvar{v}_3$\nodepart{lower}$2$};
\node (v4) at (5,0) {$\mvar{v}_4$\nodepart{lower}$3$};
\node (v5) at (7,0) {$\mvar{v}_5$\nodepart{lower}$2$};
\node (v6) at (8.5,0) {$\mvar{v}_6$\nodepart{lower}$1$};
\node (v7) at (10,0) {$\mvar{v}_7$\nodepart{lower}$0$};

\draw[myptr] (v1) to[bend left=30] (v2);
\draw[myptrdotted] (v2) to[bend left=30] (v1);
\draw[myptr] (v2) to (v3);
\draw[myptr] (v3) to (v4);
\draw[myptrdotted] (v4) to (v5);
\draw[myptr] (v5) to (v6);
\draw[myptr] (v6) to (v7);
\end{tikzpicture}
\end{minipage}

\begin{minipage}{0.98\linewidth}\centering
\begin{tikzpicture}
\tikzset{every node/.style = {draw, circle split, align=center, inner sep=1, outer sep=0}}

\node (v1) at (0,0) {$\mvar{v}_1$\nodepart{lower}$3$};
\node (v2) at (1.5,0) {$\mvar{v}_2$\nodepart{lower}$3$};
\node (v3) at (3.5,0) {$\mvar{v}_3$\nodepart{lower}$2$};
\node (v4) at (5.5,0) {$\mvar{v}_4$\nodepart{lower}$3$};
\node (v5) at (7.5,0) {$\mvar{v}_5$\nodepart{lower}$2$};
\node (v6) at (9,0) {$\mvar{v}_6$\nodepart{lower}$1$};
\node (v7) at (10.5,0) {$\mvar{v}_7$\nodepart{lower}$0$};

\draw[myptrdouble] (v1) to (v2);
\draw[myptrdotted] (v2) to (v3);
\draw[myptr] (v3) to (v4);
\draw[myptrdotted] (v4) to (v5);
\draw[myptr] (v5) to (v6);
\draw[myptr] (v6) to (v7);
\end{tikzpicture}
\end{minipage}

\begin{minipage}{0.98\linewidth}\centering
\begin{tikzpicture}
\tikzset{every node/.style = {draw, circle split, align=center, inner sep=1, outer sep=0}}

\node (v1) at (0,0) {$\mvar{v}_1$\nodepart{lower}$3$};
\node (v2) at (1.5,0) {$\mvar{v}_2$\nodepart{lower}$3$};
\node (v3) at (3.5,0) {$\mvar{v}_3$\nodepart{lower}$2$};
\node (v4) at (5.5,0) {$\mvar{v}_4$\nodepart{lower}$3$};
\node (v5) at (7,0) {$\mvar{v}_5$\nodepart{lower}$2$};
\node (v6) at (9,0) {$\mvar{v}_6$\nodepart{lower}$1$};
\node (v7) at (10.5,0) {$\mvar{v}_7$\nodepart{lower}$0$};

\draw[myptrdouble] (v1) to (v2);
\draw[myptrdotted] (v2) to (v3);
\draw[myptr] (v3) to (v4);
\draw[myptr] (v4) to (v5);
\draw[myptrdotted] (v5) to (v6);
\draw[myptr] (v6) to (v7);
\end{tikzpicture}
\end{minipage}

Calculating new dependability levels yields the following results. 
Though only the edge $(v_3, v_4)$ was added, the vast increase in the level at for example $v_1$ indicates that this simple additionally permitted flow has large impact on other areas of the policy. 
\begin{minipage}{0.98\linewidth}\centering
\begin{tikzpicture}
\tikzset{every node/.style = {draw, circle split, align=center, inner sep=1, outer sep=0}}

\node (v1) at (0,0) {$\mvar{v}_1$\nodepart{lower}$7$};
\node (v2) at (1.5,0) {$\mvar{v}_2$\nodepart{lower}$7$};
\node (v3) at (3,0) {$\mvar{v}_3$\nodepart{lower}$4$};
\node (v4) at (5,0) {$\mvar{v}_4$\nodepart{lower}$3$};
\node (v5) at (6.5,0) {$\mvar{v}_5$\nodepart{lower}$2$};
\node (v6) at (8,0) {$\mvar{v}_6$\nodepart{lower}$1$};
\node (v7) at (9.5,0) {$\mvar{v}_7$\nodepart{lower}$0$};

\draw[myptrdouble] (v1) to (v2);
\draw[myptr] (v2) to (v3);
\draw[myptr] (v3) to (v4);
\draw[myptr] (v4) to (v5);
\draw[myptr] (v5) to (v6);
\draw[myptr] (v6) to (v7);
\end{tikzpicture}
\end{minipage}

\end{example}

\paragraph*{Dependability Non-Refl}
As the previous example may have hinted, it may be unexpected that a host can depend on itself.
The invariant template can also be expressed to exclude accesses from a host to itself. 
\begin{IEEEeqnarray*}{l C l}
\sinvar\ (V,\, E)\ \nP & \ \equiv \ & (\forall v \in V.\ \ \mdef{card}\ \left(\left(\mfun{succ\mhyphen{}tran}\ (V,\, E)\ v\right) \setminus \lbrace v \rbrace \right) \leq (\nP\ v))
\end{IEEEeqnarray*}

Apart from reflexive accesses, this version of the dependability templates behaves as the original version. 

\begin{example}
\textbf{Example. }%
The default Dependability invariant also counts an access from a node to itself. 
Hence, the host in the illustration below (left) needs a dependability level of at least one. 
With the version of the Dependability invariant which excludes reflexive accesses, the node can have the default dependability level (illustrated below, right). 
\begin{minipage}{0.48\linewidth}\centering
\begin{tikzpicture}
\tikzset{every node/.style = {draw, circle split, align=center, inner sep=1, outer sep=0}}
\node (v) at (0,0) {$\mvar{v}$\nodepart{lower}$1$};
\draw[myptr] (v) to[loop above] (v);
\end{tikzpicture}
\end{minipage}
\hspace*{\fill}
\begin{minipage}{0.48\linewidth}\centering
\begin{tikzpicture}
\tikzset{every node/.style = {draw, circle split, align=center, inner sep=1, outer sep=0}}
\node (v) at (0,0) {$\mvar{v}$\nodepart{lower}$0$};
\draw[myptr] (v) to[loop above] (v);
\end{tikzpicture}
\end{minipage}
\end{example}

\section{Domain Hierarchy}
\label{sinvar:domhierarchy}
The domain hierarchy template mirrors hierarchical access control structures. 
\begin{figure}[htb]
  \centering
  		\includegraphics[width=0.65\linewidth]{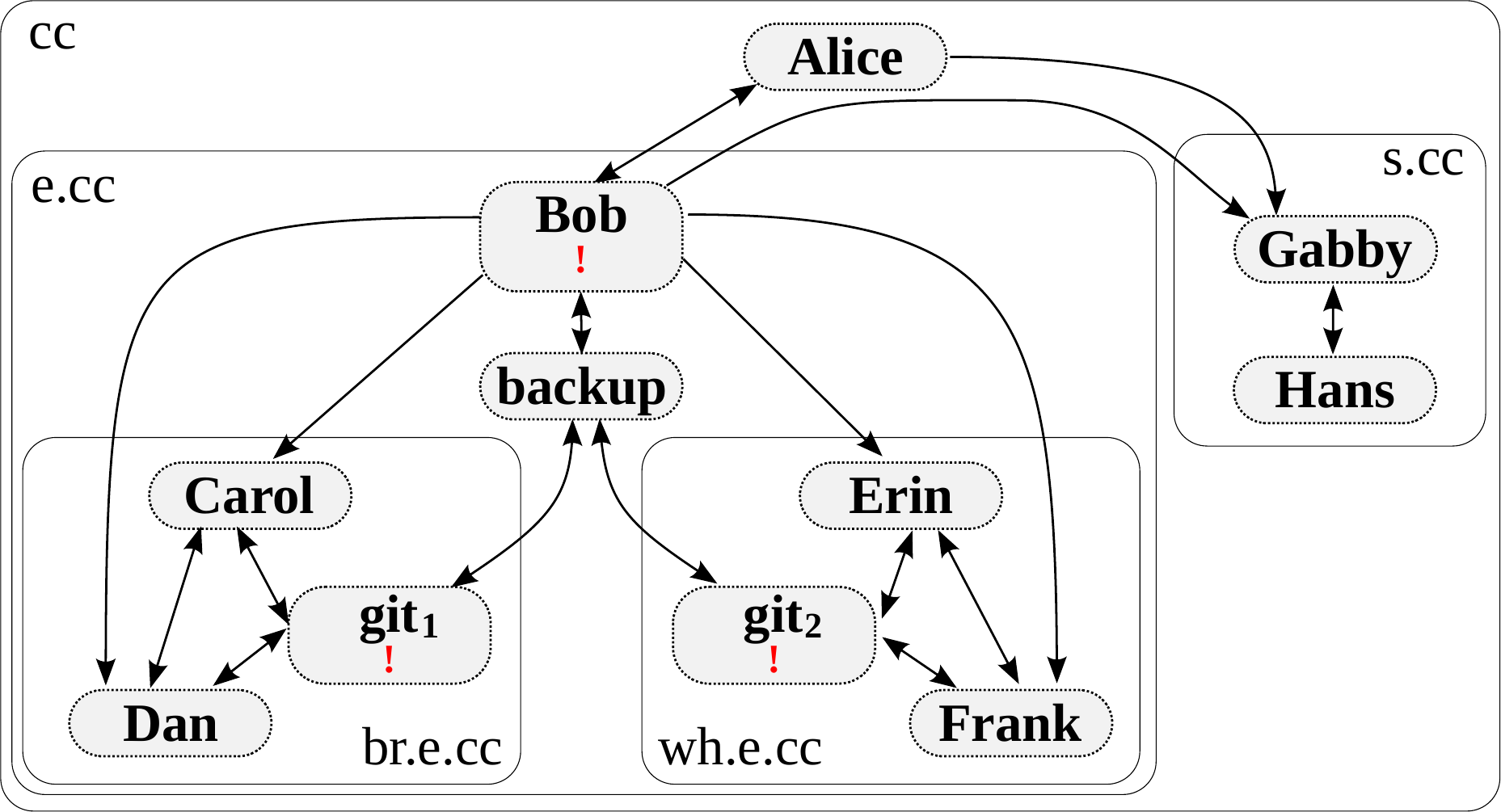}
		\caption{Policy and host mapping of \textbf{cc}.}
		\label{fig:domhierarchynetworkcc}
%
%
%
\end{figure}

\begin{figure}[htb]
  \centering
	\begin{Large}
	\begin{tikzpicture}
	\node at (0,0) {$\textnormal{\textbf{cc}}$};
	\node[rotate=45,anchor=north]  at (-0.75,-0.2) {$\sqsubseteq$};
	\node[rotate=-45,anchor=north] at (+0.75,-0.2) {$\sqsupseteq$};
	\node at (-1,-0.8) {$\textnormal{\textbf{e}}$};
	\node at (1,-0.8) {$\textnormal{\textbf{s}}$};
	\node[rotate=45,anchor=north]  at (-1.7,-1.0) {$\sqsubseteq$};
	\node[rotate=-45,anchor=north] at (-0.2,-1.0) {$\sqsupseteq$};
	\node at (-2,-1.6) {$\textnormal{\textbf{br}}$};
	\node at (0,-1.6) {$\textnormal{\textbf{wh}}$};
	\end{tikzpicture}
	\end{Large}
  		\caption{Organizational structure of \textbf{cc}.}
  		\label{fig:domain_hierarchy_example_cc}
\end{figure}
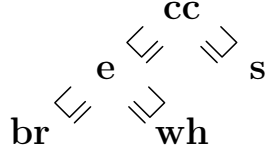

It is best introduced by example. 
The tiny car company (\textit{cc}) consists of the two sub-departments engineering (\textit{e}) and sales (\textit{s}). 
The engineering department itself consists of the brakes (\textit{br}) and the wheels (\textit{wh}) department. 
This tree-like organizational structure is illustrated in Figure~\ref{fig:domain_hierarchy_example_cc}. 
We denote a position by the fully qualified domain name, 
\eg \textit{wh.e.cc} uniquely identifies the wheels department. 
Let `$\sqsubseteq$' denote the `\emph{is below or at the same hierarchy level}' relation, \eg $\mathit{wh}.e.\mathit{cc} \sqsubseteq \mathit{wh}.e.\mathit{cc}$, $\mathit{wh}.e.\mathit{cc} \sqsubseteq e.\mathit{cc}$, and $\mathit{wh}.e.\mathit{cc} \sqsubseteq \mathit{cc}$. 
However, $ \mathit{wh}.e.\mathit{cc} \not\sqsubseteq \mathit{br}.e.\mathit{cc}$ and $\mathit{br}.e.\mathit{cc} \not\sqsubseteq \mathit{wh}.e.\mathit{cc}$. 
The `$\sqsubseteq$' relation denotes a partial order\footnote{instantiation domainNameDept :: order}. 
The company's command structures are strictly hierarchical, \ie commands are either exchanged in the same department or travel from higher departments to their sub-departments. 
Formally, the receiver's level $\sqsubseteq$ sender's level. 
%
%
%
For a host $v$, let $\mdef{level}\ (\nP\ v)$ map to the fully qualified domain name of $v$'s department. 
For example in Figure~\ref{fig:domhierarchynetworkcc}, $\mdef{level}\ (\nP\ \mathit{Bob}) = e.\mathit{cc}$. 

As in many real-world applications of a mathematical model, exceptions exist. 
Those are depicted by exclamation marks in Figure~\ref{fig:domhierarchynetworkcc}. 
For example, Bob as head of engineering is in a trusted position. 
This means he can operate as if he were in the position of Alice. 
This implies that he can communicate on par with Alice, which also implies that he might send commands to the sales department. 
We model such exceptions by assigning each host a trust level. 
This trust level specifies up to which position in the hierarchy this host may act. 
For example, Bob in \textit{e.cc} with a trust level of $1$ can act as if he were in \textit{cc}, which means he has the same command power as Alice. 
Let $\mdef{trust}\ (\nP{}\ v)$ map to $v$'s trust level. 
We define a function $\mdef{chop}$ which takes two parameters: a domain name called $\mvar{level}$ and a natural number called $\mvar{trust}$. 
It returns the domain name $\mvar{level}$ with $\mathit{trust}$ sub-domains chopped off. 
For example, $\mfun{chop}\ \mathit{br.e.cc}\ 1 = \mathit{e.cc}$ and $\mathit{chop}\ \mathit{br.e.cc}\ 2 = \mathit{cc}$. 
With this, the security invariant template can be formalized as follows. 
%
\begin{IEEEeqnarray*}{l C l}
\sinvar\ (V,\, E)\ \nP \ & \ \equiv \ & \forall (s, r) \in E. \ \ \mdef{level}\ (\nP\ r) \; \sqsubseteq \; \mfun{chop}\ \bigl((\mdef{level}\ (\nP\ s))\ (\mdef{trust}\ (\nP\ s))\bigr)
\end{IEEEeqnarray*}

In the Domain Hierarchy, the default attribute is\footnote{interpretation DomainHierarchyNG: SecurityInvariant-ACS} a special value $\bot$ with a trust of zero and which is at the lowest point in the hierarchy, \ie $\forall\; l.\ \bot \sqsubseteq l$. 
Finally, it is worth mentioning that the $\sqsubseteq$-relation forms a lattice\footnote{instantiation domainName :: lattice}, which is a desirable structure for security classes~\cite{denning1976lattice}.

\section{NoRefl}
\label{sinvar:norefl}
Entities in a policy, called hosts throughout this thesis, may correspond to several network entities. 
For example, what occurs as a single host in the policy may correspond in the implementation to a complete network subnet or another group of physical or virtual hosts. 
As long as there is a one-to-one mapping between policy entities and entities in the implementation, reflexive policy rules (\ie rules of the form $(v, v)$) correspond to in-host communication and may be disregarded from a network point of view. 
However, if a host in the policy corresponds to a group of hosts in the implementation, a reflexive policy rule dictates for the implementation whether the hosts in the group may communicate among themselves. 
Details will be discussed in Section~\ref{sec:reflexiverulesdeplay}.

\begin{example}
\textbf{Example. }%
In the illustration below, let $v_1$ and $v_2$ be nodes in the security policy. 
In the network, let $v_1$ and $v_2$ correspond to 3 physical machines each. 
The reflexive policy rule $(v_2, v_2)$ decides that the machines which correspond to $v_2$ in the policy (illustrated on the right) can communicate with each other, while the machines corresponding to $v_1$ cannot.  


\begin{minipage}{0.48\linewidth}\centering
\begin{tikzpicture}
\node[anchor=west] (hdr1) at (-1.5,1) {Policy:};
\node (v1) at (1,0) {$\mvar{v}_1$};

\node[anchor=west] (hdr2) at (-1.5,-1) {Implementation:};
\node[server] (a1) at (0,-3.5) {$\mvar{a}_1$};
\node[server] (a2) at (2,-3.5) {$\mvar{a}_2$};
\node[server] (a3) at (1,-2) {$\mvar{a}_3$};
\end{tikzpicture}
\end{minipage}
\hspace*{\fill}
\begin{minipage}{0.48\linewidth}\centering
\begin{tikzpicture}
\node[anchor=west] (hdr1) at (-1.5,1) {Policy:};
\node (v2) at (1,0) {$\mvar{v}_2$};

\draw[myptr] (v2) to[loop above] (v2);

\node[anchor=west] (hdr2) at (-1.5,-1) {Implementation:};
\node[server] (b1) at (0,-3.5) {$\mvar{b}_1$};
\node[server] (b2) at (2,-3.5) {$\mvar{b}_2$};
\node[server] (b3) at (1,-2) {$\mvar{b}_3$};

\draw[myptrdouble] (b1) to (b2);
\draw[myptrdouble] (b2) to (b3);
\draw[myptrdouble] (b1) to (b3);
\end{tikzpicture}
\end{minipage}
\bigskip
\bigskip
\end{example}

Effectively, hosts in the policy can be lifted to role descriptions by having them represent several physical hosts. 
A host in the policy can now be understood as a role. 
In a physical network implementation, many physical hosts may have the same role. 
They inherit the access rights of their role. 
Reflexive edges in the policy hence encode whether hosts of the same role may communicate with each other. 

Note that this understanding of roles does not correspond to RBAC~\cite{rbac}. 
With our simple lifting, physical hosts cannot inherit multiple roles or switch them dynamically, but roles can have several host attributes. 


The decision whether reflexive flows are allowed can be encoded with the following security invariant template. 
We distinguish between two host attributes. 
Either reflexive flows are allowed or they are denied: $\Psi = \lbrace \mconstr{Refl},\ \mconstr{NoRefl} \rbrace$. 
By default, reflexive flows are denied, $\bot = \mconstr{NoRefl}$. 
Then, the security invariant template can be formalized as follows. 
For all reflexive flows, the corresponding host must have the $\mconstr{Refl}$ attribute. 

\begin{IEEEeqnarray*}{l C l}
\sinvar\ (V,\, E)\ \nP &\ \equiv \ & \forall (s,r) \in E.\ \ s = r \longrightarrow \nP\ s = \mconstr{Refl}
\end{IEEEeqnarray*}

Noteworthy, this invariant can be interpreted as both, ACS and IFS strategy.\footnote{NoRefl-SecurityInvariant-IFS} 
Recalling that the strategy was defined by whether the offending host is the sender or the receiver (Definition~\ref{def:offenders}), and for this special invariant where sender and receiver are equal, the peculiarity becomes obvious. 
We chose it to be an access control strategy. 
This will have the additional advantage, later when translating to a stateful policy, that the requirements for access control security strategies are less strict (cf.\ Chapter~\ref{chap:esss14}), which makes the resulting stateful policy more permissive and hence easier to implement in real-world scenarios.

\section{NonInterference}
\label{sinvar:noninterference}
Parts of the following paragraph about the non-interference model in the programming languages literature have been previously published in the author's master's thesis~\cite{cornythesis}. 

\paragraph*{History of NonInterference for IFS in Programming Languages}
Information flow strategies define valid information channels between subjects~\cite{eckert2013}.
It originated from programming language research. The concept of \emph{Information Flow Security} (IFS) is to control the flow of information within programs. The original idea behind IFS is that a programmer labels confidential and non-confidential information. The compiler subsequently verifies that no information flow from confidential to non-confidential channels occurs. The confidential information must not interfere with non-confidential information. 
Originally, non-interference was introduced by Goguen and Meseguer~\cite{goguen1982} describing that a subject's possible actions must be invisible to another subject to fulfill non-interference. However, this model was recognized incorrect and several efforts were made to correct the model~\cite{csl-92-2}. A satisfactory model finally was constructed featuring the ``complete sequence of actions performed subsequent to a given action''~\cite{csl-92-2}. The now prevalent~\cite{Broberg:2009:FSD:1554339.1554352} IFS model can be described as follows: An observer with a low security level observes the public output (the output the observer can legally obtain with her security level) of a run of a system. When the same observer now observes a similar run, which differs to the first run only in actions of higher security levels, the observer must observe exactly the same output. Intuitively, this model compares any two runs of a program which only differ in their secret input and states for the program to be secure, the observable public output must be equal. 
This model has the downside that it cannot publish any computation result based on private data. Consider for example a login function which checks whether a user-supplied password matches the user's deposited password. We expect that a system can tell any unauthenticated user that her login attempt failed. It is thus necessary that the return value of the login function is public information. As the function is working with the private database of user passwords, the return value of this function is yet private. This renders IFS unsuitable or impractical for many applications. Another scenario is a statistical computation which aggregates many private user records and performs an anonymous statistical aggregation which can be published without any privacy concerns. However, this scenario can also not be modeled by traditional IFS. Myers and Liskov~\cite{Myers:1997:DMI:269005.266669} address these issues by presenting a new IFS model. The model is called decentralized as each user may impose individual restrictions on her data. In their model, users can declassify data -- that is, removing restrictions -- only if they are or indirectly act for the owner of the data. The authors admit that their model  ``does not work well in large, networked systems, where varying levels of trust exist among nodes in the network''~\cite{Myers:1997:DMI:269005.266669}. 
As the presented models focus mainly on actions which are invoked on data, Broberg and Sands~\cite{Broberg:2009:FSD:1554339.1554352} describe these models as flow insensitive. They present a flow sensitive technique (flow locks) and introduce a new representation of an attacker's knowledge.

\bigskip

\paragraph*{Our Security Invariant Template}
Our NonInterference invariant template for computer networks formalizes the requirement that two hosts must not interact with each other. 
It is a very strict information flow security strategy. 
There are only two possible host attributes: $\Psi = \lbrace \mconstr{Interfering}, \mconstr{Unrelated} \rbrace$. 
All hosts which are marked as $\mconstr{Interfering}$ must not be able to interfere with each other, by any means. 
The invariant is very strict and by default, all hosts are assumed as $\bot = \mconstr{Interfering}$. 
It is not $\Phi$-structured and we are not aware of an efficient implementation for the offending flows. 

For a starting node $v$, and a graph $G$, we define all nodes which are reachable from $v$, assuming all edges in the graph were bidirectional. 
\begin{definition}[reachable hosts in an undirected graph]
\begin{IEEEeqnarray*}{lcl}
\mfun{undirected\mhyphen{}reachable}\ (V,\, E)\ v & \ \equiv \ & \left( \mfun{succ\mhyphen{}tran}\ (V,\, E \cup \lbrace (r,s)  \mid  (s,r) \in E \rbrace)\ v\right) \setminus \lbrace v \rbrace
\end{IEEEeqnarray*}
\end{definition}

The definition excludes $v$ from the result, otherwise, the invariant can never be fulfilled because any interfering host would interfere with itself. 
With this, the invariant can be formalized as follows: For all $\mconstr{Interfering}$ hosts, they must only be able to reach hosts which are not $\mconstr{Interfering}$. 

\begin{IEEEeqnarray*}{l C l}
\sinvar\ (V,\, E)\ \nP \ & \ \equiv \ & \forall v \in \lbrace v' \in V  \mid  \nP\ v' = \mconstr{Interfering} \rbrace. \ \ \\
  & \IEEEeqnarraymulticol{2}{r}{\qquad\qquad\qquad \lbrace \nP\ v'  \mid  v' \in \mfun{undirected\mhyphen{}reachable}\ G\ v \rbrace \subseteq \lbrace \mconstr{Unrelated} \rbrace}
\end{IEEEeqnarray*}

\begin{example}
\textbf{Example. }%
For example, assume $V = \lbrace v_1, v_2, v_3, v_4 \rbrace$ and let $E = \lbrace (v_1,v_2),\allowbreak{} (v_1,v_3),\allowbreak{} (v_2,v_3),\allowbreak{} (v_3,v_4) \rbrace$. 
Assume both $v_1$ and $v_4$ are $\mconstr{Interfering}$ and $v_2$ and $v_3$ are $\mconstr{Unrelated}$.  
This setting is similar to the example of ``Comm.\ With'' in Section~\ref{sinvar:commwith}: Any combination of edges on any path from $v_1$ to $v_4$ which disconnects $v_1$ and $v_4$ may be an offending flow. 
The set of offending flows is $\lbrace\lbrace(v_1, v_2), (v_1, v_3)\rbrace, \lbrace(v_1, v_3), (v_2, v_3)\rbrace, \lbrace(v_3, v_4)\rbrace\rbrace$. 

\begin{minipage}{0.29\linewidth}\centering
\begin{tikzpicture}
\node (v1) at (0,0) {$\mvar{v}_1$};
\node (v2) at (1.5,0) {$\mvar{v}_2$};
\node (v3) at (2.5,0) {$\mvar{v}_3$};
\node (v4) at (3.5,0) {$\mvar{v}_4$};

\draw[myptrdotted] (v1) to (v2);
\draw[myptrdotted] (v1) to[bend right=30] (v3);
\draw[myptr] (v2) to (v3);
\draw[myptr] (v3) to (v4);
\end{tikzpicture}
\end{minipage}
\hspace*{\fill}
\begin{minipage}{0.29\linewidth}\centering
\begin{tikzpicture}
\node (v1) at (0,0) {$\mvar{v}_1$};
\node (v2) at (1,0) {$\mvar{v}_2$};
\node (v3) at (2.5,0) {$\mvar{v}_3$};
\node (v4) at (3.5,0) {$\mvar{v}_4$};

\draw[myptr] (v1) to (v2);
\draw[myptrdotted] (v1) to[bend right=30] (v3);
\draw[myptrdotted] (v2) to (v3);
\draw[myptr] (v3) to (v4);
\end{tikzpicture}
\end{minipage}
\hspace*{\fill}
\begin{minipage}{0.29\linewidth}\centering
\begin{tikzpicture}
\node (v1) at (0,0) {$\mvar{v}_1$};
\node (v2) at (1,0) {$\mvar{v}_2$};
\node (v3) at (2,0) {$\mvar{v}_3$};
\node (v4) at (3.5,0) {$\mvar{v}_4$};

\draw[myptr] (v1) to (v2);
\draw[myptr] (v1) to[bend right=30] (v3);
\draw[myptr] (v2) to (v3);
\draw[myptrdotted] (v3) to (v4);
\end{tikzpicture}
\end{minipage}

However, NonInterference is a stricter invariant than ``Comm.\ With''. 
If direction of the flow $(v_3, v_4)$ is reversed, the new policy can be visualized as follows. 
The ``Comm.\ With'' (example in Section~\ref{sinvar:commwith}) invariant is fulfilled for the new policy.

\begin{minipage}{0.98\linewidth}\centering
\begin{tikzpicture}
\node (v1) at (0,0) {$\mvar{v}_1$};
\node (v2) at (1,0) {$\mvar{v}_2$};
\node (v3) at (2,0) {$\mvar{v}_3$};
\node (v4) at (3,0) {$\mvar{v}_4$};

\draw[myptr] (v1) to (v2);
\draw[myptr] (v1) to[bend right=30] (v3);
\draw[myptr] (v2) to (v3);
\draw[myptr] (v4) to (v3);
\end{tikzpicture}
\end{minipage}

However, NonInterference interprets the graph as if it were undirected and still concludes that $v_1$ and $v_4$ are interfering. 
The offending flows are illustrated below. 

\begin{minipage}{0.29\linewidth}\centering
\begin{tikzpicture}
\node (v1) at (0,0) {$\mvar{v}_1$};
\node (v2) at (1.5,0) {$\mvar{v}_2$};
\node (v3) at (2.5,0) {$\mvar{v}_3$};
\node (v4) at (3.5,0) {$\mvar{v}_4$};

\draw[myptrdotted] (v1) to (v2);
\draw[myptrdotted] (v1) to[bend right=30] (v3);
\draw[myptr] (v2) to (v3);
\draw[myptr] (v4) to (v3);
\end{tikzpicture}
\end{minipage}
\hspace*{\fill}
\begin{minipage}{0.29\linewidth}\centering
\begin{tikzpicture}
\node (v1) at (0,0) {$\mvar{v}_1$};
\node (v2) at (1,0) {$\mvar{v}_2$};
\node (v3) at (2.5,0) {$\mvar{v}_3$};
\node (v4) at (3.5,0) {$\mvar{v}_4$};

\draw[myptr] (v1) to (v2);
\draw[myptrdotted] (v1) to[bend right=30] (v3);
\draw[myptrdotted] (v2) to (v3);
\draw[myptr] (v4) to (v3);
\end{tikzpicture}
\end{minipage}
\hspace*{\fill}
\begin{minipage}{0.29\linewidth}\centering
\begin{tikzpicture}
\node (v1) at (0,0) {$\mvar{v}_1$};
\node (v2) at (1,0) {$\mvar{v}_2$};
\node (v3) at (2,0) {$\mvar{v}_3$};
\node (v4) at (3.5,0) {$\mvar{v}_4$};

\draw[myptr] (v1) to (v2);
\draw[myptr] (v1) to[bend right=30] (v3);
\draw[myptr] (v2) to (v3);
\draw[myptrdotted] (v4) to (v3);
\end{tikzpicture}
\end{minipage}

\end{example}

\section{Policy Enforcement Point}
\label{sinvar:secgateway}
Hosts may belong to a certain domain. 
Sometimes, a pattern where intra-domain communication between domain members must be approved by a central instance is required. 

\begin{example}
\textbf{Example. }%
Let several virtual machines belong to the same domain and a secure hypervisor manage intra-domain communication. 
As another example, inter-device communication of slave devices in the same domain is controlled by a central master device. 
\end{example}

We call such a central instance an application-level `policy enforcement point' and present a template for this architecture. 
Five host roles are distinguished: $\Psi = \lbrace \mconstr{PolEnforcePoint},\allowbreak{} \mconstr{PolEnforcePointyIN},\allowbreak{} \mconstr{DomainMember},\allowbreak{} \mconstr{AccessibleMember},\allowbreak{} \mconstr{Unassigned} \rbrace$. 
A $\mconstr{PolEnforcePoint}$, a policy enforcement point accessible from the outside ($\mconstr{PolEnforcePointIN}$), a $\mconstr{DomainMember}$, 
a less-restricted $\mconstr{AccessibleMember}$ which is accessible from the outside world, and a default value $\bot = \mconstr{Unassigned}$ that reflects `none of these roles'. 
The following table implements the access control restrictions. 
The role of the sender (snd), role of the receiver (rcv), the result (rslt), and an explanation are given. 

\begin{table}[h!tbp]
\centering
\begin{tabular}[0.99\linewidth]{ @{}l@{ \ }l@{ }|c@{ \ }p{16em} }
	\toprule
	snd & rcv & rslt & explanation \\
	\midrule
 $\mconstr{PolEnforcePoint}$ & $\_$ & \cmark & Can send to the world. \\
 $\mconstr{PolEnforcePointIN}$ & $\_$ & \cmark & --- \grqq ---\\
 $\mconstr{DomainMember}$ & $\mconstr{DomainMember}$ & \xmark & Must not communicate directly. May communicate via $\mconstr{PolEnforcePoint}$ or $\mconstr{PolEnforcePointIN}$. \\
 $\mconstr{DomainMember}$ & $\_$ & \cmark & No restrictions for direct access to outside world. Outgoing accesses are not within the invariant's scope. \\
 $\mconstr{AccessibleMember}$ & $\mconstr{DomainMember}$ & \xmark & Must be approved. \\
 $\mconstr{AccessibleMember}$ & $\_$ & \cmark & No further restrictions, accessible members are also accessible among each other. \\
 $\mconstr{Unassigned}$ & $\mconstr{Unassigned}$ & \cmark & No restrictions. \\
 $\mconstr{Unassigned}$ & $\mconstr{PolEnforcePointIN}$ & \cmark & Accessible from outside. \\
  $\mconstr{Unassigned}$ & $\mconstr{PolEnforcePoint}$ & \xmark & Not accessible from outside. \\
 $\mconstr{Unassigned}$ & $\mconstr{AccessibleMember}$ & \cmark & Directly accessible from the outside. \\
 $\mconstr{Unassigned}$ & $\mconstr{DomainMember}$ & \xmark & Protected from outside world. \\
 \bottomrule%
\end{tabular}%
\end{table}

This template is minimalistic in that it only restricts accesses to members (from other members or the outside world), whereas accesses from members to the outside world are unrestricted. 
It can be implemented by a simple table lookup. 
In-host communication is allowed by adding $s \neq r$. 
%
\begin{IEEEeqnarray*}{lCl}
\sinvar\ (V,\, E)\ \nP \ & \ \equiv \ & \ \forall (s,\, r) \in E,\ s \neq r. \ \ \mdef{table}\ (\nP\ s)\ (\nP\ r)
\end{IEEEeqnarray*}

\begin{example}
\textbf{Example. }%
In a cloud environment, for brevity, let $\mvar{VMM}$ denote a secure hypervisor. 
The hypervisor has VM-specific security policies installed and acts as a broker between virtual machines. 
The $\mvar{VMM}$ is configured as $\mconstr{PolEnforcePoint}$. 
The VMs $\mvar{secure\mhyphen{}vm}_1$ and $\mvar{secure\mhyphen{}vm}_2$ run security-critical tasks and accesses between them and to them is mediated by the $\mvar{VMM}$. 
They are assigned the $\mconstr{DomainMember}$ host attribute. 
In contrast, $\mvar{public\mhyphen{}vm}_1$ and $\mvar{public\mhyphen{}vm}_2$ do not have special security requirements and are not mediated by the $\mvar{VMM}$. 
They are assigned the $\mconstr{AccessibleMember}$ host attribute, which makes them also directly accessible from the $\mvar{INET}$. 
Still, the two public VMs have slightly more access rights than an arbitrary host in the $\mvar{INET}$:
$\mvar{public\mhyphen{}vm}_1$ and $\mvar{public\mhyphen{}vm}_2$ may access the $\mvar{VMM}$ and hence the secure VMs if the $\mvar{VMM}$ permits it. 
The corresponding maximum policy for this setting is visualized below. 

\begin{minipage}{0.98\linewidth}\centering
\begin{tikzpicture}
\node (hypervisor) at (0,0) {$\mvar{VMM}$};
\node (securevm1) at (-3,1.5) {$\mvar{secure\mhyphen{}vm}_1$};
\node (securevm2) at (-1.5,3) {$\mvar{secure\mhyphen{}vm}_2$};
\node (publicvm1) at (1.5,3) {$\mvar{public\mhyphen{}vm}_1$};
\node (publicvm2) at (3,1.5) {$\mvar{public\mhyphen{}vm}_2$};
\node (INET) at (0,-2) {$\mvar{INET}$};

\draw[myptr] (hypervisor) to (INET);
\draw[myptr] (securevm1) to (publicvm1);
\draw[myptr] (securevm1) to (publicvm2);
\draw[myptr] (securevm1) to (INET);
\draw[myptr] (securevm2) to (publicvm1);
\draw[myptr] (securevm2) to (publicvm2);
\draw[myptr] (securevm2) to (INET);

\draw[myptrdouble] (hypervisor) to[loop right] (hypervisor);
\draw[myptrdouble] (securevm1) to (hypervisor);
\draw[myptrdouble] (securevm1) to[loop left] (securevm1);
\draw[myptrdouble] (securevm2) to (hypervisor);
\draw[myptrdouble] (securevm2) to[loop left] (securevm2);
\draw[myptrdouble] (publicvm1) to (hypervisor);
\draw[myptrdouble] (publicvm1) to[loop right] (publicvm1);
\draw[myptrdouble] (publicvm2) to (hypervisor);
\draw[myptrdouble] (publicvm2) to (publicvm1);
\draw[myptrdouble] (publicvm2) to[loop right] (publicvm2);
\draw[myptrdouble] (INET) to (publicvm1);
\draw[myptrdouble] (INET) to (publicvm2);
\draw[myptrdouble] (INET) to[loop below] (INET);
\end{tikzpicture}
\end{minipage}

This invariant template only controls how $\mconstr{DomainMember}$s can be accessed. 
It does not restrict the information flow from $\mconstr{DomainMember}$s. 
Therefore, the secure VMs may freely access anyone but may not be accessed by anyone, except the $\mvar{VMM}$. 

As an additional security requirement, let the secure VMs have confidential data. 
This is formalized with the Simplified Bell-LaPadula with Trust invariant (Section~\ref{sinvar:blptrust}). 
The $\mvar{VMM}$ is trusted and may declassify this data (security level $\mdef{unclassified}$ and trusted). 
This prevents that the secure VMs may initiate any connections to the outside by themselves and must have everything approved by the $\mvar{VMM}$.

\begin{minipage}{0.98\linewidth}\centering
\begin{tikzpicture}
\node (hypervisor) at (0,0) {$\mvar{VMM}$};
\node (securevm1) at (-3,1.5) {$\mvar{secure\mhyphen{}vm}_1$};
\node (securevm2) at (-1.5,3) {$\mvar{secure\mhyphen{}vm}_2$};
\node (publicvm1) at (1.5,3) {$\mvar{public\mhyphen{}vm}_1$};
\node (publicvm2) at (3,1.5) {$\mvar{public\mhyphen{}vm}_2$};
\node (INET) at (0,-2) {$\mvar{INET}$};

\draw[myptr] (hypervisor) to (INET);

\draw[myptrdouble] (hypervisor) to[loop right] (hypervisor);
\draw[myptrdouble] (securevm1) to (hypervisor);
\draw[myptrdouble] (securevm1) to[loop left] (securevm1);
\draw[myptrdouble] (securevm2) to (hypervisor);
\draw[myptrdouble] (securevm2) to[loop left] (securevm2);
\draw[myptrdouble] (publicvm1) to (hypervisor);
\draw[myptrdouble] (publicvm1) to[loop right] (publicvm1);
\draw[myptrdouble] (publicvm2) to (hypervisor);
\draw[myptrdouble] (publicvm2) to (publicvm1);
\draw[myptrdouble] (publicvm2) to[loop right] (publicvm2);
\draw[myptrdouble] (INET) to (publicvm1);
\draw[myptrdouble] (INET) to (publicvm2);
\draw[myptrdouble] (INET) to[loop below] (INET);
\end{tikzpicture}
\end{minipage}
\end{example}


\section{Sink}
\label{sinvar:sink}
Some hosts may be information sinks. 
That means, no information must leave those hosts. 
The information flow security strategy presented in this section formalizes the notion of information sinks. 

\begin{example}
\textbf{Example. }%
Assume logging information should not leave a central log server. 
\texttt{syslog} messages are usually sent via UDP (RFC~5426~\cite{rfc5426}), which enables a purely unidirectional channel, even without TCP acknowledgements. 
In cyber-physical systems, some actuators without sensors may also only receive commands and do not transmit any answers. 
They can also be considered information sinks. 
\end{example}

An information sink may not be limited to a single host but a pool of hosts. 
Though no information must leave the pool, it may be desirable that the pool can cooperate. 

For this security invariant, we distinguish between three types of hosts. 
The type of host attributes is defined as $\Psi = \lbrace \mconstr{Sink},\allowbreak \mconstr{SinkPool},\allowbreak \mconstr{Unassigned} \rbrace$. 
This models strict information $\mconstr{Sink}$s where no information must leave this host, 
pools of information sinks ($\mconstr{SinkPool}$) which may collaborate but no information must leave the pool, 
and $\mconstr{Unassigned}$ which does not impose any restrictions. 
The default value is $\bot = \mconstr{Unassigned}$. 

The following table formalizes the meaning of the individual host attributes. 

\begin{table}[h!tbp]
\centering
\begin{tabular}[0.99\linewidth]{ @{}l@{ \ }l@{ }|c@{ \ }p{24em} }
	\toprule
	snd & rcv & rslt & explanation \\
	\midrule
 	$\mconstr{Sink}$ & $\_$ & \xmark & No data must leave a $\mconstr{Sink}$ \\
 	$\mconstr{SinkPool}$ & $\mconstr{SinkPool}$ & \cmark & The pool can communicate with its members \\
 	$\mconstr{SinkPool}$ & $\mconstr{Sink}$ & \cmark & The pool can also send to individual sinks (but not the other way round) \\
 	$\mconstr{SinkPool}$ & $\_$ & \xmark & The pool cannot send data to the outside \\
 	$\mconstr{Unassigned}$ & $\_$ & \cmark & Everything else is not constrained \\
 	\bottomrule%
\end{tabular}%
\end{table}

With the table, the security invariant can be expressed as: 
\begin{IEEEeqnarray*}{lCl}
\sinvar\ (V,\, E)\ \nP \ & \ \equiv \ & \ \forall (s,\, r) \in E,\ s \neq r. \ \ \mdef{table}\ (\nP\ s)\ (\nP\ r)
\end{IEEEeqnarray*}

\begin{example}
\textbf{Example. }%
In this example, we model the information flow of trade secrets in a simplified SCADA factory network. 
This example is not concerned with the access control permissions, but only with information flow. 
The hosts in the example are a $\mvar{supervisor}$ for the complete production process, 
two $\mvar{control}$lers which control the production units, and two production units (called $\mvar{robot}$s). 
In addition, there is a host that represents the Internet. 

The supervisor may both access the Internet and send commands to the control units. 
The control units drive the production units. 
Since each control unit may send commands to both production units, the control units might need to synchronize for this task. 
The graph for the described scenario can be visualized as follows.

\begin{minipage}{0.98\linewidth}\centering
\begin{tikzpicture}

\node (supervisor) at (0,0) {$\mvar{supervisor}$\nodepart{lower}$\bot$};
\node (inet) at (3,0) {$\mvar{INET}$\nodepart{lower}$\bot$};
\node (control1) at (-1.5,-1.5) {$\mvar{control}_1$\nodepart{lower}$\mconstr{SinkPool}$};
\node (control2) at (1.5,-1.5) {$\mvar{control}_2$\nodepart{lower}$\mconstr{SinkPool}$};
\node (robot1) at (-1.5,-3) {$\mvar{robot}_1$\nodepart{lower}$\mconstr{Sink}$};
\node (robot2) at (1.5,-3) {$\mvar{robot}_2$\nodepart{lower}$\mconstr{Sink}$};

\draw[myptr] (supervisor) to (control1);
\draw[myptr] (supervisor) to (control2);
\draw[myptr] (control1) to (robot1);
\draw[myptr] (control1) to (robot2);
\draw[myptr] (control2) to (robot1);
\draw[myptr] (control2) to (robot2);

\draw[myptrdouble] (supervisor) to (inet);
\draw[myptrdouble] (control2) to (control1);
\end{tikzpicture}
\end{minipage}

Let the trade secrets be the production process of a product, \ie how the robots are driven. 
The main concern is that these secrets must not leak. 
The secret production steps are only encoded in the $\mvar{control}$ units, not the $\mvar{supervisor}$. 
To prevent information leakage, the $\mvar{control}$ units are information sinks, however, they need to cooperate, hence, they are labeled as $\mconstr{SinkPool}$. 
The $\mvar{robot}$s only need to receive commands and must not leak any information, therefore, they are just $\mconstr{Sink}$s. 
The $\mvar{supervisor}$ and $\mvar{INET}$ may be left unconfigured, \ie set to $\bot$. 
The policy visualized above fulfills this requirement. 
However, the following (maximum) policy also fulfills the formalized requirement.

\begin{minipage}{0.98\linewidth}\centering
\begin{tikzpicture}

\node (supervisor) at (0,0) {$\mvar{supervisor}$\nodepart{lower}$\bot$};
\node (inet) at (3,0) {$\mvar{INET}$\nodepart{lower}$\bot$};
\node (control1) at (-1.5,-1.5) {$\mvar{control}_1$\nodepart{lower}$\mconstr{SinkPool}$};
\node (control2) at (1.5,-1.5) {$\mvar{control}_2$\nodepart{lower}$\mconstr{SinkPool}$};
\node (robot1) at (-1.5,-3) {$\mvar{robot}_1$\nodepart{lower}$\mconstr{Sink}$};
\node (robot2) at (1.5,-3) {$\mvar{robot}_2$\nodepart{lower}$\mconstr{Sink}$};

\draw[myptr] (inet) to (robot1);
\draw[myptr] (inet) to (robot2);
\draw[myptr] (inet) to (control1);
\draw[myptr] (inet) to (control2);
\draw[myptr] (supervisor) to (robot1);
\draw[myptr] (supervisor) to (robot2);
\draw[myptr] (supervisor) to (control1);
\draw[myptr] (supervisor) to (control2);
\draw[myptr] (control1) to (robot1);
\draw[myptr,shorten <=-1ex] (control1) to (robot2);
\draw[myptr,shorten <=-1ex] (control2) to (robot1);
\draw[myptr] (control2) to (robot2);

\draw[myptrdouble] (inet) to[loop above] (inet);
\draw[myptrdouble] (supervisor) to (inet);
\draw[myptrdouble] (supervisor) to[loop above] (supervisor);
\draw[myptrdouble] (robot1) to[loop left] (robot1);
\draw[myptrdouble] (robot2) to[loop right] (robot2);
\draw[myptrdouble] (control1) to[loop left] (control1);
\draw[myptrdouble] (control2) to (control1);
\draw[myptrdouble] (control2) to[loop right] (control2);
\end{tikzpicture}
\end{minipage}

The picture shows that the control units and robots do not leak any information. 
However, any node, in particular $\mvar{INET}$---though it cannot receive an answer---may send arbitrary commands to the robots. 
This is because the \emph{Sink} invariant is an information flow security invariant and does not restrict access control. 
\end{example}

\section{Subnets}
\label{sinvar:subnets}
This invariant template formalizes the logical partition of a network into different segments. 
We call each segment a subnet. 

A host can either be a member of a specific subnet, can be a border router between subnets, or may not be part of the formalized security goal. 
Many different subnets can be formalized with one instance of this invariant. 
The type for host attributes is formalized as 
$\Psi = \lbrace \mconstr{Subnet}\ \mathbb{N},\allowbreak{}\ \mconstr{BorderRouter}\ \mathbb{N},\allowbreak{}\ \mconstr{Unassigned} \rbrace$. 
The types $\mconstr{Subnet}$ and $\mconstr{BorderRouter}$ have as parameter a natural number. 
This number indicates the specific subnet. 
By default, a host carries the $\bot = \mconstr{Unassigned}$ attribute.

\begin{example}
\textbf{Example. }%
Let $\nP\ v_1 = \mconstr{Subnet}\ 8$, $\nP\ v_2 = \mconstr{Subnet}\ 8$, $\nP\ v_3 = \mconstr{Subnet}\ 42$, and $\nP\ v_4 = \bot$. 
Then $v_1$ and $v_2$ are in the same subnet, $v_3$ is in a different subnet, and $v_4$ is in a completely different network segment (which is out of scope for the security requirement that is formalized with $\nP{}$). 
\end{example}

All subnets in this model are identified by a natural number. 
Hence, two subnets with a different number are distinct by definition. 
This means that this invariant does not permit different subnets to have overlapping IP address ranges. 
Since all entities are distinct and a subnet only consists of the entities in it, overlapping IP ranges cannot occur.\footnote{%
In general, we did not model IP address ranges yet. 
So far, entities are polymorphic over type \mbox{$\mathcal{V}$}. 
For example, $\mathcal{V}$ can be the set of all IPv4 addresses and entities are thus individual IP addresses. 
But since all elements of \mbox{$\mathcal{V}$} must be pairwise distinct, representing entities by overlapping IP ranges is not possible. 
One abusive workaround would be to define $\mathcal{V}$ as a set of strings and encode (possibly overlapping) IP addresses in CIDR notation into these strings. 
While at the abstraction level of our policy, two different strings represent different policy entities, soundness problems on the network level occur if a firewall does not interpret them as strings but overlapping IP ranges. 
We discuss the relation between policy entities and sets of IP addresses in detail in Part~\ref{part:existing-configs}: 
Overlapping IP address ranges with different policy actions (allow or deny) bring up the question whether the intersection should be allowed or denied. 
In contrast to our policy with only positive rules where order does not matter, we discuss the first-matching semantics of firewalls in Chapter~\ref{chap:fm15}. 
When entities correspond to IP addresses, we consider IP address spoofing in Chapter~\ref{chap:nospoof}. 
In Chapter~\ref{chap:networking16}, when inferring a policy from a network-level firewall ruleset, entities will correspond to sets of IP addresses. 
Theorem~\ref{thm:servicematrix} will conclude that the IP address ranges of all such computed entities are indeed disjoint.
Therefore, it is possible to have entities with overlapping IP address ranges in a policy (for example with the string workaround), but one needs to verify that the network-level connectivity enforced by a firewall corresponds to the desired policy. 
Our tool \fffuu{} (Section~\ref{sec:ifip:fffuu}) is suitable for this task. 
}

Border routers may connect different subnets by communicating with each other. 
The design decision for this invariant is that they are extremely restricted in their access rights. 
A core idea is that a host in a specific subnet must not be reachable from ``unconfigured'' (\ie $\bot$) parts of the network, not even indirectly.
Also, a host must not be reachable from other subnets, also, not even indirectly. 
Since border routers can communicate with each other, to fulfill the design goals, a border router must not be allowed to establish a connection to a member of a subnet. 

\begin{example}
\textbf{Example. }%
Let $\nP\ v_1 = \mconstr{Subnet}\ 1$, $\nP\ v_2 = \mconstr{Subnet}\ 2$, $\nP\ v_3 = \mconstr{BorderRouter}\ 1$, and $\nP\ v_4 = \mconstr{BorderRouter}\ 2$. 
The border routers $v_3$ and $v_4$ may communicate. 
The host $v_2$ may send packets to its border router $v_4$. 
Since $v_1$ and $v_2$ are in different subnets, $v_2$ should not be able to reach $v_1$. 
The possible path $v_2 \longrightarrow v_4 \longrightarrow v_3 \longrightarrow v_1$ is prohibited by disallowing that border routers can establish connections to subnet hosts -- including hosts of their own subnet. 

Note that $v_1$ could set up a connection to $v_3$. 
If this is a stateful connection which also permits replies from $v_3$ back to $v_1$ (cf.\ Chapter~\ref{chap:esss14}), a channel between $v_1$ and $v_2$ is established. 
Since this requires the consent of $v_1$ which needs to set up the stateful connection, $v_2$ is still not permitted to access $v_1$ on its own and this setting does not contradict the intentions of the invariant template. 
\end{example}

\noindent
Similar to previous templates, the invariant template is formalized with the help of a table. %
\begin{IEEEeqnarray*}{lCl}
\sinvar\ (V,\, E)\ \nP \ & \ \equiv \ & \ \forall (s,\, r) \in E. \ \ \mathtt{table}\ (\nP\ s)\ (\nP\ r)
\end{IEEEeqnarray*}
The access rights of the different host attributes are formalized as follows. 

\begin{table}[h!tbp]
\centering
\begin{tabular}[0.99\linewidth]{ @{}l@{ \ }l@{ }|c@{ \ }p{16em} }
	\toprule
	snd & rcv & rslt & explanation \\
	\midrule
 	$\mconstr{Subnet}\ s_1$ & $\mconstr{Subnet}\ s_2$ & $s_1 = s_2$ & Two hosts in a subnet can communicate if they are in the same subnet. \\
 	$\mconstr{Subnet}\ s_1$ & $\mconstr{BorderRouter}\ s_2$ & $s_1 = s_2 $ & A host in a subnet can communicate with its border router.  \\
 	$\mconstr{Subnet}\ s_1$ & $\mconstr{Unassigned}$ & \cmark & No restriction for communication outside of the subnet. \\
 	$\mconstr{BorderRouter}\ s_1$ & $\mconstr{Subnet}\ s_2$ & \xmark & A router should not establish connections to any hosts. This is due to the design decision that hosts must not be (even indirectly) accessible from the outside. Considering the generic requirements for invariants, it would be possible to change the condition from \xmark\ to $(s_1 = s_2)$, but this would violate this specific invariant's design decision. \\
 	$\mconstr{BorderRouter}\ s_1$ & $\mconstr{BorderRouter}\ s_2$ & \cmark & Border routers may communicate with each other. \\
 	$\mconstr{BorderRouter}\ s_1$ & $\mconstr{Unassigned}$ & \cmark & No restriction \\
 	$\mconstr{Unassigned}$ & $\mconstr{Unassigned}$ & \cmark & No restriction \\
 	$\mconstr{Unassigned}$ & $\_$ & \xmark & Must not set up connection to subnet or border routers \\
 	\bottomrule%
\end{tabular}%
\end{table}

A violation of the invariant template occurs exactly iff\footnote{SINVAR\_Subnets.violating-configurations-exhaust} 
$\bot$ is trying to connect to a non-$\bot$ host, 
or a host is trying to connect to a host or a border router of a different subnet, 
or a border router is trying to access a host.

\begin{example}
\textbf{Example. }%
Let the hosts $v_{11}$, $v_{12}$, $v_{13}$ be members of subnet $1$ and let $v_\mathrm{1b}$ be the border router of subnet $1$. 
Likewise, let the hosts $v_{21}$, $v_{22}$, $v_{23}$ be members of subnet $2$ and let $v_\mathrm{2b}$ be the border router of subnet $2$. 
Let $v_\mathrm{3b}$ be the border router of subnet $3$ and let $v_\mathrm{o}$ be an arbitrary $\bot$ host. 
The maximum policy for this setting can be visualized as follows. 

\begin{minipage}{0.98\linewidth}\centering
\begin{tikzpicture}
\node (v11) at (0,0) {$v_{11}$};
\node (v12) at (-1.5,-1) {$v_{12}$};
\node (v13) at (0,-2) {$v_{13}$};
\node (v1b) at (2,-1) {$v_\mathrm{1b}$};

\node (v21) at (6,0) {$v_{21}$};
\node (v22) at (7.5,-1) {$v_{22}$};
\node (v23) at (6,-2) {$v_{23}$};
\node (v2b) at (4,-1) {$v_\mathrm{2b}$};

\node (v3b) at (3,0) {$v_\mathrm{3b}$};

\node (vo) at (3,1) {$v_\mathrm{o}$};

\draw[myptr] (v11) to (v1b);
\draw[myptr] (v11) to (vo);
\draw[myptr] (v12) to (v1b);
\draw[myptr] (v12) to (vo);
\draw[myptr] (v13) to (v1b);
\draw[myptr] (v13) to (vo);
\draw[myptr] (v1b) to (vo);
\draw[myptr] (v21) to (v2b);
\draw[myptr] (v21) to (vo);
\draw[myptr] (v22) to (v2b);
\draw[myptr] (v22) to (vo);
\draw[myptr] (v23) to (v2b);
\draw[myptr] (v23) to (vo);
\draw[myptr] (v2b) to (vo);
\draw[myptr] (v3b) to (vo);

\draw[myptrdouble] (v11) to[loop above] (v11);
\draw[myptrdouble] (v12) to (v11);
\draw[myptrdouble] (v12) to[loop left] (v12);
\draw[myptrdouble] (v13) to (v11);
\draw[myptrdouble] (v13) to (v12);
\draw[myptrdouble] (v13) to[loop below] (v13);
\draw[myptrdouble] (v1b) to[loop below] (v1b);
\draw[myptrdouble] (v21) to[loop above] (v21);
\draw[myptrdouble] (v22) to (v21);
\draw[myptrdouble] (v22) to[loop right] (v22);
\draw[myptrdouble] (v23) to (v21);
\draw[myptrdouble] (v23) to (v22);
\draw[myptrdouble] (v23) to[loop below] (v23);
\draw[myptrdouble] (v2b) to (v1b);
\draw[myptrdouble] (v2b) to[loop below] (v2b);
\draw[myptrdouble] (v3b) to (v1b);
\draw[myptrdouble] (v3b) to (v2b);
\draw[myptrdouble] (v3b) to[loop below] (v3b);
\draw[myptrdouble] (vo) to[loop above] (vo);
\end{tikzpicture}
\end{minipage}

It can be seen that the hosts in each subnet have full connectivity among themselves. 
Everyone can access the outside-world host $v_\mathrm{o}$. 
The border routers have full connectivity among themselves but cannot access hosts in a subnet. 
Even transitively, hosts of one subnet cannot access hosts of other subnets. 
\end{example}

With this invariant template, it is impossible for some host in a subnet to be somehow accessible by the outside world. 
However, it is often desirable that a host is accessible, but only for connection which are routed over a border router. 
For this, we have extended the template\footnote{SINVAR\_Subnets2.thy} with two additional host attributes: 
\begin{itemize}
	\item A $\mconstr{BorderRouter'}$ which can access members of its own subnet. 
	\item An $\mconstr{InboundRouter}$ which is accessible by anyone, regardless of the subnet, but which cannot access hosts of a subnet. 
\end{itemize}
With this setting, a $\bot$ host can set up connections to an $\mconstr{InboundRouter}$, which can set up connections to a subnet's $\mconstr{BorderRouter'}$, which can set up connections to a host in this subnet. 

However, configuring this invariant template may get very complex. 
In addition, since this template may make assertions about a global subnet structure, it may no longer encode exactly one security requirement but may collect several. 
This may break modularity. 
Therefore, we decided to develop a simpler template which only considers one subnet. 
This will be formalized with the next invariant template. 
Statements about several subnets can be expressed with several instances of the template. 

\section{SubnetsInGW}
\label{sinvar:subnetsingw}
This invariant template encodes the following security goal for one subnet or group of hosts: 
$\mconstr{Member}$s of a subnet may freely communicate with each other but are not accessible from the outside world. 
It may be possible from the outside world to connect to a $\mconstr{Member}$ over an $\mconstr{InboundGateway}$. 
The template considers only one single subnet. 
The host attributes are $\Psi = \lbrace \mconstr{Member},\ \allowbreak{}\mconstr{InboundGateway},\ \allowbreak{}\mconstr{Unassigned} \rbrace$ where 
$\bot = \mconstr{Unassigned}$. 

With the following table, the invariant template is formalized as follows. 
\begin{table}[h!tbp]
\centering
\begin{tabular}[0.99\linewidth]{ @{}l@{ \ }l@{ }|c@{ \ }p{16em} }
	\toprule
	snd & rcv & rslt & explanation \\
	\midrule
 	$\mconstr{Member}$ & $\_$ & \cmark & Accesses to members are restricted, but not the other way round. \\
 	$\mconstr{InboundGateway}$ & $\_$ & \cmark & No restrictions, may even connect to $\mathit{Member}$s. \\
 	$\mconstr{Unassigned}$ & $\mconstr{Unassigned}$ & \cmark & No restrictions. \\
 	$\mconstr{Unassigned}$ & $\mconstr{InboundGateway}$ & \cmark & This is the only way to indirectly access $\mathit{Member}$s. \\
 	$\mconstr{Unassigned}$ & $\mconstr{Member}$ & \xmark & Direct access is prohibited. \\
 	\bottomrule%
\end{tabular}%
\end{table}

\begin{IEEEeqnarray*}{lCl}
\sinvar\ (V,\, E)\ \nP \ & \ \equiv \ & \ \forall (s,\, r) \in E. \ \ \mdef{table}\ (\nP\ s)\ (\nP\ r)
\end{IEEEeqnarray*}

With this invariant, a screened subnet architecture (sometimes called DMZ or demilitarized zone architecture) can be built. 
A $\mconstr{Member}$ is in the protected internal network and not accessible by external hosts. 
The servers in the DMZ are labeled $\mconstr{InboundGateway}$. 
With this setting, a relaxed screened subnet architecture is encoded because the servers in the DMZ can access the internal machines. 
This is sometimes desirable. 
To build an actual DMZ architecture where the servers in the DMZ must not access the internal hosts, the DMZ servers may simply be set to $\bot$.

\begin{example}
\textbf{Example. }%
We emulate the example of Section~\ref{sinvar:subnets} (Subnets) with the help of this invariant. 
Let $V = \lbrace v_{11}, v_{12}, v_{13}, v_{1b}, v_{21}, v_{22}, v_{23}, v_{2b}, v_{3b}, v_\mathrm{o} \rbrace$. 
Referring to the previous example, we will call the nodes $v_\mathrm{*b}$ border routers. 
We split the security goals into five configured security invariants. 
First, let the hosts $v_{11}$, $v_{12}$, $v_{13}$ be $\mconstr{Member}$s and let $v_\mathrm{1b}$ be the $\mconstr{InboundGateway}$. 
Likewise, we create a second configured security invariant for the hosts $v_{21}$, $v_{22}$, $v_{23}$, and $v_\mathrm{2b}$. 
These two invariants would allow the border routers to be globally accessible. 
To restrict accesses for each border router, we attach an ACL (using the Communication Partners template) to each router. 
We create one instance for each router. 
The ACLs specify that the border router is only accessible by the other routers and the members of its subnet. 
The maximum policy for this setting can be visualized as follows. 

\begin{minipage}{0.98\linewidth}\centering
\begin{tikzpicture}
\node (v11) at (0,0) {$v_{11}$};
\node (v12) at (-1.5,-1) {$v_{12}$};
\node (v13) at (0,-2) {$v_{13}$};
\node (v1b) at (2,-1) {$v_\mathrm{1b}$};

\node (v21) at (6,0) {$v_{21}$};
\node (v22) at (7.5,-1) {$v_{22}$};
\node (v23) at (6,-2) {$v_{23}$};
\node (v2b) at (4,-1) {$v_\mathrm{2b}$};

\node (v3b) at (3,0) {$v_{3b}$};

\node (vo) at (3,1) {$v_\mathrm{o}$};

\draw[myptr] (v11) to (vo);
\draw[myptr] (v12) to (vo);
\draw[myptr] (v13) to (vo);
\draw[myptr] (v1b) to (vo);
\draw[myptr] (v21) to (vo);
\draw[myptr] (v22) to (vo);
\draw[myptr] (v23) to (vo);
\draw[myptr] (v2b) to (vo);
\draw[myptr] (v3b) to (vo);

\draw[myptrdouble] (v11) to[loop above] (v11);
\draw[myptrdouble] (v12) to (v11);
\draw[myptrdouble] (v12) to[loop left] (v12);
\draw[myptrdouble] (v13) to (v11);
\draw[myptrdouble] (v13) to (v12);
\draw[myptrdouble] (v13) to[loop below] (v13);
\draw[myptrdouble] (v1b) to (v11);
\draw[myptrdouble] (v1b) to (v12);
\draw[myptrdouble] (v1b) to (v13);
\draw[myptrdouble] (v1b) to[loop below] (v1b);
\draw[myptrdouble] (v21) to[loop above] (v21);
\draw[myptrdouble] (v22) to (v21);
\draw[myptrdouble] (v22) to[loop right] (v22);
\draw[myptrdouble] (v23) to (v21);
\draw[myptrdouble] (v23) to (v22);
\draw[myptrdouble] (v23) to[loop below] (v23);
\draw[myptrdouble] (v2b) to (v1b);
\draw[myptrdouble] (v2b) to (v21);
\draw[myptrdouble] (v2b) to (v22);
\draw[myptrdouble] (v2b) to (v23);
\draw[myptrdouble] (v2b) to[loop below] (v2b);
\draw[myptrdouble] (v3b) to (v1b);
\draw[myptrdouble] (v3b) to (v2b);
\draw[myptrdouble] (v3b) to[loop below] (v3b);
\draw[myptrdouble] (vo) to[loop above] (vo);
\end{tikzpicture}
\end{minipage}

It can be seen that the resulting policy is almost equal to the one of Section~\ref{sinvar:subnets}, with the only difference that border routers are now allowed to access their subnet members. 
\end{example}

\section{Simple Tainting}
\label{sinvar:taint}
While previous invariants primarily focused on security in terms of access control and information flow, we will now present two invariants which focus on privacy. 
We will start with motivating background information and related work. 
Afterwards, we present our invariant formalization. 
Naturally, these invariants are information flow security strategies. 
Afterwards, we analyze the invariant and show that static taint analysis is as expressive as the Bell-LaPadula model.

\paragraph*{Background} 
\label{par:backgroundtainting}
Recently, dynamic taint analysis \cite{schwartz2010all} has been used successfully in the Android world to enhance user privacy~\cite{enck2010taindroid,enck2014taintdroid,droiddisintegrator2016}. 
Based on those ideas, the two invariants presented in this section and the following section formalize static taint analysis. 
We demonstrate that coarse-grained taint analysis is also applicable to the analysis and auditing of distributed architectures, can be done completely static (preventing runtime failures), while providing strong formal guarantees. 

We base our understanding of privacy on the operationalization performed by Pfitzmann and Rost~\cite{Datenschutzschutzziele} and further elaborated on by Bock and Rost~\cite{Bock2011}.
Their proposal has been adapted by the 
European Union Agency for Network and Information Security (ENISA)~\cite{ENISA2014} and by the German Standardized Data Protection Model~\cite{SDM}, showing wide acceptance of their approach. 
In summary, this set of related work bases the understanding of privacy upon the data protection goals of \emph{unlinkability}, \emph{transparency}, \emph{intervenability}, and \emph{data minimization}. 
A detailed discussion of these aspects can be found in the extended version of our paper~\cite{maltitz2016arxivprivacylong}. 
Our model of static taint analysis was designed to make these aspects of privacy more tangible. 
In brief, we address unlinkability by making it possible to see whether two taint labels are ever assigned to the same entity. 
In addition, the taint labels of an entity are a measure to promote data minimization. 
Transparency and intervenability are user-facing protection goals and our model provides a first step towards this by making the relevant information explicit. 


\begin{example}
\textbf{Example. }%
We introduce the concepts of taint analysis by a simple, fictional example:
A house, equipped with a smart meter to measure its energy consumption. The owner also provides location information via her smartphone to allow the system to turn off the lights when she leaves the home.
Once every month, the aggregated energy consumption is sent over the Internet to the energy provider for billing.

\begin{minipage}{0.98\linewidth}\centering
    \resizebox{0.99\textwidth}{!}{%
    \begin{small}
	\begin{tikzpicture}
	\node[MyRoundedBox, text width=6em] (smartphone) at (0,-1) {$\mvar{Smartphone}$ \\ $\lbrace \mdef{location} \rbrace$};
	\node[MyRoundedBox, text width=6em] (building) at (0,1) {$\mvar{Building}$\\ $\lbrace \mdef{energy} \rbrace$};
	\node[MyRoundedBox, text width=8em] (collect) at (4,0) {$\mvar{Smart Home Box}$\\ $\lbrace \mdef{energy}, \mdef{location} \rbrace$};
	\node[MyRoundedBox, text width=9em] (anon) at (8,0) {$\mvar{Anonymizer}$\\ $\mdef{untaints:}\; \lbrace \mdef{location} \rbrace$};
	\node[MyRoundedBox, text width=6em] (cloud) at (12,0) {$\mvar{Cloud}$\\ $\lbrace \mdef{energy} \rbrace$};

	\path[myptr] (smartphone) edge (collect);
	\path[myptr] (building) edge (collect);
	\path[myptr] (collect) edge (anon);
	\path[myptr] (anon) edge (cloud);
	\end{tikzpicture}
    
    \end{small}
	}
\end{minipage}	

We are interested in the privacy implications of this setup and perform a taint tracking analysis.
The system architecture is visualized in the above figure. 
The $\mvar{Building}$ produces information about its energy consumption.
Therefore, we label the $\mvar{Building}$ as taint source and assign it the $\mdef{energy}$ label.
Likewise, the $\mvar{Smartphone}$ tracks the $\mdef{location}$ of its owner.
Both data is sent to the $\mvar{Smart Home Box}$.
Since the $\mvar{Smart Home Box}$ aggregates all data, it is assigned the set $\lbrace \mdef{energy}, \mdef{location} \rbrace$ of taint labels.
The user wants to transmit only the $\mdef{energy}$ information, not her $\mdef{location}$ to the energy provider's $\mvar{Cloud}$.
Therefore, the $\mvar{Anonymizer}$ filters the information and removes all $\mdef{location}$ information.
We call this process \emph{untainting}.
Let `$\cdot \setminus \cdot$' denote the minus operation on sets.
With the $\mvar{Anonymizer}$ operating correctly, since $\lbrace \mdef{energy}, \mdef{location} \rbrace \setminus \lbrace \mdef{location} \rbrace = \lbrace \mdef{energy} \rbrace$, only $\mdef{energy}$-related information ends up in the energy provider's $\mvar{Cloud}$.
\end{example}

\medskip

We now formalize a simplified version of the security invariant template, based on the above ideas. 
For clarity, it does not provide the untainting feature. 
This feature will be added explicitly in Section~\ref{sinvar:taintuntaint}. 

The host attributes are sets of taint labels. 
For simplicity, we model taint labels as strings: $\Psi = \mathit{string}\ \mathit{set}$. 
For example, $\nP\ \mvar{Smart Home Box} = \lbrace \mdef{energy}, \mdef{location} \rbrace$. 
By default, an entity does not have any taint label, \ie $\bot = \emptyset$. 

Intuitively, information flow security according to the taint model can be understood as follows. 
Information leaving a node $v$ is tainted with $v$'s taint labels, hence every receiver $r$ must have the respective taint labels to receive the information. 
In other words, for every node $v$ in the policy, all nodes $r$ which are reachable from $v$ must have at least $v$'s taint labels. 
Representing reachability by the transitive closure (\ie $\mfun{succ\mhyphen{}tran}$), the invariant can be formalized as follows: 
\begin{IEEEeqnarray*}{lCl}
\sinvar\ (V,\, E)\ \nP \ & \ \equiv \ & \ {\forall v \in V.}\ {\forall r \in \mfun{succ\mhyphen{}tran}\ (V,\, E)\ v.}\ \ \nP\ v \subseteq \nP\ r
\end{IEEEeqnarray*}


\paragraph*{Analysis: Tainting vs.\ Bell-LaPadula Model}
The Bell-LaPadula model is the traditional, de-facto standard model for label-based information flow security. 
The question arises whether we can justify our taint model using BLP. 
We will compare our model to the simple Bell-LaPadula model (Section~\ref{sinvar:blptsimple}). 

We need to give names to the invariant templates which have always been called $\sinvar$. 
We call the tainting invariant template $\mfun{tainting}$ and we call the Bell-LaPadula template $\mfun{blp}$.

Inspired by BLP, we show an alternative definition for our $\mfun{tainting}$ invariant: 
\begin{lemma}[Localized Definition of Tainting]
\label{lem:tainting:localized}
\begin{IEEEeqnarray*}{c}
\mfun{tainting}\ (V,E)\ t = {\forall (v_1, v_2) \in E.}\ \ t\ v_1 \subseteq t\ v_2
\end{IEEEeqnarray*}
\end{lemma}

Lemma~\ref{lem:tainting:localized} also provides\footnote{SINVAR-Tainting.sinvar-preferred-def} a computational efficient formula, which only iterates over all edges and never needs to compute a transitive closure.

We now show\footnote{tainting-iff-blp} that one $\mfun{tainting}$ invariant is equal to BLP invariants for every taint label. 
We define a function $\mfun{project}\ a\ \mvar{Ts}$, which translates a set of taint labels $\mvar{Ts}$ to a security level depending on whether $a$ is in the set of taint labels. 
Formally, $\mfun{project}\ a\ \mvar{Ts} \equiv \mctrl{if}\ a \in \mvar{Ts}\ \mctrl{then}\ \mdef{confidential}\ \mctrl{else}\ \mdef{unclassified}$. 
Using function composition, the term $\mfun{project}\; a \ \circ\ \nP$ is a function which first looks up the taint labels of a node and projects them afterwards.

\begin{theorem}[Tainting and Bell-LaPadula Equivalence]
\label{thm:taint-iff-blp}
\begin{IEEEeqnarray*}{c}
\mfun{tainting}\ G\ \nP \longleftrightarrow {\forall a.}\ \mfun{blp}\ G\ (\mfun{project}\ a \ \circ\ \nP)
\end{IEEEeqnarray*}
\end{theorem}

In the context of privacy, the `$\rightarrow$'-direction of our theorem shows that one $\mfun{tainting}$ invariant guarantees individual privacy according to Bell-LaPadula for each taint label. 
This implies that every user of a system can obtain her personal privacy guarantees. 
This is one step towards transparency and intervenability. 

The `$\leftarrow$'-direction shows that $\mfun{tainting}$ is as expressive as the simple Bell-LaPadula model. 
This justifies the theoretic foundations w.r.t.\ the well-studied BLP model.
These findings are in line with Denning's lattice interpretation~\cite{denning1976lattice};  
%
however, to the best of our knowledge, we are the first to discover and formally prove this connection in the presented context. 

The theorem can be generalized\footnote{tainting-iff-blp-extended} for arbitrary (but finite) sets of taint labels~$A$. 
The $\mfun{project}$ function then maps to a numeric value of a security level by taking the cardinality of the intersection of $A$ with $\mvar{Ts}$. 
For example, if we want to project $\lbrace \mdef{location}, \allowbreak{}\mdef{temp} \rbrace$, then $\lbrace \mdef{name} \rbrace$ is $\mdef{unclassified}$, 
$\lbrace \mdef{name}, \mdef{location}, \allowbreak{}\mdef{zodiac} \rbrace$ is $\mdef{confidential}$, and $\lbrace \mdef{name}, \allowbreak{}\mdef{location}, \allowbreak{}\mdef{zodiac}, \allowbreak{}\mdef{temp} \rbrace$ is $\mdef{secret}$. 

\section{Tainting}
\label{sinvar:taintuntaint}
We continue our efforts of the previous section to formalize an invariant template for privacy based on taint analysis. 

Real-world application requires the need to untaint information, for example, when data is encrypted or properly anonymized. 
This was also demonstrated in the example of the previous section. 

We extend the host attributes to carry both taint labels and untainting information. 
The taint labels now consist of two components: the labels a node taints and the labels it untaints: 
$\Psi = \mathit{string}\ \mathit{set} \ \ \times \ \ \mathit{string}\ \mathit{set}$.
We define the helper accessor which allows us to conveniently access the individual taint label sets. 
Let $\mdef{taints}$ return the first element of the tuple and let $\mdef{untaints}$ return the second element of the tuple.\footnote{Actually, in our implementation \texttt{SINVAR-TaintingTrusted}, we create a new type \texttt{taints-raw} to avoid accidental type errors and confusion of taint labels with other tuples. }

We extend the simple $\mfun{tainting}$ invariant to support untainting: 
\begin{IEEEeqnarray*}{lCl}
\sinvar\ (V,E)\ \nP & \ \equiv \ &  {\forall (v_1, v_2) \in E.}\ \ \mdef{taints}\ (\nP\ v_1) \setminus \mdef{untaints}\ (\nP\ v_1) \subseteq \mdef{taints}\ (\nP\ v_2)
\end{IEEEeqnarray*}

To abbreviate a node's labels, we will write $X \text{---} Y$, where $X$ corresponds to the $\mdef{taints}$ and $Y$ corresponds to the $\mdef{untaints}$. 
In the example of the previous section, we have $\nP\ \mvar{Anonymizer} = \lbrace\mdef{energy}\rbrace \text{---} \lbrace \mdef{location} \rbrace$.

We impose the type constraint that $Y \subseteq X$, \ie $\mdef{untaints}\ \mvar{Ts} \subseteq \mdef{taints}\ \mvar{Ts}$.\footnote{taints-wellformedness} 
We implemented the datatype such that $X \text{---} Y$ is extended to $X \cup Y \text{---} Y$. 
Regarding previous section's example, this merely appears to be a very convenient abbreviation. 
In particular, $\nP\ \mvar{Anonymizer}$ now corresponds to $\lbrace \mdef{energy}, \mdef{location} \rbrace \text{---} \lbrace \mdef{location} \rbrace$, for which the improved tainting invariant holds and which also corresponds to our intuitive understanding of untainting.  
However, this is a fundamental requirement for the overall soundness of the invariant. 
Without the constraint, there can be dead $\mdef{untaints}$, \ie $\mdef{untaints}$ which can never have any effect. 
This would violate the uniqueness property required by the secure default parameters and can cause further problems in pathological corner cases. 
%
Yet, with this type constraint, all insights obtained for the simple model now follow analogously for this model. 

\paragraph*{Analysis: Tainting vs.\ Bell-LaPadula Model}
We now compare our improved model to the Bell-LaPadula model with trust (Section~\ref{sinvar:blptrust}). 

We give names to the invariant templates which have always been called $\sinvar$. 
We call the tainting invariant template $\mfun{tainting'}$ and we will call the Bell-LaPadula template $\mfun{blp'}$. 

In the context of Bell-LaPadula, a trusted entity is allowed to declassify information, \ie receive information of any security level and redistribute with its own level (which may be lower than the level of the received information). 
This concept is comparable to untainting. 

Recall that in Bell-LaPadula, $\mdef{trust}$ extracts the trusted flag from an entity's attributes and $\mfun{level}$ extracts the security level. 

Our insights about the equality follow analogously to the simple tainting invariant.\footnote{tainting-iff-blp-trusted} 
Let $\mfun{project}\ a\ (X \text{---} Y)$ be a function which translates $\mdef{taints}$ ($X$) and $\mdef{untaints}$ ($Y$) labels to security levels and trust of the Bell-LaPadula model. 
The function translates the security $\mdef{level}$ as follows: $\mctrl{if}\ a \in (X\ \setminus Y) \ \mctrl{then}\ \mdef{confidential}\ \mctrl{else}\ \mdef{unclassified}$. 
It translates the $\mdef{trust}$ flag as follows: $a \in Y$. 

\begin{theorem}[Tainting and Bell-LaPadula Equivalence]
\begin{IEEEeqnarray*}{c}
\mfun{tainting}'\ G\ t \longleftrightarrow {\forall a.}\ \mfun{blp}'\ G\ (\mfun{project}\ a \circ t)
\end{IEEEeqnarray*}
\end{theorem}

Similarly to the version without trust, the theorem can be generalized for arbitrary (but finite) sets of taint labels.\footnote{tainting-iff-blp-trusted-extended}

\section{System Boundaries} 
\label{sec:meta:systemboundaries}
Our formalism provides a number of useful analyses. 
For example, given a security invariant specification, it allows to compute all permitted flows which is invaluable for validating a given specification. 
However, our formalism might lack knowledge about architectural constraints which leads to the computation of an unrealistic amount of flows. 
For example, several logical entities may correspond to individual programs which are running on the same machine. 
Some programs may only communicate via IPC and are not externally reachable. 
We want to provide this knowledge to our formalism.


Therefore, we model systems with clearly defined boundaries. 
We define \emph{internal} components as nodes which are only accessible from inside the system. 
We define \emph{passive system boundaries} to be boundaries which only accept incoming connections. 
Analogously, \emph{active system boundaries} are boundaries which only establish outgoing connections. 
A \emph{boundary} may be both. 

A security invariant in our formalism must either be an access control strategy or information flow security strategy. 
An access control invariant restricts accesses \emph{from} the outside, an information flow invariant restricts leakage \emph{to} the outside. 
However, internal components of a system require both: 
they should neither be accessible from components outside of the system nor leak data to outside components. 
We overcame this limitation of our framework by constructing a model for system boundaries 
 which is internally translated to two invariants: an access control invariant (SubnetsInGW, Section~\ref{sinvar:subnetsingw}) and an information flow invariant (Bell-LaPadula, Section~\ref{sinvar:blptrust}). 
We have integrated the concept of system boundaries into our formalization and proven that the two configured invariants which are generated by a specification of a system boundary yield exactly the desired behavior.

\chapter{Evaluation \& Case Studies}
\label{sec:forte14:evalandcasestudy}

\paragraph*{Abstract}
We present a tool which implements the presented theory and evaluate our theory and tool in the case study of a cabin data network and at the example of an imaginary factory network.

\medskip

\section{Stand-Alone Tool: \topos{}}
\label{sec:forte14:impl}

\paragraph*{Version 1: Scala Implementation}
Typically, a network or policy administrator does not want to depend on a theorem prover and does not want to learn the formal language of a theorem prover. 
Therefore, we build a prototypical stand-alone tool to demonstrate applicability of our theory. 
Our tool is a stand-alone Java \texttt{jar} file and only requires a JVM to run; no external dependencies exist. 
The stand-alone tool also serves to emphasize the following:
\begin{itemize}[noitemsep]
	\item Our tool does not depend on Isabelle/HOL being installed on a user's system. 
	\item No automated provers are run in the background. 
	      Hence, our algorithms do not rely on unpredictable heuristics to solve proof obligations, which is known to fail (timeout) at runtime for certain problems; 
	      our algorithms directly solve the problems.
	\item A typical user does not need to prove anything to use our tool. 
	\item Our tool runs on common, off-the-shelf hardware, without special requirements or dependencies.
\end{itemize}
We call our prototypical tool \topos{}. 
The Scala implementation supports most of the features presented in the previous chapters. 
Its core reasoning logic consists of code generated by Isabelle/HOL. 
This guarantees the correctness of all results computed by \topos{}'s core~\cite{isabelle2012code,haftmann2010code}. 
Of course, only the core is generated by Isabelle; errors in the user interface cannot be prevented by this approach. 
In addition, the code generation only guarantees partial correctness. 
This means, termination is not guaranteed; however, if there is a result, it is correct. 
Our empirical evaluation shows that the code always terminates for reasonably-sized input. 


\paragraph*{Version 2: Isabelle/ML Implementation}
However, we realized that the tool's Scala code was diverging from the theory files over time. 

The Scala tool was based on code originally written for the author's master's thesis~\cite{cornythesis}. 
%
For the preliminary prototype of the master's thesis, about 5000 lines of theory were used to generate 2000 lines of Scala code. 
In total, the Scala tool consisted of about 6500 lines of code. 
Consequently, there was an amount of more than three times unverified code involved than there was verified code.
During the Ph.D.\ thesis, the theory (and consequently the generated code) and the manually-written code of the Scala tool grew. 
However, the ratio of about more than three times unverified code vs.\ verified code remained. 

For this Ph.D.\ thesis, the theory has been completely reworked and the Scala tool diverged more and more. 
We decided that the tool and the theories must be better kept in sync. 
Therefore, we decided to re-implement the tool completely in Isabelle. 
This has the nice advantage that the thousand lines of unverified Scala code could be replaced by only few hundred lines of ML~\cite{paulson96ML} code. 
Some lines of unverified code will always remain for the visualization. 
For our Isabelle re-implementation, they are narrowed down to the visualization with graphviz~\cite{graphviz}. 
This choice allowed to vastly improve the ratio between verified and unverified code. 
In total, after refactoring and a large cleanup, there are approximately 12000 lines of theory (excluding examples) and only about 400 lines of unverified ML code (only used for the visualization). 
This means the ratio of unverified vs.\ verified code has improved from approximately $300\%$ to approximately $3\%$. 
An additional advantage to implement our tool within Isabelle is that it is automatically maintained in the AFP~\cite{Network_Security_Policy_Verification-AFP}. 
Furthermore, the Isabelle/ML implementation is feature complete.

The Isabelle-based version of \topos{} can be used similarly to the Scala-based version. 
The main difference is that a user must start Isabelle. 
The commands which could be invoked from the Scala tool can be invoked within Isabelle.
Otherwise, the same properties as for the Scala-based tool are provided. 
In particular, a user does not need to prove anything at runtime. 
For a user without a formal background, Isabelle can be used as an interactive text editor (not as a theorem prover).

The results presented in this Ph.D.\ thesis are based on the Isabelle version of \topos{}. 
For the original paper where the case study of Section~\ref{sec:forte14:case-study} appeared first, the case study has been conducted with the Scala-based version of \topos{}. 
For this Ph.D.\ thesis, we have re-checked the case study with the Isabelle version of \topos{}.

\subsection{Computational Complexity} %
\label{subsec:toposcomputationcomplex}
\begin{figure}%
\centering%
	\includegraphics[width=0.4\linewidth]{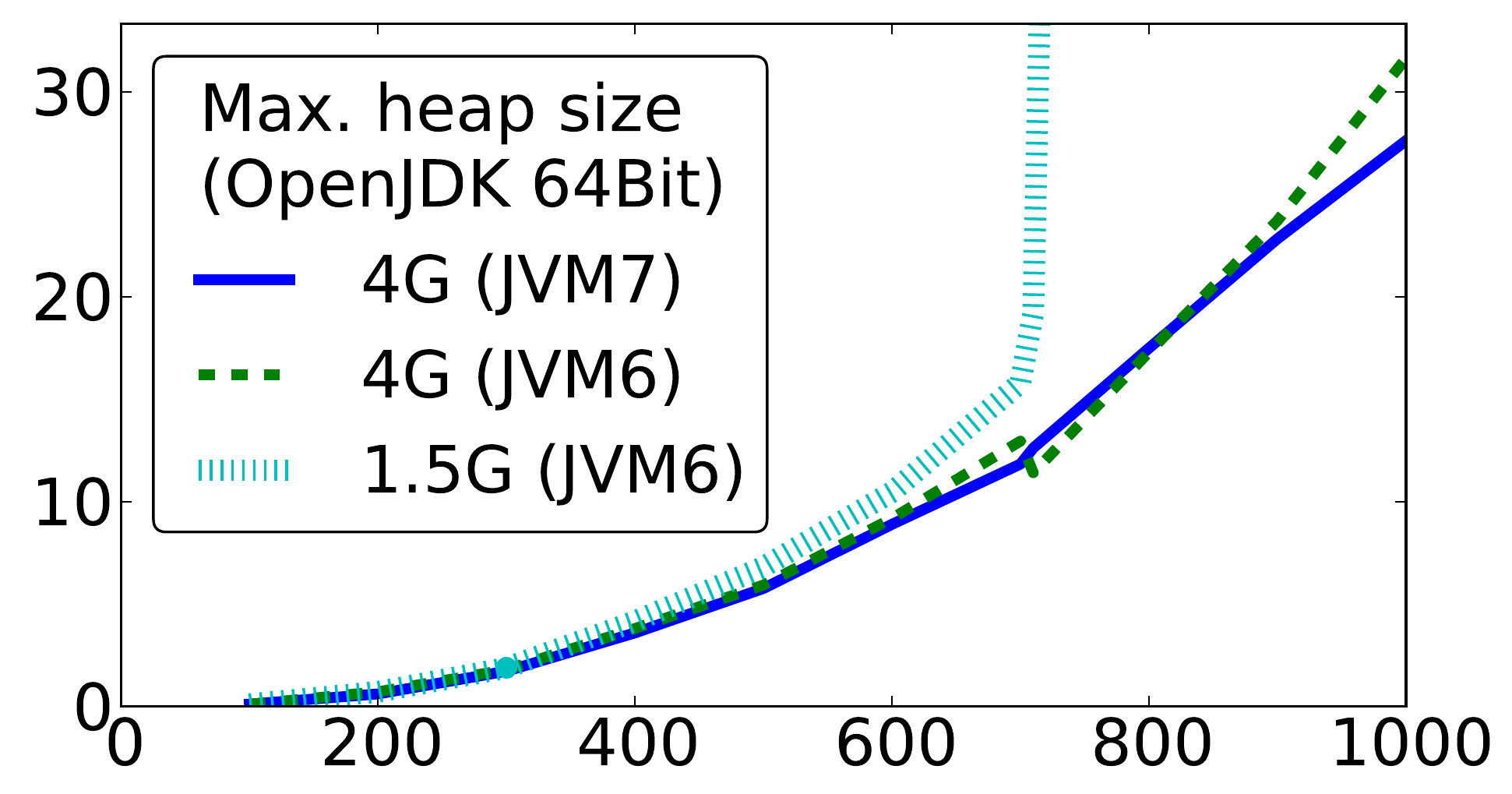}%
  \caption{Runtime of the policy construction algorithm for $100$ $\Phi$-structured invariants on an i7-2620M CPU (2.70GHz), Java Virtual Machine. X-axis: $\vert V \vert$, Y-axis: runtime in minutes.}
  \label{fig:toposforte:benchmark}
\end{figure}
\topos{} performs linear in the number of security invariants and quadratic in the number of hosts for $\Phi$-structured invariants. 
For scenarios with less than $100$ hosts, it responds interactively in less than $10$ seconds. 
With the grouping approach presented in~\cite{ou2005mulval}, 
we argue that often, setups with more than $100$ nodes are unlikely as large groups of identical hosts can be pooled into a small representative set, \eg instead of modeling $1000$ identical workstation PCs, only $2$ representatives of this large set are necessary for reasoning. 
For the presented case study (Section~\ref{sec:forte14:case-study}), all results are immediately available. 
Therefore, we provide an interactive convenient tool for designing and conceptualizing a network.

A benchmark of the automated policy construction, the most expensive algorithm, is presented in Figure~\ref{fig:toposforte:benchmark}. 
The benchmark is based on the Scala-based version of \topos{} to exclude overhead induced by Isabelle and to control the JVM heap size. 
For $\vert V \vert$ hosts, $\vert V \vert^2 / 4$ flows were created. 
With reasonable memory consumption, policies with up to $250$k flows can be processed in less than half an hour. 

\topos{} contains a lot of machine-generated code that is not optimized for performance but correctness. 
However, the overall theoretical and practical performance is sufficient for real-world usage. 
During our work with \eads{}, we never encountered any performance issues.

\section{Case Study: A Cabin Data Network}
\label{sec:forte14:case-study}
In this section, we present a slightly more complex scenario: a policy for a cabin data network for the general civil aviation. 
This example was chosen as security is very important in this domain and it provides a challenging interaction of different security invariants. 
It is a small imaginary toy example, developed in collaboration with \eads{}. 
To make it self-contained and accessible to readers without aeronautical background knowledge, it does not obey aeronautical standards (such as ARINC specification 664P5 \cite{arincdomains}). 
However, the scenario is plausible, \ie a real-world scenario may be similar. 
During our research, we also evaluated real-world scenarios in this domain. 
With this experience, we try to present a small, simplified, self-contained, plausible toy scenario that, however, preserves many real-world snares.

\medskip

\noindent The network consists of the following hosts.
\begin{description}
	\item[$\mvar{CC}$] The Cabin Core Server, a server that controls essential aircraft features, such as air conditioning and the wireless and wired telecommunication of the crew.
	\item[$\mvar{C1}$, $\mvar{C2}$] Two mobile devices for the crew to help them organize, \eg communicate, make announcements.
	\item[$\mvar{Wifi}$] A wifi hotspot that allows passengers to access the Internet with their own devices. Explicitly listed as it might also be responsible for billing passenger's Internet access.
	\item[$\mvar{IFEsrv}$] The In-Flight Entertainment server with movies, Internet access, etc. Master of the IFE displays.
	\item[$\mvar{IFE1}$, $\mvar{IFE2}$] Two In-Flight Entertainment displays, mounted at the back of passenger seats. They provide movies and Internet access. Thin clients, everything is streamed from the IFE server.
	\item[$\mvar{P1}$, $\mvar{P2}$] Two passenger-owned devices, \eg laptops, smartphones.
	\item[$\mvar{SAT}$]	A satellite uplink to the Internet.
\end{description}

\medskip

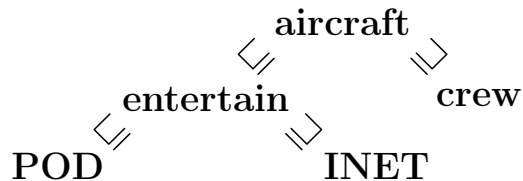
\begin{figure}[!htp]
       \centering
  			\begin{Large}
  			\begin{tikzpicture}
  			\node at (0,0) {$\textnormal{\textbf{aircraft}}$};
  			\node[rotate=45,anchor=north]  at (-1.4,-0.2) {$\sqsubseteq$};
  			\node[rotate=-45,anchor=north] at (+1.4,-0.2) {$\sqsupseteq$};
  			\node at (-1.8,-1.0) {$\textnormal{\textbf{entertain}}$};
  			\node at (1.8,-1.0) {$\textnormal{\textbf{crew}}$};
  			\node[rotate=45,anchor=north]  at (-3.3,-1.2) {$\sqsubseteq$};
  			\node[rotate=-45,anchor=north] at (-0.25,-1.2) {$\sqsupseteq$};
  			\node at (-3.75,-1.9) {$\textnormal{\textbf{POD}}$};
  			\node at (.45,-1.9) {$\textnormal{\textbf{INET}}$};
  			\end{tikzpicture}
  			\end{Large}
		\caption{Security domains of the cabin data network.}
		\label{fig:cabinnetworkhierarchy}
\end{figure}

\noindent The following three security invariants are specified.%
\begin{description}
\item[Security Invariant 1, Domain Hierarchy.] Four different security domains exist in the aircraft, cf.\ Figure~\ref{fig:cabinnetworkhierarchy}. 
They separate the \textbf{crew} domain, the \textbf{entertain}ment domain, the passenger-owned devices (\textbf{POD}) domain and the Internet (\textbf{INET}) domain. 
All devices belong to a domain. 

The following devices are in the \textbf{entertain} domain: $\mvar{IFEsrv}$, $\mvar{IFE1}$, $\mvar{IFE1}$. 

The Cabin Core Server, which is located in the \textbf{crew} domain, may also send to the entertain domain. 
Hence, it is trusted. 
Possible use cases include: Stewards coordinate food distribution or an announcement from the crew is send to the In-Flight Entertainment system (via $\mvar{CC}$) and distributed there to the IFE displays.

The $\mvar{Wifi}$ is located in the \textbf{POD} domain to be reachable by PODs. It is trusted to send to the entertain domain. 
Possible use cases include: A passenger subscribes a film from the IFE server to her notebook or establishes connections to the Internet. 

In the \textbf{INET} domain, the $\mvar{SAT}$ is isolated to prevent accesses from the Internet into the aircraft. 
%
\begin{IEEEeqnarray*}{l "c" l}
\mvar{CC} & \mapsto & \left(\mdef{level}:\ \mathit{crew}.\mathit{aircraft},\ \ \mdef{trust}:\ 1\right)\\
\mvar{C1} & \mapsto & \left(\mdef{level}:\ \mathit{crew}.\mathit{aircraft},\ \ \mdef{trust}:\ 0\right)\\
\mvar{C2} & \mapsto & \left(\mdef{level}:\ \mathit{crew}.\mathit{aircraft},\ \ \mdef{trust}:\ 0\right)\\
\mvar{IFEsrv} & \mapsto & \left(\mdef{level}:\ \mathit{entertain}.\mathit{aircraft},\ \ \mdef{trust}:\ 0\right)\\
\mvar{IFE1} & \mapsto & \left(\mdef{level}:\ \mathit{entertain}.\mathit{aircraft},\ \ \mdef{trust}:\ 0\right)\\
\mvar{IFE2} & \mapsto & \left(\mdef{level}:\ \mathit{entertain}.\mathit{aircraft},\ \ \mdef{trust}:\ 0\right)\\
\mvar{SAT} & \mapsto & \left(\mdef{level}:\ \mathit{INET}.\mathit{entertain}.\mathit{aircraft},\ \ \mdef{trust}:\ 0\right)\\
\mvar{Wifi} & \mapsto & \left(\mdef{level}:\ \mathit{POD}.\mathit{entertain}.\mathit{aircraft},\ \ \mdef{trust}:\ 1\right)\\
\mvar{P1} & \mapsto & \left(\mdef{level}:\ \mathit{POD}.\mathit{entertain}.\mathit{aircraft},\ \ \mdef{trust}:\ 0\right)\\
\mvar{P2} & \mapsto & \left(\mdef{level}:\ \mathit{POD}.\mathit{entertain}.\mathit{aircraft},\ \ \mdef{trust}:\ 0\right)\\
\end{IEEEeqnarray*}

\item[Security Invariant 2, Policy Enfrocement Point.]
\begin{sloppypar}The IFE displays are thin clients and strictly bound to their server. Peer to peer communication is prohibited. 
The Policy Enforcement Point template directly provides the respective access control restrictions.\end{sloppypar} 
%
\begin{IEEEeqnarray*}{l "c" l}
\mvar{IFEsrv} & \mapsto & \mdef{PolEnforcePointIN}\\
\mvar{IFE1} & \mapsto & \mdef{DomainMember}\\
\mvar{IFE2} & \mapsto & \mdef{DomainMember}\\
\end{IEEEeqnarray*}

\item[Security Invariant 3, Bell-LaPadula with Trust.]
Requirement 1 and 2 encode access control restrictions. 
Invariant 3 defines information flow restrictions by labeling confidential information sources according to the Bell-LaPadula model with trust. 
To protect the passenger's privacy when using the IFE displays, it is undesirable that the IFE displays communicate with anyone, except for the $\mvar{IFEsrv}$. 
Therefore, the IFE displays are marked as confidential. 
Note that requirement 2 dictates what can access the IFE displays and this requirement dictates what the IFE displays can send out. 
The $\mvar{IFEsrv}$ is considered a central trusted device. 
To enable passengers to surf the Internet on the IFE displays by forwarding the packets to the Internet or forward announcements from the crew, it must be allowed to declassify any information to the default (\ie $\mdef{unclassified}$) security level. 
Finally, the crew communication is considered more critical than the convenience features, therefore, $\mvar{CC}$, $\mvar{C1}$, and $\mvar{C2}$ are considered secret. 
As the $\mvar{IFEsrv}$ is trusted, it can receive and forward announcements from the crew.

\begin{IEEEeqnarray*}{l "c" l}
\mvar{CC} & \mapsto & \left(\mdef{level}:\ \mdef{secret},\ \ \mdef{trust}:\ \mdef{False}\right)\\
\mvar{C1} & \mapsto & \left(\mdef{level}:\ \mdef{secret},\ \ \mdef{trust}:\ \mdef{False}\right)\\
\mvar{C2} & \mapsto & \left(\mdef{level}:\ \mdef{secret}\,\ \ \mdef{trust}:\ \mdef{False}\right)\\
\mvar{IFE1} & \mapsto & \left(\mdef{level}:\ \mdef{confidential},\ \ \mdef{trust}:\ \mdef{False}\right)\\
\mvar{IFE2} & \mapsto & \left(\mdef{level}:\ \mdef{confidential},\ \ \mdef{trust}:\ \mdef{False}\right)\\
\mvar{IFEsrv} & \mapsto & \left(\mdef{level}:\ \mdef{unclassified},\ \ \mdef{trust}:\ \mdef{True}\right)\\
\end{IEEEeqnarray*}

\end{description}

This case study illustrates that this complex scenario can be divided into three security invariants that can be represented with the help of the previously presented templates. 
It also reveals that very few host attributes must be manually specified; the automatically added secure default attributes complete the configuration. 

\begin{figure}[!htp]
  \centering
       \centering
  		\includegraphics[width=0.65\linewidth]{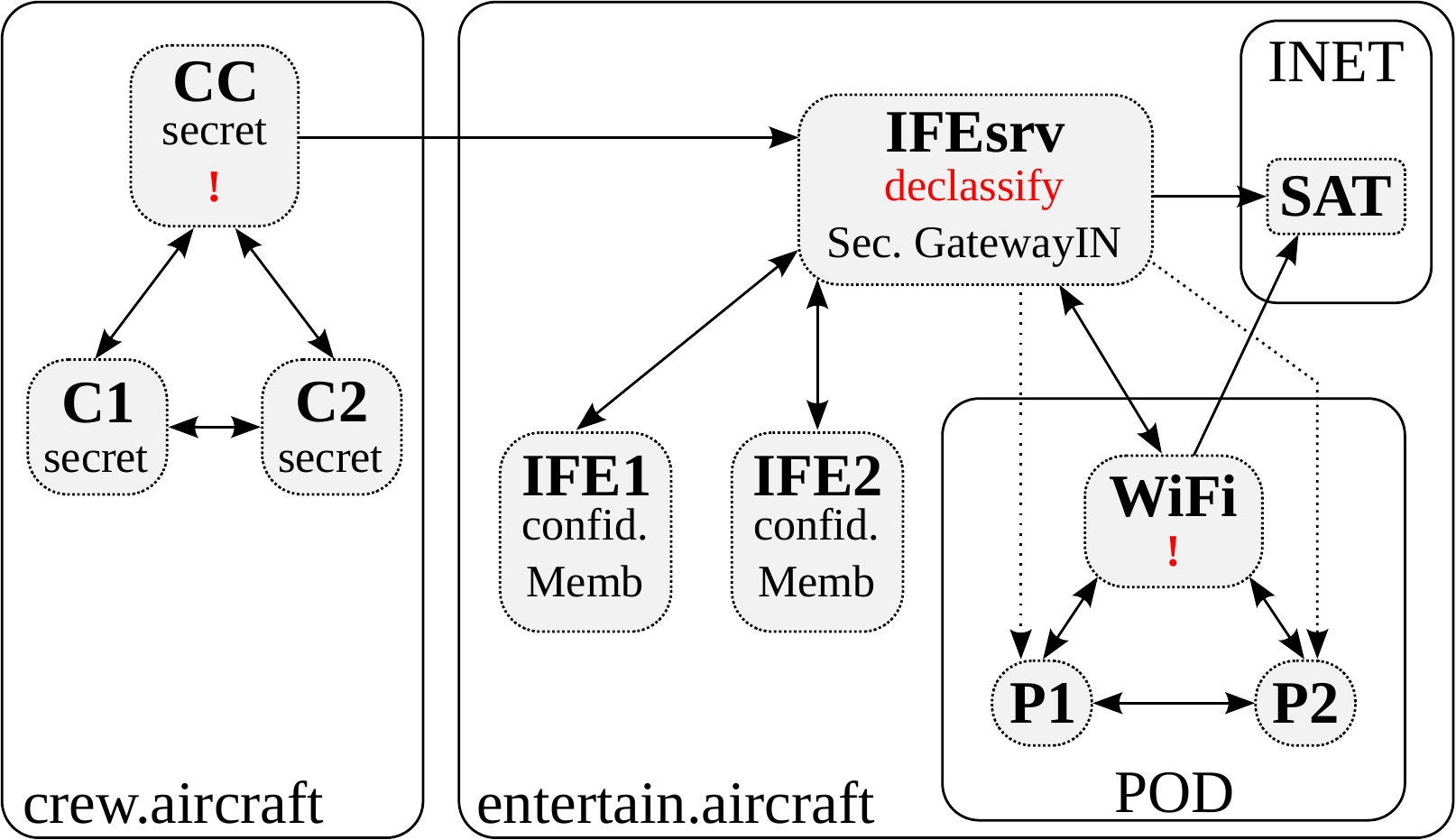}
  		\caption{Cabin network policy and hosts' attributes.}
  		\label{fig:cabinnetwork}
\end{figure}

The policy is illustrated in Figure~\ref{fig:cabinnetwork}. 
The different domains are illustrated and trusted devices according to the Domain Hierarchy are marked with an exclamation mark. 
The \emph{declassify} host attribute belongs to the Bell-LaPadula invariant and corresponds to a trusted host with the $\mdef{unclassified}$ security level. 

\paragraph*{Verifying the Policy}
Our tool verified that all security invariants are fulfilled. 

\paragraph*{Analyzing the Invariant Specification}
Focusing in \topos{}'s analysis capabilities, the following results were obtained. 
With the automated policy construction algorithm, an alternative policy was calculated. 
In Figure~\ref{fig:cabinnetwork}, the solid edges combined with the dashed edges\footnote{combined with all reflexive edges, \ie in-host communication} correspond to the uniquely defined policy with the maximum number of allowed flows. 
The solid lines were given by the policy, the dashed lines were calculated from the invariants. 
These `diffs' are computed and visualized automatically by \topos{}. 
They provide the end user with helpful feedback regarding `\emph{what do my invariants require?}' vs. `\emph{what does my policy specify?}'.
This results in a feedback loop we used extensively during our research to refine the policy and the invariants. 
It provides a `feeling' for the invariants. 

In this case study, two insights were obtained from this analysis. 
First, according to the security invariants, the IFE server could possibly connect to the passenger-owned devices directly. 
Presenting the question to an engineer, this would raise the question about whether this is actually acceptable. 
If this were not acceptable, it would indicate a bug in the specified security invariants. 
In this scenario however, it is acceptable and impossible by hardware constraints anyway. 
The second insight is that -- disregarding the two flows from the IFE server to the passenger-owned devices -- the policy only inferred from the invariants and the policy designed by hand coincide exactly. 
This is a strong indicator that the specified invariants `mean' the right thing.

\subsection{End-User Feedback Session}
\label{sec:forte14:userstudyresults}
\begin{table*}[t]
\centering
\begin{scriptsize}
\begin{tabular}{ @{}l r l r@{}c@{}r@{}c@{}r r@{}c@{}r@{}c@{}r r@{}c@{}r@{}c@{}r r@{}c@{}r@{}c@{}r@{}}
	\toprule
	Experience   & Participants       & Complexity           & \multicolumn{5}{r}{Valid} & \multicolumn{5}{r}{Violations} & \multicolumn{5}{r}{Missing} & \multicolumn{5}{r@{}}{Errors}\\
	\midrule
	Expert       & \studyExperEntries & medium but tricky    & \studyExperTopoValid & \studyExperTopoInvalid & \studyExperTopoMissing & \studyExperTopoError \\
	Intermediate & \studyInterEntries & medium               & \studyInterTopoValid & \studyInterTopoInvalid & \studyInterTopoMissing & \studyInterTopoError \\
	Novice       & \studyNovicEntries & medium               & \studyNovicTopoValid & \studyNovicTopoInvalid & \studyNovicTopoMissing & \studyNovicTopoError \\
	Total        & \studyTotalEntries & medium but tricky    & \studyTotalTopoValid & \studyTotalTopoInvalid & \studyTotalTopoMissing & \studyTotalTopoError \\
	\midrule
	\multicolumn{23}{@{}l@{}}{\footnotesize{Legend: median/arithmetic mean/std deviation}}\\
 \bottomrule
\end{tabular}
  \vskip-1ex
  \caption{Results of user feedback session.} 
  \label{tab:userstudy}
\end{scriptsize}
\end{table*}

\begin{table*}[t]
\centering
\begin{tabular}{ @{}l r@{}c@{}r@{}c@{}r r@{}c@{}r@{}c@{}r r r@{}}
	\toprule
	Experience   & \multicolumn{5}{r}{utility tool} & \multicolumn{5}{r}{utility idea} & acceptance idea & acceptance tool \\
	\midrule
	Expert       & \studyExperUtilityTool  & \studyExperUtilityIdea & \studyExperHelps & \studyExperWoulduse \\
	Intermediate & \studyInterUtilityTool  & \studyInterUtilityIdea & \studyInterHelps & \studyInterWoulduse \\
	Novice       & \studyNovicUtilityTool  & \studyNovicUtilityIdea & \studyNovicHelps & \studyNovicWoulduse \\
	Total        & \studyTotalUtilityTool  & \studyTotalUtilityIdea & \studyTotalHelps &  \studyTotalWoulduse \\
	\midrule
	\footnotesize{Utility measure:} & \multicolumn{12}{@{}p{24em}@{}}{\footnotesize{0) counter-productive, 1) more counter-productive than helpful, 2) neutral, 3) helpful, 4) extremely helpful}}\\
	\footnotesize{Legend:} & \multicolumn{12}{@{}p{24em}@{}}{\footnotesize{median/arithmetic mean/std deviation}}\\
 \bottomrule
\end{tabular}
  \vskip-1ex
  \caption{User's thoughts about the prototypical tool. } 
  \label{tab:userstudythoughts}
\end{table*}
\begin{figure}[pht]%
\centering%
	\fbox{\includegraphics[trim=1.5cm 1.5cm 1.5cm 1cm,clip=true,width=0.98\linewidth]{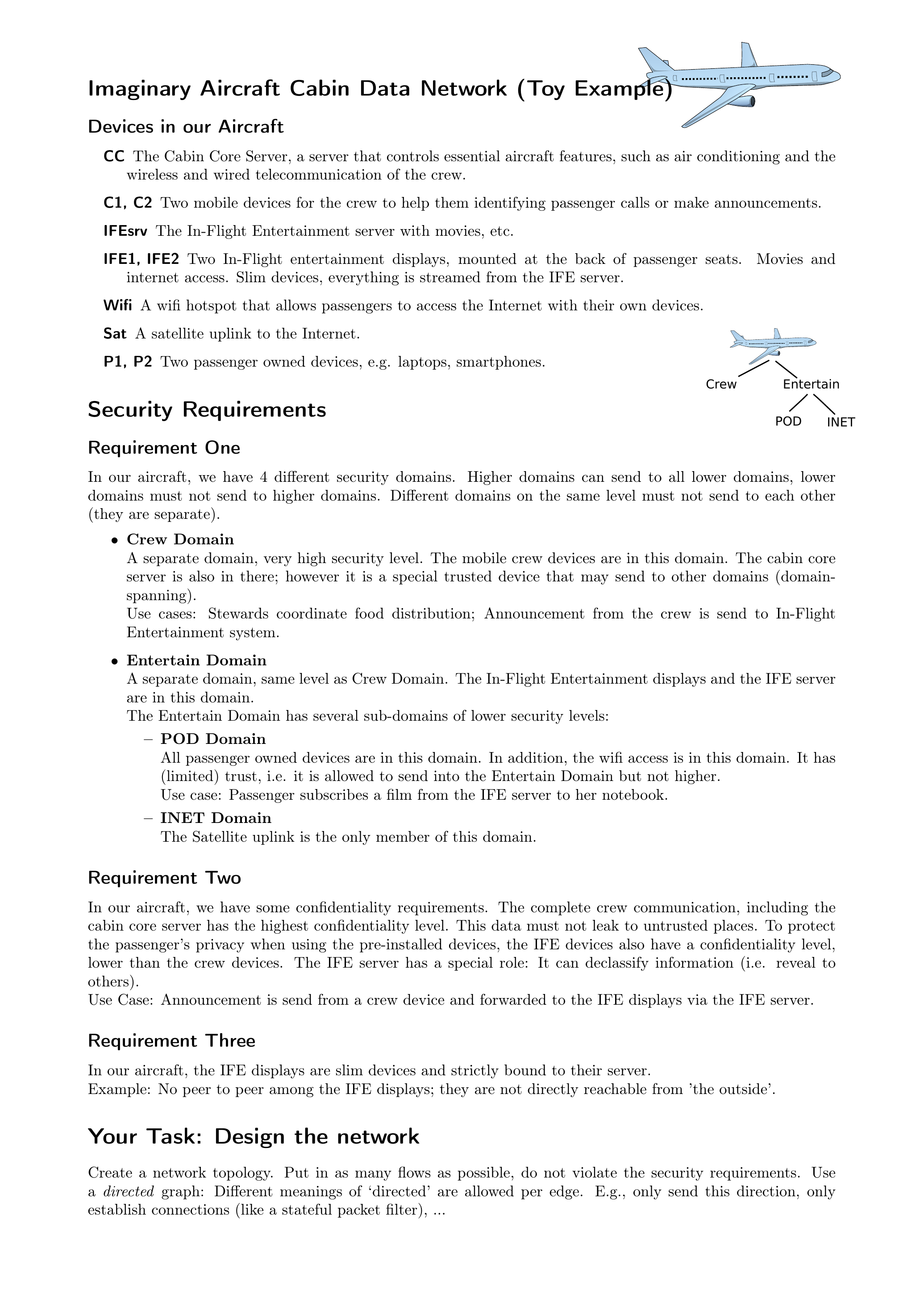}}%
  \caption{Scenario and task description as handed out to the study participants.}
  \label{fig:forte14:eadsfragebogen}
\end{figure}
%
%
%
%
To estimate the scenario's complexity, we asked some network professionals to design the scenario's policy. 
In total, \studyTotalEntries{} volunteered to participate in our study. 
No private or behavioral data was collected during this short study. 
The scenario description of Figure~\ref{fig:forte14:eadsfragebogen} was handed out to the participants. 
It was emphasized that no security requirement must be violated, but the participants should try to put the maximum number of flows in the network to fulfill as much as possible of the use cases. 
Therefore, the task was to maximize the allowed flows without violating any security invariant. 
This is a purely technical task. 
Afterwards, we asked the participants to rate the complexity of this exercise. 
The questions and answers were pre-formulated. 

The results are illustrated in Table~\ref{tab:userstudy}. 
It shows the perceived complexity of the task and the number of valid, violating, and missing flows the participants defined. 
We define the error count as the number of invalid plus the number of missing flows. 
Surprisingly, even expert network administrators made errors (both missing flows and security violations) when designing the policy. 

Afterwards, we presented our prototypical Scala tool to the participants and asked the participants about their thoughts about our tool. 
The results are summarized in Table~\ref{tab:userstudythoughts}. 
Note that this part of the user feedback session is neither a controlled experiment nor a scientific study: the conditions were not randomized, it suffers from demand bias, and there was no control group. 
The main evaluation of this work are the formal correctness proofs. 
Our only goal of this part of the user feedback session was to collect a rough feedback and some user's first thoughts. 

The overall feedback was that our tool is downright helpful (\studyTotalUtilityToolNotbl{}). 
We also introduced the idea behind our tool and admitted that the user interface of our prototype can be vastly improved. 
We asked the participants to judge the idea behind our tool.
An experienced participant raised concern that special training for novice administrators is necessary. 
However, the overall judgment about the idea was very positive and it was considered remarkably helpful (\studyTotalUtilityIdeaNotbl{}). 
In addition, \studyTotalHelps{} of the participants consider that our idea might help to manage large networks over a long period with many responsible persons. 
The positive feedback and recurring question we received during the user field study about where, when, and how expensive to obtain our tool was very motivating.
Our tool's graphical feedback was also much appreciated. 
Finally, \studyTotalWoulduse{} of the participants would want to use our tool\footnote{or a competing product, we asked to assume that an intuitive user interface is available} for similar tasks.

A detailed scenario description, the host attribute mappings, and raw data are available~\cite{userstudycabinnetwork2013}. 

\paragraph*{Related Study}
Johnson \etal\cite{johnson2010policytemplates} propose to split the process of policy authoring into three separate user roles. 
Though their policy framework and language differs from ours, their concepts can be abstracted to our system. 
They define \emph{policy element authors}, who have domain knowledge and define the necessary elements a policy can use. 
For our scenario, definitions such as the cabin core server, the in-flight entertainment system, etc.\ would be defined by this user role. 
A \emph{template author} is an experienced user who defines templates. 
For our scenario, the policy enforcement point invariant template and other templates would be defined by this user role. 
Finally, \emph{policy authors} instantiate the templates to create the actual policies. 
In our scenario, this user role would correspond to the participants of our end-user feedback session. 

In a user study with 20 experienced participants, Johnson \etal evaluate how well users can abstract over concrete policies by developing templates. 
Similar to our user feedback session, no control group was involved and demand bias was probably introduced. 
Nevertheless, the study shows that most users can successfully create templates. 
This hints that also our approach of splitting policies into generic templates and template instantiation may contribute to user-friendliness. 
The feature most used by the study participants was a policy preview feature. 
These results are in line with our user feedback session where the participants valued the automated policy construction in combination with the graph visualization. 
It shows that users in general value feedback about the meaning of a policy statement they have written. 

In the study by Johnson \etal almost half of the participants were concerned that templates may permit a policy which is too permissive. 
We believe that the principles our system is built upon prevent this issue: 
First, we built our security invariant templates with the monotonicity principle of ``prohibiting more is more or equally secure'' (Definition~\ref{def:securityinvariantmonotonicity}). 
In addition, the composability of several invariants provides the same guarantee (Section~\ref{subsec:forte:composability}). 
Second, the secure default parameter provides the incentive that adding more information makes the secure default parameter more secure (Definition~\ref{def:secure_default}).

\section{Example: Imaginary Factory Network}
\label{sec:example:imaginary-factory-network}
In this section, we give an example of an imaginary factory network. 
The example was chosen to show the interplay of several security invariants and to demonstrate their configuration effort. 
The specified security invariants deliberately include some minor specification problems. 
These problems will be used to demonstrate the inner workings of the algorithms and to visualize why some computed results will deviate from the expected results. 
At this point, we also try to outline the big picture of this thesis by including some results of the following chapters in this example. 

\subsection{Scenario Description}
The described scenario is an imaginary factory network. 
It consists of sensors and actuators in a cyber-physical system. 
The on-site production units of the factory are completely automated and there are no humans in the production area. 
Sensors are monitoring the building.
The production units are two robots which manufacture the actual goods. 
The robots are controlled by two control systems. 

\smallskip

\noindent
The network consists of the following hosts which are responsible for monitoring the building. 
\begin{description}
	\item[$\mvar{Statistics}$] A server which collects, processes, and stores all data from the sensors. 
	\item[$\mvar{SensorSink}$] A device which receives and collects data from the $\mvar{PresenceSensor}$, $\mvar{Webcam}$, $\mvar{TempSensor}$, and $\mvar{FireSensor}$. It sends the data to the $\mvar{Statistics}$ server. 
	\item[$\mvar{PresenceSensor}$] A sensor which detects whether a human is in the building. 
	\item[$\mvar{Webcam}$] A camera which monitors the building indoors. 
	\item[$\mvar{TempSensor}$] A sensor which measures the temperature in the building. 
	\item[$\mvar{FireSensor}$] A sensor which detects fire and smoke. 
\end{description}

\noindent
The following hosts are responsible for the production line. 
\begin{description}
	\item[$\mvar{MissionControl1}$] An automation device which drives and controls the robots. 
	\item[$\mvar{MissionControl2}$] An automation device which drives and controls the robots. It contains the logic for a secret production step, carried out only by $\mvar{Robot2}$. 
	\item[$\mvar{Watchdog}$] Regularly checks the health and technical readings of the robots. 
	\item[$\mvar{Robot1}$] Production robot unit 1. 
	\item[$\mvar{Robot2}$] Production robot unit 2. Performs a secret production step. 
	\item[$\mvar{AdminPc}$] A human administrator can log into this machine to supervise or troubleshoot the production. 
\end{description}

\noindent
We model one additional special host. 
\begin{description}
	\item[$\mvar{INET}$] A symbolic host which represents all hosts which are not part of this network.
\end{description}

\noindent
The security policy is visualized below. 
\bigskip

\noindent
\resizebox{0.99\linewidth}{!}{
\begin{tikzpicture}
\node (Statistics) at (0,0) {$\mvar{Statistics}$};
\node (SensorSink) at (0,-2) {$\mvar{SensorSink}$};
\node (PresenceSensor) at (-4,-4) {$\mvar{PresenceSensor}$};
\node (Webcam) at (-1.5,-4) {$\mvar{Webcam}$};
\node (TempSensor) at (1.5,-4) {$\mvar{TempSensor}$};
\node (FireSensor) at (4,-4) {$\mvar{FireSensor}$};
\node (MissionControl1) at (8,-2) {$\mvar{MissionControl1}$};
\node (MissionControl2) at (12,-2) {$\mvar{MissionControl2}$};
\node (Watchdog) at (5,-2) {$\mvar{Watchdog}$};
\node (Robot1) at (8,-4) {$\mvar{Robot1}$};
\node (Robot2) at (12,-4) {$\mvar{Robot2}$};
\node (AdminPc) at (10,0) {$\mvar{AdminPc}$};
\node (X) at (3.5,0) {$\mvar{INET}$};

\draw[myptr] (PresenceSensor) to (SensorSink);
\draw[myptr] (Webcam) to (SensorSink);
\draw[myptr] (TempSensor) to (SensorSink);
\draw[myptr] (FireSensor) to (SensorSink);
\draw[myptr] (SensorSink) to (Statistics);
\draw[myptr] (MissionControl1) to (Robot1);
\draw[myptr] (MissionControl1) to (Robot2);
\draw[myptr] (MissionControl2) to (Robot2);
\draw[myptr] (AdminPc) to (MissionControl2);
\draw[myptr] (AdminPc) to (MissionControl1);
\draw[myptr] (Watchdog) to (Robot1);
\draw[myptr] (Watchdog) to (Robot2);
\end{tikzpicture}%
}
\bigskip

The idea behind the policy is the following. 
The sensors on the left can all send their readings in a unidirectional fashion to the sensor sink, which forwards the data to the statistics server. 
In the production line, on the right, all devices will set up stateful connections. 
This means, once a connection is established, packet exchange can be bidirectional. 
This makes sure that the watchdog will receive the health information from the robots, the mission control machines will receive the current state of the robots, and the administrator can actually log into the mission control machines. 
The policy should only specify who is allowed to set up the connections. 
We will elaborate on the stateful implementation in Section~\ref{sec:example-factory:stateful}. 

\subsection{Specification of Security Invariants}
%
Several security invariants are specified. 

\begin{description}
\item[Security Invariant 1, BLP Basic. ]
	The sensors in the building may record any employee. 
	Due to privacy requirements, the sensor readings, processing, and storage of the data are treated with a high security level. 
	The presence sensor does not allow do identify an individual employee, hence produces less critical data, hence has a lower level. 
	\begin{IEEEeqnarray*}{l "c" l}
	\mvar{Statistics}     & \mapsto & 3\\
	\mvar{SensorSink}     & \mapsto & 3\\
	\mvar{PresenceSensor} & \mapsto & 2\\
	\mvar{Webcam}         & \mapsto & 3
	\end{IEEEeqnarray*}

\item[Security Invariant 2, BLP Basic. ]
	The production process is a corporate trade secret. 
	The mission control devices have the trade secrets in their program. 
	The important and secret step is done by $\mvar{MissionControl2}$. 
	\begin{IEEEeqnarray*}{l "c" l}
	\mvar{MissionControl1} & \mapsto & 1\\
	\mvar{MissionControl2} & \mapsto & 2\\
	\mvar{Robot1}            & \mapsto & 1\\
	\mvar{Robot2}            & \mapsto & 2
	\end{IEEEeqnarray*}
	
	Note that Invariant 1 and Invariant 2 are two distinct specifications. 
	They specify individual security goals independent of each other. 
	For example, in Invariant 1, $\mvar{MissionControl2}$ has the default security level $\bot = 0$ and in Invariant 2, $\mvar{PresenceSensor}$ has security level $\bot$. 
	Consequently, both cannot interact.

\item[Security Invariant 3, BLP Trusted. ]
	Monitoring the building while also ensuring privacy of the employees is an important goal for the company. 
	While the presence sensor only collects the single-bit information whether a human is present, the webcam allows identifying individual employees. 
	The data collected by the presence sensor is classified as secret while the data produced by the webcam is top secret. 
	The sensor sink only has the \emph{secret} security level, hence it is not allowed to process the data generated by the webcam. 
	However, the sensor sink aggregates all data and only distributes a statistical average which does not allow identifying individual employees. 
	It does not store the data over long periods. 
	Therefore, it is marked as trusted and may thus receive the webcam's data. 
	The statistics server, which archives all the data, is considered top secret. 
	\begin{IEEEeqnarray*}{l "c" l}
	\mvar{Statistics}     & \mapsto & \left(\mdef{level}:\ \mdef{topsecret},\ \ \mdef{trust}:\ \mdef{False}\right)\\
	\mvar{SensorSink}     & \mapsto & \left(\mdef{level}:\ \mdef{secret},\ \ \mdef{trust}:\ \mdef{True}\right)\\
	\mvar{PresenceSensor} & \mapsto & \left(\mdef{level}:\ \mdef{secret},\ \ \mdef{trust}:\ \mdef{False}\right)\\
	\mvar{Webcam}         & \mapsto & \left(\mdef{level}:\ \mdef{topsecret},\ \ \mdef{trust}:\ \mdef{False}\right)\\
	\end{IEEEeqnarray*}

\item[Security Invariant 4, Communication Partners. ]\begin{sloppypar}
	$\mvar{Robot2}$ carries out a mission-critical production step. 
	For its integrity, it must be made sure that $\mvar{Robot2}$ only receives packets from $\mvar{Robot1}$, the two mission control devices and the watchdog.
	\begin{IEEEeqnarray*}{l "c" l}
	\mvar{Robot2}            & \mapsto & \mconstr{Master}\ [\mvar{Robot1}, \mvar{MissionControl1}, \\
	                         &         & \phantom{\mconstr{Master}\ [}  \mvar{MissionControl2}, \mvar{Watchdog}] \\
	\mvar{MissionControl1}   & \mapsto & \mconstr{Care}\\
	\mvar{MissionControl2}   & \mapsto & \mconstr{Care}\\
	\mvar{Watchdog}          & \mapsto & \mconstr{Care}
	\end{IEEEeqnarray*}\end{sloppypar}
	
	Note that $\mvar{Robot1}$ is in the access list of $\mvar{Robot2}$, but it does not have the $\mconstr{Care}$ attribute. 
	This means, $\mvar{Robot1}$ can never access $\mvar{Robot2}$. 
	A tool could automatically detect such inconsistencies and emit a warning. 
	However, a tool should only emit a warning---not an error---because this setting could be intentional and desirable. 
	
	In our factory, this setting is currently desirable: 
	Three months ago, $\mvar{Robot1}$ had an irreparable hardware error and needed to be removed from the production line. 
	When removing $\mvar{Robot1}$ physically, all its host attributes were also deleted. 
	The access list of $\mvar{Robot2}$ was not changed. 
	It was planned that $\mvar{Robot1}$ will be replaced and later will have the same access rights again. 
	A few weeks later, a replacement for $\mvar{Robot1}$ arrived. 
	The replacement is also called $\mvar{Robot1}$. 
	The new robot arrived neither configured nor tested for the production. 
	After carefully testing $\mvar{Robot1}$, $\mvar{Robot1}$ has been given back the host attributes for the other security invariants. 
	Despite the ACL entry of $\mvar{Robot2}$, when $\mvar{Robot1}$ was added to the network, because of its missing $\mconstr{Care}$ attribute, it was not given automatically access to $\mvar{Robot2}$. 
	This prevented that $\mvar{Robot1}$ would accidentally impact $\mvar{Robot2}$ without being fully configured. 
	In our scenario, once $\mvar{Robot1}$ will be fully configured, tested, and verified, it will be given back the $\mconstr{Care}$ attribute. 
	
	In general, this design choice of the invariant template prevents that a newly added host may inherit access rights due to stale entries in access lists. 
	At the same time, it does not force administrators to clean up their access lists because a host may only be removed temporarily and wants to be given back its access rights later on. 
	Note that managing access lists scales quadratically in the number of hosts. 
	In contrast, the $\mconstr{Care}$ attribute can be considered as a Boolean flag which allows to temporarily enable or disable the access rights of a host locally without touching the carefully constructed access lists of other hosts. 
	It also prevents that new hosts which have the name of hosts removed long ago (but where stale access rights were not cleaned up) accidentally inherit their access rights. 
	
	This design of the invariant template was motivated by the requirements for the secure default parameter.

\item[Security Invariant 5, Domain Hierarchy. ]
	The production line is designed according to a strict command hierarchy. 
	On top of the hierarchy are control terminals which allow a human operator to intervene and supervise the production process. 
	On the level below, one distinguishes between supervision devices and control devices. 
	The watchdog is a typical supervision device whereas the mission control devices are control devices. 
	Directly below the control devices are the robots. 
	This is the structure that is necessary for the example. 
	However, the company defined a few more sub-departments for future use. 
	The full domain hierarchy tree is visualized below. 
	
	\begin{tikzpicture}
		\node at (0,0) {$\textnormal{\textbf{ControlTerminal}}$};
		\node[rotate=45,anchor=north]  at (-1.95,-0.2) {$\sqsubseteq$};
		\node[rotate=-45,anchor=north] at (+1.95,-0.2) {$\sqsupseteq$};
		\node at (-3,-0.8) {$\textnormal{\textbf{ControlDevices}}$};
		\node at (3,-0.8) {$\textnormal{\textbf{Supervision}}$};
		\node[rotate=45,anchor=north]  at (-5,-1.0) {$\sqsubseteq$};
		\node[rotate=90,anchor=east] at (-3,-0.97) {$\sqsubseteq$};
		\node[rotate=-45,anchor=north] at (-1,-1.0) {$\sqsupseteq$};
		\node at (-6,-1.6) {$\textnormal{\textbf{Robots}}$};
		\node at (-3,-1.6) {$\textnormal{\textbf{OtherStuff}}$};
		\node at (0,-1.6) {$\textnormal{\textbf{Domain3}}$};
		\node[rotate=90,anchor=east]  at (3,-0.97) {$\sqsubseteq$};
		\node[rotate=-45,anchor=north] at (4.5,-1.0) {$\sqsupseteq$};
		\node at (3,-1.6) {$\textnormal{\textbf{S}}_\mathbf{1}$};
		\node at (4.8,-1.6) {$\textnormal{\textbf{S}}_\mathbf{2}$};
	\end{tikzpicture}
	
	Apart from the watchdog, only the following linear part of the tree is used: $\textnormal{\textbf{Robots}} \sqsubseteq \textnormal{\textbf{ControlDevices}} \sqsubseteq \textnormal{\textbf{ControlTerminal}}$. 
	Because the watchdog is in a different domain, it needs a trust level of $1$ to access the robots it is monitoring. 
	%
	\begin{IEEEeqnarray*}{lcl}
	\mvar{MissionControl1} & \ \mapsto \ & \left(\mdef{level}:\ \mathit{ControlTerminal}.\mathit{ControlDevices},\ \             \mdef{trust}:\ 0\right)\\
	\mvar{MissionControl2} & \mapsto & \left(\mdef{level}:\ \mathit{ControlTerminal}.\mathit{ControlDevices},\ \                 \mdef{trust}:\ 0\right)\\
	\mvar{Watchdog}        & \mapsto & \left(\mdef{level}:\ \mathit{ControlTerminal}.\mathit{ControlDevices}.\mathit{Robots},\ \ \mdef{trust}:\ 0\right)\\
	\mvar{Robot1}            & \mapsto & \left(\mdef{level}:\ \mathit{ControlTerminal}.\mathit{Supervision},\ \ \mdef{trust}:\ 1\right)\\
	\mvar{Robot2}            & \mapsto & \left(\mdef{level}:\ \mathit{ControlTerminal}.\mathit{ControlDevices}.\mathit{Robots},\ \ \mdef{trust}:\ 0\right)\\
	\mvar{AdminPc}         & \mapsto & \left(\mdef{level}:\ \mathit{ControlTerminal},\ \                                         \mdef{trust}:\ 0\right)
	\end{IEEEeqnarray*}

\item[Security Invariant 6, Policy Enforcement Point. ]
	The sensors should not communicate with each other; all accesses must be mediated by the sensor sink. 
	%
	\begin{IEEEeqnarray*}{l "c" l}
	\mvar{SensorSink}     & \mapsto & \mconstr{PolEnforcePointIN}\\
	\mvar{PresenceSensor} & \mapsto & \mconstr{DomainMember}\\
	\mvar{Webcam}         & \mapsto & \mconstr{DomainMember}\\
	\mvar{TempSensor}     & \mapsto & \mconstr{DomainMember}\\
	\mvar{FireSensor}     & \mapsto & \mconstr{DomainMember}
	\end{IEEEeqnarray*}

\item[Security Invariant 7, Sink. ]
	The actual control program of the robots is a corporate trade secret. 
	The control commands must not leave the robots. 
	Therefore, they are declared information sinks. 
	In addition, the control command must not leave the mission control devices. 
	However, the two devices could possibly interact to synchronize and they must send their commands to the robots. 
	Therefore, they are labeled as sink pools. 
	%
	\begin{IEEEeqnarray*}{l "c" l}
	\mvar{MissionControl1} & \mapsto & \mconstr{SinkPool}\\
	\mvar{MissionControl2} & \mapsto & \mconstr{SinkPool}\\
	\mvar{Robot1}            & \mapsto & \mconstr{Sink}\\
	\mvar{Robot2}            & \mapsto & \mconstr{Sink}
	\end{IEEEeqnarray*}

\item[Security Invariant 8, Subnets. ]\begin{sloppypar}
	The sensors, including their sink and statistics server are located in their own subnet and must not be accessible from elsewhere. 
	Also, the administrator's PC is in its own subnet. 
	The production units (mission control and robots) are already isolated by the DomainHierarchy and are not added to a subnet explicitly. 
	%
	\begin{IEEEeqnarray*}{l "c" l}
	\mvar{Statistics}      & \mapsto & \mconstr{Subnet}\ 1\\
	\mvar{SensorSink}      & \mapsto & \mconstr{Subnet}\ 1\\
	\mvar{PresenceSensor}  & \mapsto & \mconstr{Subnet}\ 1\\
	\mvar{Webcam}          & \mapsto & \mconstr{Subnet}\ 1\\
	\mvar{TempSensor}      & \mapsto & \mconstr{Subnet}\ 1\\
	\mvar{FireSensor}      & \mapsto & \mconstr{Subnet}\ 1\\
	\mvar{AdminPc}         & \mapsto & \mconstr{Subnet}\ 4
	\end{IEEEeqnarray*}
	\end{sloppypar}

\item[Security Invariant 9, SubnetsInGW. ]
	The statistics server is further protected from external accesses. 
	Another, smaller subnet is defined with the only member being the statistics server. 
	The only way it may be accessed is via that sensor sink. 
	%
	\begin{IEEEeqnarray*}{l "c" l}
	\mvar{Statistics}      & \mapsto & \mconstr{Member}\\
	\mvar{SensorSink}      & \mapsto & \mconstr{InboundGateway}
	\end{IEEEeqnarray*}

\item[Security Invariant 10, NonInterference. ]
	Finally, there is a final constraint. 
	The fire sensor is managed by an external company and has a built-in GSM module to call the fire fighters in case of an emergency. 
	This additional, out-of-band connectivity is not modeled. 
	However, the contract defines that the company's administrator must not interfere in any way with the fire sensor. 
	%
	\begin{IEEEeqnarray*}{l "c" l}
	\mvar{Statistics}      & \mapsto & \mconstr{Unrelated}\\
	\mvar{SensorSink}      & \mapsto & \mconstr{Unrelated}\\
	\mvar{PresenceSensor}  & \mapsto & \mconstr{Unrelated}\\
	\mvar{Webcam}          & \mapsto & \mconstr{Unrelated}\\
	\mvar{TempSensor}      & \mapsto & \mconstr{Unrelated}\\
	\mvar{FireSensor}      & \mapsto & \mconstr{Interfering}\\ 
	\mvar{MissionControl1} & \mapsto & \mconstr{Unrelated}\\
	\mvar{MissionControl2} & \mapsto & \mconstr{Unrelated}\\
	\mvar{Watchdog}        & \mapsto & \mconstr{Unrelated}\\
	\mvar{Robot1}            & \mapsto & \mconstr{Unrelated}\\
	\mvar{Robot2}            & \mapsto & \mconstr{Unrelated}\\
	\mvar{AdminPc}         & \mapsto & \mconstr{Interfering}\\ 
	\mvar{INET}            & \mapsto & \mconstr{Unrelated}
	\end{IEEEeqnarray*}
	
    As discussed in Section~\ref{sinvar:noninterference}, this invariant is very strict and rather theoretical. 
    It is not $\Phi$-structured and may produce an exponential number of offending flows. 
    Therefore, we exclude it by default from our algorithms for now. 
\end{description}

\subsection{Policy Verification}
The given policy fulfills all the specified security invariants. 
Also, including invariant 10 (NonInterference), the policy fulfills all security invariants. 

The question, \textit{``how good are the specified security invariants?''} remains. 
Therefore, we use the algorithm from Section~\ref{subsec:policyconstruction} to generate a policy. 
Then, we will compare our manually-specified policy with the automatically generated one. 
If we exclude the NonInterference invariant from the policy construction, we know that the resulting policy must be maximal. 
Therefore, the computed policy reflects the view of the specified security invariants and, thus, gives a direct feedback whether the specified security invariants express the right thing. 
By maximality of the computed policy and monotonicity, we know that our manually-specified policy must be a subset of the computed policy. 
This allows comparing the manually-specified policy to the policy implied by the security invariants: 
If there are too many flows which are allowed according to the computed policy but which are not in our manually-specified policy, we can conclude that our security invariants are not strict enough. 

We visualize this comparison below. 
The solid edges correspond to the manually-specified policy. 
The dashed edges correspond to the flows which would be additionally permitted by the computed policy. 
\bigskip

\noindent
\resizebox{0.99\linewidth}{!}{
\begin{tikzpicture}
\node (Statistics) at (0,0) {$\mvar{Statistics}$};
\node (SensorSink) at (0,-2) {$\mvar{SensorSink}$};
\node (PresenceSensor) at (-4,-4) {$\mvar{PresenceSensor}$};
\node (Webcam) at (-1.5,-4) {$\mvar{Webcam}$};
\node (TempSensor) at (1.5,-4) {$\mvar{TempSensor}$};
\node (FireSensor) at (4,-4) {$\mvar{FireSensor}$};
\node (MissionControl1) at (8,-2) {$\mvar{MissionControl1}$};
\node (MissionControl2) at (12,-2) {$\mvar{MissionControl2}$};
\node (Watchdog) at (5,-2) {$\mvar{Watchdog}$};
\node (Robot1) at (8,-4) {$\mvar{Robot1}$};
\node (Robot2) at (12,-4) {$\mvar{Robot2}$};
\node (AdminPc) at (10,0) {$\mvar{AdminPc}$};
\node (X) at (3.5,0) {$\mvar{INET}$};

\draw[myptr] (PresenceSensor) to (SensorSink);
\draw[myptr] (Webcam) to (SensorSink);
\draw[myptr] (TempSensor) to (SensorSink);
\draw[myptr] (FireSensor) to (SensorSink);
\draw[myptr] (SensorSink) to (Statistics);
\draw[myptr] (MissionControl1) to (Robot1);
\draw[myptr] (MissionControl1) to (Robot2);
\draw[myptr] (MissionControl2) to (Robot2);
\draw[myptr] (AdminPc) to (MissionControl2);
\draw[myptr] (AdminPc) to (MissionControl1);
\draw[myptr] (Watchdog) to (Robot1);
\draw[myptr] (Watchdog) to (Robot2);

\draw[myptrdotted] (Statistics) to[loop left] (Statistics);
\draw[myptrdotted] (SensorSink) to[loop left] (SensorSink);
\draw[myptrdotted] (SensorSink) to[bend left=15] (Webcam);
\draw[myptrdotted] (PresenceSensor) to[loop below] (PresenceSensor);
\draw[myptrdotted] (Webcam) to[loop below] (Webcam);
\draw[myptrdotted] (TempSensor) to[loop below] (TempSensor);
\draw[myptrdotted] (TempSensor) to (X);
\draw[myptrdotted] (FireSensor) to[loop below] (FireSensor);
\draw[myptrdotted] (FireSensor) to (X);
\draw[myptrdotted] (MissionControl1) to[loop above] (MissionControl1);
\draw[myptrdotted] (MissionControl1) to (MissionControl2);
\draw[myptrdotted] (MissionControl2) to[loop above] (MissionControl2);
\draw[myptrdotted] (Watchdog) to (MissionControl1);
\draw[myptrdotted] (Watchdog) to[bend left=30] (MissionControl2);
\draw[myptrdotted] (Watchdog) to[loop above] (Watchdog);
\draw[myptrdotted] (Watchdog) to (X);
\draw[myptrdotted] (Robot1) to[loop below] (Robot1);
\draw[myptrdotted] (Robot2) to[loop below] (Robot2);
\draw[myptrdotted] (AdminPc) to (Watchdog);
\draw[myptrdotted] (AdminPc) to (Robot1);
\draw[myptrdotted] (AdminPc) to[loop right] (AdminPc);
\draw[myptrdotted] (AdminPc) to (X);
\draw[myptrdotted] (X) to[loop left] (X);

\end{tikzpicture}%
}
\bigskip

The comparison reveals that the following flows would be additionally permitted. 
We will discuss whether this is acceptable or if the additional permissions indicates that we probably forgot to specify a security goal. 
\begin{itemize}
	\item All reflexive flows, \ie all hosts can communicate with themselves. 
		  Since each host in the policy corresponds to one physical entity, there is no need to explicitly prohibit or allow in-host communication. 
	\item The $\mvar{SensorSink}$ may access the $\mvar{Webcam}$. 
		  Both share the same security level, there is no problem with this possible information flow. 
		  Technically, a bi-directional connection may even be desirable, since this allows the sensor sink to influence the video stream, \eg request a lower bit rate if it is overloaded. 
    \item Both the $\mvar{TempSensor}$ and the $\mvar{FireSensor}$ may access the Internet. 
    	      No security level or other privacy concerns are specified for them. 
    	      This may raise the question whether this data is indeed public. 
    	      It is up to the company to decide that this data should also be considered confidential. 
    \item $\mvar{MissionControl1}$ can send to $\mvar{MissionControl2}$. 
    	      This may be desirable since it was stated anyway that the two may need to cooperate. 
    	      Note that the opposite direction is definitely prohibited since the critical and secret production step only known to $\mvar{MissionControl2}$ must not leak. 
    \item The $\mvar{Watchdog}$ may access $\mvar{MissionControl1}$, $\mvar{MissionControl2}$, and the $\mvar{INET}$. 
    		  While it may be acceptable that the watchdog which monitors the robots may also access the control devices, 
    		  it should raise a concern that the watchdog may freely send data to the Internet. 
    		  Indeed, the watchdog can access devices which have corporate trade secrets stored but it was never specified that the watchdog should be treated confidentially. 
    		  Note that in the current setting, the trade secrets will never leave the robots. 
    		  This is because the policy only specifies a unidirectional information flow from the watchdog to the robots; the robots will not leak any information back to the watchdog. 
    		  This also means that the watchdog cannot actually monitor the robots. 
    		  Later, when implementing the scenario, we will see that the simple, hand-waving argument from the beginning that ``\textit{the watchdog connects to the robots and the robots send back their data over the established connection}'' will not work because of this possible information leak. 
    \item The $\mvar{AdminPc}$ is allowed to access the $\mvar{Watchdog}$, $\mvar{Robot1}$, and the $\mvar{INET}$. 
    		  Since this machine is trusted anyway, our fictional company does not see a problem with this. 
\end{itemize}

\subsection{Outlook: About NonInterference}
\label{subsec:example:factory:noninterference}
The NonInterference template was deliberately selected for our scenario as one of the `problematic' and rather theoretical invariants. 
Our framework allows to specify almost arbitrary invariant templates. 
We concluded that all non-$\Phi$-structured invariants which may produce an exponential number of offending flows are problematic for practical use. 
This includes ``Comm.\ With'' (Section~\ref{sinvar:commwith}), ``Not Comm.\ With'' (Section~\ref{sinvar:notcommwith}), Dependability (Section~\ref{sinvar:dependability}), and NonInterference (Section~\ref{sinvar:noninterference}). 
In this section, we discuss the consequences of the NonInterference invariant for automated policy construction. 
We will conclude that, though we can solve all technical challenges, said invariants are---due to their inherent ambiguity---not very well suited for automated policy construction. 

The computed maximum policy does not fulfill invariant 10 (NonInterference). 
This is because the fire sensor and the administrator's PC may be indirectly connected over the Internet. 

Since the NonInterference template may produce an exponential number of offending flows, it is infeasible to try our automated policy construction algorithm with it. 
We have tried to do so on a machine with \SI{128}{\giga\byte} of memory but after a few minutes, the computation ran out of memory. 
On said machine, we were unable to run our policy construction algorithm with the NonInterference invariant for more than five hosts. 

In Chapter~\ref{chap:hilbertsoffending}, we will improve the policy construction algorithm. 
The new algorithm instantly returns a solution for this scenario with a very small memory footprint. 

However, it is an inherent property of the NonInterferance template (and similar templates), that the set of offending flows is not uniquely defined. 
Consequently, since several solutions are possible, even our new algorithm may not be able to compute one maximum solution. 
It would be possible to construct some maximal solution, however, this would require to enumerate all offending flows, which is infeasible. 
Therefore, our algorithm can only return some (valid but probably not maximal) solution for non-$\Phi$-structured invariants. 

As a human, we know the scenario and the intention behind the policy. 
Probably, the best solution for policy construction with the NonInterferance property would be to restrict outgoing edges from the fire sensor. 
If we consider the policy above which was constructed without NonInterference, if we cut off the fire sensor from the Internet, we get a valid policy for the NonInterference property. 
Unfortunately, an algorithm does not have the information of which flows we would like to cut first and the algorithm needs to make some choice. 
In this example, the algorithm decides to isolate the administrator's PC from the rest of the world. 
This is also a valid solution. 
We could change the order of the elements to tell the algorithm which edges we would rather sacrifice than others. 
This may help but requires some additional input. 
The author personally prefers to construct only maximum policies with $\Phi$-structured invariants and afterwards fix the policy manually for the remaining non-$\Phi$-structured invariants. 
Though our new algorithm gives better results and returns instantly, the very nature of invariant templates with an exponential number of offending flows tells that these invariants are problematic for automated policy construction.

\subsection{Outlook: Stateful Implementation}
\label{sec:example-factory:stateful}
In this section, we will implement the policy and deploy it in a network. 
This requires discussing packet flow on the network level, which is usually bidirectional for TCP. 
However, our security policy is on the connection level (Def.\ \ref{def:securitypolicy}) and includes unidirectional flows. 
As the scenario description stated, all devices in the production line should establish stateful connections which allows -- once the connection is established -- packets to travel in both directions. 
This is necessary for the watchdog, the mission control devices, and the administrator's PC to actually perform their task. 

We compute a stateful implementation. 
We will elaborate on the criteria and the algorithms for this in Chapter~\ref{chap:esss14}. 
Below, the stateful implementation is visualized. 
It consists of the policy as visualized above. 
In addition, dashed edges visualize where answer packets are permitted. 
\bigskip

\noindent
\resizebox{0.99\linewidth}{!}{
\begin{tikzpicture}
\node (Statistics) at (0,0) {$\mvar{Statistics}$};
\node (SensorSink) at (0,-2) {$\mvar{SensorSink}$};
\node (PresenceSensor) at (-4,-4) {$\mvar{PresenceSensor}$};
\node (Webcam) at (-1.5,-4) {$\mvar{Webcam}$};
\node (TempSensor) at (1.5,-4) {$\mvar{TempSensor}$};
\node (FireSensor) at (4,-4) {$\mvar{FireSensor}$};
\node (MissionControl1) at (8,-2) {$\mvar{MissionControl1}$};
\node (MissionControl2) at (12,-2) {$\mvar{MissionControl2}$};
\node (Watchdog) at (5,-2) {$\mvar{Watchdog}$};
\node (Robot1) at (8,-4) {$\mvar{Robot1}$};
\node (Robot2) at (12,-4) {$\mvar{Robot2}$};
\node (AdminPc) at (10,0) {$\mvar{AdminPc}$};
\node (X) at (3.5,0) {$\mvar{INET}$};

\draw[myptr] (PresenceSensor) to (SensorSink);
\draw[myptr] (Webcam) to (SensorSink);
\draw[myptr] (TempSensor) to (SensorSink);
\draw[myptr] (FireSensor) to (SensorSink);
\draw[myptr] (SensorSink) to (Statistics);
\draw[myptr] (MissionControl1) to (Robot1);
\draw[myptr] (MissionControl1) to (Robot2);
\draw[myptr] (MissionControl2) to (Robot2);
\draw[myptr] (AdminPc) to (MissionControl2);
\draw[myptr] (AdminPc) to (MissionControl1);
\draw[myptr] (Watchdog) to (Robot1);
\draw[myptr] (Watchdog) to (Robot2);

\draw[myptrdotted] (SensorSink) to[bend left=15, shorten <=0.6em,shorten >=0.2em] (Webcam);
\draw[myptrdotted] (Statistics) to[bend left=15, shorten <=0.6em,shorten >=0.2em] (SensorSink);
\end{tikzpicture}%
}
\bigskip

As can be seen, only the flows between $\mvar{SensorSink} \leftrightarrow \mvar{Webcam}$ and $\mvar{Statistics} \leftrightarrow \mvar{SensorSink}$ are allowed to be stateful. 
This setup cannot be practically deployed because the watchdog, the mission control devices, and the administrator's PC also need to set up stateful connections. 
Previous section's discussion already hinted at this problem. 
The reason why the desired stateful connections are not permitted is due to information leakage. 
In detail: Security Invariant 2 (trade secrets) and Security Invariant 7 (robots information sink) are responsible. 
Both invariants prevent that any data leaves the robots and the mission control devices. 
To verify this suspicion, the two invariants are removed and the stateful flows are computed again. 
The result is visualized below. 
\bigskip


\noindent
\resizebox{0.99\linewidth}{!}{
\begin{tikzpicture}
\node (Statistics) at (0,0) {$\mvar{Statistics}$};
\node (SensorSink) at (0,-2) {$\mvar{SensorSink}$};
\node (PresenceSensor) at (-4,-4) {$\mvar{PresenceSensor}$};
\node (Webcam) at (-1.5,-4) {$\mvar{Webcam}$};
\node (TempSensor) at (1.5,-4) {$\mvar{TempSensor}$};
\node (FireSensor) at (4,-4) {$\mvar{FireSensor}$};
\node (MissionControl1) at (8,-2) {$\mvar{MissionControl1}$};
\node (MissionControl2) at (12,-2) {$\mvar{MissionControl2}$};
\node (Watchdog) at (5,-2) {$\mvar{Watchdog}$};
\node (Robot1) at (8,-4) {$\mvar{Robot1}$};
\node (Robot2) at (12,-4) {$\mvar{Robot2}$};
\node (AdminPc) at (10,0) {$\mvar{AdminPc}$};
\node (X) at (3.5,0) {$\mvar{INET}$};

\draw[myptr] (PresenceSensor) to (SensorSink);
\draw[myptr] (Webcam) to (SensorSink);
\draw[myptr] (TempSensor) to (SensorSink);
\draw[myptr] (FireSensor) to (SensorSink);
\draw[myptr] (SensorSink) to (Statistics);
\draw[myptr] (MissionControl1) to (Robot1);
\draw[myptr] (MissionControl1) to (Robot2);
\draw[myptr] (MissionControl2) to (Robot2);
\draw[myptr] (AdminPc) to (MissionControl2);
\draw[myptr] (AdminPc) to (MissionControl1);
\draw[myptr] (Watchdog) to (Robot1);
\draw[myptr] (Watchdog) to (Robot2);

\draw[myptrdotted] (SensorSink) to[bend left=15, shorten <=0.6em,shorten >=0.2em] (Webcam);
\draw[myptrdotted] (Statistics) to[bend left=15, shorten <=0.6em,shorten >=0.2em] (SensorSink);
\draw[myptrdotted] (Robot1) to[bend left=15, shorten <=0.6em,shorten >=0.2em] (MissionControl1);
\draw[myptrdotted] (Robot2) to[bend left=10, shorten <=0.6em,shorten >=0.2em] (MissionControl1);
\draw[myptrdotted] (Robot2) to[bend left=15, shorten <=0.6em,shorten >=0.2em] (MissionControl2);
\draw[myptrdotted] (MissionControl2) to[bend left=15, shorten <=0.6em,shorten >=0.2em] (AdminPc);
\draw[myptrdotted] (MissionControl1) to[bend left=15, shorten <=0.6em,shorten >=0.2em] (AdminPc);
\draw[myptrdotted] (Robot1) to[bend left=10, shorten <=0.6em,shorten >=0.2em] (Watchdog);
\draw[myptrdotted] (Robot2) to[bend left=10, shorten <=0.6em,shorten >=0.2em] (Watchdog);
\end{tikzpicture}%
}
\bigskip

This stateful policy could be transformed into a fully functional implementation. 
However, there would be no security invariants specified which protect the trade secrets. 
Without those two invariants, the invariant specification is too permissive. 
For example, if we recompute the maximum policy, we can see that the robots and mission control can leak any data to the Internet. 
Even without the maximum policy, in the stateful policy above, it can be seen that $\mvar{MissionControl1}$ can exfiltrate information from robot 2, once it establishes a stateful connection. 


Therefore, the two invariants are not removed but repaired. 
The goal is to allow the watchdog, administrator's pc, and the mission control devices to set up stateful connections without leaking corporate trade secrets to the outside. 

First, we repair invariant 2. 
On the one hand, the watchdog should be able to send packets both to $\mvar{Robot1}$ and to $\mvar{Robot2}$. 
$\mvar{Robot1}$ has a security level of $1$ and $\mvar{Robot2}$ has a security level of $2$. 
Consequently, in order to be allowed to send packets to both, $\mvar{Watchdog}$ must have a security lvel not higher than $1$. 
On the other hand, the $\mvar{Watchdog}$ should be able to receive packets from both. 
By the same argument, it must have a security level of at least $2$. 
Consequently, it is impossible to express the desired meaning in the simple BLP template. 
There are only two solutions to the problem: Either the company installs one watchdog for each security level, or the watchdog must be trusted. 
We decide for the latter option and upgrade the template to the Bell-LaPadula model with trust. 
We define the watchdog as trusted entity with a security level of $1$. 
This means, it can receive packets from and send packets to both robots but it cannot leak information to the outside world. 
We do the same for the $\mvar{AdminPc}$.

Then, we repair invariant 7. 
We realize that the following set of hosts forms one big pool of devices which must all somehow interact but where information must not leave the pool: 
The administrator's PC, the mission control devices, the robots, and the watchdog. 
Therefore, all those devices are configured to be in the same $\mconstr{SinkPool}$. 

The computed stateful policy with the repaired invariants is visualized below. 
\bigskip

\noindent
\resizebox{0.99\linewidth}{!}{
\begin{tikzpicture}
\node (Statistics) at (0,0) {$\mvar{Statistics}$};
\node (SensorSink) at (0,-2) {$\mvar{SensorSink}$};
\node (PresenceSensor) at (-4,-4) {$\mvar{PresenceSensor}$};
\node (Webcam) at (-1.5,-4) {$\mvar{Webcam}$};
\node (TempSensor) at (1.5,-4) {$\mvar{TempSensor}$};
\node (FireSensor) at (4,-4) {$\mvar{FireSensor}$};
\node (MissionControl1) at (8,-2) {$\mvar{MissionControl1}$};
\node (MissionControl2) at (12,-2) {$\mvar{MissionControl2}$};
\node (Watchdog) at (5,-2) {$\mvar{Watchdog}$};
\node (Robot1) at (8,-4) {$\mvar{Robot1}$};
\node (Robot2) at (12,-4) {$\mvar{Robot2}$};
\node (AdminPc) at (10,0) {$\mvar{AdminPc}$};
\node (X) at (3.5,0) {$\mvar{INET}$};

\draw[myptr] (PresenceSensor) to (SensorSink);
\draw[myptr] (Webcam) to (SensorSink);
\draw[myptr] (TempSensor) to (SensorSink);
\draw[myptr] (FireSensor) to (SensorSink);
\draw[myptr] (SensorSink) to (Statistics);
\draw[myptr] (MissionControl1) to (Robot1);
\draw[myptr] (MissionControl1) to (Robot2);
\draw[myptr] (MissionControl2) to (Robot2);
\draw[myptr] (AdminPc) to (MissionControl2);
\draw[myptr] (AdminPc) to (MissionControl1);
\draw[myptr] (Watchdog) to (Robot1);
\draw[myptr] (Watchdog) to (Robot2);

\draw[myptrdotted] (SensorSink) to[bend left=15, shorten <=0.6em,shorten >=0.2em] (Webcam);
\draw[myptrdotted] (Statistics) to[bend left=15, shorten <=0.6em,shorten >=0.2em] (SensorSink);
\draw[myptrdotted] (Robot1) to[bend left=10, shorten <=0.6em,shorten >=0.2em] (MissionControl1);
\draw[myptrdotted] (Robot2) to[bend left=10, shorten <=0.6em,shorten >=0.2em] (MissionControl2);
\draw[myptrdotted] (MissionControl2) to[bend left=15, shorten <=0.6em,shorten >=0.2em] (AdminPc);
\draw[myptrdotted] (MissionControl1) to[bend left=15, shorten <=0.6em,shorten >=0.2em] (AdminPc);
\draw[myptrdotted] (Robot1) to[bend left=10, shorten <=0.6em,shorten >=0.2em] (Watchdog);
\draw[myptrdotted] (Robot2) to[bend left=10, shorten <=0.6em,shorten >=0.2em] (Watchdog);
\end{tikzpicture}%
}
\bigskip

It can be seen that all connections which should be stateful are now indeed stateful. 
In addition, it can be seen that $\mvar{MissionControl1}$ cannot set up a stateful connection to $\mvar{Robot2}$. 
This is because $\mvar{MissionControl1}$ was never declared a trusted device and the confidential information in $\mvar{MissionControl2}$ and $\mvar{Robot2}$ must not leak. 

The improved invariant definition even produces a better (\ie stricter) maximum policy. 

\subsection{Outlook: Iptables Implementation}
In this section, we serialize the stateful policy to an iptables firewall ruleset. 
Our policy graph only contains positive (\ie allow) rules. 
This means, the order in which the rules are installed is irrelevant. 
Therefore, we set the default policy (\texttt{-P}) to \texttt{DROP} and iterate over all edges in the policy and emit an \texttt{ACCEPT} iptables rule.

\begin{figure*}[hptb]
\begin{minipage}{\linewidth}
\footnotesize
\begin{Verbatim}[commandchars=\\\{\},codes={\catcode`$=3\catcode`^=7}]
iptables -P FORWARD DROP
iptables -A FORWARD -i $\$\mathit{PresenceSensor\_iface}$ -s $\$\mathit{PresenceSensor\_ipv4}$ $\hfill\hookleftarrow$
               -o $\$\mathit{SensorSink\_iface}$ -d $\$\mathit{SensorSink\_ipv4}$ -j ACCEPT 
iptables -A FORWARD -i $\$\mathit{Webcam\_iface}$ -s $\$\mathit{Webcam\_ipv4}$ $\hfill\hookleftarrow$
               -o $\$\mathit{SensorSink\_iface}$ -d $\$\mathit{SensorSink\_ipv4}$ -j ACCEPT 
iptables -A FORWARD -i $\$\mathit{TempSensor\_iface}$ -s $\$\mathit{TempSensor\_ipv4}$ $\hfill\hookleftarrow$
               -o $\$\mathit{SensorSink\_iface}$ -d $\$\mathit{SensorSink\_ipv4}$ -j ACCEPT 
iptables -A FORWARD -i $\$\mathit{FireSensor\_iface}$ -s $\$\mathit{FireSensor\_ipv4}$ $\hfill\hookleftarrow$
               -o $\$\mathit{SensorSink\_iface}$ -d $\$\mathit{SensorSink\_ipv4}$ -j ACCEPT 
iptables -A FORWARD -i $\$\mathit{SensorSink\_iface}$ -s $\$\mathit{SensorSink\_ipv4}$ $\hfill\hookleftarrow$
               -o $\$\mathit{Statistics\_iface}$ -d $\$\mathit{Statistics\_ipv4}$ -j ACCEPT 
iptables -A FORWARD -i $\$\mathit{MissionControl1\_iface}$ -s $\$\mathit{MissionControl1\_ipv4}$ $\hfill\hookleftarrow$
               -o $\$\mathit{Robot1\_iface}$ -d $\$\mathit{Robot1\_ipv4}$ -j ACCEPT 
iptables -A FORWARD -i $\$\mathit{MissionControl1\_iface}$ -s $\$\mathit{MissionControl1\_ipv4}$ $\hfill\hookleftarrow$
               -o $\$\mathit{Robot2\_iface}$ -d $\$\mathit{Robot2\_ipv4}$ -j ACCEPT 
iptables -A FORWARD -i $\$\mathit{MissionControl2\_iface}$ -s $\$\mathit{MissionControl2\_ipv4}$ $\hfill\hookleftarrow$
               -o $\$\mathit{Robot2\_iface}$ -d $\$\mathit{Robot2\_ipv4}$ -j ACCEPT 
iptables -A FORWARD -i $\$\mathit{AdminPc\_iface}$ -s $\$\mathit{AdminPc\_ipv4}$ $\hfill\hookleftarrow$
               -o $\$\mathit{MissionControl2\_iface}$ -d $\$\mathit{MissionControl2\_ipv4}$ -j ACCEPT 
iptables -A FORWARD -i $\$\mathit{AdminPc\_iface}$ -s $\$\mathit{AdminPc\_ipv4}$ $\hfill\hookleftarrow$
               -o $\$\mathit{MissionControl1\_iface}$ -d $\$\mathit{MissionControl1\_ipv4}$ -j ACCEPT 
iptables -A FORWARD -i $\$\mathit{Watchdog\_iface}$ -s $\$\mathit{Watchdog\_ipv4}$ $\hfill\hookleftarrow$
               -o $\$\mathit{Robot1\_iface}$ -d $\$\mathit{Robot1\_ipv4}$ -j ACCEPT
iptables -A FORWARD -i $\$\mathit{Watchdog\_iface}$ -s $\$\mathit{Watchdog\_ipv4}$ $\hfill\hookleftarrow$
               -o $\$\mathit{Robot2\_iface}$ -d $\$\mathit{Robot2\_ipv4}$ -j ACCEPT
# SensorSink -> Webcam (answer)
iptables -I FORWARD -m state --state ESTABLISHED -i $\$\mathit{SensorSink\_iface}$ -s $\$\mathit{SensorSink\_ipv4}$ $\hfill\hookleftarrow$
               -o $\$\mathit{Webcam\_iface}$ -d $\$\mathit{Webcam_ipv4}$ -j ACCEPT
# Statistics -> SensorSink (answer)
iptables -I FORWARD -m state --state ESTABLISHED -i $\$\mathit{Statistics\_iface}$ -s $\$\mathit{Statistics\_ipv4}$ $\hfill\hookleftarrow$
               -o $\$\mathit{SensorSink\_iface}$ -d $\$\mathit{SensorSink_ipv4}$ -j ACCEPT
# Robot1 -> MissionControl1 (answer)
iptables -I FORWARD -m state --state ESTABLISHED -i $\$\mathit{Robot1\_iface}$ -s $\$\mathit{Robot1\_ipv4}$ $\hfill\hookleftarrow$
               -o $\$\mathit{MissionControl1\_iface}$ -d $\$\mathit{MissionControl1_ipv4}$ -j ACCEPT
# Robot2 -> MissionControl2 (answer)
iptables -I FORWARD -m state --state ESTABLISHED -i $\$\mathit{Robot2\_iface}$ -s $\$\mathit{Robot2\_ipv4}$ $\hfill\hookleftarrow$
               -o $\$\mathit{MissionControl2\_iface}$ -d $\$\mathit{MissionControl2_ipv4}$ -j ACCEPT
# MissionControl2 -> AdminPc (answer)
iptables -I FORWARD -m state --state ESTABLISHED $\hfill\hookleftarrow$
               -i $\$\mathit{MissionControl2\_iface}$ -s $\$\mathit{MissionControl2\_ipv4}$ $\hfill\hookleftarrow$
               -o $\$\mathit{AdminPc\_iface}$ -d $\$\mathit{AdminPc_ipv4}$ -j ACCEPT
# MissionControl1 -> AdminPc (answer)
iptables -I FORWARD -m state --state ESTABLISHED $\hfill\hookleftarrow$
               -i $\$\mathit{MissionControl1\_iface}$ -s $\$\mathit{MissionControl1\_ipv4}$ $\hfill\hookleftarrow$
               -o $\$\mathit{AdminPc\_iface}$ -d $\$\mathit{AdminPc_ipv4}$ -j ACCEPT
# Robot1 -> Watchdog (answer)
iptables -I FORWARD -m state --state ESTABLISHED -i $\$\mathit{Robot1\_iface}$ -s $\$\mathit{Robot1\_ipv4}$ $\hfill\hookleftarrow$
               -o $\$\mathit{Watchdog\_iface}$ -d $\$\mathit{Watchdog_ipv4}$ -j ACCEPT
# Robot2 -> Watchdog (answer)
iptables -I FORWARD -m state --state ESTABLISHED -i $\$\mathit{Robot2\_iface}$ -s $\$\mathit{Robot2\_ipv4}$ $\hfill\hookleftarrow$
               -o $\$\mathit{Watchdog\_iface}$ -d $\$\mathit{Watchdog_ipv4}$ -j ACCEPT
\end{Verbatim}
\end{minipage}%
\caption{iptables ruleset}
\label{fig:example-factory:iptables}
\end{figure*}

The translation for each rule is straightforward. 
Each rule has a sender and a receiver, which we can translate to iptables. 
To prevent IP spoofing, we assume that each device is connected to its own interface. 
Therefore, for the sender of a rule, we match on the input interface (\texttt{-i}) and source IP address (\texttt{-s}). 
For the receiver, we match on the output interface (\texttt{-o}) and destination IP address (\texttt{-d}). 
For the rules in the policy which are marked stateful, we additionally match on the \texttt{ESTABLISHED} state. 
The resulting ruleset can be seen in Figure~\ref{fig:example-factory:iptables}. 

We predict that more packets will be send in the `answer-direction' than the `connection-setup-direction'. 
For example, the watchdog will only send one monitoring command and afterwards, a robot will regularly send back health information. 
For performance reasons, we want the \texttt{ESTABLISHED} rules to be on top of the ruleset. 
To make the translation slightly more interesting, we will mix \texttt{-A} (append rule to the back) and \texttt{-I} (insert rule on top). 
We will later verify that the serialized iptables ruleset indeed reflects the desired policy. 

To deploy the scenario, we assign each device an IP address according to Table~\ref{tab:example-factory:ipmapping}. 
Now, the ruleset can be loaded by the Linux kernel. 
We use the results of Part~\ref{part:existing-configs} to verify the correctness of the generated ruleset. 
Therefore, we load the ruleset into our analysis tool (Part~\ref{part:existing-configs}) and compute a service matrix for an arbitrary service.

\begin{table}[h!bt]
\centering
\caption{IP Mapping}
\label{tab:example-factory:ipmapping}
\begin{tabular}[0.99\linewidth]{ l @{\hspace*{0.8em}} l }%
	\toprule
	Variable            & IP Address  \\
	\midrule
	$\$\mathit{Statistics\_ipv4}$ & $10.0.0.1$\\
	$\$\mathit{SensorSink\_ipv4}$ & $10.0.0.2$\\
	$\$\mathit{PresenceSensor\_ipv4}$ & $10.0.1.1$\\
	$\$\mathit{Webcam\_ipv4}$ & $10.0.1.2$\\
	$\$\mathit{TempSensor\_ipv4}$ & $10.0.1.3$\\
	$\$\mathit{FireSensor\_ipv4}$ & $10.0.1.4$\\
	$\$\mathit{MissionControl1\_ipv4}$ & $10.8.1.1$\\
	$\$\mathit{MissionControl2\_ipv4}$ & $10.8.1.2$\\
	$\$\mathit{Watchdog\_ipv4}$ & $10.8.8.1$\\
	$\$\mathit{Robot1\_ipv4}$ & $10.8.2.1$\\
	$\$\mathit{Robot2\_ipv4}$ & $10.8.2.2$\\
	$\$\mathit{AdminPc\_ipv4}$ & $10.8.0.1$\\
 \bottomrule%
\end{tabular}
\end{table}

\begin{figure*}[ht!bp]
\resizebox{0.99\linewidth}{!}{
\begin{tikzpicture}
\node (ip10001) at (0,0) {$10.0.0.1$};
\node (ip10002) at (0,-2) {$10.0.0.2$};
\node (ip10011) at (-1.5,-4) {$\lbrace 10.0.1.1 .. 10.0.1.4 \rbrace$};
\node (ip10881) at (5,-2) {$10.8.8.1$};
\node (ip10812) at (12,-2) {$10.8.1.2$};
\node (ip10811) at (8,-2) {$10.8.1.1$};
\node (ip10821) at (8,-4) {$10.8.2.1$};
\node (ip10822) at (12,-4) {$10.8.2.2$};
\node (ip10801) at (10,0) {$10.8.0.1$};
\node (ip0000)[align=center,text width=8cm,cloud, draw,cloud puffs=10,cloud puff arc=120, aspect=2, inner sep=-3em,outer sep=0] at (5,2) {$\lbrace 0.0.0.0 .. 10.0.0.0 \rbrace \cup \lbrace 10.0.0.3 .. 10.0.1.0 \rbrace \cup \lbrace 10.0.1.5 .. 10.8.0.0 \rbrace \cup \lbrace 10.8.0.2 .. 10.8.1.0 \rbrace \ \cup $ \\ $ \lbrace 10.8.1.3 .. 10.8.2.0 \rbrace \cup \lbrace 10.8.2.3 .. 10.8.8.0 \rbrace \cup \lbrace 10.8.8.2 .. 255.255.255.255 \rbrace $};

\draw[myptr] (ip10881) to (ip10822);
\draw[myptr] (ip10881) to (ip10821);
\draw[myptr] (ip10812) to (ip10822);
\draw[myptr] (ip10811) to (ip10822);
\draw[myptr] (ip10811) to (ip10821);
\draw[myptr] (ip10801) to (ip10812);
\draw[myptr] (ip10801) to (ip10811);
\draw[myptr] (ip10011) to (ip10002);
\draw[myptr] (ip10002) to (ip10001);
\end{tikzpicture}%
}
\caption{Analysis of firewall rules of Figure~\ref{fig:example-factory:iptables}. New connections.}
\label{fig:examplefactory:fffuuNEW}
\end{figure*}
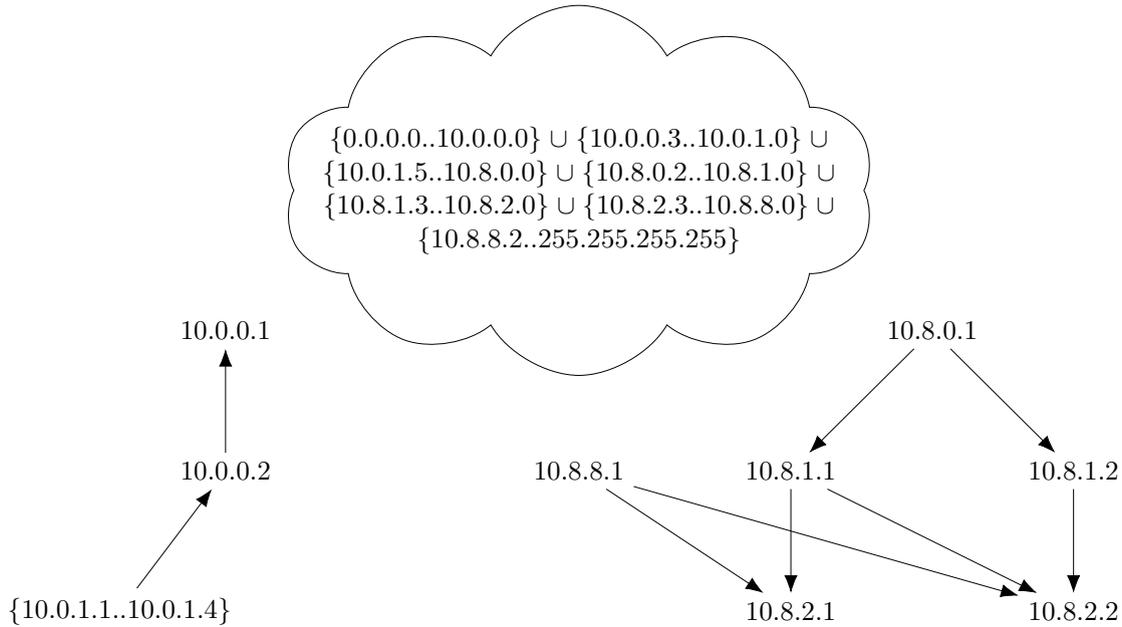

The resulting matrix for \texttt{NEW} packets is visualized in Figure~\ref{fig:examplefactory:fffuuNEW}. 
The graph shows who is allowed to set up connections with whom.

Mapping back the IP addresses to the names in the policy, we see that the iptables ruleset indeed corresponds to our desired policy. 
It can be seen that our analysis tool has pooled all sensors into one node because they all have the same access rights. 
The node at the top with the complicated IP range specification corresponds to all IP addresses which are not used in our factory: the $\mathit{INET}$.

\begin{figure*}[ht!bp]
\resizebox{0.99\linewidth}{!}{
\begin{tikzpicture}
\node (ip10012) at (0,0) {$\lbrace 10.0.0.1 \rbrace \cup \lbrace 10.0.1.2 \rbrace$};
\node (ip10002) at (0,-2) {$10.0.0.2$};
\node (ip10013) at (-1.5,-4) {$\lbrace 10.0.1.1 \rbrace \cup \lbrace 10.0.1.3 \rbrace \cup \lbrace 10.0.1.4 \rbrace$};
\node (ip10881) at (5,-2) {$10.8.8.1$};
\node (ip10812) at (12,-2) {$10.8.1.2$};
\node (ip10811) at (8,-2) {$10.8.1.1$};
\node (ip10821) at (8,-4) {$10.8.2.1$};
\node (ip10822) at (12,-4) {$10.8.2.2$};
\node (ip10801) at (10,0) {$10.8.0.1$};
\node (ip0000)[align=center,text width=8cm,cloud, draw,cloud puffs=10,cloud puff arc=120, aspect=2, inner sep=-3em,outer sep=0] at (5,2) {$ \lbrace 0.0.0.0 .. 10.0.0.0 \rbrace \cup \lbrace 10.0.0.3 .. 10.0.1.0 \rbrace \cup \lbrace 10.0.1.5 .. 10.8.0.0 \rbrace \cup \lbrace 10.8.0.2 .. 10.8.1.0 \rbrace \ \cup $ \\ $ \lbrace 10.8.1.3 .. 10.8.2.0 \rbrace \cup \lbrace 10.8.2.3 .. 10.8.8.0 \rbrace \cup \lbrace 10.8.8.2 .. 255.255.255.255\rbrace $};

\draw[myptr] (ip10881) to (ip10822);
\draw[myptr] (ip10881) to (ip10821);
\draw[myptr] (ip10822) to (ip10881);
\draw[myptr] (ip10822) to (ip10812);
\draw[myptr] (ip10821) to (ip10881);
\draw[myptr] (ip10821) to (ip10811);
\draw[myptr] (ip10812) to (ip10822);
\draw[myptr] (ip10812) to (ip10801);
\draw[myptr] (ip10811) to (ip10822);
\draw[myptr] (ip10811) to (ip10821);
\draw[myptr] (ip10811) to (ip10801);
\draw[myptr] (ip10801) to (ip10812);
\draw[myptr] (ip10801) to (ip10811);
\draw[myptr] (ip10013) to (ip10002);
\draw[myptr] (ip10012) to (ip10002);
\draw[myptr] (ip10002) to (ip10012);
\end{tikzpicture}%
}
\caption{Analysis of firewall rules of Figure~\ref{fig:example-factory:iptables}. Established connections.}
\label{fig:examplefactory:fffuuESTABLISHED}
\end{figure*}
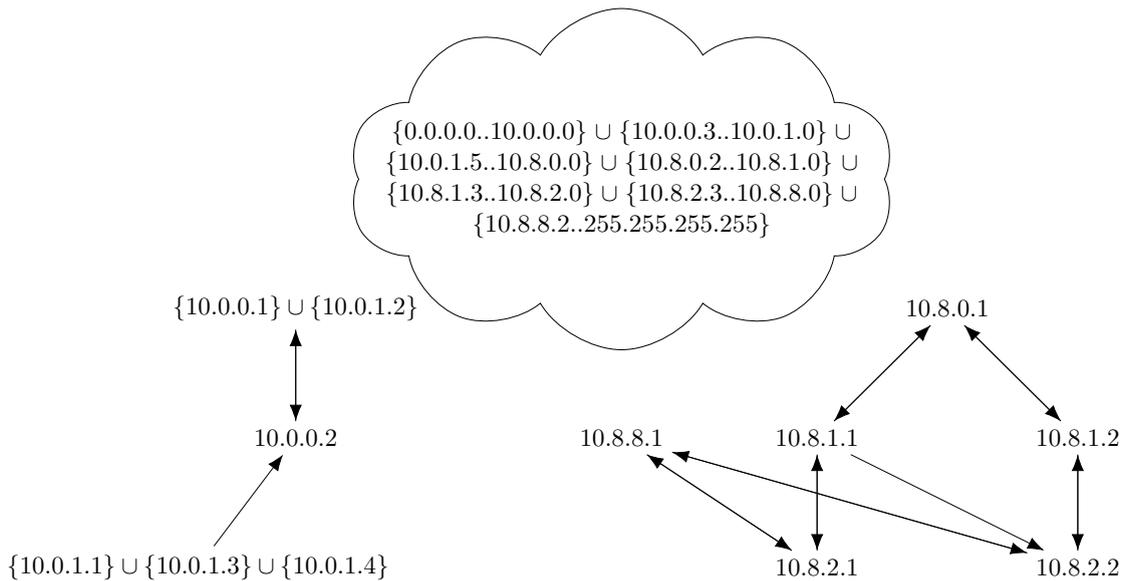

Next, we verify that all \texttt{ESTABLISHED} connections are implemented as desired. 
The connectivity matrix is visualized in Figure~\ref{fig:examplefactory:fffuuESTABLISHED}.

It can be seen that the statistics server and webcam are condensed into one node. 
Comparing to the stateful policy, they indeed have the same access rights for stateful connections. 
The policy does not show a reflexive rule for this node, hence, they still cannot communicate directly. 
All other sensors are only allowed unidirectional information flows, as specified by the policy. 
All other edges (except for mission control one to robot two) are bidirectional. 
This corresponds to the desired connectivity structure for established connections. 

Consequently, we have verified that the iptables implementation exactly corresponds to the desired stateful policy.

\section{Related Work}
\label{sec:forte14:relatedwork}
In a field study with 38 participants, Hamed and Al-Shaer discovered that ``even expert administrators can make serious mistakes when configuring the network security policy''~\cite{netsecconflicts}. 
Our user feedback session extends this finding as we discovered that even expert administrators can make serious mistakes when \emph{designing} the network security policy.

Noteworthy, Ou~\etal\cite{ou2005mulval} summarize a monotonicity property that is very similar to ours but from the opposite point of view: ``gaining privileges does not hurt an attacker[\dots]''. 

In the context of developing and implementing network protocols, Wang \etal\cite{wang2009formally} propose a framework called ``Formally Verifiable Networking''. 
They demonstrate the use of a theorem prover and leverage that this allows formalizing and verifying a specification and generate executable code out of certain specifications. 
The authors highlight that this approach enables two compatible ways to get a protocol implementation: 
First, a user can specify the protocol and (manually) verify the specification in the theorem prover and finally generate an executable code (if the specification allows it). 
The second way starts with an NDlog (Network Datalog) implementation of the protocol, which is translated into the theorem prover which can then be (manually) verified.
Our approach provides an analogue advantage of two compatible ways to get a policy. 
First, a user can specify the security invariants and automatically derive a policy from it (cf.\ Section~\ref{subsec:policyconstruction}). 
Second, a user can define a policy and verify that it corresponds to the security invariants; the verification is automatic. 
Consequently, our approach never requires a manual proof from the user. 



Cuppens \etal\cite{Cuppens2005orbacxmlfirewall} propose a policy language to administrate firewalls. 
They follow the traditional approach of the policy community to differentiate between subjects, objects, and actions. 
Because they are very accurate about their definitions, we can show that their classification can be simplified to our graph-based model. 
First, we assume that we do not match on the content of a network packet. 
This assumption can be justified since a firewall is not a deep packet inspection system. 
In addition, for example, the payload of a packet may be encrypted and a firewall cannot decrypt it. 
Cuppens \etal model their subjects as the machines in the network. 
This corresponds to the vertices in our graph. 
Next, Cuppens \etal model the actions as the allowed services a machine may use, characterized by the protocol and ports. 
Our graph can be viewed as the projection for a single service or the overall access matrix for the universe of all services.\footnote{Note: Our approach may need several graphs to represent different access rights for different services.} 
With this view, an action is reduced to the `send' permission, which is a singleton and hence exactly corresponds to an edge in our graph. 
Finally, Cuppens \etal model objects as the packets in the network, which are only characterized by their receiver. 
Since the receiver is always a machine in the network, the objects also corresponds to the subjects, which correspond to the vertices in our graph. 
Consequently, several graphs (according to our model) are as expressive as a triplet of subject, action, object in Cuppens' model. 
It is unclear which model is `better'. 
The advantages of our approach are that it isolates exactly one aspect and is hence simpler, can be easily visualized, and provides elegant algorithms to work with.

In their inspiring work, Guttman and Herzog~\cite{guttman05rigorous} describe a formal modeling approach for network security management. 
They suggest algorithms to verify whether configurations of firewalls and IPsec gateways fulfill certain security goals. 
These comparatively low-level security goals may state that a certain packet's path only passes certain areas or that packets between two networked hosts are protected by IPsec's ESP confidentiality header. 
This allows reasoning on a lower abstraction level at the cost of higher manual specification and configuration effort. 
Header space analysis~\cite{kazemian2012HSA} allows checking static network invariants such as no-forwarding-loops or traffic-isolation on the forwarding and middleboxes plane. 
It provides a common, protocol-agnostic framework and algebra on the packet header bits.


Firmato~\cite{bartal1999firmato} was designed to ease management of firewalls. 
A firewall-in\-de\-pen\-dent entity relationship model is used to specify the security policy.
With the help of a model compiler, such a model can be translated to firewall configurations. 
Ethane~\cite{ethane07} is a link layer security architecture which evolved to the network operating system NOX~\cite{gude2008nox}. 
They implement high-level security policies and propose a secure binding from host names to network addresses. 
In the long term, we consider \topos{} a valuable add-on on top of such systems for policy verification. 
For example, it could warn the administrator that a recent network policy change violates a security invariant, maybe defined years ago.

NetCore~\cite{netcore12} is a language for packet-forwarding policies in software defined networks that abstracts from low-level hardware details. 
However, compared to the abstract graph utilized in our work, it can be considered rather technical and low-level. 
However, from an abstract point of view, both describe the network topology. 
Guha \etal\cite{machineverifiednetworkcontrollers13} present a verified compiler to translate NetCore to a SDN controller. 
The authors use the Coq proof assistant~\cite{coqmanual} to verify the correctness of their SDN controller, demonstrate its suitable performance, and uncover bugs in other non-machine-verified controllers. 
As Guha \etal provide formally verified means to translate a network topology (NetCore) to real hardware and we provide formally verified means to verify the intention behind the topology (graph), we see great potential to expect provably correct networks from the abstract human intent down to the low-level hardware in the near future.

Expressive policy specification languages, such as Ponder~\cite{ponder2001}, were proposed. 
Positive authorization policies (only a small aspect of Ponder) are roughly comparable to our policy graph. 
The authors note that \eg negative authorization policies (deny-rules) can create conflicts. 
Policy constraints can be checked at compile time.
Bera \etal\cite{policy2010berapolicyformalenterprise} present a policy specification language (SPSL) with allow and deny policy rules. 
With this, a conflict-free policy specification is constructed. 
Conflict-free Boolean formulas of this policy description and the policy implementation in the security mechanisms (router ACL entries) are checked for equality using a SAT solver.
One unique feature covered is hidden service access paths, \eg http might be prohibited in zone1 but zone1 can ssh to zone2 where http is allowed. 
Craven \etal\cite{policy2009expressivedynamic} focus on policies in dynamic systems and their analysis. 
These papers require specification of the verification goals and security goals and can thus benefit from our contributions. 

This work's modeling concept is very similar to the Attribute Based Access Control (ABAC) model~\cite{abac2005}, though the underlying formal objects differ. 
ABAC distinguishes subjects, resources, and environments. 
Attributes may be assigned to each of these entities, similar to our host mappings. 
The ABAC policy model consists of positive rules which grant access based on the assigned attributes, comparably to security invariant templates. 
Therefore, our insights and contributions are also applicable to the ABAC model.

\paragraph*{Analogy to Software Architectures}
%
%
%
Finally, we want to show parallels between our work and scientific results from the field of software development and software engineering. 
Roughly speaking, a software architecture can be abstractly understood as a high level specification of a software and its documentation. 
We will use the definition that ``[a]rchitecture defines the components of a system and their dependencies''~\cite{juergens2009softwarecrchitectures}. 
An architecture is realized by an actual program, \ie code. 
For our analogy, we will equate a software architecture with security requirements and we equate code with a security policy. 

\begin{figure}[h!tp]
	\centering
	\centering
	\fbox{\includegraphics[width=0.98\linewidth]{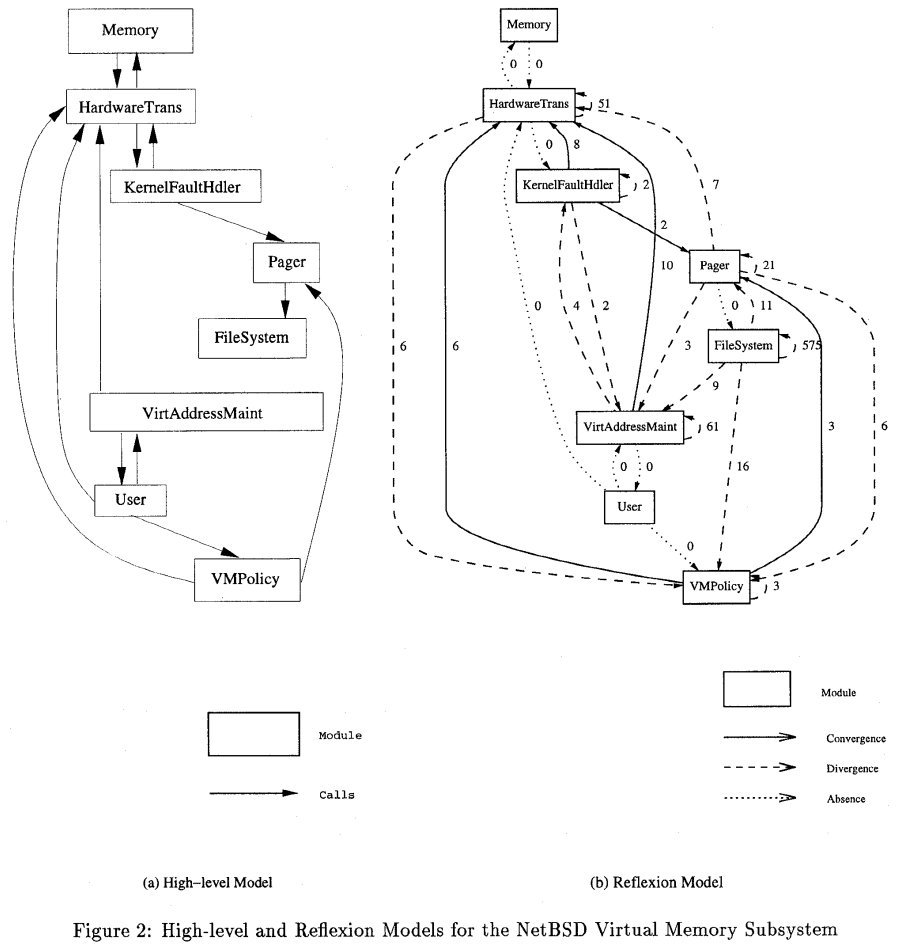}}
	\caption{By Gail C.\ Murphy, David Notkin, and Kevin Sullivan (1995)~\cite{Murphy1995architectures}}
	\label{fig:softwarearchitectures:murphy95}
\end{figure}

In software development, researchers identified the problem that the specified software architecture and the actual code diverge over time~\cite{Perry1992softwarearchitectures}. 
In fact, ``[o]ne problem with high-level models is that they are always inaccurate with respect to the system's source code''~\cite{Murphy1995architectures}. 
In real-world case studies, it was shown that documented architectures may vastly diverge from the actual code over time~\cite{juergens2009softwarecrchitectures} and that ``documentation becomes a dead artifact that is used very infrequently''~\cite{juergens2009softwarecrchitectures}.  
The same insight may also apply to an informal, textual representation of security requirements.
 
In the field of software architectures, researches have built tools to visualize the divergence of a specified architecture and the actual implementation~\cite{juergens2009softwarecrchitectures,Murphy1995architectures}. 
An example can be found in Figure~\ref{fig:softwarearchitectures:murphy95}. 
Analogously, our method allows to visualize differences between formalized security invariants and an actual policy. 
Visualizations which are generated by \topos{} look remarkably similar to visualizations of divergence in software architectures. 
For example in Figure~\ref{fig:softwarearchitectures:murphy95}, in the context of our analogy, a \emph{convergence} would correspond to a flow which is specified in the policy and allowed by the security invariants. 
A \emph{divergence} would correspond to a flow which is present in the policy but prohibited by the security invariants. 
An \emph{absence} would correspond to a flow which is not specified in the policy but would be accepted by the computed maximum policy. 

Additionally, by the same means as architecture consistency checkers help to uncover errors and architecture drift, by formalizing security requirements with our approach, it can always be ensured that the requirements and the actual policy stay consistent. 

Finally, it has been proposed~\cite{beller2012arhitecturestrict} to consider architecture strictness, where strictness refers to a measure which reflects to which degree components may access each other. 
For example, in a very strict system, no component may access another. 
This corresponds to our monotonicity principle which could be translated in the context of this analogy as follows: Increasing the strictness of a system does not decrease its security.

\section{Conclusion}
\label{sec:forte14:conclusion}
After several hundred thousand changed lines of formal theory, our simple, yet powerful, model landscape emerged. 
Representing policies as graphs makes them visualizable. 
Describing security invariants as total Boolean-valued functions is both expressive and accessible to formal analysis. 
Representing host mappings as partial configurations is end-user-friendly, transforming them to total functions makes them handy for the design of templates. 
With this simple model, we discovered important universal insights on security invariants. 
In particular, the transformation of host mappings and a simple sanity check which guarantees that security policy violations can always be resolved. 
This provides deep insights about how to express verification goals. 
The full formalization in the Isabelle/HOL theorem prover provides high confidence in the correctness.

  \chapter{Improved Policy Construction}
\label{chap:hilbertsoffending}

\paragraph*{Abstract}
The previous chapters show that our policy construction algorithm is only practically usable for $\Phi$-structured invariant temaplates. 
In this chapter, we improve the policy construction algorithm to cope with arbitrary invariants. 

\medskip

\section{Introduction}
The algorithm presented Section~\ref{subsec:policyconstruction} constructs a security policy which fulfills all security invariants. 
Summarizing the algorithm with the help of Lemma~\ref{lem:generate_valid_topology_def_alt}, for a list $\mvar{Ms}$ of configured security invariants, the algorithm simply removes all offending flows: 
\begin{IEEEeqnarray*}{c}
  \mfun{delete\mhyphen{}edges}\ G \left(\bigcup \left(\bigcup m_c \in \mdef{set}\ \mvar{Ms}.\ \mfun{set\mhyphen{}offending\mhyphen{}flows}\ m_c\ G\right)\right)
\end{IEEEeqnarray*}

The main concern with this algorithm is that it needs to construct the complete set of offending flows. 
This can be done efficiently for $\Phi$-structured invariants. 
However, as has been shown in Section~\ref{sec:forte14:model-library}, there are some invariant templates which have a different structure and where the size of the set of offending flows can grow infeasible large. 
This is for example the case for NonInterference (Section~\ref{sinvar:noninterference}), Comm.\ With (Section~\ref{sinvar:commwith}), Not Comm.\ With (Section~\ref{sinvar:notcommwith}), and Dependability (Section~\ref{sinvar:dependability}). 

In this chapter, we present a new algorithm for policy construction. 
The main idea of the new algorithm is presented by the following formula. 
\begin{IEEEeqnarray*}{c}
  \mfun{delete\mhyphen{}edges}\ G  \left(\bigcup m_c \in \mdef{set}\ \mvar{Ms}.\ \mctrl{if}\ m_c\ G\ \mctrl{then}\ \emptyset\ \mctrl{else}\ \varepsilon\ (\mfun{set\mhyphen{}offending\mhyphen{}flows}\ m_c\ G)  \right)
\end{IEEEeqnarray*}

Here, $\varepsilon$ corresponds to Hilbert's $\varepsilon$-operator. 
We use this operator in a simplified setting.\footnote{In Isabelle, a more generic version of Hilbert's $\varepsilon$ operator is available as one of the core axioms of HOL. The operator can be used with the keyword $\mathtt{SOME}$. In the context of this thesis, we can use it in a simplified fashion where we always mean $\varepsilon\ X \ \equiv \ \mdef{SOME}\ x.\ \, x \in X$.} 
Its first argument is a set and it returns an element from the set. 
The element which $\varepsilon$ returns is chosen non-deterministically: $\varepsilon$ is an \emph{indefinite} operator.
For example, if $(\varepsilon\ \lbrace 1, 2, 3 \rbrace) = x$, then $x = 1 \vee x = 2 \vee x = 3$. 
Nothing can be said if $\varepsilon$ is applied to the empty set. 
Because $\neg\; m_c\ G$ ensures that the set of offending flows is always defined,\footnote{configured-SecurityInvariant.defined-offending'} $\varepsilon$ can be safely applied here. 

Comparing the two formulas on this page, the main difference is the following:
The first one removes all offending flows whereas the second one only removes one arbitrary member of the set of offending flows per invariant.\footnote{Note that $\mfun{set\mhyphen{}offending\mhyphen{}flows}$ is a set of sets and consequently a member of the set of offending flows is a set of flows.} 
Consequently, if the set of offending flows has more than one member, the second algorithm may result in a better (here: more permissive but still sound) result.

The improved algorithm is implemented as follows.
\begin{IEEEeqnarray*}{lCl}%
  \IEEEeqnarraymulticol{3}{l}{\mfun{generate\mhyphen{}valid\mhyphen{}topology2}\ :: (\mathcal{G} \Rightarrow \mathbb{B})\ \textnormal{list} \Rightarrow \mathcal{G} \Rightarrow \mathcal{G}}\\
  \mfun{generate\mhyphen{}valid\mhyphen{}topology2}\ []\ G & = & G\\
  \mfun{generate\mhyphen{}valid\mhyphen{}topology2}\ (m_c \lstcons \mvar{Ms})\ G & = & \\
  \IEEEeqnarraymulticol{3}{l}{\qquad\qquad \mctrl{if}\ m_c\ G\ }\\
  \IEEEeqnarraymulticol{3}{l}{\qquad\qquad \mctrl{then}\ }\\
  \IEEEeqnarraymulticol{3}{l}{\qquad\qquad \qquad \mfun{generate\mhyphen{}valid\mhyphen{}topology2}\ \mvar{Ms}\ G\ }\\
  \IEEEeqnarraymulticol{3}{l}{\qquad\qquad \mctrl{else}}\\
  \IEEEeqnarraymulticol{3}{l}{\qquad\qquad \qquad \mfun{delete\mhyphen{}edges}\ \left(\mfun{generate\mhyphen{}valid\mhyphen{}topology2}\ \mvar{Ms}\ G\right) \qquad\qquad}\\
  \IEEEeqnarraymulticol{3}{r}{ \left( \varepsilon \left(\mfun{set\mhyphen{}offending\mhyphen{}flows}\ m_c\ G\right)\right)}
\end{IEEEeqnarray*}

Analogously to $\mfun{generate\mhyphen{}valid\mhyphen{}topology}$, there is an alternative, equivalent definition:
\begin{lemma}
\label{lem:generate_valid_topology2_def_alt}
\begin{IEEEeqnarray*}{c}
  \mfun{generate\mhyphen{}valid\mhyphen{}topology2}\ \mathit{Ms}\ G = \\
  \mfun{delete\_edges}\ G  \left(\bigcup m_c \in \mdef{set}\ \mvar{Ms}.\ \mctrl{if}\ m_c\ G\ \mctrl{then}\ \emptyset\ \mctrl{else}\ \varepsilon\ (\mfun{set\mhyphen{}offending\mhyphen{}flows}\ m_c\ G)  \right)
\end{IEEEeqnarray*}
\end{lemma}

The algorithm is sound\footnote{generate-valid-topology-SOME-sound}, as shown by the following Theorem.  

\begin{theorem}[Soundness of $\varepsilon$ Policy Construction]
\label{thm:generate-policy-2-sound}
Let $M$ be a set of configured security invariants. This means, each element of $M$ is of type $\mathcal{G} \Rightarrow \mathbb{B}$, is monotonic, and always has defined offending flows. Then
\begin{IEEEeqnarray*}{c}
{\forall m_c \in M.} \ \ m_c\ \left(\mfun{generate\mhyphen{}valid\mhyphen{}topology2} \ M\ G\right)
\end{IEEEeqnarray*}
\end{theorem}

The new algorithm can compute a superset of the policy which can be computed by the old algorithm. 

\begin{lemma}
\label{lem_generate-policy-2-superset}
Let $\mathit{edges}$ extract the edges from a graph. Then
\begin{IEEEeqnarray*}{c}
      \mdef{edges}\ (\mfun{generate\mhyphen{}valid\mhyphen{}topology}\ M\ G) \subseteq \mdef{edges}\ (\mfun{generate\mhyphen{}valid\mhyphen{}topology2}\ M\ G)
\end{IEEEeqnarray*}
\end{lemma}

This justifies the claim that the new algorithm is better: It may compute a more permissive policy (Lemma~\ref{lem_generate-policy-2-superset}), but the results are still sound (Theorem~\ref{thm:generate-policy-2-sound}). 
Consequently, when $\mfun{generate\mhyphen{}valid\mhyphen{}topology}$ can generate a maximum policy for $\Phi$-structured invariants, the policy computed by $\mfun{generate\mhyphen{}valid\mhyphen{}topology2}$ must also be maximal.

\paragraph*{Problems of the $\varepsilon$ operator}
As of Isabelle 2016, none of the automated solvers could solve the following example automatically: $(\varepsilon\ \lbrace 1, 2, 3 \rbrace) = x \longrightarrow x = 1 \vee x = 2 \vee x = 3$. 
Its proof required one manual step. 
We have chosen the $\varepsilon$ operator because it is the most illustrative way to present the ideas behind the algorithm in this chapter. 

However, this operator is complicated to work with. 
In addition, due to its indefinite choice, the final (deterministic) executable implementation will not use $\varepsilon$.

\section{Computing One Member of the Set of Offending Flows}
To implement $\mfun{generate\mhyphen{}valid\mhyphen{}topology2}$, one needs to select one member of the set of offending flows. 
For this algorithm to be efficient also for non-$\Phi$-structured invariant templates where the size of the set of offending flows can grow exponentially, it is important not to compute the complete set of offending flows. 
In this section, we present a deterministic algorithm which computes exactly one member of the set of offending flows without constructing the whole set.

A member of the set of offending flows $F$ has to fulfill three properties (cf.\ Definition~\ref{def:set_offending_flows_def}). 
Here, we summarize them for a configured security invariant $m_c$:
\begin{enumerate}
	\item $\neg\, m_c\ G$
	\item $m_c\ (V,\, E \setminus F)$
	\item $\forall (s,r) \in F.\ \neg\, m_c\ \left(V,\, \left(E \setminus F\right) \cup \left\lbrace \left( s,r \right) \right\rbrace \right)$
\end{enumerate}
The first property states that, independent of $F$, the invariant must be violated; otherwise, there are no offending flows. 
The second property states that after removing the flows $F$ from the policy, the security invariant is no longer violated, \ie $F$ can `fix' the policy.  
Finally, the third property states that every individual flow in $F$ must be responsible for the violation of $m_c$. 

In this section, we present an algorithm, given $m_c$ and $G$, it will calculate one such $F$ which fulfills all three properties. 
It does not rely on an efficient implementation for $\mfun{set\mhyphen{}offending\mhyphen{}flows}$ (as we could assume for $\Phi$-structured invariants). 
For this section, we will assume that the first property ($\neg\, m_c\ G$) holds. 
Otherwise, the complete $\mfun{set\mhyphen{}offending\mhyphen{}flows}$ is trivially the empty set.

The algorithm needs an over-approximation of $F$ to start with. 
By the term over-approximation, we mean a set such that property two already holds but property three may not hold. 

\begin{example}
\textbf{Example. }%
We showed that any well-formed security invariant should be fulfilled for the deny-all policy, \ie $m_c\ (V, \emptyset)$, cf.\ Theorem~\ref{thm:no-edges-validity}. 
We require this property for any well-formed security invariant. 
Consequently, for $G = (V,E)$, the complete set of edges $E$ is an over-approximation which fulfills the second property. 
\end{example}

We will call the over-approximation to start the algorithm with $\mvar{fs}$. 
It is written in lower case because $\mvar{fs}$ must be a (finite) list. 
In addition, the set of $\mvar{fs}$ must be a subset of the edges of $G$ and $\mvar{fs}$ must be distinct. 
As has been shown by the $\varepsilon$ operator, it is enough to compute one arbitrary member of the set of offending flows. 
In contrast to sets where the order of the elements does not matter, the order in the list $\mvar{fs}$ determines which member of the offending flows is computed.

The algorithm takes four parameters. 
The first parameter is the configured security invariant $m_c$. 
The second parameter, $\mvar{fs}$ is an over-approximation of the offending flows. 
The third parameter, $\mvar{keeps}$, corresponds to the flows which will be returned after minimizing. 
The fourth parameter is the security policy $G$. 
\begin{IEEEeqnarray*}{lCl}%
  \IEEEeqnarraymulticol{3}{l}{\mfun{minimalize\mhyphen{}offending\mhyphen{}overapprox}\ }\\	
  \IEEEeqnarraymulticol{3}{r}{\qquad\qquad \qquad :: (\mathcal{G} \Rightarrow \mathbb{B}) \Rightarrow (\mathcal{V}\times\mathcal{V})\ \textnormal{list} \Rightarrow (\mathcal{V}\times\mathcal{V})\ \textnormal{list} \Rightarrow \mathcal{G} \Rightarrow (\mathcal{V}\times\mathcal{V})\ \textnormal{list}}\\
  \mfun{minimalize\mhyphen{}offending\mhyphen{}overapprox}\ \_\ []\ \mvar{keeps}\ \_ & = & \mvar{keeps}\\
  \mfun{minimalize\mhyphen{}offending\mhyphen{}overapprox}\ m_c\ (f \lstcons \mvar{fs})\ \mvar{keeps}\ G & = & \\
  \IEEEeqnarraymulticol{3}{l}{\qquad\qquad \mctrl{if}\ m_c\ \left(\mfun{delete\mhyphen{}edges}\ G\ \left(\mvar{fs} \lstapp \mvar{keeps}\right)\right) }\\
  \IEEEeqnarraymulticol{3}{l}{\qquad\qquad \mctrl{then} }\\
  \IEEEeqnarraymulticol{3}{l}{\qquad\qquad \qquad \mfun{minimalize\mhyphen{}offending\mhyphen{}verapprox}\ m_c\ \mvar{fs}\ \mvar{keeps}\ G}\\
  \IEEEeqnarraymulticol{3}{l}{\qquad\qquad \mctrl{else} }\\
  \IEEEeqnarraymulticol{3}{l}{\qquad\qquad \qquad \mfun{minimalize\mhyphen{}offending\mhyphen{}overapprox}\ m_c\ \mvar{fs}\ \left(f \lstcons \mvar{keeps}\right)\ G}\\
\end{IEEEeqnarray*}

\paragraph*{Idea of the Algorithm}
The first and the fourth parameter are fixed and do not change during a run on the algorithm. 
The algorithm iterates over its second parameter $\mvar{fs}$ and stores intermediate results in its third parameter $\mvar{keeps}$. 
For each flow $f$ in $\mvar{fs}$, it checks whether it is necessary and responsible for the violation of $m_c$. 
Therefore, the algorithm checks whether $m_c$ is valid if $\mvar{fs}$ together with the $\mvar{keeps}$ but without $f$ is removed. 
If this is the case, the violation of $m_c$ can be fixed without $f$, consequently, $f$ is not responsible for the violation and can be removed. 
Otherwise, $f$ is responsible for a violation and is saved in the $\mvar{keeps}$ and will be part of the final result. 

\medskip

There are many constraints for $\mvar{keeps}$ which can be found in the formalization. 
They are necessary for the correctness proof. 
Setting $\mvar{keeps} = []$ fulfills all constraints and is the only way we will ultimately call the algorithm. 
Lemma~\ref{lem_minimalize-offending-overapprox-gives-some-offending-flow} proves\footnote{lemma minimalize-offending-overapprox-gives-some-offending-flow} correctness of the algorithm: If called with the right set of parameters, it returns one member of the set of offending flows. 

\begin{lemma}
\label{lem_minimalize-offending-overapprox-gives-some-offending-flow}
Assume $\neg\, m_c\ G$, and $G = (V,E)$. Let $\mvar{Es}$ be a distinct list which corresponds to the set $E$. Then 
\begin{IEEEeqnarray*}{c}
  \mdef{set}\ \left(\mfun{minimalize\mhyphen{}offending\mhyphen{}overapprox}\ m_c\ \mvar{Es} \ []\ G\right)\\
  \in\\
  \mfun{set\mhyphen{}offending\mhyphen{}flows}\ m_c\ G
\end{IEEEeqnarray*}
\end{lemma}

The runtime of $\mfun{minimalize\mhyphen{}offending\mhyphen{}overapprox}$ is $\vert E \vert$ times the runtime of $m_c$. 
If used for policy construction, it will call $m_c$ exactly $\vert V \vert ^2$ times. 
Hence, if $m_c$ can be computed in polynomial time, $\mfun{minimalize\mhyphen{}offending\mhyphen{}overapprox}$ is also in polynomial time. 

With $\mfun{minimalize\mhyphen{}offending\mhyphen{}overapprox}$, an executable, efficient, and deterministic implementation of $\mfun{generate\mhyphen{}valid\mhyphen{}topology2}$ is obtained. 
We will call it $\mfun{generate\mhyphen{}valid\mhyphen{}topology3}$. 
It is implemented as follows. 
%
\begin{IEEEeqnarray*}{lCl}%
  \IEEEeqnarraymulticol{3}{l}{\mfun{generate\mhyphen{}valid\mhyphen{}topology3}\ :: (\mathcal{G} \Rightarrow \mathbb{B})\ \textnormal{list} \Rightarrow \mathcal{G} \Rightarrow \mathcal{G}}\\
  \mfun{generate\mhyphen{}valid\mhyphen{}topology3}\ []\ G & = & G\\
  \mfun{generate\mhyphen{}valid\mhyphen{}topology3}\ (m_c \lstcons \mvar{Ms})\ G & = & \\
  \IEEEeqnarraymulticol{3}{l}{\qquad\qquad \mctrl{if}\ m_c\ G\ }\\
  \IEEEeqnarraymulticol{3}{l}{\qquad\qquad \mctrl{then}\ }\\
  \IEEEeqnarraymulticol{3}{l}{\qquad\qquad \qquad \mfun{generate\mhyphen{}valid\mhyphen{}topology3}\ \mvar{Ms}\ G\ }\\
  \IEEEeqnarraymulticol{3}{l}{\qquad\qquad \mctrl{else}}\\
  \IEEEeqnarraymulticol{3}{l}{\qquad\qquad \mfun{delete\mhyphen{}edges}\ \left(\mfun{generate\mhyphen{}valid\mhyphen{}topology3}\ \mvar{Ms}\ G\right) \qquad\qquad}\\
  \IEEEeqnarraymulticol{3}{r}{\qquad\qquad \left( \mfun{minimalize\mhyphen{}offending\mhyphen{}overapprox}\ m_c\ (\mdef{edges}\ G)\ []\ G \right)}
\end{IEEEeqnarray*}

\begin{sloppypar}
The same ideas as applied in Theorem~\ref{thm:generate-policy-2-sound} can be applied to show that $\mfun{generate\mhyphen{}valid\mhyphen{}topology3}$ is sound.\footnote{generate-valid-topology-some-sound} 
\end{sloppypar}

In Section~\ref{subsec:example:factory:noninterference}, we have already presented that our improved algorithm can immediately compute a policy, even with the `problematic' NonInterference invariant. 
It also showed that a user may influence the result of $\mfun{generate\mhyphen{}valid\mhyphen{}topology3}$ by reordering the edges: The edges which are listed first are preferred. 
Therefore, our framework now supports any kinds of invariant templates.

  \newcommand{\setoffending}{\mfun{set\mhyphen{}offending\mhyphen{}flows}}
\newcommand{\getoffending}{\mfun{get\mhyphen{}offending\mhyphen{}flows}}
\newcommand{\getIFS}{\mfun{getIFS}}
\newcommand{\getACS}{\mfun{getACS}}
\newcommand{\Tau}{\mathit{T}}

\chapter{Directed Security Policies: A Stateful Network Implementation}
\label{chap:esss14}


This chapter is an extended version of the following paper~\cite{diekmann2014EPTCS}:
\begin{itemize}
	\item Cornelius Diekmann, Lars Hupel, and Georg Carle. \emph{Directed Security Policies: A Stateful Network Implementation}. In Engineering Safety and Security Systems, volume 150 of Electronic Proceedings in Theoretical Computer Science, pages 20-34, Singapore, May 2014. Open Publishing Association.
\end{itemize}

\noindent
The following improvements and new contributions were added:
\begin{itemize}
	\item This work has been evaluated with the cabin data network scenario (Section~\ref{sec:esss:casestudycabinnnetwork}). 
\end{itemize}

\paragraph*{Statement on author's contributions}
For the original paper, the author of this thesis provided major contributions for the ideas, requirement specification, formalization, realization, implementation, and proof of the algorithms. 
He researched related work, evaluated, and conducted the case study. 
All improvements with regard to the paper are the work of the author of this thesis. 

\medskip

\paragraph*{Abstract}
A security policy describes the communication relationship between networked entities. 
The security policy defines rules, for example that $A$ can connect to $B$. 
In the previous chapters, the policy was represented as a directed graph on the connection level. 
This policy should be implemented in a network, for example by firewalls, such that $A$ can establish a connection to $B$ and all packets belonging to established connections are allowed. 
We call this a stateful implementation. 
This stateful implementation is usually required for a network's functionality, but it introduces the backflow from $B$ to $A$, which might contradict the security policy. 
We derive compliance criteria for a policy and its stateful implementation and present a fast algorithm to translate a security policy to a stateful policy. 

\medskip

\section{Introduction}
Large systems with high requirements for security and reliability, such as SCADA or enterprise landscapes, no longer exist in isolation but are internetworked~\cite{hansen2012research}. 
Uncontrolled information leakage and access control violations may cause severe financial loss -- as demonstrated by Stuxnet -- and may even harm people if critical infrastructure is attacked. 
Hence, network security is crucial for system security. 

A central task of a network security policy is defining the network's desired connectivity structure and hence decreasing its attack surface against access control breaches and information leakage.
A security policy defines, among others, rules determining which host is allowed to communicate with which other hosts.
One of the most prominent security mechanisms to enforce a policy are network firewalls.
For adequate protection by a firewall, its ruleset is critical~\cite{bishop2003computer,bartal1999firmato}.
For example, let $A$ and $B$ be sets of networked hosts identified by their IP addresses.
Let $A \rightarrow B$ denote a policy rule describing that $A$ is allowed to communicate with $B$.
Several solutions from the fields of formal testing~\cite{brucker2008modelfwisabelle} to formal verification~\cite{fireman2006} can guarantee that a firewall actually implements the policy $A \rightarrow B$.
However, to the best of our knowledge, one subtlety between firewall rules and policy rules remains unsolved: 
For different scenarios, there are diverging means with different protection for translating the connection-level rule $A \rightarrow B$ to network-level firewall rules. 
We will exemplify this by two scenarios. 

\paragraph{Scenario 1}
Let $A$ be a workstation in some local network and $B$ represent a hosts in the Internet.
The policy rule $A \rightarrow B$ can be justified as follows:
The workstation can access the Internet, but the hosts in the Internet cannot access the workstation, \ie the workstation is protected from attacks from the Internet.
This policy can be translated to \eg the Linux iptables firewall~\cite{iptables} as illustrated in Figure~\ref{tab:intro:statefuliptables}.
The first rule allows $A$ to establish a new connection to $B$.
The second rule allows any communication over established connections in both directions, a very common practice.
For example, $A$ can request a website and the answer is transmitted back to $A$ over the established connection.
Finally, the last rule drops all other packets.
In particular, no one can establish a connection to $A$; hence $A$ is protected from malicious accesses from the Internet.

\begin{figure*}[htb]
\begin{minipage}{\linewidth}
\small
\begin{Verbatim}[commandchars=\\\{\},codes={\catcode`$=3\catcode`^=7}]
iptables -A INPUT -s $\mathit{\$A}$ -d $\mathit{\$B}$ -m conntrack --ctstate NEW -j ACCEPT
iptables -A INPUT -m conntrack --ctstate ESTABLISHED -j ACCEPT
iptables -A INPUT -j DROP
\end{Verbatim}
\end{minipage}%
  \caption{Stateful implementation of $A \rightarrow B$ in Scenario~1}%
  \label{tab:intro:statefuliptables}%
\end{figure*}

\begin{figure*}[htb]
\begin{minipage}{\linewidth}
\small
\begin{Verbatim}[commandchars=\\\{\},codes={\catcode`$=3\catcode`^=7}]
iptables -A INPUT -s $\mathit{\$A}$ -d $\mathit{\$B}$ -j ACCEPT
iptables -A INPUT -j DROP
\end{Verbatim}
\end{minipage}%
  \caption{Stateless implementation of $A \rightarrow B$ in Scenario~2}%
  \label{tab:intro:statelessiptables}%
\end{figure*}

\paragraph{Scenario 2}
In a different scenario, the same policy rule $A \rightarrow B$ has to be translated to a completely different set of firewall rules.
Assume that $A$ is a smart meter recording electrical energy consumption data, which is in turn sent to the provider's billing gateway $B$.
There, smart meter records of many customers are collected.
That data must not flow back to any customer, as this could be a violation of other customers' privacy.
For example, under the assumption that $B$ sends packets back to $A$, a malicious customer could try to infer the energy consumption records of their neighbors with a timing attack.
In Germany, the requirement for unidirectional communication of smart meters is even standardized by a federal government agency~\cite{bsi2013smartmeter}.
The corresponding firewall rules for this scenario can be written down as shown in Figure~\ref{tab:intro:statelessiptables}.
The first rule allows packets from $A$ to $B$, whereas the second rule discards all other packets.
No connection state is established; hence no packets can be sent from $B$ to $A$.

\bigskip

These two firewall rulesets were created from the same security policy rule \mbox{$A \rightarrow B$}.
The first implementation treats ``$\rightarrow$'' as ``can initiate connections to'', whereas the second implementation treats ``$\rightarrow$'' as ``can send packets to''.
The second implementation appears to be simpler and more secure, and the firewall rules are justifiable more easily by the policy.
However, this firewall configuration is undesirable in many scenarios as it might affect the desired functionality of the network.
For example, surfing the web is not possible as no responses (\ie websites) can be transferred back to the requesting host.

A decision must be made whether to implement a policy rule $A \rightarrow B$ in the stateful (Figure~\ref{tab:intro:statefuliptables}) or in the stateless fashion (Figure~\ref{tab:intro:statelessiptables}).
The stateful fashion bears the risk of undesired side effects by allowing packet flows that are opposite to the security policy rule.
In particular, this could introduce information leakage.
On the other hand, the stateless fashion might impair the network's functionality. 
Hence, stateful flows are preferable for network operation, but are undesirable with regard to security. 
In this chapter, we tackle this problem by maximizing the number of policy rules that can be made stateful without introducing security issues.

We can see that even if a well-specified security policy exists, its implementation by a firewall configuration remains a manual and hence error-prone task.
A 2012 survey~\cite{sherry2012making} of 57 enterprise network administrators confirms that a ``majority of administrators stated misconfiguration as the most common cause of failure'' \cite{sherry2012making}.
A study~\cite{databreach2009src} conducted by Verizon from 2004 to 2009 and the United States Secret Service during 2008 and 2009 reveals that data leaks are often caused by configuration errors~\cite{databreach2009}.

In this chapter, we answer the following questions:
\begin{itemize}
	\item What conditions can be checked to verify that a stateful policy implementation complies with the directed network security policy rules?
	\item When can a policy rule $A \rightarrow B$ be upgraded to allow a stateful connection between $A$ and $B$?
\end{itemize}

Our results apply not only to firewalls but to any network security mechanisms that shape network connectivity.

The outline of this chapter is as follows. 
Section~\ref{sec:esss:example} presents a guiding example. 
Section~\ref{sec:esss:model} formalizes the key concepts of directed policies, security requirements, and stateful policies. 
Section~\ref{sec:esss:requiremensstateful} discusses the requirements for a stateful policy to comply with a directed policy. 
Section~\ref{sec:esss:algorithm} presents an algorithm to automatically derive a stateful policy. 
Sections~\ref{sec:esss:computationalcomplexity} and \ref{sec:esss:casestudy} evaluate our work: 
Section~\ref{sec:esss:computationalcomplexity} discusses the computational complexity of the algorithm, and Section~\ref{sec:esss:casestudy} presents a large real-world case study.

\section{Example}
\label{sec:esss:example}

\begin{figure*}[htb]
  \centering
  \hspace*{\fill}%
  \begin{subfigure}[t]{0.48\textwidth}
       \centering
       \captionsetup{width=0.80\textwidth}
       \includegraphics[width=1.0\textwidth]{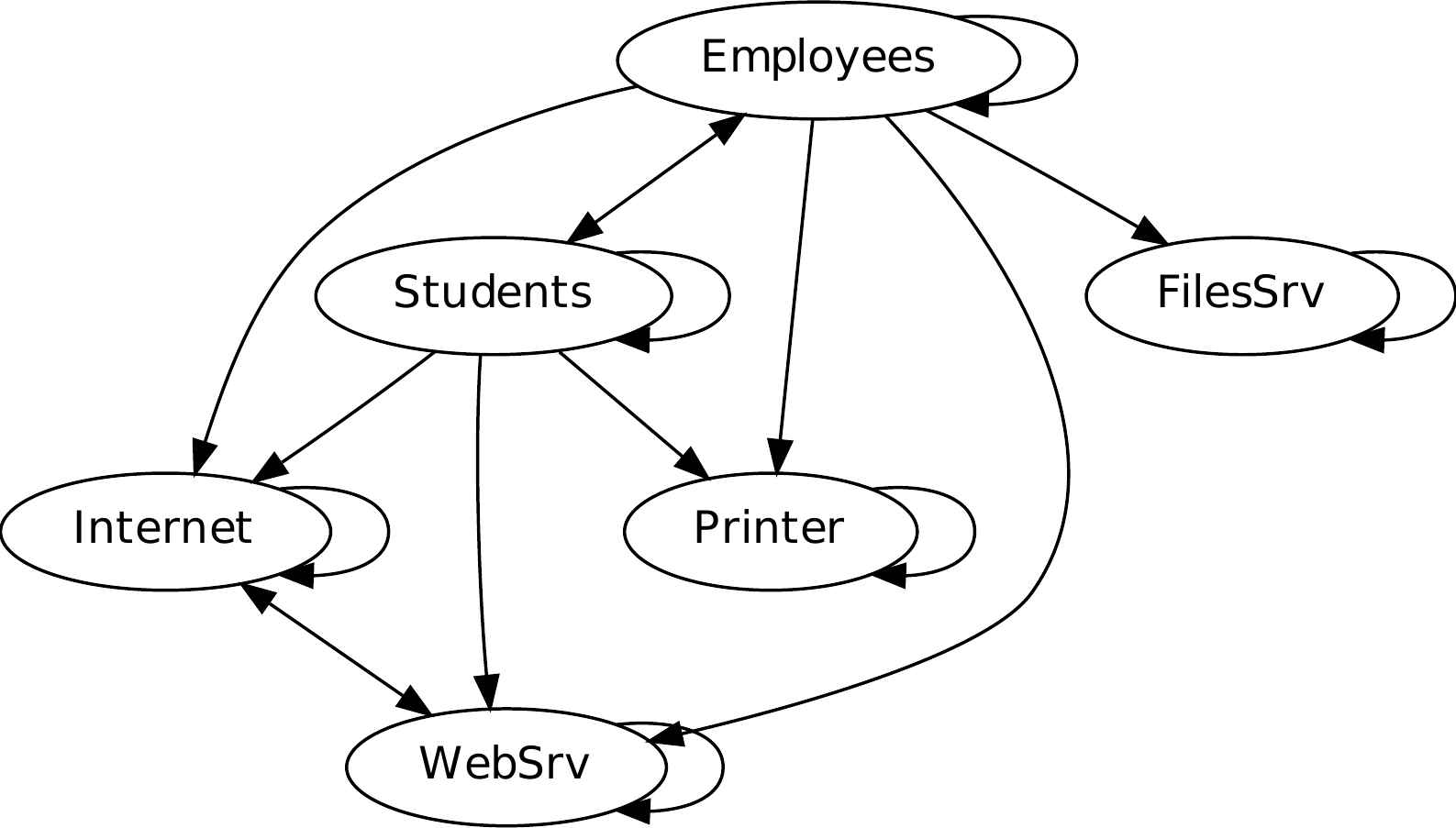}
  \caption{Network security policy}
  \label{fig:intro:policygraph}
  \end{subfigure}%
  \hspace*{\fill}%
  \begin{subfigure}[t]{0.48\textwidth}
       \centering
       \captionsetup{width=0.80\textwidth}
       \includegraphics[width=1.0\textwidth]{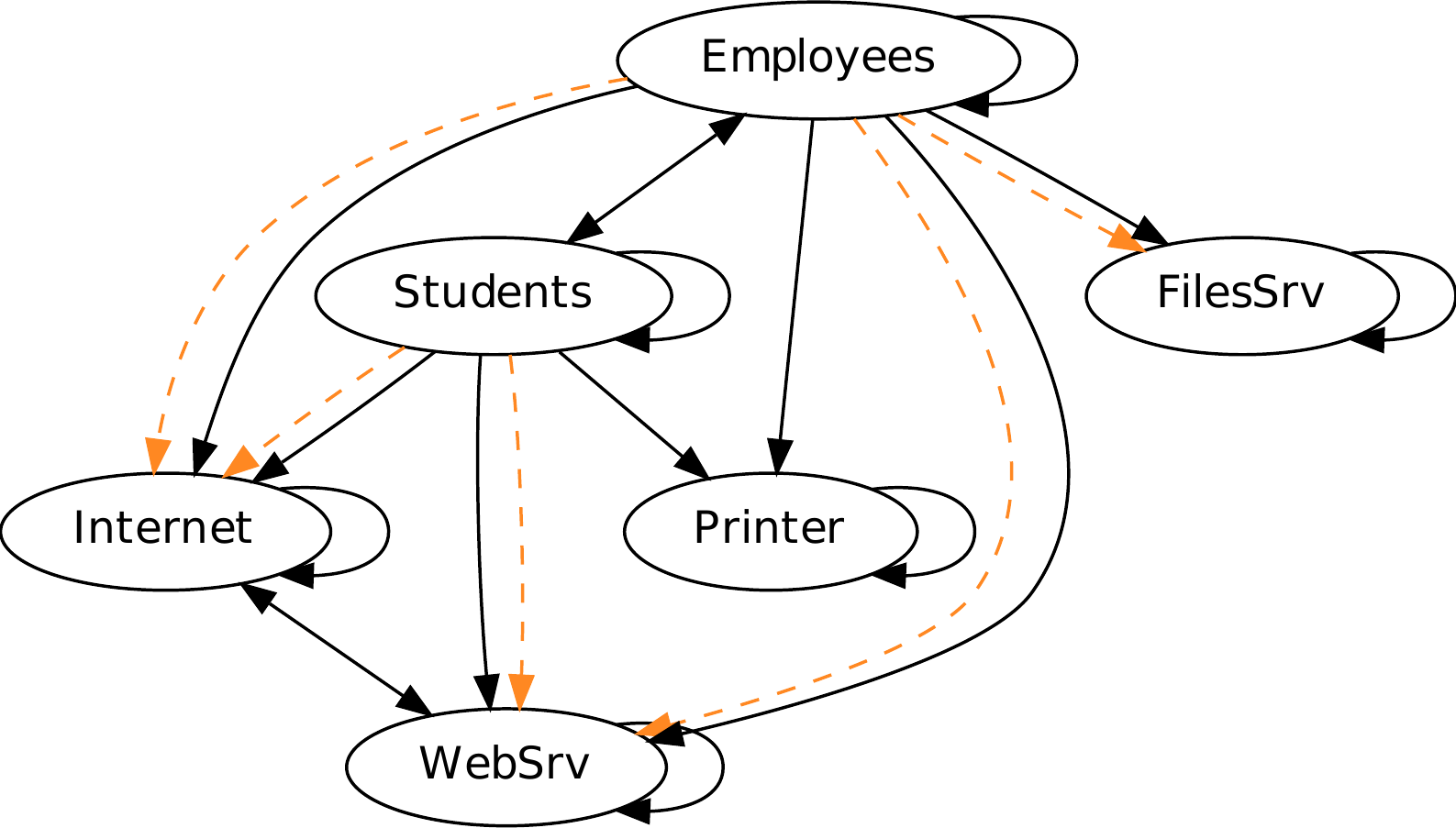}
  \caption{Stateful implementation}
  \label{fig:intro:statefulpolicygraph}
  \end{subfigure}%
  \hspace*{\fill}%
  \caption{The network security policy and its stateful implementation}
  \label{fig:intro}
\end{figure*}

We introduce a network -- from a hypothetical university department -- to illustrate the problem with a complete example and outline the solution before we describe its formalization in the next section.

The network (depicted in Figure \ref{fig:intro}) consists of the following participants: the students, the employees, a printer, a file server, a web server, and the Internet. 
The network security policy rules are depicted in Figure~\ref{fig:intro:policygraph} as a directed graph. 
A security policy rule $A \rightarrow B$ is denoted by an edge from $A$ to $B$. 
We have formalized the security invariants with the help of our security invariant template library. 
For brevity, we omit the formal configuration details since for this chapter, we will only need the distinction between information flow strategies and access control strategies. 
For this chapter, we will always write `$\cdot \rightarrow \cdot$' to visually denote a policy rule and its direction. 
The security policy is designed to fulfill the following security invariants:

\begin{description}
  \item[Access Control Invariants]
    The printer is only accessible by the employees and students; as policy, $\mvar{employees} \rightarrow \mvar{printer}$ and $\mvar{students} \rightarrow \mvar{printer}$. 
    The file server is only accessible by employees, formally \mbox{$\mvar{employees} \rightarrow \mvar{fileSrv}$}.
    The students and the employees are in a joint subnet that allows collaboration between them but protects against accesses from \eg the Internet or a compromised web or file server. 

  \item[Information Flow Invariants]
    The file server stores confidential data that must not leak to untrusted parties.
    Only the employees have the necessary security level to receive data from the file server.
    The employees are also trustworthy, \ie they may declassify and reveal any data received by the file server.
    The printer is an information sink.
    Confidential data (such as an exam) might be printed by an employee.
    No other network participants, in particular no students, are allowed to retrieve any information from the printer that might allow them to draw conclusions about the printed documents. 
    This can be formalized by the policy ``$\mvar{*} \rightarrow \mvar{printer}$'' and ``$\mvar{printer} \nrightarrow \mvar{*}$''. 
\end{description}

Note that Figure~\ref{fig:intro} are screenshots of our tool \topos{}. 

\subsection*{Stateful Policy Implementation}
Considering Figure~\ref{fig:intro:policygraph}, it is desirable to allow stateful connections from the employees and students to the Internet and the web server.
Figure~\ref{fig:intro:statefulpolicygraph} depicts the stateful policy implementation, where the additional dashed edges represent flows that are allowed to be stateful, \ie answers in the opposite direction are allowed.
Only strict stateless unidirectional communication with the printer is necessary.
The students and employees can, as already defined by the policy, bidirectionally interact with each other. 
Hence stateful semantics are not necessary for these flows.

In this chapter, we specify conditions to verify that the stateful policy implementation (\eg Figure~\ref{fig:intro:statefulpolicygraph}) complies with the directed security policy (\eg Figure~\ref{fig:intro:policygraph}).
We present an efficiently computable condition and formally prove that it implies several complex compliance conditions.
Finally, we present an algorithm that automatically computes a stateful policy from the directed policy and the security invariants.
We formally prove the algorithm's correctness and that it can always compute a maximal possible set of stateful flows with regard to access control and information flow security strategies.

\section{Formal Model}
\label{sec:esss:model}
In Chapter~\ref{chap:forte14}, we introduced our formal model. 
We presented security invariant templates and how configured security invariants can be derived from them. 
In this chapter, we only work with configured security invariants. 
For brevity, we will simply say `security invariant' for a \emph{configured} security invariant which was derived from a template. 
We first repeat the core definitions, simplified for configured security invariant templates. 

\paragraph*{Network Security Policy Rules}
We represent the network security policy's access rules as directed graph $G = (V,\, E)$. 
The type of all graphs is denoted by $\mathcal{G}$. 
For example, the policy that only consists of the rule that $A$ can send to $B$, denoted by $A \rightarrow B$, is represented by the graph $G = (\lbrace A, B \rbrace, \ \lbrace (A, B) \rbrace)$. 
An edge in the graph corresponds to a permitted flow in the network. 
We call this policy a \emph{directed policy}. 
In Section~\ref{subsec:esss:statefulpolicy}, we will introduce the notion of a stateful policy. 

We consider only syntactically well-formed graphs. 
A graph is \emph{syntactically well-formed}\footnote{FiniteGraph.wf-graph} if all nodes in the edges are also listed in the set of vertices. 
In addition, since we represent finite networks, we require that $V$ is a finite set. 
This does not prevent creating nodes that represent collections of arbitrary many hosts, \eg the node $\mvar{Internet}$ in Figure~\ref{fig:intro:policygraph} represents arbitrarily many hosts.

\paragraph*{Network Security Invariants}
A security invariant $m$ specifies whether a given policy $G$ fulfills its security requirements. 
As we focus on the network security policy's access rules which specify which hosts are allowed to communicate with which other hosts, we do not take availability or resilience requirements into account. 
Instead, we deal with only the traditional security invariants that follow the principle ``prohibiting more is more or equally secure''. 
We call this principle \emph{monotonicity}. 
To allow arbitrary network security invariants, almost any total function $m$ of type $\mathcal{G} \Rightarrow \mathbb{B}$ can be used to specify a network security requirement (as long as the proof obligations for the corresponding template imposed by Chapter~\ref{chap:forte14} can be discharged). 
 
We distinguish between the two security strategies that $m$ is set to fulfill: 
\emph{Information flow security strategies} (IFS) prevent data leakage; 
\emph{Access control strategies} (ACS) are used to prevent illegal or unauthorized accesses. 

\begin{definition}[Configured Security Invariant]
\label{def:securityinvariant}
A network security invariant $m$ is a total function \mbox{$\mathcal{G}  \Rightarrow\: \mathbb{B}$} with a security strategy (either IFS or ACS) satisfying the following conditions: 
\begin{itemize}
		\item {If no communication exists in the network, the security invariant must be fulfilled: $m \ (V,\ \emptyset) $}
		
		\item {Monotonicity: $m \ (V,E) \; \wedge \; E' \subseteq E \Longrightarrow m \ (V,E')$} 
	\end{itemize}
\end{definition}

If there is a security violation for $m$ in $G$, there must be at least one set $F \subseteq E$ such that the security violation can be remedied by removing $F$ from $E$.\footnote{Since $m \ (V,\ \emptyset) $, it is obvious that such a set always exists.} 
We call $F$ offending flows. 
$F$ is \emph{minimal} if all flows $(s, r) \in F$ contribute to the security violation. 
For $m$, the set of all minimal offending flows can be defined. 
The definition $\setoffending$ describes a set of sets, containing all minimal candidates for $F$. 
\begin{IEEEeqnarray*}{l} \setoffending \ m \ G \equiv \\
\qquad \bigl\lbrace F \subseteq E \mid \neg\, m \ G \ \wedge \ m \ (V,\ E \setminus F)\ \ \wedge \ 
         \forall (s,r) \in F.\ \neg\, m \ (V,\, (E \setminus F) \cup \lbrace (s,r) \rbrace)  \bigr\rbrace
\end{IEEEeqnarray*}

The offending flows inherit $m$'s monotonicity property.\footnote{offending-flows-union-mono} 
\begin{lemma}[Monotonicity of Offending Flows]
\label{lemma:mono-union-offending-flows}
\begin{IEEEeqnarray*}{l}
E' \subseteq E \Longrightarrow \bigcup \setoffending \ m \ (V, E') \subseteq \bigcup \setoffending \ m \ (V, E) 
\end{IEEEeqnarray*}
\end{lemma}

If there is an upper bound for the offending flows, it can be narrowed.\footnote{Un-set-offending-flows-bound-minus-subseteq} 
This is a key insight which will be used for the efficient implementation of the algorithms in this chapter. 
\begin{lemma}[Narrowed Upper Bound of Offending Flows]
\label{lemma-upperboundsubstract}
{Let $E'$ be a set of edges. 
If the offending flows are bounded, \ie if \ $\bigcup \setoffending \ m \ (V, E) \subseteq X$ holds, then
$\bigcup \setoffending \ m \ (V, E \setminus E') \subseteq X \setminus E'$.}
\end{lemma}
\begin{proof}
\begin{sloppypar}
From Lemma~\ref{lemma:mono-union-offending-flows}, we have $\bigcup \setoffending \ m \ (V, E \setminus E') \subseteq \bigcup \setoffending \ m \ (V, E)$. 
This implies that $\left( \bigcup \setoffending \ m \ (V, E \setminus E') \right) \setminus E' \subseteq \left( \bigcup \setoffending \ m \ (V, E) \right) \setminus E'$.
Since the set of offending flows only returns subsets of the graph's edges, the left hand side can be simplified: 
$\bigcup \setoffending \ m \ (V, E \setminus E') \subseteq \left( \bigcup \setoffending \ m \ (V, E) \right) \setminus E'$. 
From the assumption, it follows that $\left( \bigcup \setoffending \ m \ (V, E) \right) \setminus E' \subseteq X \setminus E'$.
We finally obtain
\begin{IEEEeqnarray*}{l}
\bigcup \setoffending \ m \ (V, E \setminus E') \subseteq \left( \bigcup \setoffending \ m \ (V, E) \right) \setminus E' \subseteq X \setminus E'
\end{IEEEeqnarray*}
by transitivity.
\end{sloppypar}
\end{proof}

We now define some helper functions for a set of configured security invariants. 

\begin{definition}[Configured Security Invariants]
\label{def:securityinvarinatlist}
We call a finite list of security invariants $M = [m_1, m_2, ..., m_k]$ a network's security invariants. 
The functions $\getIFS \ M$ (and $\getACS \ M$) return all $m \in M$ with an IFS (and ACS, respectively) security strategy.
Additionally, we abbreviate all sets of offending flows for all security invariants with $\getoffending \ M \ G = \bigcup_{m \in M} \setoffending \ m \ G$. 
Similarly to $\setoffending$, it denotes a set of sets.
\end{definition}

\subsection{Stateful Policy Implementation}
\label{subsec:esss:statefulpolicy}
We define a stateful policy similarly to a directed policy. 
\begin{definition}[Stateful Policy]
A stateful policy $\Tau = (V,\, E_\tau,\, E_\sigma)$ is a triple consisting of the networked hosts $V$, the flows $E_\tau$, and the stateful flows $E_\sigma \subseteq E_\tau$. 
\end{definition}

The meaning of $E_\sigma$ is that these flows are allowed to be stateful. 
We consider the stateful flows $E_\sigma$ as ``upgraded'' flows, hence $E_\sigma \subseteq E_\tau$. 
This means that if $(s, r) \in E_\sigma$, flows in the opposite direction, \ie $(r, s)$ may exist. 
For a set of edges $X$, we define the \emph{backflows} of $X$ as $\overleftarrow{X} = \lbrace (r,s) \mid (s,r) \in X \rbrace$. 
Hence, the semantics of $E_\sigma$ can be described as that both the flows $E_\sigma$ and $\overleftarrow{E_\sigma}$ may exist. 
We define a mapping that translates a stateful policy $\Tau$ to a directed policy $G$ as 
$\alpha \ \Tau \equiv (V,\, E_\tau \cup E_\sigma \cup \overleftarrow{E_\sigma})$. 

\begin{example}
\textbf{Example. }%
The ultimate goal is to translate a directed policy $G = (V,\, E)$ to a stateful implementation $\Tau = (V,\, E_\tau,\, E_\sigma)$ that contains as many stateful flows $E_\sigma$ as possible without introducing security flaws. 
The trivial choice is $\Tau_{\mathrm{triv}} = (V,\, E,\, \emptyset)$. 
It fulfills all security invariants because $\alpha \ \Tau_{\mathrm{triv}} = G$. 
Since $E_\sigma = \emptyset$, it does not maximize the stateful flows. 
\end{example}

\medskip
Before discussing requirements for the compliance of $\Tau$ and $G$, we first have to define the requirements for a \emph{syntactically well-formed} stateful security policy.\footnote{wf-stateful-policy, stateful-policy-compliance} 
All nodes mentioned in $E_\tau$ and $E_\sigma$ must be listed in $V$. 
The flows $E_\tau$ must be allowed by the directed policy, hence $E_\tau \subseteq E$, which also implies $E_\sigma \subseteq E$ by transitivity. 
%
%
The nodes in $\Tau$ are equal to the nodes in $G$. 
This implies that $E_\tau$ and $E_\sigma$ are finite\footnote{wf-stateful-policy.finite-$\ast$}. 
In the rest of this chapter, we always assume that $\Tau$ is syntactically well-formed. 

From these conditions, we conclude that $\Tau$ and $G$ are similar and $\Tau$ syntactically introduces neither new hosts nor flows. 
Semantically, however, $\alpha \ \Tau$ adds $\overleftarrow{E_\sigma}$, which might introduce new flows. 
Hence, the edges of $\alpha \ \Tau$ need not be a subset of $G$'s edges (nor vice versa).

\section{Requirements for Stateful Policy Implementation}
\label{sec:esss:requiremensstateful}
We assume that $G$ is a \emph{valid policy}. 
In addition to being syntactically well-formed, that means that all security invariants must be fulfilled, \ie $ \forall m \in M. \ m \ G$. 
We derive requirements to verify that a stateful policy $\Tau$ is a proper stateful implementation of $G$ without introducing security flaws.  

\subsection{Requirements for Information Flow Security Compliance}
Information leakages are critical and can occur in subtle ways. 
For example, the widely used transport protocol TCP detects data loss by sending acknowledgment packets. 
If $A$ establishes a TCP connection to $B$, then even if $B$ sends no payload, arbitrary information can be transmitted to $A$, \eg via timing channels, TCP sequence numbers, or retransmits. 
Therefore, we treat information flow security requirements carefully: 
When considering backflows, all information flow security invariants must still be fulfilled. 
\begin{IEEEeqnarray}{l}
\label{eq-ifs:fulfilled}
 \forall m \in \getIFS \  M. \ \ m \ (\alpha \ \Tau)
\end{IEEEeqnarray}

\begin{example}
\textbf{Example. }%
For our simple BLP example on page~\pageref{example:forte14:blp}, this means that no TCP connection can be established between hosts of different security levels. 
\end{example}

\subsection{Requirements for Access Control Strategies}
In contrast, the requirements for access control invariants can be slightly relaxed: 
If $A$ accesses $B$, $A$ might expect an answer from $B$ for its request. 
If $B$'s answer is transmitted via the connection that $A$ established, $B$ does not access $A$ on its own initiative. 
Only the expected answer is transmitted back to $A$. 
If $A$'s software contains no vulnerability which $B$ could exploit with its answer, no access violation occurs.\footnote{Note that we make an important assumption here. This assumption is justified as we only work on the network level and do not consider the application level, which is also the correct abstraction for network administrators when configuring network security mechanisms. It also implies that, as always, vulnerable applications with access to the Internet can cause severe damage.} 
This behavior is widely deployed in many private and enterprise networks by the standard policy that internal hosts can access the Internet and receive replies, but the Internet cannot initiate connections to internal hosts. 

Therefore, we can formulate the requirement for ACS compliance. 
Access control violations caused by stateful backflows can be tolerated.
However, \emph{negative side effects} must not be introduced by permitting these backflows. 
First, we present an example of a negative side effect.
Second, we derive a requirement for verifying the lack of side effects. 

\begin{example}
\textbf{Example. }%
We examine a building automation network. 
Let $B$ be the master controller, $A$ a door locking mechanism, and $C$ a log server that records who enters and who leaves the building. 
The controller $B$ decides when the door should be opened and what to log. 
%
The directed policy is described by $G = (\lbrace A, B, C \rbrace,\  \lbrace (B, A), (B, C)\rbrace)$. 
The only security invariant $m$ is that $A$ is not allowed to transitively access $C$. 
Let~$\rightarrow^*$ denote the transitive closure of $\rightarrow$. 
Then, $m$ prohibits $A \rightarrow^* C$, but it does not prohibit $C \rightarrow^* A$. 
In this scenario, that means that the physically accessible locking mechanism must not tamper with the integrity of the log server.

Setting $E_\sigma = \lbrace (B,A) \rbrace$ gives $\Tau = (\lbrace A, B, C \rbrace,\ \lbrace (B, A), (B, C)\rbrace,\ \lbrace (B,A) \rbrace)$, and hence $\alpha \ \Tau = (\lbrace A, B, C \rbrace, \lbrace (B, A), (B, C), (A, B)\rbrace)$. 
This attempt results in a negative side effect. 
We compute the offending flows for $m$ of $\alpha\ \Tau$ as $\lbrace \lbrace(B, C)\rbrace,\  \lbrace(A, B)\rbrace \rbrace = \lbrace \lbrace(B, C)\rbrace,\  \overleftarrow{E_\sigma} \rbrace$.
Clearly, a violation occurs in $\overleftarrow{E_\sigma}$. 
Additionally, there is a side effect: the flow from $B$ to $C$ could now cause a violation. 
Applied to our scenario, this means that in case the locking mechanism sends forged data to the controller, that data could end up in the log.  
This is a negative side effect. 
Hence $(B,A)$ cannot securely be made stateful. 
For completeness, note that because $A$ is just a simple physical actor which only executes $B$'s commands, there is no need for bidirectional communication. 
On the other hand, $(B,C)$ can be made stateful without side effects. 
\end{example}

We formalize the requirement of ``no negative side effects'' as follows: The violations caused by any subset of the backflows are at most these backflows themselves. 
\begin{IEEEeqnarray}{l}
\label{eq-acs:subsets}
\forall X \subseteq \overleftarrow{E_\sigma}.\ \forall F \in \getoffending \ (\getACS \ M) \ (V,\, E_\tau \cup E_\sigma \cup X). \ F \subseteq X
\end{IEEEeqnarray}
In particular, all offending access control violations are at most the stateful backflows. This is directly implied by the previous requirement by choosing $X$ to be $\overleftarrow{E_\sigma}$ (recall the definition of $\alpha$).  
\begin{IEEEeqnarray}{l}
\label{eq-acs:allset}
\bigcup \getoffending \ (\getACS \ M) \ (\alpha \ \Tau) \subseteq \overleftarrow{E_\sigma}
\end{IEEEeqnarray}
Also, considering all backflows individually, they cause no side effects, \ie the only violation added is the backflow itself. 
 \begin{IEEEeqnarray}{l}
\label{eq-acs:singletonset}
\forall (r, s) \in \overleftarrow{E_\sigma}. \\
\qquad \bigcup \getoffending \ (\getACS \ M) \ (V,\, E_\tau \cup E_\sigma \cup \lbrace (r, s) \rbrace)  \subseteq \lbrace (r, s) \rbrace \nonumber
\end{IEEEeqnarray} 
It is obvious that (\ref{eq-acs:subsets}) implies both (\ref{eq-acs:allset}) and (\ref{eq-acs:singletonset}).\footnote{stateful-policy-compliance.compliant-stateful-ACS-only-state-violations-union,\newline stateful-policy-compliance.compliant-stateful-ACS-no-state-singleflow-side-effect} 
The condition of (\ref{eq-acs:subsets}) is imposed on all subsets, thus ruling out all possible undesired side effects. 

However, translating (\ref{eq-acs:subsets}) to executable code results in exponential runtime complexity, because it requires iterating over all subsets of $\overleftarrow{E_\sigma}$. 
This is infeasible for any large set of stateful flows. 
In this chapter, we contribute a new formula\footnote{stateful-policy-compliance.compliant-stateful-ACS}, which implies (\ref{eq-acs:subsets}) and hence (\ref{eq-acs:allset}) and (\ref{eq-acs:singletonset}). 
It has a comparably low computational complexity and thus enables writing executable code for the automated verification of stateful and directed policies. 
\begin{IEEEeqnarray}{l}
\label{eq-acs:theallimplyACSformula}
\bigcup \getoffending \ (\getACS \ M) \ (\alpha \ \Tau ) \ \subseteq\ \overleftarrow{E_\sigma} \setminus E_\tau
\end{IEEEeqnarray}
Obviously, the runtime complexity of (\ref{eq-acs:theallimplyACSformula}) is significantly lower than (\ref{eq-acs:subsets}) (see \mbox{Section~\ref{sec:esss:computationalcomplexity}}).
The formula also bears great resemblance to (\ref{eq-acs:allset}). 
We explain the intention of (\ref{eq-acs:theallimplyACSformula}) and prove that it implies (\ref{eq-acs:subsets}). 

Note that $\overleftarrow{E_\sigma} \setminus E_\tau = \overleftarrow{\lbrace(s, r) \in E_\sigma  \mid  (r, s) \notin E_\tau \rbrace}$ \footnote{backflows-filternew-flows-state}, which means that it represents the backflows of all flows that are not already in $E_\tau$. 
In other words, it represents only the newly added backflows. 
For example, consider the flows between students and employees in Figure~\ref{fig:intro:statefulpolicygraph}: 
no stateful flows are necessary as bidirectional flows are already allowed by the policy, and the newly added backflows are represented by the dashed edges. 
Therefore, (\ref{eq-acs:theallimplyACSformula}) requires that all introduced violations are only due to the newly added backflows. 
This requirement is sufficient to imply (\ref{eq-acs:subsets}).\footnote{stateful-policy-compliance.compliant-stateful-ACS-no-side-effects}

\begin{theorem}[Efficient ACS Compliance Criterion]
For ACS, verifying that all introduced violations are only due to the newly added backflows is sufficient to verify the lack of side effects. 
Formally, $(\ref{eq-acs:theallimplyACSformula})~\Longrightarrow~(\ref{eq-acs:subsets})$. 
\end{theorem}
\begin{proof}
We assume (\ref{eq-acs:theallimplyACSformula}) and show (\ref{eq-acs:subsets}) for an arbitrary but fixed $X \subseteq \overleftarrow{E_\sigma}$. 
We need to show that $\forall F \in \getoffending \ (\getACS \ M) \ (V,\, E_\tau \cup E_\sigma \cup X). \ F \subseteq X$. 
We split $\overleftarrow{E_\sigma}$ into $\overleftarrow{E_\sigma} \ \setminus \ E_\tau$ and $\overleftarrow{E_\sigma} \setminus ( \overleftarrow{E_\sigma} \ \setminus \ E_\tau )$. 
Likewise, we can split $X$ into $X_1 \subseteq \overleftarrow{E_\sigma} \ \setminus \ E_\tau$ and $X_2 \subseteq \overleftarrow{E_\sigma} \setminus ( \overleftarrow{E_\sigma} \ \setminus \ E_\tau )$. 
Hence, $X_2 \subseteq E_\tau$ and immediately $E_\tau \cup X_2 = E_\tau$. 
This simplifies the goal as $X_2$ disappears from the edges: %
\vskip-18pt
\begin{IEEEeqnarray*}{l}%
  \forall F \in \getoffending \ (\getACS \ M) \ (V,\, E_\tau \cup E_\sigma \cup X_1). \ F \subseteq X
\end{IEEEeqnarray*}%
\vskip-2pt%
\noindent We show an even stricter version of the goal since $X = X_1 \cup X_2$. %
\vskip-18pt%
\begin{IEEEeqnarray*}{l}%
  \forall F \in \getoffending \ (\getACS \ M) \ (V,\, E_\tau \cup E_\sigma \cup X_1). \ F \subseteq X_1
\end{IEEEeqnarray*}
\vskip-2pt%
\noindent This directly follows\footnote{stateful-policy-compliance.compliant-stateful-ACS-no-side-effects-filternew-helper} by using Lemma~\ref{lemma-upperboundsubstract} and subtracting $(\overleftarrow{E_\sigma} \ \setminus \ E_\tau) \setminus X_1$ from (\ref{eq-acs:theallimplyACSformula}). 
\end{proof}

\section{Automated Stateful Policy Construction}
\label{sec:esss:algorithm}
In this section, we present algorithms to calculate a stateful implementation of a directed policy for a given set of security invariants using (\ref{eq-ifs:fulfilled}) and (\ref{eq-acs:theallimplyACSformula}). 

Instead of a set, the algorithms' last parameter is a list because the order of the elements matters. 
We use list notation as described in Chapter~\ref{chap:introisablle}. 
Since lists can be easily converted to finite sets, we make this conversion implicit for brevity. 
For example, for a list $a$, we will write the stateful policy as $(V,\, E,\, a)$, where $a$ is implicitly converted to a finite set. 

\subsection{Information Flow Security Strategies}
We start by presenting an algorithm which selects stateful edges in accordance to the IFS security invariants. 
The algorithm filters a given list of edges for edges which fulfill (\ref{eq-ifs:fulfilled}). 
It also takes as input the directed policy $G$, the security invariants $M$, and a list of edges as accumulator $a$. 
\begin{IEEEeqnarray*}{lcl}
\mfun{filterIFS} \ G \ M \ a \;\textnormal{\texttt{[]}} & \ = \ \ & a\\
\mfun{filterIFS} \ G \ M \ a \ (e \lstcons \mvar{es}) & \ = \ \ & \mctrl{if} \;\;
  \forall m \in \getIFS \ \  M. \ m \ \left(\alpha \ \left(V, E, e \lstcons a \right)\right)
   \;\; \mctrl{then} \\
\ & & \quad \mfun{filterIFS} \ G \ M \ (e \lstcons a) \ \mvar{es}\\
\ & & \mctrl{else} \\
\ & & \quad \mfun{filterIFS} \ G \ M \ a \ \mvar{es}
\end{IEEEeqnarray*}
The accumulator, initially empty, returns the result in the end. 
It is the current set of selected stateful flows. 
The algorithm is designed such that (\ref{eq-ifs:fulfilled}) always holds for $\Tau = (V, E, a)$. 
It simply iterates over all elements $e$ of the input list and checks whether the formula also holds if $e$ is added to $a$. 
If so, $e$ is added to the accumulator; otherwise, $a$ is left unchanged. 

Depending on the security invariants, multiple results are possible with this filtering criterion. 
The algorithm deterministically returns one solution. 
Users can influence the choice of edges that they want to be stateful by arranging the input list such that the preferred edges are listed first. 
If only one arbitrary solution is desired, lists and finite sets are interchangeable.

The algorithm is sound\footnote{filter-IFS-no-violations-correct} and complete.\footnote{filter-IFS-no-violations-maximal-allsubsets} 

\begin{lemma}[\texttt{filterIFS} Soundness]
If the directed policy $G = (V,\, E)$ is valid, 
then for any list $X \subseteq E$, the stateful policy $\Tau = (V,\, E,\, \mfun{filterIFS}\ G \  M \;\textnormal{\texttt{[]}} \ X)$ fulfills (\ref{eq-ifs:fulfilled}). 
\end{lemma}


\begin{lemma}[\texttt{filterIFS} Completeness]
\label{lem:esss:ifscomplete}
For $G = (V,\, E)$, 
let $E_\sigma = \mfun{filterIFS} \ G \ M \ E$. 
Then, no non-empty subset can be added to $E_\sigma$ without violating (\ref{eq-ifs:fulfilled}). %
\begin{IEEEeqnarray*}{l}
\forall X \subseteq E \setminus E_\sigma, \  X \neq \emptyset. \ 
              \neg \forall m \in \getIFS \ M \ \left(\alpha \ (V,\, E,\, E_\sigma \cup X)\right)
\end{IEEEeqnarray*}
\end{lemma}

\subsection{Access Control Strategies}
The algorithm \texttt{filterACS} follows the same principles as \texttt{filterIFS}. 
%
\begin{IEEEeqnarray*}{lcl}
\mfun{filterACS} \ G \ M \ a \;\textnormal{\texttt{[]}} & \ = \ \ & a\\
\mfun{filterACS} \ G \ M \ a \ (e \lstcons \mvar{es}) & \ = \ \ & \\
\IEEEeqnarraymulticol{3}{l}{\qquad\quad \mctrl{if}}\\
\IEEEeqnarraymulticol{3}{l}{\qquad\quad \quad
        e \notin \overleftarrow{E} \ \wedge \  \left(\forall F \in \getoffending \ (\getACS \ M) \ (\alpha\ (V,\, E,\, e :: a)). \ F \subseteq \overleftarrow{e \lstcons a}\right)
}\\
\IEEEeqnarraymulticol{3}{l}{\qquad\quad \mctrl{then}}\\
\IEEEeqnarraymulticol{3}{l}{\qquad\quad \quad \mfun{filterACS} \ G \ M \ (e \lstcons a) \ \mvar{es}}\\
\IEEEeqnarraymulticol{3}{l}{\qquad\quad \mctrl{else}}\\
\IEEEeqnarraymulticol{3}{l}{\qquad\quad \quad \mfun{filterACS} \ G \ M \ a \ \mvar{es}}
\end{IEEEeqnarray*}
As previously, the order of the elements in the list influences the choice of calculated stateful edges. 
Edges listed first are preferred. 
The algorithm is sound\footnote{filter-compliant-stateful-ACS-correct} and complete.\footnote{filter-compliant-stateful-ACS-maximal-allsubsets} 

\begin{lemma}[$\mfun{filterACS}$ Soundness]
If the directed policy $G = (V,\, E)$ is valid, 
then for any list $X \subseteq E$, the stateful policy $\Tau = (V,\, E,\, \mfun{filterACS}\ G \  M \;\textnormal{\texttt{[]}} \ X)$ fulfills (\ref{eq-acs:theallimplyACSformula}). 
\end{lemma}

To show that $\mfun{filterACS}$ computes a maximal solution, we must first identify the candidates that $\mfun{filterACS}$ might overlook. 
Flows that are already bidirectional need not be stateful. 
As illustrated in the example of Figure~\ref{fig:intro:statefulpolicygraph}, no added value is created if stateful connections between students and employees were allowed as no communication restrictions exist between these groups in the first place. 
Hence only $E \setminus \overleftarrow{E}$ is considered. 
\begin{lemma}[$\mfun{filterACS}$ Completeness]
\label{lem:esss:acscomplete}
For $G = (V,\, E)$, 
let $E_\sigma = \mfun{filterACS} \ G \ M \ E$. 
Then, no non-empty subsets $X \subseteq E \setminus ( E_\sigma \cup \overleftarrow{E})$ can be added to $E_\sigma$ without violating (\ref{eq-acs:theallimplyACSformula}). %
\begin{IEEEeqnarray*}{l}
\forall X \subseteq E \setminus ( E_\sigma \cup \overleftarrow{E} ), \  X \neq \emptyset. \\ 
\qquad \neg \left( \bigcup \getoffending \ (\getACS \ M) \ (\alpha \ (V,\, E,\, E_\sigma \cup X) ) \ \subseteq \ \overleftarrow{E_\sigma \cup X} \setminus E\right)
\end{IEEEeqnarray*}
\end{lemma}

\subsection{IFS and ACS Combined}
Finally, we combine the previous section's algorithms to derive algorithms which compute a solution that satisfies all requirements of a stateful policy.

The first algorithm\footnote{generate-valid-stateful-policy-IFSACS} simply chains $\mfun{filterIFS}$ and $\mfun{filterACS}$. 
\begin{IEEEeqnarray*}{l}
\mfun{generate1} \ G \ M \ e = \left(V,\, E,\, \mfun{filterACS} \ G \ M \  (\mfun{filterIFS} \ G \ M \ e)\right)
\end{IEEEeqnarray*}

The second algorithm\footnote{generate-valid-stateful-policy-IFSACS-2} takes the intersection of $\mfun{filterIFS}$ and $\mfun{filterACS}$. 
\begin{IEEEeqnarray*}{l}
\mfun{generate2} \ G \ M \ e = \left(V,\, E,\, \left(\mfun{filterACS} \ G \ M \ e \right) \cap  \left(\mfun{filterIFS} \ G \ M \ e\right) \right)
\end{IEEEeqnarray*}

Both algorithms are sound.\footnote{generate-valid-stateful-policy-IFSACS-stateful-policy-compliance,\newline generate-valid-stateful-policy-IFSACS-2-stateful-policy-compliance} 
It remains unclear whether both are equal in the general case. 
Furthermore, it is difficult to prove (or disprove) their completeness, because both algorithms work on almost arbitrary functions $M$. 
However, we have formal proofs for the completeness of $\mfun{filterACS}$ and $\mfun{filterIFS}$ and the structure of $\mfun{generate1}$ and $\mfun{generate2}$ suggest completeness. 
In our experiments, $\mfun{generate1}$ and $\mfun{generate2}$ always calculated the same maximal solution, at least for $\Phi$-structured invariants. 

\begin{theorem}[$\mfun{generate}\{1,2\}$ Soundness]
\label{thm:esss:generatesound}
The algorithms $\mfun{generate1}$ and $\mfun{generate2}$ calculate a stateful policy that fulfills both IFS and ACS requirements. 
\end{theorem}

\begin{example}
\textbf{Example. }%
Recall our running example. 
We illustrate how Figure~\ref{fig:intro:statefulpolicygraph} can be calculated from Figure~\ref{fig:intro:policygraph} and the security invariants. 
All ACS invariants impose only local---in contrast to transitive---access restrictions. 
Therefore, the ACS invariants lack side effects and \texttt{filterACS} selects all flows (excluding already bidirectional ones). 
The invariant that the file server stores confidential data also introduces no restrictions:  
Both $\mvar{filesSrv} \rightarrow \mvar{employees}$ and $\mvar{employees} \rightarrow \mvar{filesSrv}$ are allowed and since the employees are trusted, they can further distribute the data. 
Therefore, $\mfun{filterIFS}$ applied on only this invariant correctly selects all flows. 
Up to this point, the network's functionality is maximized. 
However, since the printer is classified as information sink, it must not leak any data. 
Therefore, $\mfun{filterIFS}$ applied to this invariant selects all but the flows to the printer. 
Ultimately, both $\mfun{generate}$ algorithms compute\footnote{Impl\_\allowbreak{}List\_\allowbreak{}Playground\_\allowbreak{}ChairNetwork\_\allowbreak{}statefulpolicy\_\allowbreak{}example.thy} the same maximal stateful policy, illustrated in Figure~\ref{fig:intro:statefulpolicygraph}. 
The soundness and completeness of the running example is hence formally proven. 
The case study in Section~\ref{sec:esss:casestudy} will focus on performance and feasibility in a large real-world example. 
\end{example}

\section{Computational Complexity}
\label{sec:esss:computationalcomplexity}
The computational complexity of all presented formulae depends on the computational complexity of the security invariants $m \in M$. 
As we allow almost any function $m$ as security invariant, the computational complexity can be arbitrarily large. 
However, most of the security invariants we use in our daily business check a property over all flows in the network ($\Phi$-structured). 
Thus, the computational complexity of $m$ is linear in the number of edges, \ie $\BigO{}(\vert E \vert)$. 
The trivial computational complexity of $\setoffending$ is in \mbox{$\BigO{}(2^{\vert E \vert} \cdot \vert E \vert^2)$}, since it iterates over all subsets of $E$. 
However, given the $\Phi$-structure of the security invariants we primarily use, we showed\footnote{BLP-offending-set, CommunicationPartners-offending-set, ...} that the offending flows for our security invariants are uniquely defined. 
They can be computed in \mbox{$\BigO{}(\vert E \vert)$}. 
The result is a singleton set whose inner set size is also in $\BigO{}(\vert E \vert)$. 
We present the computational complexity of our formulae and algorithms in this section for security invariants and offending flows with the mentioned complexity.\footnote{The computational complexity results are not formalized in Isabelle/HOL, because in its present state, there is no support for reasoning about asymptotic runtime behavior.} 
Our solution is not limited to these security invariants, but the computational complexity increases for more expensive security invariants.

We assume that set inclusion can be computed with the hedge union algorithm~\cite{hedgeuniontr} in $\BigO{}(k_i + k_j)$ for sets of size $k_{\{i,j\}}$. 
Since $E_\tau$ and $E_\sigma$ are bounded by $E$, set inclusion is in $\BigO{}(\vert E \vert)$.

Verifying information flow compliance, \ie (\ref{eq-ifs:fulfilled}), can be computed in $\BigO{}(\vert E \vert \cdot \vert M \vert)$. 
Hence, for a constant number of security invariants, the computational complexity is linear in the number of policy rules. 

To verify access control compliance, we first note that (\ref{eq-acs:subsets}) is in $\BigO{}(2^{\vert E \vert} \cdot \vert E \vert \cdot \vert M \vert)$ which is infeasible for a large policy. 
However, we provide (\ref{eq-acs:theallimplyACSformula}), which implies (\ref{eq-acs:subsets}), and can be computed in $\BigO{}(\vert E \vert \cdot \vert M \vert)$. 
Hence, for a constant number of security invariants, the computational complexity is linear in the number of policy rules. 

The $\mfun{filter}$ and $\mfun{generate}\{1,2\}$ algorithms only add $\BigO{}(\vert E \vert)$ to the complexity. 
Hence, for a constant number of security invariants, computing a stateful policy implementation from a directed policy is quadratic in the number of policy rules, which is feasible even for large policies with thousands of rules.

\section{Case Study Cabin Network Revisited}
\label{sec:esss:casestudycabinnnetwork}
\begin{figure}[htbp]
  \centering
       \centering
  		\includegraphics[width=0.65\linewidth]{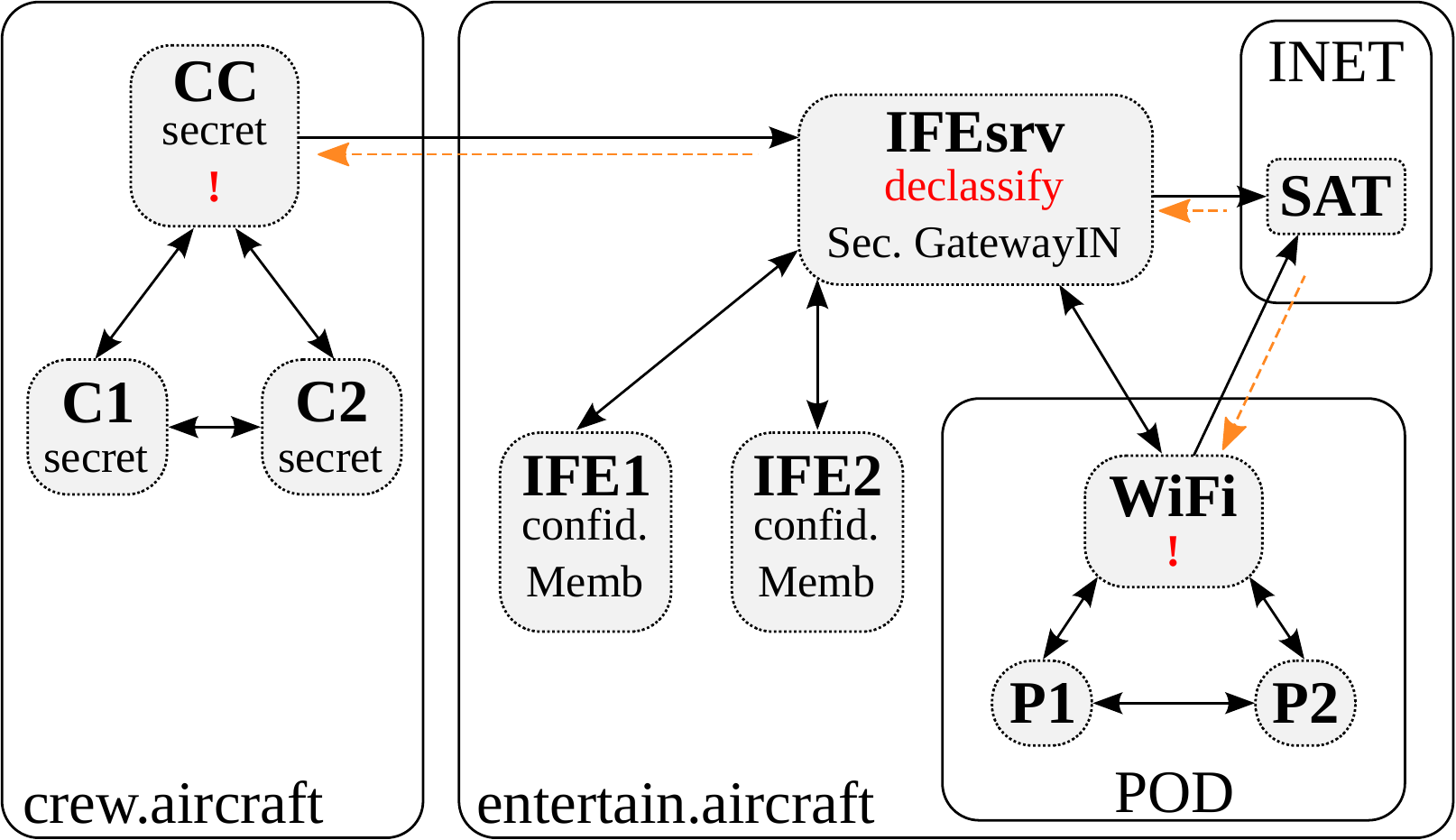}
  		\caption{Cabin network policy, host attributes, and stateful flows.}
  		\label{fig:cabinnetworkstateful}
\end{figure}

%
%
The security invariants and policy for a cabin data network for the general civil aviation have been presented and analyzed in Section~\ref{sec:forte14:case-study}. 
The policy and the host attributes have been visualized in Figure~\ref{fig:cabinnetwork}. 
In this section, we compute the stateful flows for the presented scenario. 
The stateful flows are visualized as dotted arrows in Figure~\ref{fig:cabinnetworkstateful}.

In the original policy, most of the allowed flows are already bi-directional. 
The only unidirectional flows are from the cabin core server to the in-flight entertainment system server and the IFE server and the WiFi accessing the SAT uplink to the Internet. 
The figure shows that those flows may also be stateful. 
The reason is as follows: 

The security invariants only specify one IFS invariant: 
The Bell-LaPadula (with trust) invariant. 
The IFEsrv has a lower security level than the cabin core server, so it can send answers to it without an IFS violation. 
Note that the IFEsrv is trusted, which was required in the original (non-stateful) policy because it needed to receive data from the CC and further distribute it. 
Answers from the IFEsrv to the CC cause a violation of the ACS Domain Hierarchy invariant, but there is no further negative side effect. 
The Policy Enforcement Point invariant is not violated by these answers. 
Consequently, the IFEsrv may send answers to the CC. 

For the same reasons, the SAT uplink may send answers to the IFEsrv and the WiFi. 
All devices have the same security level (though IFEsrv is additionally trusted, it has the lowest security level). 
The Policy Enforcement Point invariant is not affected by such answers. 
Only locally-contained violations of the ACS Domain Hierarchy occur. 
This corresponds exactly to the formalized requirement of accepting expected answers if the connection to the Internet was initialized by the device itself. 

The stateful policy shows that the case study can now be implemented in a fully functional network.

\section{Case Study TUM i8 Firewall}
\label{sec:esss:casestudy}
In a study, Wool~\cite{firwallerr2004} analyzed 37 firewall rulesets from telecommunications, financial, energy, media, automotive, and many other kinds of organization, collected in 2000 and 2001. 
The maximum observed ruleset size was 2671, and the average ruleset size was 144. 
Wool's study ``indicates that there are no good high-complexity rulesets''~\cite{firwallerr2004}. 
If in a scenario complicated rulesets are unavoidable, formal verification to assert their correctness is advisable. 

In this section, we analyze the firewall ruleset of TUM's Chair of Network Architectures and Services. 
With approximately 2983 rules as of November 2013, this firewall configuration can be considered representatively large. 
Almost all rules are stateful, hence the firewall generally allows all established connections and only controls who is allowed to initiate a connection. 
We publish our complete data set, 
allowing others to reproduce our results and reuse the raw data for their research. 

\begin{figure}[h!tbp]
  \centering
  		\includegraphics[width=\linewidth]{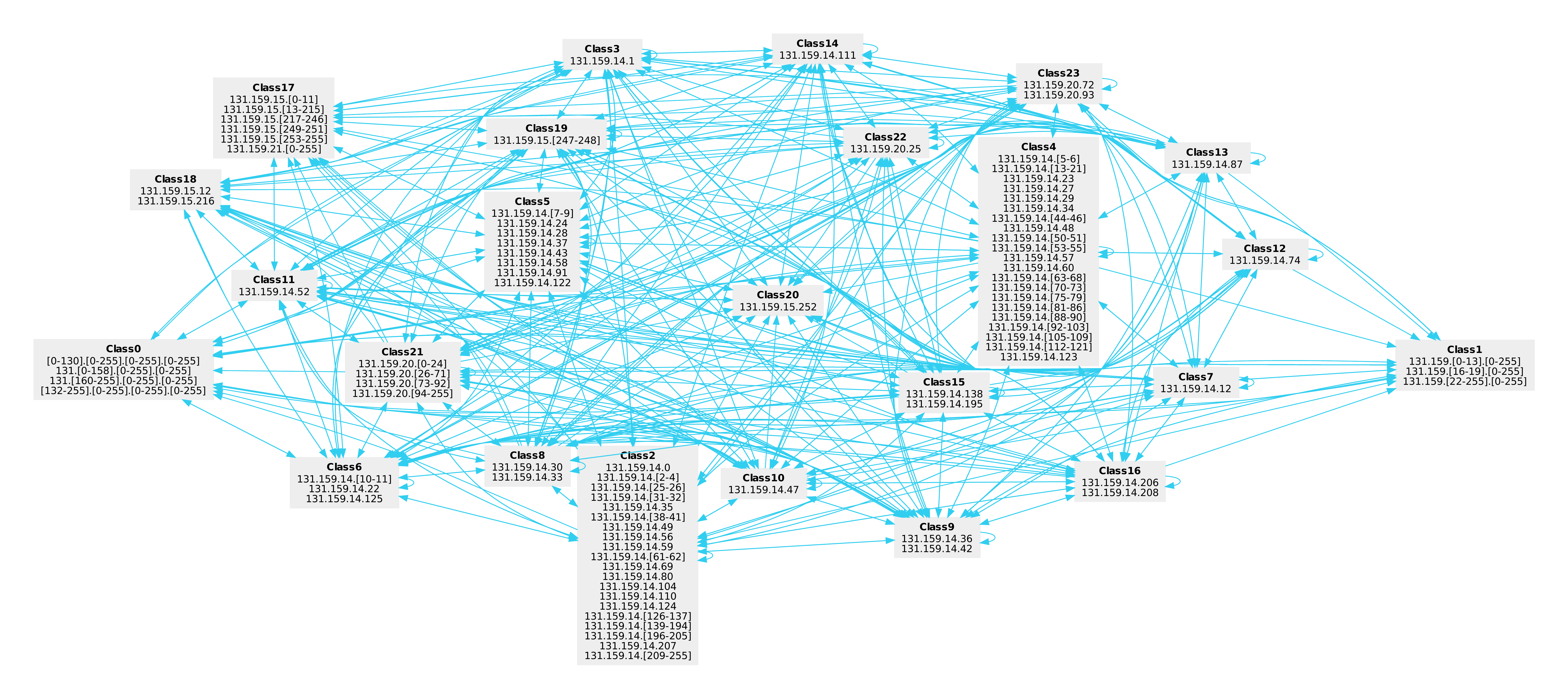}
  \caption{SSH landscape of the TUM Chair of Network Architectures and Services}
  \label{fig:eval:i8sshlandscape}
\end{figure}

As there is no written formal security policy for our network, we reverse-engineered the security policy and invariants with the help of our system administrator. 
The firewall contains rules per IP range that permit the services which are accessible from some IP range. 
Most rules are similar to rule one in Figure~\ref{tab:intro:statefuliptables}. 
We regard the firewall rules about which hosts can initiate a connection as security policy. 
It is not unusual that the implementation is also the documentation~\cite[\S1]{cspfirewall}. 
We verify that the so derived security policy, \ie which hosts can initiate connections, corresponds to the stateful implementation, \ie all connections are stateful. 

In order to prepare the firewall rules as graph, we used \emph{ITval}~\cite{marmorstein2005itval} to first partition the IP space into classes with equivalent access rights~\cite{marmorstein2006firewall} which form the nodes of our policy. 
For each of these classes, we selected representatives and queried ITval for ``which hosts can this representative connect to'' and ``which hosts can connect to this representative''. 
This method is also suggested by Marmorstein~\cite{marmorstein2006firewall}. 
The resulting IP ranges were mapped back to the classes. 
This generates the edges of the security policy graph.\footnote{We will reuse this idea and elaborate on it in Chapter~\ref{chap:networking16}. }  
We asserted that these two queries result in the same graph. 
For brevity, we restrict our attention to the SSH landscape, \ie TCP port 22. 
The full data set is publicly available. 
The SSH landscape results in  a security policy with 24 nodes (sets of IP ranges with equal access rights) and 496 edges (permissions to establish SSH connections). 
The resulting graph is shown in Figure~\ref{fig:eval:i8sshlandscape}. 

A detailed discussion with our system administrator indicated that the graphical representation of the computed graph contains helpful information. 
It reveals that the computed policy does not exactly correspond to the firewall's configuration.
At first, we could not clearly identify the cause for this discrepancy. 
In the following years, when writing our own, fully-verified tool with similar goals, we figured out that ITval contains several bugs which lead to erroneous results; the details will be presented in Part~\ref{part:existing-configs} of this thesis. 
However, this graph obtained by ITval provides a sufficient approximation of our security policy. 
Here, we will only use it for a performance evaluation, therefore, its correctness is not crucial but only its size. 
In fact, the actual, true graph is significantly smaller than the one we use in this chapter. 
For future work, we planned to generate the graph using the approach by Tongaonkar, Niranjan, and Sekar~\cite{tongaonkar2007inferring}, which, later on, we unfortunately could not reproduce because no code is publicly available. 
In the long term, we see the need for formally verified means of translating network device configurations, such as firewall rulesets, SDN flow tables, routing tables, and vendor specific access control lists to formally accessible objects, such as graphs. 
Part~\ref{part:existing-configs} of this thesis will present a tool for this.

After having constructed the security policy, we implemented our security invariants. 
They state that our IP ranges form a big set of mostly collaborating hosts. 
As a general rule, internal hosts are protected from accesses from the outside world, 
but there are many exceptions. 

No IFS invariants exist and our ACS invariants cause no side effects. 
Note that we are evaluating neither the quality of our security policy nor the quality of our security invariants, but the quality of the stateful implementation in this large real-world scenario. 
As expected, our \texttt{generate$\{$1,2$\}$} algorithms identify all unidirectional flows as upgradable to stateful. 
This shows that the standard practice to declare (almost) all rules as stateful, combined with common simple invariants does not introduce security issues. 
For our invariants, our algorithms always generate a graph $\Tau$ such that $\alpha \ \Tau = (V,\ E \cup \overleftarrow{E})$. 
This means that in this scenario, we have a formal justification that all directed policy rules correspond to their stateful implementation, without any security concern. 
This maximizes the network's functionality without introducing security risks and is thus the optimal solution. 

This statement can be generalized to all networks without IFS invariants and without side effects in the ACS invariants. 
We provide formal proofs for both $\mfun{generate}\{1,2\}$ algorithms.\footnote{generate-valid-stateful-policy-IFSACS-noIFS-noACSsideeffects-imp-fullgraph,\newline generate-valid-stateful-policy-IFSACS-2-noIFS-noACSsideeffects-imp-fullgraph} 
Due to its simplicity, universality, and convenient implications for everyday use, we state this result explicitly. 

\begin{corollary}
\label{esss:corollary1}
If there are no information flow security invariants and all access control invariants of a directed policy lack side effects, a security policy can be smoothly implemented as stateful policy, without any security issues concerning state. 
\end{corollary}

Our algorithms return this result, \ie $\alpha \ (\mfun{generate}\ G\ M\ E) = (V,\ E \cup \overleftarrow{E})$. 
If there are information flow security invariants or access control invariants with side effects, our algorithms also handle these problems.

All results can be computed interactively on today's standard hardware. 
The graph preparation, which needs to be done only once, takes several seconds. 
Our \texttt{generate} algorithms take a few seconds. 
This shows the practical low computational complexity for a large real-world study.

\section{Related Work}
\label{sec:esss:related}
In the research field of firewalls, several successful approaches to ease management~\cite{bartal1999firmato} and uncovering errors~\cite{fireman2006} exist. 
Pozo \etal\cite{cspfirewall} propose that a network security policy should exist in an informal language. 
A translation from the informal language to a formalized policy with an information content comparable to the directed policy in this work must be present. 
The same model for firewall rules and security policy is used. 
The authors model services, \ie ports, explicitly but ignore the direction of packets in their firewall model and are hence vulnerable to several attacks, such as spoofing~\cite{wool2004use}. 
Constraint Satisfaction Problem (CSP) solving techniques are used to test compliance of the security policy and the firewall ruleset. 
Using Logic Programming with Priorities (LPP), Bandara \etal\cite{bandara2009using} build a framework to detect firewall anomalies and generate anomaly-free firewall configurations from a security policy. 
The authors explicitly point out the need for solving the stateful firewall problem.

Brucker~\etal\cite{brucker2008modelfwisabelle,brucker.ea:formal-fw-testing:2014} provide a formalization of simple firewall policies in Isabelle/HOL and rewrite rules to simplify them. 
With this, they introduce HOL-TestGen/FW, a tool to generate test cases for conformance testing of a firewall ruleset, \ie that the firewall under test implements its ruleset correctly. 
The authors augment their work~\cite{brucker2013modelfwisabelle} with user-friendly high-level policies. 
This also allows the verification of a network specification with regard to these high-level policies. 

Guttman~\etal\cite{guttman05rigorous,Guttman:1997:FilteringPostures} focus on distributed network security mechanisms, such as firewalls, filtering routers, and IPsec gateways. 
Security goals centered on the path of a packet through the network can be verified against the distributed network security mechanisms configuration. 

Using formal methods, network vulnerability analysis reasons about complete networks, including the services and client software running in the network. 
Using model checking~\cite{modelchecking2000} or logic programming~\cite{ou2005mulval}, network vulnerabilities can be discovered or the absence of vulnerabilities can be shown. 
One potential drawback of these methods is that the set of vulnerabilities must be known for the analysis, which can be an advantage for postmortem network intrusion analysis, but is also a downside when trying to estimate a network's future vulnerability.

Kazemian \etal\cite{kazemian2012HSA} present a method for the packet forwarding plane to identify problems such as reachability issues, forwarding loops, and traffic leakage. 
Considering the individual packet bits, the header space is represented by a $\langle\mathit{maximum\ packet\ header\ size\ in\ bits}\rangle$-dimensional space. 
An efficient algebra on the header space is provided which enables checking of the named use cases. 


\section{Conclusion}
\label{sec:esss:conclusion}

Stateful firewall rules are commonly used to enforce network security policies. 
Due to these state-based rules, flows opposite to the security policy rules might be allowed. 
On the one hand, we argued that under presence of side effects or information flow invariants, a naive stateful implementation might break security invariants. 
On the other hand, declaring certain firewall rules to be stateless might impair the functionality of the network. 
This problem domain has often been overlooked in previous work. 

Verifying that a stateful firewall ruleset is compliant with the security policy and its invariants is computationally expensive. 
In this work, we discovered a linear-time method and contribute algorithms for verifying and also for computing stateful rulesets. 
We demonstrated that these algorithms are fast enough for reasonably large networks, while provably maintaining soundness and completeness. 


\chapter[\emph{topoS}: Synthesis of Secure Network Configurations]{Demonstrating \emph{topoS}: Theorem-Prover-Based Synthesis of Secure Network Configurations}
\label{chap:mansdnnfv}

This chapter is an extended version of the following paper~\cite{diekmann2015topos}:
\begin{itemize}
	\item \begin{sloppypar}Cornelius Diekmann, Andreas Korsten, and Georg Carle. \emph{Demonstrating topoS: Theorem-Prover-Based Synthesis of Secure Network Configurations}. In 2nd International Workshop on Management of SDN and NFV Systems, manSDN/NFV, Barcelona, Spain, November 2015.\end{sloppypar} 
\end{itemize}

\noindent
The following improvements were added:
\begin{itemize}
	\item The iptables implementation of our case study has been verified with \fffuu{} (cf.\ Part~\ref{part:existing-configs}). 
	\item Support for microservice architectures built on top of docker has been added. 
	\item The related work section was updated and extended.
\end{itemize}

\paragraph*{Statement on author's contributions}
For the original paper, the author of this thesis provided major contributions for the ideas, realization, implementation, and proof of the translation process and the \topos{} tool. 
He researched related work, and conducted the evaluation. 
Andreas Korsten contributed to the OpenFlow implementation and deployment. 
A prototypical, incomplete demonstrator of the serialization to OpenVPN has been previously presented in the author's master's thesis~\cite{cornythesis}.
This demonstrator did not consider state, nor information flow security, nor identify the necessary assumptions. 
For this thesis, we reused the idea of a central OpenVPN router but have re-implemented and re-evaluated the complete setup based on the new translation process. 
All improvements listed above are the work of the author of this thesis. 

\medskip

\paragraph*{Abstract}
We combine the results of the previous chapters to present the big picture of the translation from high-level security goals to low-level configurations of security mechanism. 
All results of this Part are combined and we present our tool \topos{} which automatically synthesizes low-level network configurations from high-level security goals.
The automation and a feedback loop help to prevent human errors.
Except for a last serialization step, \topos{} is formally verified with Isabelle/HOL, which prevents implementation errors.
In a case study, we demonstrate \topos{} by example. 
The complete transition from high-level security goals to firewall, SDN, and docker configurations is presented. 

\medskip


\section{Introduction}
Network-level access control is a fundamental security mechanism in almost every network. 
Unfortunately, configuring network-level access control devices still is a challenging, manual, and thus error-prone task~\cite{fwviz2012,fireman2006,ZhangAlShaer2007flip}. 
%
It is a known and unsolved problem for over a decade that ``corporate firewalls are often enforcing poorly written rule sets''~\cite{firwallerr2004}. 
Also, ``access list conflicts dominate the misconfiguration errors made by administrators''~\cite{netsecconflicts}. 
A recent study confirms that this problem persists as a ``majority of administrators stated misconfiguration as the most common cause of failure''~\cite{sherry2012making}.
In addition, not only is implementing a policy error-prone, but also developing it is challenging, even for experienced administrators~\cite{diekmann2014forte}.


We demonstrate our tool \topos{}: a constructive, top-down greenfield approach for network security management.
\topos{} translates high-level security goals to network security device configurations. 
The automatic translation steps prevent manual translation errors. 
Furthermore, \topos{} visualizes the results of all translation steps to help the administrator uncover specification errors.
In addition, since all intermediate transformation steps are formally verified, the correctness of \topos{} itself is guaranteed~\cite{Network_Security_Policy_Verification-AFP}. 
\topos{} is built on top of the results of this thesis, previously presented to the formal methods community~\cite{diekmann2014forte,diekmann2014EPTCS}, combines these results in a novel way, and transfers the knowledge to the network management community.
The automated tool \topos{} is the main technical contribution of this chapter.
%

We first give a short overview of \topos{} in Section~\ref{sec:sdnnfv:topos}.
Then, in Section~\ref{sec:sdnnfv:casestudy}, we present \topos{} in detail with the help of a case study. 
We discuss limitations and advantages in Section~\ref{sec:sdnnfv:discussion}, present related work in Section~\ref{sec:sdnnfv:related}, and conclude in Section~\ref{sec:sdnnfv:conclusion}.



\section{Overview of \topos{}}
\label{sec:sdnnfv:topos}

The security requirements of networks are usually scenario-specific.
Our tool \topos{} helps to configure a network according to these needs.
It takes as input the high-level security requirements and synthesizes low-level configurations for security device, \eg netfilter/iptables firewall rules or OpenFlow flow table entries.
%
%
%
It operates according to the following four-step process: 
\begin{enumerate}[label=\Alph*.]
	\item\label{topos:step:a} Formalize high-level security goals%
	\begin{enumerate}
		\item Categorize security goals
		\item Add scenario-specific knowledge
		\item $\mathbf{\star}$ Auto-complete information
	\end{enumerate}
	\item\label{topos:step:b} $\mathbf{\star}$ Construct security policy
	\item\label{topos:step:c} $\mathbf{\star}$ Construct stateful policy
	\item\label{topos:step:d} $\mathbf{\star}$ Serialize security device configurations
\end{enumerate}

All steps annotated with an asterisk are supported by \topos{}.
As the $\mathbf{\star}$-steps illustrate, once the security goals are specified, the process is completely automatic.
Between the automated steps, manual refinement is possible but requires re-verification.
This allows human intervention, while avoiding human error.

We will illustrate the process in a case study: 
Section~\ref{sub:securitygoals} presents the formalization of the security goals, illustrated in Figure~\ref{fig:runexsecinvars}. 
Figure~\ref{fig:secpol} corresponds to the derived security policy.
In Figure~\ref{fig:secpolrefined}, the administrator made some manual changes (which were accepted by the system since the changes do not violate the formalized security goals).
Finally, Figure~\ref{fig:statefulpol} corresponds to the stateful policy.
The resulting security device configuration will be illustrated for different devices, \eg Figure~\ref{fig:centralrawfirewall}, Figure~\ref{fig:vpnfirewall} and Figure~\ref{fig:openflowrules}.

The automated intermediate $\mathbf{\star}$-steps are proven correct for all inputs.
The proofs are verified with \mbox{Isabelle/HOL}~\cite{isabelle2016}. 
%
Thus, it is guaranteed that \topos{} performs correct transformations~\cite{Network_Security_Policy_Verification-AFP}.
As a side note, since the transformations are proven correct once and for all for all inputs, neither has a user to prove anything manually to use \topos{}, nor is Isabelle/HOL required to run \topos{}.

We did not verify the final step (\ie serialization of security device configurations) in general since it is merely syntactic rewriting of the result of the previous step (\cf Section~\ref{sec:sdnnfv:statefultocongif}). 
In addition, since our policy only allows positive rules, it is guaranteed to be without any conflicts. 
Nevertheless, we verified the correctness of the iptables implementation of our case study with \fffuu{} (cf.\ Part\ref{part:existing-configs}). 

We will present the steps \emph{A} to \emph{D} in the following section. 
For the sake of brevity, we only present them by example. 
Mathematical background has been presented in detail in the previous chapters. 
In this chapter, we focus on its interoperability and discuss how the underlying assumptions can be fulfilled in a real-world network. 
Further details, the correctness proofs, and the interplay of the individual steps can be found in the accompanying formalization and implementation of \topos{}. 

\section{\topos{} by Example}
\label{sec:sdnnfv:casestudy}
In this section, we demonstrate \topos{} with a small case study.
The scenario was chosen because it is minimal and comprehensible, but also realistic and contains many important aspects.
It runs live and is publicly available.\footnote{\url{http://otoro.net.in.tum.de/goals2config/} and \url{http://amygdala.ip4.net.in.tum.de/fcgi/}. The configurations are also archived at \url{https://github.com/diekmann/topoS/blob/master/thy/Network_Security_Policy_Verification/Examples/Distributed_WebApp.SDN_deployed.txt}}

\begin{figure}[h!tbp]
	\centering
  		\includegraphics[width=0.6\linewidth]{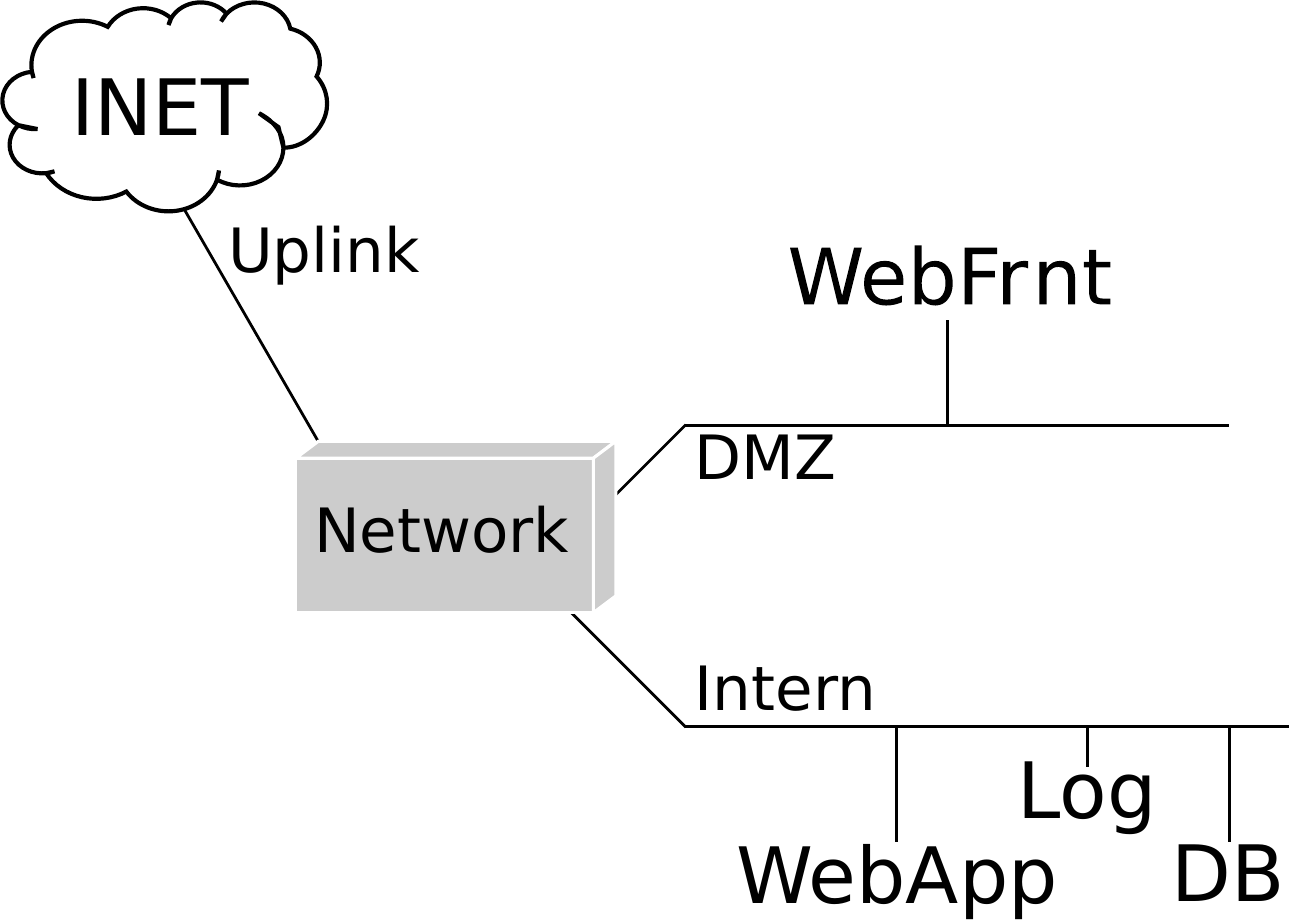}
  		\vskip-5pt
		\caption{Network Schematic} 
		\label{fig:netschematic}
\end{figure}

The case study is schematically illustrated in Figure~\ref{fig:netschematic}. 
The setup hosts a news aggregation web application, accessible from the Internet (\emph{INET}). 
It consists of a web application backend server (\mbox{$\mvar{WebApp}$}) and a frontend server (\mbox{$\mvar{WebFrnt}$}). 
The $\mvar{Web\-App}$ is connected to a database ($\mvar{DB}$) and actively retrieves data from the Internet. 
All servers send their logging data to a central, protected log server ($\mvar{Log}$).

We implemented the scenario to utilize several different protocols.
A custom backend, the $\mvar{WebApp}$ was written in \texttt{python}. 
The $\mvar{WebFrnt}$ runs \texttt{lighttpd}. 
It serves static web pages directly and retrieves dynamic websites from the $\mvar{WebApp}$ via \texttt{FastCGI}. 
All components send their \texttt{syslog} messages via UDP (RFC~5426~\cite{rfc5426}) to $\mvar{Log}$.

\subsection{Formalizing Security Goals}
\label{sub:securitygoals}
The security goals are expressed as security invariants over the network's connectivity structure.
%
%
An invariant consists of a generic part (the semantics, formalized as security invariant template) and scenario-specific information (formalized as host attributes). 
The generic part defines the type and general meaning. 
Our generic invariants currently defined are summarized in \mbox{Table~\ref{tab:invariantemplates}}. 

To construct a scenario-specific invariant, a generic invariant is instantiated with scenario-specific knowledge. 
This is done by specifying host attributes (\cf Chapter~\ref{chap:forte14}). 
These invariants and the list of entities ($\mvar{INET}$, \mbox{$\mvar{WebApp}$}, \mbox{$\mvar{WebFrnt}$}, $\mvar{DB}$, $\mvar{Log}$) is the only input needed. 
For this scenario, the following four invariants are expressed, formalized in Figure~\ref{fig:runexsecinvars}. 

\begin{enumerate}
\item First, as illustrated in Figure~\ref{fig:netschematic}, $\mvar{DB}$, $\mvar{Log}$, and $\mvar{WebApp}$ are considered internal hosts. 
We use the \emph{SubnetsInGW} invariant template (Section~\ref{sinvar:subnetsingw}) for this. 
Internal hosts are labeled with the $\mconstr{Member}$ attribute. 
The $\mvar{WebFrnt}$ must be accessible from outside, it is a classical DMZ member. 
Therefore, we label it as $\mconstr{InboundGateway}$. 

\item Next, it is expressed that the logging data must not leave the log server. 
Therefore, using the \emph{Sink} invariant (Section~\ref{sinvar:sink}), $\mvar{Log}$ is classified as information sink.

\item Using the \emph{Bell-LaPadula} invariant (Section~\ref{sinvar:blptrust}), it is specified that $\mvar{DB}$ contains confidential information. 
Since it sends its log data to the log server, this log server is also assigned the security level $\mconstr{confidential}$. 
Finally, the $\mvar{Web\-App}$ is allowed to retrieve data from the $\mvar{DB}$ and to publish it to the $\mvar{WebFrnt}$. 
Therefore, the $\mvar{Web\-App}$ is trusted and allowed to declassify the data.

\item Finally, an access control list specifies that only $\mvar{Web\-App}$ may access the $\mvar{DB}$. 
We use the \emph{Communication Partners} template from Section~\ref{sinvar:commpartners}. 
\end{enumerate}




\begin{figure}[bht]
\centering
\fbox{
	\begin{minipage}{0.8\linewidth}
	\input{content/sdnnfv15_secrequsfig}
	\end{minipage}
}
\vskip-5pt
\caption{Security Invariants (Case Study)}
\label{fig:runexsecinvars}
\end{figure}

In this example, several hosts do not have attributes assigned for all invariants. 
It is sufficient to supply an incomplete host attribute specification, since they are automatically and securely completed 
by \topos{}.
Chapter~\ref{chap:forte14} discusses the details.\footnote{The security of the auto-completion is guaranteed w.r.t.\ the provided information, \ie the auto-completion can never lead to an unnoticed security problem, given enough information is provided. 
For example, information-leakage is always uncovered, given all confidential data sources are specified.
However, if an administrator forgets to label a confidential data source, information leakage can occur.
It is trivially possible to design explicit whitelisting invariants which auto-complete to some `\emph{deny}' property. 
On the downside, this requires lots of manual configuration effort, which is avoided by the invariants utilized in this chapter.
Roughly speaking the auto-completion fulfills: ``\textit{the more information provided, the more secure the whole system}''.}
Once the invariants are specified, their management scales well in the face of changes:
When a new host is added to the network, issues are handled by the auto-completion: 
either, the new host causes a violation, which is consequently uncovered, or it can be added without any further changes.
%
Invariants are composable and modular by design, helping structured representation and archiving of knowledge.
In the worst case, inconsistent security invariants may be specified accidentally. 
This only results in an overly strict security policy being computed, which can be identified in the following step.


It has been shown that a special class of invariants, called $\Phi$-struc\-tured, exhibits several nice mathematical properties (\cf Table~\ref{tab:invariantemplates}). 
A $\Phi$-struc\-tured invariant asserts a predicate for every policy rule.  
This predicate must only depend on the sender, receiver, and their host attributes. 
In particular, these invariants and their derived algorithms are very efficiently computable. 
It is also due to the $\Phi$-struc\-tured invariants that a maximum-permissive security policy is uniquely defined.

\subsection{Constructing the Security Policy}
A network's end-to-end connectivity structure, \ie a global access control matrix, corresponds to the \emph{security policy}. 
Here, we utilize the textbook definition that a policy consists of the \emph{rules} which ensure that the network is in a secure state. 
In contrast, the security goals are expressed as \emph{invariants} over the policy and reside on a higher level of abstraction.

\begin{figure}[htb]
	\centering
	\begin{minipage}{0.29\linewidth}\centering
  		\includegraphics[width=0.99\textwidth]{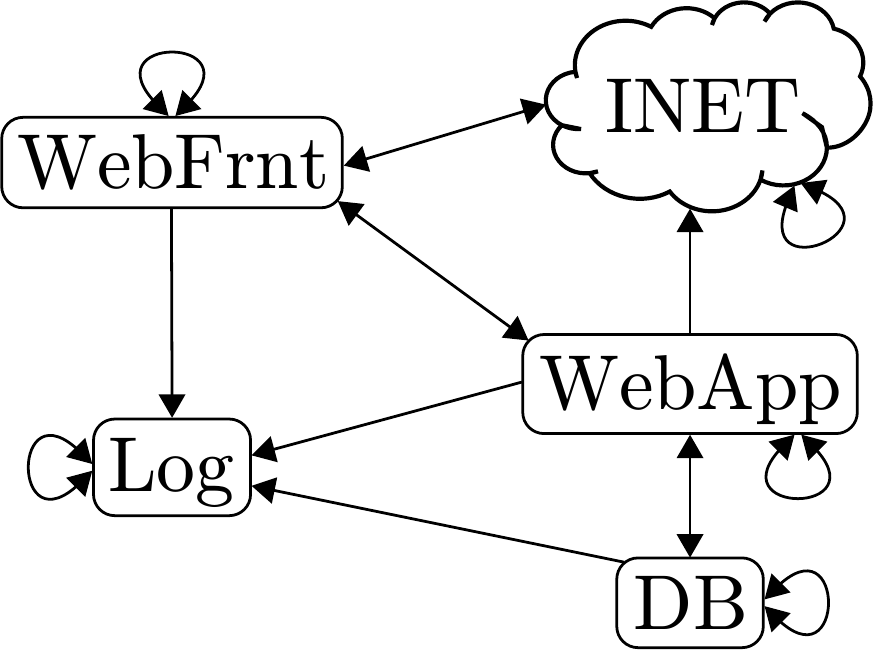}
  		\vskip-5pt
		\caption[Security Policy computed]{Security Policy\mbox{ (computed)}} 
		\label{fig:secpol}
	\end{minipage}
	\hspace*{\fill}
	\begin{minipage}{0.29\textwidth}\centering
  		\includegraphics[width=0.99\textwidth]{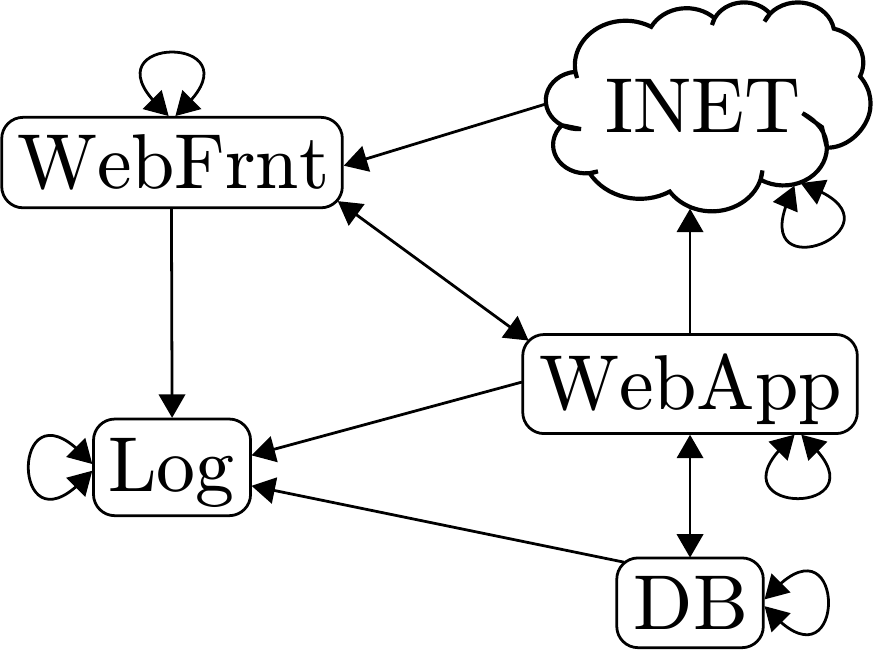}
  		\vskip-5pt
		\caption[Security Policy refined]{Security Policy\mbox{ (manually refined)}}
		\label{fig:secpolrefined}
	\end{minipage}
	\hspace*{\fill}
	\begin{minipage}{0.29\linewidth}\centering
  		\includegraphics[width=0.99\textwidth]{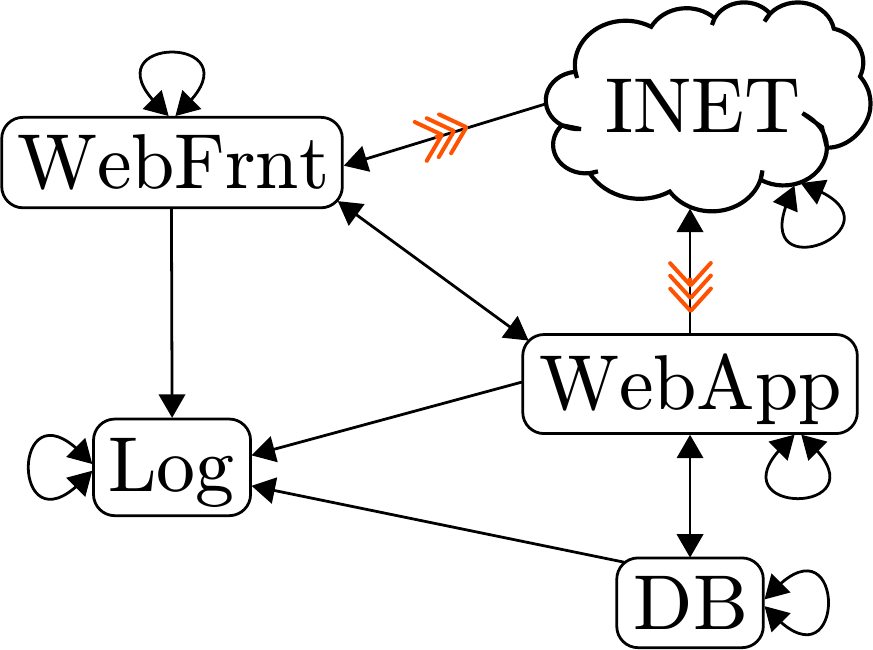}
  		\vskip-5pt
		\caption[Stateful Policy]{Stateful Policy\mbox{\quad (computed)}}
		\label{fig:statefulpol}
	\end{minipage}
\end{figure}

Graphically, a policy can be illustrated as a directed graph. 
The policy of the case study, illustrated in Figure~\ref{fig:secpol}, was automatically computed from the security invariants.


The algorithm to transform a set of security invariants into a policy starts with the allow-all policy and iteratively removes undesired rules.
This is always possible if (and only if, Theorem~\ref{thm:no-edges-validity}) the invariants hold for the deny-all policy; a static requirement which is only to be proven once for a generic invariant. 
The algorithm is sound. 
It is also complete for the invariants utilized in this example (and for $\Phi$-struc\-tured invariants in general, Theorem~\ref{thm:generate_valid_topology_phi_complete}). 

In our example, the administrator decides to manually refine the policy: there is no need for the web frontend to connect to the Internet. 
Therefore, this flow is prohibited. 
After this manual refinement, the security invariants are re-verified. 
This refined policy is shown in Figure~\ref{fig:secpolrefined}.

\subsection{Constructing the Stateful Policy}
\label{sec:sdnnfv:statefulpolicy}
The derived policy may appear adequate from a theoretical point of view but has one major problem when it comes to implementation:
The $\mvar{WebApp}$ can connect to the Internet, but the policy does not specify whether the Internet may answer this request (same for the $\mvar{WebFrnt}$ after manual refinement). 
Obviously, for this scenario, answers should be permitted; otherwise, no one would be able to use the service. 
In contrast, the Log server uses the \texttt{syslog} protocol over UDP (RFC~5426~\cite{rfc5426}). 
This protocol uses a unidirectional UDP channel and it is explicitly specified for security reasons that this is the only way the communication with the log server is permitted.

Therefore, it must be distinguished between stateful and purely unidirectional rules. 
We extend the security policy to additionally specify whether a flow might be stateful (\ie answers to requests are allowed).
Note that a flow with the stateful attribute might allow packets in the opposite direction of the policy rule and thus potentially violate security invariants.
Defining the following two consistency criteria, the stateful attributes can be computed automatically (\cf Chapter~\ref{chap:esss14}):
\begin{enumerate}
	\item No information flow violation must occur.
	\item No access control side effects must be introduced.
\end{enumerate}
To compute the stateful policy, not only a single rule but a set $S$ is to be upgraded to stateful rules. 
However, the interaction of the rules and answer paths of $S$ must not introduce negative implications. 
Therefore, in particular to verify lack of side effects, all security policies derived from upgrading all \emph{subsets} of $S$ must be verified. 
A naive approach would require exponential complexity. 
We proved that this can be done more efficiently, particularly in linear time for $\Phi$-structured invariants (Chapter~\ref{chap:esss14}). 
This insight provides an algorithm for computing the stateful policy from the security policy and the invariants. 
It is proven sound (Theorem~\ref{thm:esss:generatesound}) and complete w.r.t.\ the two criteria individually (Lemma~\ref{lem:esss:ifscomplete}, Lemma~\ref{lem:esss:acscomplete}). 
Multiple solutions for a stateful policy may exist; a user may set preferences.

For the case study, this results in a policy where the Internet can set up connections to the web frontend, likewise, the web backend can set up connections to the Internet. 
However, the logging channels are purely unidirectional UDP (stateful connections would introduce an information flow violation).
We will call this the stateful policy.
It is illustrated in Figure~\ref{fig:statefulpol}.

\subsection{Serializing Security Device Configurations} 
\label{sec:sdnnfv:statefultocongif}
By now, the network of Figure~\ref{fig:netschematic} was considered a black box.
In this section, the stateful policy is serialized to configurations for real network (security) devices.
Though the serialization step is merely syntactic rewriting of the stateful policy, care must be taken to correctly transfer the semantics. 
Therefore, all assumptions of the stateful policy must be taken into account. 
We first define two types of entities and discuss the assumptions afterwards.

We differentiate between two types of entities.
\begin{description}
	\item[Policy Entities]
		Entities relevant for the use case and required for specification of functional requirements.
		\mbox{E.\hairspace{}g}.\ all entities in Figure~\ref{fig:secpol}.
	\item[Network Infrastructure Entities]
		Not required for the description of the high-level function of a system.
		E.\hairspace{}g.\ Switches, routers, middleboxes.
		For example, everything in the network box in Figure~\ref{fig:netschematic}.
\end{description}
%
%
%
%
One policy entity may correspond to several entities in the network. 
We will call them representatives.
For example, when deploying the case study with load-balancing and redundancy, \emph{Web\-App} might be a set of backend servers; the symbolic name \emph{Web\-App} is translated to the IP range of these servers.
Hence, a one-to-many mapping lifts policy identifiers to roles of network representatives. 

\medskip

To serialize security mechanism configurations, 
\topos{} must fulfill the following three assumptions. 
\begin{description}
	\item[Structure] The enforced network connectivity structure must exactly coincide with the policy. This requires that the links are confidential and integrity protected. 
	\item[Authenticity] The policy's entities must match their network representation (\eg IP/MAC addresses). In particular, no impersonation or spoofing must be possible.
	\item[State] The stateful connection handling must match the stateful policy's semantics.
\end{description}

\subsubsection{Reflexive policy rules}
\label{sec:reflexiverulesdeplay}
When it comes to implementation, we discuss one peculiarity: Reflexive policy rules.
A reflexive rule, \eg $A \rightarrow A$, means that the policy entity $A$ can communicate with itself. 
As can be seen in the case study, all entities can communicate with themselves. 
Translating reflexive policy rules requires special care. 
For a one-to-one mapping of policy entities and network representatives, reflexive rules correspond to in-entity communication, which can obviously be ignored. 
For example, if $A$ maps to $10.0.0.4$, in-host communication is out of the scope of network access control. 
But a policy entity may also correspond to several network representatives. 
Such a scenario might occur if deployed with load balancing, \eg $\mvar{Web\-App}$ may correspond a set of backend servers. 
In this case, enforcing network access control may be necessary. 
For example, if $A$ maps to $10.0.0.0/24$ and the policy does not permit $A \rightarrow A$, the hosts in the $10.0.0.0/24$ subnet must not be able to access each other.

For the sake of brevity, we only present a one-to-one mapping between policy entities and their network representatives in this chapter.
We present four different possibilities to implement the policy.

\subsubsection{Central, Directly-Attached Firewall}
\label{deploy:iptables}
We assume that all entities are directly connected to a central firewall via an individual cable. 
An attacker does not have access to the cables.  
The firewall has an individual interface for each entity and one interface as the uplink to the Internet. 
\begin{figure*}[ht!bp]
\begin{minipage}{\linewidth}
\footnotesize
\begin{Verbatim}[commandchars=\\\{\},codes={\catcode`$=3\catcode`^=7}]
-A FORWARD -i $\mathit{\$WebFrnt\_iface}$ -s $\mathit{\$WebFrnt\_ipv4}$ -o $\mathit{\$WebFrnt\_iface}$ -d $\mathit{\$WebFrnt\_ipv4}$ -j ACCEPT
-A FORWARD -i $\mathit{\$WebFrnt\_iface}$ -s $\mathit{\$WebFrnt\_ipv4}$ -o $\mathit{\$Log\_iface}$ -d $\mathit{\$Log\_ipv4}$ -j ACCEPT
-A FORWARD -i $\mathit{\$WebFrnt\_iface}$ -s $\mathit{\$WebFrnt\_ipv4}$ -o $\mathit{\$WebApp\_iface}$ -d $\mathit{\$WebApp\_ipv4}$ -j ACCEPT
-A FORWARD -i $\mathit{\$DB\_iface}$ -s $\mathit{\$DB\_ipv4}$ -o $\mathit{\$DB\_iface}$ -d $\mathit{\$DB\_ipv4}$ -j ACCEPT
-A FORWARD -i $\mathit{\$DB\_iface}$ -s $\mathit{\$DB\_ipv4}$ -o $\mathit{\$Log\_iface}$ -d $\mathit{\$Log\_ipv4}$ -j ACCEPT
-A FORWARD -i $\mathit{\$DB\_iface}$ -s $\mathit{\$DB\_ipv4}$ -o $\mathit{\$WebApp\_iface}$ -d $\mathit{\$WebApp\_ipv4}$ -j ACCEPT
-A FORWARD -i $\mathit{\$Log\_iface}$ -s $\mathit{\$Log\_ipv4}$ -o $\mathit{\$Log\_iface}$ -d $\mathit{\$Log\_ipv4}$ -j ACCEPT
-A FORWARD -i $\mathit{\$WebApp\_iface}$ -s $\mathit{\$WebApp\_ipv4}$ -o $\mathit{\$WebFrnt\_iface}$ -d $\mathit{\$WebFrnt\_ipv4}$ -j ACCEPT
-A FORWARD -i $\mathit{\$WebApp\_iface}$ -s $\mathit{\$WebApp\_ipv4}$ -o $\mathit{\$DB\_iface}$ -d $\mathit{\$DB\_ipv4}$ -j ACCEPT
-A FORWARD -i $\mathit{\$WebApp\_iface}$ -s $\mathit{\$WebApp\_ipv4}$ -o $\mathit{\$Log\_iface}$ -d $\mathit{\$Log\_ipv4}$ -j ACCEPT
-A FORWARD -i $\mathit{\$WebApp\_iface}$ -s $\mathit{\$WebApp\_ipv4}$ -o $\mathit{\$WebApp\_iface}$ -d $\mathit{\$WebApp\_ipv4}$ -j ACCEPT
-A FORWARD -i $\mathit{\$WebApp\_iface}$ -s $\mathit{\$WebApp\_ipv4}$ -o $\mathit{\$INET\_iface}$ -d $\mathit{\$INET\_ipv4}$ -j ACCEPT
-A FORWARD -i $\mathit{\$INET\_iface}$ -s $\mathit{\$INET\_ipv4}$ -o $\mathit{\$WebFrnt\_iface}$ -d $\mathit{\$WebFrnt\_ipv4}$ -j ACCEPT
-A FORWARD -i $\mathit{\$INET\_iface}$ -s $\mathit{\$INET\_ipv4}$ -o $\mathit{\$INET\_iface}$ -d $\mathit{\$INET\_ipv4}$ -j ACCEPT
-I FORWARD -m state --state ESTABLISHED -i $\mathit{\$INET\_iface}$ -s $\mathit{\$INET\_ipv4}$ $\hfill\hookleftarrow$
             -o $\mathit{\$WebApp\_iface}$ -d $\mathit{\$WebApp\_ipv4}$ -j ACCEPT
-I FORWARD -m state --state ESTABLISHED -i $\mathit{\$WebFrnt\_iface}$ -s $\mathit{\$WebFrnt\_ipv4}$ $\hfill\hookleftarrow$
             -o $\mathit{\$INET\_iface}$ -d $\mathit{\$INET\_ipv4}$ -j ACCEPT
-P FORWARD DROP
\end{Verbatim}
\end{minipage}%
\caption{Central Firewall Rules (can be loaded with \texttt{iptables-restore})}
\label{fig:centralrawfirewall}
\end{figure*}

\begin{table}[!hbt]
\centering
\caption{Variable Assignment}
\label{tab:mansdncentralfw:ipmapping}
\small
\begin{tabular}[0.99\linewidth]{ l @{\hspace*{0.8em}} p{20em} }%
	\toprule
	Variable            & Value  \\
	\midrule
	$\$\mathit{WebFrnt\_iface}$ & \texttt{webfrnt}\\
	$\$\mathit{WebFrnt\_ipv4}$  & 10.0.0.1\\
	$\$\mathit{Log\_iface}$     & \texttt{log}\\
	$\$\mathit{Log\_ipv4}$      & 10.0.0.2\\
	$\$\mathit{DB\_iface}$      & \texttt{db}\\
	$\$\mathit{DB\_ipv4}$       & 10.0.0.3\\
	$\$\mathit{WebApp\_iface}$  & \texttt{app}\\
	$\$\mathit{WebApp\_ipv4}$   & 10.0.0.4\\
	$\$\mathit{INET\_iface}$    & \texttt{inet}\\
	$\$\mathit{INET\_ipv4}$     & 0.0.0.0/5, 8.0.0.0/7, 11.0.0.0/8, 12.0.0.0/6,\newline 16.0.0.0/4, 32.0.0.0/3, 64.0.0.0/2, 128.0.0.0/1\\
 \bottomrule%
\end{tabular}
\end{table}

\topos{} generates the firewall template shown in Figure~\ref{fig:centralrawfirewall}. 
To load the rules, we need to set the variables. 
In our shell, we \texttt{export} the interface names and IP addresses of our servers as given in Table~\ref{tab:mansdncentralfw:ipmapping}. 
The value for $\$\mathit{INET\_ipv4}$ deserves some explanation. 
It corresponds to the set of all IPv4 addresses excluding the 10.0.0.0/8 range. 
It was computed with our fully verified IP address library~\cite{IP_Addresses-AFP}. 

Technically, without additional modules, iptables only allows to match on one source or destination IP range in CIDR notation. 
However, the \texttt{iptables} userland command allows to specify several CIDR ranges as syntactic sugar, as we did for $\$\mathit{INET\_ipv4}$. 
Therefore, our configuration can be loaded without complaints. 
Internally, \texttt{iptables} translates these rules into several rules which match on one CIDR range at most. 
For this example, over 100 rules are internally created. 

\begin{figure*}[ht!bp]
\begin{minipage}{\linewidth}
	\small
\begin{Verbatim}[commandchars=\\\{\},codes={\catcode`$=3\catcode`^=7}]
webfrnt = [10.0.0.1]
log = [10.0.0.2]
db = [10.0.0.3]
inet = all_but_those_ips [10.0.0.0/8]
app = [10.0.0.4]
\end{Verbatim}
\end{minipage}%
\caption{Interface configuration for \fffuu{}}
\label{fig:centralrawfirewall:ipassmt}
\end{figure*}

To verify that the generated firewall rules provide spoofing protection and to validate our bold setting of $\$\mathit{INET\_ipv4}$, we specify the valid IP ranges per interface. 
Our tool \fffuu{}, which we will use now to verify our configurations, will be presented in detail in Part~\ref{chap:existing:overview} of this thesis. 
Figure~\ref{fig:centralrawfirewall:ipassmt} shows this specification with the syntax of our \fffuu{} tool. 
With this, our tool can immediately verify spoofing protection (\cf Chapter~\ref{chap:nospoof}) of the loaded iptables rules. 
This fulfills the required authenticity~(\cmark) assumption. 


\begin{figure*}[ht!bp]
\centering
\begin{tikzpicture}
\node[align=center,text width=6cm,cloud, draw,cloud puffs=10,cloud puff arc=120, aspect=2, inner sep=-3em,outer sep=0] (a) at (5,0) { $\{0.0.0.0 .. 9.255.255.255\} \cup \{11.0.0.0 .. 255.255.255.255\}$ };
\node (b) at (5,-3) { $\{10.0.0.4\}$ };
\node (c) at (5,-6) { $\{10.0.0.3\}$ };
\node (d) at (0,-5) { $\{10.0.0.2\}$ };
\node (e) at (0,-1) { $\{10.0.0.1\}$ };
\node (f) at (5,-8) { $\{10.0.0.0\} \cup \{10.0.0.5 .. 10.255.255.255\}$ };

\draw[myptr] (a) to[out=330,in=300,looseness=3] (a);
\draw[myptr] (a) to (e);
\draw[myptr] (b) to (a);
\draw[myptr] (b) to[loop right] (b);
\draw[myptr] (b) to (c);
\draw[myptr] (b) to (d);
\draw[myptr] (b) to (e);
\draw[myptr] (c) to (b);
\draw[myptr] (c) to[loop right] (c);
\draw[myptr] (c) to (d);
\draw[myptr] (d) to[loop below] (d);
\draw[myptr] (e) to (b);
\draw[myptr] (e) to (d);
\draw[myptr] (e) to[loop above] (e);
\end{tikzpicture}%
\caption{Reconstructed Security Policy (Structure of Figure~\ref{fig:centralrawfirewall})}
\label{fig:mansdn:centralwf:servicematrix}
\end{figure*}
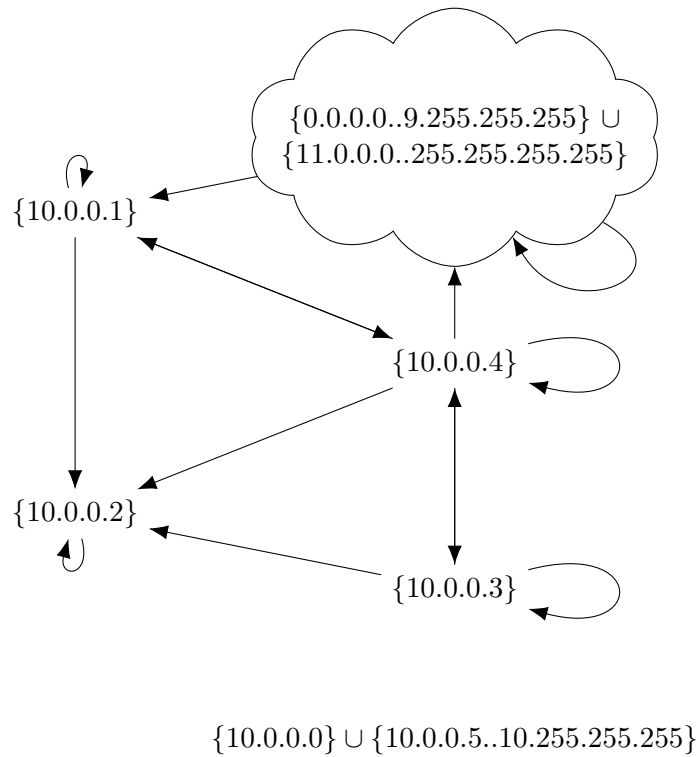

We utilize our tool \fffuu{} to compute a service matrix, \ie given the iptables configuration, we reconstruct the security policy. 
The result is shown in Figure~\ref{fig:mansdn:centralwf:servicematrix}. 
We can see that the computed graph is identical to Figure~\ref{fig:secpolrefined}. 
Hence, we have verified that our iptables configuration satisfies the structure (\cmark) assumption. 
Analogously to Section~\ref{sec:example-factory:stateful}, the stateful semantics (\cmark) can be verified.

\subsubsection{Firewall \& Central VPN Server}
\label{deploy:openvpn}
All entities connect to a central OpenVPN server which enforces the policy. 
Entities are bound to their policy name with \mbox{X.509} certificates. 
Every entity sets up a layer~3 (\texttt{tun}) VPN connection with the server.
The server authenticates entities by their certificate and centrally assigns IP addresses. 
IP spoofing over the tunnel is prevented by the server. 
This provides authenticity~(\cmark).
%
\begin{figure*}[ht!bp]
\begin{minipage}{\linewidth}
\small
\begin{Verbatim}[commandchars=\\\{\},codes={\catcode`$=3\catcode`^=7}]
-A FORWARD -i tun0 -s $\mathit{\$WebFrnt\_ipv4}$ -o tun0 -d $\mathit{\$Log\_ipv4}$ -j ACCEPT
-A FORWARD -i tun0 -s $\mathit{\$WebFrnt\_ipv4}$ -o tun0 -d $\mathit{\$WebApp\_ipv4}$ -j ACCEPT
-A FORWARD -i tun0 -s $\mathit{\$DB\_ipv4}$ -o tun0 -d $\mathit{\$Log\_ipv4}$ -j ACCEPT
-A FORWARD -i tun0 -s $\mathit{\$DB\_ipv4}$ -o tun0 -d $\mathit{\$WebApp\_ipv4}$ -j ACCEPT
-A FORWARD -i tun0 -s $\mathit{\$WebApp\_ipv4}$ -o tun0 -d $\mathit{\$WebFrnt\_ipv4}$ -j ACCEPT
-A FORWARD -i tun0 -s $\mathit{\$WebApp\_ipv4}$ -o tun0 -d $\mathit{\$DB\_ipv4}$ -j ACCEPT
-A FORWARD -i tun0 -s $\mathit{\$WebApp\_ipv4}$ -o tun0 -d $\mathit{\$Log\_ipv4}$ -j ACCEPT
-A FORWARD -i tun0 -s $\mathit{\$WebApp\_ipv4}$ -o eth0 -d $\mathit{\$INET\_ipv4}$ -j ACCEPT
-A FORWARD -i eth0 -s $\mathit{\$INET\_ipv4}$ -o tun0 -d $\mathit{\$WebFrnt\_ipv4}$ -j ACCEPT
-I FORWARD -m state --state ESTABLISHED  $\hfill\hookleftarrow$
             -i eth0 -s $\mathit{\$INET\_ipv4}$ -o tun0 -d $\mathit{\$WebApp\_ipv4}$  -j ACCEPT
-I FORWARD -m state --state ESTABLISHED   $\hfill\hookleftarrow$
             -i tun0 -s $\mathit{\$WebFrnt\_ipv4}$ -o eth0 -d $\mathit{\$INET\_ipv4}$  -j ACCEPT
-P FORWARD DROP 
\end{Verbatim}
\end{minipage}%
\caption{VPN Server Firewall Rules (can be loaded with \texttt{iptables-save})}
\label{fig:vpnfirewall}
\end{figure*}

%
%
Firewalling is applied at the server; the stateful policy is directly translated to iptables rules, shown in Figure~\ref{fig:vpnfirewall}. 
With this, the stateful semantics (\cmark) and structure (\cmark) are enforced.




\subsubsection{SDN}
With complete control over the network, as is the case with data centers, a Software-Defined Network (SDN) may be used to implement the policy.
Usually, a data center is a flat layer~2 network~\cite{bari2013data} and we need to contain layer~2 broadcasting and layer~2 attacks. 
For this, an entity's switch port\footnote{virtual hypervisor switch port or physical Top of Rack (ToR) switch port} must be known. 
%
\begin{figure*}[ht!bp]
\begin{minipage}{\linewidth}
\small
\begin{Verbatim}[commandchars=\\\{\},codes={\catcode`$=3\catcode`^=7}]
\# ARP Request
  in\_port=$\mathit{\$port\_src}$ dl\_src=$\mathit{\$mac\_src}$ dl\_dst=ff:ff:ff:ff:ff:ff  $\hfill\hookleftarrow$
  arp arp\_sha=$\mathit{\$mac\_src}$ arp\_spa=$\mathit{\$ip4\_src}$ arp\_tpa=$\mathit{\$ip4\_dst}$  $\hfill\hookleftarrow$
  priority=40000 action=mod\_dl\_dst:$\mathit{\$mac\_dst}$,output:$\mathit{\$port\_dst}$
\medskip
\# ARP Reply
  dl\_src=$\mathit{\$mac\_dst}$ dl\_dst=$\mathit{\$mac\_src}$  $\hfill\hookleftarrow$
  arp arp\_sha=$\mathit{\$mac\_dst}$ arp\_spa=$\mathit{\$ip4\_dst}$ arp\_tpa=$\mathit{\$ip4\_src}$ $\hfill\hookleftarrow$
  priority=40000 action=output:$\mathit{\$port\_src}$
\medskip
\# IPv4 one-way
  in\_port=$\mathit{\$port\_src}$ dl\_src=$\mathit{\$mac\_src}$ ip nw\_src=$\mathit{\$ip4\_src}$ nw\_dst=$\mathit{\$ip4\_dst}$  $\hfill\hookleftarrow$
  priority=40000 action=mod\_dl\_dst:$\mathit{\$mac\_dst}$,output:$\mathit{\$port\_dst}$
\medskip
\# if $\mathit{src}$ (respecively $\mathit{dst}$) is INET, 
\# replace $\mathit{\$ip4\_src}$ (respectively $\mathit{\$ip4\_dst}$) with * and decrease the priority
\end{Verbatim}
\end{minipage}%
\caption{OpenFlow Flow Table Template}%
\label{fig:openflowrules}
\end{figure*}

We deploy all entities as virtual machines on top of the hypervisor xen~\cite{xen2003}. 
All machines are executed on the same host. 
To interconnect the virtual machines, one Open vSwitch~\cite{pfaff2009openvswitch}, deployed at the hypervisor, provides connectivity.

We install OpenFlow rules which prevent MAC, IP, and ARP spoofing. 
Figure~\ref{fig:openflowrules} illustrates a template for generating a stateless rule from $\mathit{src}$ to $\mathit{dst}$. 
The rules can be loaded with \texttt{ovs-vsctl set-fail-mode $\mathit{\$switch}$ secure \&\& ovs-ofctl add-flows}. 
The first rule allows ARP requests.
Note that we rewrite the layer~2 broadcast addresses directly to the immediate receiver's address.
Rule two allows the ARP responses.
Both rules ensure that only valid ARP queries and responses are sent and received in the network.\footnote{
  For the sake of simplicity, this implementation is designed such that it gets along without an SDN controller. 
  This introduces a small hidden information flow channel (structure \xmark): the ARP responses.
  For example in Figure~\ref{fig:statefulpol}, $\mvar{Log}$ may use a timing channel or the ARP \texttt{OPER} field to exfiltrate information. 
  However, the side-channel is easily removed when an SDN controller answers all ARP requests (structure \cmark); all necessary information is present.
  }
The third rule allows IPv4 traffic.
For stateful rules, the opposite direction of Figure~\ref{fig:openflowrules}, \ie $\mathit{src}$ and $\mathit{dst}$ swapped, is added.
Any unmatched packets are dropped. 
With this set of rules, a mapping of policy identifiers to MAC and IP addresses is enforced. 
Also, correct address resolution is enforced (authenticity~\cmark). %
Without the ARP information leak, the desired connectivity structure (\cmark) is enforced.
The setup does not provide stateful handling (\xmark) by default. 
However, a network firewall or SDN firewall app can provide the desired state (\cmark) handling. 
As of version 2.5.0, Open vSwitch also supports the \texttt{ct} action for connection tracking~\cite{openvswitch250releasenotes}, providing stateful semantics (\cmark).




\subsubsection{Microservices, Containers, and Docker}
\label{deploy:docker}
Microservices are on the rise~\cite{microservicesontherise2015blog}.  
To build lightweight, efficient, distributed applications, container technology~\cite{lxc2016web} is commonly used. 
One popular software for container management and deployment is docker~\cite{dockerweb}. 
As of September 2016, over 40 thousand questions are tagged with `docker' on stackoverflow~\cite{stackoverflow}.

%
%
%
%

\begin{figure*}[ht!bp]
\begin{minipage}{\linewidth}
\footnotesize
\begin{Verbatim}[commandchars=\\\{\},codes={\catcode`$=3\catcode`^=7}]
*nat
:PREROUTING ACCEPT [0:0]
:INPUT ACCEPT [0:0]
:OUTPUT ACCEPT [0:0]
:POSTROUTING ACCEPT [0:0]
:DOCKER - [0:0]
-A PREROUTING -m addrtype --dst-type LOCAL -j DOCKER
-A OUTPUT ! -d 127.0.0.0/8 -m addrtype --dst-type LOCAL -j DOCKER
-A POSTROUTING -s 10.0.0.0/8 ! -o br-b74b417b331f -j MASQUERADE
-A POSTROUTING -s 172.17.0.0/16 ! -o docker0 -j MASQUERADE
-A DOCKER -i br-b74b417b331f -j RETURN
-A DOCKER -i docker0 -j RETURN
COMMIT
*filter
:INPUT ACCEPT [0:0]
:FORWARD ACCEPT [0:0]
:OUTPUT ACCEPT [0:0]
:DOCKER - [0:0]
:DOCKER-ISOLATION - [0:0]
-A FORWARD -j DOCKER-ISOLATION
-A FORWARD -o br-b74b417b331f -j DOCKER
-A FORWARD -o br-b74b417b331f -m conntrack --ctstate RELATED,ESTABLISHED -j ACCEPT
-A FORWARD -i br-b74b417b331f ! -o br-b74b417b331f -j ACCEPT
-A FORWARD -o docker0 -j DOCKER
-A FORWARD -o docker0 -m conntrack --ctstate RELATED,ESTABLISHED -j ACCEPT
-A FORWARD -i docker0 ! -o docker0 -j ACCEPT
-A FORWARD -i docker0 -o docker0 -j ACCEPT
-A FORWARD -i br-b74b417b331f -o br-b74b417b331f -j DROP
-A DOCKER-ISOLATION -i docker0 -o br-b74b417b331f -j DROP
-A DOCKER-ISOLATION -i br-b74b417b331f -o docker0 -j DROP
-A DOCKER-ISOLATION -j RETURN
COMMIT
\end{Verbatim}
\end{minipage}%
\caption{Default \texttt{iptables} firewall used by docker with \texttt{mynet}}
\label{fig:dockerfirewall:default}
\end{figure*}

\begin{figure*}[ht!bp]
\begin{minipage}{\linewidth}
\footnotesize
\input{content/sdnnfv_dockerfirewall_tuned_fig}
\end{minipage}%
\caption{Adapted \texttt{iptables} firewall for docker (only \texttt{filter} table shown). }
\label{fig:dockerfirewall:tuned}
\end{figure*}

\begin{figure*}[ht!bp]
\centering
\begin{tikzpicture}
\node[align=center,text width=15.5em, cloud, draw,cloud puffs=10,cloud puff arc=120, aspect=2, inner sep=-3em,outer sep=0] (a) at (5,1) { $\{0.0.0.0 .. 9.255.255.255\} \cup \{11.0.0.0 .. 255.255.255.255\}$ };
\node (b) at (5,-3) { $\{10.0.0.4\}$ };
\node (c) at (5,-6) { $\{10.0.0.3\}$ }; 
\node (d) at (0,-5) { $\{10.0.0.2\}$ };
\node (e) at (0,-1) { $\{10.0.0.1\}$ };
\node (f) at (4,-8) { $\{10.0.0.0\} \cup \{10.0.0.5 .. 10.255.255.255\}$ };

\draw[myptr] (a) to[out=330,in=310,looseness=3] (a);
\draw[myptr] (a) to (b);
\draw[myptr] (a) to[bend left] (c);
\draw[myptr] (a) to (d);
\draw[myptr] (a) to (e);
\draw[myptr] (a) to[bend right] (f);
\draw[myptr] (b) to (a);
\draw[myptr] (b) to[loop right] (b);
\draw[myptr] (b) to (c);
\draw[myptr] (b) to (d);
\draw[myptr] (b) to (e);
\draw[myptr] (b) to[bend right] (f);
\draw[myptr] (c) to (b);
\draw[myptr] (c) to[loop right] (c);
\draw[myptr] (c) to (d);
\draw[myptr] (d) to[loop below] (d);
\draw[myptr] (e) to (b);
\draw[myptr] (e) to (d);
\draw[myptr] (e) to[loop above] (e);
\end{tikzpicture}%
\caption{Reconstructed Security Policy (Structure of Figure~\ref{fig:dockerfirewall:tuned})}
\label{fig:mansdn:docker:servicematrix}
\end{figure*}

Docker itself does not provide security out of the box~\cite{dockersecurity,nccdocker2016security}. 
The security provided by containerization heavily depends on the specific docker and container setup. 
For example, if a container has the privilege to access raw sockets, the container can spoof arbitrary IP addresses or perform ARP spoofing~\cite{bui15arxivsockersec}. 
Hence, in a generic case, docker does not provide authenticity (\xmark). 
Consequently, since a malicious, privileged container can perform ARP spoofing, no guarantees about the structure (\xmark) and, hence, the stateful semantics (\xmark) can be given. 

Securing docker is out of the scope of this thesis. 
For this evaluation, we assume that docker is set up securely enough~\cite{dockersecurity} such that the firewall rules which docker generates by default are effective. 
Hence, we assume that we can refer to the identity of a container by referring to its IP address (assumption: authenticity \cmark). 
Our goal is to enforce the desired connectivity structure with the correct stateful handling. 
\medskip

This case study was built using Docker version 1.12.1, API version: 1.24. 
We want to deploy our application where each entity is realized as an individual container. 
For the sake of example, we use the Docker-provided \texttt{busybox} container images.\footnote{\url{https://hub.docker.com/_/busybox/}} 
We want to use the same IPs as in Table~\ref{tab:mansdncentralfw:ipmapping}. 

The complete application (all containers) should be isolated from other containers on the host. 
Therefore, we create a new docker network for the desired IP range. 
We also disable inter-container communication (icc) for this network. 
The network is called \texttt{mynet} and created with the following command. 

\medskip
\noindent
\begin{minipage}{\linewidth}
\footnotesize
\begin{Verbatim}[commandchars=\\\{\},codes={\catcode`$=3\catcode`^=7}]
# docker network create -d bridge --subnet=10.0.0.0/8 --gateway=10.42.0.100 $\hfill\hookleftarrow$
\qquad --opt="com.docker.network.bridge.enable_icc=false" mynet
\end{Verbatim}
\end{minipage}%
\medskip

Containers which are started in this network cannot connect to each other but can reach the Internet. 
Docker generates a default firewall, which is shown in Figure~\ref{fig:dockerfirewall:default}. 
The interface \verb~br-b74b417b331f~ is the bridge interface which was generated for \texttt{mynet}. 
We now add our custom firewall filtering rules to this setting. 
The \texttt{nat} table remains unchanged. 
We can almost directly add the rules of Figure~\ref{fig:centralrawfirewall}. 
The \verb~DOCKER-ISOLATION~ chain isolates the different docker networks from each other. 
We add our rules afterwards. 
The result is shown in Figure~\ref{fig:dockerfirewall:tuned}. 
For simplicity, we approximate the Internet-facing interface by stating that it is not our \texttt{mynet}-bridge interface. 

The question about the correctness of this ruleset remains. 
We have empirically tested the connectivity of the deployed containers. 
Yet, this does not provide a formal guarantee of its correctness. 

Similarly to Subsection~\ref{deploy:iptables}, we utilize our tool \fffuu{} to compute a service matrix. 
The `overhead' of the additional docker rules does not pose a problem to \fffuu{}. 
The result is shown in Figure~\ref{fig:mansdn:docker:servicematrix}. 
This graph is almost identical to the graph in Figure~\ref{fig:mansdn:centralwf:servicematrix}, except for two peculiarities: 
This time, we did not specify the IP range of the Internet. 
$\mvar{WebFrnt}$ can connect to the Internet, consequently, $\mvar{WebFrnt}$ can also connect to the unused IP range starting at 10.0.0.5. 
Second, we did not configure which containers or ports are publicly accessible. 
This is a docker configuration option and not part of the firewall setup. 
Therefore, the firewall setting allows all connection attempts from the Internet, but it depends on the actual docker configuration whether the containers are actually reachable. 
Hence, this setup---with all its assumptions about a secure and correct setup of docker---can enforce the desired structure (\cmark). 
The stateful semantics is provided (\cmark), which is confirmed by \fffuu{} (not visualized in the Figure). 


\paragraph*{Comparison to Container Links}
Our presented solution of configuring the iptables firewall of the host system is close to a solution supported natively by docker. 
Container links provide means to automatically install iptables rules when inter-container communication is desired between two containers~\cite{dockerlegacylinks}. 
When containers are started with the \verb~--link~ option, docker adds iptables rules similar to our setup. 
With the help of \topos{}, the necessary \verb~--link~ parameters could be generated. 
However, we decided not to follow this direction for the following reasons: 
Container links are considered legacy by the current version of docker~\cite{dockerlegacycontainerlinks}. 
In addition, the links are always bidirectional, thus no stateful semantics can be provided. 
Our approach of setting up an individual network is currently the recommended approach~\cite{dockerlegacycontainerlinks}.

\section{Discussion}
\label{sec:sdnnfv:discussion}
We discuss the limitations and advantages of \topos{}. 
\subsection{Limitations}
The process supported by \topos{} currently has two main limitations. 
First, it is completely static. 
For example, the mapping of policy entity names to their network representatives is done statically and manually. 
In general, \emph{naming} is a complex (but orthogonal) issue. 
This information should usually be managed by a resource and account management system or directory service.
%
Second, only one security device as backend is currently supported. 
However, related work suggests that this gap can be easily bridged, \eg by translating to a one big switch abstraction~\cite{monsanto2013composingonebigswitch,icfp2015smolkanetkatcompiler}.
Many networks additionally employ a variety of heterogeneous, vendor-specific middleboxes. 
In future work, it might be worth investigating to which extent additional low-level device features (\eg DHCP, IPv6, timeouts for stateful rules, ...) should be configurable on each abstraction layer.

\subsection{Advantages}
The presented process provides three novel advantages. 
First, it \emph{bridges several abstraction levels} in a uniform way. 
The intermediate results are well-specified, which allows manual intervention, visualization, and adding features.
Second, the theoretical background is \emph{completely formally verified}. 
Thus, \topos{} is more than an academic prototype but a highly trustworthy tool. 
In addition, \topos{}'s library can be reused, extended, exported to several languages, and adapted to fit the needs of other frameworks.
Finally, with the formal background, \topos{} is a first step towards high-assurance certification. 

Third, \topos{} can scale to large networks w.r.t.\ theory~(\emph{i}), computational complexity~(\emph{ii}), and management complexity~(\emph{iii}). 
The theoretical foundation (\emph{i}) scales to arbitrary networks. 
The computational complexity (\emph{ii}) depends on the type of security invariants. 
New invariants with arbitrary computational complexity can be developed for \topos{}. 
However, we found that usually only $\Phi$-structured invariants are needed, which implies the following computational complexity: $\mathrm{O}(\vert \mathit{invariants} \vert \cdot \vert \mathit{entities} \vert^2)$. 
An evaluation of \topos{}'s most expensive step has been presented in Section~\ref{subsec:toposcomputationcomplex}. 
Finally, (\emph{iii}) the complexity of managing an invariant (with exception for the ACL invariants) is linear in the number entities. 
Due to the auto-completion, it is actually better than linear.
Thus, the management complexity of \topos{} is roughly linear in the number of invariants and entities.

\section{Related Work}
\label{sec:sdnnfv:related}
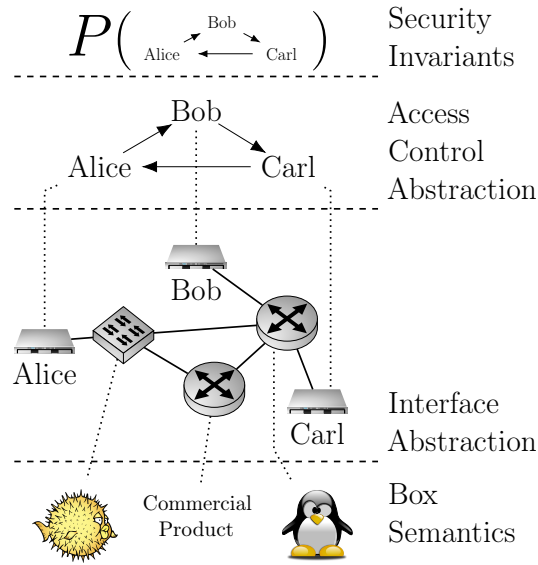
\begin{figure}[!htb]
	\centering%
  			\centering%
  			\resizebox{0.5\linewidth}{!}{%
  				\Large
  				\begin{tikzpicture}
  				\node[text width=0.3\linewidth, align=center] (iheardyoulikerecursion) at (0,1.5) {
  					\resizebox{.99\linewidth}{!}{
  						\scalebox{2}{\Huge{$P($}}\Large\begin{tikzpicture} [baseline=-3ex]
  						\node[anchor=south] (bi) at (0,0) {Bob};
  						\node[anchor=east] (ai) at (-.9,-.6) {Alice};
  						\node[anchor=west] (ci) at (+.9,-.6) {Carl};
  						
  						\draw[myptr] (ai) to (bi);
  						\draw[myptr] (bi) to (ci);
  						\draw[myptr,shorten >=1ex] (ci) to (ai.east); 
  						\end{tikzpicture}\scalebox{2}{\Huge{$)$}}%
  					}
  				};
  				
  				\def\sbordery{+.9}
  				\draw [thick,dashed] (-3,\sbordery)--(3,\sbordery);
  				\node[anchor=south west, align=left, text width=6em] at (3.,\sbordery) {Security\newline{} Invariants};

  				\node[anchor=south] (b) at (0,0) {Bob};
  				\node[anchor=east] (a) at (-.9,-.6) {Alice};
  				\node[anchor=west] (c) at (+.9,-.6) {Carl};
  				
  				\draw[myptr,shorten >=-1ex,shorten <=-1ex] (a) to (b);
  				\draw[myptr,shorten >=-1ex,shorten <=-1ex] (b) to (c);
  				\draw[myptr] (c) to (a);
  				
  				\def\abordery{-1.3}
  				\draw [thick,dashed] (-3,\abordery)--(3,\abordery);
  				\node[anchor=south west, align=left, text width=6em] at (3.,\abordery) {Access\newline{} Control\newline{} Abstraction};
  				
  				\node[server,label=below:Alice] (ra) at (-2.5,-3.5) {};
  				\node[server,label=below:Bob] (rb) at (0,-2.1) {};
  				\node[server,label=below:Carl] (rc) at (2.,-4.5) {};
  				\node[switch] (aux1) at (-1.2,-3.4) {};
  				\node[router] (aux2) at (1.5,-3.2) {};
  				\node[router] (aux3) at (.3,-4.3) {};
  				
  				\draw[thick] (ra) to (aux1);
  				\draw[thick] (rb) to (aux2);
  				\draw[thick] (rc) to (aux2);
  				\draw[thick] (aux1) to (aux2);
  				\draw[thick] (aux2) to (aux3);
  				\draw[thick] (aux1) to (aux3);

  				\draw [dotted,thick] (a)--(-2.5,-1)--(ra);
  				\draw [dotted,thick] (b)--(rb);
  				\draw [dotted,thick] (c)--($(2.,-1)+(1ex,0)$)--($(rc.north)+(1ex,0)$);
  				
  				\def\ibordery{-5.5}
  				\draw [thick,dashed] (-3,\ibordery)--(3,\ibordery);
  				\node[anchor=south west, align=left, text width=6em] at (3.,\ibordery) {Interface\newline{} Abstraction};
  				
  				\node[anchor=north] (l1) at (-2.,-5.8) {\includegraphics[height=2.5em]{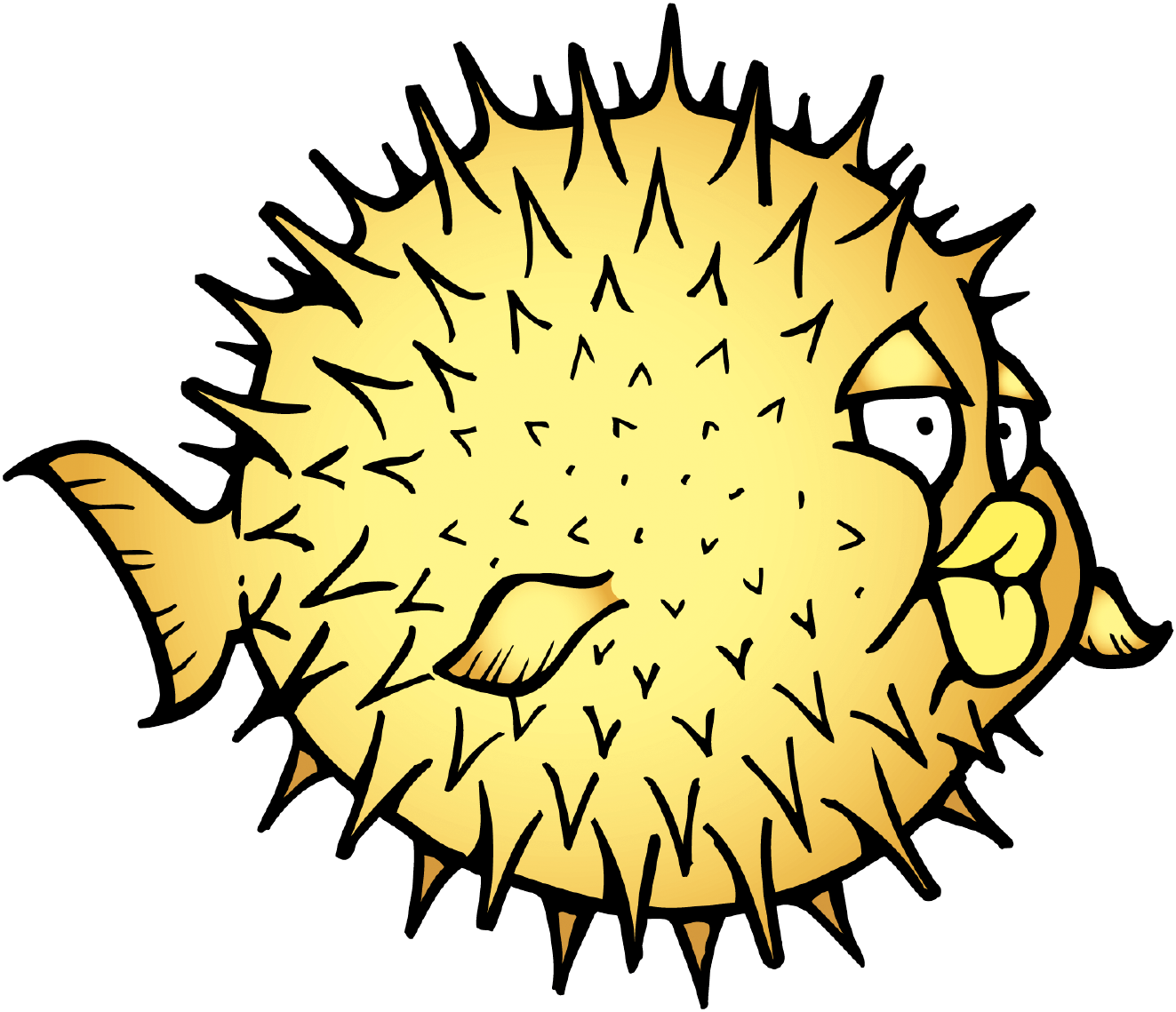}};
  				\draw [dotted,thick] (aux1)--(l1);

  				\node[anchor=north, align=center, text width=5em] (l3) at (0,-5.9) {\small{\parbox[c]{5em}{\centering Commercial Product}}};
  				\draw [dotted,thick] (aux3)--(l3);

  				\node[anchor=north] (l2) at (2.,-5.8) {\includegraphics[height=2.5em]{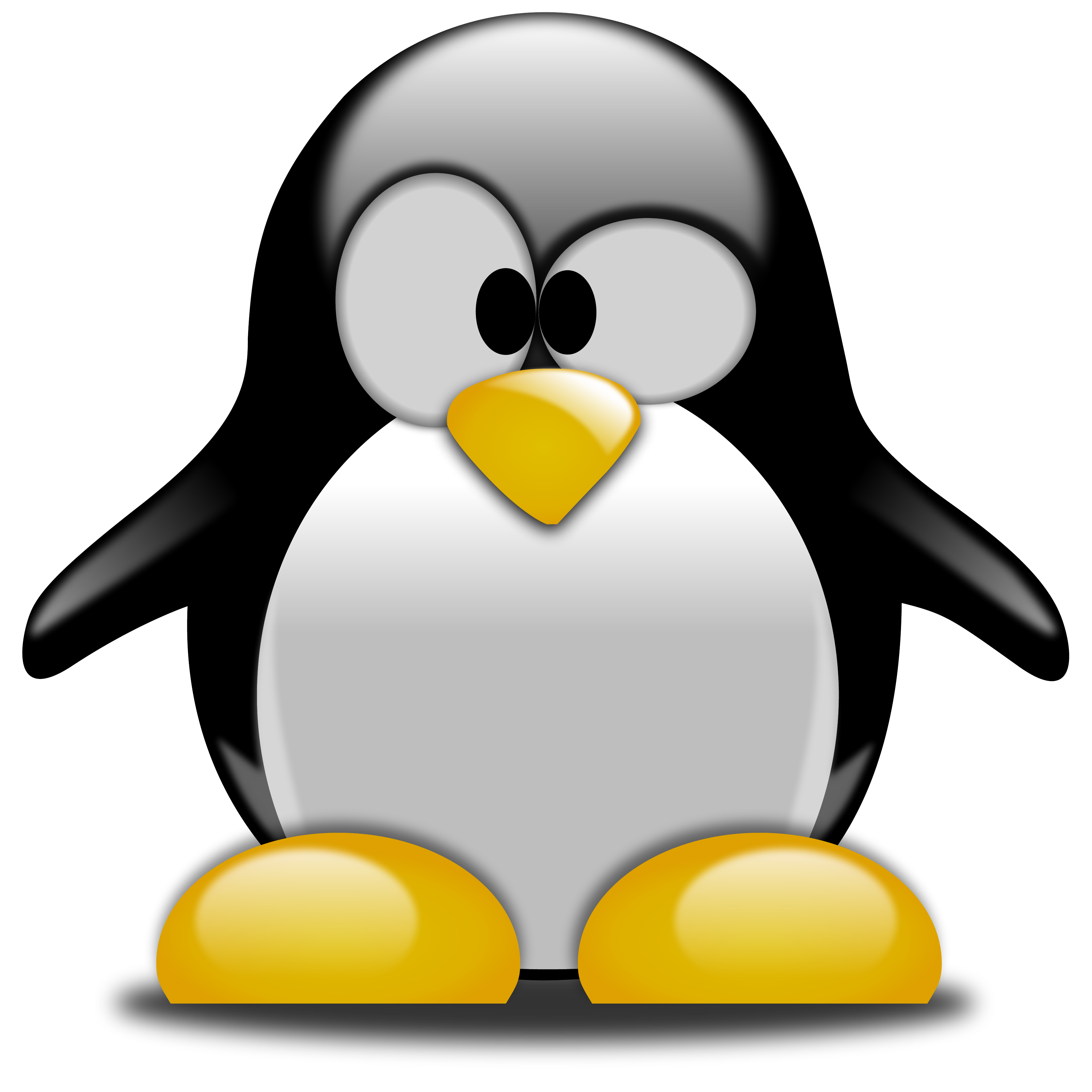}};
  				\draw [dotted,thick] ($(aux2.south)+(-1ex,0)$)--($(1.5,-5.3)+(-1ex,0)$)--(l2);
  				
  				\def\bbordery{-7}
  				\node[anchor=south west, align=left, text width=6em] at (3.,\bbordery) {Box\newline{} Semantics};
  				
  			\end{tikzpicture}}%
		\caption{Four Layer Abstractions}%
		\label{fig:4layerabstraction}%
\end{figure}

To discuss related work, we first define four management abstraction layers to subsequently classify related work.
%
%
Each abstraction layer is responsible for an individual problem domain. 
We illustrate the four layers in Figure~\ref{fig:4layerabstraction}. 
The layers have well-defined interfaces, thus, it is possible to combine solutions of individual problems. 
The discussion of related work will exhibit that this work combines perfectly with other related work focusing on different aspects. 
%
%

We propose the following four layers of abstractions.
\begin{description}
	\item[Security Invariants]
		Defines the high-level security goals.
		Representable as predicates.
		For example, Figure~\ref{fig:runexsecinvars}.
	\item[Access Control Abstraction]
		Defines the allowed accesses between policy entities.
		Representable as global access control matrix.
		For example, Figure~\ref{fig:secpol}.
	\item[Interface Abstraction]
		Defines a model of the complete network topology.
		Representable as a graph, packets are forwarded between the network entity's interfaces.
	\item[Box Semantics]
		Describes the semantics (\ie behavior) of individual network boxes.
		Usually, the semantics are vendor-specific (\eg iptables, Cisco ACLs, Snort IDS). 
\end{description}

The main difference between the interface abstraction and the box semantics is that latter describes the behavior of only one network entity, whereas the former describes the interconnection of many, possibly different, network boxes. 
In the presented case study, both coincide since only one enforcement device is considered.

With regard to Figure~\ref{fig:intro:securitycomponents}, security invariants are on the abstraction layer of security requirements, the access control abstraction corresponds to the security policy, and the interface abstraction as well as the box semantics correspond to security mechanisms.

In Figure~\ref{fig:relatedworkabstractionlayers}, we summarize how related work bridges the abstraction layers.
Related work may bridge these layers vertically or work horizontally on artifacts at one layer.
A direct arrow from the access control abstraction to the box semantics (and vice versa) means that the solution only applies to a single enforcement box. Solutions such as Firmato and Fireman achieve more and are thus listed multiple times.

\newcommand{\boxSemanticsToAccessControlAbstraction}[0]{Fireman~\cite{fireman2006}; \mbox{\fffuu{} (Chap.\ \ref{chap:networking16})\phantom{;}}}
\newcommand{\InterfaceAbstractionToAccessControl}[0]{Fireman~\cite{fireman2006}; HSA~\cite{kazemian2012HSA}; Anteater~\cite{Mai2011anteater}; \mbox{ConfigChecker}~\cite{alshaer2009configchecker}; VeriFlow~\cite{khurshid2013veriflow}} 
\newcommand{\InterfaceAbstractionMAPSAccessControl}[0]{Xie~\cite{xie2005static}; Lopes~\cite{lopes2013msrnetworkverificationprogram}}
\newcommand{\BoxSemanticsMAPSInterfaceAbstraction}[0]{HSA~\cite{kazemian2012HSA}; Anteater~\cite{Mai2011anteater}; Config\-Checker~\cite{alshaer2009configchecker}}
\newcommand{\AccessControlToInterfaceAbstraction}[0]{one big\phantom{;} switch~\cite{monsanto2013composingonebigswitch}; Firmato~\cite{bartal1999firmato}; FLIP~\cite{ZhangAlShaer2007flip}; \mbox{FortNOX}~\cite{Porras2012FortNOX}; Merlin~\cite{soule2014merlin}; Kinetic~\cite{kinetic2015}; \mbox{PBM~\cite{dinesh2002policybased}\phantom{;}}}
\newcommand{\AccessControlBISinvars}[0]{\topos{} \mbox{step \emph{B}}} 
\newcommand{\AccessControlToBoxSemantics}[0]{Firmato~\cite{bartal1999firmato}; FLIP~\cite{ZhangAlShaer2007flip}; NetKAT~\cite{icfp2015smolkanetkatcompiler}; Mignis~\cite{mignis2014};\newline Or-BAC~\cite{orbacnetwork04}; \topos{} step \emph{C}+\emph{D}}
\newcommand{\InterfaceAbstractionToBoxSemantics}[0]{RCP~\cite{Caesar2005rcp}; \mbox{OpenFlow}~\cite{mckeown2008openflow}; Merlin~\cite{soule2014merlin}; \mbox{optimized one}\phantom{;} \mbox{big switch}~\cite{Kang2013onebigswitchabstraction}; NetKAT~\cite{icfp2015smolkanetkatcompiler}; VeriFlow~\cite{khurshid2013veriflow}; FML~\cite{hinrichs2009practical}\phantom{;}}
\newcommand{\BoxSemantics}[0]{\mbox{Iptables Semantics}~\cite{diekmann2015fm}} 

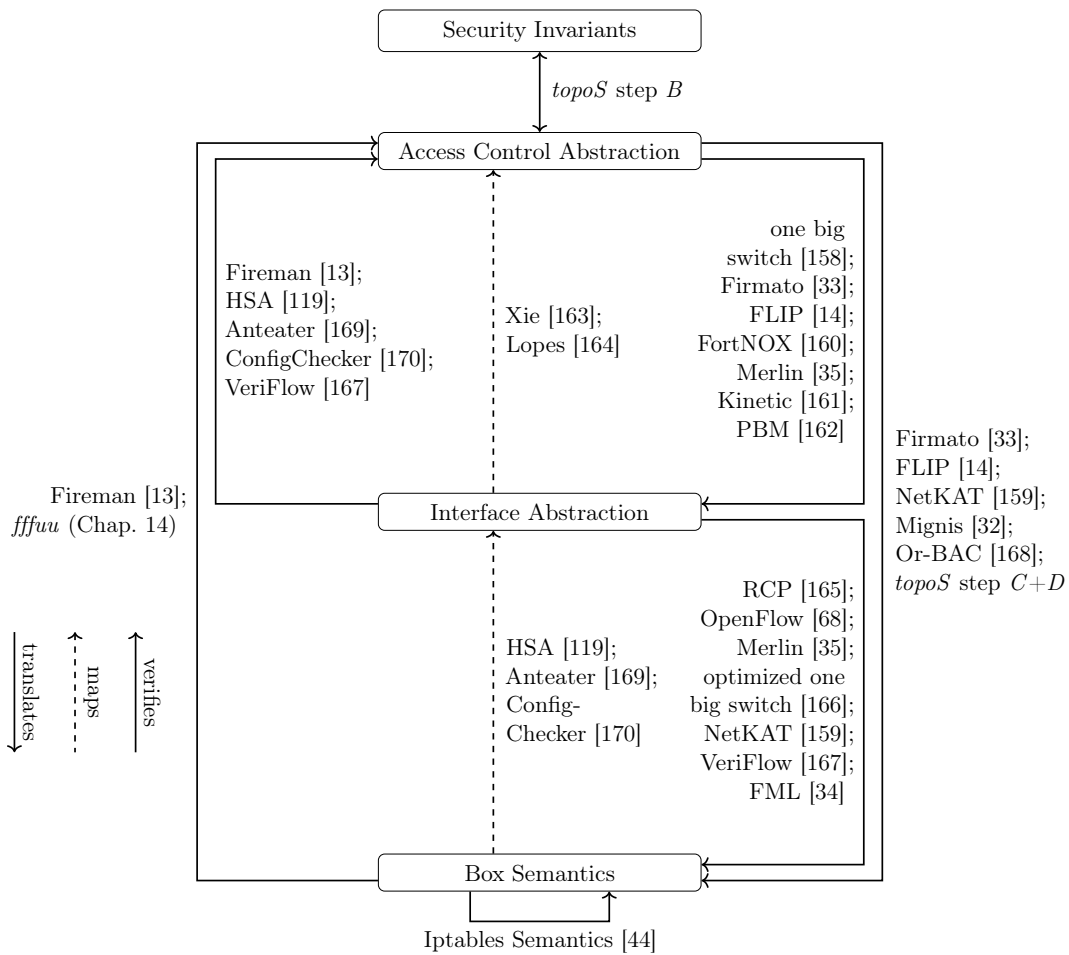
\begin{figure}[!htb]
\centering
    \resizebox{0.99\textwidth}{!}{%
 	\begin{tikzpicture}
	 \node [MyRoundedBox, text width=13em](abs1) at (0,0) {Security Invariants};
	 \node [MyRoundedBox, text width=13em](abs2) at (0,-2) {Access Control Abstraction};
	 \node [MyRoundedBox, text width=13em](abs3) at (0,-8) {Interface Abstraction};
	 \node [MyRoundedBox, text width=13em](abs4) at (0,-14) {Box Semantics};
	 \draw [thick,-to,<->] (abs1.south)--(abs2.north);
	 \node [anchor=west, align=left, text width=7em] at ($(abs1)!0.5!(abs2) + (+0.2em,0)$) {\AccessControlBISinvars};
	 
	 \draw [thick,-to,->] ($(abs2.east) + (0,-0.8ex)$)--($(abs2.east) + (+7em,-0.8ex)$)--($(abs3.east) + (+7em,+0.8ex)$)--($(abs3.east) + (0,+0.8ex)$);
	 \node [anchor=east, align=right, text width=7em] at ($(abs2)!0.5!(abs3) + (+13.8em,0)$) {\AccessControlToInterfaceAbstraction};

	 \draw [thick,-to,dashed,->] ($(abs3.north) + (-2em,0)$)--($(abs2.south) + (-2em,0)$);
	 \node [anchor=west, align=left, text width=7em] at ($(abs3)!0.5!(abs2) + (-1.8em,0)$) {\InterfaceAbstractionMAPSAccessControl};
	 
	 \draw [thick,-to,->] ($(abs3.east) + (0,-0.8ex)$)--($(abs3.east) + (+7em,-0.8ex)$)--($(abs4.east) + (+7em,+0.8ex)$)--($(abs4.east) + (0,+0.8ex)$);
	 \node [anchor=east, align=right, text width=7em] at ($(abs3)!0.5!(abs4) + (+13.8em,0)$) {\InterfaceAbstractionToBoxSemantics};
	 \draw [thick,-to,->] ($(abs2.east) + (0,+0.8ex)$)--($(abs2.east) + (+7.8em,+0.8ex)$)--($(abs4.east) + (+7.8em,-0.8ex)$)--($(abs4.east) + (0,-0.8ex)$);
	 \node [anchor=west, align=left, text width=8em] at ($(abs2)!0.5!(abs4) + (+15.0em,0)$) {\AccessControlToBoxSemantics};

	 \draw [thick,-to,dashed,->] ($(abs4.north) + (-2em,0)$)--($(abs3.south) + (-2em,0)$);
	 \node [anchor=west, align=left, text width=7em] at ($(abs4)!0.5!(abs3) + (-1.8em,0)$) {\BoxSemanticsMAPSInterfaceAbstraction};	 
	 
	 \draw [thick,-to,->] ($(abs3.west) + (0,+0.8ex)$)--($(abs3.west) + (-7em,+0.8ex)$)--($(abs2.west) + (-7em,-0.8ex)$)--($(abs2.west) + (0,-0.8ex)$);
	 \node [anchor=west, align=left, text width=10em] at ($(abs2)!0.5!(abs3) + (-13.9em,0)$) {\InterfaceAbstractionToAccessControl};
	 
	 \draw [thick,-to,->] ($(abs4.west) + (0,-0.8ex)$)--($(abs4.west) + (-7.8em,-0.8ex)$)--($(abs2.west) + (-7.8em,+0.8ex)$)--($(abs2.west) + (0,+0.8ex)$);
	 \node [anchor=east, align=right, text width=9em] at ($(abs2)!0.5!(abs4) + (-15.0em,0)$) {\boxSemanticsToAccessControlAbstraction};

	 \draw [thick,-to,->] ($(abs4.south) + (-3em,0)$)--($(abs4.south) + (-3em,-3ex)$)--($(abs4.south) + (+3em,+-3ex)$)--($(abs4.south) + (+3em,0)$);
	 \node [anchor=north, align=center, text width=10em] at ($(abs4.south)+ (0,-3ex)$) {\BoxSemantics};

	 \draw[thick,-to,->] (-6,-10) to node[above,sloped]{translates} (-6,-12);
	 \draw[thick,-to,dashed,<-] (-5,-10) to node[above,sloped]{maps} (-5,-12);
	 \draw[thick,-to,<-] (-4,-10) to node[above,sloped]{verifies} (-4,-12);
	 \end{tikzpicture}
	}
		\caption{Four Layer Abstraction in Related Work}
		\label{fig:relatedworkabstractionlayers}
\end{figure}



\begin{figure}[h!tb]
	\centering
  		\includegraphics[width=0.8\linewidth]{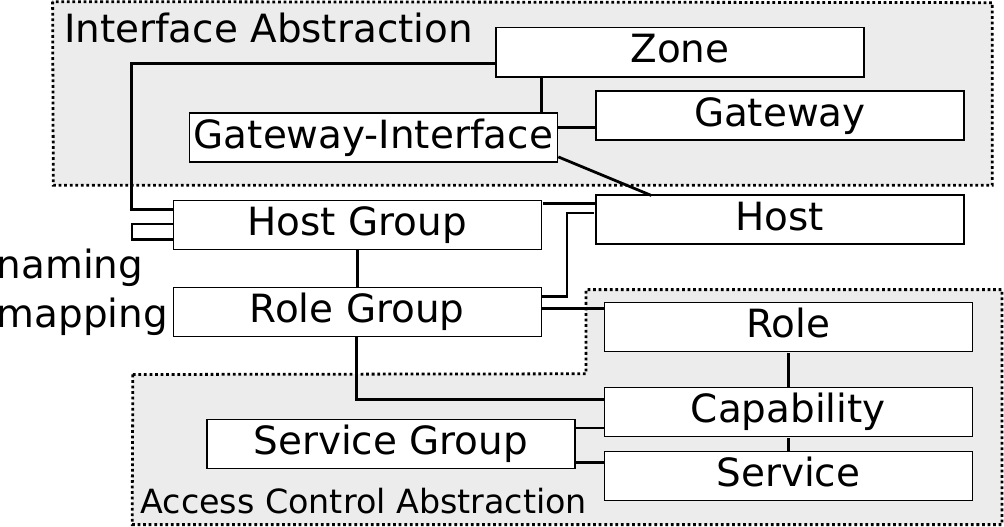}
  		\vskip-5pt
		\caption{Firmato ERM}
		\label{fig:firmatoerm}
\end{figure}

Firmato~\cite{bartal1999firmato} is the work closest related to \topos{}. 
It defines an entity relationship model 
to structure network management and compile firewall rules from it, illustrated in Figure~\ref{fig:firmatoerm}. 
Firmato focuses on roles, which correspond to policy entities in our model. 
A role has positive capabilities and is related to other roles, which can be used to derive an access control matrix. 
Zones, Gateway-Interfaces, and Gateways define the network topology, which corresponds to the interface abstraction. 
As illustrated in Figure~\ref{fig:firmatoerm}, the abstraction layers identified in this work can also be identified in Firmato's model. 
The Host Groups, Role Groups, and Hosts definitions provide a mapping from policy entities to network entities, which is Firmato's approach to the naming problem. 
With close correspondence in the underlying concepts to Firmato, Cuppens \etal\cite{Cuppens2005orbacxmlfirewall} propose a firewall configuration language based on Or-BAC~\cite{orbac}. 
Similar to Firmato (with more support for negative capabilities) is FLIP~\cite{ZhangAlShaer2007flip}, which is a high-level language with focus on \emph{service} management (\eg allow/deny HTTP). 
Essentially, both FLIP and Firmato enhance the access control abstraction horizontally by including layer four port management and traverse it vertically by serializing to firewall rules.

FML~\cite{hinrichs2009practical} is a flow-based declarative language to define, among others, access control policies in a DATALOG-like language. 
Comparably to our directed (stateless) policy, FML operates on unidirectional network flows. 
FML solves the naming problem by assuming that all entities are authenticated with IEEE 802.1X~\cite{IEEE8021X}.

Policy-based management (PBM)~\cite{dinesh2002policybased} was introduced to simplify network administration. 
Similar to our work, it proposes different levels of abstraction and describes how to translate between them. 
Policy-based management defines a generic information model~\cite{rfc3060,rfc3460} which is not limited to access control, 
however, we focus our discussion solely on access control and security. 
In a central policy repository, global policy rules are stored. 
Policy decision points retrieve these rules and interpret them. 
Using our terminology, this step translates the access control abstraction to the interface abstraction. 
A policy decision point forwards decisions to policy enforcement points, implementing the translation from the interface abstraction to box semantics. 
This last step may be very device-specific~\cite{hinrichs99pbmcisco} and is not the core focus of PBM. 
%
%
%
While PBM was built on the idea of specifying business-level abstractions in terms of requirements~\cite{dinesh2002policybased}, 
the IETF specified a rule-based policy repository~\cite{rfc3060,rfc3460}, which restricts storing high-level requirements that cannot easily be expressed as rules. 
In strong contrast to our work, where we focus on specifying higher-level requirements (\ie security invariants), the IETF Policy Framework working group focused on the specification of lower-level policies~\cite[\S\hairspace{}2.1. Policy Scope]{rfc3060}. 
This can also be witnessed in many languages which were developed over the years~\cite{Han2012surveypbm} since, in particular when it comes to security, they usually only provide access control abstractions. 
%
%
%
%
%

Mignis~\cite{mignis2014} is a declarative language to manage Netfilter/iptables firewalls. 
It focuses on packet filtering and NAT. 
The policy language is restrictive to avoid the usual policy conflicts. 
In particular, it only allows to write blacklisting-style policies, additionally NAT and filtering cannot arbitrarily be mixed to provide consistent policies.\footnote{The required assumptions for the translation impose many restrictions \cite[\S\hairspace{}5]{mignis2014}.} 
Apart from our work (Part~\ref{part:existing-configs}), it is the only work we are aware of which provides a formal semantics of an iptables firewall. 
The authors describe a semantics of the packet filtering and NAT behavior of iptables. 
Their semantics only models the aspects of iptables which are required for their policy language. 
While it supports NAT (which our semantics of Part~\ref{part:existing-configs} does not), it does not support user-defined chains or arbitrary match conditions.\footnote{Mignis supports match conditions but imposes additional assumptions on them to ensure that they are consistent with NAT. For example, the \texttt{iprange} module is not allowed. Without inspecting all matching modules manually, there is no generic way to assure whether a filter condition is compatible with Mignis.} 
Though supporting NAT and stateful filtering, advanced iptables features such as \texttt{NOTRACK} or \texttt{connmark} are not considered.

The ``One Big Switch'' Abstraction~\cite{monsanto2013composingonebigswitch,Kang2013onebigswitchabstraction,reich2013modular} allows to manage a network as if it were only one, central big switch. 
This effectively allows solutions which only support to manage one device to be applied to a complete network, consisting of multiple switches. 
With regard to Figure~\ref{fig:relatedworkabstractionlayers}, any solutions which support translating the access control abstraction to box semantics can also be applied to translate from the access control abstraction to the interface abstraction by translating to the ``One Big Switch''. 


NetKAT \cite{anderson2014netKATsemantics,icfp2015smolkanetkatcompiler} is a SDN programming language with well-defined semantics. 
It features an efficient compiler for local, global, and virtual programs to flow table entries~\cite{icfp2015smolkanetkatcompiler}. 
Among others, it allows to implement the ``One Big Switch'' Abstraction~\cite{icfp2015smolkanetkatcompiler}. 


Craven \etal\cite{craven2011policyrefinement} present a generalized (not network-specific) process to translate access control policies, enhanced with several aspects, to enforceable device-specific policies; the implementation requires a model repository of box semantics and their interplay.
Pahl delivers a data-centric, network-specific approach for managing and implementing such a repository, further focusing on things~\cite{Pahl2015IM}. 


As illustrated in Figure~\ref{fig:relatedworkabstractionlayers},
Fireman~\cite{fireman2006} is a counterpart to Firmato.
It verifies firewall rules against a global access policy.
In addition, Fireman provides verification on the same horizontal layer (\ie finding shadowed rules or inter-firewall conflicts, which do not affect the resulting end-to-end connectivity but are still most likely an implementation error).
Abstracting to its uses, one may call rcc~\cite{Feamster2005rcc} the fireman for BGP. 
\fffuu{} (presented in Part~\ref{part:existing-configs}) is unique as it not only verifies rules, but also translates them back to the access control abstraction. 


Header Space Analysis \allowbreak{}\mbox{(HSA)~\cite{kazemian2012HSA}}, Anteater \cite{Mai2011anteater}, and Config\-Checker~\cite{alshaer2009configchecker} verify several horizontal safety properties on the interface abstraction, such as absence of forwarding loops.
By analyzing reachability~\cite{xie2005static,lopes2013msrnetworkverificationprogram,Mai2011anteater,alshaer2009configchecker,kazemian2012HSA}, horizontal consistency of the interface abstraction with an access control matrix can also be verified. 
Verification of incremental changes to the interface abstraction can be done in real-time with VeriFlow~\cite{khurshid2013veriflow}, which can also prevent installation of violating rules. 
These models of the interface abstraction have many commonalities: %
The network boxes in all models are stateless and the network topology is a graph, connecting the entity's interfaces.
A function models packet traversal at a network box. 
%
Verification of incremental changes to the interface abstraction can be done in real-time by NetPlumber~\cite{kazemian2013realtimehsa} and VeriFlow~\cite{khurshid2013veriflow}.
The latter can also prevent installation of violating rules. 
These models could be considered as a giant (extended) finite state machine (FSM), where the state of a packet is an ($\textbf{interface}\times\textbf{packet}$) pair and the network topology and forwarding function represent the state transition function~\cite{lopes2013msrnetworkverificationprogram,zhang2012erificationswitching}.
In contrast to arbitrary state machines, it is believed that those derived from networks are comparatively well-behaved~\cite{zhang2012erificationswitching}.
Anteater~\cite{Mai2011anteater} differs in that interface information is implicit and packet modification is represented by relations over packet histories.


Most analysis tools make simplifying assumptions about the underlying network boxes.
Diekmann \etal\cite{diekmann2015fm} (\cf Part~\ref{part:existing-configs}) present simplification of iptables firewalls to make complex real-world firewalls available for tools with simplifying assumptions. 
Essentially, the authors horizontally simplify the box semantics.



%
FortNOX~\cite{Porras2012FortNOX} horizontally enhances the access control abstraction as it assures that rules by security apps are not overwritten by other apps.
Technically, it hooks up at the access control/interface abstraction translation. 
Kinetic~\cite{kinetic2015,kineticwebsite} is an SDN language which lifts static policies (as constructed by \topos{}) to dynamic policies.
To accomplish this, an administrator can define a simple FSM which dynamically (triggered by network events) switches between static policies. 
In addition, the FSM can be verified with a model checker.

Features are horizontally added to the interface abstraction:
a routing policy allows specifying \emph{paths} of network traffic~\cite{Kang2013onebigswitchabstraction}. 
Merlin~\cite{soule2014merlin,soule2014merlinsconferenceversion} additionally supports bandwidth assignments and network function chaining. 
Both translate from a global policy to local enforcement and Merlin provides a feature-rich language for interface abstraction policies.
Conceptually similar (but with a completely different implementation), RCP~\cite{Caesar2005rcp} allows to logically centralize routing while remaining compatible with existing routers. 


Layer two and layer three network architectures with focus on access control have been developed. 
In the terminology of this thesis, these architectures correspond to (distributed) security mechanisms, located at the interface abstraction. 
SANE~\cite{casado2006sane} and Ethane~\cite{ethane07} focus on centrally-managed enterprise networks. 
SANE presents a new architecture where each packet needs a Kerberos-like ticket~\cite{rfc4120} to be forwarded by the network, basically implementing capability-based source routing. 
In contrast, Ethane is compatible with existing Ethernet networks implementing access control in each switch, programmed in an OpenFlow-like fashion~\cite{mckeown2008openflow}. 
PBS~\cite{schulzrinne2013pbs,schulzrinne2010pbs} is an architecture developed for the context of preventing DoS attacks while being compatible with the current Internet. 
Its use-case requires distributed management, permissions are granted by receivers of a flow. 
PBS routers are configured with soft-state to allow only authorized flows using in-band signaling based on IPsec~\cite{rfc4301}. 
Both SANE~\cite{casado2006sane} and Ethane~\cite{ethane07} with their centralized approach towards policy management could serve as a backend for \topos{}. 
In fact, OpenFlow is a successor of Ethane and we have demonstrated applicability in this chapter.

We discuss further related work in detail in Chapter~\ref{chap:relatedworktable}.

\section{Conclusion}
\label{sec:sdnnfv:conclusion}
We presented \topos{}, a fully verified tool to manage network-level access control.
It was demonstrated by example; nevertheless, the correctness proofs are universally valid and \topos{} is applicable to any larger network.
The example demonstrates that a traditional network segmentation into \emph{internal} and \emph{DMZ} cannot cope with complex security goals and the traditional thought model of structure by IP ranges is no longer appropriate. 
In contrast, \topos{} only requires the high-level security goals and can automatically translate them to low-level configurations, such as firewall rules or SDN flow table entries.
%
%
%
During the translation, all intermediate results are well-defined, accessible, and can be visualized.
This provides feedback and allows manual refinement of them, including manual optimizations on lower abstraction layers.
After manual refinement, re-verification is run to avoid human error.
For the first time, the complete, automated, and verified transition from high-level security goals to Firewall, SDN, and Microservice configurations was presented.

 \part{Understanding Existing Configurations}
 \label{part:existing-configs}

 \chapter{Overview}
 \label{chap:existing:overview}
 Part~\ref{part:greenfield} focused mainly on the consistency between {security requirements} and the {security policy}. 
 It showed how to compute a policy from the requirements and how to verify an existing policy. 
 In this part, we focus on existing configurations of security mechanisms in order to derive the actual policy they are implementing. 

 \begin{center}
    \resizebox{0.99\textwidth}{!}{%
 	\begin{tikzpicture}
	 \node [MyRoundedBox, fill=LightGrey](sinvar) at (0,0) {Security Requirements};
	 \node [MyDoubleArrow](arr1) at (sinvar.east) {};
	 \node [MyRoundedBox, fill=LightYellow](policy) at (arr1.east) {Security Policies};
	 \node [MySingleLeftArrow](arr2) at (policy.east) {};
	 \node [MyRoundedBox, fill=LightYellow](mechanism) at (arr2.west) {Security Mechanisms};
	 \end{tikzpicture}
	 }
 \end{center}

 We focus on the Linux/netfilter iptables firewall. 
 We present our fully automated tool \fffuu{} which can translate a low-level iptables configuration to a policy graph. 

 \begin{center}
    \resizebox{0.99\textwidth}{!}{%
 	\begin{tikzpicture}
	 \node [MyRoundedBox, fill=LightGrey](sinvar) at (0,0) {Security Requirements};
	 \node [MyDoubleArrow](arr1) at (sinvar.east) {};
	 \node [MyRoundedBox, fill=LightYellow](policy) at (arr1.east) {Security Policies};
	 \node [MySingleLeftArrow](arr2) at (policy.east) {};
	 \node [MyRoundedBox, fill=LightYellow](mechanism) at (arr2.west) {\texttt{Linux/netfilter iptables}};
	 \draw [shorten <=-2cm,shorten >=-19ex,-latex] ($(mechanism) - (0,2.5em)$)--($(policy) - (0,2.5em)$);
	 \node [anchor=north] at ($(arr2) - (0,2.5em)$) {\fffuu{}};
	 \end{tikzpicture}
	 }
 \end{center}

 This part is structured as follows. 
 First, in Chapter~\ref{chap:fm15}, we formalize the filtering behavior of iptables and provide algorithms to transform rulesets to a simpler structure. 
 In Chapter~\ref{chap:nospoof}, we provide an algorithm to verify spoofing protection of a ruleset. 
 Having excluded IP address spoofing, we can afterwards refer to entities in our policy by their IP addresses. 
 Chapter~\ref{chap:networking16} will then present a method to partition the complete IPv4 and IPv6 address space into classes with equal access rights. 
 This yields the vertices in our policy graph. 
 With this, we compute service matrices which correspond exactly to the required policy graph for a fixed service. 
 We will demonstrate that this solves real-world problems.

\section*{Availability}
\label{sec:fm15:availability}
Our Isabelle/HOL theory files with the formalization and the referenced correctness proofs and our tool \fffuu{} are available at 
\begin{center}
\url{https://github.com/diekmann/Iptables_Semantics} \ \ and \ \ the AFP~\cite{IP_Addresses-AFP,Simple_Firewall-AFP,Iptables_Semantics-AFP,Routing-AFP}
\end{center}
It is the first fully machine-verified iptables analysis tool. 

The raw data of the analyzed firewall rulesets can be found at
\begin{center}
\url{https://github.com/diekmann/net-network}
\end{center}
To the best of our knowledge, this is the largest, publicly available collection of real-world iptables firewall rulesets.

  \chapter[Semantics-Preserving Simplification of Firewall Rule Sets]{Semantics-Preserving Simplification of Real-World Firewall Rule Sets}
\label{chap:fm15}

This chapter is an extended version of the following paper~\cite{diekmann2015fm}:
\begin{itemize}
	\item Cornelius Diekmann, Lars Hupel, and Georg Carle. \emph{Semantics-Preserving Simplification of Real-World Firewall Rule Sets}. In 20th International Symposium on Formal Methods, pages 195-212, Oslo, Norway, June 2015. Springer.
\end{itemize}

\noindent
The following major improvements and new contributions were added:
\begin{itemize}
	\item We provide a clarifying discussion of how stateful match conditions can be represented in our (seemingly) stateless model (cf.\ paragraph~\ref{par:fm15:stateful-limitations}, \nameref{par:fm15:stateful-limitations}). 
	\item We prove defindness of our semantics (Theorem~\ref{thm:bigstep-defined}). 
	\item We add support for the \iptaction{GOTO} action (which is not discussed in this chapter for brevity but can be found in the accompanying Isabelle formalization). 
	\item We present how our work could be applied to OpenFlow (Section~\ref{sec:outlook:verifyopenflow}).  
\end{itemize}

\paragraph*{Statement on author's contributions}
For the original paper, the author of this thesis developed the presented semantics. 
The author of this thesis provided major contributions for the ideas, formalization, algorithms, and proofs. 
He researched related work, and conducted the evaluation. 
All improvements with regard to the paper are the work of the author of this thesis.

\medskip

\paragraph*{Abstract}
We found that none of the available tools could handle typical, real-world \iptables{} rulesets. 
This is due to the complex chain model used by \iptables{}, but also due to the vast amount of possible match conditions that occur in real-world firewalls, many of which are not understood by academic and open source tools. 
We provide a formal big step semantics of \iptables{} filtering behavior. 
Then, we provide algorithms to transform complex firewall rulesets to a simple list model. 
Though we do not present our own fully-verified analysis algorithms yet, these transformations already enable existing analysis tools to understand real-world firewall rules. 

\medskip

\section{Introduction}
\label{sec:fm15:introduction}
Firewalls are a fundamental security mechanism for computer networks.
Several firewall solutions, ranging from open source~\cite{iptables,nftables,bsdpf} to commercial~\cite{ciscofirewallaccesslists,hpaclfirewalls}, exist.
Operating and managing firewalls is challenging as rulesets are usually written manually. 
While vulnerabilities in the firewall software itself are comparatively rare, it has been known for over a decade~\cite{firwallerr2004} that many firewalls enforce poorly written rulesets.
However, the prevalent methodology for configuring firewalls has not changed.
Consequently, studies regularly report insufficient quality of firewall rulesets \cite{fireman2006,netsecconflicts,ZhangAlShaer2007flip,databreach2009src,nelson2010margrave,sherry2012making,fwviz2012,diekmann2014forte}.

Therefore, several tools~\cite{marmorstein2005itval,marmorstein2006firewall,fireman2006,tongaonkar2007inferring,cspfirewall,nelson2010margrave,fwviz2012,fwbuilder} have been developed to ease firewall management and reveal configuration errors.
However, when we tried to analyze real-world firewalls with the publicly available tools, none of them could handle our firewall rules. 
We found that the firewall model of the available tools is too simplistic.

In this chapter, we address the following fundamental problem: Many tools do not understand real-world firewall rules. 
To solve the problem, we transform and simplify the rules such that they are understood by the respective tools.

\begin{figure*}[htb]
\begin{minipage}{\linewidth}
\footnotesize
\begin{Verbatim}[commandchars=\\\{\},codes={\catcode`$=3\catcode`^=7}]
Chain INPUT (policy ACCEPT)
target      prot source         destination  
DOS_PROTECT all  0.0.0.0/0      0.0.0.0/0  
ACCEPT      all  0.0.0.0/0      0.0.0.0/0   state RELATED,ESTABLISHED
DROP        tcp  0.0.0.0/0      0.0.0.0/0   tcp dpt:22
DROP        tcp  0.0.0.0/0      0.0.0.0/0   multiport dports 21,873,5005,5006,80, $\hfill\hookleftarrow$
                                                             548,111,2049,892
DROP        udp  0.0.0.0/0      0.0.0.0/0   multiport dports 123,111,2049,892,5353
ACCEPT      all  192.168.0.0/16 0.0.0.0/0  
DROP        all  0.0.0.0/0      0.0.0.0/0  

Chain DOS_PROTECT (1 references)
target      prot source         destination  
RETURN      icmp 0.0.0.0/0      0.0.0.0/0   icmptype 8 limit: avg 1/sec burst 5
DROP        icmp 0.0.0.0/0      0.0.0.0/0   icmptype 8
RETURN      tcp  0.0.0.0/0      0.0.0.0/0   tcp flags:0x17/0x04 limit: avg 1/sec burst 5
DROP        tcp  0.0.0.0/0      0.0.0.0/0   tcp flags:0x17/0x04
RETURN      tcp  0.0.0.0/0      0.0.0.0/0   tcp flags:0x17/0x02 limit: avg 10000/sec $\hfill\hookleftarrow$
                                                                       burst 100
DROP        tcp  0.0.0.0/0      0.0.0.0/0   tcp flags:0x17/0x02
\end{Verbatim}
\end{minipage}%
\caption{Linux \iptables{} ruleset of a Synology NAS (network attached storage) device}%
  \label{fig:firewall:synology}%
\end{figure*}

To demonstrate the problem by example, we decided to use \emph{ITVal}~\cite{marmorstein2005itval} because it natively supports \iptables{}~\cite{iptables}, is open source, and supports calls to user-defined chains. 
However, ITVal's firewall model is representative of the model used by the majority of tools; therefore, the problems described here also apply to a vast range of other tools.
Firewall models used in related work are surveyed in Section~\ref{sec:fm15:firewall-models}.
For this example, we use the firewall rules in Figure~\ref{fig:firewall:synology}, taken from an NAS device.
The ruleset reads as follows: 
First, incoming packets are sent to the user-defined chain \texttt{DOS\_PROTECT}, where some rate limiting is applied. 
Afterwards, the firewall allows all packets which belong to already established connections. 
This is generally considered good practice.
Then, some services, identified by their ports, are blocked.
Finally, the firewall allows all packets from the local network 192.168.0.0/16 and discards all other packets.
%
%
We used ITVal to partition the IP space into equivalence classes (\ie ranges with the same access rights)~\cite{marmorstein2006firewall}.
The expected result is a set of two IP ranges: the local network 192.168.0.0/16 and the ``rest''.
However, ITVal erroneously only reports one IP range: the universe.
Removing the first two rules (in particular the call in the \texttt{DOS\_PROTECT} chain) lets ITVal compute the expected result.

We identified two main problems which prevent tools from ``understanding'' real-world firewalls. 
First, calling and returning from custom chains, due to the possibility of complex nested chain calls. 
Second, more seriously, most tools do not understand the firewall's match conditions.
In the above example, the rate limiting is not understood.
The problem of unknown match conditions cannot simply be solved by implementing the rate limiting feature for the respective tool.
The major reason is that the underlying algorithm might not be capable of dealing with this special case.
Additionally, firewalls, such as \iptables{}, support numerous match conditions and several new ones are added in every release. 
As of version 1.6.0 (Linux kernel 4.10, early 2017), \iptables{} supports more than 60 match conditions with over 200 individual options. 
Firewalls, such as \iptables{}, support numerous match conditions and several new ones are added in every release. 
We expect even more match conditions for nftables~\cite{nftables} in the future since they can be written as simple userspace programs~\cite{whylovenftables2014blog}.
%
%
Therefore, it is virtually impossible to write a tool which understands all possible match conditions.

In this chapter, we build a fundamental prerequisite to enable tool-supported analysis of \emph{real-world} firewalls: 
We present several steps of semantics-preserving ruleset simplification, which lead to a ruleset that is ``understandable'' to subsequent analysis tools:  
First, we unfold all calls to and returns from user-defined chains.
This process is exact and valid for arbitrary match conditions.
Afterwards, we process unknown match conditions.
For that, we embed a ternary-logic semantics into the firewall's semantics.
Due to ternary logic, all match conditions not understood by subsequent analysis tools can be treated as always yielding an unknown result.
In a next step, all unknown conditions can be removed.
This introduces an over- and underapproximation ruleset, called upper/lower closure. 
Guarantees about the original ruleset dropping/allowing a packet can be given by using the respective closure ruleset.

To summarize, we provide the following contributions: 
\begin{enumerate}
  \item A formal semantics of \iptables{} packet filtering (Section~\ref{sec:fm15:iptables-semantics}),
  \item Chain unfolding: transforming a ruleset in the complex chain model to a ruleset in the simple list model (Section~\ref{sec:fm15:unfolding}),
  \item An embedded semantics with ternary logic, supporting arbitrary match conditions, introducing a lower/upper closure of accepted packets (Section~\ref{sec:fm15:ternary}), and
  \item Normalization and translation of complex logical expressions to an \iptables{}-compatible format, discovering a meta-logical firewall algebra (Section~\ref{sec:fm15:normalization}).
\end{enumerate}

We evaluate applicability on large real-world firewalls in Section~\ref{sec:fm15:evaluation}.
All proofs are machine-verified with Isabelle~\cite{isabelle2016}. 
Therefore, the correctness of all obtained results only depends on a small and well-established mathematical kernel and the \iptables{} semantics (Figure~\ref{fig:semantics}).

\section{Firewall Models in the Literature and Related Work}
\label{sec:fm15:firewall-models}
Packets are routed through the firewall and the firewall needs to decide whether to allow or deny a packet.
A firewall ruleset determines the firewall's filtering behavior.
The firewall inspects its ruleset for each single packet to determine the action to apply to the packet.
The ruleset can be viewed as a list of rules; usually it is processed sequentially and the first matching rule is applied.

The literature agrees on the definition of a single firewall rule.
It consists of a predicate (the match expression) and an action.
If the match expression applies to a packet, the action is performed.
%
%
%
%
Usually, a packet is scrutinized by several rules. 
Zhang~\etal\cite{ZhangAlShaer2007flip} specify a common format for packet filtering rules.
The action is either ``allow'' or ``deny'', which directly corresponds to the firewall's filtering decision.
The ruleset is processed strictly sequentially.
Yuan \etal\cite{fireman2006} call this the \emph{simple list model}.
ITVal also supports calls to user-defined chains as an action.
This allows ``jumping'' within the ruleset without having a final filtering decision yet.
This is called the \emph{complex chain model}~\cite{fireman2006}.
%
%

In general, a packet header is a bitstring which can be matched against~\cite{zhang2012qbfsatfirewalls}.
Zhang~\etal\cite{ZhangAlShaer2007flip} support matching on the following packet header fields: 
IP source and destination address, protocol, and port on layer 4.
This model is commonly found in the literature~\cite{cspfirewall,bartal1999firmato,ZhangAlShaer2007flip,fireman2006,brucker2008modelfwisabelle,brucker.ea:formal-fw-testing:2014}.
%
%
ITVal extends these match conditions with flags (\eg \texttt{TCP SYN}) and connection states (\texttt{INVALID}, \texttt{NEW}, \texttt{ESTABLISHED}, \texttt{RELATED)}.
The state matching is treated as just another match condition.%
\footnote{Firewalls can be stateful or stateless. 
 Most firewalls nowadays are stateful, which means the firewall remembers and tracks information of previously seen packets, \eg the TCP connection a packet belongs to and the state of this connection. 
 ITVal does not track the state of connections. Match conditions on connection states are treated exactly the same as matches on a packet header. 
  In general, focusing on rulesets and not firewall implementation, matching on \iptables{} conntrack states is exactly as matching on any other (stateless) condition. 
 However, internally, not only the packet header is consulted but also the current connection tables. 
 Note that existing firewall analysis tools also largely ignore state~\cite{nelson2010margrave}.
 In our semantics, we also model stateless matching. } 
This model is similar to Margrave's model for IOS~\cite{nelson2010margrave}.
When comparing these features to the simple firewall in Figure~\ref{fig:firewall:synology}, it becomes obvious that none of these tools supports that firewall.

We are not the first to propose simplifying firewall rulesets to enable subsequent analysis. 
Brucker \etal\cite{brucker.ea:formal-fw-testing:2014,brucker2010firewalltransform,UPF_Firewall-AFP} provide algorithms to transform firewall rulesets to enable the testing of a firewall. 
They prove correctness of their transformations in Isabelle/HOL. 
Unfortunately, their supported firewall model is very limited, even more limited than the model used by the tools presented in the previous paragraph. 
They support the simple list model and are restricted to match only on networks, ports, and protocols. 
Instead of natively supporting IP addresses, they support the notion of networks which imposes additional constraints: 
While it is natural for IP address ranges in a ruleset to overlap (\eg 10.0.0.0/8 and 10.42.42.0/24), their notion of networks requires that all different networks in a ruleset must be distinct~\cite{brucker2010firewalltransform,brucker.ea:formal-fw-testing:2014}. 
Their work is only evaluated on three very small real-world rulesets (less than 15 rules each) of the same institute.

We are not aware of any tool which uses a model fundamentally different than those described in the previous paragraphs. 
Our model enhances existing work in that we use ternary logic to support arbitrary match conditions.
To analyze a large \iptables{} firewall, the authors of Margrave~\cite{nelson2010margrave} translated it to basic Cisco IOS access lists~\cite{ciscofirewallaccesslists} by hand.
With our simplification, we can automatically remove all features not understood by basic Cisco IOS.
This enables translation of any \iptables{} firewall to a basic Cisco access lists which is guaranteed to drop no more packets than the original \iptables{} firewall. 
This opens up all tools available only for Cisco IOS access lists, \eg Margrave \cite{nelson2010margrave} and Header Space Analysis~\cite{kazemian2012HSA}.%
\footnote{Note that the other direction is considered easy~\cite{ciscotoiptables}, because basic Cisco IOS access lists have ``no nice features''~\cite{ciscononicefeatures}. Also note that there also are \emph{Advanced} Access Lists. }

%


\section{Semantics of \iptables{}}
\label{sec:fm15:iptables-semantics}



We formalized the semantics of a subset of \iptables{}.
The semantics focuses on access control, which is done in the \texttt{INPUT}, \texttt{OUTUT}, and \texttt{FORWARD} chain of the \texttt{filter} table. 
Thus packet modification (\eg NAT) is not considered (and also not allowed in these chains).

Match conditions, \eg \texttt{source 192.168.0.0/24} and \texttt{protocol TCP}, are called \emph{primitives}.
A primitive matcher $\gamma$ decides whether a packet matches a primitive.
Formally, based on a set $X$ of primitives and a set of packets $P$, a primitive matcher $\gamma$ is a binary relation over $X$ and $P$.
The semantics supports arbitrary packet models and match conditions, hence both remain abstract in our definition.

In one firewall rule, several primitives can be specified.
Their logical connective is conjunction, for example $\verb~src 192.168.0.0/24~ \ \mathit{and} \  \verb~tcp~$.
Disjunction is omitted because it is neither needed for the formalization nor supported by \iptables{}; this is consistent with the model by Jeffrey and Samak~\cite{fwmodelchecking2009jeffry}.
Primitives can be combined in an algebra of \emph{match expressions} $M_X$:
\begin{IEEEeqnarray*}{c}
    \mathit{mexpr} \quad = \quad  x \quad \mctrl{for}\ x \in X 
                   \quad | \quad  \mconstr{\neg}\, \mathit{mexpr} 
                   \quad | \quad \mathit{mexpr} \ \mconstr{\wedge}\ \mathit{mexpr} 
                   \quad | \quad \mconstr{True}
\end{IEEEeqnarray*}
For a primitive matcher $\gamma$ and a match expression $m \in M_X$, we write \mbox{$\matches{m}$} if a packet $p \in P$ matches $m$, essentially lifting $\gamma$ to a relation over $M_X$ and $P$, with the connectives defined as usual.
With completely generic $P$, $X$, and $\gamma$, the semantics can be considered to have access to an oracle which understands all possible match conditions.

Furthermore, we support the following \emph{actions}, modeled closely after \iptables{}: $\iptaction{Accept}$, $\iptaction{Reject}$, $\iptaction{Drop}$, $\iptaction{Log}$, $\iptaction{Empty}$, $\iptaction{Call}\ c$ $\text{for a chain $c$}$, and $\iptaction{Return}$.
A \emph{rule} can be defined as a tuple $(m,\,a)$ for a match expression $m$ and an action $a$.
A list (or sequence) of rules is called a \emph{chain}.
For example, the beginning of the \verb~DOS_PROTECT~ chain in Figure~\ref{fig:firewall:synology} is $[(\verb~icmp~ \wedge \verb~icmptype 8 limit:~\,\dots,\: \iptaction{Return}),\: \dots]$
.

A set of chains associated with a name is called a \emph{ruleset}.
Let $\Gamma$ denote the mapping from chain names to chains.
For example, $\Gamma \ \verb~DOS_PROTECT~$ returns the contents of the \verb~DOS_PROTECT~ chain.
We assume that $\Gamma$ is well-formed that means, if a $\iptaction{Call}\ c$ action occurs in a ruleset, then the chain named $c$ is defined in $\Gamma$.
This assumption is justified as the Linux kernel only accepts well-formed rulesets.


\begin{figure}[htb]
\centering
  \begin{mathpar}
    \textsc{Skip}\quad\inferrule{ }{\bigstep{[]}{t}{t}} \and
    \textsc{Accept}\quad\inferrule{\matches{m}}{\bigstep{[(m,\:\iptaction{Accept})]}{\undecided}{\allow}} \and
    \textsc{Drop}\quad\inferrule{\matches{m}}{\bigstep{[(m,\:\iptaction{Drop})]}{\undecided}{\deny}} \and
    \textsc{Reject}\quad\inferrule{\matches{m}}{\bigstep{[(m,\:\iptaction{Reject})]}{\undecided}{\deny}} \and
    \textsc{NoMatch}\quad\inferrule{\nmatches{m}}{\bigstep{[(m,\:a)]}{\undecided}{\undecided}} \and
    \textsc{Decision}\quad\inferrule{t \neq \undecided}{\bigstep{\mvar{rs}}{t}{t}} \and
    \textsc{Seq}\quad\inferrule{\bigstep{\mvar{rs}_1}{\undecided}{t} \\ \bigstep{\mvar{rs}_2}{t}{t'}}{\bigstep{\mvar{rs}_1 \lstapp \mvar{rs}_2}{\undecided}{t'}} \and
    \textsc{CallResult}\quad\inferrule{\matches{m} \\ \\ \bigstep{\Gamma\ c}{\undecided}{t}}{\bigstep{[\left(m,\: \iptaction{Call}\ c\right)]}{\undecided}{t}}\and
    \textsc{CallReturn}\quad\inferrule{\matches{m} \\ \Gamma\; c = \mvar{rs}_1 \lstapp (m',\: \iptaction{Return}) \lstcons \mvar{rs}_2  \\\\ \matches{m'} \\ \bigstep{\mvar{rs}_1}{\undecided}{\undecided}}{\bigstep{[\left(m,\: \mathtt{Call}\ c\right)]}{\undecided}{\undecided}} \and
    \textsc{Log}\quad\inferrule{\matches{m}}{\bigstep{[(m,\:\iptaction{Log})]}{\undecided}{\undecided}} \and
    \textsc{Empty}\quad\inferrule{\matches{m}}{\bigstep{[(m,\:\iptaction{Empty})]}{\undecided}{\undecided}}
  \end{mathpar}
\vskip1ex
  \hfill (for any primitive matcher $\gamma$ and any well-formed ruleset $\Gamma$)
  \caption{Big Step semantics for \iptables{}}
  \label{fig:semantics}
\end{figure}

The semantics of a firewall w.r.t.\ to a given packet $p$, a background ruleset $\Gamma$, and a primitive matcher $\gamma$ can be defined as a relation over the currently active chain and the state before and the state after processing this chain.
The semantics is specified in Figure~\ref{fig:semantics}.\footnote{inductive iptables-bigstep} 
On the left side of the turnstile ($\vdash$), the constants $\Gamma$,$\gamma$,$p$ are written. 
The expression $\bigstep{\mvar{rs}}{t}{t'}$ states that starting with state $t$, after processing the chain $\mvar{rs}$, the resulting state is $t'$.
For a packet $p$, our semantics focuses on firewall filtering decisions.
Therefore, only the following three states are necessary:	
The firewall may allow ($\allow$) or deny ($\deny$) the packet, or it may not have come to a decision yet~($\undecided$). 

We will now discuss the most important rules. 
The \textsc{Accept} rule describes the following: if the packet $p$ matches the match expression $m$, then the firewall with no filtering decision ($\undecided$) processes the singleton chain $[(m,\:\iptaction{Accept})]$ by switching to the allow state.
Both the \textsc{Drop} and \textsc{Reject} rules deny a packet; the difference is only in whether the firewall generates some informational message, which does not influence filtering. 
The \textsc{NoMatch} rule specifies that if the firewall has not come to a filtering decision yet, it can process any non-matching rule without changing its state.
The \textsc{Decision} rule specifies that as soon as the firewall made a filtering decision, it does not change its decision.
The \textsc{Seq} rule specifies that if the firewall has not come to a filtering decision and it processes the chain $\mvar{rs}_1$ which results in state $t$ and starting from $t$ processes the chain $\mvar{rs}_2$ which results in state $t'$, then both chains can be processed sequentially, ending in state $t'$. 
The \textsc{CallResult} rule specifies that if a matching $\iptaction{Call}$ to a chain named ``$c$'' occurs, the resulting state $t$ is the result of processing the chain $\Gamma\; c$.
Likewise, the \textsc{CallReturn} rule specifies that if processing a prefix $\mvar{rs}_1$ of the called chain does not lead to a filtering decision and directly afterwards, a matching $\iptaction{Return}$ rule occurs, the called chain is processed without result.%
\footnote{The semantics gets stuck if a $\iptaction{Return}$ occurs on top-level.
However, this is not a problem since we make sure that this cannot happen.
\iptables{} specifies that a $\iptaction{Return}$ on top-level in a built-in chain is allowed and in this corner case, the chain's default policy is executed. 
To comply with this behavior, we always start analysis of a ruleset as follows: 
$[(\mdef{True},\: \iptaction{Call}\ \mvar{start\mhyphen{}chain}),\allowbreak{}\: (\mdef{True},\: \mvar{default\mhyphen{}policy})]$, where the start chain is one of \iptables{}' built-in \texttt{INPUT}, \texttt{FORWARD}, or \texttt{OUTPUT} chains with a default policy of either $\iptaction{Accept}$ or $\iptaction{Drop}$. }
The \textsc{Log} rule does not influence the filtering behavior.
Similarly, the \textsc{Empty} rule does not result in a filtering decision.
An \textsc{Empty} rule, \ie a rule without an action, occurs if \iptables{} only updates its internal state, \eg updating packet counters. 

\paragraph{Model Limitations and Stateful Matchers}
\label{par:fm15:stateful-limitations}
Our primitive matcher is completely stateless: $\gamma :: \left(\mathit{X} \Rightarrow \mathit{packet} \Rightarrow \mathbb{B}\right)$. 
However, \iptables{} also allows stateful operations, such as marking a packet and matching on the marking later on. 

The documentation of \iptables{} distinguishes between match extensions and target extensions. 
Ideally, almost all match extensions can be used as if they were stateless. 
Anything which performs an action should be implemented as target extension, \ie action. 
For example, marking a packet with \texttt{CONNMARK} is an action. 
Matching on a \texttt{CONNMARK} marking is a match condition. 
Our semantics does not support the \texttt{CONNMARK} action. 
This is not a problem since usually, new \texttt{CONNMARK} markings are not set in the \texttt{filter} table. 
However, it is possible to match on existing markings. 
Since our primitive matchers and packets are completely generic, this case can be represented within our model:
Instead of keeping an internal \texttt{CONNMARK} state, an additional ``ghost field'' must be introduced in the packet model. 
Since packets are immutable, this field cannot be set by a rule but the packet must be given to the firewall with the final value of the ghost field already set.
Hence, an analysis must be carried out with the correct value in the ghost fields when the packet is given to the \texttt{filter} table. 
We admit that this model is very unwieldy in general. 
However, for one of the most used stateful modules of iptables, namely connection state tracking with \texttt{conntrack} and \texttt{state}, this model has been proven to be very convenient.\footnote{Semantics-Stateful.thy} 
We will elaborate on stateful connection tracking (which can be fully supported by our semantics) in Section~\ref{subsec:ifip:ctstate}. 
When later embedding into the more practical ternary semantics, all further stateful primitives (such as \texttt{CONNMARK}) can be considered ``unknown'' and are correctly abstracted by these semantics.

What if a match extension maintains an internal state and changes its behavior on every invocation? 
Ideally, due to usability, \iptables{} match extensions should not exhibit this behavior; however, the \texttt{recent} module or the \texttt{connbytes} exhibit similar behavior. 
This means, the tautology in Boolean logic ``$a \wedge \neg a = \mdef{False}$'' does not hold if $a$ is a module which updates an internal state and its matching behavior after every invocation. 
Therefore, one might argue that our \iptables{} model can only be applied to stateless match conditions. 
If we add some state $\sigma$ and updated state $\sigma'$ to the match condition, the formula ``$a_\sigma \wedge \neg a_{\sigma'}$'' now correctly represents stateful match conditions. 
Therefore, it is only wrong to perform equality operations on stateful match conditions but not to model stateful match conditions with a specific, fixed state. 
In our implementation, we immediately embed everything in ternary logic and treat all stateful primitives as ``unknown''; 
more precisely, we only perform simplification on primitives which are definitely stateless. 
This prevents this error and yields ``$a \wedge \neg a = \mdef{Unknown}$'', which is a correct model since we do not know about a potential internal state of some arbitrary match condition $a$. 
To additionally convince the reader about the soundness of our approach, it would be possible to adapt the parser to give a unique ID to every primitive which is not known to be stateless. 
This unique ID represents the internal state of that particular match condition on that particular position in a ruleset. 
It prevents equality operations between multiple invocations of a stateful match condition. 
It does not change any of our algorithms since we immediately embed all these primitives in ternary logic and treat them as ``unknown''. 

For future work, if we want to consider \eg the \texttt{raw} or \texttt{mangle} table with its extended set of actions or OpenFlow with its full set of actions, a semantics needs to be designed with a mutable packet model. 


\paragraph*{Analysis and Use of the Semantics}

The subsequent parts of this chapter are all based on these semantics.
Whenever we provide a procedure $P$ to operate on chains, we proved that the firewall's filtering behavior is preserved, formally:%
\begin{IEEEeqnarray*}{c}
\bigstep{P \ \mvar{rs}}{t}{t'} \quad \text{\textit{iff}} \quad \bigstep{\mvar{rs}}{t}{t'}
\end{IEEEeqnarray*}%
All our proofs are machine-verified with Isabelle.
Therefore, once the reader is convinced of the semantics as specified in Figure~\ref{fig:semantics}, the correctness of all subsequent theorems follows automatically -- without any hidden assumptions or limitations. 

The rules in Figure~\ref{fig:semantics} are designed such that every rule can be inspected individually.
However, considering all of them together, it is not immediately clear whether the result depends on the order of their application to a concrete ruleset and packet.
Theorem~\ref{thm:bigstep-deterministic} states that the semantics is deterministic, \ie only one uniquely defined outcome is possible.\footnote{iptables-bigstep-deterministic}

\begin{theorem}[Determinism]%
\label{thm:bigstep-deterministic}%
\begin{IEEEeqnarray*}{c}
\text{If} \quad \bigstep{\mvar{rs}}{s}{t} \quad \text{and} \quad \bigstep{\mvar{rs}}{s}{t'} \quad \text{then} \quad t=t'
\end{IEEEeqnarray*}
\end{theorem}

Next, we show that the semantics are actually defined, \ie there is always a decision for any packet and ruleset.\footnote{semantics-bigstep-defined} 
We assume that the ruleset does not have an infinite loop and that all chains which are called exist in the background ruleset. 
These conditions are checked by the Linux kernel and can thus be assumed. 
The way we start any analysis (namely $[(\mdef{True},\: \iptaction{Call}\ \mvar{start\mhyphen{}chain}),\allowbreak{}\: (\mdef{True},\: \mvar{default\mhyphen{}policy})]$, where the  default policy is either $\iptaction{Accept}$ or $\iptaction{Drop}$) directly implies that we cannot have a top-level $\iptaction{Return}$. 
In addition, we assume that only the actions defined in Figure~\ref{fig:semantics} occur in the ruleset; our parser rejects everything else.

\begin{theorem}[Defined]%
\label{thm:bigstep-defined}%
If the caller-callee relation is well-founded and finite (\ie there are no infinite calling loops) and $\Gamma$ is well-formed (\ie all chain names which are called are defined) and there is no $\iptaction{Return}$ on top-level and all actions are supported by the semantics, then
\begin{IEEEeqnarray*}{c}
\exists t.\ \ \bigstep{\mvar{rs}}{s}{t}
\end{IEEEeqnarray*}
\end{theorem}

To also assert \emph{empirically} that we only allow analysis of iptables rulesets which are defined according to our semantics, we always check the preconditions of Theorem~\ref{thm:bigstep-defined} at runtime when our tool loads a ruleset. 
First, we can statically verify that $\Gamma$ is well-formed by verifying that all chain names which are referenced in an action are defined and that no unsupported actions occur. 
Next, our tool verifies that there are no infinite loops by unfolding the ruleset (see next section) only a finite but sufficiently large number of times and abort if the ruleset is not in the proper form afterwards. 
For all real-world firewalls we have analyzed, these conditions were almost never violated. 
They were only violated for a negligible quantity of firewalls which used very special iptables actions\footnote{For example setting \iptaction{CONNMARK} in the \texttt{filter} table where it was not necessary, redirecting packets to userspace with \iptaction{NFQUEUE} where we do not know how the userspace application handles them, or specialized loggings such as \iptaction{NFLOG} which is technically equivalent to the \iptaction{LOG} semantics and could directly be supported.} not supported by our semantics or special hand-crafted firewalls which deliberately violate a property and which are also rejected by the Linux kernel. 

\section{Custom Chain Unfolding}
\label{sec:fm15:unfolding}
In this section, we present algorithms to convert a ruleset from the complex chain model to the simple list model.

The function $\mfun{pr}$ (``process return'') iterates over a chain.
If a $\iptaction{Return}$ rule is encountered, all subsequent rules are amended by adding the $\iptaction{Return}$ rule's negated match expression as a conjunct.
Intuitively, if a $\iptaction{Return}$ rule occurs in a chain, all following rules of this chain can only be reached if the $\iptaction{Return}$ rule does not match.
\begin{IEEEeqnarray*}{lCl}
  \mfun{add\mhyphen{}match} \ m' \ \mvar{rs} & = & [(m \wedge m',\,a).\ (m,\,a) \leftarrow \mvar{rs}] \\
  \mfun{pr} \ [] & = & [] \\
  \mfun{pr} \ ((m,\, \iptaction{Return}) \lstcons \mvar{rs}) & = & \mfun{add\mhyphen{}match} \ (\neg m) \ (\mfun{pr} \ \mvar{rs})\\
  \mfun{pr} \ ((m,\, a) \lstcons \mfun{rs}) & = & (m,\, a) \lstcons \mfun{pr} \ \mvar{rs}
\end{IEEEeqnarray*}

The function $\mfun{pc}$ (``process call'') iterates over a chain, unfolding one level of $\iptaction{Call}$ rules.
If a $\iptaction{Call}$ to the chain $c$ occurs, the chain itself (\ie $\Gamma \ c$) is inserted instead of the $\iptaction{Call}$.
However, $\iptaction{Return}$s in the chain need to be processed and the match expression for the original $\iptaction{Call}$ needs to be added to the inserted chain.
\begin{IEEEeqnarray*}{lCl}
  \mfun{pc} \ [] & = & []\\
  \mfun{pc} \ ((m,\, \iptaction{Call}\ c) \lstcons \mvar{rs}) & = & \mfun{add\mhyphen{}match} \  m \ \left(\mfun{pr} \ \left(\Gamma \ c\right)\right) \lstapp \mfun{pc} \ \mvar{rs} \\
  \mfun{pc} \ ((m,\, a) \lstcons \mvar{rs}) & \ = \ & (m,\, a) \lstcons \mfun{pc} \ \mvar{rs}
\end{IEEEeqnarray*}

The procedure $\mfun{pc}$ can be applied arbitrarily many times and preserves the semantics.\footnote{unfolding-n-sound-complete} 
It is sound and complete.
\begin{theorem}[Soundness and Completeness]%
\label{thm:unfolding-sound-complete}%
\begin{IEEEeqnarray*}{c}
   \bigstep{\mfun{pc}^\mvar{n} \ \mvar{rs}}{t}{t'} \quad  \text{iff} \quad  \bigstep{\mvar{rs}}{t}{t'}
\end{IEEEeqnarray*}
\end{theorem}

In each iteration, the algorithm unfolds one level of $\iptaction{Call}$s.
The algorithm needs to be applied until the result no longer changes.
Note that the semantics allows non-terminating rulesets; however, the only rulesets that are interesting for analysis are the ones actually accepted by the Linux kernel.\footnote{The relevant check is in \texttt{mark\_source\_chains}, file \url{source/net/ipv4/netfilter/ip_tables.c} of the Linux kernel version 3.2.} 
%
Since it rejects rulesets with loops, both our algorithm and the resulting ruleset are guaranteed to terminate.

\begin{corollary}
Every ruleset (with only $\iptaction{Accept}$, $\iptaction{Drop}$, $\iptaction{Reject}$, $\iptaction{Log}$, $\iptaction{Empty}$, $\iptaction{Call}$, $\iptaction{Return}$ actions) accepted by the Linux kernel can be unfolded completely while preserving its filtering behavior.
\end{corollary}

In addition to unfolding calls, the following transformations applied to any ruleset preserve the semantics:
\begin{itemize}
  \item 
    Replacing $\iptaction{Reject}$ actions with $\iptaction{Drop}$ actions,\footnote{iptables-bigstep-rw-Reject} 
  \item 
    Removing $\iptaction{Empty}$ and $\iptaction{Log}$ rules,\footnote{iptables-bigstep-rm-LogEmpty} 
  \item 
    Simplifying match expressions which contain $\mdef{True}$ or $\neg\,\mdef{True}$.\footnote{unfold-optimize-ruleset-CHAIN} 
   \item 
   For some given primitive matcher, specific optimizations may also be performed,\footnote{unfold-optimize-ruleset-CHAIN}  \eg rewriting \texttt{src 0.0.0.0/0} to $\mdef{True}$.
\end{itemize}

Therefore, after unfolding and optimizing, a chain which only contains $\iptaction{Allow}$ or $\iptaction{Drop}$ actions is left.
In the subsequent sections, we require this as a precondition.
As an example, recall the firewall in Figure~\ref{fig:firewall:synology}.
Its \verb~INPUT~ chain after unfolding and optimizing is listed in Figure~\ref{fig:firewall:synology-unfolded}.
%
%
Observe that the computed match expressions are beyond iptable's expressiveness.
An algorithm to normalize the rules to an \iptables{}-compatible format will be described in Section~\ref{sec:fm15:normalization}.

\begin{figure}[t]%
\begin{IEEEeqnarray*}{l}%
[           \left(\neg\,\left(\verb~icmp~ \wedge \verb~icmptype 8 limit:~\dots\right) \,\wedge \verb~icmp~ \wedge \verb~icmptype 8~,\, \iptaction{Drop}\right),\;\\
\phantom{[} (\neg\,\left(\verb~icmp~ \wedge \verb~icmptype 8 limit:~\dots\right) \,\wedge \neg\,\left(\verb~tcp~ \wedge \verb~tcp flags:0x17/0x04 limit:~ \dots\right) \,\wedge \\
\phantom{[( } \verb~tcp~ \wedge \verb~tcp flags:0x17/0x04~,\,  \iptaction{Drop}),\; \dots, \;
\phantom{[} \left(\verb~src 192.168.0.0/16~,\, \iptaction{Accept}\right),\; \dots ]%
\end{IEEEeqnarray*}%
\vskip-10pt 
\caption{Unfolded Synology Firewall}%
  \label{fig:firewall:synology-unfolded}%
\end{figure}

\section{Unknown Primitives}
\label{sec:fm15:ternary}
As we argued earlier, it is infeasible to support all possible primitives of a firewall.
Suppose a new firewall module is created which provides the \verb~ssh_blacklisted~ and \verb~ssh_innocent~ primitives. 
The former applies if an IP address has had too many invalid SSH login attempts in the past; 
the latter is the opposite of the former.
Since we made up these primitives, no existing tool will support them. 
However, a new version of \iptables{} could implement them or they can be provided as third-party kernel modules. 
Therefore, our ruleset transformations must take unknown primitives into account. 
To achieve this, we lift the primitive matcher $\gamma$  to ternary logic, adding $\mdef{Unknown}$ as matching outcome. 
We embed this new ``approximate'' semantics into the semantics described in the previous sections.
Thus, it becomes easier to construct matchers tailored to the primitives supported by a particular tool.

\subsection{Ternary Matching}

Logical conjunction and negation on ternary values are as before, with these additional rules for $\mdef{Unknown}$ operands (commutative cases omitted):
\begin{IEEEeqnarray*}{c}
    \mdef{True}  \wedge  \mdef{Unknown} = \mdef{Unknown} \ \quad
    \mdef{False}  \wedge  \mdef{Unknown} = \mdef{False} \ \quad
                      \neg \: \mdef{Unknown} = \mdef{Unknown}
\end{IEEEeqnarray*}
These rules correspond to Kleene's 3-valued logic~\cite{kleene1952introduction} and are well-suited for firewall semantics: 
The first equation states that, if one condition matches, the final result only depends on the other condition.
The next equation states that a rule cannot match if one of its conditions does not match.
Finally, by negating an unknown value, no additional information can be inferred.

\begin{sloppypar}
We demonstrate this by example: 
the two rulesets
  $\left[(\texttt{ssh\_blacklisted},\, \iptaction{Drop})\right]$
and 
$ \left[(\mdef{True},\, \iptaction{Call}\ c)\right]$
  where $\Gamma\, c\, = [(\mathtt{ssh\_innocent},\, \iptaction{Return}),\, (\mdef{True},\, \iptaction{Drop})]$
have exactly the same filtering behavior. 
After unfolding, the second ruleset collapses to
 $ \left[(\neg\; \texttt{ssh\_innocent},\, \iptaction{Drop})\right]$. 
Both the \texttt{ssh\_blacklisted} and the \texttt{ssh\_innocent} primitives are $\mdef{Unknown}$ to our matcher. 
Thus, since both rulesets have the same filtering behavior, a packet matching $\mdef{Unknown}$ in the first ruleset should also match $\neg\;\mdef{Unknown}$ in the second ruleset. 
\end{sloppypar}


%

\subsection{Closures}
In the ternary semantics, it may be unknown whether a rule applies to a packet.
Therefore, the matching semantics are extended with an \emph{``in-doubt''-tactic}.
This tactic is consulted if the result of a match expression is $\mdef{Unknown}$.
It decides whether a rule applies.

We introduce the \emph{in-doubt-$\mathit{allow}$} and \emph{in-doubt-$\mathit{deny}$} tactics.
The first tactic forces a match if the rule's action is $\iptaction{Accept}$ and a mismatch if it is $\iptaction{Drop}$.
The second tactic behaves in the opposite manner.
Note that an unfolded ruleset is necessary, since no behavior can be specified for $\iptaction{Call}$ and $\iptaction{Return}$ actions.\footnote{The final decision ($\allow$ or $\deny$) for $\iptaction{Call}$ and $\iptaction{Return}$ rules depends on the called/calling chain.}

We denote the exact Boolean semantics with ``$\Rightarrow$'' and embedded ternary semantics with an arbitrary tactic $\alpha$ with ``$\Rightarrow_\alpha$''.
In particular, $\alpha = \mathit{allow}$ for \emph{in-doubt-$\mathit{allow}$} and $\alpha = \mathit{deny}$ analogously.

``$\Rightarrow$'' and ``$\Rightarrow_\alpha$'' are related to the in-doubt-tactics as follows: 
considering the set of all accepted packets, \emph{in-doubt-$\mathit{allow}$} is an overapproximation, whereas \emph{in-doubt-$\mathit{deny}$} is an underapproximation.
In other words, if ``$\Rightarrow$'' accepts a packet, then ``$\Rightarrow_{\operatorname{allow}}$'' also accepts the packet.
Thus, from the opposite perspective, the \emph{in-doubt-$\mathit{allow}$} tactic can be used to guarantee that a packet is certainly dropped.
Likewise, if ``$\Rightarrow$'' denies a packet, then ``$\Rightarrow_{\operatorname{deny}}$'' also denies this packet.
Thus, the \emph{in-doubt-$\mathit{deny}$} tactic can be used to guarantee that a packet is certainly accepted.

For example, the unfolded firewall of Figure~\ref{fig:firewall:synology} contains rules which drop a packet if a limit is exceeded.
If this rate limiting is not understood by $\gamma$, the \emph{in-doubt-$\mathit{allow}$} tactic will never apply this rule, while with the \emph{in-doubt-$\mathit{deny}$} tactic, it is applied universally.

We say that the Boolean and the ternary matchers agree iff they return the same result or the ternary matcher returns $\mdef{Unknown}$.
Interpreting this definition, the ternary matcher may always return $\mdef{Unknown}$ and the Boolean matcher serves as an oracle which knows the correct result.
Note that we never explicitly specify anything about the Boolean matcher; therefore the model is universally valid, \ie the proof holds for an arbitrary oracle.

If the exact and ternary matcher agree, then the set of all packets allowed by the \emph{in-doubt-$\mathit{deny}$} tactic is a subset of the packets allowed by the exact semantics, which in turn is a subset of the packets allowed by the \emph{in-doubt-$\mathit{allow}$} tactic.\footnote{FinalAllowClosure} 
Therefore, we call all packets accepted by $\Rightarrow_{\operatorname{deny}}$ the \emph{lower closure}, \ie the semantics which accepts at most the packets that the exact semantics accepts.
Likewise, we call all packets accepted by $\Rightarrow_{\operatorname{allow}}$ the \emph{upper closure}, \ie the semantics which accepts at least the packets that the exact semantics accepts.
Every packet which is not in the upper closure is guaranteed to be dropped by the firewall.

\begin{theorem}[Lower and Upper Closure of Allowed Packets]%
\label{thm:FinalAllowClosure}%
\begin{IEEEeqnarray*}{c}
  \left\lbrace p.\ \bigstepapprox{\mvar{rs}}{\undecided}{\allow}{\operatorname{deny}} \right\rbrace \\
  \subseteq\\
  \left\lbrace p.\ \bigstep{\mvar{rs}}{\undecided}{\allow} \right\rbrace \\
  \subseteq\\
  \left\lbrace p.\ \bigstepapprox{\mvar{rs}}{\undecided}{\allow}{\operatorname{allow}} \right\rbrace
\end{IEEEeqnarray*}
\end{theorem}

The opposite holds for the set of denied packets.\footnote{FinalDenyClosure} 

For the example in Figure~\ref{fig:firewall:synology}, we computed the closures (without the \texttt{RELATED\hskip-1.2pt{},\allowbreak{}ESTABLISHED} rule, see Section~\ref{sec:fm15:establishedrule}) and a ternary matcher which only understands IP addresses and layer 4 protocols.
The lower closure is the empty set since rate limiting could apply to any packet.
The upper closure is the set of packets originating from $192.168.0.0/16$.

\subsection{Removing Unknown Matches}
In this section, as a final optimization, we remove all unknown primitives.
We call this algorithm $\mfun{pu}$ (``process unknowns'').
For this step, the specific ternary matcher and the choice of the in-doubt-tactic must be known.

In every rule, top-level unknown primitives can be rewritten to $\mdef{True}$ or $\neg\, \mdef{True}$. 
For example, let $m_u$ be a primitive which is unknown to $\gamma$. Then, for in-doubt-allow, $(m_u,\, \iptaction{Accept})$ is equal to $(\mdef{True},\, \iptaction{Accept})$ and $(m_u,\, \iptaction{Drop})$ is equal to $(\neg\,\mdef{True},\, \iptaction{Drop})$. 
Similarly, negated unknown primitives and conjunctions of (negated) unknown primitives can be rewritten.

Hence, the base cases of $\mfun{pu}$ are straightforward. 
However, the case of a negated conjunction of match expressions requires some care. 
The following equation represents the De Morgan rule, specialized to the in-doubt-allow tactic. 
\begin{IEEEeqnarray*}{l}
\mfun{pu} \ (\neg\,(m_1 \wedge m_2),\; a) \quad = \quad
  \begin{cases}
    \mdef{True} & \qquad\hfill \mctrl{if}\ \mfun{pu} \ (\neg\,m_1,\; a) = \mdef{True}  \phantom{\neg\,} \\
    \mdef{True} & \qquad\hfill \mctrl{if}\ \mfun{pu}  \ (\neg\,m_2,\; a) = \mdef{True} \phantom{\neg\,} \\
    \mfun{pu} \ (\neg\,m_2,\; a) & \qquad\hfill \mctrl{if}\ \mfun{pu}  \ (\neg\,m_1,\; a) = \neg\,\mdef{True} \\
    \mfun{pu} \ (\neg\,m_1,\; a) & \qquad\hfill \mctrl{if}\ \mfun{pu}  \ (\neg\,m_2,\; a) = \neg\,\mdef{True} \\
    \multicolumn{2}{l}{ \neg\left(\neg\,\mfun{pu}\; \left(\neg\,m_1,\, a\right) \wedge \neg\,\mfun{pu}\; \left(\neg\,m_2,\, a\right)\right)  \qquad \mctrl{otherwise} }
  \end{cases}
\end{IEEEeqnarray*}

A note about the notation: The algorithm explicitly tests for `$\cdot = \mdef{True}$', since in this context, $\mdef{True}$ is the syntactic base case of a match expression $M_X$.  

The $\neg\;\mdef{Unknown} = \mdef{Unknown}$ equation is responsible for the complicated nature of the De Morgan rule.
Fortunately, we machine-verified all our algorithms.\footnote{transform-remove-unknowns-upper, transform-remove-unknowns-lower} 
For example, during our research, we wrote a seemingly simple (but incorrect) version of $\mfun{pu}$ and everybody agreed that the algorithm looks correct.
In the early empirical evaluation, with yet unfinished proofs, we did not observe our bug.
Only because of the failed correctness proof did we realize that we introduced an equation that only holds in Boolean logic.

\begin{theorem}[Soundness and Completeness]%
\begin{IEEEeqnarray*}{c}
\bigstepapprox{[\textnormal{\mfun{pu}} \ r.\ \ r \leftarrow \ \mvar{rs}]}{t}{t'}{\operatorname{allow}} \quad \text{iff} \quad \bigstepapprox{\mvar{rs}}{t}{t'}{\operatorname{allow}}
\end{IEEEeqnarray*}
\end{theorem}
\begin{theorem}%
Algorithm $\mfun{pu}$ removes all unknown primitive match expressions. 
\end{theorem}

An algorithm for the in-doubt-deny tactic (with the same equation for the De Morgan case) can be specified in a similar way.
Thus, $\Rightarrow_{\alpha}$ can be treated as if it were defined only on Boolean logic with only known match expressions.

\begin{sloppypar}
As an example, we examine the ruleset of the upper closure of Figure~\ref{fig:firewall:synology} (without the \texttt{RELATED\hskip-1.2pt{},\allowbreak{}ESTABLISHED} rule, see Section~\ref{sec:fm15:establishedrule}) for a ternary matcher which only understands IP addresses and layer 4 protocols.
The ruleset is simplified to $[(\verb~src 192.168.0.0/16~,\, \iptaction{Accept}),\allowbreak{}\, (\mdef{True},\, \iptaction{Drop})]$. 
ITVal can now directly compute the correct results on this ruleset.
\end{sloppypar}

\subsection[The RELATED,ESTABLISHED Rule]{The \texttt{RELATED\hskip-1.2pt{},ESTABLISHED} Rule}
\label{sec:fm15:establishedrule}
Since firewalls process rules sequentially, the first rule has no dependency on any previous rules.
Similarly, rules at the beginning have very low dependencies on other rules.
Therefore, firewall rules in the beginning can be inspected manually, whereas the complexity of manual inspection increases with every additional preceding rule.

It is good practice~\cite{iptablesperfectruleset} to start a firewall with an \texttt{ESTABLISHED} (and sometimes \texttt{RELATED}) rule.
This also happens in Figure~\ref{fig:firewall:synology} after the rate limiting.
The \texttt{ESTABLISHED} rule usually matches most of the packets~\cite{iptablesperfectruleset},\footnote{We revalidated this observation in September 2014 and found that in our firewall, which has seen more than 15 billion packets (19+ Terabyte data) since the last reboot, more than 95\% of all packets matched the first \texttt{RELATED\hskip-1.2pt{},\allowbreak{}ESTABLISHED} rule. } which is important for performance; however, when analyzing the filtering behavior of a firewall, it is important to consider how a connection can be brought to this state.
Therefore, we remove this rule and only focus on the connection setup. 

The \texttt{ESTABLISHED} rule essentially allows packet flows in the opposite direction of all subsequent rules (cf.\ Chapter~\ref{chap:esss14}). 
Unless there are special security requirements (which is not the case in any of our analyzed scenarios), the \texttt{ESTABLISHED} rule can be excluded when analyzing the connection setup~(Corollary~\ref{esss:corollary1}).\footnote{%
The same can be concluded for reflexive ACLs in Cisco's IOS Firewall~\cite{ciscofirewallaccesslists}.} 
If the \texttt{ESTABLISHED} rule is removed and in the subsequent rules, for example, a primitive \texttt{state NEW} occurs, our ternary matcher returns \texttt{Unknown}.
The closure procedures handle these cases automatically, without the need for any additional knowledge.

In Section~\ref{subsec:ifip:ctstate}, we will describe our improvements which will enable support for conntrack state. 
There will no longer be the need to manually exclude rules. 
In short, we will fully support matches on conntrack state such as \texttt{ESTABLISHED} or \texttt{NEW}. 
The observation and argument of this section remains: for access control analysis, we focus on \texttt{NEW} packets. 

\section{Normalization}
\label{sec:fm15:normalization}
Ruleset unfolding may result in non-atomic match expressions, \eg $\neg\:(a \wedge b)$.
\iptables{} only supports match expressions in \emph{Negation Normal Form} (NNF).\footnote{
Since match expressions do not contain disjunctions, any match expression in NNF is trivially also in \emph{Disjunctive Normal Form} (DNF). }
There, a negation may only occur before a primitive, not before compound expressions.
For example, $\neg\:(\verb~src ~\mathit{ip}) \:\wedge\: \mathtt{tcp}$ is a valid NNF formula, whereas $\neg\:\left(\left(\verb~src ~\mathit{ip}\right) \:\wedge\: \mathtt{tcp}\right)$ is not. 
The reason is that \iptables{} rules are usually specified on the command line and each primitive is an argument to the \texttt{iptables} command, for example $\verb~! --src ~\mathit{ip}\verb~  -p tcp~$. 
We normalize match expressions to NNF, using the following observations:

The De Morgan rule can be applied to match expressions, splitting one rule into two.
For example, $\left[\left(\neg\;(\verb~src ~\mathit{ip} \;\wedge\; \verb~tcp~),\, \iptaction{Allow}\right)\right]$ and $[(\neg\;\verb~src ~\mathit{ip},\; \iptaction{Allow}),\allowbreak \, (\neg\; \verb~tcp~,\, \iptaction{Allow})]$ are equivalent. 
%
This introduces a ``meta-logical'' disjunction consisting of a sequence of consecutive rules with a shared action.
For example, $[(m_1,\, a),\; (m_2,\, a)]$ is equivalent to $[(m_1 \vee m_2,\, a)]$. 

For sequences of rules with the same action, a distributive law akin to common Boolean logic holds.
For example, the conjunction of the two rulesets $[(m_1,\, a),\allowbreak{}\; (m_2,\, a)]$ and $[(m_3,\, a),\allowbreak{}\; (m_4,\, a)]$ is equivalent to the ruleset $[({m_1 \wedge m_3},\allowbreak{}\, a),\allowbreak{}\; ({m_1 \wedge m_4},\, a),\; (m_2 \wedge m_3,\, a),\; (m_2 \wedge m_4,\, a)]$.
This can be illustrated with a situation where $a = \iptaction{Accept}$ and a packet needs to pass two firewalls in a row.

We can now construct a procedure which converts a rule with a complex match expression to a sequence of rules with match expressions in NNF. 
It is independent of the particular primitive matcher and the in-doubt tactic used. 
The algorithm $\mfun{n}$ (``normalize'') of type $M_X \Rightarrow M_X\ \mathit{list}$ is defined as follows:
\begin{IEEEeqnarray*}{lcl}
  \mfun{n} \ \mdef{True} & \ = \ & [\mdef{True}]\\
  \mfun{n} \ (m_1 \wedge m_2) & \ = \ & [x \wedge y.\ \ x \leftarrow \mfun{n}\ m_1,\; y \leftarrow \mfun{n}\ m_2]\\
  \mfun{n} \ (\neg\;(m_1 \wedge m_2)) & \ = \ & \mfun{n}\ (\neg m_1) \ \lstapp \ \mfun{n}\ (\neg m_2)\\
  \mfun{n} \ (\neg\neg m) & \ = \ & \mfun{n}\ m\\
  \mfun{n} \ (\neg \mdef{True}) & \ = \ & []\\
  \mfun{n} \ x & \ = \ & [x] \qquad \text{ for } x \in X\\ 
  \mfun{n} \ (\neg x) & \ = \ & [\neg x] \qquad \text{ for } x \in X
\end{IEEEeqnarray*}
The second equation corresponds to the distributive law, the third to the De Morgan rule.
For example, $\mfun{n}\ \left(\neg\,(\verb~src~\;\mvar{ip} \wedge \verb~tcp~)\right) = \left[\neg\,\verb~src~\;\mvar{ip},\, \neg\, \verb~tcp~\right]$.
The fifth rule states that non-matching rules can be removed completely.

The unfolded ruleset of Figure~\ref{fig:firewall:synology-unfolded}, which consists of $9$ rules, can be normalized to a ruleset of $20$ rules (due to distributivity).
In the worst case, normalization can cause an exponential blowup.
Our evaluation shows that this is not a problem in practice, even for large rulesets. 
This is because rulesets are usually managed manually, which naturally limits their complexity to a level processible by state-of-the-art hardware. 

\begin{theorem}
  \mfun{n} always terminates, all match expressions in the returned list are in NNF, and their conjunction is equivalent to the original expression.\footnote{normalized-nnf-match-normalize-match}
\end{theorem}

We show soundness and completeness w.r.t.\ arbitrary $\gamma$, $\alpha$, and primitives.\footnote{normalize-match-correct}
Hence, it also holds for the Boolean semantics.
In general, proofs about the ternary semantics are stronger (the ternary primitive matcher can simulate the Boolean matcher).\footnote{$\beta_\mathrm{magic}$-approximating-bigstep-fun-iff-iptables-bigstep, LukassLemma}

\begin{theorem}[Soundness and Completeness]%
\label{thm:n-sound-complete-ternary}%
  \begin{IEEEeqnarray*}{c}
    \bigstepapprox{[(m',\, a).\ \ m' \leftarrow\mfun{n}\ m]}{t}{t'}{\alpha} 
    \quad \text{iff} \quad  \bigstepapprox{[(m,\, a)]}{t}{t'}{\alpha}
  \end{IEEEeqnarray*}
\end{theorem}


After having been normalized by $\mfun{n}$, the rules can mostly be fed back to \iptables{}.
For some specific primitives, \iptables{} imposes additional restrictions, \eg that at most one primitive of a type may be present in a single rule. 
For our evaluation, we only need to solve this issue for IP address ranges in CIDR notation~\cite{rfc4632}. 
We introduced and verified another transformation which computes intersection of IP address ranges, which returns at most one range.
This is sufficient to process all rulesets we encountered during evaluation. 
In the following chapters, we show how to support more primitives; the evaluation in this chapter only focuses on IP addresses.

\section{Evaluation}
\label{sec:fm15:evaluation}
In this section, we demonstrate the applicability of our ruleset preprocessing. 
Usually, network administrators are not inclined towards publishing their firewall ruleset because of potential negative security implications. 
For this evaluation, we have obtained approximately $20\textnormal{k}$ real-world rules and the permission to publish them.
In addition to the running example in Figure~\ref{fig:firewall:synology} (a small real-world firewall), we tested our algorithms on four other real-world firewalls.
We put focus on the third ruleset, because it is one of the largest and the most interesting one.

For our analysis, we wanted to know how the firewall partitions the IPv4 space.
Therefore, we used a matcher $\gamma$ which only understands source/destination IP addresses and the layer 4 protocols TCP and UDP.
Our algorithms do not require special processing capabilities, they can be executed within seconds on a common off-the-shelf \SI{4}{\giga\byte} RAM laptop.

\paragraph*{Ruleset 1}
is taken from a Shorewall~\cite{shorewall} firewall, running on a home router, with around 500 rules.
We verified that our algorithms correctly unfold, preprocess, and simplify this ruleset.
We expected to see, in both the upper and lower closure, that the firewall drops packets from private IP ranges.
However, we could not see this in the upper closure and verified that the firewall does indeed not block such packets if their connection is in a certain state.
The administrator of the firewall confirmed this issue and is currently investigating it.

\paragraph*{Ruleset 2}
is taken from a small firewall script found online~\cite{iptablesringofsaturn}.
Although it only contains about 50 rules,  we found that it contains a serious mistake. 
We assume the author accidentally confused \iptables{}' \texttt{-I} (insert at top) and \texttt{-A} (append at tail) options.
We saw this after unfolding, as the firewall allows nearly all packets at the beginning.
Subsequent rules are shadowed and cannot apply.
However, these rules come with a documentation of their intended purpose, such as ``drop reserved addresses'', which highlights the error.   
We verified the erroneous behavior by installing the firewall on our systems.
The author is currently investigating this issue.
Thus, our unfolding algorithm alone can provide valuable insights.

\paragraph*{Ruleset 3 \& 4}
are taken from the main firewall of our lab (Chair of Network Architectures and Services).
One snapshot was taken 2013 with 2800 rules and one snapshot was taken 2014, containing around 4000 rules.
It is obvious that these rulesets have historically grown.
About ten years ago, these two rulesets would have been the largest real-world rulesets ever analyzed in academia~\cite{firwallerr2004}.

We present the analysis results of the 2013 version of the firewall.
Details can be found in the additional material. 
We removed the first three rules. 
The first rule was the \texttt{ESTABLISHED} rule, as discussed in Section~\ref{sec:fm15:establishedrule}.
Our focus was put on the second rule when we calculated the lower closure:  
this rule was responsible for the lower closure being the empty set.
Upon closer inspection of this rule, we realized that it was `dead', \ie it can never apply.
We confirmed this observation by changing the target to a $\iptaction{Log}$ action on the real firewall and could never see a hit of this rule for months.
Due to our analysis, this rule could be removed.
%
The third rule performed SSH rate limiting (a $\iptaction{Drop}$ rule).
We removed this rule because we had a very good understanding of it.
Keeping it would not influence correctness of the upper closure, but lead to a smaller lower closure than necessary.

First, we tested the ruleset with the well-maintained Firewall Builder~\cite{fwbuilder}.
The original ruleset could not be imported by Firewall Builder due to $22$ errors, caused by unknown match expressions.
Using the calculated upper closure, Firewall Builder could import this ruleset without any problems.

Next, we tested ITVal's IP space partitioning query~\cite{marmorstein2006firewall}.
On our original ruleset with 2800 rules, ITVal completed the query with around \SI{3}{\giga\byte} of RAM in around \SI{1}{\minute}.
Analyzing ITVal's debug output, we found that most of the rules were not understood correctly due to unknown primitives.
Thus, the results were spurious.
We could verify this as 127.0.0.0/8, obviously dropped by our firewall, was grouped into the same class as the rest of the Internet.
In contrast, using the upper and lower closure ruleset, ITVal correctly identifies 127.0.0.0/8 as its own class.

We found another interesting result about the ITVal tool: 
The (optimized) upper closure ruleset only contains around 1000 rules and the lower closure only around 500 rules.
Thus, we expected that ITVal could process these rulesets significantly faster.
However, the opposite is the case: ITVal requires more than 10 times the resources (both CPU and RAM, we had to move the analysis to a $>$~\SI{40}{\giga\byte} RAM cluster) to finish the analysis of the closures.
We assume that this is due to the fact that ITVal now understands \emph{all} rules. 
Yet, Chapter~\ref{chap:networking16} will reveal that ITVal still computes wrong results.

\section{Outlook: Verifying OpenFlow Rules}
\label{sec:outlook:verifyopenflow}
OpenFlow~\cite{ofspec10,ofspec15} is a standard for configuring OpenFlow-enabled switches. 
It is usually referred to in the context of  Software-Defined Networking (SDN) and is a hot topic in network management and operations since almost ten years (cf.\ Section~\ref{sec:intro:problemclassification-and-sdn}). 
In this section, we elaborate on our decision to focus on iptables instead of OpenFlow and describe how our results could also contribute to the verification of low-level OpenFlow rulesets. 


This chapter, and in general the complete Part~\ref{part:existing-configs} of this thesis, focuses on the analysis of iptables instead of OpenFlow for several reasons: 
  Despite OpenFlow 1.0~\cite{ofspec10} being available for over five years, compared to iptables, OpenFlow is a relatively young and not very wide-spread technology. 
  In contrast, the iptables firewall is wide-spread, real-world approved, supports a large amount of features, and is in productive use for over a decade. 
  There are also decade-old configurations which utilize a vast amount of features, which are no longer fully understood by the administrator~\cite{diekmanngithubnetnetwork}. 
  As of July 2016, the popular network of stackoverflow~\cite{stackoverflow} serverfault\footnote{\url{https://serverfault.com/}} counts more than a hundred times more questions related to iptables than OpenFlow and the superuser\footnote{\url{https://superuser.com/}} network (also a part of the stackoverflow network) counts even a thousand times more questions related to iptables than OpenFlow. 
  %
  %
  Over the years, iptables has evolved into a system with an enormous amount of (legacy) features. 
  Compared to this, OpenFlow is a relatively young and tidy technology. 
  But we anticipate to see a similar feature-creep over the years, considering \eg Nicira extensions~\cite{openflowniciraextensions2016github} or attempts to enhance OpenFlow with generic FPGAs to add ``exotic funcionality [sic]''~\cite{byma2015openflowcloudfpga}. 
  In a broader context, by extending OpenFlow or one of its stateful, more feature-rich, proposed successors~\cite{bianchi2014openstate}, many iptables features have been already reimplemented on top of it~\cite{2016arXiv161102853P}. 
  
  Our declared goal was to provide scientific methods to understand challenging configurations (as observed in iptables) and evaluate our methodology on complex, real-world, legacy-grown systems. 
  The insights we obtained can also be applied to OpenFlow. 
  In particular, a large portion of Part~\ref{part:existing-configs} focuses on match conditions, \eg abstracting over unknowns, optimizing, rewriting, normalizing, or even replacing interfaces by IP addresses.  
  Our work on match condition can be directly reused in the context of OpenFlow. 
  
  However, iptables is not OpenFlow. 
  In particular, the OpenFlow standard defines a vast amount of actions which can be performed for a packet. 
  In contrast, iptables \texttt{filter}ing primarily uses the two actions \iptaction{ACCEPT} and \iptaction{DROP}. 
  This is because a firewall clearly separates filtering from other network functions, such as packet rewriting. 
  OpenFlow implementations tend to easily mixes those. 
  We have shown how to deal with unknown match conditions but unknown actions are an unsolved problem. 
  We have discussed what would be required for a full OpenFlow semantics . 
  In particular, a mutable packet model (cf.\ paragraph~\ref{par:fm15:stateful-limitations}, \nameref{par:fm15:stateful-limitations}) would be necessary, which our methods do not support. 
  However, there is no technical need for OpenFlow switches to mix packet filtering with other operations. 
  For example, the pipelined OpenFlow Router architecture constructed by Nelson \etal\cite[cf.\ \S\hairspace{}3, Fig.\ 3]{nelson2015exodus} clearly separates packet filtering from packet forwarding and rewriting. 
  In general, using pipeline processing as specified in recent OpenFlow standards~\cite{ofspec15} might be a step forward to separate filtering from forwarding and rewriting. 
  This may also help compilers which produce OpenFlow rules and suffer from a large blow-up which is induced by a cross product over several tables to join rules for different actions into one table~\cite{icfp2015smolkanetkatcompiler}. 
  Such a filtering table implemented by OpenFlow rules without unspecified behavior could be analyzed by our presented methods.

\section{Conclusion}
This work was motivated by the fact that we could not find any tool which helped analyzing our lab's and other firewall rulesets.
Though much related work about firewall analysis exists, all academic firewall models are too simplistic to be applicable to those real-world rulesets.
With the transformations presented in this chapter, they became processable by existing tools.
With only a small amount of manual inspection, we found previously unknown issues in four real-world firewalls.

We introduced an approximation to reduce even further the complexity of real-world firewalls for subsequent analysis.
In our evaluation, we found that the approximation is good enough to provide meaningful results.
In particular, using further tools, we were finally able to provide our administrator with a plausible answer to the question of how our firewall partitions the IP space.

Our transformations can be extended for different firewall configurations. 
A user must only provide a primitive matcher for the firewall match conditions she wishes to support.
Since we use ternary logic, a user can specify ``unknown'' as matching outcome, which makes definition of new primitive matchers very easy.
The resulting firewall ruleset conforms to the simple list model in Boolean logic (\ie the common model found in the literature).

Future work includes increasing the accuracy of the approximation by providing more feature-rich primitive matchers and directly implementing firewall analysis algorithms in Isabelle to formally verify them.
Another planned application is to assist firewall migration between different vendors and migrating legacy firewall systems to new technologies.
In particular, such a migration can be easily prototyped by installing a new firewall in chain with the legacy firewall such that packets need to pass both systems: with the assumption that users only complain if services no longer work, the formal argument in this chapter proves that the new firewall with an upper closure ruleset operates without user complaints. 
A new fast firewall with a lower closure ruleset allows bypassing a slow legacy firewall, probably removing a network bottleneck, without security concerns.

%

\chapter{Certifying Spoofing-Protection of Firewalls}
\label{chap:nospoof}

This chapter is an extended version of the following paper~\cite{diekmann2015cnsm}:
\begin{itemize}
	\item Cornelius Diekmann, Lukas Schwaighofer, and Georg Carle. \emph{Certifying Spoofing-Protection of Firewalls}. In 11th International Conference on Network and Service Management, CNSM, Barcelona, Spain, November 2015.
\end{itemize}

\noindent
The following improvements were added:
\begin{itemize}
	\item One new evaluation section (Section~\ref{subsec:cnsm:eval:japanfw}) was added to the accepted version of the paper which also made it into the published, camera-ready version. 
	\item Two new examples were added to the evaluation. 
	\item Spoofing protection was integrated into the \fffuu{} tool. 
	\item We discuss why the simple firewall model (cf.\ Chaper~\ref{chap:networking16}) is explicitly not used here. 
\end{itemize}

\paragraph*{Statement on author's contributions}
For the original paper, the author of this thesis provided major contributions for the ideas, mathematical formalization, realization, implementation, and proof of the presented algorithm. 
He researched related work, and conducted the evaluation. 
All improvements are the work of the author of this thesis. 

\medskip

\paragraph*{Abstract}
In this chapter, we present an algorithm to certify IP spoofing protection of firewall rulesets. 
It fits the following into the big picture of this thesis: 
Once the spoofing protection of a firewall configuration is verified, interfaces and IP addresses can be related. 
Overall, it allows to identify and name entities by their IP address.  
This will be necessary to construct service matrices in the following chapter. 

\medskip

\section{Introduction}
In firewalls, it is good practice and sometimes an essential security feature to prevent IP address spoofing attacks \cite{rfc2827,rfc3704,tolarisblogdisablerpfilter,bartal1999firmato,marmorstein2005itval,wool2004use}.
The Linux kernel offers reverse path filtering~\cite{kernelrpfilter}, which can conveniently prevent such attacks. 
However, in many scenarios (\eg asymmetric routing~\cite{iptablesperfectruleset}, failover~\cite{tolarisblogdisablerpfilter}, zone-spanning interfaces, or multihoming~\cite{rfc3704}), reverse path filtering must be disabled or is too coarse-grained (\eg for filtering special purpose addresses~\cite{rfc6890}).
In these cases, spoofing protection has to be provided by the firewall which is configured by the administrator.
Unfortunately, writing firewall rules is prone to human error, a fact known for over a decade~\cite{firwallerr2004}.

For Linux netfilter/iptables firewalls~\cite{iptables}, we present an algorithm to certify spoofing protection of a ruleset.
It provides the following contributions; a unique set of features in their combination:
\begin{itemize}
	\item It is formally and machine-verifiably proven sound with Isabelle/HOL. 
	\item It supports the largest subset of iptables features compared to any other firewall analysis system. 
	\item It was tested on the largest (publicly-available) firewall which was ever analyzed in academia. 
	\item It terminates within a second for thousands of rules. 
\end{itemize}

We chose iptables because it provides one of the most complex firewall semantics widely deployed~\cite{fireman2006,pozo2009model}.
Of course, our algorithm is also applicable to similar or less complex (\eg Cisco PIX) firewall systems.

\section{Related Work}
There are several popular static firewall analysis tools.
The Firewall Policy Advisor~\cite{alshaer2004firewallpolicyanomaly} discovers inconsistencies between pairs of rules (\eg one rule completely overshadowing the other) in distributed firewall setups.
A similar tool is FIREMAN~\cite{fireman2006}. 
It can discover inconsistencies within rulesets or between firewalls and verify setups against administrator-defined policies.
To represent sets of packets, Binary Decision Diagrams (BDDs) are used.
It does not support matching on interfaces in a rule.
Since interfaces can be strings of arbitrary length, they may not be ideal for the encoding in BDDs and adding support to FIREMAN might be complicated or deteriorate its performance.

%

Margrave~\cite{nelson2010margrave} can be used to query (distributed) firewalls.
It is well suited to troubleshoot and to debug firewall configurations or to show the impact of ruleset edits.
For certain queries, it can also find scenarios, \eg it can show concrete packets which violate a security policy.
This scenario finding is sound but not complete.
Its query language as well as Margrave's internal implementation is based on first-order logic. 
A similar tool (relying on BDDs), with a different query language and focus on complete networks, is ConfigChecker~\cite{alshaer2011configcheckershort,alshaer2009configchecker}.
ITVal~\cite{marmorstein2005itval} can also be used to query firewalls.
It can also describe the firewall in terms of equivalent IP address spaces~\cite{marmorstein2006firewall}, which does not require the administrator to pose specific queries.


None of these tools can directly verify spoofing protection.
Nor are the tools themselves formally verified, which limits confidence in their results. 
In addition, the tools only support a limited subset of real-world firewalls (cf.\ Chapter~\ref{chap:fm15}). 
If the firewall under analysis exceeds this feature set, the tools either produce erroneous results or cannot continue. 

Jeffrey and Samak~\cite{fwmodelchecking2009jeffry} analyze the complexity of firewall analysis and conclude that most questions (for variable packet models) are NP-complete.
They show that SAT solvers usually outperform BDD-based implementations.

\section{Mathematical Background}

\paragraph*{Iptables Semantics}
To define and prove correctness of our algorithm, one must first agree on the semantics (\ie behavior) of an iptables firewall.
We rely on our semantics specified in Chapter~\ref{chap:fm15}. 
The semantics are well-suited for the tables where spoofing protection is usually implemented: the \texttt{filter} table or the \texttt{raw} table; given that only the following actions are used: 
$\iptaction{Accept}$, $\iptaction{Drop}$, $\iptaction{Reject}$, $\iptaction{Log}$, calling to and $\iptaction{Return}$ing from user-defined chains, as well as the ``empty'' action.

In the semantics, matching a packet against match conditions (\eg IP source or destination addresses) is mathematically defined with a ``magic oracle'' which understands all possible matches.
Obviously, semantics with a ``magic oracle'' cannot be expressed in terms of executable code.
Nevertheless, the algorithm we present in this chapter is both executable and proven sound w.\,r.\,t.\ these semantics.

\section{Spoofing Protection -- Mathematically}
To define spoofing protection, two data sets are required:
The firewall ruleset $\mvar{rs}$ and the IP addresses assignment $\mvar{ipassmt}$.
$\mvar{ipassmt}$ is a mapping from interfaces to an IP address range.
Usually, it can be obtained by \texttt{ip route}.
We will write $\mvar{ipassmt}[i]$ to get the corresponding IP range of interface $i$.
For the following examples, we assume 
\begin{IEEEeqnarray*}{l}
\mvar{ipassmt} = [\mathtt{eth0} \mapsto \lbrace 192.168.0.0/24 \rbrace ]
\end{IEEEeqnarray*}

\begin{definition}[Spoofing Protection]
\label{def:nospoofstrict}
A firewall ruleset provides \emph{spoofing protection} if for all interfaces $i$ specified in $\mvar{ipassmt}$, all packets from $i$ which are accepted by the ruleset have a source IP address contained in $\mvar{ipassmt}[i]$.
\end{definition}

For example, using pseudo iptables syntax (which omits the chain), the following ruleset implements spoofing protection:\medskip

\begin{minipage}{.9\linewidth}
\small
\begin{Verbatim}[commandchars=\\\{\},codes={\catcode`$=3\catcode`^=7}]
-i eth0 ! --src 192.168.0.0/24 -j DROP
-j ACCEPT
\end{Verbatim}
\end{minipage}%
\bigskip


\paragraph*{Spoofing Protection with Unknowns}
iptables supports numerous match conditions and new ones may be added in future.
It is practically infeasible to support all match conditions in a tool (cf.\ Chapter~\ref{chap:fm15}).
As we do not want our algorithm to abort if an unknown match occurs, we will refine Definition~\ref{def:nospoofstrict} to incorporate unknown matches.
This is motivated by the following examples.

We assume that \verb~--foo~ is a match condition which is unknown to us.
Therefore, we cannot guarantee that\medskip

\begin{minipage}{.9\linewidth}
	\small
\begin{Verbatim}[commandchars=\\\{\},codes={\catcode`$=3\catcode`^=7}]
-i eth0 ! --src 192.168.0.0/24 --foo -j DROP
-j ACCEPT
\end{Verbatim}
\end{minipage}\bigskip

implements spoofing protections since \verb~--foo~ could prevent certain spoofed packets from being dropped.
Also, the following ruleset might neither implement spoofing protection since ~\verb~--foo~ might allow spoofed packets:\medskip

\begin{minipage}{.9\linewidth}
	\small
\begin{Verbatim}[commandchars=\\\{\},codes={\catcode`$=3\catcode`^=7}]
--foo -j ACCEPT
-i eth0 ! --src 192.168.0.0/24 -j DROP
-j ACCEPT
\end{Verbatim}
\end{minipage}\bigskip

However, the following ruleset definitely implements spoofing protection; 
Independently of the meaning of \verb~--foo~ and \verb~--bar~, it is guaranteed that no spoofed packets are allowed:\medskip

\begin{minipage}{.9\linewidth}
	\small
\begin{Verbatim}[commandchars=\\\{\},codes={\catcode`$=3\catcode`^=7}]
--foo -j DROP
-i eth0 ! --src 192.168.0.0/24 -j DROP
--bar -j ACCEPT
\end{Verbatim}
\end{minipage}\bigskip

\noindent
This motivates Definition~\ref{def:nospoofpotentially}.

\begin{definition}[Certifiable Spoofing Protection]
\label{def:nospoofpotentially}
A firewall ruleset provides \emph{spoofing protection} if for all interfaces $i$ specified in $\mvar{ipassmt}$, all packets from $i$ which are \emph{potentially} accepted by the ruleset have a source IP address in the IP range of $i$.
\end{definition}

This new definition is stricter than the original one: \mbox{Definition~\ref{def:nospoofpotentially}} implies \mbox{Definition~\ref{def:nospoofstrict}}\footnote{approximating-imp-booloan-semantics-nospoofing}, which directly follows from Theorem~\ref{thm:FinalAllowClosure}. 
Therefore the new definition is \emph{sound} and can be used to prove that the last example implements spoofing protection.

However, depending on the meaning of \verb~--foo~, some of the previous examples might also implement spoofing protection.
This cannot be shown with Definition~\ref{def:nospoofpotentially}, so the new definition is not \emph{complete}.
However, as long as we anticipate unknown matches to occur, it is impossible to obtain completeness.

\section{Spoofing Protection -- Executable}
A straight forward spoofing protection proof of a firewall ruleset using Definition~\ref{def:nospoofpotentially} would require iterating over all packets, which is obviously infeasible.
We present an efficient executable algorithm to certify spoofing protection. 
It does not rely on BDDs or an external SAT/SMT solver. 

We assume the ruleset to be certified is preprocessed.
For this, we rely on the semantics-preserving ruleset simplification of Chapter~\ref{chap:fm15} which rewrites a ruleset to a semantically equivalent ruleset where only $\iptaction{Accept}$ and $\iptaction{Drop}$ actions occur. 
A preprocessed ruleset always has an explicit deny-all or allow-all rule at the end; it can never be empty. 
This rule corresponds to a chain's default policy. 

It would also be possible writing the algorithm for the simple firewall model which will be presented in Chapter~\ref{chap:networking16}. 
We chose not to do so since the simple model does not support negated interfaces, which are essential for spoofing protection. 
The big picture of the thesis is as follows: 
Once we have verified spoofing protection, we can rewrite matches on interfaces with matches in IP addresses. 
Thus, we need to check for spoofing protection first and can afterwards translate to a simpler firewall model which does not support matches on (negated) interfaces. 
These details are presented in the following chapter. 

\medskip

We call our algorithm $\mfun{sp}$ (``spoofing protection''). 
It certifies spoofing protection for one interface $i$ in $\mvar{ipassmt}$.
Using $\mfun{sp}$ to certify all $i \in \mvar{ipassmt}$ for a ruleset implies spoofing protection according to Definition~\ref{def:nospoofpotentially}.

We assume the global, static, and fixed parameters of $\mfun{sp}$ are an interface $i$ and the $\mvar{ipassmt}$. 
Then, $\mfun{sp}$ has the following type signature: 
\begin{IEEEeqnarray*}{c}
\mathit{rule}\ \mathit{list} \ \Rightarrow \ \mathit{ipaddr}\ \mathit{set} \ \Rightarrow \ \mathit{ipaddr}\ \mathit{set} \Rightarrow \mathbb{B}
\end{IEEEeqnarray*}
The first parameter ($\mathit{rule}\ \mathit{list}$) is the preprocessed firewall ruleset.
To recapitulate, a rule is a tuple $(m, a)$, where $m$ is the match condition and $a$ the action.
The action is either $\iptaction{Accept}$ or $\iptaction{Drop}$.
For a packet $p$, there is a predicate $\matches{m}$ which tells whether the packet $p$ matches the match condition $m$.
%
The second parameter ($\mathit{ipaddr}\ \mathit{set}$) is the set of potentially allowed source IP addresses for $i$. 
The third parameter ($\mathit{ipaddr}\ \mathit{set}$) is the set of definitely denied source IP addresses for $i$. 
The last parameter ($\mathbb{B}$) is a Boolean, which is true if spoofing protection could be certified.

Before we present the algorithm, we first present its correctness theorem (which will be proven later).\footnote{no-spoofing-iface, and ultimately spoofing protection according to Definition~\ref{def:nospoofstrict}: no-spoofing-executable-set}
\begin{theorem}[$\mfun{sp}$ sound]
\label{thm:spsound}
For any ruleset $\mvar{rs}$, if 
\begin{IEEEeqnarray*}{lcl}
\forall i \in \mvar{ipassmt}.\ \ \mfun{sp}\ \mvar{rs}\ \lbrace\rbrace \ \lbrace\rbrace
\end{IEEEeqnarray*}
then $\mvar{rs}$ provides spoofing protection according to Definition~\ref{def:nospoofpotentially}. 
\end{theorem}

The algorithm $\mfun{sp}$, presented in \mbox{Figure~\ref{algo:sp}}, is implemented recursively.
It iterates over the firewall ruleset.
\begin{figure*}
\begin{IEEEeqnarray*}{lcl}
	\mfun{sp} \ [] \ A \ D & \ = \ & (A \setminus D) \subseteq \bigcup \mvar{ipassmt}[i] \\
    \mfun{sp} \ ((m, \iptaction{Accept}) \lstcons \mvar{rs}) \ A \ D & \ = \ & \\
    \IEEEeqnarraymulticol{3}{r}{
      \mfun{sp} \ rs \ (A \cup \lbrace \mvar{ip}  \mid  
	     \exists \textnormal{$p$ from interface $i$ with src address $\mvar{ip}$}.\ \matches{m} \rbrace ) \ D 
	}\\
    \mfun{sp} \ ((m, \iptaction{Drop}) \lstcons rs) \ A \ D & \ = \ & \\
    \IEEEeqnarraymulticol{3}{r}{\qquad\qquad
      \mfun{sp} \ rs \ A \ \left(D \cup \left(\left\lbrace \mvar{ip}  \mid  
         \forall \textnormal{$p$ from interface $i$ with src address $\mvar{ip}$}.\ \matches{m} \right\rbrace \setminus A\right)\right)
     }%
%
\end{IEEEeqnarray*}
  \caption{Our algorithm to certify spoofing protection.}
  \label{algo:sp}
\end{figure*}

The base case is for an empty ruleset.
Here, $A$ and $D$ are the set of allowed/denied source IP addresses.
The firewall provides spoofing protection if the set of potentially allowed sources minus the set of definitely denied sources is a subset of the allowed IP range.

The two recursive calls collect these sets $A$ and $D$.
If the rule is an $\iptaction{Accept}$ rule, the set $A$ is extended with the set of sources possibly accepted in this rule.
If the rule is a $\iptaction{Drop}$ rule, the set $D$ is extended with the set of sources definitely denied in this rule, excluding any sources which were potentially accepted earlier.

As Theorem~\ref{thm:spsound} already states, $\mfun{sp}$ can be started with any ruleset and the empty set for $A$ and $D$.

We will now describe how the $\mathit{ipaddr}\ \mathit{set}$ operations are implemented.
In general, we symbolically represent a set of IP addresses as a set of IP range intervals.
Since IP ranges are commonly expressed in CIDR notation (\eg $\mathit{a.b.c.d/n}$), the interval datatype proves to be very efficient.

Next, the set $\lbrace \mvar{ip}  \mid  \exists \textnormal{$p$ from interface $i$ with src address $\mvar{ip}$}.\allowbreak\ \matches{m}\rbrace$ requires executable code.
Obviously, a straight-forward implementation which tests the existence of any packet is infeasible.
We provide an over-approximation for this set, the correctness proof confirms that this approach is sound.
First we check that $i$ matches all input interfaces specified in the match expression $m$.
If this is not the case, the set is obviously empty.
Otherwise, we collect the intersection of all matches on source IPs in $m$.
If no source IPs are specified in $m$, then $m$ matches any source IP and we return the universe.

The set $\lbrace \mvar{ip}  \mid  \forall \textnormal{$p$ from interface $i$ with src address $\mvar{ip}$}.\allowbreak\ \matches{m}\rbrace$ can be computed similarly.
However, we need to return an under-approximation here.
First, we check that $i$ matches, otherwise the set is empty.
Next, we remove all matches on input interfaces and source IPs from $m$.
If the remaining match expression is not unconditionally true, then we return the empty set.
Otherwise, we return the intersection of all source IP addresses specified in $m$, or the universe if $m$ does not restrict source IPs.

Note that after preprocessing we always have an explicit allow-all or deny-all rule at the end of the firewall ruleset.
Thus, $ A \cup D $ will always hold the universe after consuming the last rule.

\section{Evaluation -- Mathematically}
We outline the main idea of the correctness proof. 
Since $\mfun{sp}$ operates on a fixed interface, we define certifiable spoofing protection for a fixed interface $i$.
Showing Definition~\ref{def:nospoofpotentiallyoneiface} for all interfaces is equivalent to Definition~\ref{def:nospoofpotentially}.

\begin{definition}
\label{def:nospoofpotentiallyoneiface}
\begin{IEEEeqnarray*}{c}
\forall p \in \left\lbrace \mathit{p}  \mid  \textnormal{$p$ from $i$ and potentially accepted by the firewall} \right\rbrace.\ \ 
  \mdef{src\mhyphen{}ip}\ p \in \bigcup \mvar{ipassmt}[i]
\end{IEEEeqnarray*}
\end{definition}

The correctness proof of $\mfun{sp}$ is done by induction over the firewall ruleset.
Theorem~\ref{thm:spsound} does not lend itself to induction, since it features two empty sets which would generate unusable induction hypotheses.
To obtain a strong induction hypothesis, we generalize.
The ruleset is split into two parts: $\mvar{rs}_1$ and $\mvar{rs}_2$.
We assume that the algorithm correctly iterated over $\mvar{rs}_1$. 
For this lemma, we use the following notation:
\vskip-2.2em
\begin{IEEEeqnarray*}{l}
 A_\mathrm{exact} = \lbrace \mvar{ip}  \mid  \exists p.\ \textnormal{$p$ from $i$ with src $\mvar{ip}$ and accepted by $\mvar{rs}_1$} \rbrace \\
 D_\mathrm{exact} = \lbrace \mvar{ip}  \mid  \forall p.\ \textnormal{$p$ from $i$ with src $\mvar{ip}$ and denied by $\mvar{rs}_1$} \rbrace 
\end{IEEEeqnarray*}

\begin{lemma}
\label{lemma:nospoofgeneralized}
If 
 $A_\mathrm{exact} \subseteq A$ and $D \subseteq D_\mathrm{exact}$ and $\mfun{sp} \ \mvar{rs}_2 \ A \ D$ then Definition~\ref{def:nospoofpotentiallyoneiface} holds for $\mvar{rs}_1 \lstapp \mvar{rs}_2$
\end{lemma}
\begin{proof}
The proof is done by induction over $\mvar{rs}_2$ for arbitrary $\mvar{rs}_1$, $A$, and $D$.

Base case (\ie $\mvar{rs}_2 = []$): From $\mfun{sp} \ [] \ A \ D$ we conclude $(A \setminus D) \subseteq \bigcup \mvar{ipassmt}[i]$.
Since $A$ is an over-approximation and $D$ an under-approximation: $A_\mathrm{exact} \setminus D_\mathrm{exact} \subseteq A \setminus D$.
Since no IP address can be both accepted and denied we get $A_\mathrm{exact} \setminus D_\mathrm{exact} = A_\mathrm{exact}$.
From transitivity we conclude $A_\mathrm{exact} \subseteq \bigcup \mvar{ipassmt}[i]$, which implies spoofing protection for that interface according to Definition~\ref{def:nospoofpotentiallyoneiface}.

The two induction steps (one for $\iptaction{Accept}$ and one for $\iptaction{Drop}$ rules) follow from the induction hypothesis.
The over- and under- approximations were carefully constructed, such that the subset relations continue to hold.
The executable implementations of these sets also respect the subset relation; hence, the induction hypothesis solves these cases.
\end{proof}

\begin{proof}[Proof of Theorem~\ref{thm:spsound}]
Lemma~\ref{lemma:nospoofgeneralized} can be instantiated where $\mvar{rs}_1$ is the empty ruleset and $A$ and $D$ are the empty set.
For this particular choice, it is easy to see that the preconditions hold.
Thus, for any $\mvar{rs}$, we conclude $\mfun{sp} \ \mvar{rs} \ \lbrace\rbrace \ \lbrace\rbrace$ implies Definition~\ref{def:nospoofpotentiallyoneiface}.
Since this holds for arbitrary interfaces, we conclude Definition~\ref{def:nospoofpotentially}.
\end{proof}

Thus, our algorithm is proven \emph{sound}. 
This means, if the algorithm certifies a ruleset, then this ruleset is guaranteed to implement spoofing protection.

Note that our algorithm only certifies; debugging a ruleset in case of a certification failure remains manual. 
To debug, the proof of Lemma~\ref{lemma:nospoofgeneralized} suggest to consider the first rule where $(A \setminus D) \subseteq \bigcup \mvar{ipassmt}[i]$ is violated.


The algorithm is \emph{not complete}.
This means, there may be rulesets which implement spoofing protection but cannot be certified by the algorithm.
This is bought by the approximations and by the support for unknown match conditions.
For example, the following ruleset cannot be certified:\medskip

\begin{minipage}{.9\linewidth}
	\small
\begin{Verbatim}[commandchars=\\\{\},codes={\catcode`$=3\catcode`^=7}]
-i eth0 ! --src 192.168.0.0/24 --foo -j DROP
-i eth0 ! --src 192.168.0.0/24 ! --foo -j DROP
-j ACCEPT
\end{Verbatim}
\end{minipage}\bigskip

This is a reasonable decision for a completely unknown \verb~--foo~, since it might update an internal state and the mathematical equation ``$\verb~foo~ \vee \neg\;\verb~foo~ = \mdef{True}$'' may not hold.
However, if \verb~foo~ is replaced by the known and stateless match condition \verb~--protocol tcp~, the ruleset can be shown to correctly implement spoofing protection.
The algorithm, however, cannot certify it, since it does not track this match condition.
However, this is a made-up and bad-practice example, and we never encountered such special cases in any real-world ruleset.
The evaluation ---in which vast amounts of unknowns occurred--- shows that the algorithm certified all rulesets which included spoofing protection and correctly failed only for those rulesets which did not (correctly) implement it.
Thus, the incompleteness is primarily a theoretical limitation.

\section{Evaluation -- Empirically}
Often, firewalls start with an \verb~ESTABLISHED~ rule.
A packet can only match this rule if it belongs to a connection which has been accepted by the firewall previously.
Hence, the \verb~ESTABLISHED~ rule does not contribute to the access control policy for connection setup enforced by the firewall~\cite{fireman2006}.
Likewise, spoofed packets can only be allowed by the \verb~ESTABLISHED~ rule if they are allowed by any of the subsequent ACL rules.
Therefore, as done in the previous chapter (Section~\ref{sec:fm15:establishedrule}), we either exclude this rule from our analysis or only consider packets of state \verb~NEW~.

We tested the algorithm on several real-word rulesets~\cite{diekmanngithubnetnetwork}. 
Most of them either did not provide spoofing protection or had an obvious spoofing protection and could thus be certified.
In this Section, we present the results of certifying four rulesets. 
The first two rulesets~\cite{fwbuilder2011userguide,cybercity2005spoofing} are simple examples and were found as the first two results of a goolge query for ``\texttt{iptables spoofing}''. 
They primarily serve as example. 
Afterwards, we will put our main focus in this section on a large and interesting real-world ruleset.

First of all, for all rulesets, our algorithm was extremely fast:
Once the ruleset is preprocessed (few seconds) the certification algorithm only takes fractions of seconds for rulesets with several thousand rules.
We omit a detailed performance evaluation since these orders of magnitude are sufficient for a static/offline analysis system. 
Certification runs of our algorithm, in particular for the larger rulesets, were usually faster than reloading the ruleset on the firewall system itself.

\subsection{Firewall Builder Documentation}

\begin{figure*}[htb]
\centering
\begin{minipage}{0.7\linewidth}
\small
\begin{Verbatim}[commandchars=\\\{\},codes={\catcode`$=3\catcode`^=7}]
*filter
:INPUT ACCEPT [16:6016]
:FORWARD ACCEPT [0:0]
:OUTPUT ACCEPT [16:6016]
:In_RULE_0 - [0:0]
-A INPUT -s 192.0.2.1/32 -i eth0 -j In_RULE_0
-A INPUT -s 192.168.1.1/32 -i eth0 -j In_RULE_0
-A INPUT -s 192.168.1.0/24 -i eth0 -j In_RULE_0
-A FORWARD -s 192.0.2.1/32 -i eth0 -j In_RULE_0
-A FORWARD -s 192.168.1.1/32 -i eth0 -j In_RULE_0
-A FORWARD -s 192.168.1.0/24 -i eth0 -j In_RULE_0
-A In_RULE_0 -j LOG --log-prefix "RULE 0 -- DENY " --log-level 6
-A In_RULE_0 -j DROP
COMMIT
\end{Verbatim}
\end{minipage}%
  \caption{Spoofing Protection Ruleset from the Firewall Builder Documentation}%
  \label{tab:cnsm:eval:firewallbuilder}%
\end{figure*}

Our first example is taken from the Firewall Builder~\cite{fwbuilder} user guide \cite[14.2.6. Anti-spoofing rules]{fwbuilder2011userguide}. 
Their example defines the IP addresses of the firewall as 192.168.1.1 and 192.0.2.1 and network behind as 192.168.1.0/24. 
Consequently, on the external-facing interface (\texttt{eth0}), if any of these IP addresses occurs, it must be spoofed. 
For spoofing protection, these IP addresses must not occur. 
Let $\mdef{UNIV}$ denote the universe of all IPv4 addresses, then $\mvar{ipassmt} = [\mathtt{eth0} \mapsto \mdef{UNIV} \setminus \lbrace 192.168.1.1,\ 192.0.2.1,\ 192.168.1.0\ ..\ 192.168.1.255 \rbrace ]$. 
Our implementation allows to conveniently express this as \texttt{eth0 = all\_but\_those\_ips [192.168.1.1, 192.0.2.1, 192.168.1.0/24]}. 
Aside, our tool warns that the $\mvar{ipassmt}$ is not complete, \ie it does not cover the complete IPv4 address space. 
This is mainly because the firewall only defines spoofing protection on the external-facing interface and does not consider (in this example) outgoing spoofing from the internal interface. 
The firewall rules are shown in Figure~\ref{tab:cnsm:eval:firewallbuilder} and our tool immediately certifies spoofing protection for \texttt{eth0} for both the \texttt{FORWARD} and the \texttt{INPUT} chain. 

\smallskip

Spoofing protection can be considered a rather low-level property of a firewall. 
Considering the big picture of this thesis, once we have certified spoofing protection, we can abstract over such low-level properties. 
For example for this ruleset, we may now assume that we have spoofing protection. 
If we now simplify the ruleset under this assumption, the ruleset is effectively only one rule: Accept all.
This high-level view and the respective simplifications which are built on top of spoofing protection will be presented in the following chapter.

\subsection{Blog Post}

\begin{figure*}[htb]
\centering
\begin{minipage}{0.7\linewidth}
\small
\begin{Verbatim}[commandchars=\\\{\},codes={\catcode`$=3\catcode`^=7}]
*filter
:INPUT ACCEPT [77:24937]
:FORWARD ACCEPT [0:0]
:OUTPUT ACCEPT [78:25303]
-A INPUT -s 202.54.10.20/32 -i eth1 -j DROP
-A INPUT -s 192.168.1.0/24 -i eth1 -j DROP
-A INPUT -s 0.0.0.0/8 -i eth1 -j DROP
-A INPUT -s 127.0.0.0/8 -i eth1 -j DROP
-A INPUT -s 10.0.0.0/8 -i eth1 -j DROP
-A INPUT -s 172.16.0.0/12 -i eth1 -j DROP
-A INPUT -s 192.168.0.0/16 -i eth1 -j DROP
-A INPUT -s 224.0.0.0/3 -i eth1 -j DROP
-A OUTPUT -s 202.54.10.20/32 -o eth1 -j DROP
-A OUTPUT -s 192.168.1.0/24 -o eth1 -j DROP
-A OUTPUT -s 0.0.0.0/8 -o eth1 -j DROP
-A OUTPUT -s 127.0.0.0/8 -o eth1 -j DROP
-A OUTPUT -s 10.0.0.0/8 -o eth1 -j DROP
-A OUTPUT -s 172.16.0.0/12 -o eth1 -j DROP
-A OUTPUT -s 192.168.0.0/16 -o eth1 -j DROP
-A OUTPUT -s 224.0.0.0/3 -o eth1 -j DROP
COMMIT
\end{Verbatim}
\end{minipage}%
  \caption{Spoofing Protection Ruleset from ``Linux Iptables Avoid IP Spoofing And Bad Addresses Attacks'' Blog Post}%
  \label{tab:cnsm:eval:cyber}%
\end{figure*}

The next example is taken from a Linux blog which explains spoofing protection with iptables~\cite{cybercity2005spoofing}. 
With 202.54.10.20 beeing the IP address of the machine which should be protected, the author lists the following IP ranges as ``bad'': 
0.0.0.0/8, 127.0.0.0/8, 10.0.0.0/8, 172.16.0.0/12, 192.168.0.0/16, 224.0.0.0/3, and the ``LAN IP range'' 192.168.1.0/24. 
Note that the last IP range is already included in the fifth IP range, which causes no problem to our tool. 
The author defines \texttt{eth1} as the Internet-facing interface. 
As in the previous example, we can define $\mvar{ipassmt}$ as \texttt{eth0 = all\_but\_those\_ips [\textnormal{\textit{``bad'' IP ranges}}]}. 
The firewall rules are generated by a shell script, ultimately, the rules as shown in Figure~\ref{tab:cnsm:eval:cyber} are generated. 
Our tool immediately certifies spoofing protection for the \texttt{INPUT} chain.

Additionally, the shell script loads the very same rules with \verb~-o eth1~ instead of \verb~-i eth1~ for the \texttt{OUTPUT} chain. 
Therefore, we also want to check spoofing protection for the \texttt{OUTPUT} chain. 
For outbound spoofing protection, it must be assured that the machine uses only its assigned IP address, \ie $\mvar{ipassmt} = [\mathtt{eth1} \mapsto \lbrace 202.54.10.20 \rbrace ]$. 
The check for spoofing protection immediately fails. 
The \texttt{OUTPUT} chain prevents that the machine spoofs with the set of ``bad'' IP ranges. 
However, it does not prevent that the machine spoofs its source IP address with some valid IP address which is not defined as ``bad''. 
More severe, as the very first rule of the \texttt{OUTPUT} chain, the ruleset prevents outgoing packets from the only valid IP address: The address of the machine itself.  
This means the machine is not able to send any (non-spoofed) packets to the Internet. 
Reading the comment section of the blog post reveals that three users also noticed that their machine loses outbound Internet connectivity if they run the shell script.


\subsection{Firewall of Our Lab}
We present the certification of a firewall with about 4800 rules, connecting about 20 VLANs. 
Every VLAN has its own interface. 
An excerpt of the ruleset and the $\mvar{ipassmt}$ is shown in Figure~\ref{fig:spoofing:ipassmti8}.

\begin{figure*}[htb]
\begin{minipage}{.55\linewidth}
\footnotesize%
Old rules (without spoofing protection):
\begin{Verbatim}[commandchars=\\\{\},codes={\catcode`$=3\catcode`^=7}]
-A FORWARD -p tcp -m state --state NEW $\hfill\hookleftarrow$
               -m tcp --dport 22 \dots $\hfill\hookleftarrow$
               -m recent --update \dots $\hfill\hookleftarrow$
               --name ratessh --rsource $\hfill\hookleftarrow$
               -j LOG_RECENT_DROP
\dots
-A FORWARD -s 131.159.14.0/25 -i eth1.96 -j mac_96
-A FORWARD -i eth1.96 -j ranges_96
-A FORWARD -d 131.159.14.0/25 -o eth1.96 $\hfill\hookleftarrow$
               -j filter_96
-A FORWARD -d 224.0.0.0/4 -o eth1.96 -j filter_96
-A FORWARD -s 131.159.14.128/26 -i eth1.108 $\hfill\hookleftarrow$
               -j mac_108
-A FORWARD -i eth1.108 -j ranges_108
-A FORWARD -d 131.159.14.128/26 -o eth1.108 $\hfill\hookleftarrow$
               -j filter_108
-A FORWARD -d 224.0.0.0/4 -o eth1.108 -j filter_108
\dots
-A mac_96 -s 131.159.14.92/32 $\hfill\hookleftarrow$
              -m mac --mac-source \dots -j RETURN
-A mac_96 -s 131.159.14.92/32 -j DROP
\dots
-A ranges_96 -s 131.159.14.0/25 -j RETURN
-A ranges_96 -j LOG_DROP
\dots
\end{Verbatim}

\vspace*{6ex}

Improved rules (with spoofing protection):
\begin{Verbatim}[commandchars=\\\{\},codes={\catcode`$=3\catcode`^=7}]
-A FORWARD -i eth1.110 -j NOTFROMHERE
-A FORWARD -p tcp -m state --state NEW $\hfill\hookleftarrow$
               -m tcp --dport 22 \dots $\hfill\hookleftarrow$
               -m recent --update \dots $\hfill\hookleftarrow$
               --name ratessh --rsource $\hfill\hookleftarrow$
               -j LOG_RECENT_DROP
\dots
-A FORWARD -s 131.159.14.0/25 -i eth1.96 -j mac_96
-A FORWARD -i eth1.96 -j ranges_96
-A FORWARD -s 131.159.14.128/26 -i eth1.108 $\hfill\hookleftarrow$
               -j mac_108
-A FORWARD -i eth1.108 -j ranges_108
\dots
-A FORWARD -d 131.159.14.0/25 -o eth1.96 $\hfill\hookleftarrow$
               -j filter_96
-A FORWARD -d 224.0.0.0/4 -o eth1.96 -j filter_96
-A FORWARD -d 131.159.14.128/26 -o eth1.108 $\hfill\hookleftarrow$
               -j filter_108
-A FORWARD -d 224.0.0.0/4 -o eth1.108 -j filter_108
\dots
-A mac_96 -s 131.159.14.92/32 $\hfill\hookleftarrow$
              -m mac --mac-source \dots -j RETURN
-A mac_96 -s 131.159.14.92/32 -j DROP
\dots
-A ranges_96 -s 131.159.14.0/25 -j RETURN
-A ranges_96 -j LOG_DROP
\dots
-A NOTFROMHERE -s 131.159.14.0/23 -j LOG_DROP
\dots
\end{Verbatim}
\end{minipage}%
\hfill%
\begin{minipage}{.38\linewidth}%
\footnotesize%
Specified $\mvar{ipassmt}$:
\begin{Verbatim}[commandchars=\\\{\},codes={\catcode`$=3\catcode`^=7}]
eth0 = [0.0.0.0-255.255.255.255]
eth1.96 = [131.159.14.3/25]
eth1.108 = [131.159.14.129/26]
eth1.109 = [131.159.20.11/24]
eth1.110 = all_but_those_ips [
  131.159.14.0/23,
  131.159.20.0/23,
  192.168.212.0/23,
  188.95.233.0/24,
  188.95.232.192/27,
  188.95.234.0/23,
  192.48.107.0/24,
  188.95.236.0/22,
  185.86.232.0/22
  ]
eth1.116 = [131.159.15.131/26]
eth1.152 = [131.159.15.252/28]
eth1.171 = [131.159.15.2/26]
eth1.173 = [131.159.21.252/24]
eth1.1010 = [131.159.15.227/28]
eth1.1011 = [131.159.14.194/27]
eth1.1012 = [131.159.14.238/28]
eth1.1014 = [131.159.15.217/27]
eth1.1016 = [131.159.15.66/26]
eth1.1017 = [131.159.14.242/28]
eth1.1111 = [192.168.212.4/24]
eth1.97 = [188.95.233.2/24]
eth1.1019 = [188.95.234.2/23]
eth1.1020 = [192.48.107.2/24]
eth1.1023 = [188.95.236.2/22]
eth1.1025 = [185.86.232.2/22]
eth1.1024 = all_but_those_ips [
  131.159.14.0/23,
  131.159.20.0/23,
  192.168.212.0/23,
  188.95.233.0/24,
  188.95.232.192/27,
  188.95.234.0/23,
  192.48.107.0/24,
  188.95.236.0/22,
  185.86.232.0/22
  ]
\end{Verbatim}
\end{minipage}%
  \caption{Firewall rules without spoofing protection and improved rules}%
  \label{fig:spoofing:ipassmti8}%
\end{figure*}

Trying the certification, it immediately fails.
Responsible was a work-around rule which should only have existed temporarily but was forgotten.
This rule is now on the administrator's ``things to do the right way'' list and we exclude it for further evaluation.

Certifying spoofing protection for the first VLAN interface succeeds instantly.
However, trying to certify all other VLANs fails.
The reason is an error in the ruleset.
For every VLAN $n$, the firewall defines three custom chains: \texttt{mac\_}$n$, \texttt{ranges\_}$n$, and \texttt{filter\_}$n$.
The \texttt{mac\_}$n$ chain verifies that for hosts with registered MAC addresses and static IP addresses, nobody (with a different MAC address) steals the IP address.
This chain is primarily to avoid manual IP assignment errors.
Next, the \texttt{ranges\_}$n$ chain should prevent outgoing spoofing.
Finally, the \texttt{filter\_}$n$ chain allows packets for certain registered services.
The main error was that a spoofed packet from VLAN $m$ could be accepted by \texttt{filter\_}$n$ before it had to pass the \texttt{ranges\_}$m$ check.
The discovery of this error also discovered that the \texttt{mac\_}$m$ chains were not working reliably.
We verified these findings by sending and receiving such spoofed packet via the real firewall.

Finally, we fixed the firewall by moving all \texttt{mac\_}$n$ and \texttt{ranges\_}$n$ chains before any \texttt{filter\_}$n$ chains.
The certification for all but one\footnote{This one VLAN where the certification fails is only for internal testing purposes and deliberately features no spoofing protection} internal VLAN interfaces succeeds.

Next, the interfaces attached to the Internet are certified.
The IP address range was defined as the universe of all IPs, excluding the IPs owned by the institute. 
Here, certification failed in a first run.
Responsible were some ssh rate limiting rules.
These rules were originally designed to prevent too many outgoing ssh connections.
However, since spoofing protection did not apply to them, an attacker could exploit them for a DOS attack against the internal network:
The attacker floods the firewall with ssh TCP SYN packets with spoofed internal addresses.
This exhausts the ssh limit of the internal hosts and it is no longer possible for them to establish new ssh connections.
This flaw was fixed and certification subsequently succeeds.

The improved and certified firewall ruleset is now in production use. 

\subsection{Remote Firewall}
\label{subsec:cnsm:eval:japanfw}
After this, another administrator got interested and wanted to implement spoofing protection for his firewall. 
To complicate matters, he was in Japan and the firewall in Germany. 
It was a key requirement that he would not lock himself out. 
Our tool could certify both: His proposed changes to the firewall correctly enforce spoofing protection and he will not lose ssh access. 
To provide a \emph{sound} guarantee for the latter, we applied the same idea as in our algorithm, but in reverse: the ruleset is abstracted to a stricter version (\ie a version that blocks more packets) and we consequently certify that it still allows \verb~NEW~ and \verb~ESTABLISHED~ ssh packets from the Internet.

\section{Conclusion}
We present an easy-to-use algorithm. 
It is fast enough to be run on every ruleset update.
It discovered real problems in a large, production-use firewall.
Both, the theoretical algorithm as well as the executable code are proven sound, hence if the algorithm certifies a firewall, the ruleset is \emph{proven} to implement spoofing protection correctly.

  \chapter{\fffuu{}: Verified iptables Firewall Analysis}
\label{chap:networking16}
  
This chapter is an extended version of the following paper~\cite{diekmann2016networking}:
\begin{itemize}
	\item Cornelius Diekmann, Julius Michaelis, Maximilian Haslbeck, and Georg Carle, \emph{Verified iptables Firewall Analysis}. In IFIP Networking 2016, Vienna, Austria, May 2016.
\end{itemize}

\noindent
The following major improvements were added: 
\begin{itemize}
 	\item Performance improvement of about 10x; correctness proven. 
 	\item Identified and fixed a bug in the port matching semantics, large rework of the theory. 
 	\begin{itemize}
	 	\item Generic result: Bug either affects any firewall analysis tool or tool is limited in its expressiveness (Section~\ref{par:ifip:portsbug}).
 	\end{itemize}
 	\item Added ultimate service matrix correctness theorem (Theorem~\ref{thm:servicematrix}). 
 	\item Performed the evaluation again several times (performance measurements, identify effects of bug). 
 	\item Stand-alone haskell tool \fffuu{} (Section~\ref{sec:ifip:fffuu}), which gives another few orders of magnitude speedup. 
	\item IPv6 support for the translation, simple firewall, and presented algorithms. Featuring probably the first formally verified IPv6 RFC 5952~\cite{rfc5952} pretty printer~\cite{IP_Addresses-AFP}. 
	\item A formalization of longest prefix routing~\cite{Routing-AFP} to enable output port rewriting. 
 	\item Additional evaluation. 
\end{itemize}

  \paragraph*{Statement on author's contributions}
  For the original paper, the author of this thesis provided major contributions for the overall idea. 
  He invented and implemented the simple firewall model and provided major contributions for the ideas, realization, implementation, and proof of the translation from the real-world iptables firewall model to the simplified model. 
  For this, he also enhanced the model with state. 
  He researched related work, collected the data, and conducted the evaluation. 
  He is the main author of the Haskell tool \fffuu{}. 
  Maximilian Haslbeck implemented (as a part of his Bachelor's thesis) a first prototype of the IP address space partitioning algorithm as well as the service matrix algorithm and showed its correctness. 
  The author of this thesis later on improved the performance of the algorithm and showed the final service matrix theorem (based on Haslbeck's previous work). 
  Julius Michaelis provided major contributions for the CIDR split algorithm and its correctness proof and contributed to the word interval type which we use to store and operate on IPv4 addresses, which was generalized by the author of this thesis to support arbitrary machine words which is necessary for IPv6 support. 
  Julius Michaelis formalized longest prefix routing and implemented the output port rewriting. 
  All other improvements with regard to the paper are the work of the author of this thesis. 

  \medskip

  \paragraph*{Abstract}
  In this chapter, we present our fully verified iptables firewall ruleset analysis framework. 
  The core analysis capability is the translation of an iptables ruleset to a policy, \ie we present a method to compute graphs which partition the complete IPv4 and IPv6 address space and show the allowed accesses between partitions for a fixed service. 
  Such a graph can be visualized for manual inspection or verified with the methods of Part~\ref{part:greenfield}. 
  This allows uncovering scenario-specific firewall errors. 
  Internally, we are working with a simplified firewall model and a core contribution is the translation of complex real-world iptables firewall rules into this model. 
  A real-world evaluation demonstrates the applicability of our tool.

  \medskip

\section{Introduction}
Firewall rulesets are inherently difficult to manage. 
It is a well-studied but unsolved problem that many rulesets show several configuration errors~\cite{firwallerr2004,wool2010firewall,diekmann2015fm}.
Tools were designed to help uncover configuration errors and verify a ruleset. 
In this thesis, we focus on tools for the static of rulesets. 
They have the benefit that the analysis can be carried out offline, without any negative effects on the network. 
In contrast to testing, static analysis can achieve a full coverage (\eg the results hold for all packets) and thus are able to uncover all errors and give strong guarantees for the absence of certain classes of errors. 
However, in practice, static ruleset analysis tools fail for various reasons:
They do not support the vast amount of firewall features, they require the administrator to learn a complex query language which might be more complex than the firewall language itself, the analysis algorithms do not scale to large firewalls, and the output of the verification tools itself cannot be trusted.


To overcome these issues and to foster static analysis and verification of real-world firewall rulesets, we present the first fully verified and large-scale tested Linux/netfilter iptables firewall analysis and verification tool. 
In detail, our contributions are: 
\begin{itemize}
\item A simple firewall model, designed for mathematical beauty and ease of static analysis  (Section~\ref{sec:models})
\item A series of translation steps to translate real-world firewall rulesets into this simple model (Section~\ref{sec:translating-primitives})
\item Static and automatic firewall analysis methods, based on the simple model, featuring
	\begin{itemize}
	\item IP address space partitioning (Section~\ref{sec:ip-partition})
	\item Minimal service matrices (Section~\ref{sec:matrix})
	\end{itemize}
\item Full formal and machine-verifiable proof of correctness (Section~Availability)
\item Evaluation on large real-world data set (Section~\ref{sec:evaluation})
\end{itemize}



The Linux iptables firewall is wide-spread, has evolved over a long time, and is well-known for its vast amount of features. 
In addition, in production networks, huge, complex, and legacy firewall rulesets have evolved over time. 
Therefore, iptables poses a particular challenge. 
Naturally, our methodology can also be applied to firewalls with simpler semantics, or younger technology with yet fewer features, \eg Cisco IOS Access Lists or filtering OpenFlow flow tables. 

In contrast to work which considers generic firewall configuration errors~\cite{firwallerr2004,wool2010firewall}, the goal of our IP address space partitioning and service matrices is to uncover scenario-specific misconfiguration. 
For example, our tool will not warn that inbound HTTP is allowed to more than 256 IPs, however, it will generate an overview of which IPs may establish and receive HTTP connections. 
This overview can then be inspected by an administrator. 
For a large webhoster, having more than 256 IPs where inbound HTTP is allowed may be perfectly fine, whereas in an airplane, even the possibility of a single HTTP connection from the coffee machine to the avionics may raise concern.\footnote{This is a made up example.}

We outline related work in Section~\ref{sec:related}. 
The real-world and simplified firewall models are presented in Section~\ref{sec:models}. 
We detail on the translation between these models in Section~\ref{sec:translating-primitives}. 
Afterwards, we present the IP address space partitioning (Section~\ref{sec:ip-partition}) and service matrices (Section~\ref{sec:matrix}). 
In Section~\ref{sec:evaluation}, we evaluate our algorithms on a large set of real-world iptables rulesets.

\section{Related Work}
\label{sec:related}
As introduced in Section~\ref{sec:fm15:iptables-semantics}, we call the features a firewall can use to match on packets \emph{primitives}. 
For example, among others, iptables supports the following primitives: src IP address, layer 4 port, inbound interface, conntrack state, entries and limits in the \texttt{recent} list,~\dots 
While our previous work in Chapter~\ref{chap:fm15} focused on the high-level semantics of iptables, this chapter focuses on primitives. 

Popular tools for static firewall analysis include FIREMAN~\cite{fireman2006}, Capretta \etal\cite{capretta2007coqfwconflicts}, and the Firewall Policy Advisor~\cite{alshaer2004firewallpolicyanomaly}. 
They support the following primitives: IP addresses, ports, and protocol.
This corresponds to (a subset of) our simple firewall model, hence, these tools would not be applicable to most firewalls from our evaluation. 
The tools focus on detecting conflicts between rules and can consequently not offer service matrices. 

The work most similar to our IP address space partitioning is ITVal~\cite{marmorstein2005itval}: 
It supports a large set of iptables features and can compute an IP address space partition~\cite{marmorstein2006firewall}. 
Unfortunately, ITVal only supports IPv4, is not formally verified, and its implementation has several errors. 
For example, ITVal produces spurious results if the number of significant bits in IP addresses in CIDR notation~\cite{rfc4632} is not a multiple of 8. 
This appears to be a coding error in the translation to ITVal's internal data structure which is easy to fix for non-negated IP address ranges. 
ITval does not consider logical negations which may occur when $\iptaction{RETURN}$ing prematurely from user-defined chains, which leads to wrong interpretation of complement sets. 
It does not support abstracting over unknown primitives but simply ignores them, which also leads to spurious results. 
Those are problems of the internal logic and are thus non-trivial to repair. 
For rulesets with more than 1000 rules, ITVal requires tens of gigabytes of RAM. 
It does not support IPv6. 
Finally, ITVal neither proves the soundness nor the minimality of its IP address range partitioning.
Nevertheless, ITVal demonstrates the need for and the use of IP address range partitioning and has demonstrated that its implementation works well on rulesets which do not trigger the aforementioned errors.
Building on the ideas of ITVal (but with a different algorithm), we overcome all presented issues.


Exodus~\cite{nelson2015exodus} translates existing device configurations to a simpler model, similar to our translation step. 
It translates router configurations to a high-level SDN controller program, which is implemented on top of OpenFlow. 
Exodus supports many Cisco IOS features.
The translation problem solved by Exodus is comparable to this paper's problem of translating to a simple firewall model:
OpenFlow 1.0 only supports a limited set of features (comparable to our simple firewall) whereas IOS supports a wide range of features (comparable to iptables);
A complex language is ultimately translated to a simple language, which is the `hard' direction.

Complementary to our verification tool, and well-suited for debugging, is Margrave~\cite{nelson2010margrave}.
It can be used to query firewalls and to troubleshoot configurations or to show the impact of ruleset edits.
Margrave can find scenarios, \ie it can show concrete packets which violate a security policy.
Our framework does not show such information. 
Margrave's query language (which a potential user is required to learn to use this tool) is based on first-order logic. 



\section{Firewall Semantics}
\label{sec:models}

First, we present a very simple firewall model. 
This model was designed to feature nice mathematical properties but it is too simplistic to mirror the real world. 
Afterwards, we will compare it to our model for real-world firewalls of Chapter~\ref{chap:fm15}. 
Section~\ref{sec:translating-primitives} will show how rulesets can be translated between these two models. 
This preprocessing step converts firewall rulesets from the real-world model to the simple model, which greatly simplifies all future static firewall analysis. 

\subsection{Simple Firewall}
We will write firewall rules as tuple $(m,\; a)$, where $m$ is a match expression and $a$ is the action the firewall performs if $m$ matches for a packet.
The firewall has two possibilities for the filtering decision: it may accept ($\allow$) the packet or deny ($\deny$) the packet. 
There is also an intermediate state ($\undecided$) in which the firewall did not come to a filtering decision. 
Note that iptables firewalls always have a default policy and the $\undecided$ case cannot occur as final decision.

The simple model is a simple recursive function. 
The first parameter is the ruleset the firewall iterates over, the second parameter is the packet.
%
%
\begin{IEEEeqnarray*}{llllcl}
\mfun{simple\mhyphen{}fw}\ \ & []\ \                                       & p & \ = \ & \undecided \\
\mfun{simple\mhyphen{}fw} & ((m,\; \iptaction{Accept})\lstcons{}\mathit{rs})\ & p & \ = \ & \mctrl{if}\ \mfun{match} \ m\ p\ \mctrl{then} \ \allow \ \mctrl{else} \  \mfun{simple\mhyphen{}fw} \ \mathit{rs}\  p\\
\mfun{simple\mhyphen{}fw} & ((m,\; \iptaction{Drop})\lstcons{}\mathit{rs})\   & p & \ = \ & \mctrl{if}\ \mfun{match} \ m\ p\ \mctrl{then} \ \deny \ \mctrl{else} \  \mfun{simple\mhyphen{}fw} \ \mathit{rs}\  p
\end{IEEEeqnarray*}

A function $\mfun{match}$ tests whether a packet $p$ matches the match condition $m$.\footnote{Note that this is not the same function as in Chapter~\ref{chap:fm15} since this simple $\mfun{match}$ function does not require parameter $\gamma$. Basically, it already has the primitive matcher hard-coded in it. } 
The match condition is an 7-tuple, consisting of the following primitives: 
\begin{IEEEeqnarray*}{c}
	\left(\mathrm{in},\ \mathrm{out},\ \mathrm{src},\ \mathrm{dst},\ \mathrm{protocol},\ \mathrm{src\ ports},\ \mathrm{dst\ ports}\right)
\end{IEEEeqnarray*}
In contrast to iptables, negating matches is not supported. 
In detail, the following primitives are supported:
\begin{itemize}
	\item in/out interface, including support for the `\texttt{+}' wildcard
	\item src/dst IP address range in CIDR notation, \eg 192.168.0.0/24
	\item protocol ($\mconstr{Any}$, tcp, udp, icmp, or any numeric protocol identifier)
	\item src/dst interval of ports, \eg 0:65535
\end{itemize}

For example, we obtain an empty match (a match that does not apply to any packet) \emph{iff} an end port is greater than the start port.\footnote{empty-match} 
%
The match which matches any packet is constructed by setting the interfaces to ``\texttt{+}'', the ips to 0.0.0.0/0, the ports to 0:65535 and the protocol to $\mconstr{Any}$.\footnote{simple-match-any}
%

We require that all match conditions are well-formed, \ie it is only allowed to match on ports (other than the universe 0:65535) if the protocol is tcp, udp, or sctp.

With this type of match expression, it is possible to implement a function $\mfun{conj}$ which takes two match expressions $m_1$ and $m_2$ and returns exactly one match expression being the conjunction of both.\footnote{simple-match-and-correct}
\begin{theorem}[Conjunction of two simple match expressions]%
\label{thm:conj}%
\begin{IEEEeqnarray*}{c}
\mfun{match} \ m_1\ p\ \wedge \mfun{match} \ m_2\ p\ \longleftrightarrow \mfun{match} \ (\mfun{conj}\ m_1 \ m_2) \ p\
\end{IEEEeqnarray*}%
\end{theorem}%
Computing the conjunction of the individual match expressions for port intervals and single protocols is straightforward. 
The conjunction of two intervals in CIDR notation is either empty or the smaller of both intervals.
The conjunction of two interfaces is either empty if they do not share a common prefix, otherwise it is the longest of both interfaces (non-wildcard interfaces dominate wildcard interfaces).

The $\mfun{conj}$ of two well-formed matches is again well-formed.\footnote{simple-match-and-valid} 

The type of match expressions was carefully designed such that the conjunction of \emph{two} match expressions is only \emph{one} match expression. 
If features were added to the match expression, for example negated interfaces, this would no longer be possible.
Of the features most commonly found in our iptables firewall rulesets~\cite{diekmanngithubnetnetwork}, we only found that it would further be possible to add TCP flags to the match expression without violating the aforementioned conjunction property. 
Considering common features of firewalls in general~\cite{pozo2009model}, it would probably be possible to enhance the ICMP support of our model.

\subsection{Semantics of Iptables}
\label{sec:real-world-semantics}
We now repeat the important aspects of our model of a real-world iptables firewall from Chapter~\ref{chap:fm15}. 
Most firewall analysis is concerned with the access control rules of a firewall, therefore the model focuses on the \texttt{filter} table. 
This implies, packet modification (\eg NAT, which must not occur in this table) is not considered in this work. 
The model supports the following common actions: 
$\iptaction{Accept}$, $\iptaction{Drop}$, $\iptaction{Reject}$, $\iptaction{Log}$, $\iptaction{Goto}$, $\iptaction{Call}$ing to and $\iptaction{Return}$ing from user-defined chains, as well as the ``empty'' action.
%
%
%
The model is defined as an inductive predicate with the following syntax:
\begin{IEEEeqnarray*}{c}
\bigstep{\mathit{rs}}{s}{t}
\end{IEEEeqnarray*}
The ruleset of the firewall is $\mathit{rs}$ and the packet under examination is $p$. 
The states $s$ and $t$ are in $\lbrace\allow,\, \deny,\, \undecided \rbrace$.
The starting state of the firewall is $s$, usually \undecided.
The filtering decision after processing $\mathit{rs}$ is $t$, usually $\allow$ or $\deny$.
User-defined chains are stored in $\Gamma$, which corresponds to the background ruleset.
A primitive matcher $\gamma$ (a boolean function which takes a primitive and the packet as parameters) decides whether a certain primitive matches for a packet. 
Note that the model and all algorithms on top of it are proven correct for an arbitrary $\gamma$, hence, this model supports \emph{all} iptables matching features. 
Obviously, there is no executable code for an arbitrary $\gamma$. 
However, the algorithms which transform rulesets are executable. 

We make use of these algorithms, in particular: 
An algorithm which unfolds all calls to and returns from user-defined chains and rewriting of further actions. 
This leaves a ruleset where only the following actions occur: $\iptaction{Accept}$ and $\iptaction{Drop}$.\footnote{rewrite-Goto-chain-safe and unfold-optimize-common-matcher-univ-ruleset-CHAIN} 
Thus, a large step for translating the real-world model to the simple firewall model is already accomplished. 
Translating the match expressions remains.
The real-world model allows a match expression to be an arbitrary propositional logic expression. 
However, iptables only accepts match expressions in \emph{negation normal form} (NNF). 
A Boolean formula is in NNF \emph{iff} all occurring negations are on primitives, 
\ie there are no nested negated expressions. 
For example, iptables can load \texttt{-s\ 10.0.0.0/8 !\;-p\ tcp} but not \texttt{!\;(-s\ 10.0.0.0/8 -p\ tcp)}. 
However, such negated expressions may occur as a result of the unfolding algorithm. 
We already presented an algorithm to translate a ruleset to a ruleset where all match conditions are in NNF.\footnote{NNF normalizing may create additional rules. } 
However, there is an additional constraint imposed by iptables, not solved by the algorithm: A primitive must only occur at most once. 
This problem will be addressed in this Chapter. 

We have implemented a subset of $\gamma$, namely for all primitives supported by the simple firewall and some further primitives, detailed in Section~\ref{sec:translating-primitives}. 
Chapter~\ref{chap:fm15} provides an algorithm to abstract over all `unknown' primitives which are not understood by our subset implementation of $\gamma$. 
This algorithm leads to an approximation of the ruleset. 
It can either be an overapproximation which results in a more permissive ruleset, or an underapproximation, which results in a stricter ruleset.
For the sake of example, we will only consider the overapproximation in this chapter, the underapproximation is analogous and can be found in our formalization.

Since stateful firewalls usually accept all packets which belong to an \texttt{ESTABLISHED} connection, the interesting access control rules in a ruleset only apply to \texttt{NEW} packets. 
We only consider \texttt{NEW} packets, \ie \texttt{{-}{-}ctstate NEW} and \texttt{{-}{-}syn} for TCP packets.
Our first goal is to translate a ruleset from the real-world model to the simple model. 
We have proven\footnote{new-packets-to-simple-firewall-overapproximation, new-packets-to-simple-firewall-underapproximation} that the set of new packets accepted by the simple firewall is a superset (overapproximation) of the packets accepted by the real-world model. 
This is a core contribution and we detail on the translation in the following section.

\begin{theorem}[Translation to simple firewall model]
\label{thm:to-simple-upper}
\begin{IEEEeqnarray*}{c}
  \left\lbrace p.\ \ \mfun{new}\ p \, \wedge \, \bigstep{\mathit{rs}}{\undecided}{\allow}\right\rbrace \\
  \subseteq\\
  \left\lbrace p.\ \ \mfun{new}\ p \, \wedge \,  \mfun{simple\mhyphen{}fw}\ (\mfun{translate\mhyphen{}oapprox}\ \mathit{rs})\ p = \allow \right\rbrace 
\end{IEEEeqnarray*}
\end{theorem}

Any packet dropped by the translated, overapproximated simple firewall ruleset is guaranteed to be dropped by the real-world firewall, for arbitrary $\gamma$, $\Gamma$, $\mathit{rs}$.
Similar guarantees for certainly accepted packets can be given by considering the translated underapproximation. 
Given the simple and carefully designed model of the $\mfun{simple\mhyphen{}fw}$, it is much easier to write algorithms to analyze and verify the translated rulesets. 


\begin{example}
\textbf{Example. }%
%
%
We consider a \texttt{FORWARD} chain with a default policy of \iptaction{DROP} and a user-defined chain \texttt{foo}.
\smallskip

\begin{minipage}{.8\linewidth}
\small
\begin{Verbatim}[commandchars=\\\{\},codes={\catcode`$=3\catcode`^=7}]
-P FORWARD DROP
-A FORWARD -s 10.0.0.0/8 -j foo
-A foo ! -s 10.0.0.0/9 -j DROP
-A foo -p tcp -j ACCEPT
\end{Verbatim}
\end{minipage}
\smallskip

This ruleset, though it only consist of three rules and a default policy, is complicated to analyze.
Our translation algorithm translates it to the simple firewall model, where the ruleset becomes remarkably simple. 
We use $*$ to denote a wildcard: 
\begin{IEEEeqnarray*}{rCcCCCCCcCCCCClcl}
( & * & , & * & , & \texttt{10.128.0.0/9} & , & * & , & *            & , & * & , & * & ) & \hspace*{1em} & \iptaction{DROP} \\
( & * & , & * & , & \texttt{10.0.0.0/8}   & , & * & , & \texttt{TCP} & , & * & , & * & ) &               & \iptaction{ACCEPT} \\
( & * & , & * & , & *            & , & * & , & *            & , & * & , & * & ) &               & \iptaction{DROP}
\end{IEEEeqnarray*}
No over- or underapproximation occurred since all primitives could be translated. 
Note the 10.128.0.0/9 address.
\end{example}

\section{Translating Primitives}
\label{sec:translating-primitives}
A firewall has the same behavior for two rulesets $\mathit{rs}_1$ and $\mathit{rs}_2$\phantom{,} \emph{iff} for all packets, the firewall computes the same filtering decision for $\mathit{rs}_1$ and $\mathit{rs}_2$. 
Formally, 
\begin{IEEEeqnarray*}{c}
\forall p\ s\ t.\ \ \bigstep{\mathit{rs}_1}{s}{t} \ \ \longleftrightarrow \ \  \bigstep{\mathit{rs}_2}{s}{t}
\end{IEEEeqnarray*}%
In this section, we present algorithms to transform an arbitrary $\mathit{rs_1}$ to $\mathit{rs_2}$ without changing the behavior of the firewall.\footnote{All lemmas and results of the following subsections ultimately yield Theorem~\ref{thm:to-simple-upper} and are referenced in its proof.} 
In the resulting $\mathit{rs_2}$, all primitives will be normalized such that the translation to the $\mfun{simple\mhyphen{}fw}$ is obvious. 
We continue by describing the normalization of all common primitives found in iptables rulesets.

\subsection{IPv4 Addresses}
According to Nelson~\cite{nelson2010margrave}, ``[m]odeling IP addresses efficiently is challenging.'' 
First, we present a datatype to efficiently perform set operations on intervals of machine words, \eg \SI{32}{\bit} integers. 
We will use this type for IPv4 addresses, but we have generalized to machine words of arbitrary length, \eg IPv6 addresses or L4 ports. 
For the sake of brevity, we will present our formalization at the example of IPv4. 
We call our datatype a word interval ($\mathit{wi}$), and $\mconstr{WI}\ \mathit{start}\ \mathit{end}$ describes the interval with $\mathit{start}$ and $\mathit{end}$ inclusive. 
The $\mconstr{Union}$ of two $\mathit{wi}$s is defined recursively. 
\begin{IEEEeqnarray*}{l}
  \mctrl{datatype}\ \mathit{wi} = \mconstr{WI}\ \mathit{word}\ \mathit{word} \ | \ \mconstr{Union}\ \mathit{wi} \ \mathit{wi}
\end{IEEEeqnarray*}
Let $\mfun{set}$ denote the interpretation into mathematical sets, then $\mathit{wi}$ has the following semantics: 
$\mfun{set}\ (\mconstr{WI}\ \mathit{start}\ \mathit{end}) = \lbrace \mathit{start} .. \mathit{end} \rbrace$ and 
$\mfun{set}\ \mconstr{Union}\ \mathit{wi}_1\ \mathit{wi}_2 = \mfun{set}\ (\mathit{wi}_1) \cup \mfun{set}\ (\mathit{wi}_2)$.

An IP address in CIDR notation or IP addresses specified by \eg \texttt{-m iprange} can be translated to one $\mconstr{WI}$. 
We have implemented and proven the common set operations:
`$\cup$', `$\lbrace\rbrace$', `$\setminus$', `$\cap$', `$\subseteq$', and `$=$'.
These operations are linear in the number of $\mconstr{Union}$-constructors.
The result is optimized by merging adjacent and overlapping intervals and removing empty intervals. 
We can also represent `$\mconstr{UNIV}$' (the universe of all IP addresses). 
Since most rulesets use IP addresses in CIDR notation or intervals in general, the $\mathit{wi}$ datatype has proven to be very efficient.
Recall that the intersection of two intervals, constructed from addresses in CIDR notation, is either empty or the smaller of both intervals.\footnote{ipcidr-conjunct-correct}

The datatype $\mathit{wi}$ is an internal representation and for the simple firewall, the result needs to be represented in CIDR notation.
For this direction, one $\mconstr{WI}$ may correspond to several CIDR ranges.
We describe an algorithm to $\mfun{split}$ off one CIDR range from an arbitrary word interval $r$.
The output is a CIDR range and $r^\prime$, the remainder after splitting off this CIDR range. 
$\mfun{split}$ is implemented as follows: 
Let $\mathit{a}$ be the lowest element in $r$.
If this does not exist, then $r$ corresponds to the empty set and the algorithm terminates.
Otherwise, we construct the list of CIDR ranges $[a/0, a/1, ..., a/32]$.
The first element in the list which is well-formed (\ie all bits after the network prefix must be zero) and which is a subset of $r$ is the wanted element.
Note that this element always exists. 
It is subtracted from $r$ to obtain $r^\prime$.
To convert $r$ completely to a list of CIDR ranges, this is applied recursively until it yields no more results.
This algorithm is guaranteed to terminate and the resulting list in CIDR notation corresponds to the same set of IP addresses as represented by $r$.\footnote{cidr-split-prefix} 
Formally, 
$\bigcup \mfun{map}\ \mfun{set}\ (\mfun{split}\ r) = \mfun{set}\ r$. 

\begin{sloppypar}
For example, 
$\mfun{split}\ (\mconstr{WI}\; \textnormal{10.0.0.0}\; \textnormal{10.0.0.15}) = [\textnormal{10.0.0.0/28}]$ and 
$\mfun{split}\ (\mconstr{WI}\ \textnormal{10.0.0.1}\ \textnormal{10.0.0.15}) = [\textnormal{10.0.0.1/32}, \textnormal{10.0.0.2/31}, \textnormal{10.0.0.4/30}, \textnormal{10.0.0.8/29}]$. 
\end{sloppypar}

With the help of these functions, arbitrary IP address ranges can be translated to the format required by the simple firewall. 
The following is applied to matches on src and dst IP addresses: 
First, the IP match expression is translated to a word interval. 
If the match on an IP range is negated, we compute $\mdef{UNIV} \setminus \mathit{wi}$. 
All matches in one rule can be joined to a single word interval, using the $\cap$ operation. 
The resulting word interval is translated to a set of non-negated CIDR ranges. 
Using the NNF normalization, at most one match on an IP range in CIDR notation remains.
We have proven that this process preserves the firewall's filtering behavior. 

\begin{sloppypar}
We conclude with a simple, synthetic worst-case example. 
The evaluation shows that this worst-case does not prevent successful analysis: 
\texttt{-m iprange {-}{-}src-range 0.0.0.1-255.255.255.254}. 
Translated to the simple firewall, this one range blows up to 62 ranges in CIDR notation.
A similar blowup may occur for negated IP ranges. 
\end{sloppypar}


\begin{sloppypar}
Note that, while pretty printing IPv4 addresses in dotecimal notation (\ie \verb~<dotnum> ::= <snum> "." <snum> "." <snum> "." <snum>~~\cite{rfc780}) is simple, 
pretty printing IPv6 addresses is non-trivial~\cite{rfc5952} and our implementation contains the first formally, machine-verified IPv6 pretty printer~\cite{IP_Addresses-AFP}. 
\end{sloppypar}

\subsection{Conntrack State}
\label{subsec:ifip:ctstate}
%
%
If a packet $p$ is matched against the stateful match condition \texttt{ESTABLISHED}, conntrack looks up $p$ in its state table. 
When the firewall comes to a filtering decision for $p$, if the packet is not dropped and the state was \texttt{NEW}, the conntrack state table is updated such that the flow of $p$ is now \texttt{ESTALISHED}. 
Similarly, other conntrack states are handled. 

We present an alternative model for this behavior:
Before the firewall starts processing the ruleset for $p$, the conntrack state table is consulted for the state of the connection of $p$.
This state is added as a (phantom) tag to $p$.
Therefore, ctstate can be modeled as just another header field of $p$. 
When processing the ruleset, it is not necessary to inspect the conntrack table but only the virtual state tag of the packet.
After processing, the state table is updated accordingly.

We have proven that both models are equivalent.\footnote{Semantics\_Stateful.thy} 
The latter model is simpler for analysis purposes since the conntrack state can be considered an ordinary packet field.\footnote{This holds because the semantics does not modify a packet during filtering. }  

In Theorem~\ref{thm:to-simple-upper}, we are only interested in \texttt{NEW} packets. 
In contrast to previous work, there is no longer the need to manually exclude \texttt{ESTABLISHED} rules from a ruleset. 
The alternative model allows us to consider only \texttt{NEW} packets: all state matches can be removed (by being pre-evaluated for an arbitrary \texttt{NEW} packet) from the ruleset without changing the filtering behavior of the firewall.

\subsection{Layer 4 Ports}
Translating singleton ports or intervals of ports to the simple firewall is straightforward. 
A challenge remains for negated port ranges and the \texttt{multiport} module. 
However, the word interval type is also applicable to \SI{16}{\bit} machine words and solves these challenges. 
%
%
%
%
For ports, there is no need to translate an interval back to CIDR notation.\footnote{ 
As a side note, OpenFlow (technically, the Open vSwitch) defines CIDR-like matching for L4 ports. 
With the small change of converting ports to CIDR-like notation, our simple firewall can be directly converted to OpenFlow and we have the first (almost) fully verified translation of iptables rulesets to SDN.} 

\paragraph*{Bug in the Original Paper}
\label{par:ifip:portsbug}
We made a serious mistake~\cite{iptablesgithub113issue} when specifying the semantics of matches on ports (which also made it into the camera ready paper~\cite{diekmann2016networking}). 
Fortunately, the mistake only occurs in corner cases and did not affect the published evaluation.\footnote{However, we have seen rulesets in the wild which triggered the bug, hence, it is not purely of academic nature.} 
The error is now fixed. 

Since we have proven the correctness of all our algorithms and checked all assumptions, the bug did not exist in the code. 
The bug exists in the model. 
More precisely, we formalized a semantics of ports which does not correspond to reality. 
We defined the datatype of a source port match as follows: 
\begin{IEEEeqnarray*}{l}
  \mctrl{datatype}\ \mathit{src\mhyphen{}ports} = \mconstr{SrcPorts}\ \ \mlq{}16\ \mathit{word} \times 16\ \mathit{word}\mrq{}
\end{IEEEeqnarray*}

This datatype describes a source port match as an interval of \SI{16}{\bit} port numbers. 
The match semantics for a packet were defined such that the source port of the packet must be in the interval. 
For example, packet $p$ matches $\mconstr{SrcPorts}\ a\ b$ iff $m.\mathrm{src\mhyphen{}port} \in \lbrace a .. b \rbrace$. 
We defined $\mconstr{DstPorts}$ analogously. 

With these semantics, we can construct a corner case which describes why these semantics do not correspond to reality. 
We consider the following firewall. 
\bigskip

\begin{minipage}{.9\linewidth}
\small
\begin{Verbatim}[commandchars=\\\{\},codes={\catcode`$=3\catcode`^=7}]
*filter
:FORWARD ACCEPT [0:0]
:CHAIN - [0:0]
-A FORWARD -j CHAIN
-A CHAIN -p tcp -m tcp --sport 22 -j RETURN
-A CHAIN -p udp -m udp --dport 80 -j RETURN
-A CHAIN -j DROP
COMMIT
\end{Verbatim}
\end{minipage}
\bigskip

The firewall in \texttt{iptables-save} format shows the \texttt{filter} table, which consists of the two chains \texttt{FORWARD} and \texttt{CHAIN}. 
The \texttt{FORWARD} chain is built-in and has a default policy if \texttt{ACCEPT} here. 
Starting at the \texttt{FORWARD} chain, any packet which is processed by this firewall is directly sent to the user-defined chain \texttt{CHAIN} first. 
A packet can only \texttt{RETURN} if it is a tcp packet with source port 22 or a udp packet destination port 80. 
All other packets are dropped. 
Hence, this firewall expresses in a complicated way the following policy: \textit{``Drop everything which is not tcp src port 22 or udp dst port 80''}. 
This ruleset, though it does not have an obvious use, was artificially constructed to demonstrate our bug. 
Our tool has `simplified' the ruleset the following:

\begin{IEEEeqnarray*}{rCcCCCCCcCCCCClcl}
( & * & , & * & , & * & , & * & , & *            & , & 0:21\; & ,\ & 0:79 & ) & \hspace*{1em} & \iptaction{DROP} \\
( & * & , & * & , & * & , & * & , & *            & , & 0:21\; & ,\ & 81:65535 & ) & \hspace*{1em} & \iptaction{DROP} \\
( & * & , & * & , & * & , & * & , & *            & , & 23:65535\; & ,\ & 0:79 & ) & \hspace*{1em} & \iptaction{DROP} \\
( & * & , & * & , & * & , & * & , & *            & , & 23:65535\; & ,\ & 81:65535 & ) & \hspace*{1em} & \iptaction{DROP} \\
( & * & , & * & , & * & , & * & , & * & , & * & , & * & ) &               & \iptaction{ACCEPT} \\
\end{IEEEeqnarray*}

Given our semantics, the simplification is correct. 
In reality, this simple firewall is wrong for various reasons. 
First, it is not well-formed, \ie it tries to match on ports without specifying a protocol. 
Second, it has mixed up udp and tcp ports. 

The problem lies in our semantics of $\mconstr{SrcPorts}$ and $\mconstr{DstPorts}$. 
Basically, there is no such a thing as `ports'. 
Yet, there exist tcp ports, udp ports, sctp ports, \dots

We have fixed the issue by including the protocol in the match for a port: 
\begin{IEEEeqnarray*}{l}
  \mctrl{datatype}\ \mathit{src\mhyphen{}ports} = \mconstr{SrcPorts}\ \ \mlq{}8\ \mathit{word}\mrq{}\ \ \ \mlq{}16\ \mathit{word} \times 16\ \mathit{word}\mrq{}
\end{IEEEeqnarray*}

The $8\ \mathit{word}$ corresponds to the \texttt{protocol} field in IPv4~\cite{rfc791}, respectively the \texttt{Next Header} field in IPv6~\cite{rfc2460}, identifying protocols by their assigned numbers~\cite{rfc1700,rfc3232}. 
It does not allow a wildcard. 
The semantics defines that the protocol of a packet must be the same as specified in the datatype and that the source port must be in the interval (as in the first definition). 

With the corrected semantics, our tool computes the correct and expected result: 

\begin{IEEEeqnarray*}{rCcCCCCCcCCCCClcl}
( & * & , & * & , & * & , & * & , & \;\texttt{UDP}\; & , & * & ,\ & 0:79 & ) & \hspace*{1em} & \iptaction{DROP} \\
( & * & , & * & , & * & , & * & , & \texttt{UDP} & , & * & ,\ & 81:65535 & ) & \hspace*{1em} & \iptaction{DROP} \\
( & * & , & * & , & * & , & * & , & \texttt{TCP} & , & 0:21\; & ,\ & * & ) & \hspace*{1em} & \iptaction{DROP} \\
( & * & , & * & , & * & , & * & , & \texttt{TCP} & , & 23:65535\; & ,\ & * & ) & \hspace*{1em} & \iptaction{DROP} \\
( & * & , & * & , & * & , & * & , & * & , & * & , & * & ) &               & \iptaction{ACCEPT} \\
\end{IEEEeqnarray*}

The negation of a match on ports is the interesting corner case to which the presented problems can be reduced to. 
We will illustrate the issue by a simpler example. 
Assuming we have one rule which tries to accept every packet which is not udp destination port 80.\footnote{Note that this cannot be expressed directly in one rule with iptables. In the example, we used the semantics of \iptaction{RETURN} to construct a compound negated match expression. } 
For simplicity, we assume we have one rule as follows: 
\verb~! (-p udp --dport 80) -j ACCEPT~. 
Semantically, to unfold this negation, the rule matches either everything which is not udp or everything which is udp but not destination port 80. 
It can be expressed with the following two rules: 
\verb~! -p udp -j ACCEPT~ followed by \verb~-p udp ! --dport80 -j ACCEPT~. 
We use this strategy in our tool to unfold the negation of matches on ports. 
Note the type dependencies which occur: Negating one rule that matches on ports yields both a rule which matches on protocols and one rule which matches on ports. 

This example also shows that any tool which reduces match conditions to a flat bit vector is either buggy (it loses the protocol which belongs to a match on ports) or cannot support complicated negations. 
This includes tools which reduce firewall analysis to SAT~\cite{fwmodelchecking2009jeffry} or BDDs~\cite{fireman2006,alshaer2011configcheckershort}. 
It may probably also affect ITVal~\cite{marmorstein2005itval} which relies on multi-way decision diagrams (MDD). 
This was also the case for our $\bigstep{\mathit{rs}}{s}{t}$ semantics with the buggy $\gamma$ described in this paragraph. 
Our simple firewall model does not allow complicated negations and we have proven that the match conditions are always well-formed, hence, the presented class of errors cannot occur there.

\subsection{TCP Flags}
Iptables can match on a set of L4 flags. 
To match on flags, a $\mathit{mask}$ selects the corresponding flags and $c$ declares the flags which must be present. 
For example, the match \texttt{{-}{-}syn} is a synonym for $\mathit{mask}=\mathtt{SYN,RST,ACK,FIN}$ and $c=\mathtt{SYN}$. 
For a set $f$ of flags in a packet, matching can be formalized as $(f \cap \mathit{mask}) = c$. 
If $c$ is not a subset of $\mathit{mask}$, the expression cannot match; we call this the empty match. 
We proved that two matches $(\mathit{mask}_1, c_1)$ and $(\mathit{mask}_2, c_2)$ are equal if and only if 
$( \mctrl{if}\ c_1 \subseteq \mathit{mask}_1 \wedge c_2 \subseteq \mathit{mask}_2 \
          \mctrl{then}\ c_1 = c_2 \wedge \mathit{mask}_1 = \mathit{mask}_2 \ \mctrl{else}\ (\neg c_1 \subseteq \mathit{mask}_1) \wedge (\neg c_2 \subseteq \mathit{mask}_2))$ holds. 
We also proved that the conjunction of two matches is exactly 
$( \mctrl{if}\ c_1 \subseteq \mathit{mask}_1 \wedge c_2 \subseteq \mathit{mask}_2 \wedge \mathit{mask}_1 \cap \mathit{mask}_2 \cap c_1 = \mathit{mask}_1 \cap \mathit{mask}_2 \cap c_2\  \mctrl{then}\ (\mathit{mask}_1 \cup \mathit{mask}_2,\, c_1 \cup c_2) \ \mctrl{else}\ \mathtt{empty})$. 
If we assume \texttt{{-}{-}syn} for a packet, we can remove all matches which are equal to \texttt{{-}{-}syn} and add the \texttt{{-}{-}syn} match as conjunction to all other matches on flags and remove empty matches. 
Some matches on flags may remain, \eg $\mathtt{URG}$, which need to be abstracted over later.

\subsection{Interfaces}
The simple firewall model does not support negated interfaces, \eg \texttt{!\;{-}i eth+}.
Therefore, they must be removed. 
We first motivate the need for abstracting over negated interfaces. 

For whitelisting scenarios, one might argue, that negated interfaces is bad practice anyway. 
This is because new (virtual) interfaces might be added to the system at runtime and a match on negated interfaces might now also include these new interfaces. 
Therefore, it can be argued that negated interfaces correspond to blacklisting, which is not recommended for most firewalls.
However, the main reason why negated interfaces are not supported by our model is of technical nature:
Let $\mfun{set}$ denote the set of interfaces that match an interface expression.
For example, $\mfun{set}\ \texttt{eth0} = \lbrace\texttt{eth0}\rbrace$ and $\mfun{set}\ \texttt{eth+}$ is the set of all interfaces that start with the prefix \texttt{eth}.
If the match on $\texttt{eth+}$ is negated, then it matches all strings in the complement set: $\mdef{UNIV} \setminus (\mfun{set}\ \texttt{eth+})$.
The simple firewall model requires that a conjunction of two primitives is again at most one primitive. 
This can obviously not be achieved with such sets. 
In addition, working with negated interfaces can cause great confusion. 
Note that the interface match condition `\texttt{+}' matches any interfaces.
Also note that $\textnormal{`\texttt{+}'} \in \mdef{UNIV} \setminus (\mfun{set}\ \texttt{eth+})$.
In the second equation, `\texttt{+}' is not a wildcard character but the name of an interface.
The confusion introduced by negated interfaces becomes more apparent when one realizes that `\texttt{+}' can occur as both wildcard character and normal character.
Therefore, it is not possible to construct an interface match condition which matches exactly on the interface `\texttt{+}', because a `\texttt{+}' at the end of an interface match condition is interpreted as wildcard.\footnote{We strongly discourage the use of ``\texttt{ip link set eth0 name +}'' in production. Please fix your container startup scripts with untrusted input now!} 
While technically, the Linux kernel would allow to match on `\texttt{+}' as a normal character~\cite{kernelmatchinterface}, the \texttt{iptables} userland command does not permit to construct such a match~\cite{xtablesmtach}.

\subsubsection*{Correlating with IP Ranges}
\label{sec:ifcerewrite}
Later, in Section~\ref{sec:ip-partition}, we will compute an IP address space partition. 
For best clarity, this partition must not be `polluted' with interface information. 
Therefore, for the partition, we will assume that no matches on interfaces occur in the ruleset. 
In this subsection, we describe a method to get rid of both, negated and non-negated interfaces while preserving their relation to IP address ranges. 

Input Interfaces are usually assigned an IP range of valid source IPs which are expected to arrive on that interface. 
Let $\mathit{ipassmt}$ be a mapping from interfaces to an IP address range.
This information can be obtained by \texttt{ip route} and \texttt{ip addr}. 
We will write $\mathit{ipassmt}[i]$ to get the corresponding IP range of interface $i$.
For the following examples, we assume 
\begin{IEEEeqnarray*}{l}%
\mathit{ipassmt} = [\mathtt{eth0} \mapsto \lbrace \textnormal{10.8.0.0/16} \rbrace ]
\end{IEEEeqnarray*}%
The goal is to rewrite input interfaces with the corresponding source IP range.
For example, we would like to replace all occurrences of \texttt{{-}i eth0} with \texttt{{-}s 10.8.0.0/16}.
This idea can only be sound if there are no spoofed packets; we only expect packets with a source IP of \texttt{10.8.0.0/16} to arrive at \texttt{eth0}. 
Once we have assured that the firewall blocks spoofed packets, we can assume in a second step that there are no spoofed accepted packets left. 
By default, the Linux kernel offers reverse path filtering, which blocks spoofed packet automatically.
In this case we can assume that no spoofed packets occur.
In some complex scenarios, reverse path filtering needs to be disabled and spoofed packets should be blocked manually with the help of the firewall ruleset.
In the previous Chapter~\ref{chap:nospoof}, we presented an algorithm to verify that a ruleset correctly blocks spoofed packets. 
This algorithm is integrated in our framework, proven sound, works on the same $\mathit{ipassmt}$ and does not need the simple firewall model (\ie supports negated interfaces).
If some interface $i$ should accept arbitrary IP addresses (essentially not providing spoofing protection), it is possible to set $\mathit{ipassmt}[i] = \mdef{UNIV}$. 
Therefore, we can verify spoofing protection according to $\mathit{ipassmt}$ at runtime and afterwards continue with the assumption that no spoofed packets occur.

Under the assumption that no spoofed packets occur, we will now present two algorithms to relate an input interface $i$ to $\mathit{ipassmt}[i]$. 
Both approaches are valid for negated and non-negated interfaces. 
Approach one provides better results but requires stronger assumptions (which can be checked at runtime), whereas approach two is applicable without further assumptions.

\paragraph*{Approach One}
In general, it is considered bad practice~\cite{firwallerr2004,wool2004use} to have zone-spanning interfaces. 
Two interfaces are zone-spanning if they share a common, overlapping IP address range. 
Mathematically, absence of zone-spanning interfaces means that for any two interfaces in $\mathit{ipassmt}$, their assigned IP range must be disjoint.
Our tool emits a warning if $\mathit{ipassmt}$ contains zone-spanning interfaces. 
If absence of zone-spanning interfaces is checked, then all input interfaces can be replaced by their assigned source IP address range. 
This preserves exactly the behavior of the firewall.
The idea is that in this case a bidirectional mapping between input interfaces and source IPs exists.
Interestingly, our proof does not need the assumption that $\mathit{ipassmt}$ maps to the complete IP universe.

\paragraph*{Approach Two}
Unfortunately, though considered bad practice, we found many zone-spanning interfaces in many real-world rulesets and hence cannot apply the previous algorithm. 
First, we proved that correctness of the described rewriting algorithm implies lack of zone-spanning interfaces.\footnote{iface-replace-needs-ipassmt-disjoint}
This leads to the conclusion that it is impossible to perform rewriting without this assumption.
Therefore, we present an algorithm which adds the IP range information to the ruleset (without removing the interface match), thus constraining the match on input interfaces to their IP range.
The algorithm computes the following: 
Whenever there is a match on an input interface $i$, the algorithm looks up the corresponding IP range of that interface and adds \texttt{{-}s $\mathit{ipassmt}[i]$} to the rule. 
To prove correctness of this algorithm, no assumption about zone-spanning interfaces is needed, $\mathit{ipassmt}$ may only be defined for a subset of the interfaces, and the range of $\mathit{ipassmt}$ may not cover the complete IP universe. 
Consequently, there is no need for a user to specify $\mathit{ipassmt}$, but having it may yield more accurate results.

\paragraph*{Output Port Rewriting}
Our presented approaches for input interface rewriting can be generalized to also support output interface (\texttt{-o}) rewriting. 
The very core idea is to replace a match on an output interface by the corresponding IP address range which is determined by the system's routing table. 
To do this, we parse the routing table, map it to a relation (which provides a structure which is independent of its order), and compute the inverse of the relation. 
This ultimately provides a mapping for each interface and its corresponding IP address range. 

This computed mapping is very similar to the $\mathit{ipassmt}$. 
In fact, we found it to be a helpful debugging tool to compare the inverse routing relation to an $\mathit{ipassmt}$. 
For convenience, we also provide a function to compute an $\mathit{ipassmt}$ from a routing table. 

Essentially, computing the inverse routing relation semantically is the same behavior as found in strict reverse path filtering~\cite{rfc3704}. 
We have formally proven\footnote{Routing/rpf-strict-correct} this observation. 

Because a routing table may change frequently, even triggered by external malicious routing advertisements, by default, we refrain from output port rewriting in this work. 
It is not applied for Table~\ref{tab:somenumbers}; however, we additionally show how the results for firewall \textbf{D} in Section~\ref{sec:evaluation} will improve with its help. 

\subsection{Abstracting Over Primitives}
Some primitives cannot be translated to the simple model.
Chapter~\ref{chap:fm15} already provides the function $\mfun{pu}$ which removes all unknown match conditions. 
This leads to an approximation and is the main reason for the `$\subseteq$' relation in Theorem~\ref{thm:to-simple-upper}.
We found that we can also rewrite any known primitive \emph{at any time} to an unknown primitive.
This can be used to apply additional knowledge during preprocessing. 
For example, since we understand flags, we know that the following condition is false, hence rules using it can be removed: \texttt{{-}{-}syn $\wedge$ {-}{-}tcp-flags RST,ACK RST}. 
After this optimization, all remaining flags can be treated as unknowns and abstracted over afterwards.
This allows to easily add additional knowledge and optimization strategies for further primitive match conditions without the need to adapt any algorithm which works on the simple firewall model.
We proved soundness of this approach:  
The `$\subseteq$' relation in Theorem~\ref{thm:to-simple-upper} is preserved.

\section{IP Address Space Partition}
\label{sec:ip-partition}
In the following sections, we will work on rulesets translated to the $\mfun{simple\mhyphen{}fw}$ model. 
In this section, we will compute a partition of the IP address space. 
Our algorithms works for IP addresses of arbitrary length, in particular IPv4 and IPv6. 
For the sake of example, we will present only IPv4 here. 
All IP addresses in the same partition must show the same behavior w.r.t\ the firewall ruleset. 
We do not require that the partition is minimal. 
Therefore, the following would be a valid solution: 
$\left\lbrace \left\lbrace \textnormal{0} \right\rbrace,\ \left\lbrace \textnormal{1} \right\rbrace,\ \dots,\ \left\lbrace \textnormal{255.255.255.255} \right\rbrace \right\rbrace$. 
However, we will need the partition as starting point for a further algorithm and a partition of size ${2}^{32}$ is too large for this purpose and infeasible for IPv6.  
In this section, we will present an algorithm to compute a partition which behaves roughly linear in the number of rules for real-world rulesets.
First, we motivate the partitioning idea with the following observation. 
\begin{lemma}
\label{thm:partset}
For an arbitrary packet $p$, we write $p(src \mapsto s)$ to fix the src IP address to $s$. 
Let $X$ be the set of all src IP matches specified in $\mathit{rs}$, \ie $X$ is a set of CIDR ranges. 
If
\begin{IEEEeqnarray*}{c}
\forall A \in X.\  B \subseteq A \vee B \cap A = \lbrace\rbrace
\end{IEEEeqnarray*}
then let $s_1 \in B$ and $s_2 \in B$ then
\begin{IEEEeqnarray*}{c}
\mfun{simple\mhyphen{}fw}\ \mathit{rs}\ p(src \mapsto s_1) \ = \ \mfun{simple\mhyphen{}fw}\ \mathit{rs}\ p(src \mapsto s_2)
\end{IEEEeqnarray*}
\end{lemma}
%
Reading the lemma backwards, it states that all packets with arbitrary source IPs picked from $B$ are treated equally by the firewall. 
Therefore, $B$ is a member of an IP address range partition. 
The condition imposed on $B$ is that for all src CIDR ranges specified in the ruleset (called $A$ in the lemma), $B$ is either a subset of the range or disjoint.
The lemma shows that this condition is sufficient for $B$, therefore we will construct an algorithm to compute $B$. 
For an arbitrary set $X$, this condition is purely set-theoretic and we can solve it independently from the firewall theory.

%
%
%
%

For simplicity, we use finite sets and lists interchangeably.
We will write an algorithm $\mfun{part}$ and reuse the common list algorithm from functional programming $\mfun{foldr}$. 
For $X$, the following algorithm computes a partition: $\mfun{foldr}\ \mfun{part}\ X\ \lbrace \mdef{UNIV} \rbrace$. 
In addition, it is guaranteed that the union of the resulting partition is equal to the universe. 
For our scenario, this means that the partitioning covers the complete IPv4 space. 
The algorithm $\mfun{part}$ is implemented as follows: 
The first parameter is a set $S \in X$, the second parameter $\mathit{TS}$ is a set of sets and corresponds to the remaining set which will be partitioned. 
In the first call $\mathit{TS} = \lbrace \mdef{UNIV} \rbrace$. 
For a fixed $S$, $\mfun{part} \ S \ \mathit{TS}$ iterates over $\mathit{TS}$ and splits the set such that the precondition of Lemma~\ref{thm:partset} holds: 
Written as recursive function: 
$\mfun{part} \ S \ (\lbrace T \rbrace \cup \mathit{TS}) = (S \cap T) \cup (T \setminus S)  \cup (\mfun{part}\ (S \setminus T)\ \mathit{TS})$

The result size of calling $\mfun{part}$ once can be up to two times the size of $\mathit{TS}$. 
This means, the partition of a complete firewall ruleset is in $O(2^{\vert\mathit{rules}\vert})$. 
However, the empirical evaluation shows that the resulting size for real-world rulesets is much better. 
This is because IP address ranges may overlap in a ruleset, but they do not overlap in the worst possible way for all pairs of rules. 
Consequently, at least one of the sets $S \cap T$ or $T \setminus S$ is usually empty and can be optimized away. 
For example, for our largest firewall, the number of computed partitions is 10 times smaller than the number of rules. 
Table~\ref{tab:somenumbers} confirms that the number of partitions is usually less than the number of rules. 

%
%
%
%

Our algorithm fulfills the assumption of Lemma~\ref{thm:partset} for arbitrary $X$. 
Because IP addresses occur as source and destination in a ruleset, we use our partitioning algorithm where $X$ is the set of all IPs found in the ruleset. 
The result is a partition where for any two IPs in the same partition, setting the src or dst of an arbitrary packet to one of the two IPs, the firewall behaves equally. 
This results in a stronger version of Lemma~\ref{thm:partset}, which holds without any assumption and also holds for both src and dst IPs simultaneously.\footnote{getParts-samefw} 
In addition, the partition covers the complete IPv4 address space.\footnote{getParts-complete} 


\section{Service Matrices}
\label{sec:matrix}
The IP address space partition may not be minimal. 
That means, two different partitions may exhibit exactly the same behavior. 
Therefore, for manual firewall verification, these partitions may be misleading. 
Marmorstein elaborates on this problem~\cite{marmorstein2006firewall}. 
ITVal's solution is to minimize the partition. 
We suggest to minimize the partition for a fixed service. 
The evaluation shows that the result is smaller and thus clearer.

A fixed service corresponds to a fixed packet with arbitrary IPs. 
For example, we can define ssh as TCP, dport 22, arbitrary sport $\geq$ 1024.
A service matrix describes the allowed accesses for a specific service over the complete IPv4 address space. 
It can be visualized as graph; for example, the ruleset of Figure~\ref{fig:ifipfffuuexample} is visualized in Figure~\ref{fig:ifipfffuuu:matrix}. 
An example of a firewall with several thousands of rules is shown in Figure~\ref{fig:tumssh}. 
For clarity, this figure uses symbolic names (\eg $\mathit{servers}$) instead of IP addresses. 
The raw IP addresses can be found in Figure~\ref{fig:tumsshwithips}. 
More complicated examples with highly fragmented IP ranges are shown in Figure~\ref{fig:eval_measrdroid:i8fw:port80} and Figure~\ref{fig:eval_measrdroid:i8fw:port80ipv6}; actually they are from the same firewall at a later point in time.  
All matrices are minimal, \ie they cannot be compressed any further. 

First, we describe when a firewall exhibits the same behavior for arbitrary source IPs $s_1, s_2$ and a fixed packet~$p$: 
%
%
%
\begin{IEEEeqnarray*}{lcr}%
\forall d.\ & \mfun{simple\mhyphen{}fw}\ \mathit{rs}\ p(src \mapsto s_1,\ dst \mapsto d) & \ = \\
	        & \mfun{simple\mhyphen{}fw}\ \mathit{rs}\ p(src \mapsto s_2,\ dst \mapsto d) &
\end{IEEEeqnarray*}%
We say the firewall shows same behavior for a fixed service if, in addition, the analogue condition holds for destination IPs.

We present a function $\mfun{groupWIs}$, which computes the minimal partition for a fixed service. 
For this, the full access control matrix for inbound and outbound connections of each partition member is generated. 
This can be done by taking arbitrary representatives from each partition as source and destination address and executing $\mfun{simple\mhyphen{}fw}$ for the fixed packet with those fixed IPs. 
The matrix is minimized by merging partitions with equal rights, \ie equal rows in the matrix. 
%
%
%
This algorithm is quadratic in the number of partitions. 
An early evaluation~\cite{diekmann2016networking} shows that it scales surprisingly well, even for large rulesets, since the number of partitions is usually small. 

The algorithm is sound\footnote{build-ip-partition-same-fw}, complete\footnote{build-ip-partition-complete}, and minimal.\footnote{build-ip-partition-same-fw-min}

\begin{theorem}[$\mfun{groupWIs}$ is Sound and Minimal]
\label{thm:groupwis}
For any two IPs in any member of $\mfun{groupWIs}$, the firewall shows the same behavior for a fixed service. 

For any two arbitrary members $A$ and $B$ in $\mfun{groupWIs}$, if we can find two IPs in $A$ and $B$ respectively where the firewall shows the same behavior for a fixed service, then $A = B$. 
\end{theorem}

\paragraph*{Improving Performance}
We assume that the ruleset has a default policy; we fall back to our slower algorithm otherwise but any simplified ruleset obtained from well-formed iptables has a default policy.\footnote{Since we can easily check at runtime whether a ruleset has a default policy, this fallback only exists that we can write down our theorems without requiring the assumption of a default policy since our faster algorithm (with default policy) and slower algorithm (without default policy) compute the same result. In practice, any ruleset has a default policy and the faster algorithm is always used.} 
The algorithm described above performs two calls (one for source IP and one for destination IP) to $\mfun{simple\mhyphen{}fw}$ for each pair of representatives in the partition. 
The algorithm is significantly slowed down by the quadratic number of calls to $\mfun{simple\mhyphen{}fw}$. 
Instead of repeatedly executing $\mfun{simple\mhyphen{}fw}$ for all representatives as source and destination address, for a fixed service and fixed source address, we can pre-compute the set of all matching destination addresses with one iteration over the ruleset. 
The same holds for the matching source addresses. 
This roughly brings down the quadratic number of calls to $\mfun{simple\mhyphen{}fw}$ to a linear number of iterations over the ruleset. 
Note that the asymptotic runtime is still quadratic. 
We have implemented this improved algorithm and proven that Theorem~\ref{thm:groupwis} and Theorem~\ref{thm:servicematrix} still hold for it. 
The empirical evaluation shows that this improvement yields a speedup of about 10x, \ie 1000\%.

\paragraph*{Final Theorem}
The function $\mfun{groupWIs}$ computes a minimal partition and the allowed accesses over representatives of the IPv4 address space. 
However, it does not directly allow drawing graphs as shown in \eg Figure~\ref{fig:tumssh}, Figure~\ref{fig:eval_measrdroid:i8fw:port80}, or Figure~\ref{fig:eval_measrdroid:i8fw:port80ipv6}. 
To draw a graph, for example with tikz~\cite{tantau2016tikz}, one first needs to print the nodes and print the edges afterwards. 
The name of the nodes (representatives) should not be printed but th IP range it actually represents. 
For example, a graph may be defined as follows: 

\begin{small}
\begin{verbatim}
\begin{tikzpicture}
  \node (a) at (-4,-4) {$\{131.159.21.0 .. 131.159.21.255\}$};
  \node (b) at (4,-4) {$\{131.159.15.240 .. 131.159.15.255\}$};
  \node (c) at (0,-6) {$\{127.0.0.0 .. 127.255.255.255\}$};
  ...
  
  \draw (a) to (b);
  \draw (c) to (a);
  \draw (c) to (b);
  \draw (c) to[loop above] (c);
  ...
\end{tikzpicture}  
\end{verbatim}
\end{small}

In this example, the node names \texttt{a}, \texttt{b}, and \texttt{c} are representatives which semantically correspond to the set of IP addresses on their right. 
The edges mean that the complete IP ranges may communicate, \eg \verb~\draw (a) to (b)~ means that the complete set 131.159.21.0/24 may establish connections to 131.159.15.240/28. 
In the final drawing, the identifiers \texttt{a}, \texttt{b}, and \texttt{c} are not shown but only their corresponding IP ranges. 
We present a final theorem which justifies the correctness of graphs which are drawn according to our method.\footnote{access-matrix} 

\begin{theorem}[Service Matrix]
\label{thm:servicematrix}
Let $V$ be the nodes of the service matrix. $V$ is a map from representatives to the IP range of the representative. Let $E$ be the edges of the service matrix. Then, %
	%
	\begin{IEEEeqnarray*}{c}%
	\bigl(\exists\ s_\mathrm{repr}\ d_\mathrm{repr}\ s_\mathrm{range}\ d_\mathrm{range}.\ \ (s_\mathrm{repr},\; d_\mathrm{repr}) \in E \ \wedge \\
	\phantom{(} V \ s_\mathrm{repr} = \mconstr{Some}\ s_\mathrm{range} \ \wedge \ s \in s_\mathrm{range} \ \wedge \\
	\phantom{(} V \ d_\mathrm{repr} = \mconstr{Some}\ d_\mathrm{range} \ \wedge \ d \in d_\mathrm{range} \bigr) \\
	\longleftrightarrow \\
	\mfun{simple\mhyphen{}fw}\ \mathit{rs}\ p(src \mapsto s,\ dst \mapsto d) \ = \ $\allow$
	\end{IEEEeqnarray*}
\end{theorem}

The theorem reads as follows: 
For a fixed connection, one can look up IP addresses (source $s$ and destination $d$ pairs) in the graph 
if and only if the firewall accepts this $(s, d)$ IP address pair for the fixed connection. 

The part which complicates the formalization is the formulation of `look up IP addresses [..] in the graph'. 
To look up source IP address $s$ in the graph, one first locates $s$ as a member in one of the IP ranges, here $s_\mathrm{range}$. 
This IP range is represented by a representative $s_\mathrm{repr}$. 
The same is done to obtain $d_\mathrm{repr}$. 
The theorem now says that $(s_\mathrm{repr},\; d_\mathrm{repr}) \in E$ iff the firewall allows packets from $s$ to $d$. 
The theorem is actually stronger because the if-and-only-if relationship in combination with the existential quantifier also implies that there is always exactly one range in which we can find $s$ and $d$ (which means that our graph always contains a complete and disjoint representation of the IP address space). 

\section{Stand-Alone Haskell Tool \fffuu{}}
\label{sec:ifip:fffuu}
We used Isabelle's code generation features~\cite{isabelle2012code,haftmann2010code} to build a stand-alone tool in Haskell. 
Since all analysis and transformation algorithms are written in Isabelle, we only needed to add parsers and user interface. 
Overall, more than 80\% of the code is generated by Isabelle, which gives a strong correctness guarantee. 

We call our tool \fffuu{}, the \emph{f}ancy \emph{f}ormal \emph{f}irewall \emph{u}niversal \emph{u}nderstander. 

\fffuu{} requires only one parameter to run, namely, an \texttt{iptables-save} dump. 
This makes it very usable. 
Optionally, one may pass an \emph{ipassmt}, change the \texttt{table} or \texttt{chain} which is loaded, pass a routing table for output port rewriting, or select the services for the service matrix.

\paragraph*{Example}
We demonstrate \fffuu{} by a small example. 
We want to infer the intention behind the ruleset shown in Figure~\ref{fig:ifipfffuuexample}. 
Though this ruleset was artificially crafted to demonstrate certain corner cases, it is based on actual rules from real-world firewalls~\cite{serverfaultmultiportnegation,diekmanngithubnetnetwork}. 
Also note that the interface name \verb~\e[31m~\includegraphics[height=1.8ex]{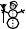}\verb~\e[0m~ with utf-8 symbols and shell escapes for color~\cite{xtermescapes} is perfectly valid. 

\begin{figure*}[ht!bp]
	\begin{minipage}{.99\linewidth}
\small
\begin{Verbatim}[commandchars=\\\{\},codes={\catcode`$=3\catcode`^=7}]
*filter
:INPUT DROP [0:0]
:FORWARD DROP [0:0]
:OUTPUT DROP [0:0]
:DOS_PROTECT - [0:0]
:GOOD~STUFF - [0:0]
-A FORWARD -j DOS_PROTECT
-A FORWARD -j GOOD~STUFF
-A FORWARD -p tcp -m multiport ! --dports 80,443,6667,6697 -m hashlimit $\hfill\hookleftarrow$
    --hashlimit-above 10/sec --hashlimit-burst 20 --hashlimit-mode srcip $\hfill\hookleftarrow$
    --hashlimit-name aflood --hashlimit-srcmask 8 -j LOG
-A FORWARD ! -i lo -s 127.0.0.0/8 -j DROP
-A FORWARD -i internal -s 131.159.21.0/24 -j ACCEPT
-A FORWARD -s 131.159.15.240/28 -d 131.159.21.0/24 -j DROP
-A FORWARD -p tcp -d 131.159.15.240/28 -j ACCEPT
-A FORWARD -i \includegraphics[height=1.8ex]{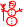} -p tcp -s 131.159.15.240/28 -j ACCEPT
-A GOOD~STUFF -i lo -j ACCEPT
-A GOOD~STUFF -m state --state ESTABLISHED -j ACCEPT
-A GOOD~STUFF -p icmp -m state --state RELATED -j ACCEPT
-A DOS_PROTECT -i eth1 -p icmp -m icmp --icmp-type 8 $\dots$ --limit 1/sec -j RETURN
-A DOS_PROTECT -i eth1 -p icmp -m icmp --icmp-type 8 -j DROP
COMMIT
\end{Verbatim}
\end{minipage}%
\caption{Example Ruleset}
\label{fig:ifipfffuuexample}
\end{figure*}

\begin{figure}
	\centering
\resizebox{\linewidth}{!}{%
	\scriptsize
	\begin{tikzpicture}
	\node[align=left] (a) at (-4,-4) {$\{131.159.21.0 .. 131.159.21.255\}$}; 
	\node[align=left] (b) at (4,-4) {$\{131.159.15.240 .. 131.159.15.255\}$}; 
	\node[align=left] (c) at (0,-6) {$\{127.0.0.0 .. 127.255.255.255\}$};
	\node[align=center,text width=5cm,cloud, draw,cloud puffs=10,cloud puff arc=120, aspect=2, inner sep=-2.5em,outer sep=0] (d) at (0,0) {$\{0.0.0.0 .. 126.255.255.255\} \cup \{128.0.0.0 .. 131.159.15.239\} \cup \{131.159.16.0 .. 131.159.20.255\} \cup \{131.159.22.0 .. 255.255.255.255\}$};
	
	\draw[myptr] (a) to[loop above] (a);
	\draw[myptr] (a) to (b);
	\draw[myptr] (a) to (c);
	\draw[myptr] (a) to (d);
	\draw[myptr] (b) to[loop above] (b);
	\draw[myptr] (b) to (c);
	\draw[myptr] (b) to (d);
	\draw[myptr] (c) to (a);
	\draw[myptr] (c) to (b);
	\draw[myptr] (c) to[loop below] (c);
	\draw[myptr] (c) to (d);
	\draw[myptr] (d) to (b);
	\end{tikzpicture}%
}
\caption{Service Matrix of Ruleset in Figure~\ref{fig:ifipfffuuexample}}
\label{fig:ifipfffuuu:matrix}
\end{figure}
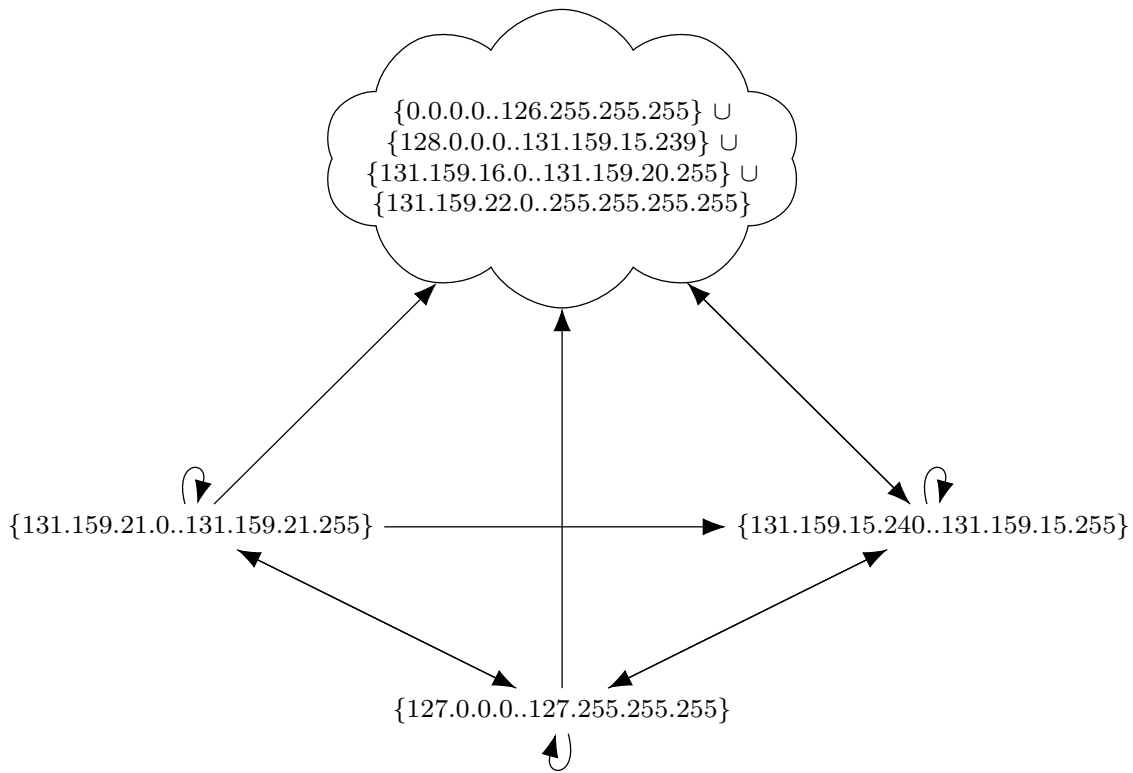

It is hard to guess what the ruleset is implementing. 
The service matrix will provide clarity. 
We dump the ruleset into \fffuu{}, not requiring any additional parameters or manual steps to compute it. 
The resulting service matrix (for arbitrary ports) is shown in Figure~\ref{fig:ifipfffuuu:matrix}. 
An arrow from one IP range to another IP range indicates that the first range may set up connections with the second. 

At the bottom, we see the localhost range of 127.0.0.0/8. 
The reflexive arrow (localhost to localhost) shows that the firewall does not block its own localhost traffic, which is usually a good sign. 
However, localhost traffic is usually not interesting for a firewall analysis since this range is usually not routed~\cite{rfc6890}. 
We will ignore it from now. 

On the top, in the cloud, we see a large set of IP addresses. 
This corresponds to the Internet. 
On the left, we see the 131.159.21.0/24 range. 
It may access the Internet and the 131.159.15.240/28 range. 
On the right, we see the 131.159.15.240/28 range, which may only access the Internet but not the 131.159.21.0/24 range. 

Gazing at the figure, we might recognize the overall architecture: The firewall simply implements the textbook version of the Demilitarized Zone (DMZ) architecture. 
Starting from the original \texttt{iptables-save} input, without the help of \fffuu{}, this would be extremely hard to uncover and verify.

\section{Evaluation}
\label{sec:evaluation}
We obtained real-world rulesets from over 15 firewalls. 
Some are central, production-critical devices. 
They are written by different authors, utilize a vast amount of different features and exhibit different styles and patterns. 
The fact that we publish the complete rulesets is an important contribution (cf.\ \cite{firwallerr2004,wool2010firewall}). 
To the best of our knowledge, this is the largest, publicly-available collection of real-world iptables rulesets. 
Note: some administrators wish to remain anonymous so we replaced their public IP addresses with public IP ranges of our institute, preserving all IP subset relationships. 

Table~\ref{tab:somenumbers} summarizes the evaluation's results. 
The first column (Fw) labels the analyzed ruleset. 
Column two (Rules) contains the number of rules (only the filter table) in the output of \texttt{iptables-save}. 
We work directly and completely on this real-world data. 
Column three describes the analyzed chain. 
Depending on the type of firewall, we either analyzed the \texttt{FORWARD} (FW) or the \texttt{INPUT} (IN) chain. 
For a host firewall, we analyzed IN; for a network firewall, \eg on a gateway or router, we analyzed FW. 
In parentheses, we wrote the number of rules after unfolding the analyzed chain. 
The unfolding also features some generic, straight-forward optimizations, such as removing rules where the match expression is $\mdef{False}$. 
Column four (Simple rules) is the number of rules when translated to the simple firewall. 
In parentheses, we wrote the number of simple firewall rules when interfaces are removed. 
This ruleset is used subsequently to compute the partitions and service matrices. 
In column five (Use), we mark whether the translated simple firewall is useful. 
We will detail on the metric later. 
Column six (Parts) lists the number of IP address space partitions. 
For comparison, we give the number of partitions computed by ITVal in parentheses. 
In Column seven (ssh) and eight (http), we give the number of partitions for the service matrices for ssh and http. 
In column nine (Time (ITVal)), for comparison, we put the runtime of the partitioning by ITVal in parentheses in seconds, minutes, or hours. 
In column ten (Time (this)), we give the overall runtime of our analysis.

When translating to the simple firewall, to accomplish support for arbitrary matching primitives, some approximations need to be performed. 
For every firewall, the first row states the overapproximation (more permissive), the second row the underapproximation (more strict).

\newcommand{\asterspce}[0]{\phantom{\textsuperscript{$\ast$}}}

\begin{table}[h!bt]
\centering
\begin{footnotesize}
\begin{tabular}{ l r   r@{\,}r   r@{\,}r  c  r r r  r r }
	\toprule
	Fw & Rules       & \multicolumn{2}{c}{\parbox[t]{\widthof{(unfolded)}}{\raggedleft{Chain\\(unfolded)}}}      & \multicolumn{2}{c}{\parbox[t]{\widthof{Simple\,rules}}{\raggedleft{Simple\,rules\\(no\,ifaces)}}}   & Use  & \parbox[t]{\widthof{(ITVal)}}{\raggedleft{Parts\\(ITVal)}} & ssh & http & \parbox[t]{3em}{\raggedleft{Time\\(ITVal)}} & \parbox[t]{3em}{\raggedleft{Time\\(this)}}\\
	\midrule
	%
	   
	A & 2784           & FW & (2376)          & 2381 & (1920)                & \cmark & 246 (1)           & 13   & 9   & \SI{3}{\hour}\textsuperscript{$\ast$} & \SI{172}{\second} \\ %
	   & -              & FW & (2376)          & 2837 & (581)                & \xmark\ \textsuperscript{r} & 522 (1)   & 1   & 1   & \SI{9}{\hour}\textsuperscript{$\ast$} &  \SI{194}{\second} \\ 

	A & 4113           & FW & (2922)          & 3114 & (2862)                & \cmark & 334 (2)           & 11   & 11   & \SI{27}{\hour}\textsuperscript{$\ast$} & \SI{302}{\second} \\
	   & -              & FW & (2922)          & 3585 & (517)                 & \xmark\ \textsuperscript{r} & 490 (1)   & 1   & 1   & \SI{8}{\hour}\asterspce{} &  \SI{320}{\second} \\ 

	A  & 4814           & FW & (4403)          & 3574 & (3144)                & \cmark & 364 (2)           & 9   & 12   & \SI{46}{\hour}\textsuperscript{$\ast$} & \SI{477}{\second} \\ %
	   & -              & FW & (4403)          & 5123 & (1601)                & \xmark\ \textsuperscript{r} & 1574 (1)   & 1   & 1   & \SI{3}{\hour}\textsuperscript{$\ast$} &  \SI{618}{\second}  \\ 
	   
	A & 4946           & FW & (4887)          & 4004 & (3570)                & \cmark & 371 (2)           & 9   & 12   & \SI{53}{\hour}\textsuperscript{$\ast$} & \SI{578}{\second} \\ 
	   & -              & FW & (4887)          & 5563 & (1613)                & \xmark\ \textsuperscript{r} & 1585 (1)   & 1   & 1   & \SI{4}{\hour}\textsuperscript{$\ast$}  &  \SI{820}{\second} \\ 

	B & 88              & FW & (40)          & 110 & (106)                & \cmark & 50 (4)           & 4   & 2   & \SI{2}{\second}\asterspce{} & \SI{3}{\second} \\ 
	   & -              & FW & (40)          & 183 & (75)                & \cmark  & 40 (1)   & 1   & 1   & \SI{1}{\second}\asterspce{} & \SI{2}{\second} \\ 

	
	%
	C  & 53             & FW & (30)                    & 29 & (12)             & \cmark  & 8 (1)        & 1  &  1 & \SI{1}{\second}\asterspce{} & \SI{1}{\second} \\ 
	   & -              & FW & (30)                    & 27 & (1)               & \cmark  & 1 (1)       & 1  &  1 & \SI{1}{\second}\asterspce{} & \SI{1}{\second} \\ 
	   & -              & IN & (49)                   & 74 & (46)               & \cmark   & 38 (1)        & 1  &  1 & \SI{1}{\second}\asterspce{} & \SI{1}{\second} \\ %
	   & -              & IN & (49)                   & 75 & (21)               & \cmark  & 6 (1)          & 1  &  1 & \SI{1}{\second}\asterspce{} & \SI{1}{\second} \\ %

	D  & 373            & FW & (2649)                 & 3482 & (166)            & \cmark      & 43 (1)        & 1   &  1  & \SI{3}{\second}\asterspce{} & \SI{22}{\second} \\ 
	   & -              & FW & (2649)                 & 16592 & (1918)          & \xmark      & 67 (1)        & 1   &  1  & \SI{33}{\minute}\textsuperscript{$\ast$} & \SI{49}{\second} \\ %

	E  & 31             & IN & (24)                    & 57 & (27)             & \cmark  & 4 (3)        & 1    & 2 & \SI{1}{\second}\asterspce{} & \SI{10}{\second} \\ 
	%
	   & -              & IN & (24)                    & 61 & (45)             & \xmark\ \textsuperscript{r} & 3 (1)  & 1    & 1  & \SI{1}{\second}\asterspce{} & \SI{1}{\second} \\ 
	
	F  & 263            & IN & (261)                    & 263 & (263)          & \cmark & 250 (3)      & 3    & 3 & \SI{2}{\minute}\asterspce{} & \SI{80}{\second} \\ 
	   & -              & IN & (261)                    & 265 & (264)          & \cmark  & 250 (3)      & 3    & 3 & \SI{3}{\minute}\asterspce{} & \SI{57}{\second} \\ 

	G  & 68             & IN & (28)                     & 20 & (20)            & \cmark & 8 (5)       & 1     & 2 & \SI{1}{\second}\asterspce{} & \SI{8}{\second} \\
	   & -              & IN & (28)                     & 19 & (19)            & \xmark & 8 (2)       & 2     & 2 & \SI{1}{\second}\asterspce{} & \SI{1}{\second} \\

	H  & 19             & FW & (20)                     & 10 & (10)           & \xmark & 9 (1)       & 1      & 1 & \SI{1}{\second}\asterspce{} & \SI{8}{\second} \\ 
	   & -              & FW & (20)                     & 8 & (8)             & \xmark\ \textsuperscript{r} & 3 (1)       & 1      & 1 & \SI{1}{\second}\asterspce{} & \SI{1}{\second} \\ 

	I  & 15             & FW & (5)                     & 4 & (4)             & \cmark  & 4 (4)       & 4      & 4 & \SI{1}{\second}\asterspce{} & \SI{8}{\second}\\ 
	   & -              & FW & (5)                     & 4 & (4)             & \cmark  & 4 (4)       & 4      & 4 & \SI{1}{\second}\asterspce{} & \SI{1}{\second} \\ 

	J  & 48             & FW & (12)                    & 5 & (5)             & \cmark  & 3 (2)       & 2      & 2 & \SI{1}{\second}\asterspce{} & \SI{6}{\second} \\ 
	   & -              & FW & (12)                    & 8 & (2)             & \cmark  & 1 (1)       & 1      & 1 & \SI{1}{\second}\asterspce{} & \SI{1}{\second} \\ 

	K  & 21             & FW & (9)                     & 7 & (6)             & \cmark  & 3 (1)       & 1      & 1 & \SI{1}{\second}\asterspce{} & \SI{12}{\second} \\ 
	   & -              & FW & (9)                     & 4 & (3)             & \cmark  & 2 (1)       & 1      & 1 & \SI{1}{\second}\asterspce{} & \SI{1}{\second} \\ 

	L  & 27             & IN & (16)                    & 19 & (19)            & \cmark  & 17 (3)      & 2      & 2 & \SI{1}{\second}\asterspce{} & \SI{1}{\second} \\ 
	   &  -             & IN & (16)                    & 18 & (18)            & \cmark  & 17 (3)     & 2      & 2 & \SI{1}{\second}\asterspce{} & \SI{1}{\second} \\ 

	M  & 80             & IN & (92)                    & 64 & (16)            & \cmark   & 2 (2)       & 1      & 2 & \SI{1}{\second}\asterspce{} & \SI{6}{\second} \\ 
	   &  -             & IN & (92)                    & 58 & (27)            & \xmark   & 11 (1)      & 1      & 1 & \SI{1}{\second}\asterspce{} & \SI{1}{\second} \\ 

	N  & 34             & FW & (14)                    & 12 & (12)            & \cmark   & 10 (6)      & 6      & 6 & \SI{2}{\second}\asterspce{} & \SI{2}{\second} \\ %
	   &  -             & FW & (14)                    & 12 & (12)            & \cmark   & 10 (6)      & 6      & 6 & \SI{2}{\second}\asterspce{} & \SI{1}{\second} \\ %

	O  & 8              & IN & (7)                    & 9 & (9)            & \cmark   & 3 (3)          & 1      & 2 & \SI{1}{\second}\asterspce{} & \SI{1}{\second} \\ %
	   &  -             & IN & (7)                    & 8 & (8)            & \cmark   & 3 (3)          & 1      & 2 & \SI{1}{\second}\asterspce{} & \SI{1}{\second} \\ %

	P  & 595            & IN & (15)                   & 8 & (8)            & \cmark   & 3 (2)          & 2      & 2 & \SI{1}{\second}\asterspce{} & \SI{6}{\second} \\ %
	   &  -             & IN & (15)                   & 9 & (9)            & \cmark   & 3 (2)          & 2      & 2 & \SI{1}{\second}\asterspce{} & \SI{6}{\second} \\ %
	   & 595            & FW & (66)                   & 64 & (64)          & \cmark   & 60 (5)         & 5      & 4 & \SI{22}{\second}\asterspce{} & \SI{6}{\second} \\ 
	   &  -             & FW & (66)                   & 63 & (63)           & \cmark  & 60 (5)         & 5      & 4 & \SI{22}{\second}\asterspce{} & \SI{7}{\second} \\ 
%

	Q  & 58             & IN & (59)                   & 65 & (65)          & \cmark   & 21 (1)         & 1      & 1 & \SI{2}{\second}\asterspce{} & \SI{2}{\second} \\ 
      &  -             & IN & (59)                   & 62 & (62)          & \cmark   & 21 (2)         & 2      & 1 & \SI{2}{\second}\asterspce{} & \SI{1}{\second} \\ 
%
	   
	
	R  & 30             & FW & (28)                   & 123 & (123)          & \cmark   & 14 (1)         & 1      & 6 & \SI{1}{\second}\asterspce{} & \SI{1}{\second} \\ %
	&  -             & FW & (28)                   & 20 & (3)          & \cmark   & 2 (2)         & 2      & 1 & \SI{1}{\second}\asterspce{} & \SI{1}{\second} \\ %
 \midrule%
 \multicolumn{12}{p{0.7\linewidth}}{\textsuperscript{$\ast$} ITVal memory consumption, in order of appearance:\newline
 		\SI{84}{\giga\byte}, \SI{96}{\giga\byte}, \SI{94}{\giga\byte}, \SI{95}{\giga\byte}, \SI{61}{\giga\byte}, \SI{98}{\giga\byte}, \SI{96}{\giga\byte}, \SI{21}{\giga\byte} }\\
 \bottomrule%
\end{tabular}%
\end{footnotesize}
\vskip+3pt%
\caption{Summary of Evaluation on Real-World Firewalls}
\label{tab:somenumbers}
\end{table}

In contrast to previous work, there is no longer the need to manually exclude certain rules from the analysis (cf.\ Section~\ref{sec:fm15:establishedrule}). 
%
For some rulesets, we do not know the interface configuration. 
For others, there were zone-spanning interfaces. 
For these reasons, as proven in Section~\ref{sec:ifcerewrite}, in the majority of cases, we could not rewrite interfaces. 
This is one reason for the differences between over- and underapproximation. 


\begin{sloppypar}
We loaded all translated simple firewall rulesets (without interfaces) with \texttt{iptables-restore}. 
This validates that our results are well-formed. 
We then used iptables directly to generate the firewall format required by ITVal (\texttt{iptables -L -n}). 
Our translation to the simple firewall is required because ITVal cannot understand the original complex rulesets and produces flawed results for them. 
\end{sloppypar}

\paragraph*{Performance}
%
The code of our tool is automatically generated by Isabelle. 
Isabelle can translate executable algorithms to SML. 
For verifiable correctness, Isabelle also generates code for many datastructures which are already in the standard library of many programming languages.
Usually, the machine-generated code by Isabelle can be quite inefficient. 
For example, lookups in Isabelle-generated dictionaries have linear lookup time, compared to constant lookup time of standard library implementations. 
In contrast, ITVal is highly optimized C++ code.
%
%
We benchmarked our tool on a commodity i7-2620M laptop with 2 physical cores and \SI{8}{\giga\byte} of RAM. 
In contrast, we executed ITVal on a server with 16 physical Xeon E5-2650 cores and \SI{128}{\giga\byte} RAM.
The runtime measured for our tool is the complete translation to the two simple firewalls, computation of partitions, and the two service matrices. 
In contrast, the ITVal runtime only consists of computing one partition. 

These benchmark settings are extremely `unfair' for our tool. 
Indeed, exporting our tool to a standalone Haskell application, replacing some common datastructures with optimized ones from the Haskell std lib, enabling aggressive compiler optimization and parallelization, and running our tool on the Xeon server, the runtime of our tool improves by orders of magnitude. 
Our stand-alone tool \fffuu{} also achieves a better runtime by orders of magnitude. 
Nevertheless, we chose the `unfair' setting to demonstrate the feasibility of running fully verified code directly in a theorem prover. 
In addition, we preserve the property of full verification; even for the results of executable code.\footnote{There are methods to improve the performance and provably preserve correctness~\cite{Lammich2013autoref,Lammich2012isabellerefine}, which are out of the scope of this thesis} 

Table~\ref{tab:somenumbers} shows that our tool outperforms ITVal for large firewalls. 
We added ITVal's memory requirements to the table if they exceeded \SI{20}{\giga\byte}. 
ITVal requires an infeasible amount of memory for larger rulesets while our tool can finish on commodity hardware. 
The overall numbers show that the runtime for our tool is sufficient for static, offline analysis, even for large real-word rulesets. 

For our daily use and convenience, we use our Haskell tool \fffuu{} which adds another order of magnitude of speedup to our numbers of Table~\ref{tab:somenumbers}. 

\paragraph*{Quality of results}
The main goal of ITVal is to compute a minimal partition while ours may not be minimal. 
Since a service matrix is more specific than a partition, a partition cannot be smaller than a service matrix. 
ITVal may produce spurious results (and it did in certain examples) while ours are provably correct. 
For firewalls $A$ and $R$, it can be seen that ITVals's results must be spurious: 
If the number of partitions calculated by ITVal is smaller than that those of a service matrix, this is an error in ITVal.
However, comparing the number of partitions for other rulesets, we can see that ITVal often computes better results. 
Our service matrices are provably minimal and can improve on ITVal's partition. 

In column five, we show the usefulness of the translated simple firewall (including interfaces). 
We deem a firewall useful if interesting information was preserved by the approximation. 
Therefore, we manually inspected the rulesest and compared it to the original. 
For the overapproximation, we focused on preserved (non-shadowed) $\iptaction{DROP}$ rules. 
For the underapproximation, we focused on preserved (non-shadowed) $\iptaction{ACCEPT}$ rules.
If the firewall features some rate-limiting for all packets in the beginning, the underapproximation is naturally a drop-all ruleset because the rate-limiting could apply to all packets. 
According to our metric, such a ruleset is of no use (but the only sound solution).  
We indicate this case with an~\textsuperscript{r}. 
The table indicates that, usually, at least one approximation per firewall is useful.


For brevity, we only elaborate on the most interesting rulesets and stories. 

\begin{figure}[!htbp]
\centering
\begin{tikzpicture}
\node (g1) at (0,0) {$\mathit{internal}$}; 
\node (g2) at (1,2) {$\mathit{servers}$}; 
\node (m) at (0.7,4) {multicast}; 
\node (rest) at (-1,3.5) {INET};
\node (l) at (2.7,0.5) {localhost}; 
\node (ip1) at (3,3.5) {$\mathit{ip}_1$}; 
//messzeugs, alte

\node (ip2) at (5,0) {$\mathit{ip}_2$}; 
//ausserhalb unseres netzes, darf rein, fuer scan

\node (ips3) at (5,2) {AS routers}; 
// spezielle netze
// backbone zu neuem as 64-127 nur router
// 128-191: netsec.net.in.tum.de
//   hosts im neuen AS vor firewall
// 188.1.239.86 DFN router, gateway, transfer net

\node (ips4) at (-0.9,2.8) {INET'}; 
 //unused?

\draw[myptr] (g1) to [loop left] (g1);
\draw[myptr] (g1) to (g2);
\draw[myptr] (g1) to (m);
\draw[myptr] (g1) to (rest);
\draw[myptr] (g1) to (ip1);
\draw[myptr] (g1) to (ip2);
\draw[myptr] (g1) to (ips3);
\draw[myptr] (g1) to (ips4);
\draw[myptr] (g1) to (l);
\draw[myptr] (g2) to (g1);
\draw[myptr] (g2) to [loop left] (g2);
\draw[myptr] (g2) to (m);
\draw[myptr] (g2) to (rest);
\draw[myptr] (g2) to (ip1);
\draw[myptr] (g2) to (ip2);
\draw[myptr] (g2) to (ips3);
\draw[myptr] (g2) to (ips4);
\draw[myptr] (g2) to (l);
\draw[myptr] (m) to (g2);
\draw[myptr] (m) to [loop above] (m);
\draw[myptr] (rest) to (g2);
\draw[myptr] (rest) to (m);
\draw[myptr] (ip1) to (g1);
\draw[myptr] (ip1) to (g2);
\draw[myptr] (ip1) to (m);
\draw[myptr] (ip1) to (rest);
\draw[myptr] (ip1) to [loop above] (ip1);
\draw[myptr] (ip1) to (ip2);
\draw[myptr] (ip1) to (ips3);
\draw[myptr] (ip1) to (ips4);
\draw[myptr] (ip1) to (l);
\draw[myptr] (ip2) to (g1);
\draw[myptr] (ip2) to (g2);
\draw[myptr] (ip2) to (m);,
\draw[myptr] (ip2) to (ip1);
\draw[myptr] (ips3) to (g1);
\draw[myptr] (ips3) to (g2);
\draw[myptr] (ips3) to (m);
\draw[myptr] (ips3) to (rest);
\draw[myptr] (ips3) to (ip1);
\draw[myptr] (ips3) to (ip2);
\draw[myptr] (ips3) to [loop above] (ips3);
\draw[myptr] (ips3) to (ips4);
\draw[myptr] (ips3) to (l);
\draw[myptr] (ips4) to (g2);
\draw[myptr] (ips4) to (m);
\draw[myptr] (ips4) to (ip1);

\end{tikzpicture}
\caption{TUM i8 ssh Service Matrix}
\label{fig:tumssh}
\end{figure}
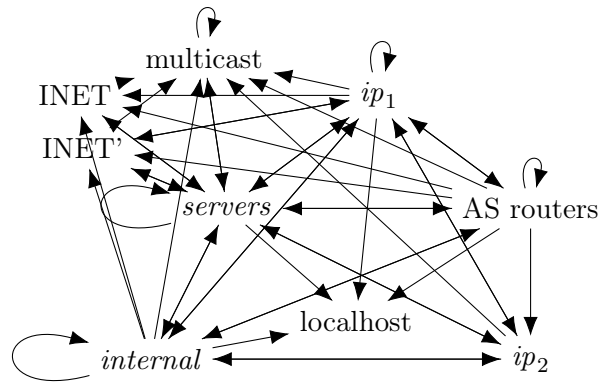

\begin{figure}[!htbp]
	\centering
	\resizebox{\linewidth}{!}{%
	\begin{footnotesize}%
	\begin{tikzpicture}
	\node[anchor=north, text badly ragged left, text width=10cm] (g1) at (-10,-10) {$\{131.159.14.0 .. 131.159.14.7\} \cup \{131.159.14.12 .. 131.159.14.21\} \cup \{131.159.14.23 .. 131.159.14.25\} \cup 131.159.14.27 \cup \{131.159.14.29 .. 131.159.14.33\} \cup \{131.159.14.38 .. 131.159.14.39\} \cup 131.159.14.41 \cup \{131.159.14.43 .. 131.159.14.51\} \cup \{131.159.14.53 .. 131.159.14.55\} \cup 131.159.14.57 \cup \{131.159.14.59 .. 131.159.14.68\} \cup \{131.159.14.70 .. 131.159.14.82\} \cup \{131.159.14.84 .. 131.159.14.103\} \cup \{131.159.14.105 .. 131.159.14.110\} \cup \{131.159.14.112 .. 131.159.14.124\} \cup \{131.159.14.126 .. 131.159.14.136\} \cup \{131.159.14.138 .. 131.159.14.139\} \cup \{131.159.14.141 .. 131.159.14.144\} \cup \{131.159.14.147 .. 131.159.14.154\} \cup \{131.159.14.157 .. 131.159.14.162\} \cup \{131.159.14.164 .. 131.159.14.168\} \cup \{131.159.14.170 .. 131.159.14.200\} \cup \{131.159.14.202 .. 131.159.14.213\} \cup \{131.159.14.215 .. 131.159.15.4\} \cup 131.159.15.6 \cup \{131.159.15.14 .. 131.159.15.15\} \cup \{131.159.15.21 .. 131.159.15.22\} \cup 131.159.15.26 \cup 131.159.15.28 \cup \{131.159.15.30 .. 131.159.15.31\} \cup \{131.159.15.33 .. 131.159.15.35\} \cup \{131.159.15.37 .. 131.159.15.38\} \cup \{131.159.15.40 .. 131.159.15.41\} \cup \{131.159.15.45 .. 131.159.15.46\} \cup \{131.159.15.49 .. 131.159.15.53\} \cup 131.159.15.55 \cup 131.159.15.57 \cup 131.159.15.59 \cup \{131.159.15.61 .. 131.159.15.68\} \cup \{131.159.15.70 .. 131.159.15.196\} \cup \{131.159.15.198 .. 131.159.15.227\} \cup \{131.159.15.229 .. 131.159.15.233\} \cup \{131.159.15.235 .. 131.159.15.246\} \cup \{131.159.15.250 .. 131.159.15.255\} \cup \{131.159.20.0 .. 131.159.20.20\} \cup \{131.159.20.22 .. 131.159.20.28\} \cup \{131.159.20.31 .. 131.159.20.35\} \cup \{131.159.20.37 .. 131.159.20.44\} \cup \{131.159.20.46 .. 131.159.20.51\} \cup \{131.159.20.53 .. 131.159.20.58\} \cup \{131.159.20.60 .. 131.159.20.62\} \cup \{131.159.20.64 .. 131.159.20.70\} \cup \{131.159.20.72 .. 131.159.20.73\} \cup \{131.159.20.75 .. 131.159.20.84\} \cup \{131.159.20.86 .. 131.159.20.96\} \cup \{131.159.20.98 .. 131.159.20.117\} \cup 131.159.20.119 \cup \{131.159.20.121 .. 131.159.20.123\} \cup \{131.159.20.125 .. 131.159.20.138\} \cup \{131.159.20.140 .. 131.159.20.149\} \cup \{131.159.20.152 .. 131.159.20.154\} \cup \{131.159.20.156 .. 131.159.20.159\} \cup \{131.159.20.161 .. 131.159.20.164\} \cup \{131.159.20.167 .. 131.159.20.179\} \cup \{131.159.20.181 .. 131.159.20.184\} \cup \{131.159.20.186 .. 131.159.20.199\} \cup \{131.159.20.201 .. 131.159.20.232\} \cup \{131.159.20.235 .. 131.159.20.255\} \cup \{185.86.232.0 .. 185.86.235.255\} \cup \{188.95.233.0 .. 188.95.233.3\} \cup \{188.95.233.5 .. 188.95.233.8\} \cup \{188.95.233.10 .. 188.95.233.255\} \cup \{192.48.107.0 .. 192.48.107.255\}$}; 
	
	\node[anchor=north, text badly ragged left, text width=10cm] (g2) at (1,2) {$\{131.159.14.8 .. 131.159.14.11\} \cup 131.159.14.22 \cup 131.159.14.26 \cup 131.159.14.28 \cup \{131.159.14.34 .. 131.159.14.37\} \cup 131.159.14.40 \cup 131.159.14.42 \cup 131.159.14.52 \cup 131.159.14.56 \cup 131.159.14.58 \cup 131.159.14.69 \cup 131.159.14.83 \cup 131.159.14.104 \cup 131.159.14.111 \cup 131.159.14.125 \cup 131.159.14.137 \cup 131.159.14.140 \cup \{131.159.14.145 .. 131.159.14.146\} \cup \{131.159.14.155 .. 131.159.14.156\} \cup 131.159.14.163 \cup 131.159.14.169 \cup 131.159.14.201 \cup 131.159.14.214 \cup 131.159.15.5 \cup \{131.159.15.7 .. 131.159.15.13\} \cup \{131.159.15.16 .. 131.159.15.20\} \cup \{131.159.15.23 .. 131.159.15.25\} \cup 131.159.15.27 \cup 131.159.15.29 \cup 131.159.15.32 \cup 131.159.15.36 \cup 131.159.15.39 \cup \{131.159.15.42 .. 131.159.15.44\} \cup \{131.159.15.47 .. 131.159.15.48\} \cup 131.159.15.56 \cup 131.159.15.58 \cup 131.159.15.60 \cup 131.159.15.69 \cup 131.159.15.197 \cup 131.159.15.228 \cup 131.159.15.234 \cup \{131.159.15.247 .. 131.159.15.249\} \cup 131.159.20.21 \cup \{131.159.20.29 .. 131.159.20.30\} \cup 131.159.20.36 \cup 131.159.20.45 \cup 131.159.20.52 \cup 131.159.20.59 \cup 131.159.20.63 \cup 131.159.20.71 \cup 131.159.20.74 \cup 131.159.20.85 \cup 131.159.20.97 \cup 131.159.20.118 \cup 131.159.20.120 \cup 131.159.20.124 \cup 131.159.20.139 \cup \{131.159.20.150 .. 131.159.20.151\} \cup 131.159.20.155 \cup 131.159.20.160 \cup \{131.159.20.165 .. 131.159.20.166\} \cup 131.159.20.180 \cup 131.159.20.185 \cup 131.159.20.200 \cup \{131.159.20.233 .. 131.159.20.234\} \cup \{131.159.21.0 .. 131.159.21.255\} \cup \{188.95.232.192 .. 188.95.232.223\} \cup 188.95.233.4 \cup 188.95.233.9 \cup \{188.95.234.0 .. 188.95.239.255\}$}; 
	
	\node (m) at (0.7,5.5) {$\{224.0.0.0 .. 239.255.255.255\}$}; 
	
	\node[align=center, text width=10cm,cloud, draw,cloud puffs=10,cloud puff arc=120, aspect=2, inner sep=-4.5em,outer sep=0] (rest) at (-11,3.5) {$\{0.0.0.0 .. 126.255.255.255\} \cup \{128.0.0.0 .. 131.158.255.255\} \cup \{131.160.0.0 .. 138.246.253.4\} \cup \{138.246.253.6 .. 185.86.231.255\} \cup \{185.86.236.0 .. 188.1.239.85\} \cup \{188.1.239.87 .. 188.95.232.63\} \cup \{188.95.232.224 .. 188.95.232.255\} \cup \{188.95.240.0 .. 192.48.106.255\} \cup \{192.48.108.0 .. 223.255.255.255\} \cup \{240.0.0.0 .. 255.255.255.255\}$}; 
	
	\node (l) at (1,-10.5) {$\{127.0.0.0 .. 127.255.255.255\}$}; 
	\node (ip1) at (5,4) {$131.159.15.54$}; 
	//messzeugs, alte
	
	\node (ip2) at (10,-9) {$138.246.253.5$}; 
	//ausserhalb unseres netzes, darf rein, fuer scan
	
	\node[anchor=north, align=center, text width=6cm] (ips3) at (9,2) {$188.1.239.86 \cup \{188.95.232.64 .. 188.95.232.191\}$}; 
	// spezielle netze
	// backbone zu neuem as 64-127 nur router
	// 128-191: netsec.net.in.tum.de
	//   hosts im neuen AS vor firewall
	// 188.1.239.86 DFN router, gateway, transfer net
	
	\node[align=center, text width=6cm,cloud, draw,cloud puffs=6,cloud puff arc=120, aspect=2, inner sep=-4em,outer sep=0] (ips4) at (-10.9,-2.8) {$\{131.159.0.0 .. 131.159.13.255\} \cup \{131.159.16.0 .. 131.159.19.255\} \cup \{131.159.22.0 .. 131.159.255.255\}$}; 

	\draw[myptr] (g1) to [loop right] (g1);
	\draw[myptr] (g1) to (g2);
	\draw[myptr] (g1) to (m);
	\draw[myptr] (g1) to (rest);
	\draw[myptr] (g1) to (ip1);
	\draw[myptr] (g1) to (ip2);
	\draw[myptr] (g1) to (ips3);
	\draw[myptr] (g1) to (ips4);
	\draw[myptr] (g1) to (l);
	\draw[myptr] (g2) to (g1);
	\draw[myptr] (g2) to [loop below] (g2);
	\draw[myptr] (g2) to (m);
	\draw[myptr] (g2) to (rest);
	\draw[myptr] (g2) to (ip1);
	\draw[myptr] (g2) to (ip2);
	\draw[myptr] (g2) to (ips3);
	\draw[myptr] (g2) to (ips4);
	\draw[myptr] (g2) to (l);
	\draw[myptr] (m) to (g2);
	\draw[myptr] (m) to [loop above] (m);
	\draw[myptr] (rest) to (g2);
	\draw[myptr] (rest) to (m);
	\draw[myptr] (ip1) to (g1);
	\draw[myptr] (ip1) to (g2);
	\draw[myptr] (ip1) to (m);
	\draw[myptr] (ip1) to (rest);
	\draw[myptr] (ip1) to [loop above] (ip1);
	\draw[myptr] (ip1) to (ip2);
	\draw[myptr] (ip1) to (ips3);
	\draw[myptr] (ip1) to (ips4);
	\draw[myptr] (ip1) to (l);
	\draw[myptr] (ip2) to (g1);
	\draw[myptr] (ip2) to (g2);
	\draw[myptr] (ip2) to (m);,
	\draw[myptr] (ip2) to (ip1);
	\draw[myptr] (ips3) to (g1);
	\draw[myptr] (ips3) to (g2);
	\draw[myptr] (ips3) to (m);
	\draw[myptr] (ips3) to (rest);
	\draw[myptr] (ips3) to (ip1);
	\draw[myptr] (ips3) to (ip2);
	\draw[myptr] (ips3) to [loop above] (ips3);
	\draw[myptr] (ips3) to (ips4);
	\draw[myptr] (ips3) to (l);
	\draw[myptr] (ips4) to (g2);
	\draw[myptr] (ips4) to (m);
	\draw[myptr] (ips4) to (ip1);
	
	\end{tikzpicture}
	\end{footnotesize}
	}
	\caption{TUM i8 ssh Service Matrix (with raw IP addresses)}
	\label{fig:tumsshwithips}
\end{figure}
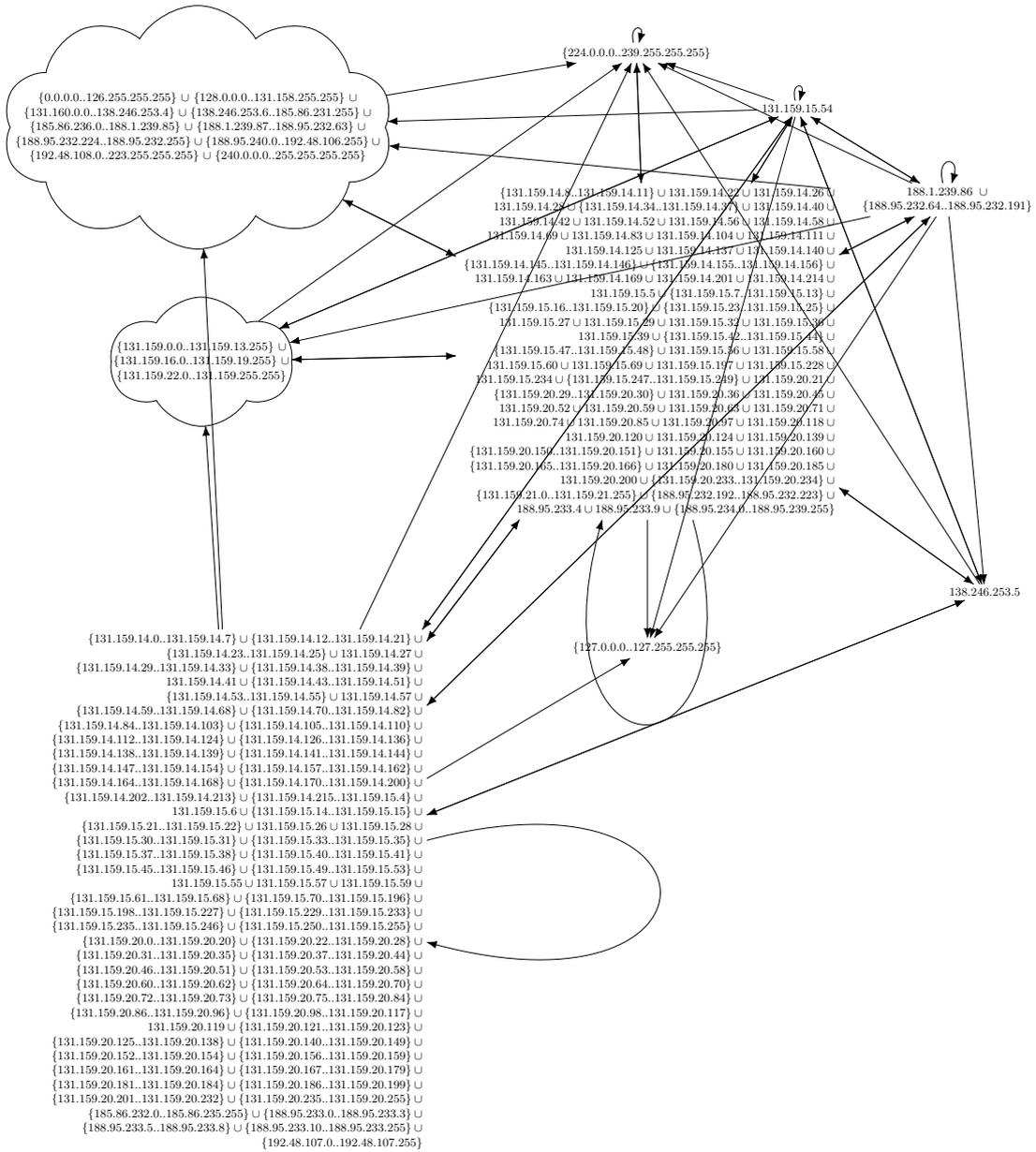

\paragraph*{Firewall A} 
This firewall is the core firewall of our lab (Chair of Network Architectures and Services). 
It has two uplinks, interconnects several VLANs, hence, the firewall matches on more than 20 interfaces. 
It has around 500 direct users and one transfer network for an AS behind it. 
The traffic is usually several Mbit/s.
The dumps are from Oct 2013, Sep 2014, May 2015, Sep 2015 and the changing number of rules indicates that it is actively managed. 
The firewall starts with some rate-limiting rules. 
Therefore, its stricter approximation assumes that the rate-limiting always applies and transforms the ruleset into a deny-all ruleset. 
The more permissive approximation abstracts over this rate-limiting and provides a very good approximation of the original ruleset. 
%
%
The ssh service matrix is visualized in Figure~\ref{fig:tumssh} and in Figure~\ref{fig:tumsshwithips} with the raw IP addresses. 
The figure can be read as follows: 
The vast majority of our IP addresses are grouped into \emph{internal} and \emph{servers}. 
Servers are reachable from the outside, internal hosts are not. 
$\mathit{ip}_1$ and $\mathit{ip}_2$ are two individual IP addresses with special exceptions.
There is also a group for the backbone routers of the connected AS. 
%
INET is the set of IP addresses which does not belong to us, basically the Internet. 
INET' is another part of the Internet. 
With the help of the service matrix, the administrator confirmed that the existence of INET' was an error caused by a stale rule. 
The misconfiguration has been fixed. 
Figure~\ref{fig:tumssh} summarizes over 4000 firewall rules and helps to easily visually verify the complex ssh setup of our firewall. 
The administrator was also interested in the kerberos-adm and ldap service matrices. 
They helped verifying the complex setup and discovered potential for ruleset cleanup. 

We have used the \fffuu{} tool further on to analyze our firewall. 
For example, in Chapter~\ref{chap:puttingtogether}, Figure~\ref{fig:eval_measrdroid:i8fw:port80} and Figure~\ref{fig:eval_measrdroid:i8fw:port80ipv6} were created from a snapshot of June 2016 and depict the service matrix for http. 
The figures show the raw IP addresses. 
It can be seen that the `two INETs' bug has been fixed. 
Note that the service matrix is minimal, \ie there is no way to compress it any further. 
The two figures reveal the intrinsic complexity of this firewall. 
However, the figures ---though complicated--- can still be visualized on one page, something which would be impossible for the thousands of rules of the actual ruleset. 
This demonstrates that our service matrices can give a suitable overview over complicated rulesets.

\paragraph*{Firewall D} 
This firewall was taken from a Shorewall system with 373 rules and 65 chains. 
It can be seen that unfolding increases the number of rules. 
This is due to linearizing the complex call structures generated by the user-defined chains. 
The transformation to the simple firewall further increases the ruleset size. 
This is, among others, due to rewriting several negated IP matches back to non-negated CIDR ranges and NNF normalization. 
However, the absolute numbers tell that this blow up is no problem for computerized analysis. 
The firewall basically wires interfaces together, \ie it heavily uses \texttt{-i} and \texttt{-o}. 
This can be easily seen in the overapproximation. 
There are also many zone-spanning interfaces. 
As we have proven, it is impossible to rewrite interface in this case. 
In addition, for some interfaces, no IP ranges are specified. 
Hence, this ruleset is more of a link layer firewall than a network layer firewall. 
Consequently, the service matrices are barely of any use.

\begin{figure*}[!htbp]
	\centering\resizebox{0.8\linewidth}{!}{
	\small%
	\begin{tikzpicture}
	\node[align=center,text width=6.5cm,cloud, draw,cloud puffs=8,cloud puff arc=120, aspect=2, inner sep=-2em,outer sep=0] (a) at (5,3) { $ \{0.0.0.0 .. 10.13.36.255\} \cup \{ 10.13.38.0 .. 10.13.42.127\} \cup \{10.13.42.192 .. 10.13.42.255\} \cup \{10.13.43.32 .. 10.13.43.255\} \cup \{10.13.45.0 .. 141.76.2.3\} \cup \{141.76.2.5 .. 255.255.254.25\} \cup \{255.255.255.1 .. 255.255.255.255\}$ };
	\node (b) at (8,-6) { $ \{$10.13.37.0 .. 10.13.37.255$\}$ $\cup$ $\{$10.13.44.0 .. 10.13.44.255$\}$ $\cup$ 255.255.255.0 };
	\node (c) at (13,3) { 141.76.2.4 }; 
	\node (d) at (14,-1) { $ \{10.13.43.16 .. 10.13.43.31\}$ }; 
	\node[align=center,text width=5cm] (e) at (9,-3) { $ \{10.13.42.128 .. 10.13.42.175\} \cup \{10.13.43.0 .. 10.13.43.15\}$ };
	\node (f) at (2,-3) { $ \{10.13.42.176 .. 10.13.42.191\}$ }; 
	
	\draw[myptr] (d) to (c);
	\draw[myptr] (d) to[loop below] (d);
	\draw[myptr] (e) to (a);
	\draw[myptr] (e) to (c);
	\draw[myptr] (e) to[loop below] (e);
	\draw[myptr] (f) to (a);
	\draw[myptr] (f) to (b);
	\draw[myptr] (f) to (c);
	\draw[myptr] (f) to (d);
	\draw[myptr] (f) to (e);
	\draw[myptr] (f) to[loop below] (f);
	\end{tikzpicture}%
	}
	\caption{Firewall D ssh Service Matrix with input and output port rewriting}
	\label{fig:ifip:sqrl2014withrouting} 
\end{figure*}
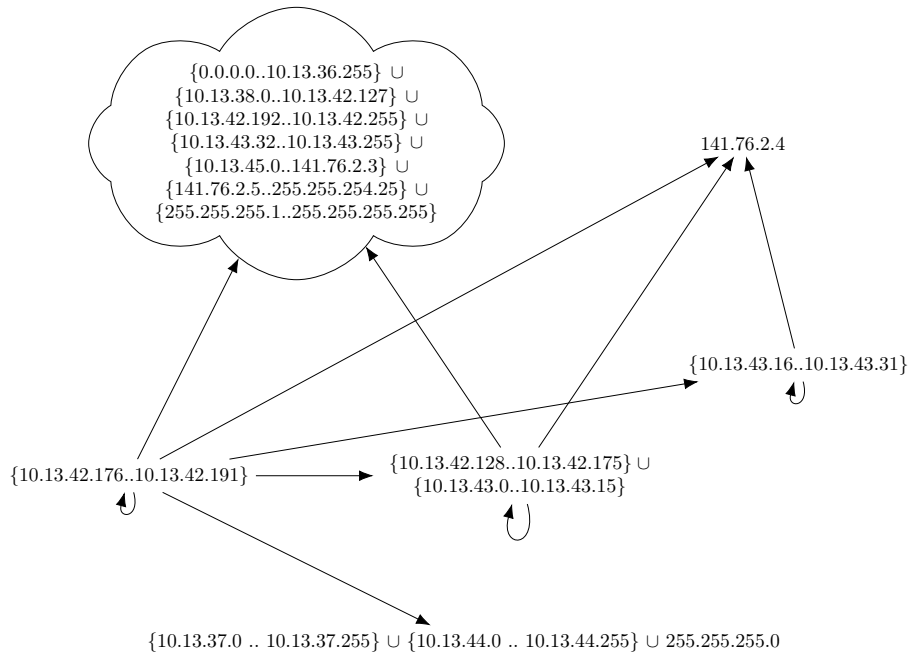

Later on, having obtained more detailed interface and routing configurations, we tried again with input and output port rewriting. 
The result is not shown in the table but visualized in Figure~\ref{fig:ifip:sqrl2014withrouting}. 
The figure now correctly summarizes the network architecture enforced by the firewall. 
It shows the general Internet, a debian update server (141.76.2.4), and four internal networks with different access rights.

\paragraph*{Firewall E}
This ruleset was taken from a NAS device from the introduction (Figure~\ref{fig:firewall:synology}). 
The ruleset first performs some rate-limiting, consequently, the underapproximation corresponds to the deny-all ruleset. 
The table lists a more recent version of the ruleset after a system update. 
Our ssh service matrix reveals a misconfiguration: ssh was accidentally left enabled after the update. 
After this discovery, the flaw was fixed. 
The service matrix for the other services provided by the NAS (not listed in the table) verifies that these services are only accessible from the local network. 
This finally yields the expected result as motivated in Section~\ref{sec:fm15:introduction}. 

\paragraph*{Firewall F}
This firewall is running on a publicly accessible server. 
The firewall first allows everything for localhost, then blocks IP addresses which have shown malicious behavior in the past and finally allows certain services. 
Since most rules are devoted to blocking malicious IPs, our IP address space partition roughly grows linear with the number of rules. 
The service matrices, however, reveal that there are actually only three classes of IP ranges: localhost, the blocked IPs, and all other IPs which are granted access to the services. 

\paragraph*{Firewall G}
For this production server, the service matrices verified that a SQL daemon is only accessible from a local network and three explicitly-defined public IP addresses.

\paragraph*{Firewall H}
This ruleset from 2003 appears to block Kazaa filesharing traffic during working hours. 
In addition, a rule drops all packets with the string ``X-Kazaa-User''. 
The more permissive abstraction correctly tells that the firewall may accept all packets for all IPs (if the above conditions do not hold). 
Hence, the firewall is essentially abstracted to an allow-all ruleset. 
According to our metric, this information is not useful. 
However, in this scenario, this information may reveal an error in the ruleset: 
The firewall explicitly permits certain IP ranges, however, the default policy is $\iptaction{ACCEPT}$ and includes all these previously explicitly permitted ranges. 
By inspecting the structure of the firewall, we suppose that the default policy should be $\iptaction{DROP}$. 
This possible misconfiguration was uncovered by the overapproximation. 
The underapproximation does not understand the string match on ``X-Kazaa-User'' in the beginning and thus corresponds to the deny-all ruleset. 
However, a manual inspection of the underapproximation still reveals an interesting error: 
The ruleset also tries to prevent MAC address spoofing for some hard-coded MAC/IP pairs. 
However, we could not see any drop rules for spoofed MAC addresses in the underapproximation. 
Indeed, the ruleset allows non-spoofed packets but forgets to drop the spoofed ones.
This firewall demonstrates the worst case for our approximations: one set of accepted packets is the universe, the other is the empty set.\footnote{Note that this ruleset is actually severely broken and no better approximation would be possible.} 
However, manual inspection of the simplified ruleset helped reveal several errors. 
This demonstrates that even if the service matrices do not contain any information, the other output of our tool may still contain interesting information.

%
%

%
\paragraph*{Firewall P} 
This is the ruleset of the main firewall of a medium-sized company. 
The administrator asked us what her ruleset was doing. 
She did not reveal her intentions to prevent that the analysis might be skewed towards the expected outcome. 
We calculated the simplified firewall rules and service matrices. 
Using the underapproximation, we could also give guarantees about the packets which are definitely allowed by the firewall. 
The administrator critically inspected the output of our tool. 
Finally, she confirmed that the firewall was working exactly as intended. 
This demonstrates: Not only finding errors but showing correctness is one of the key strengths of our tool. 

After the analysis, the administrator revealed her true intentions. 
She has previously upgraded the system to iptables. 
Her users, the company's employees, got aware of that fact. 
She received some complaints about connectivity issues and the employees were blaming the firewall.
However, the administrator was suspecting that the connectivity issues were triggered by some users who are behaving against the corporate policy, \eg sharing user accounts. 
With the help of our analysis, the administrator could reject all accusations about her firewall config and follow her initial suspicion about misbehaving employees.

A few months later, we received some final feedback that the firewall was perfect and ``users are stupid''. 

\paragraph*{Firewall R} 
This ruleset was extracted from a docker host. 
We discuss this scenario and the history of the ruleset in detail in Chapter~\ref{chap:dynamicdocker}. 
For remote management, the ruleset allows unconstrained ssh access for all machines, which can be seen by the fact that the ssh service matrix only shows one partition. 
In contrast, an advanced setup is enforced for http and the http matrix is visualized in Figure~\ref{fig:dynamicdocker:fffuu:dockermynet.topos4.1.state}. 
Being able to verify the publicly exposed http setup while neglecting the ssh maintenance setup demonstrates the advantage of calculating our access matrices for each service. 
The lower closure also exhibits one interesting detail: Except for one host which is rate limited, ssh connectivity is guaranteed. 
Ironically, ITVal segfaults on the original ruleset. 
With our processing, it terminates successfully but returns a spurious result.

\section{Conclusion}
We have demonstrated the first, fully verified, real-world applicable analysis framework for firewall rulesets. 
Our tool supports the Linux iptables firewall because it is widely used and well-known for its vast amount of features. 
It directly works on \texttt{iptables-save} output. 
We presented an algebra on common match conditions and a method to translate complex conditions to simpler ones. 
Further match conditions, which are either unknown or cannot be translated, are approximated in a sound fashion. 
This results in a translation method for complex, real-world rulesets to a simple model. 
The evaluation demonstrates that, despite possible approximation, the simplified rulesets preserve the interesting aspects of the original ones. 

Based on the simplified model, we presented algorithms to partition the IPv4 and IPv6 address space and compute service matrices. 
This allows summarizing and verifying the firewall in a clear manner. 

The analysis is fully implemented in the Isabelle theorem prover. 
No additional input or knowledge of mathematics is required by the administrator. 
Our stand-alone Haskell tool \fffuu{} can perform the analysis automatically, only requiring the following input: \texttt{iptables-save}. 

The evaluation demonstrates applicability on many real-world rulesets. 
For this, to the best of our knowledge, we have collected and published the largest collection of real-world iptables rulesets in academia. 
We demonstrated that our approach can outperform existing tools with regard to: correctness, supported match conditions, CPU time, and RAM requirements. 
Our tool helped to verify lack of or discover previously unknown errors in real-world, production rulesets.

\part{Applicability \& Conclusion}
\label{part:applicability}

\chapter{Overview}
\label{chap:partthree:overview}
In this part, we demonstrate the interplay of the tools we developed in Part~\ref{part:greenfield} and Part~\ref{part:existing-configs}. 
As the following figure illustrates, our tools \fffuu{} and \topos{} work on common abstractions and allow to translate between different levels of abstraction. 
  \begin{center}
  	\resizebox{0.99\textwidth}{!}{%
  \begin{tikzpicture}
  \node [MyRoundedBox, fill=LightYellow](sinvar) at (0,0) {Security Requirements};
  \node [MyDoubleArrow](arr1) at (sinvar.east) {};
  \node [MyRoundedBox, fill=LightYellow](policy) at (arr1.east) {Security Policies};
  \node [MyDoubleArrow](arr2) at (policy.east) {};
  \node [MyRoundedBox, fill=LightYellow](mechanism) at (arr2.east) {\texttt{iptables}};

  \path[draw,thick,-latex,shorten >=0.5cm,shorten <=0.5cm] 
  (sinvar) to[bend left]    node[anchor=south, yshift=1ex] {\topos{} Policy Construction} (policy);

  \path[draw,thick,-latex,shorten >=0.5cm,shorten <=0.5cm] 
  (policy) to[bend left]    node[anchor=south, yshift=1ex] {\topos{} Serialize} (mechanism);

  \path[draw,thick,-latex,shorten >=0.5cm,shorten <=0.5cm] 
  (mechanism) to[bend left]    node[anchor=north, yshift=-1ex] {\fffuu{} Service Matrices} (policy);

  \path[draw,thick,-latex,shorten >=0.5cm,shorten <=0.5cm] 
  (policy) to[bend left]    node[anchor=north, yshift=-1ex] {\topos{} Verification} (sinvar);
  \end{tikzpicture}%
  }
  \end{center}

This part is structured as follows. 
First, we give an overview of the applicability of our tools in Chapter~\ref{chap:dynamicdocker} by presenting a story about the dynamic management of docker container networking. 
Afterwards, in Chapter~\ref{chap:puttingtogether}, we demonstrate our tools in a real-world case study, namely a privacy evaluation of the Android MeasrDroid app. 
Having demonstrated the big picture of the individual results obtained in this thesis, in Chapter~\ref{chap:answers} we summarize our answers to the scientific questions asked at the beginning. 
In Chapter~\ref{chap:relatedworktable}, we compare this work to the state of the art. 
We summarize the applicability of our work in Chapter~\ref{chap:applicability} and conclude this thesis in Chapter~\ref{chap:conclusion}.

  \chapter{Demonstrating Dynamic Microservice Management}
\label{chap:dynamicdocker}
%
%
%

%
%
%
%

\paragraph*{Abstract}
In this chapter, we demonstrate the interplay of \topos{} and \fffuu{}. 
We present a fictional story about administrating a dynamically changing docker environment. 
From the initial design and software engineering through network operations and automation, we show how our tools help.

\section{Introduction}
\begin{figure}[!hbt]%
	\centering%
	\resizebox{\textwidth}{!}{%
		\begin{tikzpicture}
		\node (tool) at (0,0) {\Large\topos{}};
		
		\node[anchor=east, text width=0.3\linewidth, align=center] (net) at ($(tool.west) + (-1.5,2.5)$) {
			\includegraphics[width=\linewidth]{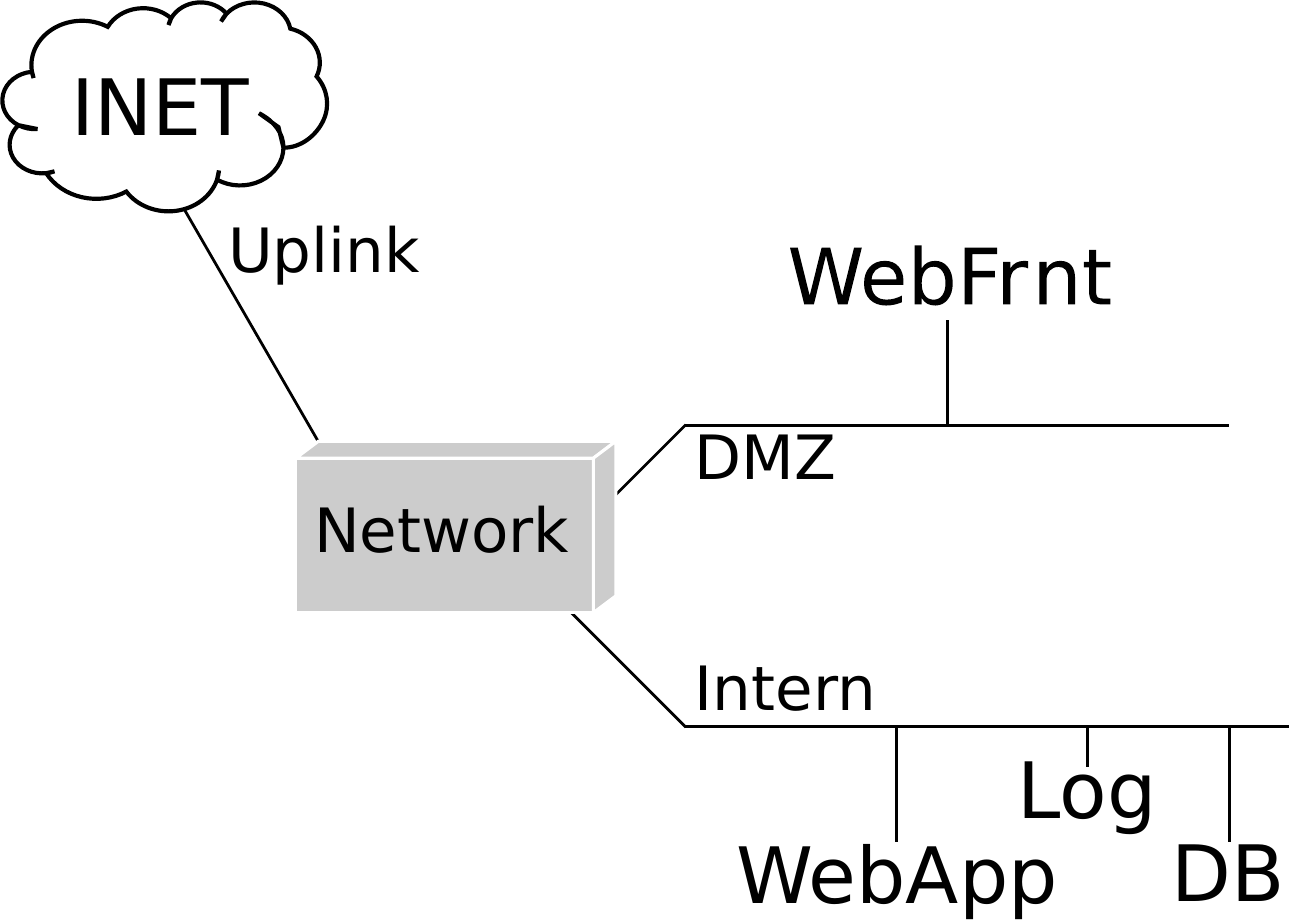}\newline
			(Figure~\ref{fig:netschematic})
		};
		\node[anchor=east, text width=0.3\linewidth, align=center] (plus) at ($(tool.west) + (-1.5,0)$) {$+$};
		\node[anchor=east, text width=0.3\linewidth, align=center] (secreqs) at ($(tool.west) + (-1.5,-2.5)$) {
			\fbox{
				\resizebox{.89\linewidth}{!}{
					\begin{minipage}{\widthof{Bell-LaPadula $\lbrace\mvar{WebApp} \mapsto \mdef{declassify} \ (\mdef{trusted})\rbrace$ }}
					\input{content/sdnnfv15_secrequsfig}
					\end{minipage}
				}
			}\newline
			(Figure~\ref{fig:runexsecinvars})
		};
		
		\path let \p1 = (plus) in let \p2 = (secreqs.east) in node[MySingleArrow,anchor=center](arr1) at ($(\x2,\y1)!0.5!(tool.west)$) {};
		
		\node[anchor=west, text width=0.45\linewidth, align=center] (result) at ($(tool.east) + (1.5,0)$) {
			\fbox{%
				\resizebox{0.96\linewidth}{!}{%
					\begin{minipage}{\widthof{\verb~-A FORWARD -o br-b74b417b331f -m conntrack --ctstate RELATED,ESTABLISHED -j ACCEPT   ~}}%
					\input{content/sdnnfv_dockerfirewall_tuned_fig}%
					\end{minipage}%
				}%
			}\newline
			(Figure~\ref{fig:dockerfirewall:tuned})};
		
		\node [MySingleArrow,anchor=center](arr2) at ($(tool.east)!0.5!(result.west)$) {};
		\end{tikzpicture}%
	}%
	\caption{Input and Output of \topos{}}%
	\label{fig:dynamicdocker:toposiface}%
\end{figure}

This chapter demonstrates applicability of our tools in a dynamic context by telling a fictional story about administrator Alice. 
Alice loves docker and she maintains the distributed web application we have presented in Chapter~\ref{chap:mansdnnfv}. 
She knows that it is best security practice when using docker to decrease the attack surface by limiting container networking~\cite{docker2015sec}. 
Hence, she takes great care about her firewall settings. 
In Figure~\ref{fig:dynamicdocker:toposiface}, we summarize how \topos{} was used in Chapter~\ref{chap:mansdnnfv} to create the initial firewall rules.

%

\begin{figure}[!thb]
	\centering%
	\resizebox{\textwidth}{!}{%
		\begin{tikzpicture}%
		\node (tool) at (0,0) {\Large\fffuu{}};
		
		\node[anchor=east, text width=0.42\linewidth, align=center] (fw) at ($(tool.west) + (-1.5,0)$) {\fbox{%
				\resizebox{0.96\linewidth}{!}{%
					\begin{minipage}{\widthof{\verb~-A FORWARD -o br-b74b417b331f -m conntrack --ctstate RELATED,ESTABLISHED -j ACCEPT   ~}}%
					\input{content/sdnnfv_dockerfirewall_tuned_fig}%
					\end{minipage}%
				}%
			}\newline
			(Figure~\ref{fig:dockerfirewall:tuned})};
		
		\node [MySingleArrow,anchor=center](arr1) at ($(fw.east)!0.5!(tool.west)$) {};
		
		\node[anchor=west, text width=0.4\linewidth, align=center] (result) at ($(tool.east) + (1.5,0)$) {
			\includegraphics[width=\linewidth]{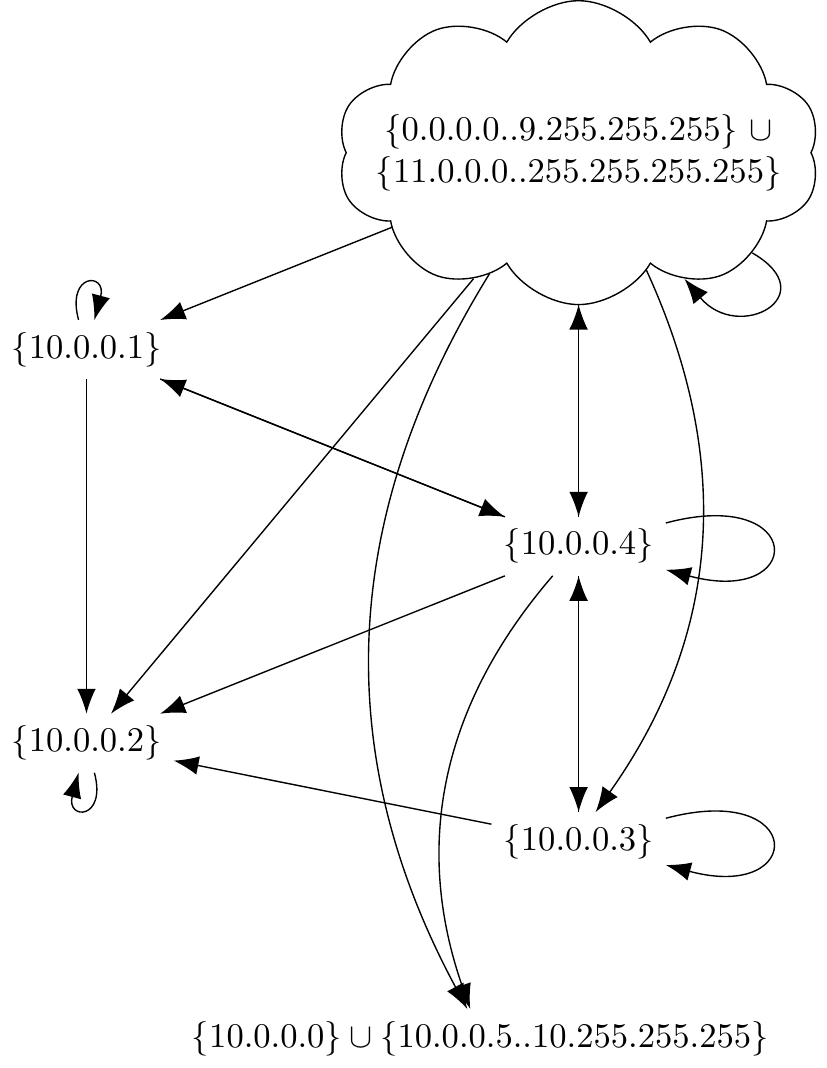}\newline
			(Figure~\ref{fig:mansdn:docker:servicematrix})};
		
		\node [MySingleArrow,anchor=center](arr2) at ($(tool.east)!0.5!(result.west)$) {};
		\end{tikzpicture}%
	}%
	\caption{Input and Output of \fffuu{}}%
	\label{fig:dynamicdocker:fffuuiface}%
\end{figure}

Throughout this thesis, we have mainly focused on \emph{static} configuration management. 
Since our tools are completely automatic, they can also be used in \emph{dynamic} contexts. 
Alice installs a cronjob which runs \texttt{iptables-save} regularly and looks for changes. 
If a change to the ruleset is discovered, \fffuu{} is run to compute an overview of the firewall policy currently enforced.  
The result is visualized with tikz and emailed to Alice. 
We visualize this process in Figure~\ref{fig:dynamicdocker:fffuuiface}. 
In this story, the requirements, setup, and configurations are frequently changed and we demonstrate the use of our tools in this dynamic context.

\section{Network Access Control in Docker}
First, Alice questions her initial decision of Chapter~\ref{chap:mansdnnfv} to operate custom firewall rules and researches whether docker can provide similar features out of the box. 

Since docker version 1.10, it is possible to create custom internal networks~\cite{docker2016blog110}. 
The \texttt{{-}{-}internal} flag protects a network from accesses from the outside. 
However, by default, all containers in an internal network can access each other. 
It is possible to configure the network such that containers cannot access each other. 
Ironically, this feature was broken in our version of docker and containers could still communicate~\cite{github2016dockerinternalicc}. 
In addition, there is no possibility to enable fine-grained access control between the containers in a network. 
A \texttt{{-}{-}link} option exists to connect two containers within a custom network, but it merely sets environment variables, it does not influence the actual IP connectivity. 

The docker design philosophy is to decouple the application developer from networking details~\cite{docker2016blognetphilosophy}. 
Only a coarse-grained network abstraction in terms of different networks is exposed to the application developer. 
The network IT team ---\ie Alice--- should manage the network. 
Since one compromised container in an internal network can attack all other containers in its network, Alice desires further network-level access control. 


Due to the lack of fine-grained access control, Alice sticks to her custom firewall rules in the docker host. 
In addition, the fact that the fix to the bug mentioned above was initially not considered a security issue by the docker developers confirms Alice's choice of running her own firewall. 
As it turns out, many administrators are looking for means to fine-tune their docker firewall because of certain shortcomings in docker~\cite{github2016dockerfwbypass}.


\section{The First Day}
At the beginning of this story, the installed iptables ruleset corresponds to the one listed in Figure~\ref{fig:dockerfirewall:tuned}. 
Over time, a ruleset evolves. 
On her first day, Alice receives a call from a friend at \url{heise.de} (193.99.144.80), who is complaining that one of her containers is pinging his webserver excessively. 
Alice knows that the web backend tests connectivity from time to time by sending one echo request to heise. 
She decides to postpone investigating the core of the problem and installs a short-term mitigation by just rate limiting all connections to heise. 
We print changes to the ruleset in unified \texttt{diff} format. 
Alice installs the following rules. 

\smallskip

\newcommand{\diffadd}[1]{\textcolor{Green}{#1}}
\newcommand{\diffdel}[1]{\textcolor{Red}{#1}}

\begin{minipage}{.95\linewidth}
	\footnotesize
	\begin{Verbatim}[commandchars=\\\{\},codes={\catcode`$=3\catcode`^=7}]
 :DOCKER-ISOLATION - [0:0]
 :MYNET - [0:0]
 -A FORWARD -j DOCKER-ISOLATION
\diffadd{+-A FORWARD -d 193.99.144.80 -m recent --set --name rateheise --rsource}
\diffadd{+-A FORWARD -d 193.99.144.80 -m recent --update --seconds 60 --hitcount 3} $\hfill\hookleftarrow$
             \diffadd{--name rateheise --rsource -j DROP}
 -A FORWARD -j MYNET
 -A FORWARD -o br-b74b417b331f -j DOCKER
 -A FORWARD -o br-b74b417b331f -m conntrack --ctstate RELATED,ESTABLISHED -j ACCEPT
\end{Verbatim}
\end{minipage}
\medskip

While this change to the ruleset is identified by her cronjob, \fffuu{} confirms that the overall access control structure of the firewall has not changed: It still corresponds to Figure~\ref{fig:mansdn:docker:servicematrix}. 

While Alice is updating the ruleset, she remembers from Figure~\ref{fig:mansdn:docker:servicematrix} that the Internet still has too many direct access rights to her internal containers. 
This is not a major problem since her docker configuration ultimately prevents unintended accesses from the Internet. 
Yet, while having opened the ruleset in her editor, Alice decides to fix this issue right away as a second line of defense. 

\smallskip

\begin{minipage}{.95\linewidth}
	\footnotesize
	\begin{Verbatim}[commandchars=\\\{\},codes={\catcode`$=3\catcode`^=7}]
 -A MYNET -i br-b74b417b331f -s 10.0.0.4 -o br-b74b417b331f -d 10.0.0.3 -j ACCEPT
 -A MYNET -i br-b74b417b331f -s 10.0.0.4 -o br-b74b417b331f -d 10.0.0.2 -j ACCEPT
 -A MYNET -i br-b74b417b331f -s 10.0.0.4 -o br-b74b417b331f -d 10.0.0.4 -j ACCEPT
\diffdel{--A MYNET -i br-b74b417b331f -s 10.0.0.4 ! -o br-b74b417b331f -j ACCEPT}
\diffdel{--A MYNET ! -i br-b74b417b331f -o br-b74b417b331f -d 10.0.0.1 -j ACCEPT}
\diffadd{+-A MYNET -i br-b74b417b331f -s 10.0.0.4 ! -o br-b74b417b331f ! -d 10.0.0.0/8 -j ACCEPT}
\diffadd{+-A MYNET ! -i br-b74b417b331f ! -s 10.0.0.0/8 -o br-b74b417b331f -d 10.0.0.1 -j ACCEPT}
 -A MYNET -i br-b74b417b331f -j DROP
\diffadd{+-A MYNET -o br-b74b417b331f -j DROP}
\diffadd{+-A MYNET -s 10.0.0.0/8 -j DROP}
\diffadd{+-A MYNET -d 10.0.0.0/8 -j DROP}
 COMMIT
	\end{Verbatim}
\end{minipage}

\medskip

After a few seconds, she receives an email with the firewall overview generated by \fffuu{}, shown in Figure~\ref{fig:dynamicdocker:fffuu:dockermynet3}. 
Comparing this figure 
  to Figure~\ref{fig:mansdn:docker:servicematrix}, it can be seen that the Internet is now appropriately constrained.

\begin{figure}[!thb]
	\centering
	\begin{minipage}[t]{0.32\linewidth}\centering
			\resizebox{\linewidth}{!}{\Large
				\begin{tikzpicture}
\node[align=center,text width=15.5em, cloud, draw,cloud puffs=10,cloud puff arc=120, aspect=2, inner sep=-3em,outer sep=0] (a) at (5,1) { $\{0.0.0.0 .. 9.255.255.255\} \cup \{11.0.0.0 .. 255.255.255.255\}$ };
\node (b) at (5,-3) { $\{10.0.0.4\}$ };
\node (c) at (5,-6) { $\{10.0.0.3\}$ }; 
\node (d) at (0,-5) { $\{10.0.0.2\}$ };
\node (e) at (0,-1) { $\{10.0.0.1\}$ };
\node (f) at (4,-8) { $\{10.0.0.0\} \cup \{10.0.0.5 .. 10.255.255.255\}$ };

\draw[myptr] (a) to[out=330,in=310,looseness=3] (a);
\draw[myptr] (a) to (b);
\draw[myptr] (a) to[bend left] (c);
\draw[myptr] (a) to (d);
\draw[myptr] (a) to (e);
\draw[myptr] (a) to[bend right] (f);
\draw[myptr] (b) to (a);
\draw[myptr] (b) to[loop right] (b);
\draw[myptr] (b) to (c);
\draw[myptr] (b) to (d);
\draw[myptr] (b) to (e);
\draw[myptr] (b) to[bend right] (f);
\draw[myptr] (c) to (b);
\draw[myptr] (c) to[loop right] (c);
\draw[myptr] (c) to (d);
\draw[myptr] (d) to[loop below] (d);
\draw[myptr] (e) to (b);
\draw[myptr] (e) to (d);
\draw[myptr] (e) to[loop above] (e);
\end{tikzpicture}%
			}
	\caption*{Copy of Figure~\ref{fig:mansdn:docker:servicematrix}}
	\end{minipage}%
	\hspace*{\fill}%
	\begin{minipage}[t]{0.32\textwidth}\centering
		\resizebox{\linewidth}{!}{\Large
		\begin{tikzpicture}
		\node[align=center,text width=15.5em, cloud, draw,cloud puffs=10,cloud puff arc=120, aspect=2, inner sep=-3em,outer sep=0] (a) at (5,1) { $\{0.0.0.0 .. 9.255.255.255\} \cup \{11.0.0.0 .. 255.255.255.255\}$ };
		\node (b) at (5,-3) { $\{10.0.0.4\}$ };
		\node (c) at (5,-6) { $\{10.0.0.3\}$ }; 
		\node (d) at (0,-5) { $\{10.0.0.2\}$ };
		\node (e) at (0,-1) { $\{10.0.0.1\}$ };
		\node (f) at (4,-8) { $\{10.0.0.0\} \cup \{10.0.0.5 .. 10.255.255.255\}$ };
		
		\draw[myptr] (a) to[out=330,in=310,looseness=3] (a);
		\draw[myptr] (a) to (e);
		\draw[myptr] (b) to (a);
		\draw[myptr] (b) to[loop right] (b);
		\draw[myptr] (b) to (c);
		\draw[myptr] (b) to (d);
		\draw[myptr] (b) to (e);
		\draw[myptr] (c) to (b);
		\draw[myptr] (c) to[loop right] (c);
		\draw[myptr] (c) to (d);
		\draw[myptr] (d) to[loop below] (d);
		\draw[myptr] (e) to (b);
		\draw[myptr] (e) to (d);
		\draw[myptr] (e) to[loop above] (e);
		\end{tikzpicture}%
		}
	\caption{Overview computed by \fffuu{}}
	\label{fig:dynamicdocker:fffuu:dockermynet3}
	\end{minipage}%
	\hspace*{\fill}%
	\begin{minipage}[t]{0.32\linewidth}\centering
	   \resizebox{\linewidth}{!}{\Large
		\begin{tikzpicture}
		\node[align=center,text width=15.5em,cloud, draw,cloud puffs=10,cloud puff arc=120, aspect=2, inner sep=-3em,outer sep=0] (a) at (5,1) { $\{0.0.0.0 .. 9.255.255.255\} \cup \{11.0.0.0 .. 255.255.255.255\}$ };
		\node (b) at (5,-3) { $\{10.0.0.4\}$ };
		\node (c) at (5,-6) { $\{10.0.0.3\}$ }; 
		\node (d) at (0,-5) { $\{10.0.0.2\}$ };
		\node (e) at (0,-1) { $\{10.0.0.1\}$ };
		\node (f) at (4,-8) { $\{10.0.0.0\} \cup \{10.0.0.5 .. 10.255.255.255\}$ };
	
		\draw[myptr] (a) to[out=330,in=310,looseness=3] (a);
		\draw[myptr] (a) to (e);
		\draw[myptr] (b) to (a);
		\draw[myptr] (b) to[loop right] (b);
		\draw[myptr] (b) to (c);
		\draw[myptr] (b) to (d);
		\draw[myptr] (b) to (e);
		\draw[myptr] (c) to (b);
		\draw[myptr] (c) to[loop right] (c);
		\draw[myptr] (c) to (d);
		\draw[myptr] (d) to[loop below] (d);
		\draw[myptr] (d) to (e);
		\draw[myptr] (e) to (b);
		\draw[myptr] (e) to (d);
		\draw[myptr] (e) to[loop above] (e);
	\end{tikzpicture}%
	}
	\caption{Overview computed by \fffuu{} after WebDev call} 
	\label{fig:dynamicdocker:fffuu:dockermynet4}
	\end{minipage}%
\end{figure}

\section{A Call from the Web Developer}

Continuing the story, Alice receives a call from her web developer. 
He requests ssh access to all containers. 
In addition, he requests that the log server (10.0.0.2) may access a status page of the web frontend (10.0.0.1) over HTTP. 
Both permissions should only be granted temporarily for debugging purposes. 
Alice sets up the appropriate firewall rules. 

\smallskip

\begin{minipage}{.95\linewidth}
	\footnotesize
	\begin{Verbatim}[commandchars=\\\{\},codes={\catcode`$=3\catcode`^=7}]
 -A FORWARD -j DOCKER-ISOLATION
 -A FORWARD -d 193.99.144.80 -m recent --set --name rateheise --rsource
 -A FORWARD -d 193.99.144.80 -m recent --update --seconds 60 --hitcount 3 $\hfill\hookleftarrow$
             --name rateheise --rsource -j DROP
\diffadd{+-A FORWARD -m state --state ESTABLISHED,RELATED -j ACCEPT}
\diffadd{+-A FORWARD -p tcp --dport 22 -j ACCEPT}
\diffadd{+-A FORWARD -s 10.0.0.2 -d 10.0.0.1 -p tcp --dport 80 -j ACCEPT}
 -A FORWARD -j MYNET
 -A FORWARD -o br-b74b417b331f -j DOCKER
 -A FORWARD -o br-b74b417b331f -m conntrack --ctstate RELATED,ESTABLISHED -j ACCEPT
	\end{Verbatim}
\end{minipage}

\medskip


While the docker container connectivity works as intended, \fffuu{} now computes two interesting service matrices.\footnote{By default, \fffuu{} only computes the service matrices for ssh and HTTP. } 
First, it visualizes that there are no longer any restrictions for ssh, \ie \fffuu{} only shows one node. 
This one node comprises the complete IPv4 address space and may access itself. 
Second, \fffuu{} presents a new HTTP service matrix, shown in Figure~\ref{fig:dynamicdocker:fffuu:dockermynet4}. 
The only difference is that the log server may access the web frontend. 
The two service matrices underline that Alice implemented her web developer's request correctly.

\section{Emergency Response \& Scaling Horizontally}
While the policy overview computed by \fffuu{} is still comprehensible for Alice, the raw firewall ruleset, now comprising 37 rules, is slowly becoming a mess. 
The ruleset partly contains unused artifacts installed by docker and Alice's hot fixes are cluttered all over it. 
While Alice is poring over about how she could clean up the rules, she receives an emergency call. 
Her manager tells her that the webservice was mentioned on reddit and that the web frontend cannot cope with the additional load. 
Alice is spawning an additional frontend container with IP address 10.0.0.42. 

Alice's firewall adheres to best practices and implements whitelisting. 
Consequently, the new container does not have any connectivity. 
Alice now is in the urgent situation to get the firewall rules set up which permit connectivity for the second frontend instance. 
Just permitting everything is not an acceptable option for security-aware Alice.

\begin{figure}[htb]
	\centering
	\begin{minipage}[b]{0.49\linewidth}\centering
		\centering
		\includegraphics[trim=0cm 1cm 0cm 0cm,clip=true,width=.9\textwidth]{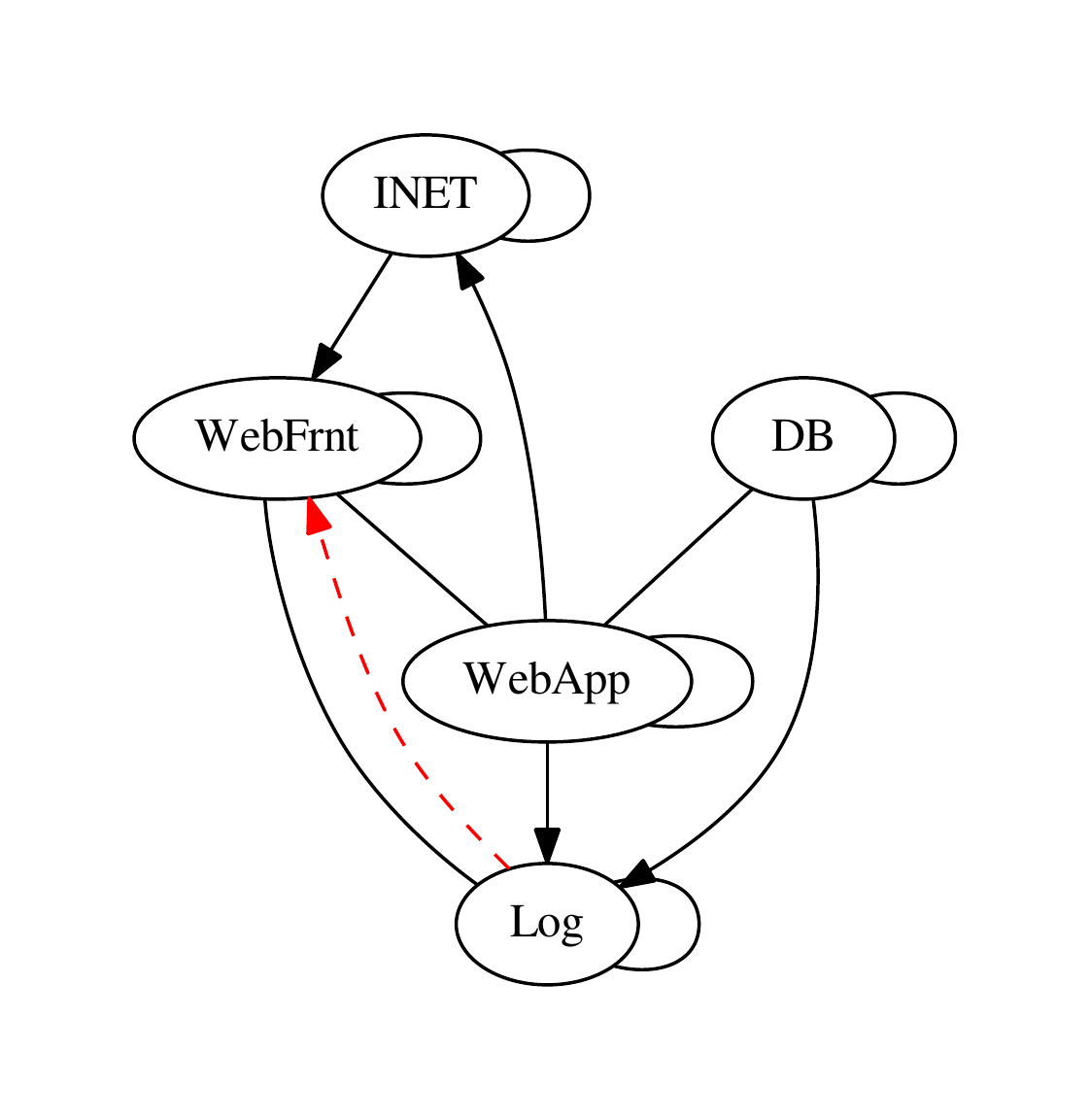}
		\vskip-1ex
		\caption{Policy Verification reveals a violation (generated by \topos{})}
		\label{fig:dynamicdocker:topos:dockermynet4}
	\end{minipage}%
	\hspace*{\fill}%
	\begin{minipage}[b]{0.49\textwidth}\centering
		\centering
		\includegraphics[trim=0cm 1cm 0cm 0cm,clip=true,width=.9\textwidth]{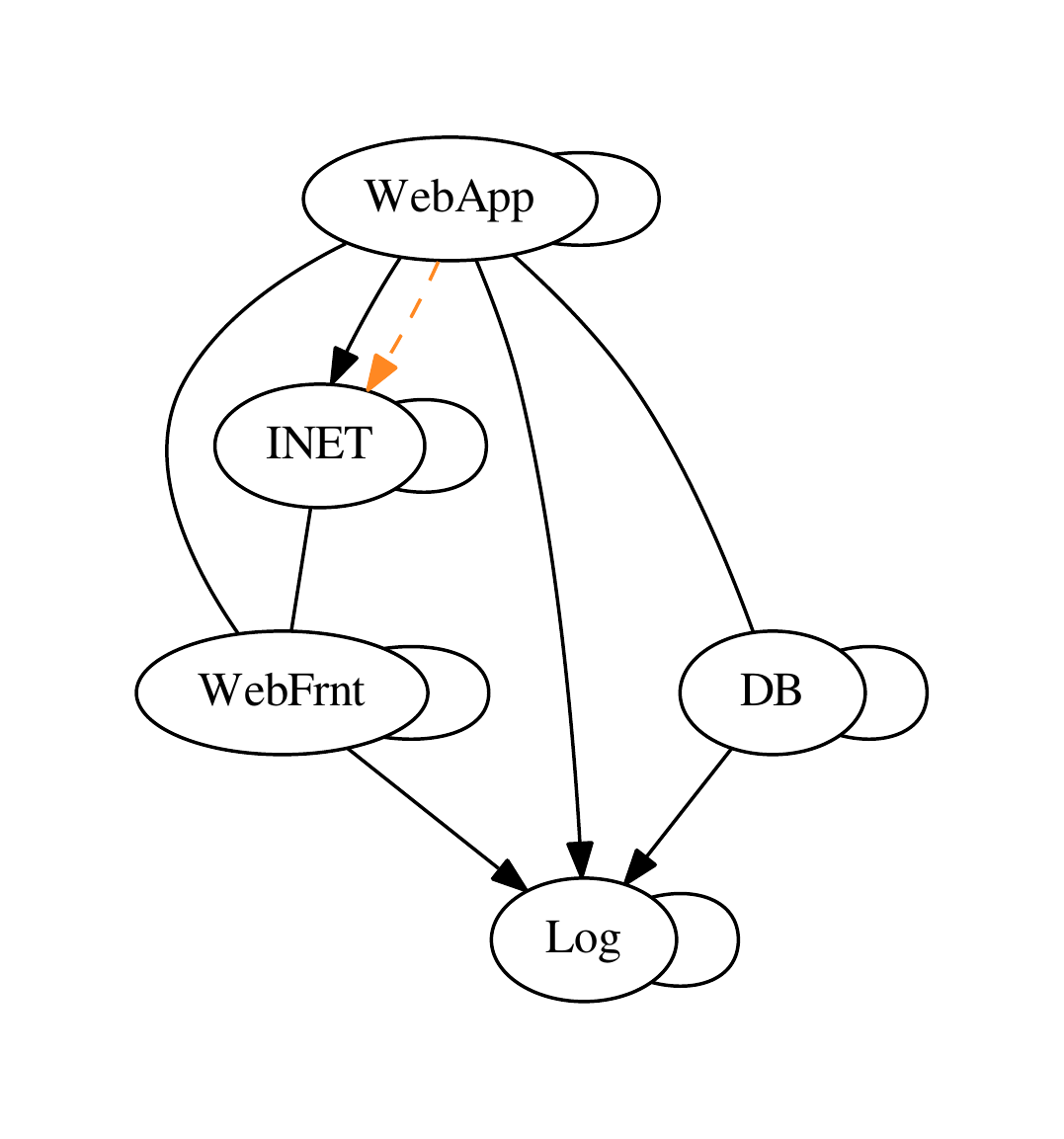}
		\vskip-1ex
		\caption{Stateful Policy (generated by \topos{})}
		\label{fig:dynamicdocker:topos:dockermynet4.stateful}
	\end{minipage}%
\end{figure}

Fortunately, Alice remembers that she has specified the security requirements with \topos{} long ago (Figure~\ref{fig:runexsecinvars}). 
First, Alice loads the policy overview for HTTP computed by \fffuu{} (Figure~\ref{fig:dynamicdocker:fffuu:dockermynet4}) into \topos{}. 
She immediately gets a visualization (Figure~\ref{fig:dynamicdocker:topos:dockermynet4}) which tells her that her policy has already diverged from the security requirements: The dashed red arrow, visualized by \topos{}, reveals a flow which is violating the security requirements. 
It corresponds to the flow installed upon the request of her web developer previously.  
Based on the security requirements, Alice should probably have rejected the request from her web developer in the very beginning. 
But this is neither the time to cast the blame nor to discuss security requirements. 
Given the time pressure and the many odd-looking docker-generated rules in her ruleset, Alice decides that her current ruleset is unsalvageable. 
Alice is not the first person to notice that the docker daemon may make surprising changes to an iptables configuration~\cite{docker2014blogufw,github2016dockerfwbypass}. 
She decides to take over complete control over the ruleset. 
Therefore, Alice prohibits docker from making any changes to the firewall by setting \texttt{{-}{-}iptables=false}.


\begin{figure*}[htbp]
	\begin{minipage}{\linewidth}
		\footnotesize
\begin{Verbatim}[commandchars=\\\{\},codes={\catcode`$=3\catcode`^=7}]
-A FORWARD -i $\mathit{\$dockerbr}$ -s $\mathit{\$WebFrnt\_ipv4}$ -o $\mathit{\$dockerbr}$ -d $\mathit{\$WebFrnt\_ipv4}$ -j ACCEPT 
-A FORWARD -i $\mathit{\$dockerbr}$ -s $\mathit{\$WebFrnt\_ipv4}$ -o $\mathit{\$dockerbr}$ -d $\mathit{\$Log\_ipv4}$ -j ACCEPT 
-A FORWARD -i $\mathit{\$dockerbr}$ -s $\mathit{\$WebFrnt\_ipv4}$ -o $\mathit{\$dockerbr}$ -d $\mathit{\$WebApp\_ipv4}$ -j ACCEPT 
-A FORWARD -i $\mathit{\$dockerbr}$ -s $\mathit{\$WebFrnt\_ipv4}$ -o $\mathit{\$INET\_iface}$ -d $\mathit{\$INET\_ipv4}$ -j ACCEPT 
-A FORWARD -i $\mathit{\$dockerbr}$ -s $\mathit{\$DB\_ipv4}$ -o $\mathit{\$dockerbr}$ -d $\mathit{\$DB\_ipv4}$ -j ACCEPT 
-A FORWARD -i $\mathit{\$dockerbr}$ -s $\mathit{\$DB\_ipv4}$ -o $\mathit{\$dockerbr}$ -d $\mathit{\$Log\_ipv4}$ -j ACCEPT 
-A FORWARD -i $\mathit{\$dockerbr}$ -s $\mathit{\$DB\_ipv4}$ -o $\mathit{\$dockerbr}$ -d $\mathit{\$WebApp\_ipv4}$ -j ACCEPT 
-A FORWARD -i $\mathit{\$dockerbr}$ -s $\mathit{\$Log\_ipv4}$ -o $\mathit{\$dockerbr}$ -d $\mathit{\$Log\_ipv4}$ -j ACCEPT 
-A FORWARD -i $\mathit{\$dockerbr}$ -s $\mathit{\$WebApp\_ipv4}$ -o $\mathit{\$dockerbr}$ -d $\mathit{\$WebFrnt\_ipv4}$ -j ACCEPT 
-A FORWARD -i $\mathit{\$dockerbr}$ -s $\mathit{\$WebApp\_ipv4}$ -o $\mathit{\$dockerbr}$ -d $\mathit{\$DB\_ipv4}$ -j ACCEPT 
-A FORWARD -i $\mathit{\$dockerbr}$ -s $\mathit{\$WebApp\_ipv4}$ -o $\mathit{\$dockerbr}$ -d $\mathit{\$Log\_ipv4}$ -j ACCEPT 
-A FORWARD -i $\mathit{\$dockerbr}$ -s $\mathit{\$WebApp\_ipv4}$ -o $\mathit{\$dockerbr}$ -d $\mathit{\$WebApp\_ipv4}$ -j ACCEPT 
-A FORWARD -i $\mathit{\$dockerbr}$ -s $\mathit{\$WebApp\_ipv4}$ -o $\mathit{\$INET\_iface}$ -d $\mathit{\$INET\_ipv4}$ -j ACCEPT 
-A FORWARD -i $\mathit{\$INET\_iface}$ -s $\mathit{\$INET\_ipv4}$ -o $\mathit{\$dockerbr}$ -d $\mathit{\$WebFrnt\_ipv4}$ -j ACCEPT 
-A FORWARD -i $\mathit{\$INET\_iface}$ -s $\mathit{\$INET\_ipv4}$ -o $\mathit{\$INET\_iface}$ -d $\mathit{\$INET\_ipv4}$ -j ACCEPT 
-I FORWARD -m state --state ESTABLISHED -i $\mathit{\$INET\_iface}$ $\hfill\hookleftarrow$
            -s $\mathit{\$INET\_ipv4}$ -o $\mathit{\$dockerbr}$ -d $\mathit{\$WebApp\_ipv4}$ -j ACCEPT 
\end{Verbatim}
\end{minipage}%
\caption{Fresh ruleset generated by \topos{}, considering only the requirements}
\label{fig:dynamicdocker:topos:dockermynet.topos4}
\end{figure*}

To start over, Alice asks \topos{} to compute a completely new ruleset for her, based only on the requirements specified before. 
\topos{} computes the stateful policy shown in Figure~\ref{fig:dynamicdocker:topos:dockermynet4.stateful}; the dashed orange flow indicates a connection with stateful semantics. 
The result is serialized to the firewall rules shown in Figure~\ref{fig:dynamicdocker:topos:dockermynet.topos4}. 
All containers are attached to the same docker bridge $\mathit{\$dockerbr} = \texttt{br-b74b417b331f}$. 
Alice only needs to fill in the IP addresses of the machines and the Internet-facing interface. 
Alice sticks to the IP addresses of her containers as listed in Table~\ref{tab:mansdncentralfw:ipmapping}, except for the web frontend which now has two active containers running. 
She sets $\mathit{\$WebFrnt\_ipv4} = \textnormal{10.0.0.1,10.0.0.42}$. 
This syntax is supported by iptables and iptables will automatically expand this syntax to several rules upon loading. 
To specify the missing variables, \ie the interface and IP range of the Internet, Alice simply defines that the Internet is `everything except her docker subnet'. 
Therefore, Alice negates her docker interface and internal docker IP range. 
\begin{samepage}
For example, the second last rule becomes the following: 

\begin{footnotesize}
\begin{Verbatim}[commandchars=\\\{\},codes={\catcode`$=3\catcode`^=7}]
-A FORWARD ! -i br-b74b417b331f ! -s 10.0.0.0/8 $\hfill\hookleftarrow$
            ! -o br-b74b417b331f ! -d 10.0.0.0/8 -j ACCEPT
\end{Verbatim}
\end{footnotesize}
\end{samepage}


Yet, for some rules, the iptables command complains that negation is not allowed with multiple source or destination IP addresses. 
For example in line four, iptables prohibits the use of \verb~! -d 10.0.0.0/8~ in combination with the two source addresses \verb~-s 10.0.0.1,10.0.0.42~ specified for $\mathit{\$WebFrnt\_ipv4}$. 
To work around this iptables limitation, Alice uses the \verb~iprange~ module to declare the IP range of the Internet. 
\begin{samepage}
For example, the fourth rule now becomes%
\begin{footnotesize}
\begin{Verbatim}[commandchars=\\\{\},codes={\catcode`$=3\catcode`^=7}]
-A FORWARD -i br-b74b417b331f -s 10.0.0.1,10.0.0.42 $\hfill\hookleftarrow$
           ! -o br-b74b417b331f -m iprange ! --dst-range 10.0.0.0-10.255.255.255 -j ACCEPT
\end{Verbatim}
\end{footnotesize}
\end{samepage}

\begin{figure}[!thb]
	\centering
	\begin{minipage}[t]{0.32\linewidth}\centering
		\resizebox{\linewidth}{!}{\Large
		\begin{tikzpicture}
		\node[align=center,text width=15.5em,cloud, draw,cloud puffs=10,cloud puff arc=120, aspect=2, inner sep=-3em,outer sep=0] (a) at (5,1) { $\{0.0.0.0 .. 9.255.255.255\} \cup \{11.0.0.0 .. 255.255.255.255\}$ };
		\node (c) at (5,-3) { $\{10.0.0.4\}$ };
		\node (d) at (5,-6) { $\{10.0.0.3\}$ }; 
		\node (e) at (0,-5) { $\{10.0.0.2\}$ };
		\node (b) at (0,-1) { $\{10.0.0.1,10.0.0.42\}$ };
		\node[align=center,text width=15.5em] (f) at (3,-8) { $\{10.0.0.0\} \cup \{10.0.0.5 .. 10.0.0.41\} \cup \{10.0.0.43 .. 10.255.255.255\} $ };
		
		\draw[myptr] (a) to[out=330,in=310,looseness=3] (a);
		\draw[myptr] (a) to (b);
		\draw[myptr] (b) to (a);
		\draw[myptr] (b) to[loop above] (b);
		\draw[myptr] (b) to (c);
		\draw[myptr] (b) to (e);
		\draw[myptr] (c) to (a);
		\draw[myptr] (c) to (b);
		\draw[myptr] (c) to[loop right] (c);
		\draw[myptr] (c) to (d);
		\draw[myptr] (c) to (e);
		\draw[myptr] (d) to (c);
		\draw[myptr] (d) to[loop right] (d);
		\draw[myptr] (d) to (e);
		\draw[myptr] (e) to[loop below] (e);
		\end{tikzpicture}%
		}
	\caption{Overview of Figure~\ref{fig:dynamicdocker:topos:dockermynet.topos4} computed by \fffuu{}} 
	\label{fig:dynamicdocker:fffuu:dockermynet.topos4}
	\end{minipage}%
	\hspace*{\fill}%
	\begin{minipage}[t]{0.32\textwidth}\centering
		\resizebox{.7\linewidth}{!}{\Large
			\begin{tikzpicture}
			\node[align=center,text width=15.5em,cloud, draw,cloud puffs=10,cloud puff arc=120, aspect=2, inner sep=-3em,outer sep=0] (a) at (5,0) { $\{0.0.0.0 .. 255.255.255.255\}$ };
			
			\draw[myptrdotted] (a) to[loop below] (a);
			
			\end{tikzpicture}%
		}
	\caption{Allowed established flows, computed by \fffuu{}} 
	\label{fig:dynamicdocker:fffuu:dockermynet.topos4.1.established}
	\end{minipage}%
	\hspace*{\fill}%
	\begin{minipage}[t]{0.32\linewidth}\centering
	   \resizebox{\linewidth}{!}{\Large
	   \begin{tikzpicture}
	   	\node[align=center,text width=15.5em,cloud, draw,cloud puffs=10,cloud puff arc=120, aspect=2, inner sep=-3em,outer sep=0] (a) at (5,1) { $\{0.0.0.0 .. 9.255.255.255\} \cup \{11.0.0.0 .. 255.255.255.255\}$ };
	   	\node (c) at (5,-3) { $\{10.0.0.4\}$ };
	   	\node (d) at (5,-6) { $\{10.0.0.3\}$ }; 
	   	\node (e) at (0,-5) { $\{10.0.0.2\}$ };
	   	\node (b) at (0,-1) { $\{10.0.0.1,10.0.0.42\}$ };
	   	\node[align=center,text width=15.5em] (f) at (3,-8) { $\{10.0.0.0\} \cup \{10.0.0.5 .. 10.0.0.41\} \cup \{10.0.0.43 .. 10.255.255.255\} $ };
	   	
	   	\draw[myptr] (a) to[out=330,in=310,looseness=3] (a);
	   	\draw[myptr] (a) to (b);
	   	\draw[myptr] (b) to (a);
	   	\draw[myptr] (b) to[loop above] (b);
	   	\draw[myptr] (b) to (c);
	   	\draw[myptr] (b) to (e);
	   	\draw[myptr] (c) to (a);
	   	\draw[myptr] (c) to (b);
	   	\draw[myptr] (c) to[loop right] (c);
	   	\draw[myptr] (c) to (d);
	   	\draw[myptr] (c) to (e);
	   	\draw[myptr] (d) to (c);
	   	\draw[myptr] (d) to[loop right] (d);
	   	\draw[myptr] (d) to (e);
	   	\draw[myptr] (e) to[loop below] (e);
	   	
	   	\draw[myptrdotted,orange] (a) to[bend left=15, shorten <=0.6em,shorten >=0.2em] (c);
	   	\end{tikzpicture} %
	   	}
	\caption{HTTP Service Matrix with state (by \fffuu{})}
	\label{fig:dynamicdocker:fffuu:dockermynet.topos4.1.state}
	\end{minipage}%
\end{figure}

%
Fortunately, \fffuu{} understands all those matching modules. 
The firewall overview is visualized in Figure~\ref{fig:dynamicdocker:fffuu:dockermynet.topos4}. 
It is remarkably similar to Figure~\ref{fig:dynamicdocker:fffuu:dockermynet3}, the last, old visualization where the ruleset was still in a good state. 
The main difference is that the web frontend is now represented by two machines and that it may establish connections to the Internet itself. 
This has been prohibited by the old policy but it does not contradict any security requirement. 
A final test confirms that the container connectivity works as expected and the two frontend instances can cope with the load. 

\section{The Logging Information Leak}
Looking at her todo list, Alice decides to install some of the old rules again. 
This time, she designs a clean ruleset and handles all of her temporary rules in a chain she calls \texttt{CUSTOM}. 
After her custom chain, she hands over control to the \topos{}-generated rules. 
Alice still has not investigated why some container is excessively pinging 193.99.144.80, so she installs the rate limiting again. 
Alice is more careful about the other temporary rules. 
\topos{} has shown her that the log server must not communicate with the web frontend, so she is not enabling this rule. 
However, she does not see a problem with the ssh exception and enables it again. 
She installs the following rules.

\smallskip

\begin{minipage}{.95\linewidth}
	\footnotesize
	\begin{Verbatim}[commandchars=\\\{\},codes={\catcode`$=3\catcode`^=7}]
 :INPUT ACCEPT [0:0]
 :FORWARD DROP [0:0]
 :OUTPUT ACCEPT [0:0] +:CUSTOM - [0:0]
\diffadd{+-A FORWARD -j CUSTOM}
\diffadd{+-A CUSTOM -d 193.99.144.80 -m recent --set --name rateheise --rsource}
\diffadd{+-A CUSTOM -d 193.99.144.80 -m recent --update --seconds 60 --hitcount 3} $\hfill\hookleftarrow$
            \diffadd{--name rateheise --rsource -j DROP}
\diffadd{+-A CUSTOM -m state --state ESTABLISHED -j ACCEPT}
\diffadd{+-A CUSTOM -p tcp -m tcp --dport 22 -j ACCEPT}
 -A FORWARD -i br-b74b417b331f -s 10.0.0.1,10.0.0.42  $\hfill\hookleftarrow$
             -o br-b74b417b331f -d 10.0.0.1,10.0.0.42 -j ACCEPT 
 -A FORWARD -i br-b74b417b331f -s 10.0.0.1,10.0.0.42  $\hfill\hookleftarrow$
             -o br-b74b417b331f -d 10.0.0.2 -j ACCEPT 
 -A FORWARD -i br-b74b417b331f -s 10.0.0.1,10.0.0.42  $\hfill\hookleftarrow$
             -o br-b74b417b331f -d 10.0.0.4 -j ACCEPT
	\end{Verbatim}
\end{minipage}

\medskip

Alice knows that it is a good practice to have a rule which allows all packets belonging to an established connection~\cite{iptablesperfectruleset}. 
She definitely needs an \verb~ESTABLISHED~ rule to make ssh work so she just copies it from a guide. 
Though, Alice wonders why \topos{} did not generate such a rule. 
Afterwards, she becomes skeptical about her decision and want's to double check. 
Asking \fffuu{} about the potential packet flows once a connection is initiated, \fffuu{} confirms that there are currently no limitations at all, not even for HTTP. 
This is visualized in Figure~\ref{fig:dynamicdocker:fffuu:dockermynet.topos4.1.established}. 
She compares this to the stateful implementation intended by \topos{}, shown in Figure~\ref{fig:dynamicdocker:topos:dockermynet4.stateful}. 
The dashed orange line indicates a flow with stateful semantics, \ie packets may flow in both directions once the connection was initiated by the web app. 
She realizes that \topos{} takes great care to enforce unidirectional information flow to the log server. 
This is due to the information sink security invariant specified in the requirements. 
Alice knows from recent news that a badly protected log server may leak information which may lead to the compromise of all her machines~\cite{rhel2016logpwn}. 
Therefore, Alice restricts her \verb~ESTABLISHED~ rule to ssh. 
She uses the \texttt{multiport} module which conveniently allows to match on source or destination port within one rule. 
She makes the following final adjustment to her ruleset. 

\smallskip

\begin{minipage}{.95\linewidth}
	\footnotesize
	\begin{Verbatim}[commandchars=\\\{\},codes={\catcode`$=3\catcode`^=7}]
 -A CUSTOM
 -A CUSTOM -d 193.99.144.80/32 -m recent --set --name rateheise --rsource
 -A CUSTOM -d 193.99.144.80/32 -m recent --update --seconds 60 --hitcount 3 $\hfill\hookleftarrow$
            --name rateheise --rsource -j DROP
\diffdel{--A CUSTOM -p tcp -m state --state ESTABLISHED -j ACCEPT}
\diffadd{+-A CUSTOM -p tcp -m state --state ESTABLISHED -m multiport --ports 22 -j ACCEPT}
 -A CUSTOM -p tcp -m tcp --dport 22 -j ACCEPT
 COMMIT
	\end{Verbatim}
\end{minipage}

\medskip

Alice runs one final verification of the implemented policy with \fffuu{}, shown in Figure~\ref{fig:dynamicdocker:fffuu:dockermynet.topos4.1.state}. 
This time, she also includes the stateful flows. 
\fffuu{} identifies only one stateful flow, visualized by an orange dashed line. 
The direction of the stateful flow is the other way round compared to Figure~\ref{fig:dynamicdocker:topos:dockermynet4.stateful}. 
This is merely an artifact of the visualization, a stateful flow is essentially bidirectional once it is established. 
Otherwise, Figure~\ref{fig:dynamicdocker:topos:dockermynet4.stateful} and Figure~\ref{fig:dynamicdocker:fffuu:dockermynet.topos4.1.state} show isomorphic graphs. 
This verifies that Alice's firewall rules are correct.


%
%

\section{Related Docker Work}
%
%
Tools to improve firewall management for docker hosts exist~\cite{github2016dockerfw,github2016dfwfw}. 
Docker-fw~\cite{github2016dockerfw} is a convenient iptables wrapper with docker-specific features, such as retrieving the IP address from a container name using the Docker API. It currently only supports the default docker bridge, but not custom networks. 
%
DFWFW~\cite{github2016dfwfw} is also a convenient tool to manage the iptables firewall in a docker host. 
It runs as a daemon and can apply changes dynamically if the docker setup changes, \eg if new containers are instantiated.

Both tools provide features that could help to make the management process with \topos{} and \fffuu{} more convenient. 
At the moment, \topos{} generates raw iptables rules but leaves the actual IP addresses to be set by the user, \eg $\mathit{\$WebFrnt\_ipv4}$ in Figure~\ref{fig:dynamicdocker:topos:dockermynet.topos4}. 
To further automate the setup, \topos{} could generate docker-fw~\cite{github2016dockerfw} rules which automatically resolve the correct IP address. 
To further automate firewall management, \topos{} could directly generate DFWFW~\cite{github2016dfwfw} configurations. 
This would mean that no manual configuration is required any longer if multiple instances of the same container are spawned.

We have tested \topos{} together with DFWFW. 
For this, we simply adapted the \topos{} serialization step to generate rules in the DFWFW configuration format. 
Since DFWFW is also built to primarily support whitelisting, the translation is straightforward. 
A rule in this format first matches on the docker network (in our scenario, we called the network \texttt{mynet}), then it allows to specify the source and destination container, it allows to specify an arbitrary string which will be added to the iptables match expression, and finally the iptables action. 
The match on the container names permits the use of Perl regular expressions. 
To allow dynamic spawning of multiple instances of a container, we wrote a regex which matches on the container name and any trailing number, \eg \texttt{webfrnt}, \texttt{webfrnt1}, \texttt{webfrnt-1}, \texttt{webfrnt200}. 
The beginning of the configuration file looks as follows: 

\smallskip

\begin{minipage}{.95\linewidth}
	\footnotesize
	\begin{Verbatim}[commandchars=\\\{\},codes={\catcode`$=3\catcode`^=7}]
\{
  "container_to_container": \{
  "rules": [
    \{
       "network": "mynet",
       "src_container": "Name =\textasciitilde \textasciicircum{}webfrnt-?\textbackslash{}\textbackslash{}d*\$",
       "dst_container": "Name =\textasciitilde \textasciicircum{}webfrnt-?\textbackslash{}\textbackslash{}d*\$",
       "filter": "",
       "action": "ACCEPT"
     \}, 
     \{
       "network": "mynet",
       "src_container": "Name =\textasciitilde \textasciicircum{}webfrnt-?\textbackslash{}\textbackslash{}d*\$",
       "dst_container": "Name =\textasciitilde \textasciicircum{}log-?\textbackslash{}\textbackslash{}d*\$",
       "filter": "",
       "action": "ACCEPT"
     \}, 
       $\dots$
	\end{Verbatim}
\end{minipage}

\medskip

Disregarding all the line breaks, it is similar to Figure~\ref{fig:dynamicdocker:topos:dockermynet.topos4}, but only the first two rules are shown. 
We tested that all containers have the necessary connectivity with this setting. 
We also tested that the firewall gets dynamically updated once we instantiate new copies of our containers and that the most obvious attempts to subvert the security policy are successfully blocked. 
We also verified the generated iptables rules with \fffuu{}. 
It reveals that the overall setting is indeed good, but it also uncovers two open issues: 
First, Internet connectivity is again unconstrained and the stateless semantics are not enforced correctly, \ie once a connection with the log server is established, bidirectional communication is permitted. 
We leave these engineering issues of fine tuning the DFWFW configuration to future work. 

\section{Comparison to Academic State-of-the-Art}
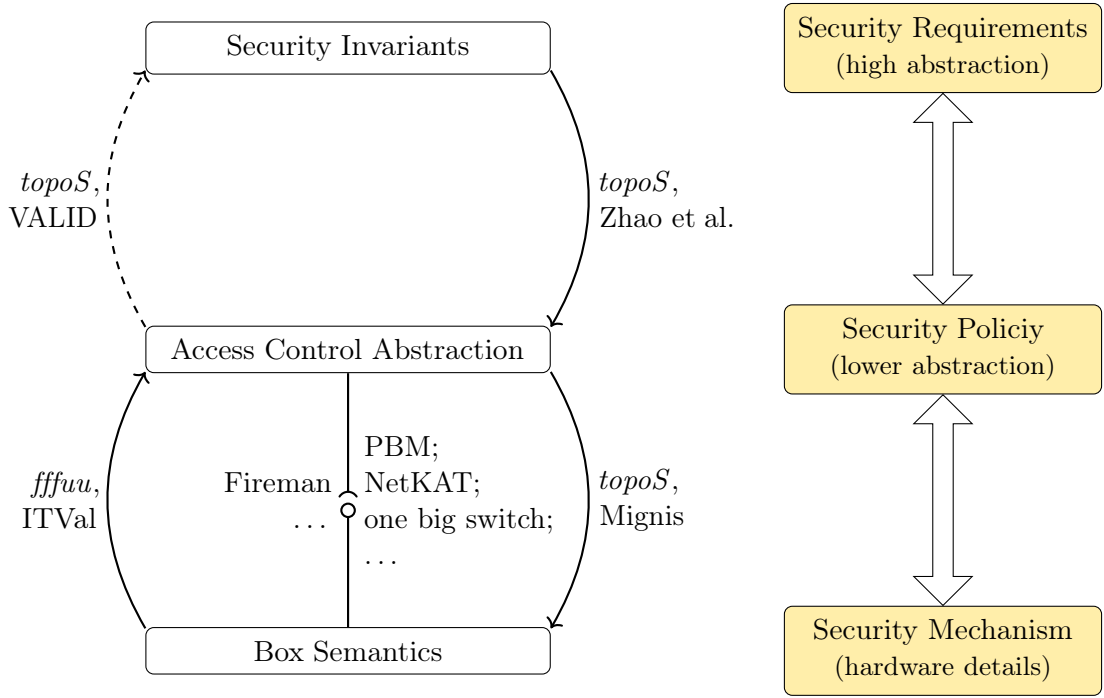
\begin{figure}[!htb]%
	\centering%
	\begin{tikzpicture}%
	\node [MyRoundedBox, text width=13em](abs1) at (0,0) {Security Invariants};
	\node [MyRoundedBox, text width=13em](abs2) at (0,-4) {Access Control Abstraction};
	\node [MyRoundedBox, text width=13em](abs4) at (0,-8) {Box Semantics};
	
	\draw [thick,-to,->] ($(abs1.east)+(0,-1.8ex)$) to[bend left] ($(abs2.east)+(0,+1.8ex)$);
	\node [anchor=west, align=left, text width=7em] at ($(abs1)!0.5!(abs2) + (+8.3em,0)$) {\topos{}, Zhao~et$\;$al.};
	
	\draw [thick,-to,dashed,->] ($(abs2.west)+(0,+1.8ex)$) to[bend left] ($(abs1.west)+(0,-1.8ex)$);
	\node [anchor=east, align=right, text width=3em] at ($(abs1)!0.5!(abs2) + (-8.3em,0)$) {\topos{}, VALID};
	
	\draw [thick,-to,->] ($(abs2.east) + (0,-1.8ex)$) to[bend left] ($(abs4.east) + (0,+1.8ex)$);
	\node [anchor=west, align=left, text width=4em] at ($(abs2)!0.5!(abs4) + (+8.3em,0)$) {\topos{}, \mbox{Mignis}};
	
	\draw [thick,-to,->] ($(abs4.west) + (0,+1.8ex)$) to[bend left] ($(abs2.west) + (0,-1.8ex)$);
	\node [anchor=east, align=right, text width=3em] at ($(abs2)!0.5!(abs4) + (-8.3em,0)$) {\fffuu{}, ITVal};
	
	\draw [thick,-to,-(] ($(abs2.south)$)--($(abs4.north)!0.5!(abs2.south) + (0,.2ex)$);
	\draw [thick,-to,o-] ($(abs4.north)!0.5!(abs2.south) + (0,-.2ex)$)--($(abs4.north)$);
	\node [anchor=west, align=left, text width=7em] at ($(abs2)!0.5!(abs4) + (+0.2em,0)$) {PBM;\newline{}\mbox{NetKAT;}\newline{}\mbox{one big switch;}\newline{}\dots};
	\node [anchor=east, align=right, text width=4em] at ($(abs2)!0.5!(abs4) + (-0.2em,0)$) {Fireman \dots};


	\node [MyRoundedBox, fill=LightYellow](req) at ($(abs1) + (15em,0)$) {Security Requirements \small{(high~abstraction)}};
	\node [MyRoundedBox, fill=LightYellow](pol) at ($(abs2) + (15em,0)$) {Security Policiy \small{(lower~abstraction)}};
	\node [MyRoundedBox, fill=LightYellow](mech) at ($(abs4) + (15em,0)$) {Security Mechanism \small{(hardware~details)}};
	\node [MyDoubleArrow, anchor=center, rotate=90, text width=5.9em](arr1) at ($(req.south)!0.5!(pol.north)$) {};
	\node [MyDoubleArrow, anchor=center, rotate=90, text width=5.9em](arr2) at ($(pol.south)!0.5!(mech.north)$) {};
	\end{tikzpicture}
	\caption{Overview of Comparison to Related Work}
	\label{fig:dockerrelated}
\end{figure}

We are not aware of any academic related work about network access control management specifically for docker. 
To the best of our knowledge, our tools are the only ones where applicability for a docker environment has been demonstrated. 
But docker is merely an application example. 
We broaden the scope and compare this work to the general state of the art of tools for helping network management and administration. 
The comparison is outlined in Figure~\ref{fig:dockerrelated}. 
The figure is aligned similar to Figure~\ref{fig:relatedworkabstractionlayers}, but we omit the Interface Abstraction. 
We add the three security components of Figure~\ref{fig:intro:securitycomponents} on the right to visualize the corresponding level of abstraction. 
The solid single arrows on the left part of the figure visualize work which is able to translate between the corresponding abstractions. 
The dashed single arrow represents work which is only capable of verification: 
Given as input an access control policy and security invariants, their conformance can be verified. 
But one cannot be derived from the other. 
As discussed Chapter~\ref{chap:introduction}, a translation from an access control policy to security invariants is not possible without guessing a policy author's intention.


Our combination of \topos{} and \fffuu{} is the only compatible, jointly designed toolset which is able to bridge all levels of abstractions in both directions out of the box. 
The fictional story reveals that this back-and-forth is useful in several scenarios. 
The story also reveals single features an administrator may wish for. 
We now compare \topos{} and \fffuu{} with related work specifically for the following use cases, considered in isolation. 

\paragraph*{Build Networks Based on a Security Requirement Specification}
Instead of writing a policy or low-level configuration by hand, the fictional story shows that it is useful to generate working network configurations directly from a scenario-specific security requirement specification. 
This is useful for the initial design and implementation, as well as for starting over in certain scenarios. 

VALID~\cite{bleikertz2011VALID} allows to express security requirements, but it cannot derive network configurations from them. 
Zhao \etal\cite{zhao2011policyremanet} also propose a framework which allows to express security requirements. 
In contrast to VALID, their framework additionally allows to derive working network configurations. 
The language proposed by Zhao \etal exposes a lot of formalism to the administrator and almost bears more resemblance to programming than it bears to specifying. 
Compared with this, \topos{} distinguishes between templates and instantiating a template. 
While defining new templates also bears resemblance to programming and is only intended for expert users, the common operation to define a specification is by instantiating templates, which only requires configurations and exposes very little formalism to the administrator. 



\paragraph*{Allow Intervention and Low-Level Control for the Administrator}
The fictional story revealed several occasions where Alice wanted to fine-tune the low-level policy by hand. 
One example was the temporary, ad-hoc permission for ssh. 
Another example was the rate limiting to \url{heise.de}. 
It may also be imaginable that Alice needs to restructure her ruleset at some point for performance reasons or to support a chain managed by \texttt{fail2ban}. 
Except for the ssh permission, those are true low-level operations which should not be achievable on a higher level of abstraction.\footnote{tautologically, higher levels of abstraction abstract over low-level details.} 

The Mignis~\cite{mignis2014} firewall configuration language allows to specify filtering policies. 
It gives the administrator the optional possibility to add arbitrary additional low-level iptables match conditions to the high-level rules. 
These additional match conditions may introduce soundness issues. 
The language does not permit the administrator to change the generated iptables rules directly, \eg reordering, restructuring, or ad-hoc changes without recompiling are not allowed. 
In contrast, \topos{} permits arbitrary changes to the generated iptables rules since \fffuu{} can be used to verify correctness of the changes.

\paragraph*{Detect Erosion and Drift of the Implemented Policy vs.\ the Specified Policy}
%
The terms \emph{erosion} and \emph{drift} are usually used for software architectures~\cite{Perry1992softwarearchitectures}. 
However, our fictional story shows that network security policies and the corresponding configurations also easily decay, become unmaintainable, and violations of the original requirements sneak in. 
In addition, being able to detect differences between a configuration and a specified policy is an important step towards understanding legacy configurations or verifying manual low-level changes, as discussed in the previous paragraph. 

To analyze the current policy enforced by an iptables ruleset, ITVal~\cite{marmorstein2005itval,marmorstein2006firewall} can be used. 
Similar to \fffuu{}, it allows to partition the complete IPv4 address range into equivalence classes. 
ITVal computes one set of equivalence classes jointly for all layer~4 ports. 
In contrast, \fffuu{} can only compute them for one selected port. 
It depends on the scenario which of the two approaches is more suitable: 
ITVal's overview is very helpful for a first, quick, overview of a firewall. 
\fffuu{}'s service matrices provide better granularity, once one knows which ports one is interested in. 
ITVal only supports IPv4 while \fffuu{} supports IPv4 and IPv6. 
In addition, ITVal is known to have bugs (Chapter~\ref{chap:networking16}) while \fffuu{} is formally proven correct. 
Ironically, ITVal segfaults for some docker rulesets of this chapter while \fffuu{} processes them without complaint. 

\paragraph*{Distributed Enforcement?}
Our work only focuses on one single, central enforcement device. 
But having only one central firewall or only one central docker host is not a satisfying scenario. 
This raises the question whether our work is useless for large installations or whether distributed enforcement is an orthogonal issue.
The Interface Abstraction---discussed in Section~\ref{sec:sdnnfv:related} and omitted in Figure~\ref{fig:dockerrelated}---corresponds to distributed enforcement. 
Figure~\ref{fig:relatedworkabstractionlayers} lists several related work which takes one centralized policy and enforces it in a distributed fashion. 
Therefore, distributed enforcement is an orthogonal issue and our tools can help to develop, verify, and maintain the centralized policy which is then enforced in a distributed fashion. 
For example, policy-based management~\cite{dinesh2002policybased} systems or the NetKAT compiler~\cite{icfp2015smolkanetkatcompiler} for SDN could be used. 
We explicitly visualize an interface boundary \tikz[baseline]{
\draw [thick,-to,-(] (0,.7ex)--($(.5,.7ex) + (-.2ex,0)$);
\draw [thick,-to,o-] ($(.5,.7ex) + (.2ex,0)$)--(1,.7ex);
} in Figure~\ref{fig:dockerrelated} to highlight that \topos{} produces an access control matrix, which is understood by many technologies for distributed enforcement. 
Therefore, \topos{} can be used as a module for access control within another system. 
%
In addition, algorithms for distributed firewall analysis (as supported by Fireman~\cite{fireman2006}) can also benefit from the pre-processing and simplification provided by \fffuu{}.

\section{Conclusion}
We showed how our tools \topos{} and \fffuu{} can be used to design, manage, and operate docker-based environments. 
Chapter~\ref{chap:mansdnnfv} shows how our toolset helps during the design phase of a setup and this chapter shows how our tools can help during daily operations. 
We demonstrated several situations in which our tools provide useful feedback, uncover bugs, and even help migrating setups. 
The duality of \topos{} and \fffuu{} in combination with their common policy abstraction makes them a powerful combination and enhances the academic state-of-the-art.

We also enhanced the state-of-the-art of docker container management by combining the dynamic docker firewall framework DFWFW with our static policy management tool \topos{}. 
While the docker firewall enforces the security policy, \topos{} generates it. 
This combination lifts \topos{} to dynamic contexts since it allows dynamic spawning and deletion of containers at runtime while still providing strong guarantees about the enforced security requirements. 
In addition, our tool \fffuu{} can verify at runtime the correct operation according to the policy.

Notably, our tools are not limited to docker environments, but also applicable to different scenarios. 
This becomes explicit as there was not a single change necessary for \fffuu{} to support the docker scenarios. 
To support DFWFW as additional backend, only few lines in the serialization component of \topos{} needed to be adapted.


%
%

%



  \chapter{Case Study: MeasrDroid Privacy Evaluation and Improvement}
\label{chap:puttingtogether}
  
This chapter is an extended version of a part of the following paper~\cite{maltitz2016fmpriv}:
\begin{itemize}
	\item Marcel von Maltitz, Cornelius Diekmann and Georg Carle, \emph{Taint Analysis for System-Wide Privacy Audits: A Framework and Real-World Case Studies}. In 1st Workshop for Formal Methods on Privacy, Limassol, Cyprus, November 2016. Note: no proceedings published. 
\end{itemize}


\paragraph*{Statement on author's contributions}
The evaluation of MeasrDroid was led by von Maltitz as domain expert on privacy. 
He provided major contributions for the modeling of MeasrDroid. 
Both authors, the author of this thesis and von Maltitz, contributed equally to the audit of the real-world MeasrDroid system. 
Most of the work for the audit was performed by the tools \topos{} and \fffuu{} which are developed by the author of this thesis.
%

    
\medskip

\paragraph*{Abstract}
  In this chapter, we put all the results of this thesis together, demonstrated by a real-world case study.  
  We audit privacy requirements of the smartphone measurement system MeasrDroid. 
  First, we use \topos{} to model the MeasrDroid architecture and to formalize and verify the privacy requirements. 
  Next, using \fffuu{}, we audit the real implementation of MeasrDroid enforced by a network firewall. 
  We uncover previously unknown bugs, use \topos{} to automatically compute a correct firewall ruleset, and deploy the improved ruleset to the real system. 
  Our audit bridges the gap from a high-level security requirement analysis to complex low-level firewall rules. 
  To the best of our knowledge, this is the first time that such an audit has been performed completely with the assurance level provided by the theorem prover Isabelle/HOL. 
  

\section{Requirements and their Formalization}
\begin{figure}[h!tb]
\centering
   \resizebox{0.999\textwidth}{!}{%
   	\scriptsize
	\begin{tikzpicture}
	\node[MyRoundedBox, text width=5em] (SensorsA) at (-8.5,2) {$\mvar{Sensors}_\mathrm{A}$\\ $\lbrace\mdef{A}\rbrace$ --- $\lbrace\rbrace$};
	\node[MyRoundedBox, text width=6em] (EncryptionA) at (-6,2) {$\mvar{Encryption}_\mathrm{A}$\\ $\lbrace\rbrace$ --- $\lbrace\mdef{A}\rbrace$};
	\node[MyRoundedBox, text width=5em, fill=white] (ClientAout) at (-3,2) {$\mvar{Client}_\mathrm{A\mhyphen{}out}$\\ $\lbrace\rbrace$ ---$\lbrace\rbrace$};
	\node[MyRoundedBox, text width=5em] (SensorsB) at (-8.5,0) {$\mvar{Sensors}_\mathrm{B}$\\ $\lbrace\mdef{B}\rbrace$ --- $\lbrace\rbrace$};
	\node[MyRoundedBox, text width=6em] (EncryptionB) at (-6,0) {$\mvar{Encryption}_\mathrm{C}$\\ $\lbrace\rbrace$--- $\lbrace\mdef{B}\rbrace$};
	\node[MyRoundedBox, text width=5em ,fill=white] (ClientBout) at (-3,0) {$\mvar{Client}_\mathrm{B\mhyphen{}out}$\\ $\lbrace\rbrace$ ---$\lbrace\rbrace$};
	\node[MyRoundedBox, text width=5em] (SensorsC) at (-8.5,-2) {$\mvar{Sensors}_\mathrm{C}$\\ $\lbrace\mdef{C}\rbrace$ --- $\lbrace\rbrace$};
	\node[MyRoundedBox, text width=6em] (EncryptionC) at (-6,-2) {$\mvar{Encryption}_\mathrm{C}$\\ $\lbrace\rbrace$  --- $\lbrace\mdef{C}\rbrace$};
	\node[MyRoundedBox, text width=5em,fill=white] (ClientCout) at (-3,-2) {$\mvar{Client}_\mathrm{C\mhyphen{}out}$\\ $\lbrace\rbrace$ ---$\lbrace\rbrace$};
	\node[MyRoundedBox,fill=white] (UploadDroid) at (0.5,0) {$\mvar{UploadDroid}$\\ $\lbrace\rbrace$ ---$\lbrace\rbrace$};
	\node[MyRoundedBox, text width=7em, fill=white] (C3POin) at (5,0) {$Data\mhyphen{}Retriever$\\ $\lbrace\rbrace$ ---$\lbrace\rbrace$};
	\node[MyRoundedBox, text width=5em] (C3PODecA) at (8.3,2) {$\mvar{Dec\mhyphen{}A}$\\ $\lbrace\mdef{A}\rbrace$ --- $\lbrace\rbrace$};
	\node[MyRoundedBox, text width=5em] (C3PODecB) at (8.3,0) {$\mvar{Dec\mhyphen{}B}$\\ $\lbrace\mdef{B}\rbrace$ --- $\lbrace\rbrace$};
	\node[MyRoundedBox, text width=5em] (C3PODecC) at (8.3,-2) {$\mvar{Dec\mhyphen{}C}$\\ $\lbrace\mdef{C}\rbrace$ --- $\lbrace\rbrace$};
	\node[MyRoundedBox, text width=7em] (C3POStorage) at (11,0) {$\mvar{Storage}$\\ $\lbrace\mdef{A}, \mdef{B}, \mdef{C}\rbrace$ --- $\lbrace\rbrace$};

\begin{pgfonlayer}{background}
	\draw[dashed] ($(SensorsA.north west) + (-1ex,1ex)$) rectangle ($(ClientAout.south east) + (-1ex,-1ex)$);
	\node[align=left] at ($(ClientAout.north) + (0,3ex)$) {Smartphone $A$};
	
	\draw[dashed] ($(SensorsB.north west) + (-1ex,1ex)$) rectangle ($(ClientBout.south east) + (-1ex,-1ex)$);
	\node[align=left] at ($(ClientBout.north) + (0,3ex)$) {Smartphone $B$};
	
	\draw[dashed] ($(SensorsC.north west) + (-1ex,1ex)$) rectangle ($(ClientCout.south east) + (-1ex,-1ex)$);
	\node[align=left] at ($(ClientCout.north) + (0,3ex)$) {Smartphone $C$};
	
	\draw[dashed] ($(UploadDroid.north west) + (2ex,1ex)$) rectangle ($(UploadDroid.south east) + (-2ex,-1ex)$);

	\draw[dashed] let \p1 = (C3POin.west) in
			      let \p2 = (C3PODecA.north) in 
			      let \p3 = (C3POStorage.east) in 
			      let \p4 = (C3PODecC.south) in ($(\x1,\y2) + (+2ex,1ex)$) rectangle ($(\x3,\y4) + (2ex,-1ex)$);
	\node[align=left] at ($(C3POStorage.north) + (+6ex,19ex)$) {$\mvar{Collect Droid}$};
\end{pgfonlayer}

	\path[myptr] (SensorsA) edge (EncryptionA);
	\path[myptr] (EncryptionA) edge (ClientAout);
	\path[myptr] (ClientAout) edge (UploadDroid);
	\path[myptr] (SensorsB) edge (EncryptionB);
	\path[myptr] (EncryptionB) edge (ClientBout);
	\path[myptr] (ClientBout) edge (UploadDroid);
	\path[myptr] (SensorsC) edge (EncryptionC);
	\path[myptr] (EncryptionC) edge (ClientCout);
	\path[myptr] (ClientCout) edge (UploadDroid);
	\path[myptr] (C3POin) edge (UploadDroid);
	\path[myptr] (C3POin) edge (C3PODecA);
	\path[myptr] (C3POin) edge (C3PODecB);
	\path[myptr] (C3POin) edge (C3PODecC);
	\path[myptr] (C3PODecA) edge (C3POStorage);
	\path[myptr] (C3PODecB) edge (C3POStorage);
	\path[myptr] (C3PODecC) edge (C3POStorage);
	\end{tikzpicture}
   }
   \caption{MeasrDroid Architecture}
  \label{fig:casestudy:measrdroid:arch}
\end{figure}
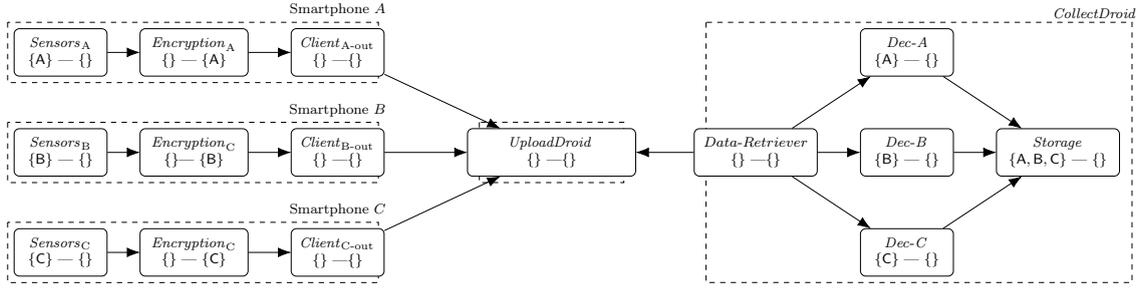

MeasrDroid~\cite{measrdroid} is a system for collecting smartphone sensor data for research purposes. 
The collected data may be privacy-critical. 
MeasrDroid's architecture is illustrated in Figure~\ref{fig:casestudy:measrdroid:arch}. 
The figure illustrates three user smartphones. 
They send their sensor readings over the Internet to $\mvar{Upload Droid}$. 
Ultimately, the data is stored and analyzed by $\mvar{Collect Droid}$, a trusted machine. 
To decrease the attack surface of this machine, it is not reachable over the Internet. 
Instead, the smartphones push the data to a server called $\mvar{Upload Droid}$.
$\mvar{Collect Droid}$ in turn regularly polls that server for new information.
Since $\mvar{Upload Droid}$ is particularly exposed, a compromise of this machine must not lead to a privacy violation of the users. 
Therefore, $\mvar{Upload Droid}$ must be completely uncritical. 
This is achieved by having the smartphones encrypt the data such that only $\mvar{Collect Droid}$ can decrypt again. 
Consequently, $\mvar{Upload Droid}$ only sees encrypted, uncritical data. 

We model the system in \topos{} and formalized the privacy requirements. 
Figure~\ref{fig:casestudy:measrdroid:arch} provides an overview of the model. 
The figure shows the architecture and the taint labels (cf.\ Section~\ref{sinvar:taintuntaint}) we assign to each entity. 
Each smartphone generates individual user data, labeled $\mdef{A}$, $\mdef{B}$, and $\mdef{C}$. 
The encryption components make the data unreadable for any unauthorized entity, effectively untainting the privacy-critical information. 
Once the data is decrypted again, the taint labels are restored. 
The $\mvar{Storage}$ is the most critical component as it aggregates all user data. 
It can be seen that $\mvar{Upload Droid}$ is completely uncritical because it does not have any labels at all. 
In addition to the Tainting invariant, the dotted lines visualize the system boundaries (cf.\ Section~\ref{sec:meta:systemboundaries}). 

\begin{figure*}[htb]
		\includegraphics[width=.99\textwidth]{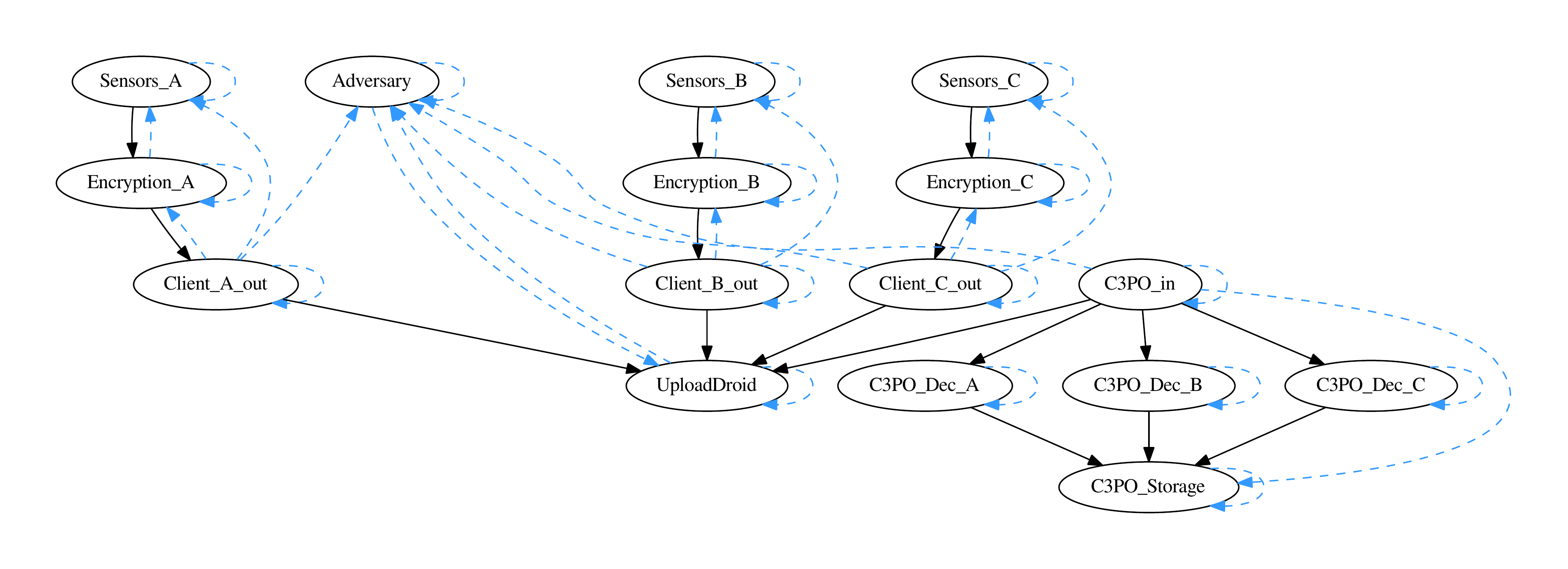}
	\caption{Analysis of the MeasrDroid architecture specification (generated by \topos{})}
	\label{fig:measrdroid:analysisbytopos}
\end{figure*}

We evaluate this system specification shown in Figure~\ref{fig:casestudy:measrdroid:arch}.\footnote{\texttt{Examples/Tainting/MeasrDroid.thy}} 
First of all, \topos{} verifies that all security invariants are fulfilled for the presented architecture. 
Adding an additional adversary node to the architecture, we use \topos{}' policy construction algorithm to visualize which interaction between the entities would be allowed. 
The visualization generated by \topos{} is shown in Figure~\ref{fig:measrdroid:analysisbytopos}. 
It is the original unmodified output of \topos{} we obtain during our analysis. 
In our Isabelle formalization, we give the entities names closer to their corresponding name in the actual implementation 
Most notably, $\mvar{Collect Droid}$ is called $\mvar{C3PO}$ and the $\mvar{Data\mhyphen{}Retriever}$ of $\mvar{Collect Droid}$ is called $\mvar{C3PO\_in}$.
We find the following flows that are allowed but not considered in our policy (architecture):  
\begin{itemize}
\item All reflexive flows, i.e. every component can interact with itself. This is acceptable. 
\item Within each smartphone, internally, arbitrary communication is possible. We cannot prevent this at a user's smartphone.
\item Every smartphone could send data to the adversary.
	It is important that this is generally allowed since we do not want to put any restrictions on the Internet connectivity of a smartphone. 
	For example, this allows the smartphone user to surf facebook, which is not a trusted component in our system. 
	The collected data is encrypted once it leaves the smartphone (via MeasrDroid). 
	Therefore, sensor data is not leaked.
\item $\mvar{C3PO\_in}$ could send data to the adversary. 
	At this point, the data is still encrypted.
	It would be possible to add an additional security invariant to make sure that $\mvar{C3PO\_in}$
	only connects to $\mvar{Upload Droid}$.
\item An adversary could send data to $\mvar{Upload Droid}$.
	Since the system shall be accessible to any smartphone connected to the Internet without authentication,
	we cannot prevent that a malicious user might send fake data.
\item The $\mvar{Upload Droid}$ could send data to the adversary.
	This does not undermine the security concept because $\mvar{Upload Droid}$ only stores encrypted data. 
	In fact, the security assumptions were from the very beginning that $\mvar{Upload Droid}$ can get compromised.
\item $\mvar{Collect Droid}$ (C3PO) could directly save data in its database without decrypting.
	This is acceptable and might potentially be used in a future version for backups.
\end{itemize}

The evaluation of the architecture provides the following high-level conclusions with respect to the overall privacy requirements: 
Data minimization can already be performed by the client application on the smartphone. 
This improves control by the data subject, the user. 
Furthermore, distributed collection better realizes unlinkability directly from the beginning.
With regard to asset identification, every smartphone is only critical for the corresponding individual while $\mvar{Collect Droid}$ is critical for all users. 
$\mvar{Upload Droid}$ is considered completely uncritical. 
$\mvar{Collect Droid}$ only has an active boundary; intrusion has to be considered as an attack vector and protection using a firewall to restrict incoming connections is vital.

\section{Auditing the Real MeasrDroid}
MeasrDroid is deployed and in productive use since 2013. 
The previous sections presented a theoretical evaluation of its architecture. 
This evaluation was not available at the time MeasrDroid was developed. 
In this section, we evaluate the real MeasrDroid implementation with regard to our findings of the previous sections.

First, we collected all physical and virtual machines which are associated with MeasrDroid. 
We found the following machines:

\begin{minipage}{.8\linewidth} 
\begin{description}
	\setlist{nolistsep}
	\item[droid0] Virtual machine
		\begin{itemize}
			\item IPv4: 131.159.15.16
			\item IPv6: 2001:4ca0:2001:13:216:3eff:fea7:6ad5
			\item Name in the model: not present
			\item Purpose: DNS server, not relevant for MeasrDroid's architecture
		\end{itemize}
	\end{description}
\end{minipage}
\vspace*{.8\baselineskip}

\begin{minipage}{.8\linewidth} 
	\begin{description}
	\setlist{nolistsep}
	\item[droid1] Virtual machine
		\begin{itemize}
			\item IPv4: 131.159.15.42
			\item IPv6: 2001:4ca0:2001:13:216:3eff:fe03:34f8
			\item Name in the model: $\mvar{Upload Droid}$
			\item Purpose: Receive data via http/https
		\end{itemize}
\end{description}
\end{minipage}
\vspace*{.8\baselineskip}

\begin{minipage}{.8\linewidth} 
	\begin{description}
		\setlist{nolistsep}
	\item[c3po] Physical, powerful machine
		\begin{itemize}
			\item IPv4: 131.159.15.52
			\item IPv6: 2001:4ca0:2001:13:2e0:81ff:fee0:f02e
			\item Name in the model: $\mvar{Collect Droid}$
			\item Purpose: Trusted collection and storage
		\end{itemize}
\end{description}
\end{minipage}
\bigskip

\begin{sloppypar}
The relevant machines are $\mvar{Upload Droid}$ at 131.159.15.42 and $\mvar{Collect Droid}$ at 131.159.15.52. 
We find that the machines do not have a firewall set up. 
All rely on the central firewall of our lab. 
\end{sloppypar}

This central firewall may be the largest, real-world, publicly available iptables firewall in the world and handles many different machines and networks. 
MeasrDroid is only a tiny fragment of it. 
We obtain a snapshot from June 2016 and make it publicly available~\cite{diekmanngithubnetnetwork}. 
The firewall is managed by several users and consists of over 5500 IPv4 rules.

MeasrDroid relies on the protocols http (port 80), https (port 443), and ssh (port 22). 
The architecture does not specify any port numbers, but our tool \fffuu{} needs a port number to compute a service matrix. 
For a full audit, the following analysis should be carried out for all port numbers used by MeasrDroid. 
For conciseness, we focus our audit on port 80. 
Notably, our theoretical analysis, in particular the model shown in Figure~\ref{fig:casestudy:measrdroid:arch}, has abstracted over concrete port number at all times.

\begin{figure}
\centering
\resizebox{\linewidth}{!}{%
\begin{footnotesize}%
\begin{tikzpicture}
\node[align=left] (a) at (0,-10) {$\{224.0.0.0 .. 239.255.255.255\}$};
\node[align=center, text width=10cm,cloud, draw,cloud puffs=10,cloud puff arc=120, aspect=2, inner sep=-3em,outer sep=0] (b) at (0,4) {$\{0.0.0.0 .. 45.56.113.32\} \cup \{45.56.113.34 .. 80.81.195.255\} \cup \{80.81.197.0 .. 81.169.253.163\} \cup \{81.169.253.165 .. 85.214.129.213\} \cup \{85.214.129.215 .. 94.186.159.97\} \cup \{94.186.159.99 .. 126.255.255.255\} \cup \{128.0.0.0 .. 131.159.13.255\} \cup \{131.159.16.0 .. 131.159.19.255\} \cup \{131.159.22.0 .. 138.246.252.255\} \cup \{138.246.254.0 .. 148.251.90.44\} \cup \{148.251.90.46 .. 185.86.231.255\} \cup \{185.86.236.0 .. 188.1.239.85\} \cup \{188.1.239.87 .. 188.95.232.63\} \cup \{188.95.232.224 .. 188.95.232.255\} \cup \{188.95.240.0 .. 192.48.106.255\} \cup \{192.48.108.0 .. 223.255.255.255\} \cup \{240.0.0.0 .. 255.255.255.255\}$};
\node[anchor=north, text badly ragged, text width=10cm] (c) at (-8,0) {$\{131.159.14.0 .. 131.159.14.10\} \cup \{131.159.14.12 .. 131.159.14.25\} \cup \{131.159.14.27 .. 131.159.14.35\} \cup \{131.159.14.37 .. 131.159.14.41\} \cup \{131.159.14.43 .. 131.159.14.46\} \cup \{131.159.14.48 .. 131.159.14.59\} \cup \{131.159.14.61 .. 131.159.14.62\} \cup \{131.159.14.64 .. 131.159.14.84\} \cup \{131.159.14.86 .. 131.159.14.124\} \cup \{131.159.14.126 .. 131.159.14.139\} \cup \{131.159.14.141 .. 131.159.14.144\} \cup \{131.159.14.147 .. 131.159.14.168\} \cup \{131.159.14.170 .. 131.159.14.203\} \cup \{131.159.14.205 .. 131.159.14.208\} \cup \{131.159.14.210 .. 131.159.14.211\} \cup 131.159.14.213 \cup \{131.159.14.217 .. 131.159.14.220\} \cup \{131.159.14.222 .. 131.159.15.3\} \cup 131.159.15.6 \cup 131.159.15.8 \cup 131.159.15.10 \cup 131.159.15.12 \cup 131.159.15.15 \cup \{131.159.15.18 .. 131.159.15.19\} \cup 131.159.15.22 \cup \{131.159.15.24 .. 131.159.15.25\} \cup 131.159.15.28 \cup 131.159.15.37 \cup 131.159.15.40 \cup 131.159.15.45 \cup \mathbf{\textcolor{red}{131.159.15.52}} \cup 131.159.15.55 \cup \{131.159.15.60 .. 131.159.15.67\} \cup \{131.159.15.69 .. 131.159.15.225\} \cup \{131.159.15.227 .. 131.159.15.228\} \cup \{131.159.15.230 .. 131.159.15.232\} \cup \{131.159.15.234 .. 131.159.15.245\} \cup \{131.159.15.249 .. 131.159.15.255\} \cup \{131.159.20.0 .. 131.159.20.41\} \cup \{131.159.20.43 .. 131.159.20.44\} \cup \{131.159.20.46 .. 131.159.20.51\} \cup \{131.159.20.53 .. 131.159.20.58\} \cup \{131.159.20.60 .. 131.159.20.71\} \cup \{131.159.20.73 .. 131.159.20.154\} \cup \{131.159.20.156 .. 131.159.20.201\} \cup \{131.159.20.203 .. 131.159.20.242\} \cup \{131.159.20.244 .. 131.159.20.255\} \cup \{188.95.233.0 .. 188.95.233.3\} \cup \{188.95.233.6 .. 188.95.233.8\} \cup \{188.95.233.10 .. 188.95.233.255\} \cup \{192.48.107.0 .. 192.48.107.255\}$};
\node[anchor=north, text badly ragged left, text width=10cm] (d) at (8,0) {$131.159.14.11 \cup 131.159.14.26 \cup 131.159.14.36 \cup 131.159.14.42 \cup 131.159.14.47 \cup 131.159.14.60 \cup 131.159.14.63 \cup 131.159.14.85 \cup 131.159.14.125 \cup 131.159.14.140 \cup \{131.159.14.145 .. 131.159.14.146\} \cup 131.159.14.169 \cup 131.159.14.204 \cup 131.159.14.214 \cup 131.159.14.221 \cup \{131.159.15.4 .. 131.159.15.5\} \cup 131.159.15.7 \cup 131.159.15.9 \cup 131.159.15.11 \cup \{131.159.15.13 .. 131.159.15.14\} \cup \{131.159.15.16 .. 131.159.15.17\} \cup \{131.159.15.20 .. 131.159.15.21\} \cup 131.159.15.23 \cup \{131.159.15.26 .. 131.159.15.27\} \cup \{131.159.15.29 .. 131.159.15.36\} \cup \{131.159.15.38 .. 131.159.15.39\} \cup \{131.159.15.41 .. \mathbf{\textcolor{red}{131.159.15.42}} .. 131.159.15.44\} \cup \{131.159.15.46 .. 131.159.15.49\} \cup 131.159.15.51 \cup \{131.159.15.53 .. 131.159.15.54\} \cup \{131.159.15.56 .. 131.159.15.59\} \cup 131.159.15.68 \cup 131.159.15.226 \cup 131.159.15.229 \cup 131.159.15.233 \cup \{131.159.15.246 .. 131.159.15.248\} \cup 131.159.20.42 \cup 131.159.20.45 \cup 131.159.20.52 \cup 131.159.20.59 \cup 131.159.20.72 \cup 131.159.20.155 \cup 131.159.20.202 \cup 131.159.20.243 \cup \{131.159.21.0 .. 131.159.21.255\} \cup \{185.86.232.0 .. 185.86.235.255\} \cup \{188.95.232.192 .. 188.95.232.223\} \cup \{188.95.233.4 .. 188.95.233.5\} \cup 188.95.233.9 \cup \{188.95.234.0 .. 188.95.239.255\}$};
\node[align=left] (e) at (-5,-16) {$188.1.239.86 \cup \{188.95.232.64 .. 188.95.232.191\}$};
\node[align=center, text width=6cm] (f) at (11,-16) {$45.56.113.33 \cup 81.169.253.164 \cup 85.214.129.214 \cup 94.186.159.98 \cup 148.251.90.45$};
\node[align=center, text width=6cm] (g) at (6,-9) {$\{138.246.253.6 .. 138.246.253.10\} \cup 138.246.253.19$};
\node[align=left] (h) at (-8,-13) {$138.246.253.5$};
\node[align=center, text width=6cm] (i) at (0,-20) {$\{138.246.253.0 .. 138.246.253.4\} \cup \{138.246.253.11 .. 138.246.253.18\} \cup \{138.246.253.20 .. 138.246.253.255\}$};
\node[align=left] (j) at (4,-13) {$131.159.15.50$};
\node[align=center, text width=6cm] (k) at (-3,-12) {$131.159.14.212 \cup \{131.159.14.215 .. 131.159.14.216\}$};
\node[align=left] (l) at (3,-18) {$131.159.14.209$};
\node[align=left] (m) at (8,-10) {$\{127.0.0.0 .. 127.255.255.255\}$};
\node[align=left] (n) at (11,-14) {$\{80.81.196.0 .. 80.81.196.255\}$};

\draw[myptr] (a) to[loop above] (a);
\draw[myptr] (a) to (d);
\draw[myptr] (b) to (a);
\draw[myptr] (b) to (d);
\draw[myptr] (c) to (a);
\draw[myptr] (c) to (b);
\draw[myptr, color=red] (c) to[loop above] (c);
\draw[myptr] ($(c.east)+(0,1ex)$) to[bend left, in=160, out=20] ($(d.west)+(0,1ex)$);
\draw[myptr] (c) to (e);
\draw[myptr] (c) to (f);
\draw[myptr] (c) to (g);
\draw[myptr] (c) to (h);
\draw[myptr] (c) to (i);
\draw[myptr] (c) to (j);
\draw[myptr] (c) to (k);
\draw[myptr] (c) to (l);
\draw[myptr] (c) to (m);
\draw[myptr] (c) to (n);
\draw[myptr] (d) to (a);
\draw[myptr] (d) to (b);
\draw[myptr,color=red] ($(d.west)+(0,-1ex)$) to[bend left, in=160, out=20] ($(c.east)+(0,-1ex)$);
\draw[myptr] (d) to[loop above] (d);
\draw[myptr] (d) to (e);
\draw[myptr] (d) to (f);
\draw[myptr] (d) to (g);
\draw[myptr] (d) to (h);
\draw[myptr] (d) to (i);
\draw[myptr] (d) to (j);
\draw[myptr] (d) to (k);
\draw[myptr] (d) to (l);
\draw[myptr] (d) to (m);
\draw[myptr] (d) to (n);
\draw[myptr] (e) to (a);
\draw[myptr] (e) to (b);
\draw[myptr] (e) to (c);
\draw[myptr] (e) to (d);
\draw[myptr] (e) to[loop above] (e);
\draw[myptr] (e) to (f);
\draw[myptr] (e) to (g);
\draw[myptr] (e) to (h);
\draw[myptr] (e) to (i);
\draw[myptr] (e) to (j);
\draw[myptr] (e) to (k);
\draw[myptr] (e) to (l);
\draw[myptr] (e) to (m);
\draw[myptr] (e) to (n);
\draw[myptr] (f) to (a);
\draw[myptr] (f) to (d);
\draw[myptr] (f) to (k);
\draw[myptr] (f) to (l);
\draw[myptr] (g) to (a);
\draw[myptr] (g) to (d);
\draw[myptr] (g) to (j);
\draw[myptr] (g) to (k);
\draw[myptr] (h) to (a);
\draw[myptr] (h) to (c);
\draw[myptr] (h) to (d);
\draw[myptr] (h) to (j);
\draw[myptr] (h) to (k);
\draw[myptr] (h) to (l);
\draw[myptr] (i) to (a);
\draw[myptr] (i) to (d);
\draw[myptr] (i) to (k);
\draw[myptr] (j) to (a);
\draw[myptr] (j) to (b);
\draw[myptr] (j) to (c);
\draw[myptr] (j) to (d);
\draw[myptr] (j) to (e);
\draw[myptr] (j) to (f);
\draw[myptr] (j) to (g);
\draw[myptr] (j) to (h);
\draw[myptr] (j) to (i);
\draw[myptr] (j) to[loop above] (j);
\draw[myptr] (j) to (k);
\draw[myptr] (j) to (l);
\draw[myptr] (j) to (m);
\draw[myptr] (j) to (n);
\draw[myptr] (k) to (a);
\draw[myptr] (k) to (b);
\draw[myptr] (k) to (c);
\draw[myptr] (k) to (d);
\draw[myptr] (k) to (e);
\draw[myptr] (k) to (f);
\draw[myptr] (k) to (g);
\draw[myptr] (k) to (h);
\draw[myptr] (k) to (i);
\draw[myptr] (k) to (j);
\draw[myptr] (k) to[loop above] (k);
\draw[myptr] (k) to (l);
\draw[myptr] (k) to (m);
\draw[myptr] (k) to (n);
\draw[myptr] (l) to (a);
\draw[myptr] (l) to (b);
\draw[myptr] (l) to (c);
\draw[myptr] (l) to (d);
\draw[myptr] (l) to (e);
\draw[myptr] (l) to (f);
\draw[myptr] (l) to (g);
\draw[myptr] (l) to (h);
\draw[myptr] (l) to (i);
\draw[myptr] (l) to (j);
\draw[myptr] (l) to (k);
\draw[myptr] (l) to[loop above] (l);
\draw[myptr] (l) to (m);
\draw[myptr] (l) to (n);
\draw[myptr] (n) to (a);
\draw[myptr] (n) to (d);
\draw[myptr] (n) to (l);
\end{tikzpicture}%
\end{footnotesize}%
}%
\caption{MeasrDroid: Main firewall -- IPv4 http connectivity matrix}
\label{fig:eval_measrdroid:i8fw:port80}
\end{figure}
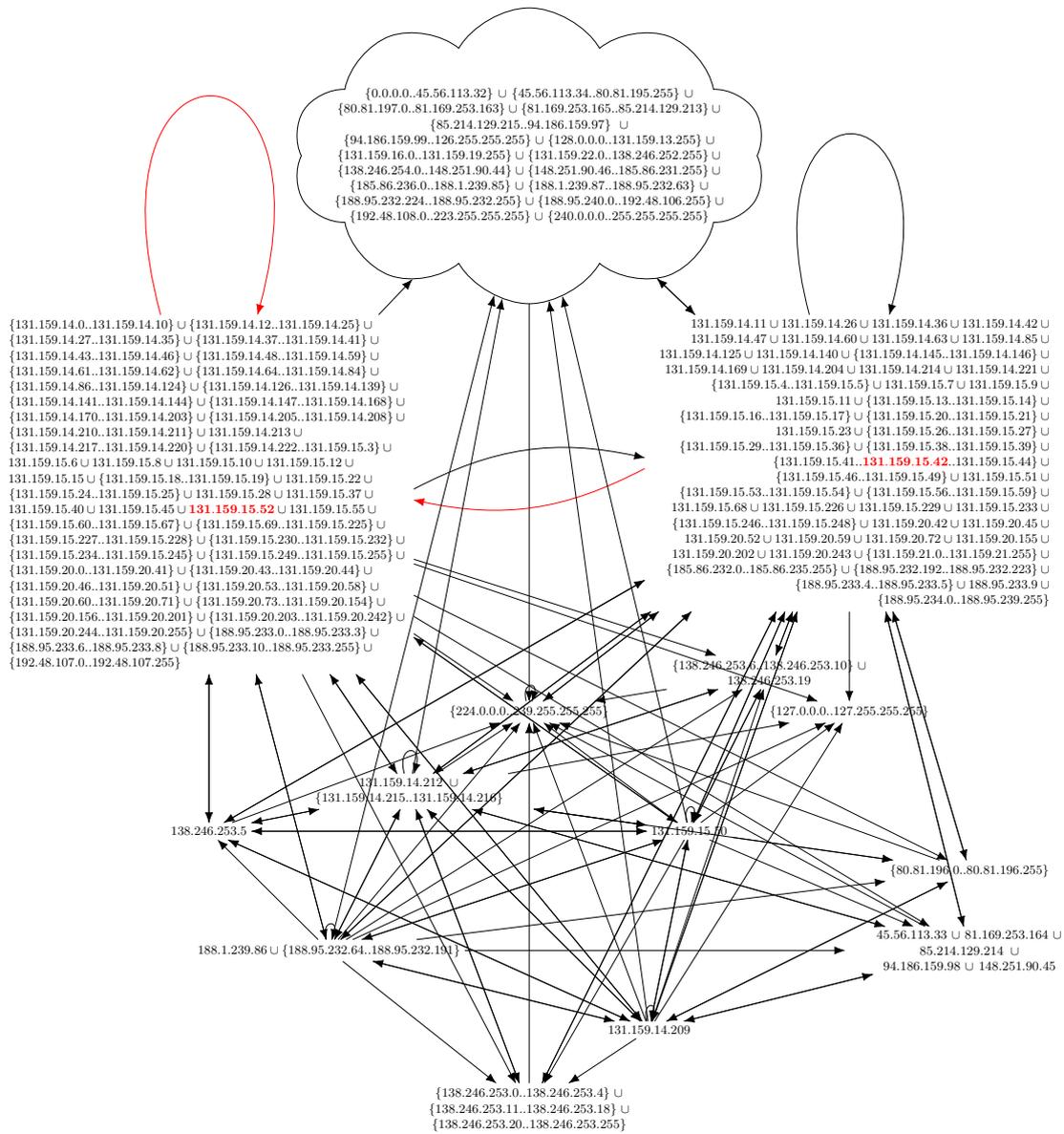

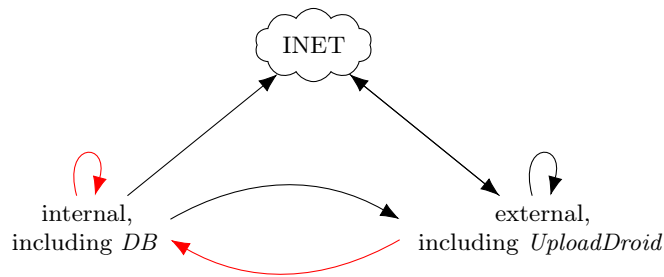
\begin{figure}
\centering
\begin{footnotesize}
\begin{tikzpicture}
\node[align=center,text width=3cm,cloud, draw,cloud puffs=10,cloud puff arc=120, aspect=2, inner sep=-3em,outer sep=0] (b) at (0,2) {INET};
\node[anchor=north, align=center,text width=2cm] (c) at (-3,0) {internal,\\ including $\mvar{DB}$};
\node[anchor=north, align=center,text width=3.5cm] (d) at (3,0) {external,\\ including $\mvar{Upload Droid}$};

\draw[myptr] (b) to (d);
\draw[myptr] (c) to (b);
\draw[myptr, color=red] (c) to[loop above] (c);
\draw[myptr] ($(c.east)+(0,+1ex)$) to[bend left] ($(d.west)+(0,+1ex)$);
\draw[myptr] (d) to (b);
\draw[myptr,color=red] ($(d.west)+(0,-1ex)$) to[bend left] ($(c.east)+(0,-1ex)$);
\draw[myptr] (d) to[loop above] (d);
\end{tikzpicture}%
\end{footnotesize}
\caption{MeasrDroid: Main firewall -- simplified connectivity matrix}
\label{fig:eval_measrdroid:i8fw:port80:simple}
\end{figure}

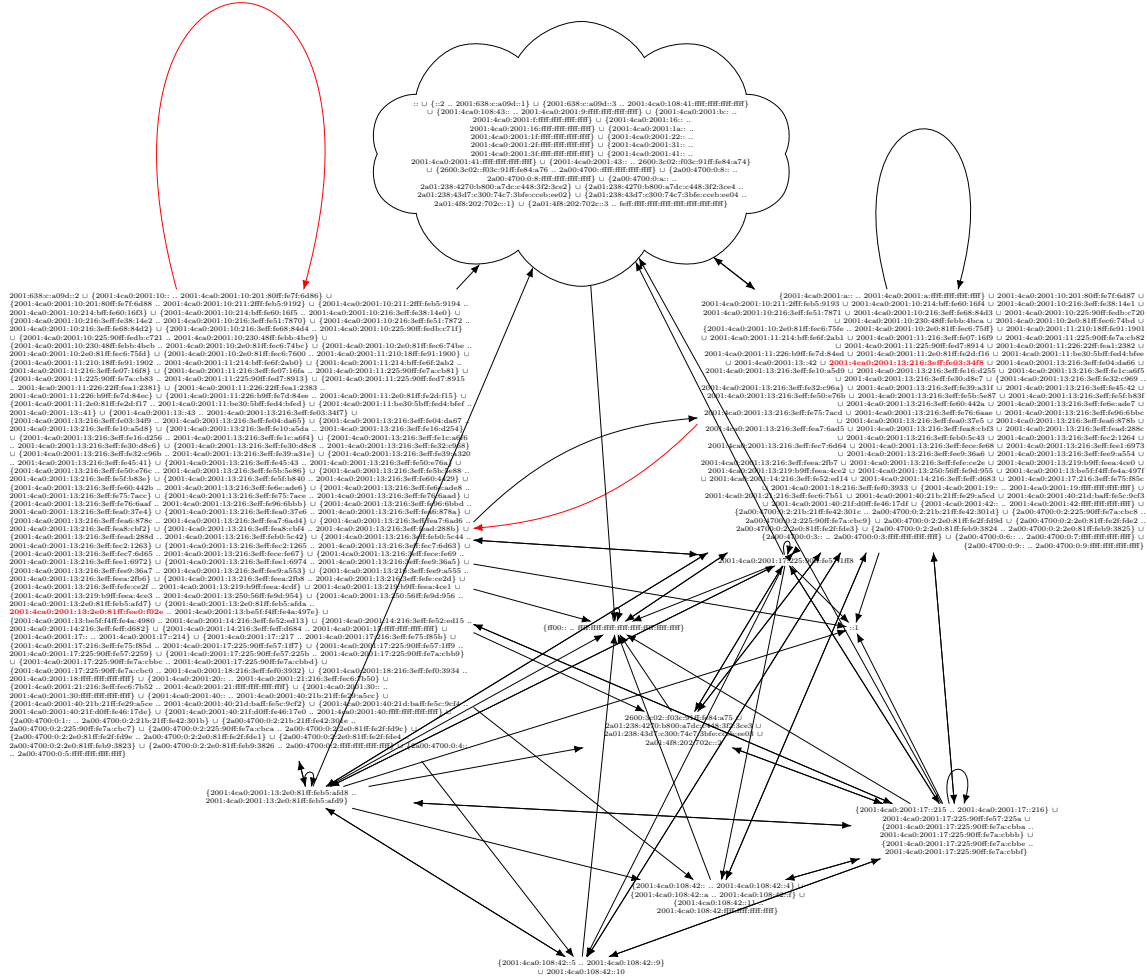
\begin{figure}
\centering
\resizebox{\linewidth}{!}{%
\tiny%
\begin{tikzpicture}
\node[align=left] (a) at (1,-10) { $\{$ff00:: .. ffff:ffff:ffff:ffff:ffff:ffff:ffff:ffff$\}$ };
\node[text badly centered, text width=6cm] (b) at (3,-13) {2600:3c02::f03c:91ff:fe84:a75 $\cup$ 2a01:238:4270:b800:a7dc:c448:3f2:3ce3 $\cup$ 2a01:238:43d7:c300:74c7:3bfe:cceb:ee03 $\cup$ 2a01:4f8:202:702c::2};
\node[anchor=north, text badly ragged left, text width=13cm] (c) at (10,0) { $\{$2001:4ca0:2001:a:: .. 2001:4ca0:2001:a:ffff:ffff:ffff:ffff$\}$  $\cup$ 2001:4ca0:2001:10:201:80ff:fe7f:6d87 $\cup$ 2001:4ca0:2001:10:211:2fff:feb5:9193 $\cup$ 2001:4ca0:2001:10:214:bff:fe60:16f4 $\cup$ 2001:4ca0:2001:10:216:3eff:fe38:14e1 $\cup$ 2001:4ca0:2001:10:216:3eff:fe51:7871 $\cup$ 2001:4ca0:2001:10:216:3eff:fe68:84d3 $\cup$ 2001:4ca0:2001:10:225:90ff:fedb:c720 $\cup$ 2001:4ca0:2001:10:230:48ff:febb:4bca $\cup$ 2001:4ca0:2001:10:2e0:81ff:fec6:74bd $\cup$  $\{$2001:4ca0:2001:10:2e0:81ff:fec6:75fe .. 2001:4ca0:2001:10:2e0:81ff:fec6:75ff$\}$  $\cup$ 2001:4ca0:2001:11:210:18ff:fe91:1901 $\cup$ 2001:4ca0:2001:11:214:bff:fe6f:2ab1 $\cup$ 2001:4ca0:2001:11:216:3eff:fe07:16f9 $\cup$ 2001:4ca0:2001:11:225:90ff:fe7a:cb82 $\cup$ 2001:4ca0:2001:11:225:90ff:fed7:8914 $\cup$ 2001:4ca0:2001:11:226:22ff:fea1:2382 $\cup$ 2001:4ca0:2001:11:226:b9ff:fe7d:84ed $\cup$ 2001:4ca0:2001:11:2e0:81ff:fe2d:f16 $\cup$ 2001:4ca0:2001:11:be30:5bff:fed4:bfee $\cup$ 2001:4ca0:2001:13::42 $\cup$ \textbf{\textcolor{red}{2001:4ca0:2001:13:216:3eff:fe03:34f8}} $\cup$ 2001:4ca0:2001:13:216:3eff:fe04:da66 $\cup$ 2001:4ca0:2001:13:216:3eff:fe10:a5d9 $\cup$ 2001:4ca0:2001:13:216:3eff:fe16:d255 $\cup$ 2001:4ca0:2001:13:216:3eff:fe1c:a6f5 $\cup$ 2001:4ca0:2001:13:216:3eff:fe30:d8c7 $\cup$  $\{$2001:4ca0:2001:13:216:3eff:fe32:c969 .. 2001:4ca0:2001:13:216:3eff:fe32:c96a$\}$  $\cup$ 2001:4ca0:2001:13:216:3eff:fe39:a31f $\cup$ 2001:4ca0:2001:13:216:3eff:fe45:42 $\cup$ 2001:4ca0:2001:13:216:3eff:fe50:e76b $\cup$ 2001:4ca0:2001:13:216:3eff:fe5b:5e87 $\cup$ 2001:4ca0:2001:13:216:3eff:fe5f:b83f $\cup$ 2001:4ca0:2001:13:216:3eff:fe60:442a $\cup$ 2001:4ca0:2001:13:216:3eff:fe6e:ade7 $\cup$ 2001:4ca0:2001:13:216:3eff:fe75:7acd $\cup$ 2001:4ca0:2001:13:216:3eff:fe76:6aae $\cup$ 2001:4ca0:2001:13:216:3eff:fe96:6bbc $\cup$ 2001:4ca0:2001:13:216:3eff:fea0:37e5 $\cup$ 2001:4ca0:2001:13:216:3eff:fea6:878b $\cup$ 2001:4ca0:2001:13:216:3eff:fea7:6ad5 $\cup$ 2001:4ca0:2001:13:216:3eff:fea8:cbf3 $\cup$ 2001:4ca0:2001:13:216:3eff:fead:288c $\cup$ 2001:4ca0:2001:13:216:3eff:feb0:5c43 $\cup$ 2001:4ca0:2001:13:216:3eff:fec2:1264 $\cup$ 2001:4ca0:2001:13:216:3eff:fec7:6d64 $\cup$ 2001:4ca0:2001:13:216:3eff:fece:fe68 $\cup$ 2001:4ca0:2001:13:216:3eff:fee1:6973 $\cup$ 2001:4ca0:2001:13:216:3eff:fee9:36a6 $\cup$ 2001:4ca0:2001:13:216:3eff:fee9:a554 $\cup$ 2001:4ca0:2001:13:216:3eff:feea:2fb7 $\cup$ 2001:4ca0:2001:13:216:3eff:fefe:ce2e $\cup$ 2001:4ca0:2001:13:219:b9ff:feea:4ce0 $\cup$ 2001:4ca0:2001:13:219:b9ff:feea:4ce2 $\cup$ 2001:4ca0:2001:13:250:56ff:fe9d:955 $\cup$ 2001:4ca0:2001:13:be5f:f4ff:fe4a:497f $\cup$ 2001:4ca0:2001:14:216:3eff:fe52:ed14 $\cup$ 2001:4ca0:2001:14:216:3eff:feff:d683 $\cup$ 2001:4ca0:2001:17:216:3eff:fe75:f85c $\cup$ 2001:4ca0:2001:18:216:3eff:fef0:3933 $\cup$  $\{$2001:4ca0:2001:19:: .. 2001:4ca0:2001:19:ffff:ffff:ffff:ffff$\}$  $\cup$ 2001:4ca0:2001:21:216:3eff:fec6:7b51 $\cup$ 2001:4ca0:2001:40:21b:21ff:fe29:a5cd $\cup$ 2001:4ca0:2001:40:21d:baff:fe5c:9cf3 $\cup$ 2001:4ca0:2001:40:21f:d0ff:fe46:17df $\cup$  $\{$2001:4ca0:2001:42:: .. 2001:4ca0:2001:42:ffff:ffff:ffff:ffff$\}$  $\cup$  $\{$2a00:4700:0:2:21b:21ff:fe42:301c .. 2a00:4700:0:2:21b:21ff:fe42:301d$\}$  $\cup$  $\{$2a00:4700:0:2:225:90ff:fe7a:cbc8 .. 2a00:4700:0:2:225:90ff:fe7a:cbc9$\}$  $\cup$ 2a00:4700:0:2:2e0:81ff:fe2f:fd9d $\cup$  $\{$2a00:4700:0:2:2e0:81ff:fe2f:fde2 .. 2a00:4700:0:2:2e0:81ff:fe2f:fde3$\}$  $\cup$  $\{$2a00:4700:0:2:2e0:81ff:feb9:3824 .. 2a00:4700:0:2:2e0:81ff:feb9:3825$\}$  $\cup$  $\{$2a00:4700:0:3:: .. 2a00:4700:0:3:ffff:ffff:ffff:ffff$\}$  $\cup$  $\{$2a00:4700:0:6:: .. 2a00:4700:0:7:ffff:ffff:ffff:ffff$\}$  $\cup$  $\{$2a00:4700:0:9:: .. 2a00:4700:0:9:ffff:ffff:ffff:ffff$\}$ };
\node[anchor=north, text badly ragged, text width=13.5cm] (d) at (-10,0) {2001:638:c:a09d::2 $\cup$  $\{$2001:4ca0:2001:10:: .. 2001:4ca0:2001:10:201:80ff:fe7f:6d86$\}$  $\cup$  $\{$2001:4ca0:2001:10:201:80ff:fe7f:6d88 .. 2001:4ca0:2001:10:211:2fff:feb5:9192$\}$  $\cup$  $\{$2001:4ca0:2001:10:211:2fff:feb5:9194 .. 2001:4ca0:2001:10:214:bff:fe60:16f3$\}$  $\cup$  $\{$2001:4ca0:2001:10:214:bff:fe60:16f5 .. 2001:4ca0:2001:10:216:3eff:fe38:14e0$\}$  $\cup$  $\{$2001:4ca0:2001:10:216:3eff:fe38:14e2 .. 2001:4ca0:2001:10:216:3eff:fe51:7870$\}$  $\cup$  $\{$2001:4ca0:2001:10:216:3eff:fe51:7872 .. 2001:4ca0:2001:10:216:3eff:fe68:84d2$\}$  $\cup$  $\{$2001:4ca0:2001:10:216:3eff:fe68:84d4 .. 2001:4ca0:2001:10:225:90ff:fedb:c71f$\}$  $\cup$  $\{$2001:4ca0:2001:10:225:90ff:fedb:c721 .. 2001:4ca0:2001:10:230:48ff:febb:4bc9$\}$  $\cup$  $\{$2001:4ca0:2001:10:230:48ff:febb:4bcb .. 2001:4ca0:2001:10:2e0:81ff:fec6:74bc$\}$  $\cup$  $\{$2001:4ca0:2001:10:2e0:81ff:fec6:74be .. 2001:4ca0:2001:10:2e0:81ff:fec6:75fd$\}$  $\cup$  $\{$2001:4ca0:2001:10:2e0:81ff:fec6:7600 .. 2001:4ca0:2001:11:210:18ff:fe91:1900$\}$  $\cup$  $\{$2001:4ca0:2001:11:210:18ff:fe91:1902 .. 2001:4ca0:2001:11:214:bff:fe6f:2ab0$\}$  $\cup$  $\{$2001:4ca0:2001:11:214:bff:fe6f:2ab2 .. 2001:4ca0:2001:11:216:3eff:fe07:16f8$\}$  $\cup$  $\{$2001:4ca0:2001:11:216:3eff:fe07:16fa .. 2001:4ca0:2001:11:225:90ff:fe7a:cb81$\}$  $\cup$  $\{$2001:4ca0:2001:11:225:90ff:fe7a:cb83 .. 2001:4ca0:2001:11:225:90ff:fed7:8913$\}$  $\cup$  $\{$2001:4ca0:2001:11:225:90ff:fed7:8915 .. 2001:4ca0:2001:11:226:22ff:fea1:2381$\}$  $\cup$  $\{$2001:4ca0:2001:11:226:22ff:fea1:2383 .. 2001:4ca0:2001:11:226:b9ff:fe7d:84ec$\}$  $\cup$  $\{$2001:4ca0:2001:11:226:b9ff:fe7d:84ee .. 2001:4ca0:2001:11:2e0:81ff:fe2d:f15$\}$  $\cup$  $\{$2001:4ca0:2001:11:2e0:81ff:fe2d:f17 .. 2001:4ca0:2001:11:be30:5bff:fed4:bfed$\}$  $\cup$  $\{$2001:4ca0:2001:11:be30:5bff:fed4:bfef .. 2001:4ca0:2001:13::41$\}$  $\cup$  $\{$2001:4ca0:2001:13::43 .. 2001:4ca0:2001:13:216:3eff:fe03:34f7$\}$  $\cup$  $\{$2001:4ca0:2001:13:216:3eff:fe03:34f9 .. 2001:4ca0:2001:13:216:3eff:fe04:da65$\}$  $\cup$  $\{$2001:4ca0:2001:13:216:3eff:fe04:da67 .. 2001:4ca0:2001:13:216:3eff:fe10:a5d8$\}$  $\cup$  $\{$2001:4ca0:2001:13:216:3eff:fe10:a5da .. 2001:4ca0:2001:13:216:3eff:fe16:d254$\}$  $\cup$  $\{$2001:4ca0:2001:13:216:3eff:fe16:d256 .. 2001:4ca0:2001:13:216:3eff:fe1c:a6f4$\}$  $\cup$  $\{$2001:4ca0:2001:13:216:3eff:fe1c:a6f6 .. 2001:4ca0:2001:13:216:3eff:fe30:d8c6$\}$  $\cup$  $\{$2001:4ca0:2001:13:216:3eff:fe30:d8c8 .. 2001:4ca0:2001:13:216:3eff:fe32:c968$\}$  $\cup$  $\{$2001:4ca0:2001:13:216:3eff:fe32:c96b .. 2001:4ca0:2001:13:216:3eff:fe39:a31e$\}$  $\cup$  $\{$2001:4ca0:2001:13:216:3eff:fe39:a320 .. 2001:4ca0:2001:13:216:3eff:fe45:41$\}$  $\cup$  $\{$2001:4ca0:2001:13:216:3eff:fe45:43 .. 2001:4ca0:2001:13:216:3eff:fe50:e76a$\}$  $\cup$  $\{$2001:4ca0:2001:13:216:3eff:fe50:e76c .. 2001:4ca0:2001:13:216:3eff:fe5b:5e86$\}$  $\cup$  $\{$2001:4ca0:2001:13:216:3eff:fe5b:5e88 .. 2001:4ca0:2001:13:216:3eff:fe5f:b83e$\}$  $\cup$  $\{$2001:4ca0:2001:13:216:3eff:fe5f:b840 .. 2001:4ca0:2001:13:216:3eff:fe60:4429$\}$  $\cup$  $\{$2001:4ca0:2001:13:216:3eff:fe60:442b .. 2001:4ca0:2001:13:216:3eff:fe6e:ade6$\}$  $\cup$  $\{$2001:4ca0:2001:13:216:3eff:fe6e:ade8 .. 2001:4ca0:2001:13:216:3eff:fe75:7acc$\}$  $\cup$  $\{$2001:4ca0:2001:13:216:3eff:fe75:7ace .. 2001:4ca0:2001:13:216:3eff:fe76:6aad$\}$  $\cup$  $\{$2001:4ca0:2001:13:216:3eff:fe76:6aaf .. 2001:4ca0:2001:13:216:3eff:fe96:6bbb$\}$  $\cup$  $\{$2001:4ca0:2001:13:216:3eff:fe96:6bbd .. 2001:4ca0:2001:13:216:3eff:fea0:37e4$\}$  $\cup$  $\{$2001:4ca0:2001:13:216:3eff:fea0:37e6 .. 2001:4ca0:2001:13:216:3eff:fea6:878a$\}$  $\cup$  $\{$2001:4ca0:2001:13:216:3eff:fea6:878c .. 2001:4ca0:2001:13:216:3eff:fea7:6ad4$\}$  $\cup$  $\{$2001:4ca0:2001:13:216:3eff:fea7:6ad6 .. 2001:4ca0:2001:13:216:3eff:fea8:cbf2$\}$  $\cup$  $\{$2001:4ca0:2001:13:216:3eff:fea8:cbf4 .. 2001:4ca0:2001:13:216:3eff:fead:288b$\}$  $\cup$  $\{$2001:4ca0:2001:13:216:3eff:fead:288d .. 2001:4ca0:2001:13:216:3eff:feb0:5c42$\}$  $\cup$  $\{$2001:4ca0:2001:13:216:3eff:feb0:5c44 .. 2001:4ca0:2001:13:216:3eff:fec2:1263$\}$  $\cup$  $\{$2001:4ca0:2001:13:216:3eff:fec2:1265 .. 2001:4ca0:2001:13:216:3eff:fec7:6d63$\}$  $\cup$  $\{$2001:4ca0:2001:13:216:3eff:fec7:6d65 .. 2001:4ca0:2001:13:216:3eff:fece:fe67$\}$  $\cup$  $\{$2001:4ca0:2001:13:216:3eff:fece:fe69 .. 2001:4ca0:2001:13:216:3eff:fee1:6972$\}$  $\cup$  $\{$2001:4ca0:2001:13:216:3eff:fee1:6974 .. 2001:4ca0:2001:13:216:3eff:fee9:36a5$\}$  $\cup$  $\{$2001:4ca0:2001:13:216:3eff:fee9:36a7 .. 2001:4ca0:2001:13:216:3eff:fee9:a553$\}$  $\cup$  $\{$2001:4ca0:2001:13:216:3eff:fee9:a555 .. 2001:4ca0:2001:13:216:3eff:feea:2fb6$\}$  $\cup$  $\{$2001:4ca0:2001:13:216:3eff:feea:2fb8 .. 2001:4ca0:2001:13:216:3eff:fefe:ce2d$\}$  $\cup$  $\{$2001:4ca0:2001:13:216:3eff:fefe:ce2f .. 2001:4ca0:2001:13:219:b9ff:feea:4cdf$\}$  $\cup$ 2001:4ca0:2001:13:219:b9ff:feea:4ce1 $\cup$  $\{$2001:4ca0:2001:13:219:b9ff:feea:4ce3 .. 2001:4ca0:2001:13:250:56ff:fe9d:954$\}$  $\cup$  $\{$2001:4ca0:2001:13:250:56ff:fe9d:956 .. 2001:4ca0:2001:13:2e0:81ff:feb5:afd7$\}$  $\cup$  $\{$2001:4ca0:2001:13:2e0:81ff:feb5:afda .. \textbf{\textcolor{red}{2001:4ca0:2001:13:2e0:81ff:fee0:f02e}} .. 2001:4ca0:2001:13:be5f:f4ff:fe4a:497e$\}$  $\cup$  $\{$2001:4ca0:2001:13:be5f:f4ff:fe4a:4980 .. 2001:4ca0:2001:14:216:3eff:fe52:ed13$\}$  $\cup$  $\{$2001:4ca0:2001:14:216:3eff:fe52:ed15 .. 2001:4ca0:2001:14:216:3eff:feff:d682$\}$  $\cup$  $\{$2001:4ca0:2001:14:216:3eff:feff:d684 .. 2001:4ca0:2001:15:ffff:ffff:ffff:ffff$\}$  $\cup$  $\{$2001:4ca0:2001:17:: .. 2001:4ca0:2001:17::214$\}$  $\cup$  $\{$2001:4ca0:2001:17::217 .. 2001:4ca0:2001:17:216:3eff:fe75:f85b$\}$  $\cup$  $\{$2001:4ca0:2001:17:216:3eff:fe75:f85d .. 2001:4ca0:2001:17:225:90ff:fe57:1ff7$\}$  $\cup$  $\{$2001:4ca0:2001:17:225:90ff:fe57:1ff9 .. 2001:4ca0:2001:17:225:90ff:fe57:2259$\}$  $\cup$  $\{$2001:4ca0:2001:17:225:90ff:fe57:225b .. 2001:4ca0:2001:17:225:90ff:fe7a:cbb9$\}$  $\cup$  $\{$2001:4ca0:2001:17:225:90ff:fe7a:cbbc .. 2001:4ca0:2001:17:225:90ff:fe7a:cbbd$\}$  $\cup$  $\{$2001:4ca0:2001:17:225:90ff:fe7a:cbc0 .. 2001:4ca0:2001:18:216:3eff:fef0:3932$\}$  $\cup$  $\{$2001:4ca0:2001:18:216:3eff:fef0:3934 .. 2001:4ca0:2001:18:ffff:ffff:ffff:ffff$\}$  $\cup$  $\{$2001:4ca0:2001:20:: .. 2001:4ca0:2001:21:216:3eff:fec6:7b50$\}$  $\cup$  $\{$2001:4ca0:2001:21:216:3eff:fec6:7b52 .. 2001:4ca0:2001:21:ffff:ffff:ffff:ffff$\}$  $\cup$  $\{$2001:4ca0:2001:30:: .. 2001:4ca0:2001:30:ffff:ffff:ffff:ffff$\}$  $\cup$  $\{$2001:4ca0:2001:40:: .. 2001:4ca0:2001:40:21b:21ff:fe29:a5cc$\}$  $\cup$  $\{$2001:4ca0:2001:40:21b:21ff:fe29:a5ce .. 2001:4ca0:2001:40:21d:baff:fe5c:9cf2$\}$  $\cup$  $\{$2001:4ca0:2001:40:21d:baff:fe5c:9cf4 .. 2001:4ca0:2001:40:21f:d0ff:fe46:17de$\}$  $\cup$  $\{$2001:4ca0:2001:40:21f:d0ff:fe46:17e0 .. 2001:4ca0:2001:40:ffff:ffff:ffff:ffff$\}$  $\cup$  $\{$2a00:4700:0:1:: .. 2a00:4700:0:2:21b:21ff:fe42:301b$\}$  $\cup$  $\{$2a00:4700:0:2:21b:21ff:fe42:301e .. 2a00:4700:0:2:225:90ff:fe7a:cbc7$\}$  $\cup$  $\{$2a00:4700:0:2:225:90ff:fe7a:cbca .. 2a00:4700:0:2:2e0:81ff:fe2f:fd9c$\}$  $\cup$  $\{$2a00:4700:0:2:2e0:81ff:fe2f:fd9e .. 2a00:4700:0:2:2e0:81ff:fe2f:fde1$\}$  $\cup$  $\{$2a00:4700:0:2:2e0:81ff:fe2f:fde4 .. 2a00:4700:0:2:2e0:81ff:feb9:3823$\}$  $\cup$  $\{$2a00:4700:0:2:2e0:81ff:feb9:3826 .. 2a00:4700:0:2:ffff:ffff:ffff:ffff$\}$  $\cup$  $\{$2a00:4700:0:4:: .. 2a00:4700:0:5:ffff:ffff:ffff:ffff$\}$ };
\node[text badly centered,text width=10cm,cloud, draw,cloud puffs=10,cloud puff arc=120, aspect=2, inner sep=-3em,outer sep=0] (e) at (0,4) {:: $\cup$  $\{$::2 .. 2001:638:c:a09d::1$\}$  $\cup$  $\{$2001:638:c:a09d::3 .. 2001:4ca0:108:41:ffff:ffff:ffff:ffff$\}$  $\cup$  $\{$2001:4ca0:108:43:: .. 2001:4ca0:2001:9:ffff:ffff:ffff:ffff$\}$  $\cup$  $\{$2001:4ca0:2001:b:: .. 2001:4ca0:2001:f:ffff:ffff:ffff:ffff$\}$  $\cup$  $\{$2001:4ca0:2001:16:: .. 2001:4ca0:2001:16:ffff:ffff:ffff:ffff$\}$  $\cup$  $\{$2001:4ca0:2001:1a:: .. 2001:4ca0:2001:1f:ffff:ffff:ffff:ffff$\}$  $\cup$  $\{$2001:4ca0:2001:22:: .. 2001:4ca0:2001:2f:ffff:ffff:ffff:ffff$\}$  $\cup$  $\{$2001:4ca0:2001:31:: .. 2001:4ca0:2001:3f:ffff:ffff:ffff:ffff$\}$  $\cup$  $\{$2001:4ca0:2001:41:: .. 2001:4ca0:2001:41:ffff:ffff:ffff:ffff$\}$  $\cup$  $\{$2001:4ca0:2001:43:: .. 2600:3c02::f03c:91ff:fe84:a74$\}$  $\cup$  $\{$2600:3c02::f03c:91ff:fe84:a76 .. 2a00:4700::ffff:ffff:ffff:ffff$\}$  $\cup$  $\{$2a00:4700:0:8:: .. 2a00:4700:0:8:ffff:ffff:ffff:ffff$\}$  $\cup$  $\{$2a00:4700:0:a:: .. 2a01:238:4270:b800:a7dc:c448:3f2:3ce2$\}$  $\cup$  $\{$2a01:238:4270:b800:a7dc:c448:3f2:3ce4 .. 2a01:238:43d7:c300:74c7:3bfe:cceb:ee02$\}$  $\cup$  $\{$2a01:238:43d7:c300:74c7:3bfe:cceb:ee04 .. 2a01:4f8:202:702c::1$\}$  $\cup$  $\{$2a01:4f8:202:702c::3 .. feff:ffff:ffff:ffff:ffff:ffff:ffff:ffff$\}$ };
\node[text badly centered,text width=6cm] (f) at (11,-16) { $\{$2001:4ca0:2001:17::215 .. 2001:4ca0:2001:17::216$\}$  $\cup$ 2001:4ca0:2001:17:225:90ff:fe57:225a $\cup$  $\{$2001:4ca0:2001:17:225:90ff:fe7a:cbba .. 2001:4ca0:2001:17:225:90ff:fe7a:cbbb$\}$  $\cup$  $\{$2001:4ca0:2001:17:225:90ff:fe7a:cbbe .. 2001:4ca0:2001:17:225:90ff:fe7a:cbbf$\}$ };
\node[align=center, text width=6cm] (g) at (6,-8) {2001:4ca0:2001:17:225:90ff:fe57:1ff8};
\node[align=left, text width=6cm] (h) at (-8,-15) { $\{$2001:4ca0:2001:13:2e0:81ff:feb5:afd8 .. 2001:4ca0:2001:13:2e0:81ff:feb5:afd9$\}$ };
\node[align=center,text width=6cm] (i) at (0,-20) { $\{$2001:4ca0:108:42::5 .. 2001:4ca0:108:42::9$\}$  $\cup$ 2001:4ca0:108:42::10};
\node[text badly centered,text width=6cm] (j) at (4,-18) { $\{$2001:4ca0:108:42:: .. 2001:4ca0:108:42::4$\}$  $\cup$  $\{$2001:4ca0:108:42::a .. 2001:4ca0:108:42::f$\}$  $\cup$  $\{$2001:4ca0:108:42::11 .. 2001:4ca0:108:42:ffff:ffff:ffff:ffff$\}$ };
\node[align=left] (k) at (8,-10) {::1};

\draw[myptr] (a) to[loop above] (a);
\draw[myptr] (a) to (c);
\draw[myptr] (b) to (a);
\draw[myptr] (b) to (c);
\draw[myptr] (b) to (f);
\draw[myptr] (b) to (g);
\draw[myptr] (c) to (a);
\draw[myptr] (c) to (b);
\draw[myptr] (c) to[loop above] (c);
\draw[myptr, color=red] ($(c.west)+(0,-1ex)$) to[bend left, in=160, out=20] ($(d.east)+(0,-1ex)$);
\draw[myptr] (c) to (e);
\draw[myptr] (c) to (f);
\draw[myptr] (c) to (g);
\draw[myptr] (c) to (h);
\draw[myptr] (c) to (i);
\draw[myptr] (c) to (j);
\draw[myptr] (c) to (k);
\draw[myptr] (d) to (a);
\draw[myptr] (d) to (b);
\draw[myptr] ($(d.east)+(0,1ex)$) to[bend left, in=160, out=20] ($(c.west)+(0,1ex)$);
\draw[myptr, color=red] (d) to[loop above] (d);
\draw[myptr] (d) to (e);
\draw[myptr] (d) to (f);
\draw[myptr] (d) to (g);
\draw[myptr] (d) to (h);
\draw[myptr] (d) to (i);
\draw[myptr] (d) to (j);
\draw[myptr] (d) to (k);
\draw[myptr] (e) to (a);
\draw[myptr] (e) to (c);
\draw[myptr] (f) to (a);
\draw[myptr] (f) to (b);
\draw[myptr] (f) to (c);
\draw[myptr] (f) to (d);
\draw[myptr] (f) to (e);
\draw[myptr] (f) to[loop above] (f);
\draw[myptr] (f) to (g);
\draw[myptr] (f) to (h);
\draw[myptr] (f) to (i);
\draw[myptr] (f) to (j);
\draw[myptr] (f) to (k);
\draw[myptr] (g) to (a);
\draw[myptr] (g) to (b);
\draw[myptr] (g) to (c);
\draw[myptr] (g) to (d);
\draw[myptr] (g) to (e);
\draw[myptr] (g) to (f);
\draw[myptr] (g) to[loop above] (g);
\draw[myptr] (g) to (h);
\draw[myptr] (g) to (i);
\draw[myptr] (g) to (j);
\draw[myptr] (g) to (k);
\draw[myptr] (h) to (a);
\draw[myptr] (h) to (b);
\draw[myptr] (h) to (c);
\draw[myptr] (h) to (d);
\draw[myptr] (h) to (e);
\draw[myptr] (h) to (f);
\draw[myptr] (h) to (g);
\draw[myptr] (h) to[loop above] (h);
\draw[myptr] (h) to (i);
\draw[myptr] (h) to (j);
\draw[myptr] (h) to (k);
\draw[myptr] (i) to (a);
\draw[myptr] (i) to (c);
\draw[myptr] (i) to (f);
\draw[myptr] (i) to (h);
\draw[myptr] (j) to (a);
\draw[myptr] (j) to (c);
\draw[myptr] (j) to (f);
\end{tikzpicture}%
}%
\caption{MeasrDroid: Main firewall -- IPv6 http connectivity matrix}
\label{fig:eval_measrdroid:i8fw:port80ipv6}
\end{figure}

The structure of the MeasrDroid architecture (cf.\ Figure~\ref{fig:casestudy:measrdroid:arch}) should be recognizable in the rules of our central firewall. 
In particular, $\mvar{Collect Droid}$ should not be reachable from the Internet, $\mvar{Upload Droid}$ should be reachable from the Internet, and $\mvar{Collect Droid}$ should be able to pull data from $\mvar{Upload Droid}$. 
This information may be hidden somewhere in the more than 5500 IPv4 and over 6000 IPv6 firewall rules. 
We use our tool \fffuu{} to extract the access control structure of the firewall. 
The result is visualized in Figure~\ref{fig:eval_measrdroid:i8fw:port80} for IPv4. 
The IPv6 structure is shown in Figure~\ref{fig:eval_measrdroid:i8fw:port80ipv6}. 
These figures may first appear highly confusing, which is due to the sheer intrinsic complexity of the access control policy enforced by the firewall. 
We have highlighted three entities in both figures. 
Because the structure and the results are similar for IPv4 and IPv6 and due to its long addresses, the IPv6 graph is even worse readable than the IPv4 graph. 
Thus, we continue our analysis only with IPv4. 
First, at the top, the IP range enclosed in a cloud corresponds to the IP range that is not used by our department, \ie the Internet. 
The large block on the left corresponds to most internal machines that are not globally accessible. 
The IP address we mark in bold red within this block belongs to $\mvar{Collect Droid}$. 
Therefore, by inspecting the arrows, we have formally verified our first auditing goal: $\mvar{Collect Droid}$ is not directly accessible from the Internet. 
The other large IP block on the right belongs to machines that are globally accessible. 
The IP address we marked in bold red within this block belongs to $\mvar{Upload Droid}$. 
Therefore, we have verified our second auditing goal: $\mvar{Upload Droid}$ should be reachable from the Internet. 
In general, it is pleasant to see that the two machines are in different access groups. 
Finally, we see that the class of IP addresses including $\mvar{Collect Droid}$ can access $\mvar{Upload Droid}$. 
Hence, we have verified our third auditing goal. 

For the sake of example, simplicity, and presentiveness, we disregard that most machines at the bottom of Figure~\ref{fig:eval_measrdroid:i8fw:port80} could attack $\mvar{Collect Droid}$.\footnote{Our method is also applicable to the complete scenario; this would only decrease clarity without contributing any new insights. We acknowledge the sheer complexity of this real-world setup with all its side-constraints.} 
Therefore, the huge access control structure at the bottom of Figure~\ref{fig:eval_measrdroid:i8fw:port80} is not related to MeasrDroid and can be ignored. 
We extract only the relevant (and simplified) parts in Figure~\ref{fig:eval_measrdroid:i8fw:port80:simple}. 
In the previous paragraph, we only presented the successful parts of the audit. 
Our audit also reveals many problems related to MeasrDroid, visualized by red arrows. 
The problems can be clearly recognized in Figure~\ref{fig:eval_measrdroid:i8fw:port80:simple}: 
\begin{itemize}
	\item $\mvar{Upload Droid}$ can connect to $\mvar{Collect Droid}$. 
	      This is a clear violation of the architecture. 
	      We have empirically verified this highly severe problem by logging into $\mvar{Upload Droid}$ and connecting to $\mvar{Collect Droid}$. 
	\item In general, most internal machines may access $\mvar{Collect Droid}$, which violates the architecture. 
	\item There are no restrictions for $\mvar{Upload Droid}$ with regard to outgoing connections. 
	      In theory, it should only passively retrieve data and never initiate connections by itself (disregarding system updates). 
	\item We uncover a special IP address with special access rights towards $\mvar{Collect Droid}$ (only shown in the full figure). 
          We find an abandoned server that has no current relevance for the MeasrDroid system. 
          As a consequence, the access rights are revoked.
\end{itemize}

Therefore, our audit could verify some core assertions about the actual implementation. 
In addition, our audit could uncover and confirm serious bugs in the implementation. 
These bugs were unknown prior to our audit and we could only uncover them with the help of our tools.

\section{Automatically Fixing Bugs}
To fix the problems our audit uncovered, we decide to install additional firewall rules at $\mvar{Collect Droid}$. 
\topos{} can automatically generate the rules for us. 
We detail the \topos{}-supported firewall configuration in the next section.

\paragraph*{A Firewall for C3PO}
\topos{} can generate a global firewall for the complete MeasrDroid architecture. 
We filter the output for rules which affect $\mvar{Collect Droid}$. 
Note that \topos{} generates a fully functional \emph{stateful} firewall for us.
For our architecture, \topos{} generates the two simple rules shown in Figure~\ref{fig:measrdroid:iptables:c3po:topos}. 

\begin{figure}[htb]%
	\footnotesize\centering
	\begin{minipage}{.9\linewidth}%
\begin{Verbatim}[commandchars=\\\{\},codes={\catcode`$=3\catcode`^=7}]
*filter
:INPUT DROP [0:0]
:FORWARD DROP [0:0]
:OUTPUT DROP [0:0]
-A OUTPUT -s 131.159.15.52 -d 131.159.15.42 -j ACCEPT
-A INPUT -m state --state ESTABLISHED -s 131.159.15.42 -d 131.159.15.52 -j ACCEPT
COMMIT
\end{Verbatim}
	\end{minipage}%
	\caption{Automatically generated \texttt{iptables} rules}
	\label{fig:measrdroid:iptables:c3po:topos}
\end{figure}

The first rule allows $\mvar{Collect Droid}$ to connect to the $\mvar{Upload Droid}$. 
The second rule allows the $\mvar{Upload Droid}$ to answer to such existing connections. 

\begin{figure}[h!tb]%
	\footnotesize\centering
	\begin{minipage}{.9\linewidth}%
\begin{Verbatim}[commandchars=\\\{\},codes={\catcode`$=3\catcode`^=7}]
*filter
:INPUT DROP [0:0]
:FORWARD DROP [0:0]
:OUTPUT DROP [0:0]
# topoS generated: 
# C3PO -> UploadDroid
-A OUTPUT -s 131.159.15.52 -d 131.159.15.42 -j ACCEPT
# UploadDroid -> C3PO (answer)
-A INPUT -m state --state ESTABLISHED -s 131.159.15.42 -d 131.159.15.52 -j ACCEPT
# custom additional rules
-A INPUT -i lo -j ACCEPT
-A OUTPUT -o lo -j ACCEPT
-A OUTPUT -p icmp -j ACCEPT
-A INPUT -p icmp -m state --state ESTABLISHED,RELATED -j ACCEPT
-A INPUT -s 131.159.20.190/24 -p tcp -m tcp --dport 22 -j ACCEPT
-A OUTPUT -m state --state ESTABLISHED -p tcp -m tcp --sport 22 -j ACCEPT
# DHCP
-A INPUT -p udp --dport 67:68 --sport 67:68 -j ACCEPT
-A OUTPUT -p udp --dport 67:68 --sport 67:68 -j ACCEPT
# ntp
-A OUTPUT -p udp --dport 123 -j ACCEPT
-A INPUT -p udp --sport 123 -j ACCEPT
# DNS
-A OUTPUT -p udp --dport 53 -j ACCEPT
-A INPUT -p udp --sport 53 -m state --state ESTABLISHED  -j ACCEPT
# further output, policy could be improved.
# Notice mails to admin, system updates, ...
-A OUTPUT -p tcp -j LOG
-A OUTPUT -p tcp -j ACCEPT
-A INPUT -p tcp -m state --state ESTABLISHED -j LOG
-A INPUT -p tcp -m state --state ESTABLISHED -j ACCEPT
COMMIT
\end{Verbatim}
	\end{minipage}%
	\caption{Manually tuned \texttt{iptables} rules}
	\label{fig:measrdroid:iptables:c3po:manual}
\end{figure}

Afterwards, we manually extend the rules to further allow some core network services such as ICMP, DHCP, DNS, NTP, and SSH for remote management. 
In addition, we allow further outgoing TCP connections from $\mvar{Collect Droid}$ (for example for system updates) but log those packets. 
The modified ruleset is illustrated in Figure~\ref{fig:measrdroid:iptables:c3po:manual}.

We analyze the modified firewall with \fffuu{} to ensure that it still conforms to the overall policy. 
%
%
%
%
%
%
%
%
\fffuu{} immediately verifies that $\mvar{Collect Droid}$ is no longer reachable from any machine (excluding localhost connection) over HTTP.  
We further verify that we do not lock ourselves out from ssh access from our internal network. 
After this verification, the firewall is deployed to the real $\mvar{Collect Droid}$ machine. 
Similarly, we implement and deploy an IPv6 firewall.

\section{Conclusion}
We used \topos{} to model the system architecture of MeasrDroid and to formalize privacy and security requirements. 
Using the built-in verification capabilities of \topos{}, we verified that our formalization corresponds to our intention and that its design guarantees the desired privacy properties. 
Using \fffuu{} to analyze whether the existing firewall correctly enforces the modeled policy, we uncovered previously unknown errors. 
In turn, we used \topos{} to generate a correct firewall for us, applied some manual low-level improvements, and used our tools again to verify that these improvements do not violate the requirements.

  \chapter{Achieved Scientific Results}
\label{chap:answers}

\begin{quote}
	\textit{Q: What is the importance of specifications and verifications to industry today and how do you think
		they will evolve over time? For example, do you believe that more automated tools will help improve
		the state of the art?}
	
	\textit{A: There is a race between the increasing complexity of the systems we build and our ability to develop
		intellectual tools for understanding that complexity. If the race is won by our tools, then systems will
		eventually become easier to use and more reliable. If not, they will continue to become harder to use and
		less reliable for all but a relatively small set of common tasks. Given how hard thinking is, if those
		intellectual tools are to succeed, they will have to substitute calculation for thought. }
\end{quote}
\hfill Interview with L.\ Lamport, (2002)~\cite{llamport2002interview}.

\bigskip

\section{Answers to the Research Questions}
In Section~\ref{sec:researchobjectives}, we asked two guiding research questions and identified four non-functional requirements that a solution must fulfill. 
In this section, we summarize our answers and elaborate on how we have fulfilled the non-functional requirements. 

\subsection*{Fulfilling the Non-Functional Requirements}

\paragraph{NF1} \textit{Can we provide automated tools for the solutions to \textbf{Q1} and \textbf{Q2}?}\medskip\newline
We contribute our fully automated tools \topos{} and \fffuu{}. 
Given the input to the tools \ie configured security invariants for \topos{} or an iptables firewall for \fffuu{}, the tools can operate completely automatic, not requiring any manual proof at runtime. 

 \begin{center}
 	\resizebox{0.99\textwidth}{!}{%
 		\begin{tikzpicture}
 		\node [MyRoundedBox, fill=LightYellow](sinvar) at (0,0) {Security Requirements};
 		\node [MyDoubleArrow](arr1) at (sinvar.east) {};
 		\node [MyRoundedBox, fill=LightYellow](policy) at (arr1.east) {Security Policies};
 		\node [MyDoubleArrow](arr2) at (policy.east) {};
 		\node [MyRoundedBox, fill=LightYellow](mechanism) at (arr2.east) {\texttt{iptables}};
 		
 		\draw [shorten <=-2cm,-latex] ($(arr1) + (0,2.5em)$)--($(mechanism) + (0,2.5em)$);
 		\node [anchor=south] at ($(policy) + (0,2.5em)$) {\topos};

 		\draw [shorten <=-2cm,shorten >=-11ex,-latex] ($(mechanism) - (0,2.5em)$)--($(policy) - (0,2.5em)$);
 		\node [anchor=north] at ($(arr2) - (0,2.5em)$) {\fffuu{}};
 		\end{tikzpicture}
 	}
 \end{center}

\paragraph{NF2} \textit{Can the correctness of the tools be justified?}\medskip\newline
We have formally and machine-checked proven the correctness of our theory with Isabelle/HOL. 
The core reasoning logic of both tools is generated by Isabelle, which guarantees partial correctness. 
In this context \emph{partial} correctness~\cite{isabelle2012code,haftmann2010code} means that the computed results are correct. 
It is not formally guaranteed that there may always be results, \eg non-termination or memory limits may abort a computation. 
Our empirical evaluation shows that the generated code both terminates and runs efficiently on commodity hardware. 
Consequently, disregarding errors in the user interface or parser, results computed by our tools are formally guaranteed to be correct. 

\paragraph{NF3} \textit{Is over-formalism exposed to the administrator?}\medskip\newline
Our tool \topos{} requires a specification of the security requirements. 
However, a user is not required to encode this specification with complicated formulas. 
Our attribute-based approach means that a user can simply assign attributes to entities. 
We have demonstrated with the Scala implementation of our tool that a simple set of JSON configuration files is sufficient. 
The secure auto completion further minimizes manual specification effort. 
Our tool \fffuu{} exposes even less formalism to its users: By default, it does not require any manual input from its users and outputs visualizable results.

\begin{figure}[!htb]
\centering
  \begin{tikzpicture}[%
	node distance=2.25cm,align=center,auto,
	state/.style={
           rectangle,
           rounded corners,
           draw=black, very thick,
           minimum height=2em,
           }
    ]

	   \node[state](q_1) at (0,0) {high-level abstraction};
	   \node[state](q_0) at (8,0) {low-level configuration}; 
	    \path[draw,thick,->,shorten >=0.5cm,shorten <=0.5cm] 
	    (q_0) to[bend left]    node {abstraction} (q_1);
	    \path[draw,thick,->,shorten >=0.5cm,shorten <=1cm] 
	    (q_1) to[bend left]    node {synthesis} (q_0);
	    \path[draw,thick,->] 
	    (q_0) edge [loop above] node {change, optimize} ()
	    (q_1) edge [loop above] node {verify} ()
	    (q_1) edge [loop below] node {modify} ()
	    (q_0) edge [loop below] node {verify} ();
	\end{tikzpicture}

  \caption{Tool-Supported Full Circle}
  \label{fig:example:intro:toolprozess}
\end{figure}
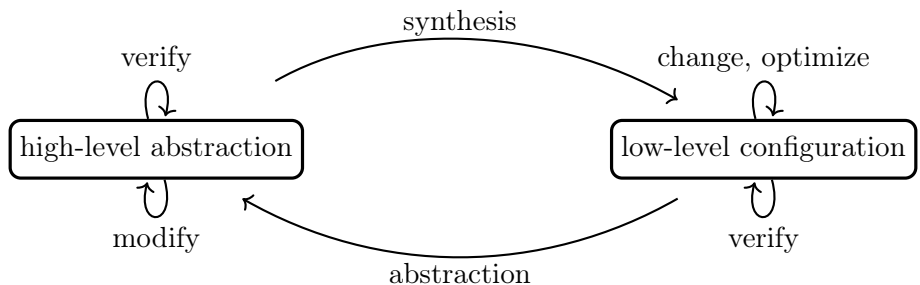

\paragraph{NF4} \textit{Are the solutions to \textbf{Q1} and \textbf{Q2} compatible?}\medskip\newline
%
%
Our methodology works on well-defined intermediate results and gives the administrator full control over them at all levels of abstraction. 
We represent policies as graphs, which is the common abstraction level where both directions presented in this thesis (synthesizing new policies vs.\ verifying existing policies) meet.  
The two directions are compatible with each other and an administrator may freely choose to which extent she wants to use our methodology and to which extent she wants to remain in full control. 
It is even possible to use our toolset in a full circle, illustrated in Figure~\ref{fig:example:intro:toolprozess}: 
Starting at an arbitrary step, one can (\textit{i})~infer the policy of a legacy configuration and (\textit{ii})~verify the policy w.r.t.\ high-level security requirements. 
Afterwards, one can (\textit{iii})~abolish the old configuration and synthesize a new, clearer configuration out of the security requirements. 
Next, one can (\textit{iv})~apply custom low-level optimizations to the synthesized policy or merge it with parts of the legacy configuration again.
Afterwards, one can (\textit{v}) verify that it still complies with the specified high-level requirements by abstracting the low-level policy to a policy graph. 
This corresponds again to steps (\textit{i}) and (\textit{ii}), which closes the circle.

\subsection*{Answers to Question 1}
\begin{center}
	\textit{``How can we design secure networks from scratch?''}
\end{center}

\paragraph{Q1.1} \textit{How can the security requirements be specified?}\medskip\newline
	We contribute a theory and generic theoretical framework which allows specifying security requirements as predicates over a policy. 
	Our formalization of security invariants is composable by construction and thus allows to modularize the security requirement specification. 
	It exposes low manual configuration overhead by the secure auto completion and our library of ready-to-use templates. 
	We present a generic method to describe offending flows and proved efficient implementations for certain invariants. 
	This enables easy debugging of a requirements specification and provides visual feedback. 
	
\paragraph{Q1.2} \textit{How can a security policy be derived from the requirements?}\medskip\newline
	The formalization of security invariants is carefully designed in many iterations such that both automatic derivation of a policy and automatic verification of policies are supported. 
	We further present enhancements that improve their performance. 
	Our evaluations show that the method scales well. 
	
\paragraph{Q1.3} \textit{How can a policy be deployed to real network security mechanisms?}\medskip\newline
	We augment our graph representation of a policy to incorporate the notion of stateful connections. 
	This extended graph can be directly serialized to several security mechanisms. 
	We identify the assumptions that a real-world setup must fulfill in order to be compatible with the theory. 
	By deploying scenarios to actual networks, we demonstrate applicability of our theory for firewalls, OpenFlow-enabled switches, and containers. 
	For the iptables firewall, we have proven the correctness of the final results. 

\subsection*{Answers to Question 2}
\begin{center}
	\textit{``How can we analyze and verify existing configurations?''}
\end{center}

\paragraph{Q2.1} \textit{What are the semantics of a security mechanism?}\medskip\newline
We contribute a formal semantics of iptables packet filtering. 
To the best of our knowledge, it is the first and only formal semantics for a complicated firewall system which is complete with regard to the matching features and the only formal semantics that supports control flow with user-defined chains. 
	
\paragraph{Q2.2} \textit{How does an entity in a security mechanism configuration correspond to an entity in a policy?}\medskip\newline
First, we present an algorithm which verifies that a firewall ruleset implements spoofing protection. 
This allows to refer to entities by their IP addresses. 
Afterwards, we present an algorithm to group sets of IP addresses by their access rights, effectively computing the vertices of the graph representation. 

	
\paragraph{Q2.3} \textit{How can a high-level policy be derived from a low-level security mechanism configuration?}\medskip\newline
First, we contribute an algorithm which transforms a complicated iptables ruleset to a simpler firewall model, abstracting over all unnecessary low-level details. 
Despite abstracting over low-level details, we have proven soundness of our approach, but not necessarily completeness due to the loss of low-level details. 
Any sound approach which converts a complex language into a simpler less expressive language is necessarily incomplete. 

Our approach enables to compute an overview of the connectivity permitted by a firewall. 
This enables sound security evaluation: the actual firewall definitely does not allow more than our simplified firewall, but due to the potential incompleteness, the actual firewall could drop even more.\footnote{w.r.t.\ our monotonicity assumption, dropping more does not decrease security} 
Vice versa, our approach also allows reasoning about the set of packets that a firewall definitely accepts. 
Despite the incompleteness, the usefulness of our approximation has been successfully evaluated in many real-world rulesets.

Second, we contribute an algorithm which partitions the complete IPv4 and IPv6 address space into equivalence classes with equal access rights for a fixed service. 
This yields a graph which represents the policy enforced by a firewall, effectively computing a minimal access control matrix. 
This graph can be visualized and further verified with the answers we contributed for \textbf{Q1}.

\paragraph{Q2.4} \textit{Can a derived high-level policy be verified w.r.t.\ a given set of security requirements?}\medskip\newline
As we already mentioned when asking the questions, question \textbf{Q1.2} is strengthened such that an answer to it must already entail an answer to this question. 
Indeed, our theory directly supports this and our answer to \textbf{NF1} shows that our methodology supports the full circle.

\section{Tackling Complexity in Network Security Management}
In Section~\ref{sec:intro:problemclassification-and-sdn}, we discussed the complexity challenges in software engineering and how they can be interpreted in the context of network security management. 
We distinguished between \emph{accidental} difficulties and \emph{essential} difficulties. 
In this section, we sum up how we have approached these difficulties in this thesis. 

A large set of related work in the field of configuration management only tackles accidental complexity. 
Many attempts have been made before to hide low-level (accidentally complex) configuration aspects with a high-level language~\cite{mignis2014,ZhangAlShaer2007flip,orbacnetwork04,hinrichs2009practical,ethane07}. 
But this cannot solve the core problem of essential difficulties.

Essential difficulties cannot be removed. 
However, in this thesis, we use the following key ideas to tackle them as good as possible:
\begin{itemize}
\item \emph{complexity}: Intrinsic complexity cannot be removed. 
But we can isolate and modularize certain aspects to only focus on one problem at once. 
For example, in Chapter~\ref{chap:forte14}, we break down the task of specifying security requirements into several isolated tasks of specifying well-defined security invariants. 
Our design ensures composability of these invariants.  
In Chapter~\ref{sec:forte14:model-library} we provide a library to further isolate the generic part of security invariants (template) from the scenario-specific configuration (host attribute mapping). 
In Chapter~\ref{chap:esss14}, we isolate stateless access control from its stateful network implementation. 
In Chapter~\ref{chap:networking16}, we completely isolate low-level firewall primitives and access control. 
\item \emph{conformity}: Part~\ref{part:existing-configs} is completely devoted to the understanding and verification existing iptables configurations. 
\item \emph{changeability}: We provide a formal proof for all our algorithms. 
 The proof, by its very nature, reveals all implicit assumption and is valid under any worst-case assumptions, thus covering all unforeseeable cases. 
\item \emph{invisibility}: We provide clear, intermediate artifacts. 
 All results of our algorithms can be presented and visualized as graph. 
\end{itemize}


\noindent
With regard to this view, the structure of this thesis can be understood as follows.
\begin{description}
	\item[Part~\ref{part:greenfield}] Breaking down the \emph{complexity} of network security management into several isolated aspects and creating \emph{visible} intermediate results. 
	\item[Part~\ref{part:existing-configs}] Ensuring \emph{conformity} with existing, legacy configurations. Separating low-level accidental complexity from inherent complexity. 
\end{description}

In 1987, it was envisioned that ``one of the most promising of the current technological efforts, and one that attacks the essence, not the accidents, of [network security configuration], is the development of approaches and tools for rapid prototyping of systems as prototyping is part of the iterative specification of requirements.''~\cite{brooks1987no}\footnote{We have strongly adapted the quote for our context. The original quote said ``software problem'' instead of ``network security configuration''.} 
Now, 30 years later, our iterative process illustrated in Figure~\ref{fig:example:intro:toolprozess}, finally enables this process and provides formally verified tools for it. 
 

  \chapter{Comparison to State-of-the-Art}
\label{chap:relatedworktable}
In this chapter, we compare the contributions of this thesis in the field of \emph{static network configuration} to the state of the art. 
In particular, we consider work that provides answers to the question \textit{``How would I want to design my network?''} and, with focus on security in terms of managing access control, work that provides answers to the question \textit{``Which machines should be allowed to speak to each other?''}.



\section{List of Criteria}
We compare languages, tools, and frameworks for static network configuration and focus on security. 
We set up a list of criteria based on the finest features found in related work. 
With regard to \textbf{NF1}, we consider only work where executable code is available or an executable implementation is described. 


We start with two quality criteria. 
Our non-functional requirement \textbf{NF2} asked \textit{``can the correctness of the tools be justified?''}. 
To motivate the importance of this question, it was recently found~\cite{icfp2015smolkanetkatcompiler} that even the well-engineered SDX setup~\cite{gupta2014sdx} had errors because of errors in the underlying tool. 
Guha \etal\cite{machineverifiednetworkcontrollers13} provide further examples where ``tools make simplifying assumptions that are routinely violated by real hardware''. 
%
%
To formally justify overall correctness of a tool, at least two artifacts are needed. 
We describe these artifacts in the following two paragraphs. 

\paragraph*{Formal Semantics}
First, a {formal semantics} must describe a model and the behavior of the targeted network mechanisms to be able to reason about a configuration. 
Ideally, for a tool which compiles a high-level language to low-level configurations, two semantics should be provided: one semantics of the targeted network elements (\eg a semantics of OpenFlow switches or the iptables firewall) and one semantics of the high-level management language. 
NetCore provides a detailed semantics of the operations and essential features of OpenFlow~\cite{machineverifiednetworkcontrollers13}. 
Both NetKAT~\cite{anderson2014netKATsemantics} and NetCore~\cite{netcore12} provide clear, explicit, formal semantics of their language. 
The presence of a formalized semantics makes assumptions explicit and allows formal reasoning.

\paragraph*{Formal Verification}
Second, when using a tool to help with security-related issues, a lot of trust is placed in the tool. 
The question about the correctness of the tool w.r.t.\ the formal semantics needs to be answered. 
The authors of the Mignis~\cite{mignis2014} tool provide a proof of correctness of their theory. 
However, the proof only exists on paper and it remains unclear whether the actual implementation is also correct. 
Remarkably, the correctness of the NetCore toolset~\cite{machineverifiednetworkcontrollers13} is proven correct with the Coq~\cite{coqmanual} proof assistant. 
This provides a machine-verified proof of correctness and constitutes our criterion of {formal verification}.

Those two criteria, namely the availability of a formal semantics and machine-verified formal correctness proofs, ensure a high quality of a tool and also demonstrate a generic scientifically sound theory.

\paragraph*{High-Level Language}
Next, we survey related work with regard to our first question \textbf{Q1} \textit{``How can we design secure networks from scratch?''} 
In particular, \textbf{Q1.1} ``\textit{How can the security requirements be specified?}'' raises the question at which level of abstraction the security requirements should be expressed. 
With regard to Figure~\ref{fig:intro:securitycomponents}, security requirements reside on a higher level of abstraction than policies. 
However, most related work discussed in this chapter only considers the abstraction level of policies, which leaves the formalization of security requirements as a policy as an error-prone manual step. 
Notably, Zhao \etal\cite{zhao2011policyremanet} try to tackle this exact problem. 
They state that
``[i]t is increasingly important to develop policy refinement that automates high level requirements into [policy] implementation in policy-based system management.''~\cite{zhao2011policyremanet}. 
With regard to Figure~\ref{fig:4layerabstraction}, Zhao \etal propose to start at the Security Invariants while most other work starts at the Access Control Abstraction. 
We call a language that directly supports specification as security invariants a \emph{High-Level Language}. 
%
%
%

\paragraph*{Built-In Verification}
Yet, given a specification in a high-level language or as (high-level) policy, the question whether a specification expresses the intended meaning remains. 
Therefore, a management language should be built with automated verification in mind. 
This means that the language should be limited in the expressiveness to allow formal reasoning but still be as expressible as necessary for real-world application~\cite{nelson2013balanceofpower}. 
Both Flowlog and NetKAT are based on this principle. 

\paragraph*{Stateful Connection Semantics}
Though we primarily focus on static configuration languages, stateful connection semantics must not be overlooked. 
For example, a simple policy such as `Internet hosts can only send packets to an internal host if the internal host initiated the connection' may be found in almost every firewall. 
McClurg \etal\cite{netkat2016state} note that most languages that have been proposed so far lack this critical feature. 
Also, Ad{\~a}o \etal\cite{mignis2014} note that a lot of firewall work is limited to stateless firewalls and therefore barely applicable to real-world. 
Both focus on explicit support for stateful connections. 

\paragraph*{Legacy Support}
``[A]dministrators quite likely depend upon the behavior of their existing configurations''~\cite{nelson2015exodus}. 
Therefore, it is important to investigate how existing configurations can be ported to a new, proposed management language. 
The Exodus system~\cite{nelson2015exodus} focuses on migrating existing, legacy configurations from Cisco devices to the Flowlog language. 
Our criterion of legacy support requires that legacy configurations can be migrated to a proposed language. 
Afterwards, for this criterion, the legacy configuration and devices can be abandoned; our criterion does not require coexistence with legacy infrastructure. 
%
%
It is even more challenging to build a hybrid approach which allows to operate legacy and new infrastructure in a hybrid way~\cite{b42013googlesdn}.


\paragraph*{Low Level Access}
Our second main research question \textbf{Q2} asks \textit{``How can we analyze and verify existing configurations?''}. 
The previous criterion of legacy support implies a feature to import an existing configuration and, probably, analyze it afterwards. 
However, once an existing configuration is imported, there is no way of going back. 
Existing configurations may contain certain low-level statements that are no longer available in the high-level language. 
Since a high-level language should usually abstract over low-level details, this is a desirable property. 
Yet, these low-level features are sometimes necessary, as can be seen in several examples of iptables~\cite{diekmanngithubnetnetwork,cloudflare2014blogbpf,serverfaultiptables}.  
To cope with this issue, the Mignis~\cite{mignis2014} firewall configuration language explicitly allows its user to add arbitrary low-level iptables match conditions to the high-level rules. 
This also gives the user low-level access. 
It adds the problem that once a user may perform low-level modifications, there arises the need to verify that these low-level modifications comply with high-level specifications.

\section{Evaluation of Related Work}
We summarize our comparison of related work in Table~\ref{tab:relatedworktable}. 
First, we summarize the evaluation criteria applied in said table. 
Afterwards, we detail on the individual related work. 
Finally, we summarize how our tools achieve the criteria in Section~\ref{sec:related:howweachieve}. 

\begin{description}
	\item[F.\ Sem (Formal Semantics)] A checkmark is awarded if a formal semantics of the targeted network elements or the language exists. 
		Two checkmarks are awarded if the semantics is formalized and at least type checked by a theorem prover.  
\end{description}
\begin{description}
	\item[F.\ Veri (Formal Verification)] A checkmark is awarded if the compilation or translation process of a tool is formally proven. 
		We award two checkmarks if the proof is machine-checked. 
\end{description}
\begin{description}
	\item[HLL (High-Level Language)] A checkmark is awarded if the language allows to start with a specification of high-level security requirements (in contrast to languages which start with a policy). 
\end{description}
\begin{description}
	\item[Veri Spec (Built-In Verification)] One checkmark is awarded if the language comes with built with built-in verification. Two checkmarks are awarded if some automatic verification without the need for a manual specification is possible. 
\end{description}
\begin{description}
	\item[ct (Stateful Connection Semantics)] A checkmark is awarded if the tool supports stateful connection semantics. 
		The abbreviation \textbf{ct} is derived from the abbreviation for conntrack. 
\end{description}
\begin{description}
	\item[Leg Sup (Legacy Support)] A checkmark is awarded if a solution provides legacy-support, \eg reading iptables or other existing configurations and making them available to the high-level abstraction. 
\end{description}
\begin{description}
    \item[LL Access (Low Level Access)] One checkmark is awarded if the language allows the administrator to perform low-level configuration tuning. 
    		Two checkmarks are awarded if it allows to check whether the low-level changes are sound. 
\end{description}

\begin{table}[hbt]
	\centering
	\caption{Comparison to related work.}
	\vskip-1ex
	\label{tab:relatedworktable}
\centering
\begin{small}
	\begin{tabular}[0.99\linewidth]{ l | l | l | l | l | l | l | l}
		\toprule
		                   & F.\ Sem          & F.\ Veri	      & HLL     & Veri Spec    & ct        & Leg Sup & LL Access \\
		\midrule           
		Flowlog \& co      & \xmark           & \xmark            & \xmark  & \cmark\cmark & \cmark    & \cmark    & \xmark       \\
		NetCore            & \cmark\cmark     & \cmark\cmark      & \xmark  & \cmark       & \xmark    & \xmark    & \xmark       \\
		NetKAT family      & \cmark\cmark     & \cmark            & \xmark  & \cmark       & \xmark    & \xmark    & \xmark       \\
		VALID              & \cmark           & \xmark            & \cmark  & \xmark       & \xmark    & \xmark    & \xmark       \\
		Zhao et al.        & \xmark           & \xmark            & \cmark  & \xmark       & \xmark    & \xmark    & \xmark       \\
		Mignis             & \cmark           & \cmark            & \xmark  & \xmark       & \cmark    & \xmark    & \cmark       \\
		\midrule
		\midrule
		\topos{} + \fffuu{}      & \cmark\cmark     & \cmark\cmark      & \cmark  & \cmark\cmark & \cmark    & \cmark    & \cmark\cmark     \\
		\bottomrule%
	\end{tabular}%
\end{small}
\end{table}


\section{Discussion of Related Work in Detail}
%

\paragraph*{Flowlog \& co}
Nelson presents the programming language Flowlog~\cite{nelson2014flowlog} for SDNs, which was preliminary presented in a workshop paper~\cite{nelson2013balanceofpower}.
Flowlog is an event-driven language with a syntax similar to SQL. 
It allows to write dynamic SDN controller programs; strictly speaking, it is not a static configuration language. 
However, Flowlog tries to compile a large amount of a program to static OpenFlow rules, handling only the bare minimum at the controller. 
Flowlog supports arbitrary state in a program via mutable tables. 
For example, it is easy to construct a stateful firewall with it~\cite{nelsongithubflowlogfirewall}. 
With the same abstraction, Flowlog allows to call external programs. 

Flowlog generates OpenFlow rules indirectly by generating a NetCore (cf.\ next paragraph) specification, which is in turn compiled to OpenFlow. 
The system does not allow an administrator to modify the generated NetCore or OpenFlow rules since this could introduce errors w.r.t.\ the Flowlog specification. 
It is not possible to assess whether an arbitrary NetCore or OpenFLow ruleset corresponds to a given Flowlog specification. 
%
The Flowlog ``compiler has only been tested, not proven correct.''~\cite[\S\hairspace{}7]{nelson2014flowlog}

Flowlog was especially designed with built-in verification and analysis in mind~\cite{nelson2013phdthesis,nelson2013balanceofpower}. 
Using the alloy analyzer~\cite{alloywebsite}, Flowlog programs can be verified or counterexamples can be found. 
The translation from a Flowlog program to an alloy specification is automatic. 
Of course, external programs cannot be translated, thus a user must describe them axiomatically. 
A user can manually formalize a verification goal as alloy statement. 
Because alloy can only perform bounded verification, a lack of a counterexample does not automatically imply that the verification goal is met. 
Some experience is required by the user to argue about the verification bounds and the resulting completeness in certain cases. 
Later on, Chimp~\cite{nelson2015chimp} was designed which automatically compares the behavior of two versions of a Flowlog program. 
Chimp does not require any manual specification of verification goals and is thus very usable. 
Due to the clever construction of the Flowlog language, a user must only provide few bounds about the size of the analyzed network and Chimp can automatically generate bounds for which the analysis is both sound and complete (for most cases, cf.\ \cite[\S\hairspace{}5]{nelson2015chimp}). 


Flowlog focuses on network programming. 
It does not support to specify security requirements directly. 
Though, an advanced user might formalize them as alloy specification and verify her Flowlog implementation against it. 
Given only a specification of security requirements, the system does not support to derive a Flowlog program automatically. 
Therefore, with regard to our criterion definition, Flowlog is not a high-level language. 

Legacy network configurations can be translated to Flowlog. 
The Exodus system~\cite{nelson2015exodus} processes a set of Cisco IOS configurations. 
It supports a remarkable subset of all features. 
The translation is not formally verified, but all assumptions and supported features are clearly stated. 
The static parts of the evaluation's configurations are statically verified (with HSA~\cite{kazemian2012HSA}) and dynamic parts are verified using Flowlog's built-in verification capabilities.

\paragraph*{NetCore}
NetCore~\cite{netcore12} is a stateless forwarding policy language with formal semantics. 
It was designed from the very beginning to feature formal semantics which allows to build verification tools on top~\cite{guha2013formalshort}. 
It contributes an OpenFlow syntax and semantics~\cite{machineverifiednetworkcontrollers13} and ``the first machine-verified SDN controller''~\cite{machineverifiednetworkcontrollers13}. 
High-level properties of NetCore programs can be proven~\cite{stewart2013computationalnetcorecoq}. 
It was also attempted to extend this verification to networks with several switches~\cite{netocre2016severalswitchesbad}. 
All the verification was carried out with the Coq~\cite{coqmanual} proof assistant. 

NetCore pioneered in its domain but ultimately, it was superseded by NetKAT (see next paragraph) because its semantics is not sound for KAT~\cite{anderson2014netKATsemantics}. 
Flowlog still uses NetCore as backend. 
We focus this short summary only on NetCore's features which are formally, machine-checked verified. 
In contrast, while NetKAT and Flowlog provide more features, they are not formally verified.

\paragraph*{NetKAT}
NetKAT provides a semantics of network packet forwarding~\cite{anderson2014netKATsemantics}. 
The semantics is based on Kleene Algebra with Tests (KAT), which allows sound and complete reasoning about statements expressed in NetKAT. 
A network topology can be encoded in NetKAT. 
In addition, forwarding, filtering, and packet modification policies can also be encoded with NetKAT. 
Consequently, NetKAT can be considered a static network programming language with well-defined semantics. 
An explicit low-level hardware semantics -- such as Featherweight OpenFlow~\cite{machineverifiednetworkcontrollers13} -- is not provided, but the behavior of OpenFlow flow table switch processing can be directly expressed with NetKAT. 

A concrete NetKAT statement (network description or network program) can be verified. 
Verification is done by checking equivalence of two statements. 
For example, one can verify the equivalence of an implemented NetKAT forwarding policy and the high-level NetKAT policy ``$A$ can reach $B$''. 
To verify that certain packets are dropped, one can check that the implemented policy for a specific packet is equivalent to the drop-all policy~\cite{anderson2014netKATsemantics}. 
A fast decision procedure exists for these kinds of queries~\cite{netkat2015coalgebraicdecision}. 
Unfortunately, a user needs to specify queries manually, which bears the chance of replicating a bug in both a NetKAT program and a NetKAT specification. 
Pre-defined queries for common verification problems exist \eg all-pair connectivity and loop freedom. 
In general, equivalence of NetKAT statements is PSPACE complete.

It is unclear whether existing, complex, legacy network configurations can be automatically translated to NetKAT. 
For the evaluation of the NetKAT decision procedure~\cite{netkat2015coalgebraicdecision}, router configurations of the Stanford backbone\footnote{Archived at \url{https://github.com/diekmann/net-network/tree/master/hassell-public_Stanford/Stanford_backbone}} have been translated to NetKAT statements. 
It is unclear which features are supported by the translation in general since this tool seems not to be present in the repository~\cite{githubNetKATcoalgebraic}. 
However, a parser from JSON to NetKAT exists~\cite{githubNetKATjsonparser}. 
The supported features are similar to OpenFlow 1.0, which we do not consider as legacy support. 
%

NetKAT features an efficient compiler for local, global, and virtual programs to flow table entries~\cite{icfp2015smolkanetkatcompiler}. 
A local program describes the behavior of individual switches. 
A global program allows to express network-wide behavior, such as paths through the network. 
A virtual program allows to express a behavior based on a virtual topology \eg considering the network as one big switch.

%
NetKAT has been extended to support stateful network programs and event-driven programming~\cite{netkat2016state}. 
A stateful NetKAT program can be considered a graph where the nodes correspond to static NetKAT programs and the edges correspond to network events, such as a packet arriving at a switch. 
Upon receipt of an event, the network configuration transitions from one static NetKAT configuration to the next. 
The NetKAT decision procedure~\cite{netkat2015coalgebraicdecision} and existing compiler~\cite{icfp2015smolkanetkatcompiler} cannot be applied to a stateful NetKAT program, but for each state a stateful NetKAT program can be transformed into a stateless NetKAT program where aforementioned tools are applicable. 
Therefore, the verification of a stateful NetKAT program requires enumerating all states. 
Compilation is also based on enumerating all states and reusing the existing compiler for the stateless NetKAT programs. 
This overall compiler setup pre-compiles OpenFlow rules for all states in advance. 
While this programming language may be suitable for certain scenarios, it is not well-suited for implementing connection tracking in a stateful firewall. 
The paper~\cite{netkat2016state} shows an example of a stateful firewall which allows host $A$ to send packets to host $B$, but only allows host $B$ to send packets to host $A$ if host $A$ has sent a packet first. 
This link layer filtering behavior does not correspond to the actual connection tracking performed by stateful firewalls. 
A common stateful firewall performs connection tracking, \ie tracking the IP-5-tuple tuple of source/destination IP addresses, protocol, and source/destination ports. 
Enumerating all possible states (here, all possible values of an IP-5-tuple) or pre-compiling all possible rules is infeasible for a state of such size.

A decision procedure for KA (Kleene Algebra) and KAT (Kleene Algebra with Tests) has been formally verified with the Coq proof assistant~\cite{braibant2010kaincoq,Pous2013katincoq}. 
Also, the semantics of NetKAT has been formalized in Coq~\cite{githubnetkatrepoissue2}. 
However, KAT tools cannot be directly applied to NetKAT because ``NetKAT extends KAT with network-specific primitives''~\cite{netkat2015coalgebraicdecision}. 
Consequently, the NetKAT toolset itself (compiler, decision procedure) is not formally verified~\cite{githubnetkatrepoissue2}. 

\paragraph*{VALID}
Bleikertz and Gro{\ss} state the need to express security requirements in a high-level language and verify policies against them. 
They propose VALID~\cite{bleikertz2011VALID} which is a specification language to express security invariants in cloud infrastructures. 
VALID has aspects in common with \topos{}, for example, it requires a network's connectivity and information flow structure in the format of a graph as input and allows to specify predicates over it. 
The language is formally defined in the AVISPA Intermediate Format (IF). 
It can verify that a given network topology conforms to specified high-level goals, but it cannot translate these goals to a network topology nor give feedback about the meaning of the specified goals. 
VALID is merely a language to express security requirements, hence, a VALID specification cannot be deployed to a network. 

\paragraph*{Zhao \etal}
Zhao \etal\cite{zhao2011policyremanet} present a policy refinement framework for network services. 
They state the need to express security requirements in high-level terms and present a high-level logic language to encode them. 
For this language, an automated translation (\emph{refinement}) procedure to low-level policies is presented. 
The low-level policies can be directly enforced by security mechanisms. 


The logic-based, abstract policy language roughly corresponds to first-order logic with set theory, orderings, and relations over an UML class diagram. 
The policy language allows to express almost unrestricted statements. 
As an example, given an UML class diagram which models all natural numbers as $\mathit{class}(\mathbb{N})$, statements about all natural numbers are possible. 
It is thinkable that number theory and even G{\"o}del's incompleteness theorem can be modeled in the language, which demonstrates its expressiveness. 
If it were possible to express enumerable but infinite UML objects, \eg $\mathit{obj}(\mdef{zero}, \mathbb{N})$, $\mathit{obj}(\mdef{one}, \mathbb{N})$, $\mathit{obj}(\mdef{two}, \mathbb{N})$, etc., arbitrary computations could be possible. 
The policy language by itself is expressive enough to make statements about any natural number. 
However, a finite UML definition for the policy domain needs to be given. 
Ultimately, this means that only statements about a finite amount of natural numbers can be given to the refinement process and the refinement process itself is decidable. 
We believe that the finiteness of the UML diagram is the only factor which prevents this mighty system from being Turing complete. 
%
The downside for a user of such a language may be that it is hard to specify and hard to read. 
The system does not provide a reader with feedback about whether the encoded high-level security requirement actually corresponds to the policy author's intention. 

In contrast, our tool \topos{} comes with a library of pre-specified logic formulas (security invariant templates). 
A policy author only needs to assign attributes but is not required to manually encode requirements as formulas. 
In addition, \topos{} gives immediate feedback to a policy author by visualizing the resulting access control policy as graph. 
This enables a policy author to verify her requirement specification or perform a \textit{``What-if?''} analysis. 

While the language is well thought out, the prototype implementation and refinement process hint at some small problems. 
The refinement process contains a non-deterministic step, which means that several different refinements might fulfill a high-level policy rule. 
Under certain circumstances, this may result in soundness issues: 
Such a refined rule may only cover one subset of subjects that are allowed/denied access, but where the high-level policy intended to allow/deny also access for a different set of subjects. 
This remark is rather theoretical, as we were unable to empirically construct such an example for the prototype implementation. 
The current prototype implementation only supports refinement of one single policy rule. 
Since rules may conflict or overlap, it is not possible to just translate a set of rules one by one individually. 
This effectively limits policies for the prototype implementation to one rule (without introducing conflict resolution or soundness issues due to the ordering of ACLs). 
The translation to a Routing+Firewall architecture (called ROFL) may result in further soundness issues since Bloom filters~\cite{bloom1970filters} are used to encode permitted subjects and a false-positive may result in granting access to a subject that should not have access. 
Testing the prototype implementation, we found a policy which produces a refinement which is not expected, but we could not figure out the reason of the problem. 

The author of this thesis wants to thank Hang Zhao for providing the source code of the prototype implementation on request and for a constructive email conversation.

\paragraph*{Mignis}
Mignis~\cite{mignis2014} presents a formal semantics of the filtering behavior and NAT behavior for the netfilter/iptables firewall. 
The semantics supports the actions \iptaction{ACCEPT}, \iptaction{DROP}, \texttt{SNAT}, and \texttt{DNAT}; it does not consider user-defined chains. 
The match conditions for matching on source/destination IP addresses and source/destination ports are modeled explicitly. 
In addition, an arbitrary predicate $\phi$ over the packet and the connection tracking state is modeled to represent arbitrary further match conditions. 
This predicate $\phi$ is always assumed to be well-behaved. 
Unfortunately, the semantics only exists on paper. 
Common questions whether the semantics is always defined or determinism remain unanswered. 
The Mignis tool is written in python~\cite{githubmignis}. 

The main focus of the Mignis tool is to present a new policy language for firewall management. 
Hence, the semantics mainly models the subset of iptables that is also supported by the Mignis language. 
The predicate $\phi$ is intended to give the user low-level access to add additional matching features to iptables rules which are generated by Mignis. 

The policy language is declarative and rules are position-independent. 
The language supports \iptaction{DROP}, \iptaction{ACCEPT}, \iptaction{DNAT}, and \iptaction{SNAT} rules. 
\iptaction{DROP} rules always take precedence over \iptaction{ACCEPT} rules, which avoids traditional firewall rule conflicts by definition. 
For each rule, the matching IP addresses and ports can be specified. 
In addition, for each rule an arbitrary predicate $\phi$ can be added to give the administrator additional low-level control over the ruleset. 

The Mignis tool decides the overall control flow of the generated iptables rules. 
This prevents an administrator from reordering rules -- a common performance optimization which is obstructed by Mignis. 
An administrator can only influence match conditions for individual rules via the $\phi$ predicates. 
Since a user cannot influence the control flow, this reduces the effective matching possibilities originally provided by iptables: 
 With one iptables rule, it is not possible to write a rule that negates several primitive match conditions.
 For example, one cannot write directly \verb~! (-p tcp --dport 80)~, which should match everything that is not tcp port 80.\footnote{For this specific example, it is possible encode the desired behavior manually by two rules: \verb~-p tcp~ and \verb~-p tcp ! --dport 80~. This manual approach is error prone and does not scale for larger, complicated expressions.} 
 However, calling a user-defined chain and \iptaction{RETURN}ing allows to logically construct exactly the desired match condition. 
 This pattern is observed quite often in real-world rulesets~\cite{diekmanngithubnetnetwork}.
Therefore, the low-level control admitted by Mignis is rather limited.

\begin{sloppypar}
The user-defined predicate $\phi$ must be well-behaved. 
For example, it must behave exactly the same for a packet before and after NATing. 
The Mignis implementation~\cite{githubmignis} checks that no $\phi$ contains strings such as \verb~--sport~ or \verb~--source-port~.  
However, this check can be too restrictive as, for example, Mignis also rejects the comment \verb~-m comment " --sports "~. 
More severely, without having an explicit whitelist which enumerates all `good' match features, it is impossible to check that a user-specified predicate is actually well-behaved. 
For example, the match extensions \verb~bpf~, \verb~string~, \verb~u32~ can all be used to match on ports and IP addresses, and \verb~iprange~, \verb~addrtype~, \verb~realm~, \verb~recent~, and \verb~set~ can be used to match on IP addresses. 
On the one hand, if Mignis allows arbitrary predicates $\phi$, it sacrifices soundness. 
On the other hand, if Mignis only allows an explicit whitelist for $\phi$, it limits its expressiveness. 
In the current implementation ``mignis cannot provide any correctness guarantee about the whole configuration if [arbitrary match extensions are used]''~\cite{githubmignisissue2}.\footnote{Original quote: ``You can write any filter you want after |, but mignis cannot provide any correctness guarantee about the whole configuration if you do that.''~\cite{githubmignisissue2}} 
\end{sloppypar}



A simple query language to verify Mignis policies is work in progress~\cite{githubmignis}. 

%
%
%

\section{Evaluation of \topos{} and \fffuu{} w.r.t.\ the Criteria}
\label{sec:related:howweachieve}
In this section, we summarize how the results of this thesis are evaluated with regard to the presented criteria. 

\begin{description}
	\setlength{\itemsep}{.1ex}
	\item[Formal Semantics]
We provide both, a low-level semantics of iptables and formalization of our high-level language as security invariants. 
Both are formalized in the Isabelle/HOL theorem prover. 
In addition, important properties of the semantics are shown, \eg determinism of the iptables semantics or definedness of the offending flows. 


\item[Formal Verification]
Our tools \topos{} and \fffuu{} are formally machine-verified by \mbox{Isabelle/HOL}. 

\item[High-Level Language]
Our approach allows to start with a high-level language to encode security requirements. 
For user-friendliness, a security requirement can be formalized by instantiating pre-defined security invariant templates with scenario-specific information. 
In addition, any higher-order logic formula which fulfills the proof obligations for a security invariant template\footnote{Definition~\ref{def:securityinvariantmonotonicity}, the invariant must be true for the deny-all policy (Theorem~\ref{thm:no-edges-validity}), and Definition~\ref{def:unique_default}} can be added to the language. 
Ultimately, requirements are expressed as a set of security invariants over a policy. 
We showed that, given only the invariants, a policy can be automatically computed. 
This makes \topos{} a true high-level language according to our criterion. 


\item[Built-In Verification]
Our tool \topos{} also provides built-in verification. 
For example, it is possible to compare formalized requirements with a manually specified policy. 
Once a specification of the requirements exists, the generic policy construction algorithm allows to provide automatic feedback about the requirements. 
This does not require any additional input. 
Among others, it can compute the allowed flows resulting from the requirements, or conversely, can visualize everything the requirements prohibit. 
The semantical difference between two requirement specifications can be automatically compared by comparing the policies constructed from these requirements. 
Our tool \fffuu{} gives feedback about an iptables ruleset without any additional input: it can visualize the overall policy as a partition over the complete IP address space. 

\item[Stateful Connection Semantics]
Both our tools \topos{} and \fffuu{} support stateful connection semantics. 

\item[Legacy Support]
Legacy support for iptables firewalls is provided by \fffuu{}, which translates an iptables ruleset to a policy. 
The resulting policy can, in turn, be loaded by \topos{}. 

\item[Low Level Access]
The other way round, any iptables ruleset generated by \topos{} can be manually low-level tuned, and \fffuu{} can compute whether the policy has changed. 
If it has changed, \topos{} in turn can verify whether the changes introduce a violation with regard to the security requirements. 

\end{description}

  \chapter{Summary of Applicability and Application}
\label{chap:applicability}

\begin{quote}
	\textit{Beware of ``the real world''. A speaker's appeal to it is always an invitation not to challenge his tacit assumptions. When a speaker refers to the real world, a useful counterploy is pointing out that he only refers to \underline{his} perception of it.}
\end{quote}
\hfill E.\ W.\ Dijkstra, A bagatelle for the left hand (1982)~\cite{EWD:EWD800}.

\section*{Applicability and Application}
\noindent
In Chapter~\ref{chap:introduction}, we asked the question
\begin{center}
\textit{``How can we provide means to help the administrator to configure secure networks or verify the security of existing network configurations?''}
\end{center}
We have split the problem statement into various sub problems and presented our answers in Chapter~\ref{chap:answers}. 
In this chapter, we summarize how our results apply to various domains. 

\section{Generic Policy Management and Reasoning}
In this thesis, we developed a generic theory for policy management, reasoning, and verification. 
Our theory is independent of enforcing mechanisms and can thus be applied in various domains, not only traditional network management. 
We demonstrated the applicability in the following areas:

\paragraph*{Designing Privacy Policies}
As shown in Chapter~\ref{chap:puttingtogether}, our framework and tool \topos{} can be used to construct and evaluate policies with regard to privacy. 

\paragraph*{Auditing}
After having constructed a privacy policy, our \fffuu{} tool can be used to perform an audit of the real implementation of the system. 
We discovered and fixed previously unknown errors. 
The results are also presented in Chapter~\ref{chap:puttingtogether}. 

\paragraph*{Microservice Management}
As shown in Chapter~\ref{chap:mansdnnfv} and Chapter~\ref{chap:dynamicdocker}, our \topos{} framework can be used to construct an access policy for microservices. 
The ruleset can be deployed on a docker host. 

\paragraph*{Hypervisor Management}
We showed that the same policy we have constructed in Chapter~\ref{chap:mansdnnfv} can also be enforced by Open vSwitch. 
Thus, instead of only relying on the weak isolation provided by docker, the complete scenario can be deployed where each service is running in its own virtual machine on top of xen. 
This provides stronger isolation and we have shown how that our approach can be easily used to manage the internal network of a hyperviser such as xen. 

\paragraph*{Cyber Physical Systems}
In Section~\ref{sec:forte14:case-study}, we showed how our tool \topos{} can help in a real-world scenario ---in collaboration with \eads--- to evaluate a cabin data network for the general civil aviation. 
Section~\ref{sec:example:imaginary-factory-network} further exemplifies the applicability for a different cyber physical system. 

\paragraph*{Software Architectures}
Section~\ref{sec:forte14:relatedwork} shows a striking parallel between our work and the field of software architectures. 
Viewing an architecture as policy, our method of policy verification is also applicable to software architectures. 
We have successfully used our method to verify the software architecture of a Munich startup (unfortunately, the results are not public). 

\section{Iptables Firewall Analysis}
The theoretical aspects of our work are ``substantial''~\cite{iptablesafpmaillarry}. 
Among others, we contribute a generic method to abstract over unknown match conditions which allowed us to build the first tool which can understand \emph{any} iptables match condition. 
For the first time, a fully machine-verified analysis tool, and in general, a tool to analyze non-trivial firewalls is presented. 
Our tool is probably also a stepping stone to bridge gaps between the formal methods community and the computer networking community.

Throughout this thesis, we demonstrated real-world applicability of our tool. 
We also use it internally in our lab~\cite{github2016insalata}. 
In addition, we have successfully solved real problems on serverfault~\cite{serverfaultmultiportnegation,serverfaultpinput,serverfaultspoofing,serverfaultwhyblocking,serverfaultwhywhitelistdns,serverfaultwhywhitelistdns,serverfaultipv6fail}, debugged real-word problems (Chapter~\ref{chap:networking16}), found real-world errors (Chapter~\ref{chap:puttingtogether}), and earned several stars on github~\cite{diekmanngithubiptablessemantics}. 
The practical applicability has also been demonstrated to hundreds of hackers~\cite{diekmann32c3firewall,diekmann1curry,diekmann2curry,diekmann3curry}.

  \section{Software-Defined Networking}
  \label{sec:contextsdn}
  %
  %
  %
  %
  Software-Defined Networking (SDN) is a hot topic in network management and operations since almost ten years (cf.\ Section~\ref{sec:intro:problemclassification-and-sdn}). 
  For its driving position in research, we discuss the results of this thesis in the context of SDN in more detail. 
  
  SDN is often considered to be equivalent to the technology OpenFlow~\cite{ofspec10,ofspec15}. 
  But SDN is more than just OpenFlow.  
  In the beginning, OpenFlow was just an ``an initial attempt''~\cite[\S\hairspace{}1]{mckeown2008openflow}, constraint by the hardware ``vendors' need for closed platforms''~\cite[\S\hairspace{}1]{mckeown2008openflow}. 
  Yet, OpenFlow is an available technology and it is widely used in research~\cite{anderson2014netKATsemantics,icfp2015smolkanetkatcompiler} and practice~\cite{openflowatgoogle2012,telekom2013kuvs,b42013googlesdn,google2016sre}. 
  In the next two subsections, we classify the contributions of this thesis with regard to the two point of views: ``SDN-is-OpenFlow'' and ``SDN as paradigm''.  
  
  \subsection{Contributions to the OpenFlow-Centric SDN Point of View}
  \paragraph*{Creating New OpenFlow Rules}
  In Part~\ref{part:greenfield} of this thesis, we show how to generate OpenFlow rules, given only a high-level specification of security requirements. 
  For the first time, the complete, automated, and verified translation from high-level security goals to an annotated graph that could be translated to OpenFlow rules was shown. 
  
  The formal guarantees only cover the transition from the security goals to the graph structure. 
  The final syntactic rewriting step to OpenFlow is not formally verified. 
  The advantage of this generic approach is that it allows to replace OpenFlow by different backend technologies. 
  For example, we also generated (and verified) iptables firewall rules. 
  
  To make our demonstrator easily accessible, reproducible, and to focus on the core contributions of our research, we chose to set up our demonstrator with only one central OpenFlow-enabled switch.      
  However, there is no technical limitation in our approach that requires this specific setup. 
  It would also be possible to serialize the output of \topos{} to a SDN programing language~\cite{icfp2015smolkanetkatcompiler,soule2014merlin,soule2014merlinsconferenceversion}, which then compiles to a distributed set of OpenFlow-enabled switches. 
  We discuss related work in Section~\ref{sec:sdnnfv:related} and illustrate how \topos{} could interact with related work in Figure~\ref{fig:relatedworkabstractionlayers}. 
  
  %

  \paragraph*{Verifying OpenFlow Rules}
  In Section~\ref{sec:outlook:verifyopenflow}, we discuss how our insights and library for firewall ruleset analysis can also contribute to the analysis of OpenFlow rules.
  
  \paragraph*{Translating Existing Technologies to OpenFlow}
  
  Based on the results of this thesis, the first fully machine-verified translation of iptables to OpenFlow rules has been demonstrated~\cite{LOFT-AFP}.

  \subsection{Contributions to SDN as Abstract Paradigm}
  The core paradigm of SDN is the separation of the control plane from the data plane and centralization of control~\cite{keown2013sdnsummerschoool}. 
  %
  In this thesis, we put the separation of control plane from data plane to the extreme by only considering static configurations.

  
  \paragraph*{Creating New SDNs}
  In Part~\ref{part:greenfield}, we propose methods to design new networks. 
  Our graph model corresponds to the SDN global network view \cite{sck11futurenetworkspastprotocols}.
  Therefore, our methodology is well-suited to implement an access control module in a network operating system~\cite{gude2008nox,reich2013modular}.


  Since our methods allow the static, offline verification of a policy, it can guarantee within the limits of the model that there are no runtime security problems. 
  Our control plane (\ie policy) can be completely separated from the data plane components (\eg OpenFlow rules, iptables rules).

  We demonstrated how \topos{} can generate iptables rules for a central OpenVPN controller. 
  This architecture of a star topology enforced by a central VPN sever with access policies implemented in software is also a SDN by definition. 
  This generalized view was well appreciated by the workshop participants~\cite{diekmann2015topos} and demonstrates the general applicability of our method. 
  The complete Part~\ref{part:greenfield} of this thesis explores network policy management independently of the underlying data plane. 

  %
  %
  %
  %

  \chapter{Conclusion}
  \label{chap:conclusion}
  \begin{quote}
  	\textit{Beware of bugs in the above code; I have only proved it correct, not tried it.}
  \end{quote}
  \hfill D.\ Knuth, Letter to Emde Boas (1977)~\cite{knuth77onlyproved}.

  \vspace*{10ex}

  \noindent
  In this thesis, we explored the application of formal methods to make our networks more secure. 
  In order to ease the daily life of network administrators and to prevent configuration errors, we presented two fully automated tools. 
  While a user of our tools needs not to worry about any formalism, the output of our tools provides a very high level of confidence due to the formal verification. 
  While we found that many theory-heavy works are not applicable to real-word data, attributable to the feature creep of most practical configuration languages, we have demonstrated large-scale real-world applicability of our tools and theory. 

  During this thesis, we roamed between two communities, hopefully narrowing the gaps between the two. 
  In the Isabelle community, we started and actively populated the networks section in the archive of formal proofs. 
  In the networks community, we demonstrated the application of rigorous formal methods to real-world problems and data. 
  Our formal approach allows to reduce complex problems to a simple structure, allowing to translate between different levels of abstraction. 
  We provide not only academic prototypical tools, but a large library of ready-to-use formally verified components for simplifying network access control. 
  

  
  \cleardoublepage
  \pagestyle{plain}
  \chapter*{Acknowledgements}
\phantomsection
\pdfbookmark[-1]{Acknowledgements}{Acknowledgements}

This thesis would not have been possible without the support of many people. 
Foremost, I want to thank all people who put great effort into this research over many years, constantly supported me, co-authored, and provided many important contributions. 
Thank you Lars Hupel, Julius Michaelis, Andreas Korsten, and Lukas Schwaighofer. 

During this research, many papers have been published. 
Now, I would like to thank all of my co-authors: Thank you Marcel von Maltitz, Stephan-A.\ Posselt, Heiko Niedermayer, Oliver Hanka, Holger Kinkelin, and Michael Schlatt. 
In addition, I want to thank all of the people who have not been listed before but who commented on the early drafts of our papers and research and gave valuable feedback: Thank you Lars Noschinski, Manuel Eberl, Lothar Braun, Fabian Immler, Benjamin Hof, and Jasmin Blanchette. 
We thank all the (anonymous) administrators who donated their firewall configs to our research. 

I would like express my gratitude to Prof.\ Dr.-Ing.\ Georg Carle for supervising this thesis, and for making this research possible altogether.

Finally, I am also very grateful to my parents Annette and Hermann Diekmann who have supported me and always believed in me.

  \cleardoublepage

  \pagestyle{fancy}

  \mathchardef\UrlBreakPenalty=10
  \mathchardef\UrlBigBreakPenalty=10

\phantomsection

\addcontentsline{toc}{part}{\bibname}


\bibliographystyle{IEEEtran}
\bibliography{literature}

\begin{thebibliography}{100}
\providecommand{\url}[1]{#1}
\csname url@samestyle\endcsname
\providecommand{\newblock}{\relax}
\providecommand{\bibinfo}[2]{#2}
\providecommand{\BIBentrySTDinterwordspacing}{\spaceskip=0pt\relax}
\providecommand{\BIBentryALTinterwordstretchfactor}{4}
\providecommand{\BIBentryALTinterwordspacing}{\spaceskip=\fontdimen2\font plus
\BIBentryALTinterwordstretchfactor\fontdimen3\font minus
  \fontdimen4\font\relax}
\providecommand{\BIBforeignlanguage}[2]{{%
\expandafter\ifx\csname l@#1\endcsname\relax
\typeout{** WARNING: IEEEtran.bst: No hyphenation pattern has been}%
\typeout{** loaded for the language `#1'. Using the pattern for}%
\typeout{** the default language instead.}%
\else
\language=\csname l@#1\endcsname
\fi
#2}}
\providecommand{\BIBdecl}{\relax}
\BIBdecl

\bibitem{EWD:EWD896}
\BIBentryALTinterwordspacing
E.~W. Dijkstra, ``{On the nature of Computing Science},'' Aug. 1984, {EWD896}.
  [Online]. Available:
  \url{http://www.cs.utexas.edu/users/EWD/ewd08xx/EWD896.PDF}
\BIBentrySTDinterwordspacing

\bibitem{burges2004sysadminbook}
M.~Burgess, \emph{Principles of Network and System Administration}, 2nd,
  Ed.\hskip 1em plus 0.5em minus 0.4em\relax Wiley, Feb. 2004, {ISBN}
  978-0-470-86807-2.

\bibitem{bellovin2009configuration}
S.~M. Bellovin and R.~Bush, ``{Configuration Management and Security},''
  \emph{IEEE Journal on Selected Areas in Communications}, vol.~27, no.~3, pp.
  268--274, Apr. 2009.

\bibitem{sck11futurenetworkspastprotocols}
S.~Shenker, M.~Casado, T.~Koponen, and N.~McKeown, ``{The Future of Networking,
  and the Past of Protocols},'' Open Networking Summit, Oct. 2011.

\bibitem{mck2012sdntame}
N.~McKeown, ``How {SDNs} will tame networks,'' Hot Interconnects, Aug. 2012,
  keynote.

\bibitem{guttman05rigorous}
J.~D. Guttman and A.~L. Herzog, ``{Rigorous Automated Network Security
  Management},'' \emph{International Journal of Information Security}, vol.~4,
  no.~1, pp. 29--48, Feb. 2005.

\bibitem{firwallerr2004}
A.~Wool, ``{A Quantitative Study of Firewall Configuration Errors},''
  \emph{IEEE Computer}, vol.~37, no.~6, pp. 62--67, Jun. 2004.

\bibitem{wool2010firewall}
A.~Wool, ``{Trends in Firewall Configuration Errors: Measuring the Holes in
  Swiss Cheese},'' \emph{IEEE Internet Computing}, vol.~14, no.~4, pp. 58--65,
  Jul. 2010.

\bibitem{netsecconflicts}
H.~Hamed and E.~Al-Shaer, ``{Taxonomy of Conflicts in Network Security
  Policies},'' \emph{IEEE Communications Magazine}, vol.~44, no.~3, pp.
  134--141, Mar. 2006.

\bibitem{sherry2012making}
J.~Sherry, S.~Hasan, C.~Scott, A.~Krishnamurthy, S.~Ratnasamy, and V.~Sekar,
  ``{Making Middleboxes Someone Else’s Problem: Network Processing as a Cloud
  Service},'' \emph{ACM SIGCOMM Computer Communication Review}, vol.~42, no.~4,
  pp. 13--24, Oct. 2012.

\bibitem{diekmann2014forte}
C.~Diekmann, S.-A. Posselt, H.~Niedermayer, H.~Kinkelin, O.~Hanka, and
  G.~Carle, ``{Verifying Security Policies using Host Attributes},'' in
  \emph{Formal Techniques for Distributed Objects, Components, and Systems:
  34th IFIP WG 6.1 International Conference (FORTE)}.\hskip 1em plus 0.5em
  minus 0.4em\relax Berlin, Germany: Springer Berlin Heidelberg, Jun. 2014, pp.
  133--148.

\bibitem{fwviz2012}
F.~Mansmann, T.~G\"{o}bel, and W.~Cheswick, ``{Visual Analysis of Complex
  Firewall Configurations},'' in \emph{9th International Symposium on
  Visualization for Cyber Security}, ser. VizSec '12.\hskip 1em plus 0.5em
  minus 0.4em\relax ACM, Oct. 2012, pp. 1--8.

\bibitem{fireman2006}
L.~Yuan, H.~Chen, J.~Mai, C.-N. Chuah, Z.~Su, and P.~Mohapatra, ``{FIREMAN: A
  Toolkit for FIREwall Modeling and ANalysis},'' in \emph{IEEE Symposium on
  Security and Privacy}, May 2006, pp. 199--213.

\bibitem{ZhangAlShaer2007flip}
B.~Zhang, E.~Al-Shaer, R.~Jagadeesan, J.~Riely, and C.~Pitcher,
  ``{Specifications of a High-level Conflict-free Firewall Policy Language for
  Multi-domain Networks},'' in \emph{12th ACM symposium on Access control
  models and technologies}, ser. SACMAT'07.\hskip 1em plus 0.5em minus
  0.4em\relax ACM, Jun. 2007, pp. 185--194.

\bibitem{databreach2009src}
\BIBentryALTinterwordspacing
{Verizon Business RISK team} and {United States Secret Service}, ``{2010 Data
  Breach Investigations Report},'' 2010. [Online]. Available:
  \url{http://www.verizonenterprise.com/resources/reports/rp_2010-DBIR-combined-reports_en_xg.pdf}
\BIBentrySTDinterwordspacing

\bibitem{nelson2010margrave}
T.~Nelson, C.~Barratt, D.~J. Dougherty, K.~Fisler, and S.~Krishnamurthi, ``{The
  Margrave Tool for Firewall Analysis},'' in \emph{24th USENIX Large
  Installation System Administration Conference (LISA)}.\hskip 1em plus 0.5em
  minus 0.4em\relax San Jose, CA: USENIX Association, Nov. 2010.

\bibitem{survery2012networktroubleshooting}
H.~Zeng, P.~Kazemian, G.~Varghese, and N.~McKeown, ``{A Survey on Network
  Troubleshooting},'' Technical Report Stanford/TR12-HPNG-061012, Jun. 2012.

\bibitem{bsigrundschutz}
\BIBentryALTinterwordspacing
\emph{{IT-Grundschutz-Kataloge}}, {BSI} Std., Rev. 15. EL Stand 2016. [Online].
  Available:
  \url{https://www.bsi.bund.de/DE/Themen/ITGrundschutz/ITGrundschutzKataloge/itgrundschutzkataloge_node.html}
\BIBentrySTDinterwordspacing

\bibitem{networkdowntime2009}
\BIBentryALTinterwordspacing
{Netcordia, Inc.}, ``{Network Downtime, the Configuration Errors},''
  whitepaper, 2009, currently unavailable; archived mirror. [Online].
  Available:
  \url{https://web.archive.org/web/20160717225153/http://www.pmi.it/file/whitepaper/000140.pdf}
\BIBentrySTDinterwordspacing

\bibitem{oppenheimer2003internet}
D.~Oppenheimer, A.~Ganapathi, and D.~A. Patterson, ``{Why do Internet services
  fail, and what can be done about it?}'' in \emph{4th USENIX Symposium on
  Internet Technologies and Systems}, ser. USITS'03, vol.~67, Seattle,
  WA.\hskip 1em plus 0.5em minus 0.4em\relax USENIX Association, Mar. 2003.

\bibitem{Cuppens2005orbacxmlfirewall}
F.~Cuppens, N.~Cuppens-Boulahia, T.~Sans, and A.~Mi{\`e}ge, \emph{{Formal
  Aspects in Security and Trust: IFIP TC1 WG1.7 Workshop on Formal Aspects in
  Security and Trust (FAST), World Computer Congress}}.\hskip 1em plus 0.5em
  minus 0.4em\relax Boston, MA: Springer US, 2005, ch. A Formal Approach to
  Specify and Deploy a Network Security Policy, pp. 203--218.

\bibitem{pozo2009model}
S.~Pozo, R.~Ceballos, and R.~M. Gasca, ``{Model-Based Development of firewall
  rule sets: Diagnosing model inconsistencies},'' \emph{Information and
  Software Technology}, vol.~51, no.~5, pp. 894--915, May 2009.

\bibitem{iptables}
\BIBentryALTinterwordspacing
{The netfilter.org project}, ``netfilter/iptables project.'' [Online].
  Available: \url{http://www.netfilter.org/}
\BIBentrySTDinterwordspacing

\bibitem{maniptablesextensions}
``{man (8) iptables-extensions},'' version 1.4.21.

\bibitem{diekmanngithubnetnetwork}
\BIBentryALTinterwordspacing
``Analyzed firewall rulesets (raw data),'' accompanying material. [Online].
  Available: \url{https://github.com/diekmann/net-network}
\BIBentrySTDinterwordspacing

\bibitem{cloudflare2014blogbpf}
\BIBentryALTinterwordspacing
M.~Majkowski, ``{BPF} -- the forgotten bytecode,'' {CloudFlare} blog, May 2014,
  retrieved May 2016. [Online]. Available:
  \url{https://blog.cloudflare.com/bpf-the-forgotten-bytecode/}
\BIBentrySTDinterwordspacing

\bibitem{serverfaultiptables}
\BIBentryALTinterwordspacing
``Over 4000 questions tagged with `iptables','' serverfault, Jun. 2016.
  [Online]. Available: \url{http://serverfault.com/questions/tagged/iptables}
\BIBentrySTDinterwordspacing

\bibitem{netcore12}
C.~Monsanto, N.~Foster, R.~Harrison, and D.~Walker, ``{A Compiler and Run-time
  System for Network Programming Languages},'' in \emph{39th annual ACM
  SIGPLAN-SIGACT symposium on Principles of programming languages}, ser. POPL
  '12.\hskip 1em plus 0.5em minus 0.4em\relax ACM, Jan. 2012, pp. 217--230.

\bibitem{nelson2014flowlog}
T.~Nelson, A.~D. Ferguson, M.~J. Scheer, and S.~Krishnamurthi, ``{Tierless
  Programming and Reasoning for Software-Defined Networks},'' in \emph{11th
  USENIX Symposium on Networked Systems Design and Implementation (NSDI)}, ser.
  NSDI'14.\hskip 1em plus 0.5em minus 0.4em\relax Seattle, WA: USENIX
  Association, Apr. 2014, pp. 519--531.

\bibitem{zhao2011policyremanet}
H.~Zhao, J.~Lobo, A.~Roy, and S.~M. Bellovin, ``{Policy Refinement of Network
  Services for MANETs},'' in \emph{12th IFIP/IEEE International Symposium on
  Integrated Network Management (IM 2011)}, Dublin, Ireland, May 2011.

\bibitem{anderson2014netKATsemantics}
C.~J. Anderson, N.~Foster, A.~Guha, J.-B. Jeannin, D.~Kozen, C.~Schlesinger,
  and D.~Walker, ``{NetKAT: Semantic Foundations for Networks},'' in \emph{41st
  ACM SIGPLAN-SIGACT Symposium on Principles of Programming Languages}, ser.
  POPL '14.\hskip 1em plus 0.5em minus 0.4em\relax San Diego, California: ACM,
  Jan. 2014, pp. 113--126.

\bibitem{mignis2014}
P.~Ad{\~{a}}o, C.~Bozzato, G.~{Dei Rossi}, R.~Focardi, and F.~L. Luccio,
  ``{Mignis: A Semantic Based Tool for Firewall Configuration},'' in \emph{27th
  Computer Security Foundations Symposium}, ser. CSF.\hskip 1em plus 0.5em
  minus 0.4em\relax IEEE, Jul. 2014, pp. 351--365.

\bibitem{bartal1999firmato}
Y.~Bartal, A.~Mayer, K.~Nissim, and A.~Wool, ``{Firmato: A Novel Firewall
  Management Toolkit},'' in \emph{Symposium on Security and Privacy}.\hskip 1em
  plus 0.5em minus 0.4em\relax IEEE, May 1999, pp. 17--31.

\bibitem{hinrichs2009practical}
T.~L. Hinrichs, N.~S. Gude, M.~Casado, J.~C. Mitchell, and S.~Shenker,
  ``{Practical Declarative Network Management},'' in \emph{1st ACM workshop on
  Research on enterprise networking}, ser. WREN'09.\hskip 1em plus 0.5em minus
  0.4em\relax ACM, Aug. 2009, pp. 1--10.

\bibitem{soule2014merlin}
\BIBentryALTinterwordspacing
R.~Soul{\'{e}}, S.~Basu, P.~J. Marandi, F.~Pedone, R.~Kleinberg, E.~G. Sirer,
  and N.~Foster, ``{Merlin: A Language for Provisioning Network Resources},''
  \emph{CoRR}, vol. abs/1407.1199, 2014. [Online]. Available:
  \url{http://arxiv.org/abs/1407.1199}
\BIBentrySTDinterwordspacing

\bibitem{policy2010berapolicyformalenterprise}
P.~Bera, S.~Ghosh, and P.~Dasgupta, ``{Policy Based Security Analysis in
  Enterprise Networks: A Formal Approach},'' \emph{IEEE Transactions on Network
  and Service Management}, vol.~7, no.~4, pp. 231--243, Dec. 2010.

\bibitem{isabelle2016}
\BIBentryALTinterwordspacing
T.~Nipkow, L.~C. Paulson, and M.~Wenzel, \emph{Isabelle/HOL: A Proof Assistant
  for Higher-Order Logic}, ser. LNCS.\hskip 1em plus 0.5em minus 0.4em\relax
  Springer, 2002, last updated 2016, vol. 2283. [Online]. Available:
  \url{http://isabelle.in.tum.de/}
\BIBentrySTDinterwordspacing

\bibitem{Network_Security_Policy_Verification-AFP}
\BIBentryALTinterwordspacing
C.~Diekmann, ``{Network Security Policy Verification},'' \emph{Archive of
  Formal Proofs}, Jul. 2016, formal proof development. [Online]. Available:
  \url{https://www.isa-afp.org/entries/Network_Security_Policy_Verification.shtml}
\BIBentrySTDinterwordspacing

\bibitem{IP_Addresses-AFP}
\BIBentryALTinterwordspacing
C.~Diekmann, J.~Michaelis, and L.~Hupel, ``{IP Addresses},'' \emph{Archive of
  Formal Proofs}, Jun. 2016, formal proof development. [Online]. Available:
  \url{http://isa-afp.org/entries/IP_Addresses.shtml}
\BIBentrySTDinterwordspacing

\bibitem{Simple_Firewall-AFP}
\BIBentryALTinterwordspacing
C.~Diekmann, J.~Michaelis, and M.~Haslbeck, ``{Simple Firewall},''
  \emph{Archive of Formal Proofs}, Aug. 2016, formal proof development.
  [Online]. Available: \url{http://isa-afp.org/entries/Simple_Firewall.shtml}
\BIBentrySTDinterwordspacing

\bibitem{Iptables_Semantics-AFP}
\BIBentryALTinterwordspacing
C.~Diekmann and L.~Hupel, ``{Iptables Semantics},'' \emph{Archive of Formal
  Proofs}, Sep. 2016, formal proof development. [Online]. Available:
  \url{http://isa-afp.org/entries/Iptables_Semantics.shtml}
\BIBentrySTDinterwordspacing

\bibitem{Routing-AFP}
\BIBentryALTinterwordspacing
J.~Michaelis and C.~Diekmann, ``{Routing},'' \emph{Archive of Formal Proofs},
  Aug. 2016, formal proof development. [Online]. Available:
  \url{http://isa-afp.org/entries/Routing.shtml}
\BIBentrySTDinterwordspacing

\bibitem{LOFT-AFP}
\BIBentryALTinterwordspacing
J.~Michaelis and C.~Diekmann, ``{LOFT -- Verified Migration of Linux Firewalls
  to SDN},'' \emph{Archive of Formal Proofs}, Oct. 2016, formal proof
  development. [Online]. Available: \url{http://isa-afp.org/entries/LOFT.shtml}
\BIBentrySTDinterwordspacing

\bibitem{diekmann2015fm}
C.~Diekmann, L.~Hupel, and G.~Carle, ``{Semantics-Preserving Simplification of
  Real-World Firewall Rule Sets},'' in \emph{Formal Methods}, Jun. 2015.

\bibitem{diekmann2016networking}
C.~Diekmann, J.~Michaelis, M.~Haslbeck, and G.~Carle, ``{Verified iptables
  Firewall Analysis},'' in \emph{IFIP Networking 2016}, Vienna, Austria, May
  2016.

\bibitem{iptablesafpmaillarry}
\BIBentryALTinterwordspacing
L.~Paulson, ``{new in the AFP: Iptables\_Semantics},'' isabelle users mailing
  list, Sep. 2016. [Online]. Available:
  \url{https://lists.cam.ac.uk/pipermail/cl-isabelle-users/2016-September/msg00023.html}
\BIBentrySTDinterwordspacing

\bibitem{iptablesafpmailgerwin}
\BIBentryALTinterwordspacing
G.~Klein, ``{AFP: LOFT},'' isabelle users mailing list, Oct. 2016. [Online].
  Available:
  \url{https://lists.cam.ac.uk/pipermail/cl-isabelle-users/2016-October/msg00055.html}
\BIBentrySTDinterwordspacing

\bibitem{diekmann32c3firewall}
\BIBentryALTinterwordspacing
C.~Diekmann, ``{Verified Firewall Ruleset Verification -- Math, Functional
  Programming, Theorem Proving, and an Introduction to Isabelle/HOL},'' {32.
  Chaos Communication Congress (32C3)}, Dec. 2015. [Online]. Available:
  \url{https://media.ccc.de/v/32c3-7195-verified_firewall_ruleset_verification}
\BIBentrySTDinterwordspacing

\bibitem{diekmann1curry}
\BIBentryALTinterwordspacing
C.~Diekmann, ``{Verified Firewall Ruleset Verification 1/2},'' {elfte Treffen
  des Curry Clubs Augsburg}, Feb. 2016. [Online]. Available:
  \url{http://curry-club-augsburg.de/posts/2016-01-05-elftes-treffen.html}
\BIBentrySTDinterwordspacing

\bibitem{diekmann2curry}
\BIBentryALTinterwordspacing
C.~Diekmann, ``{Verified Firewall Ruleset Verification 2/2},'' {zw{\"o}lfte
  Treffen des Curry Clubs Augsburg}, Mar. 2016. [Online]. Available:
  \url{http://curry-club-augsburg.de/posts/2016-02-26-zwoelftes-treffen.html}
\BIBentrySTDinterwordspacing

\bibitem{diekmann3curry}
\BIBentryALTinterwordspacing
C.~Diekmann, ``{Make Firewalls Verified Again},'' {achtzehnte Treffen des Curry
  Clubs Augsburg}, Sep. 2016. [Online]. Available:
  \url{http://curry-club-augsburg.de/posts/2016-06-29-achtzehntes-treffen.html}
\BIBentrySTDinterwordspacing

\bibitem{bishop2003compsec}
M.~Bishop, ``{What Is Computer Security?}'' \emph{IEEE Security \& Privacy},
  vol.~1, no.~1, pp. 67--69, Feb. 2003.

\bibitem{cornythesis}
C.~Diekmann, ``{Security Requirement Modeling as Configuration Management for
  Scenario-Specific Networks},'' Master's thesis, Technische Universit{\"a}t
  M{\"u}nchen, May 2013.

\bibitem{debadminhandbook2015}
R.~Hertzog and R.~Mas, \emph{{The Debian Administrator's Handbook} -- {Debian
  Jessie from Discovery to Mastery}}, 1st~ed.\hskip 1em plus 0.5em minus
  0.4em\relax \url{debian-handbook.info}, Oct. 2015, {ISBN} 979-10-91414-05-0.

\bibitem{infrastructure2007report}
{Arbor Networks}, ``Worldwide infrastructure security report,'' Sep. 2007,
  archived at
  \url{http://web.archive.org/web/20160728134338/http://www.lateledipenelope.it/public/WISP_US_12sept07.pdf}.

\bibitem{infrastructure2016report}
{Arbor Networks}, ``Worldwide infrastructure security report,'' 2016, archived
  at
  \url{https://web.archive.org/web/20160417124116/https://www.arbornetworks.com/images/documents/WISR2016_EN_Web.pdf}.

\bibitem{imc2013demystifying}
R.~Potharaju and N.~Jain, ``{Demystifying the Dark Side of the Middle: A Field
  Study of Middlebox Failures in Datacenters},'' in \emph{13th SIGCOMM
  Conference on Internet Measurement (IMC)}.\hskip 1em plus 0.5em minus
  0.4em\relax ACM, Oct. 2013.

\bibitem{ethane07}
M.~Casado, M.~J. Freedman, J.~Pettit, J.~Luo, N.~McKeown, and S.~Shenker,
  ``{Ethane: Taking Control of the Enterprise},'' in \emph{Conference on
  Applications, technologies, architectures, and protocols for computer
  communications}, ser. SIGCOMM '07.\hskip 1em plus 0.5em minus 0.4em\relax
  Kyoto, Japan: ACM, Aug. 2007, pp. 1--12.

\bibitem{autosecpolicymgnt01}
J.~Burns, A.~Cheng, P.~Gurung, S.~Rajagopalan, P.~Rao, D.~Rosenbluth, A.~V.
  Surendran, and J.~Martin, D.M., ``{Automatic Management of Network Security
  Policy},'' in \emph{DARPA Information Survivability Conference \& Exposition
  II. DISCEX}, vol.~2, Aug. 2001, pp. 12--26.

\bibitem{arxivnsfworkshopformalprivacy2016}
\BIBentryALTinterwordspacing
S.~Chong, J.~Guttman, A.~Datta, A.~Myers, B.~Pierce, P.~Schaumont, T.~Sherwood,
  and N.~Zeldovich, ``{Report on the NSF Workshop on Formal Methods for
  Security},'' Aug. 2016. [Online]. Available:
  \url{https://arxiv.org/abs/1608.00678v1}
\BIBentrySTDinterwordspacing

\bibitem{bjorner_et_al:DR:2015:5044}
\BIBentryALTinterwordspacing
N.~Bj{\o}rner, N.~Foster, P.~B. Godfrey, and P.~Zave, ``{Formal Foundations for
  Networking (Dagstuhl Seminar 15071)},'' \emph{Dagstuhl Reports}, vol.~5,
  no.~2, pp. 44--63, 2015. [Online]. Available:
  \url{http://drops.dagstuhl.de/opus/volltexte/2015/5044}
\BIBentrySTDinterwordspacing

\bibitem{databreach2009}
\BIBentryALTinterwordspacing
J.~Kirk, ``{Verizon: Data breaches often caused by configuration errors},''
  networkworld, Jul. 2010, retrieved Aug 2016. [Online]. Available:
  \url{http://www.computerworld.com/article/2519657/network-security/verizon--data-breaches-often-caused-by-configuration-errors.html}
\BIBentrySTDinterwordspacing

\bibitem{verizon2016databreach}
\BIBentryALTinterwordspacing
{Verizon and Contributing Organizations}, ``{2016 Data Breach Investigations
  Report},'' 2016, retrieved Aug 2016. [Online]. Available:
  \url{http://www.verizonenterprise.com/verizon-insights-lab/dbir/}
\BIBentrySTDinterwordspacing

\bibitem{faultyaccess2016online}
\BIBentryALTinterwordspacing
A.~Peters, ``{Faulty Access Controls Led to Morgan Stanley Data Breach: FTC},''
  americanbanker.com, Aug. 2015, retrieved Aug 2016. [Online]. Available:
  \url{http://www.americanbanker.com/news/bank-technology/faulty-access-controls-led-to-morgan-stanley-data-breach-ftc-1076098-1.html}
\BIBentrySTDinterwordspacing

\bibitem{telekom2013kuvs}
\BIBentryALTinterwordspacing
A.~Gladisch and F.-J. Westphal, ``{Carrier Landscape for SDN -- Next Level of
  Telco Industrialization?}'' 44. Treffen der VDE/ITG-Fachgruppe 5.2.4, Mobile
  Network (Function) Virtualization and Software Defined Networking, Nov. 2013,
  deutsche Telekom Innovation Labs. [Online]. Available:
  \url{http://www.ikr.uni-stuttgart.de/Content/itg/fg524/Meetings/2013-11-15-Muenchen/08_ITG524_Muenchen_Westphal.pdf}
\BIBentrySTDinterwordspacing

\bibitem{EPFL-REPORT-187131}
P.~Peresini and D.~Kostic, ``{Is the Network Turing-Complete?}'' EPFL, Tech.
  Rep. EPFL-REPORT-187131, Jun. 2013, after a personal communication with the
  author of this thesis in Sep.\ 2015, the report was updated.

\bibitem{casado2006sane}
M.~Casado, T.~Garfinkel, A.~Akella, M.~J. Freedman, D.~Boneh, N.~McKeown, and
  S.~Shenker, ``{SANE: A Protection Architecture for Enterprise Networks},'' in
  \emph{15th USENIX Security Symposium}, vol.~15, no.~10.\hskip 1em plus 0.5em
  minus 0.4em\relax Vancouver, B.C., Canada: USENIX Association, Jul. 2006.

\bibitem{mckeown2008openflow}
N.~McKeown, T.~Anderson, H.~Balakrishnan, G.~Parulkar, L.~Peterson, J.~Rexford,
  S.~Shenker, and J.~Turner, ``{OpenFlow: Enabling Innovation in Campus
  Networks},'' \emph{ACM SIGCOMM Computer Communication Review}, vol.~38,
  no.~2, pp. 69--74, Apr. 2008.

\bibitem{openflowatgoogle2012}
U.~Hoelzle, ``{OpenFlow @ Google},'' Open Networking Summit, Apr. 2012,
  keynote.

\bibitem{b42013googlesdn}
S.~Jain, A.~Kumar, S.~Mandal, J.~Ong, L.~Poutievski, A.~Singh, S.~Venkata,
  J.~Wanderer, J.~Zhou, M.~Zhu, J.~Zolla, U.~H\"{o}lzle, S.~Stuart, and
  A.~Vahdat, ``{B4: Experience with a Globally-deployed Software Defined
  WAN},'' in \emph{ACM SIGCOMM}.\hskip 1em plus 0.5em minus 0.4em\relax Hong
  Kong, China: ACM, Aug. 2013, pp. 3--14.

\bibitem{brooks1987no}
F.~P. Brooks, ``{No Silver Bullet -- Essence and Accidents of Software
  Engineering},'' \emph{IEEE computer}, vol.~20, no.~4, pp. 10--19, Apr. 1987.

\bibitem{safari2007softwareeng}
G.~Booch, R.~A. Maksimchuk, M.~W. Engle, B.~J. Young, J.~Conallen, and K.~A.
  Houston, \emph{Object-Oriented Analysis and Design with Applications}.\hskip
  1em plus 0.5em minus 0.4em\relax Addison-Wesley Professional, Apr. 2007, no.
  Third Edition.

\bibitem{techtraget2013feamstersdn}
\BIBentryALTinterwordspacing
M.~McNickle, ``{Nick Feamster}: {OpenFlow} {SDN} for access control is just the
  beginning,'' techtarget.com, Nov. 2013. [Online]. Available:
  \url{http://searchsdn.techtarget.com/news/2240208949/Nick-Feamster-OpenFlow-SDN-for-access-control-is-just-the-beginning}
\BIBentrySTDinterwordspacing

\bibitem{benson2009unraveling}
T.~Benson, A.~Akella, and D.~Maltz, ``{Unraveling the Complexity of Network
  Management},'' in \emph{6th USENIX Symposium on Networked Systems Design and
  Implementation (NSDI)}, ser. NSDI'09.\hskip 1em plus 0.5em minus 0.4em\relax
  Berkeley, CA, USA: USENIX Association, Apr. 2009, pp. 335--348.

\bibitem{cc2012p3}
\BIBentryALTinterwordspacing
{Common Criteria}, ``{Part 3: Security assurance components},'' \emph{{Common
  Criteria for Information Technology Security Evaluation}}, vol.
  CCMB-2012-09-003, no. Version 3.1 Revision 4, Sep. 2012. [Online]. Available:
  \url{http://www.commoncriteriaportal.org/files/ccfiles/CCPART3V3.1R4.pdf}
\BIBentrySTDinterwordspacing

\bibitem{afpall}
\BIBentryALTinterwordspacing
G.~Klein, T.~Nipkow, and L.~Paulson, Eds., \emph{{Archive of Formal Proofs}},
  2016. [Online]. Available: \url{https://www.isa-afp.org/}
\BIBentrySTDinterwordspacing

\bibitem{abac2005}
E.~Yuan and J.~Tong, ``{Attributed Based Access Control (ABAC) for Web
  Services},'' in \emph{IEEE International Conference on Web Services}, ser.
  ICWS, Jul. 2005.

\bibitem{samarati2001accessdickerbroken}
P.~Samarati and S.~C. de~Vimercati, \emph{{Foundations of Security Analysis and
  Design: Tutorial Lectures}}.\hskip 1em plus 0.5em minus 0.4em\relax Berlin,
  Heidelberg: Springer Berlin Heidelberg, 2001, ch. Access Control: Policies,
  Models, and Mechanisms, pp. 137--196.

\bibitem{jajodia2001flexibledatabaseaccesscontrol}
S.~Jajodia, P.~Samarati, M.~L. Sapino, and V.~S. Subrahmanian, ``{Flexible
  Support for Multiple Access Control Policies},'' \emph{ACM Transactions on
  Database Systems (TODS)}, vol.~26, no.~2, pp. 214--260, Jun. 2001.

\bibitem{mcilroy1992multilevel}
M.~D. McIlroy and J.~A. Reeds, ``{Multilevel Security in the UNIX Tradition},''
  \emph{Software: Practice and Experience}, vol.~22, no.~8, pp. 673--694, Aug.
  1992.

\bibitem{sel42013IFS}
T.~Murray, D.~Matichuk, M.~Brassil, P.~Gammie, T.~Bourke, S.~Seefried,
  C.~Lewis, X.~Gao, and G.~Klein, ``{seL4: From General Purpose to a Proof of
  Information Flow Enforcement},'' in \emph{IEEE Symposium on Security and
  Privacy}, May 2013, pp. 415--429.

\bibitem{denning1976lattice}
D.~E. Denning, ``{A Lattice Model of Secure Information Flow},''
  \emph{Communications of the ACM}, vol.~19, no.~5, pp. 236--243, May 1976.

\bibitem{bell1973secure1}
D.~E. Bell and L.~J. LaPadula, ``{Secure Computer Systems: Mathematical
  Foundations},'' MITRE Corporation, Bedford, MA, MTR-2547 Vol. I, Mar. 1973.

\bibitem{bell1973secure2}
D.~E. Bell and L.~J. LaPadula, ``{Secure Computer Systems: A Mathematical
  Model},'' MITRE Corporation, Bedford, MA, MTR-2547 Vol. II, May 1973.

\bibitem{bell1973securerefinedmodel}
D.~E. Bell, ``{Secure Computer Systems: A Refinement of the Mathematical
  Model},'' MITRE Corporation, Bedford, MA, MTR-2547 Vol. III, Dec. 1973.

\bibitem{bell1975padula3}
D.~E. Bell and L.~J. LaPadula, ``{Secure computer systems: Unified Exposition
  and MULTICS Interpretation},'' MITRE Corporation, Bedford, MA, {MTR}-2997,
  Jul. 1975.

\bibitem{blphistory}
D.~E. Bell, ``{Looking Back at the {Bell-La Padula} Model},'' in \emph{21st
  Annual Computer Security Applications Conference}, Dec. 2005, pp. 337--351.

\bibitem{bishop2003computer}
M.~Bishop, \emph{Computer Security: Art and Science}.\hskip 1em plus 0.5em
  minus 0.4em\relax Addison-Wesley, 2003, {ISBN} 0-201-44099-7.

\bibitem{eckert2013}
C.~Eckert, \emph{IT-Sicherheit: Konzepte-Verfahren-Protokolle}, 8th~ed.\hskip
  1em plus 0.5em minus 0.4em\relax Oldenbourg Verlag, 2013, {ISBN} 3486721380.

\bibitem{bellovin2004lookbackseciritytcp}
S.~M. Bellovin, ``{A Look Back at ``Security Problems in the TCP/IP Protocol
  Suite''},'' in \emph{Annual Computer Security Applications Conference}, Dec.
  2004, invited paper.

\bibitem{composable94}
L.~Gong and X.~Qian, ``{The Complexity and Composability of Secure
  Interoperation},'' in \emph{IEEE Computer Society Symposium on Research in
  Security and Privacy}, May 1994, pp. 190--200.

\bibitem{restrictiveness}
D.~McCullough, ``{A Hookup Theorem for Multilevel Security},'' \emph{IEEE
  Transactions on Software Engineering}, vol.~16, no.~6, pp. 563--568, Jun.
  1990.

\bibitem{maltitz2016fmpriv}
M.~{von Maltitz}, C.~{Diekmann}, and G.~{Carle}, ``{Taint Analysis for
  System-Wide Privacy Audits: A Framework and Real-World Case Studies},'' 1st
  Workshop for Formal Methods on Privacy, Nov. 2016, workshop without
  proceedings.

\bibitem{Transitive-Closure-AFP}
\BIBentryALTinterwordspacing
C.~Sternagel and R.~Thiemann, ``{Executable Transitive Closures of Finite
  Relations},'' \emph{Archive of Formal Proofs}, Mar. 2011, formal proof
  development. [Online]. Available:
  \url{https://www.isa-afp.org/entries/Transitive-Closure.shtml}
\BIBentrySTDinterwordspacing

\bibitem{rbac}
R.~S. Sandhu, E.~J. Coynek, H.~L. Feinsteink, and C.~E. Youmank, ``{Role-Based
  Access Control Models},'' \emph{IEEE Computer}, vol.~29, no.~2, pp. 38--47,
  Feb. 1996.

\bibitem{goguen1982}
J.~A. Goguen and J.~Meseguer, ``{Security Policies and Security Models},'' in
  \emph{IEEE Symposium on Security and Privacy}, Apr. 1982, pp. 11--20.

\bibitem{csl-92-2}
\BIBentryALTinterwordspacing
J.~Rushby, ``{Noninterference, Transitivity, and Channel-Control Security
  Policies},'' Tech. Rep., Dec. 1992. [Online]. Available:
  \url{http://www.csl.sri.com/papers/csl-92-2/}
\BIBentrySTDinterwordspacing

\bibitem{Broberg:2009:FSD:1554339.1554352}
N.~Broberg and D.~Sands, ``{Flow-Sensitive Semantics for Dynamic Information
  Flow Policies},'' in \emph{Fourth Workshop on Programming Languages and
  Analysis for Security}, ser. PLAS '09.\hskip 1em plus 0.5em minus 0.4em\relax
  ACM, Jun. 2009, pp. 101--112.

\bibitem{Myers:1997:DMI:269005.266669}
A.~C. Myers and B.~Liskov, ``{A Decentralized Model for Information Flow
  Control},'' \emph{SIGOPS Oper. Syst. Rev.}, vol.~31, no.~5, pp. 129--142,
  Oct. 1997.

\bibitem{rfc5426}
A.~Okmianski, ``{Transmission of Syslog Messages over UDP},'' RFC 5426
  (Proposed Standard), Internet Engineering Task Force, Mar. 2009.

\bibitem{schwartz2010all}
E.~J. Schwartz, T.~Avgerinos, and D.~Brumley, ``{All You Ever Wanted to Know
  About Dynamic Taint Analysis and Forward Symbolic Execution (but might have
  been afraid to ask)},'' in \emph{IEEE Symposium on Security and
  Privacy}.\hskip 1em plus 0.5em minus 0.4em\relax IEEE, May 2010, pp.
  317--331.

\bibitem{enck2010taindroid}
W.~Enck, P.~Gilbert, B.-G. Chun, L.~P. Cox, J.~Jung, P.~McDaniel, and A.~N.
  Sheth, ``{TaintDroid: An Information-Flow Tracking System for Realtime
  Privacy Monitoring on Smartphones},'' in \emph{Operating Systems Design and
  Implementation}, ser. OSDI'10.\hskip 1em plus 0.5em minus 0.4em\relax
  Vancouver, BC, Canada: USENIX Association, Oct. 2010, pp. 393--407.

\bibitem{enck2014taintdroid}
W.~Enck, P.~Gilbert, S.~Han, V.~Tendulkar, B.-G. Chun, L.~P. Cox, J.~Jung,
  P.~McDaniel, and A.~N. Sheth, ``{TaintDroid: An Information-Flow Tracking
  System for Realtime Privacy Monitoring on Smartphones},'' \emph{ACM
  Transactions on Computer Systems (TOCS)}, vol.~32, no.~2, pp. 5:1--5:29, Jun.
  2014.

\bibitem{droiddisintegrator2016}
\BIBentryALTinterwordspacing
E.~Tromer and R.~Schuster, ``{DroidDisintegrator: Intra-Application Information
  Flow Control in Android Apps (extended version)},'' in \emph{11th ACM Asia
  Conference on Computer and Communications Security}, ser. ASIA CCS '16.\hskip
  1em plus 0.5em minus 0.4em\relax Xi'an, China: ACM, May 2016, pp. 401--412,
  author's extended version referenced. [Online]. Available:
  \url{http://www.cs.tau.ac.il/~tromer/disintegrator/disintegrator.pdf}
\BIBentrySTDinterwordspacing

\bibitem{Datenschutzschutzziele}
M.~Rost and A.~Pfitzmann, ``{Datenschutz-Schutzziele – revisited},''
  \emph{Datenschutz und Datensicherheit DuD}, vol.~33, no.~6, pp. 353--358,
  Jul. 2009.

\bibitem{Bock2011}
K.~Bock and M.~Rost, ``{Privacy By Design und die Neuen Schutzziele},''
  \emph{DuD}, vol.~35, no.~1, pp. 30--35, Mar. 2011.

\bibitem{ENISA2014}
G.~Danezis, J.~Domingo-Ferrer, M.~Hansen, J.-H. Hoepman, D.~L. Metayer,
  R.~Tirtea, and S.~Schiffner, ``{Privacy and Data Protection by Design -- from
  policy to engineering},'' ENISA, Tech. Rep., Jan. 2015.

\bibitem{SDM}
\BIBentryALTinterwordspacing
{AK Technik der Konferenz der unabh{\"{a}}ngigen Datenschutzbeh{\"{o}}rden des
  Bundes und der L{\"{a}}nder}, ``{Das Standard-Datenschutzmodell},'' Konferenz
  der unabh{\"{a}}ngigen Datenschutzbeh{\"{o}}rden des Bundes und der
  L{\"{a}}nder, Darmstadt, Tech. Rep. V.0.9, Oct. 2015. [Online]. Available:
  \url{https://www.datenschutzzentrum.de/artikel/954-Handbuch-zum-Standard-Datenschutzmodell.html}
\BIBentrySTDinterwordspacing

\bibitem{maltitz2016arxivprivacylong}
\BIBentryALTinterwordspacing
M.~{von Maltitz}, C.~{Diekmann}, and G.~{Carle}, ``{Taint Analysis for
  System-Wide Privacy Audits: A Framework and Real-World Case Studies},''
  \emph{ArXiv e-prints}, vol. abs/1608.04671, Aug. 2016. [Online]. Available:
  \url{https://arxiv.org/abs/1608.04671}
\BIBentrySTDinterwordspacing

\bibitem{isabelle2012code}
F.~Haftmann and L.~Bulwahn, \emph{{Code generation from Isabelle/HOL
  theories}}, Feb. 2016.

\bibitem{haftmann2010code}
F.~Haftmann and T.~Nipkow, ``Code generation via higher-order rewrite
  systems,'' in \emph{Functional and Logic Programming}.\hskip 1em plus 0.5em
  minus 0.4em\relax Springer, 2010, pp. 103--117.

\bibitem{paulson96ML}
L.~C. Paulson, \emph{{ML for the Working Programmer}}, 2nd~ed.\hskip 1em plus
  0.5em minus 0.4em\relax Cambridge University Press, Jun. 1996, {ISBN}
  978-0521565431.

\bibitem{graphviz}
\BIBentryALTinterwordspacing
E.~R. Gansner, E.~Koutsofios, and S.~C. North, ``{Drawing Graphs with
  \emph{dot}},'' Jan. 2015. [Online]. Available:
  \url{http://www.graphviz.org/pdf/dotguide.pdf}
\BIBentrySTDinterwordspacing

\bibitem{ou2005mulval}
X.~Ou, S.~Govindavajhala, and A.~W. Appel, ``{MulVAL: A Logic-based Network
  Security Analyzer},'' in \emph{14th USENIX Security Symposium}.\hskip 1em
  plus 0.5em minus 0.4em\relax Baltimore, MD: USENIX Association, Jul. 2005,
  pp. 113--128.

\bibitem{arincdomains}
\emph{{Aircraft Data Network, Part 5, Network Domain Characteristics and
  Interconnection}}, ARINC Specification 664P5, Apr. 2005.

\bibitem{userstudycabinnetwork2013}
\BIBentryALTinterwordspacing
C.~Diekmann, O.~Hanka, S.-A. Posselt, and M.~Schlatt, ``{Imaginary Aircraft
  Cabin Data Network (Toy Example)},'' Jul. 2013. [Online]. Available:
  \url{http://www.net.in.tum.de/pub/diekmann/cabin_data_network.pdf}
\BIBentrySTDinterwordspacing

\bibitem{johnson2010policytemplates}
M.~Johnson, J.~Karat, C.~M. Karat, and K.~Grueneberg, ``{Usable Policy Template
  Authoring for Iterative Policy Refinement},'' in \emph{Symposium on Policies
  for Distributed Systems and Networks (POLICY)}.\hskip 1em plus 0.5em minus
  0.4em\relax IEEE, Jul. 2010, pp. 18--21.

\bibitem{wang2009formally}
A.~Wang, L.~Jia, C.~Liu, B.~T. Loo, O.~Sokolsky, and P.~Basu, ``{Formally
  Verifiable Networking},'' in \emph{8th ACM Workshop on Hot Topics in
  Networks}, Oct. 2009.

\bibitem{kazemian2012HSA}
P.~Kazemian, G.~Varghese, and N.~McKeown, ``{Header Space Analysis: Static
  Checking for Networks},'' in \emph{9th USENIX Symposium on Networked Systems
  Design and Implementation (NSDI)}, ser. NSDI'12.\hskip 1em plus 0.5em minus
  0.4em\relax San Jose, CA: USENIX Association, Apr. 2012, pp. 113--126.

\bibitem{gude2008nox}
N.~Gude, T.~Koponen, J.~Pettit, B.~Pfaff, M.~Casado, N.~McKeown, and
  S.~Shenker, ``{NOX: Towards an Operating System for Networks},'' \emph{ACM
  SIGCOMM Computer Communication Review}, vol.~38, no.~3, pp. 105--110, Jul.
  2008.

\bibitem{machineverifiednetworkcontrollers13}
A.~Guha, M.~Reitblatt, and N.~Foster, ``{Machine-Verified Network
  Controllers},'' in \emph{34th ACM SIGPLAN conference on Programming language
  design and implementation}, ser. PLDI '13.\hskip 1em plus 0.5em minus
  0.4em\relax Seattle, Washington, USA: ACM, Jun. 2013, pp. 483--494.

\bibitem{coqmanual}
\BIBentryALTinterwordspacing
{The Coq development team}, \emph{{The Coq proof assistant reference manual}},
  {LogiCal Project}, 2004, last updated 2016. [Online]. Available:
  \url{http://coq.inria.fr/}
\BIBentrySTDinterwordspacing

\bibitem{ponder2001}
N.~Damianou, N.~Dulay, E.~Lupu, and M.~Sloman,
  ``\BIBforeignlanguage{English}{{The Ponder Policy Specification Language}},''
  in \emph{\BIBforeignlanguage{English}{Policies for Distributed Systems and
  Networks}}, ser. LNCS.\hskip 1em plus 0.5em minus 0.4em\relax Springer Berlin
  Heidelberg, 2001, vol. 1995, pp. 18--38.

\bibitem{policy2009expressivedynamic}
R.~Craven, J.~Lobo, J.~Ma, A.~Russo, E.~Lupu, and A.~Bandara, ``{Expressive
  Policy Analysis with Enhanced System Dynamicity},'' in \emph{4th
  International Symposium on Information, Computer, and Communications
  Security}, ser. ASIACCS '09.\hskip 1em plus 0.5em minus 0.4em\relax Sydney,
  Australia: ACM, Mar. 2009, pp. 239--250.

\bibitem{juergens2009softwarecrchitectures}
M.~Feilkas, D.~Ratiu, and E.~J{\"u}rgens, ``{The Loss of Architectural
  Knowledge during System Evolution: An Industrial Case Study},'' in \emph{17th
  International Conference on Program Comprehension (ICPC)}, May 2009, pp.
  188--197.

\bibitem{Murphy1995architectures}
G.~C. Murphy, D.~Notkin, and K.~Sullivan, ``{Software Reflexion Models:
  Bridging the Gap Between Source and High-level Models},'' \emph{SIGSOFT
  Software Engineering Notes}, vol.~20, no.~4, pp. 18--28, Oct. 1995.

\bibitem{Perry1992softwarearchitectures}
D.~E. Perry and A.~L. Wolf, ``{Foundations for the Study of Software
  Architecture},'' \emph{SIGSOFT Software Engineering Notes}, vol.~17, no.~4,
  pp. 40--52, Oct. 1992.

\bibitem{beller2012arhitecturestrict}
M.~Beller and E.~J{\"u}rgens:, ``{How Strict is Your Architecture?}'' in
  \emph{14th Workshop on Software Reengineering}, Bad Honnef (Germany), 2012.

\bibitem{diekmann2014EPTCS}
C.~Diekmann, L.~Hupel, and G.~Carle, ``{Directed Security Policies: A Stateful
  Network Implementation},'' in \emph{Engineering Safety and Security Systems
  (ESSS)}, ser. Electronic Proceedings in Theoretical Computer Science, vol.
  150.\hskip 1em plus 0.5em minus 0.4em\relax Singapore: Open Publishing
  Association, May 2014, pp. 20--34.

\bibitem{hansen2012research}
\BIBentryALTinterwordspacing
A.~H.~R. Hansen, ``Protecting critical infrastructure,'' \emph{ASA Institute
  for Risk \& Innovation}, pp. 1--12, Jun. 2012. [Online]. Available:
  \url{http://anniesearle.com/web-services/Documents/ResearchNotes/ASA_ResearchNote_ProtectingCriticalInfrastructure_June2012.pdf}
\BIBentrySTDinterwordspacing

\bibitem{brucker2008modelfwisabelle}
A.~D. Brucker, L.~Br{\"u}gger, and B.~Wolff, ``{Model-based Firewall
  Conformance Testing},'' in \emph{Testing of Software and Communicating
  Systems}.\hskip 1em plus 0.5em minus 0.4em\relax Springer, 2008, pp.
  103--118.

\bibitem{bsi2013smartmeter}
\emph{{Technische Richtlinie BSI TR-03109-1} -- Anforderungen an die
  Interoperabilit\"{a}t der Kommunikationseinheit eines intelligenten
  Messsystems}, 1st~ed., Bundesamt f\"{u}r Sicherheit in der
  Informationstechnik, Mar. 2013, \url{https://www.bsi.bund.de}.

\bibitem{hedgeuniontr}
\BIBentryALTinterwordspacing
S.~Adams, ``{Implementing Sets Efficiently in a Functional Language},''
  University of Southampton, Tech. Rep. CSTR 92-10, 1992. [Online]. Available:
  \url{http://groups.csail.mit.edu/mac/users/adams/BB/}
\BIBentrySTDinterwordspacing

\bibitem{cspfirewall}
S.~Pozo, R.~Ceballos, and R.~M. Gasca, ``{CSP-Based Firewall Rule Set Diagnosis
  using Security Policies},'' in \emph{2nd International Conference on
  Availability, Reliability and Security (ARES)}.\hskip 1em plus 0.5em minus
  0.4em\relax Los Alamitos, CA, USA: IEEE, Apr. 2007, pp. 723--729.

\bibitem{marmorstein2005itval}
R.~M. Marmorstein and P.~Kearns, ``{A Tool for Automated iptables Firewall
  Analysis},'' in \emph{USENIX Annual Technical Conference, FREENIX
  Track}.\hskip 1em plus 0.5em minus 0.4em\relax USENIX Association, Apr. 2005,
  pp. 71--81.

\bibitem{marmorstein2006firewall}
R.~Marmorstein and P.~Kearns, ``{Firewall Analysis with Policy-based Host
  Classification},'' in \emph{20th USENIX Large Installation System
  Administration Conference (LISA)}, vol.~6.\hskip 1em plus 0.5em minus
  0.4em\relax Washington, D.C.: USENIX Association, Dec. 2006.

\bibitem{tongaonkar2007inferring}
A.~Tongaonkar, N.~Inamdar, and R.~Sekar, ``{Inferring Higher Level Policies
  from Firewall Rules},'' in \emph{21st USENIX Large Installation System
  Administration Conference (LISA)}, vol.~7.\hskip 1em plus 0.5em minus
  0.4em\relax Dallas, TX: USENIX Association, Nov. 2007, pp. 1--10.

\bibitem{wool2004use}
A.~Wool, ``The use and usability of direction-based filtering in firewalls,''
  \emph{Computers \& Security}, vol.~23, no.~6, pp. 459--468, 2004.

\bibitem{bandara2009using}
A.~K. Bandara, A.~C. Kakas, E.~C. Lupu, and A.~Russo, ``{Using Argumentation
  Logic for Firewall Configuration Management},'' in \emph{IFIP/IEEE
  International Symposium on Integrated Network Management}, Jun. 2009, pp.
  180--187.

\bibitem{brucker.ea:formal-fw-testing:2014}
\BIBentryALTinterwordspacing
A.~D. Brucker, L.~Br{\"u}gger, and B.~Wolff,
  ``\BIBforeignlanguage{USenglish}{{Formal Firewall Conformance Testing: An
  Application of Test and Proof Techniques}},''
  \emph{\BIBforeignlanguage{USenglish}{Software Testing, Verification \&
  Reliability (STVR)}}, vol.~25, no.~1, pp. 34--71, 2015. [Online]. Available:
  \url{https://www.brucker.ch/bibliography/abstract/brucker.ea-formal-fw-testing-2014}
\BIBentrySTDinterwordspacing

\bibitem{brucker2013modelfwisabelle}
A.~D. Brucker, L.~Br\"{u}gger, and B.~Wolff, ``{HOL-TestGen/FW: An Environment
  for Specification-based Firewall Conformance Testing},'' in
  \emph{International Colloquium on Theoretical Aspects of Computing -– ICTAC
  2013}, ser. Lecture Notes in Computer Science, Z.~Liu, J.~Woodcock, and
  H.~Zhu, Eds.\hskip 1em plus 0.5em minus 0.4em\relax Springer Berlin
  Heidelberg, 2013, vol. 8049, pp. 112--121.

\bibitem{Guttman:1997:FilteringPostures}
J.~D. Guttman, ``{Filtering Postures: Local Enforcement for Global Policies},''
  in \emph{IEEE Symposium on Security and Privacy}, Washington, DC, USA, May
  1997.

\bibitem{modelchecking2000}
R.~Ritchey and P.~Ammann, ``{Using Model Checking to Analyze Network
  Vulnerabilities},'' in \emph{IEEE Symposium on Security and Privacy}, May
  2000, pp. 156--165.

\bibitem{diekmann2015topos}
C.~Diekmann, A.~Korsten, and G.~Carle, ``{Demonstrating topoS:
  Theorem-prover-based synthesis of secure network configurations},'' in
  \emph{11th International Conference on Network and Service Management
  (CNSM)}, Barcelona, Spain, Nov. 2015, pp. 366--371.

\bibitem{bari2013data}
M.~F. Bari, R.~Boutaba, R.~Esteves, L.~Z. Granville, M.~Podlesny, M.~G.
  Rabbani, Q.~Zhang, and M.~F. Zhani, ``{Data Center Network Virtualization: A
  Survey},'' \emph{{IEEE} Communications Surveys \& Tutorials}, vol.~15, no.~2,
  pp. 909--928, Sep. 2013.

\bibitem{xen2003}
P.~Barham, B.~Dragovic, K.~Fraser, S.~Hand, T.~Harris, A.~Ho, R.~Neugebauer,
  I.~Pratt, and A.~Warfield, ``{Xen and the Art of Virtualization},''
  \emph{SIGOPS Operating Systems Review}, vol.~37, no.~5, pp. 164--177, Oct.
  2003.

\bibitem{pfaff2009openvswitch}
\BIBentryALTinterwordspacing
B.~Pfaff, J.~Pettit, K.~Amidon, M.~Casado, T.~Koponen, and S.~Shenker,
  ``{Extending Networking into the Virtualization Layer},'' in \emph{Workshop
  on Hot Topics in Networks (HotNets)}.\hskip 1em plus 0.5em minus 0.4em\relax
  ACM, Oct. 2009. [Online]. Available:
  \url{http://openvswitch.org/papers/hotnets2009.pdf}
\BIBentrySTDinterwordspacing

\bibitem{openvswitch250releasenotes}
\BIBentryALTinterwordspacing
{Open vSwitch}, ``{Release Notes v2.5.0},'' Feb. 2016. [Online]. Available:
  \url{http://openvswitch.org/releases/NEWS-2.5.0}
\BIBentrySTDinterwordspacing

\bibitem{microservicesontherise2015blog}
\BIBentryALTinterwordspacing
P.~Cal\c{c}ado, ``How we ended up with microservices,'' Sep. 2015, retrieved
  Sep 2016. [Online]. Available:
  \url{http://philcalcado.com/2015/09/08/how_we_ended_up_with_microservices.html}
\BIBentrySTDinterwordspacing

\bibitem{lxc2016web}
\BIBentryALTinterwordspacing
``{LinuxContainers.org -- Infrastructure for container projects.}'' Sep. 2016.
  [Online]. Available: \url{https://linuxcontainers.org/}
\BIBentrySTDinterwordspacing

\bibitem{dockerweb}
\BIBentryALTinterwordspacing
{Docker, Inc.}, ``docker,'' 2016. [Online]. Available:
  \url{https://www.docker.com/}
\BIBentrySTDinterwordspacing

\bibitem{stackoverflow}
\BIBentryALTinterwordspacing
``{Stack Overflow}.'' [Online]. Available: \url{http://stackoverflow.com/}
\BIBentrySTDinterwordspacing

\bibitem{dockersecurity}
\BIBentryALTinterwordspacing
{Docker Inc.}, ``Docker security,'' Sep. 2016. [Online]. Available:
  \url{https://docs.docker.com/engine/security/security/}
\BIBentrySTDinterwordspacing

\bibitem{nccdocker2016security}
\BIBentryALTinterwordspacing
J.~Hertz, ``{Abusing Privileged and Unprivileged Linux Containers},'' {NCC
  Group}, 2016. [Online]. Available:
  \url{https://www.nccgroup.trust/globalassets/our-research/us/whitepapers/2016/june/container_whitepaperpdf/}
\BIBentrySTDinterwordspacing

\bibitem{bui15arxivsockersec}
\BIBentryALTinterwordspacing
T.~Bui, ``{Analysis of Docker Security},'' \emph{CoRR}, vol. abs/1501.02967,
  Jan. 2015. [Online]. Available: \url{http://arxiv.org/abs/1501.02967}
\BIBentrySTDinterwordspacing

\bibitem{dockerlegacylinks}
\BIBentryALTinterwordspacing
{Docker, Inc.}, ``{Network Configuration},'' version v1.5. [Online]. Available:
  \url{https://docs.docker.com/v1.5/articles/networking/}
\BIBentrySTDinterwordspacing

\bibitem{dockerlegacycontainerlinks}
\BIBentryALTinterwordspacing
{Docker Inc.}, ``Legacy container links,'' Sep. 2016. [Online]. Available:
  \url{https://docs.docker.com/engine/userguide/networking/default_network/dockerlinks/}
\BIBentrySTDinterwordspacing

\bibitem{monsanto2013composingonebigswitch}
C.~Monsanto, J.~Reich, N.~Foster, J.~Rexford, and D.~Walker, ``{Composing
  Software Defined Networks},'' in \emph{10th USENIX Symposium on Networked
  Systems Design and Implementation (NSDI)}, ser. NSDI'13.\hskip 1em plus 0.5em
  minus 0.4em\relax Lombard, IL: USENIX Association, Apr. 2013, pp. 1--13.

\bibitem{icfp2015smolkanetkatcompiler}
S.~Smolka, S.~A. Eliopoulos, N.~Foster, and A.~Guha, ``{A Fast Compiler for
  NetKAT},'' in \emph{International Conference on Functional Programming
  (ICFP)}.\hskip 1em plus 0.5em minus 0.4em\relax ACM, Sep. 2015, pp. 328--341.

\bibitem{Porras2012FortNOX}
P.~Porras, S.~Shin, V.~Yegneswaran, M.~Fong, M.~Tyson, and G.~Gu, ``{A Security
  Enforcement Kernel for OpenFlow Networks},'' in \emph{First Workshop on Hot
  Topics in Software Defined Networks}, ser. HotSDN '12.\hskip 1em plus 0.5em
  minus 0.4em\relax Helsinki, Finland: ACM, Aug. 2012, pp. 121--126.

\bibitem{kinetic2015}
H.~Kim, J.~Reich, A.~Gupta, M.~Shahbaz, N.~Feamster, and R.~Clark, ``{Kinetic:
  Verifiable Dynamic Network Control},'' in \emph{12th USENIX Symposium on
  Networked Systems Design and Implementation (NSDI)}, ser. NSDI'15.\hskip 1em
  plus 0.5em minus 0.4em\relax Oakland, CA: USENIX Association, May 2015, pp.
  59--72.

\bibitem{dinesh2002policybased}
D.~C. Verma, ``{Simplifying Network Administration Using Policy-Based
  Management},'' \emph{IEEE Network}, vol.~16, no.~2, pp. 20--26, Mar. 2002.

\bibitem{xie2005static}
G.~G. Xie, J.~Zhan, D.~A. Maltz, H.~Zhang, A.~G. Greenberg,
  G.~Hj{\'{a}}lmt{\'{y}}sson, and J.~Rexford, ``{On Static Reachability
  Analysis of IP Networks},'' in \emph{24th Annual Joint Conference of the IEEE
  Computer and Communications Societies (INFOCOM)}, vol.~3.\hskip 1em plus
  0.5em minus 0.4em\relax IEEE, Mar. 2005, pp. 2170--2183.

\bibitem{lopes2013msrnetworkverificationprogram}
\BIBentryALTinterwordspacing
N.~P. Lopes, N.~Bj{\o}rner, P.~Godefroid, and G.~Varghese, ``{Network
  Verification in the Light of Program Verification},'' Tech. Rep., Sep. 2013.
  [Online]. Available:
  \url{http://research.microsoft.com/apps/pubs/default.aspx?id=201589}
\BIBentrySTDinterwordspacing

\bibitem{Caesar2005rcp}
M.~Caesar, D.~Caldwell, N.~Feamster, J.~Rexford, A.~Shaikh, and J.~van~der
  Merwe, ``{Design and Implementation of a Routing Control Platform},'' in
  \emph{2nd USENIX Symposium on Networked Systems Design and Implementation
  (NSDI)}, ser. NSDI'05.\hskip 1em plus 0.5em minus 0.4em\relax Boston, MA:
  USENIX Association, May 2005, pp. 15--28.

\bibitem{Kang2013onebigswitchabstraction}
N.~Kang, Z.~Liu, J.~Rexford, and D.~Walker, ``{Optimizing the ``One Big
  Switch'' Abstraction in Software-defined Networks},'' in \emph{9th ACM
  Conference on Emerging Networking Experiments and Technologies}, ser.
  CoNEXT.\hskip 1em plus 0.5em minus 0.4em\relax ACM, Dec. 2013, pp. 13--24.

\bibitem{khurshid2013veriflow}
A.~Khurshid, X.~Zou, W.~Zhou, M.~Caesar, and P.~B. Godfrey, ``{VeriFlow:
  Verifying Network-Wide Invariants in Real Time},'' in \emph{10th USENIX
  Symposium on Networked Systems Design and Implementation (NSDI)}, ser.
  NSDI'13.\hskip 1em plus 0.5em minus 0.4em\relax Lombard, IL: USENIX
  Association, Apr. 2013, pp. 15--27.

\bibitem{orbacnetwork04}
F.~Cuppens, N.~Cuppens-Boulahia, T.~Sans, and A.~Mi{\`e}ge, ``{A Formal
  Approach to Specify and Deploy a Network Security Policy},'' in \emph{Formal
  Aspects of Security and Trust (FAST)}.\hskip 1em plus 0.5em minus 0.4em\relax
  Springer US, Aug. 2004, pp. 203--218.

\bibitem{Mai2011anteater}
H.~Mai, A.~Khurshid, R.~Agarwal, M.~Caesar, P.~B. Godfrey, and S.~T. King,
  ``{Debugging the Data Plane with Anteater},'' in \emph{ACM SIGCOMM}, Toronto,
  Ontario, Canada, Aug. 2011, pp. 290--301.

\bibitem{alshaer2009configchecker}
E.~Al-Shaer, W.~Marrero, A.~El-Atawy, and K.~Elbadawi, ``{Network Configuration
  in A Box: Towards End-to-End Verification of Network Reachability and
  Security},'' in \emph{International Conference on Network Protocols
  (ICNP)}.\hskip 1em plus 0.5em minus 0.4em\relax IEEE, Oct. 2009, pp.
  123--132.

\bibitem{orbac}
A.~{Abou El Kalam}, R.~{El Baida}, P.~Balbiani, S.~Benferhat, F.~Cuppens,
  Y.~Deswarte, A.~Mi\`{e}ge, C.~Saurel, and G.~Trouessin, ``{Organization Based
  Access Control},'' in \emph{IEEE 4th International Workshop on Policies for
  Distributed Systems and Networks (POLICY)}, Jun. 2003, pp. 120--131.

\bibitem{IEEE8021X}
\emph{{802.1X -- Port based Network Access Control}}, IEEE Std. 802.1X-2001,
  Jun. 2001.

\bibitem{rfc3060}
B.~Moore, E.~Ellesson, J.~Strassner, and A.~Westerinen, ``{Policy Core
  Information Model -- Version 1 Specification},'' RFC 3060 (Proposed
  Standard), Internet Engineering Task Force, Feb. 2001, updated by RFC 3460.

\bibitem{rfc3460}
B.~Moore, ``{Policy Core Information Model (PCIM) Extensions},'' RFC 3460
  (Proposed Standard), Internet Engineering Task Force, Jan. 2003.

\bibitem{hinrichs99pbmcisco}
S.~Hinrichs, ``{Policy-Based Management: Bridging the Gap},'' in \emph{15th
  Computer Security Applications Conference (ACSAC)}.\hskip 1em plus 0.5em
  minus 0.4em\relax IEEE, Dec. 1999, pp. 209--218.

\bibitem{Han2012surveypbm}
W.~Han and C.~Lei, ``A survey on policy languages in network and security
  management,'' \emph{Computer Networks}, vol.~56, no.~1, pp. 477--489, Jan.
  2012.

\bibitem{reich2013modular}
J.~Reich, C.~Monsanto, N.~Foster, J.~Rexford, and D.~Walker, ``{Modular SDN
  Programming with Pyretic},'' \emph{{;login:} The USENIX Magazine}, vol.~38,
  no.~5, pp. 40--47, Oct. 2013.

\bibitem{craven2011policyrefinement}
R.~Craven, J.~Lobo, E.~Lupu, A.~Russo, and M.~Sloman, ``{Policy Refinement:
  Decomposition and Operationalization for Dynamic Domains},'' in \emph{7th
  Conference on Network and Service Management (CNSM)}, Oct. 2011, pp. 1--9.

\bibitem{Pahl2015IM}
M.-O. Pahl, ``{Data-Centric Service-Oriented Management of Things},'' in
  \emph{IFIP/IEEE International Symposium on Integrated Network Management
  (IM)}, Ottawa, Canada, May 2015.

\bibitem{Feamster2005rcc}
N.~Feamster and H.~Balakrishnan, ``{Detecting BGP Configuration Faults with
  Static Analysis},'' in \emph{2nd USENIX Symposium on Networked Systems Design
  and Implementation (NSDI)}, ser. NSDI'05.\hskip 1em plus 0.5em minus
  0.4em\relax Boston, MA: USENIX Association, May 2005, pp. 43--56.

\bibitem{kazemian2013realtimehsa}
P.~Kazemian, M.~Chan, H.~Zeng, G.~Varghese, N.~McKeown, and S.~Whyte, ``{Real
  Time Network Policy Checking Using Header Space Analysis},'' in \emph{10th
  USENIX Symposium on Networked Systems Design and Implementation (NSDI)}, ser.
  NSDI'13.\hskip 1em plus 0.5em minus 0.4em\relax Lombard, IL: USENIX
  Association, Apr. 2013, pp. 99--111.

\bibitem{zhang2012erificationswitching}
S.~Zhang, S.~Malik, and R.~McGeer, ``\BIBforeignlanguage{English}{{Verification
  of Computer Switching Networks: An Overview}},'' in
  \emph{\BIBforeignlanguage{English}{Automated Technology for Verification and
  Analysis}}, ser. LNCS.\hskip 1em plus 0.5em minus 0.4em\relax Springer, 2012,
  pp. 1--16.

\bibitem{kineticwebsite}
\BIBentryALTinterwordspacing
H.~Kim, J.~Reich, A.~Gupta, M.~Shahbaz, N.~Feamster, and R.~Clark, ``{Kinetic:
  Verifiable Dynamic Control},'' 2014, retrieved Sep 2016. [Online]. Available:
  \url{http://resonance.noise.gatech.edu/}
\BIBentrySTDinterwordspacing

\bibitem{soule2014merlinsconferenceversion}
R.~Soul{\'e}, S.~Basu, P.~J. Marandi, F.~Pedone, R.~Kleinberg, E.~G. Sirer, and
  N.~Foster, ``{Merlin: A Language for Provisioning Network Resources},'' in
  \emph{10th ACM International on Conference on Emerging Networking Experiments
  and Technologies}, ser. CoNEXT.\hskip 1em plus 0.5em minus 0.4em\relax
  Sydney, Australia: ACM, Dec. 2014, pp. 213--226.

\bibitem{rfc4120}
C.~Neuman, T.~Yu, S.~Hartman, and K.~Raeburn, ``{The Kerberos Network
  Authentication Service (V5)},'' RFC 4120 (Proposed Standard), Internet
  Engineering Task Force, Jul. 2005, updated by RFCs 4537, 5021, 5896, 6111,
  6112, 6113, 6649, 6806, 7751.

\bibitem{schulzrinne2013pbs}
S.~G. Hong and H.~Schulzrinne, ``{PBS: Signaling Architecture for Network
  Traffic Authorization},'' \emph{IEEE Communications Magazine}, vol.~51,
  no.~7, pp. 89--96, Jul. 2013.

\bibitem{schulzrinne2010pbs}
S.~G. Hong, H.~Schulzrinne, and S.~Weiland, ``{Signaling Architecture for
  Network Traffic Authorization},'' in \emph{Global Telecommunications
  Conference (GLOBECOM)}.\hskip 1em plus 0.5em minus 0.4em\relax IEEE, Dec.
  2010, pp. 1--6.

\bibitem{rfc4301}
S.~Kent and K.~Seo, ``{Security Architecture for the Internet Protocol},'' RFC
  4301 (Proposed Standard), Internet Engineering Task Force, Dec. 2005, updated
  by RFC 6040.

\bibitem{nftables}
\BIBentryALTinterwordspacing
{The netfilter.org project}, ``netfilter/nftables project.'' [Online].
  Available: \url{http://www.netfilter.org/}
\BIBentrySTDinterwordspacing

\bibitem{bsdpf}
\BIBentryALTinterwordspacing
``{PF}: The {OpenBSD} packet filter.'' [Online]. Available:
  \url{http://www.openbsd.org/faq/pf/}
\BIBentrySTDinterwordspacing

\bibitem{ciscofirewallaccesslists}
\BIBentryALTinterwordspacing
``{Cisco IOS Firewall -- Configuring IP Access Lists},'' Document ID: 23602,
  Dec. 2007. [Online]. Available:
  \url{http://www.cisco.com/c/en/us/support/docs/security/ios-firewall/23602-confaccesslists.html}
\BIBentrySTDinterwordspacing

\bibitem{hpaclfirewalls}
\BIBentryALTinterwordspacing
{Hewlett Packard}, ``{IP} firewall configuration guide,'' 2005. [Online].
  Available:
  \url{ftp://ftp.hp.com/pub/networking/software/ProCurve-SR-IP-Firewall-Config-Guide.pdf}
\BIBentrySTDinterwordspacing

\bibitem{fwbuilder}
\BIBentryALTinterwordspacing
{NetCitadel, Inc.}, ``{FirewallBuilder},'' ver. 5.1. [Online]. Available:
  \url{http://www.fwbuilder.org}
\BIBentrySTDinterwordspacing

\bibitem{whylovenftables2014blog}
E.~Leblond, ``Why you will love nftables,'' Jan. 2014,
  \url{https://home.regit.org/2014/01/why-you-will-love-nftables/}.

\bibitem{zhang2012qbfsatfirewalls}
S.~Zhang, A.~Mahmoud, S.~Malik, and S.~Narain, ``{Verification and Synthesis of
  Firewalls Using SAT and QBF},'' in \emph{Network Protocols (ICNP)}, Oct.
  2012, pp. 1--6.

\bibitem{brucker2010firewalltransform}
A.~D. Brucker, L.~Br\"{u}gger, P.~Kearney, and B.~Wolff, ``Verified firewall
  policy transformations for test case generation,'' in \emph{3rd International
  Conference on Software Testing, Verification and Validation}.\hskip 1em plus
  0.5em minus 0.4em\relax IEEE, Apr. 2010, pp. 345--354.

\bibitem{UPF_Firewall-AFP}
\BIBentryALTinterwordspacing
A.~D. Brucker, L.~Br\"{u}gger, and B.~Wolff, ``{Formal Network Models and Their
  Application to Firewall Policies},'' \emph{Archive of Formal Proofs}, Jan.
  2017, formal proof development. [Online]. Available:
  \url{http://isa-afp.org/entries/UPF_Firewall.shtml}
\BIBentrySTDinterwordspacing

\bibitem{ciscotoiptables}
\BIBentryALTinterwordspacing
B.~Renard, ``{cisco-acl-to-iptables},'' 2013, retrieved Sep 2014. [Online].
  Available: \url{http://git.zionetrix.net/?a=summary&p=cisco-acl-to-iptables}
\BIBentrySTDinterwordspacing

\bibitem{ciscononicefeatures}
\BIBentryALTinterwordspacing
M.~Gartenmeister, ``{Iptables vs.\ Cisco PIX},'' Apr. 2005. [Online].
  Available:
  \url{http://lists.netfilter.org/pipermail/netfilter/2005-April/059714.html}
\BIBentrySTDinterwordspacing

\bibitem{fwmodelchecking2009jeffry}
A.~Jeffrey and T.~Samak, ``{Model Checking Firewall Policy Configurations},''
  in \emph{Policies for Distributed Systems and Networks}.\hskip 1em plus 0.5em
  minus 0.4em\relax IEEE, Jul. 2009, pp. 60--67.

\bibitem{kleene1952introduction}
S.~C. Kleene, \emph{Introduction to Metamathematics}, ser. Bibliotheca
  Mathematica, D.~Bruijn, V.~Dantzig, and D.~Groot, Eds.\hskip 1em plus 0.5em
  minus 0.4em\relax Amsterdam: North-Holland, 1952, {ISBN} 978-0923891572.

\bibitem{iptablesperfectruleset}
\BIBentryALTinterwordspacing
J.~Engelhardt, ``Towards the perfect ruleset,'' May 2011. [Online]. Available:
  \url{http://inai.de/documents/Perfect_Ruleset.pdf}
\BIBentrySTDinterwordspacing

\bibitem{rfc4632}
V.~Fuller and T.~Li, ``{Classless Inter-domain Routing (CIDR): The Internet
  Address Assignment and Aggregation Plan},'' RFC 4632 (Best Current Practice),
  Internet Engineering Task Force, Aug. 2006.

\bibitem{shorewall}
\BIBentryALTinterwordspacing
T.~M. Eastep, ``iptables made ease -- shorewall,'' 2014. [Online]. Available:
  \url{http://shorewall.net/}
\BIBentrySTDinterwordspacing

\bibitem{iptablesringofsaturn}
\BIBentryALTinterwordspacing
``{IPTables Example Config},'' retrieved Sep 2014. [Online]. Available:
  \url{http://networking.ringofsaturn.com/Unix/iptables.php}
\BIBentrySTDinterwordspacing

\bibitem{ofspec10}
\BIBentryALTinterwordspacing
B.~Pfaff, B.~Heller, D.~Talayco, D.~Erickson, G.~Gibb, G.~Appenzeller,
  J.~Tourrilhes, J.~Pettit, K.~Yap, M.~Casado, M.~Kobayashi, N.~McKeown,
  P.~Balland, R.~Price, R.~Sherwood, and Y.~Yiakoumis, ``{OpenFlow Switch
  Specification} v1.0.0,'' Dec. 2009. [Online]. Available:
  \url{http://archive.openflow.org/documents/openflow-spec-v1.0.0.pdf}
\BIBentrySTDinterwordspacing

\bibitem{ofspec15}
\BIBentryALTinterwordspacing
A.~Nygren, B.~Pfaff, B.~Lantz, B.~Heller, C.~Barker, C.~Beckmann, D.~Cohn,
  D.~Malek, D.~Talayco, D.~Erickson, D.~McDysan, D.~Ward, E.~Crabbe,
  F.~Schneider, G.~Gibb, G.~Appenzeller, J.~Tourrilhes, J.~Tonsing, J.~Pettit,
  K.~Yap, L.~Poutievski, L.~Dunbar, L.~Vicisano, M.~Casado, M.~Takahashi,
  M.~Kobayashi, M.~Orr, N.~Yadav, N.~McKeown, N.~dHeureuse, P.~Balland,
  R.~Madabushi, R.~Ramanathan, R.~Price, R.~Sherwood, S.~Das, S.~Gandham,
  S.~Curtis, S.~Natarajan, T.~Mizrahi, T.~Yabe, W.~Ding, Y.~Yiakoumis,
  Y.~Moses, and Z.~L. Kis, ``{OpenFlow Switch Specification} v1.5.1,'' Apr.
  2015, {ONF TS-025}. [Online]. Available:
  \url{https://www.opennetworking.org/images/stories/downloads/sdn-resources/onf-specifications/openflow/openflow-switch-v1.5.1.pdf}
\BIBentrySTDinterwordspacing

\bibitem{openflowniciraextensions2016github}
\BIBentryALTinterwordspacing
{Nicira, Inc.}, ``{Nicira extensions},'' {openvswitch/ovs} repository, Jun.
  2016, revision fb8f22c186b89cd36059c37908f940a1aa5e1569. [Online]. Available:
  \url{https://github.com/openvswitch/ovs/blob/master/include/openflow/nicira-ext.h}
\BIBentrySTDinterwordspacing

\bibitem{byma2015openflowcloudfpga}
S.~Byma, N.~Tarafdar, T.~Xu, H.~Bannazadeh, A.~Leon-Garcia, and P.~Chow,
  ``{Expanding OpenFlow Capabilities with Virtualized Reconfigurable
  Hardware},'' in \emph{ACM/SIGDA International Symposium on Field-Programmable
  Gate Arrays}, ser. FPGA '15.\hskip 1em plus 0.5em minus 0.4em\relax Monterey,
  California: ACM, Feb. 2015, pp. 94--97.

\bibitem{bianchi2014openstate}
G.~Bianchi, M.~Bonola, A.~Capone, and C.~Cascone, ``{OpenState: Programming
  Platform-independent Stateful Openflow Applications Inside the Switch},''
  \emph{ACM SIGCOMM Computer Communication Review}, vol.~44, no.~2, pp. 44--51,
  Apr. 2014.

\bibitem{2016arXiv161102853P}
L.~{Petrucci}, N.~{Bonelli}, M.~{Bonola}, G.~{Procissi}, C.~{Cascone},
  D.~{Sanvito}, S.~{Pontarelli}, G.~{Bianchi}, and R.~{Bifulco}, ``{Towards a
  Stateful Forwarding Abstraction to Implement Scalable Network Functions in
  Software and Hardware},'' \emph{ArXiv e-prints}, Nov. 2016.

\bibitem{nelson2015exodus}
T.~Nelson, A.~D. Ferguson, D.~Yu, R.~Fonseca, and S.~Krishnamurthi, ``{Exodus:
  Toward Automatic Migration of Enterprise Network Configurations to SDNs},''
  in \emph{1st ACM SIGCOMM Symposium on Software Defined Networking Research},
  ser. SOSR '15, no.~13.\hskip 1em plus 0.5em minus 0.4em\relax Santa Clara,
  California: ACM, Jun. 2015, pp. 13:1--13:7.

\bibitem{diekmann2015cnsm}
C.~Diekmann, L.~Schwaighofer, and G.~Carle, ``{Certifying Spoofing-Protection
  of Firewalls},'' in \emph{11th International Conference on Network and
  Service Management (CNSM)}, Barcelona, Spain, Nov. 2015, pp. 168--172.

\bibitem{rfc2827}
P.~Ferguson and D.~Senie, ``{Network Ingress Filtering: Defeating Denial of
  Service Attacks which employ IP Source Address Spoofing},'' RFC 2827 (Best
  Current Practice), Internet Engineering Task Force, May 2000, updated by RFC
  3704.

\bibitem{rfc3704}
F.~Baker and P.~Savola, ``{Ingress Filtering for Multihomed Networks},'' RFC
  3704 (Best Current Practice), Internet Engineering Task Force, Mar. 2004.

\bibitem{tolarisblogdisablerpfilter}
\BIBentryALTinterwordspacing
T.~J. Wagner, ``Disabling reverse-path filtering in complex networks,'' Jul.
  2009, retrieved Mai 2015. [Online]. Available:
  \url{https://www.tolaris.com/2009/07/13/disabling-reverse-path-filtering-in-complex-networks/}
\BIBentrySTDinterwordspacing

\bibitem{kernelrpfilter}
\BIBentryALTinterwordspacing
{Linux Kernel Sources}, ``ip-sysctl,'' Kernel 4.0 Documentation, 2015.
  [Online]. Available:
  \url{https://www.kernel.org/doc/Documentation/networking/ip-sysctl.txt}
\BIBentrySTDinterwordspacing

\bibitem{rfc6890}
M.~Cotton, L.~Vegoda, R.~Bonica, and B.~Haberman, ``{Special-Purpose IP Address
  Registries},'' RFC 6890 (Best Current Practice), Internet Engineering Task
  Force, Apr. 2013.

\bibitem{alshaer2004firewallpolicyanomaly}
E.~S. Al-Shaer and H.~H. Hamed, ``{Discovery of Policy Anomalies in Distributed
  Firewalls},'' in \emph{Annual Joint Conference of the IEEE Computer and
  Communications Societies (INFOCOM)}, vol.~4, Mar. 2004, pp. 2605--2616.

\bibitem{alshaer2011configcheckershort}
E.~Al-Shaer and M.~Alsaleh, ``{ConfigChecker: A Tool for Comprehensive Security
  Configuration Analytics},'' in \emph{Configuration Analytics and Automation
  (SAFECONFIG)}, Oct. 2011, pp. 1--2.

\bibitem{fwbuilder2011userguide}
\BIBentryALTinterwordspacing
{NetCitadel, Inc.}, ``{Firewall Builder 5 User's Guide},'' 2011. [Online].
  Available:
  \url{http://www.fwbuilder.org/4.0/docs/users_guide5/UsersGuide5.pdf}
\BIBentrySTDinterwordspacing

\bibitem{cybercity2005spoofing}
\BIBentryALTinterwordspacing
V.~Gite, ``{Linux Iptables Avoid IP Spoofing And Bad Addresses Attacks},''
  nixCraft blog, Jun. 2005. [Online]. Available:
  \url{http://www.cyberciti.biz/tips/linux-iptables-8-how-to-avoid-spoofing-and-bad-addresses-attack.html}
\BIBentrySTDinterwordspacing

\bibitem{rfc5952}
S.~Kawamura and M.~Kawashima, ``{A Recommendation for IPv6 Address Text
  Representation},'' RFC 5952 (Proposed Standard), Internet Engineering Task
  Force, Aug. 2010.

\bibitem{capretta2007coqfwconflicts}
V.~Capretta, B.~Stepien, A.~Felty, and S.~Matwin, ``{Formal Correctness of
  Conflict Detection for Firewalls},'' in \emph{Workshop on Formal Methods in
  Security Engineering}.\hskip 1em plus 0.5em minus 0.4em\relax ACM, Nov. 2007,
  pp. 22--30.

\bibitem{rfc780}
S.~Sluizer and J.~Postel, ``{Mail Transfer Protocol},'' RFC 780, Internet
  Engineering Task Force, May 1981, obsoleted by RFC 788.

\bibitem{iptablesgithub113issue}
\BIBentryALTinterwordspacing
{diekmann/Iptables\_Semantics}, ``{Issue \#113 -- Port numbers belong to a
  specific protocol},'' github, Jul. 2016. [Online]. Available:
  \url{https://github.com/diekmann/Iptables_Semantics/issues/113}
\BIBentrySTDinterwordspacing

\bibitem{rfc791}
J.~Postel, ``{Internet Protocol},'' RFC 791 (INTERNET STANDARD), Internet
  Engineering Task Force, Sep. 1981, updated by RFCs 1349, 2474, 6864.

\bibitem{rfc2460}
S.~Deering and R.~Hinden, ``{Internet Protocol, Version 6 (IPv6)
  Specification},'' RFC 2460 (Draft Standard), Internet Engineering Task Force,
  Dec. 1998, updated by RFCs 5095, 5722, 5871, 6437, 6564, 6935, 6946, 7045,
  7112.

\bibitem{rfc1700}
J.~Reynolds and J.~Postel, ``{Assigned Numbers},'' RFC 1700 (Historic),
  Internet Engineering Task Force, Oct. 1994, obsoleted by RFC 3232.

\bibitem{rfc3232}
J.~Reynolds, ``{Assigned Numbers: RFC 1700 is Replaced by an On-line
  Database},'' RFC 3232 (Informational), Internet Engineering Task Force, Jan.
  2002.

\bibitem{kernelmatchinterface}
\BIBentryALTinterwordspacing
{Linux Kernel Sources}, ``{Linux/include/linux/netfilter/x\_tables.h},'' Kernel
  4.6. [Online]. Available:
  \url{http://lxr.free-electrons.com/source/include/linux/netfilter/x_tables.h?v=4.6#L343}
\BIBentrySTDinterwordspacing

\bibitem{xtablesmtach}
\BIBentryALTinterwordspacing
{netfilter coreteam}, ``{libxtables/xtables.c}.'' [Online]. Available:
  \url{https://git.netfilter.org/iptables/tree/libxtables/xtables.c?h=v1.6.0#n518}
\BIBentrySTDinterwordspacing

\bibitem{tantau2016tikz}
T.~Tantau and C.~Feuersaenger, ``{The TikZ and pgf packages},'' 2016,
  pgfversion 3.0.1a.

\bibitem{serverfaultmultiportnegation}
\BIBentryALTinterwordspacing
{CrazyCat}, ``{iptables multiport and negation},'' {Server Fault Question},
  Aug. 2016. [Online]. Available:
  \url{http://serverfault.com/questions/793631/iptables-multiport-and-negation/}
\BIBentrySTDinterwordspacing

\bibitem{xtermescapes}
\BIBentryALTinterwordspacing
E.~Moy, S.~Gildea, and T.~Dickey, ``{XTerm Control Sequences},'' 2016.
  [Online]. Available:
  \url{http://invisible-island.net/xterm/ctlseqs/ctlseqs.html}
\BIBentrySTDinterwordspacing

\bibitem{Lammich2013autoref}
P.~Lammich, \emph{{Automatic Data Refinement}}.\hskip 1em plus 0.5em minus
  0.4em\relax Berlin, Heidelberg: Springer Berlin Heidelberg, 2013, pp. 84--99.

\bibitem{Lammich2012isabellerefine}
P.~Lammich and T.~Tuerk, \emph{{Applying Data Refinement for Monadic Programs
  to Hopcroft's Algorithm}}.\hskip 1em plus 0.5em minus 0.4em\relax Berlin,
  Heidelberg: Springer Berlin Heidelberg, 2012, pp. 166--182.

\bibitem{docker2015sec}
\BIBentryALTinterwordspacing
A.~Mouat, \emph{{Docker Security -- Using Containers Safely in Production}},
  1st~ed.\hskip 1em plus 0.5em minus 0.4em\relax O'Reilly, 2015. [Online].
  Available: \url{https://www.openshift.com/promotions/docker-security.html}
\BIBentrySTDinterwordspacing

\bibitem{docker2016blog110}
\BIBentryALTinterwordspacing
{Docker Core Engineering}, ``{Docker 1.10: New Compose File, Improved Security,
  Networking And Much More!}'' blog, Feb. 2016, retrieved Dec 2016. [Online].
  Available: \url{https://blog.docker.com/2016/02/docker-1-10/}
\BIBentrySTDinterwordspacing

\bibitem{github2016dockerinternalicc}
\BIBentryALTinterwordspacing
C.~Diekmann, ``{Issue \#29108 -- Networking/Security: Custom net with both
  {-}{-}internal and {-}{-}icc=false does not block icc},'' github, Dec. 2016.
  [Online]. Available: \url{https://github.com/docker/docker/issues/29108}
\BIBentrySTDinterwordspacing

\bibitem{docker2016blognetphilosophy}
\BIBentryALTinterwordspacing
J.~Radhakrishnan, ``{Docker Networking Design Philosophy},'' blog, Mar. 2016,
  retrieved Dec 2016. [Online]. Available:
  \url{https://blog.docker.com/2016/03/docker-networking-design-philosophy/}
\BIBentrySTDinterwordspacing

\bibitem{github2016dockerfwbypass}
\BIBentryALTinterwordspacing
B.~Meyer, ``{Issue \#22054 -- Docker Network bypasses Firewall, no option to
  disable},'' github issue and discussion, Apr. 2016, retrieved Dec 2016.
  [Online]. Available: \url{https://github.com/docker/docker/issues/22054}
\BIBentrySTDinterwordspacing

\bibitem{docker2014blogufw}
\BIBentryALTinterwordspacing
V.~Petersson, ``{The dangers of UFW + Docker},'' blog, Nov. 2014, retrieved Nov
  2016. [Online]. Available:
  \url{http://blog.viktorpetersson.com/post/101707677489/the-dangers-of-ufw-docker}
\BIBentrySTDinterwordspacing

\bibitem{rhel2016logpwn}
\BIBentryALTinterwordspacing
I.~Duffy, ``{Azure bug bounty Pwning Red Hat Enterprise Linux},'' blog, Nov.
  2016, retrieved Nov 2016. [Online]. Available:
  \url{http://ianduffy.ie/blog/2016/11/26/azure-bug-bounty-pwning-red-hat-enterprise-linux/}
\BIBentrySTDinterwordspacing

\bibitem{github2016dockerfw}
\BIBentryALTinterwordspacing
{gdm85}, ``{Docker-fw},'' github, Mar. 2016, commit d335396. [Online].
  Available: \url{https://github.com/gdm85/docker-fw}
\BIBentrySTDinterwordspacing

\bibitem{github2016dfwfw}
\BIBentryALTinterwordspacing
I.~Rad, ``{DFWFW -- Docker Firewall Framework},'' github, Aug. 2016, commit
  94d1036. [Online]. Available: \url{https://github.com/irsl/dfwfw}
\BIBentrySTDinterwordspacing

\bibitem{bleikertz2011VALID}
S.~Bleikertz and T.~Gro{\ss}, ``{A Virtualization Assurance Language for
  Isolation and Deployment},'' in \emph{International Symposium on Policies for
  Distributed Systems and Networks (POLICY)}.\hskip 1em plus 0.5em minus
  0.4em\relax IEEE, Jun. 2011, pp. 33--40.

\bibitem{measrdroid}
{Chair of Network Architectures and Services, TUM}, ``{MeasrDroid},''
  \url{http://www.droid.net.in.tum.de/},
  \url{https://play.google.com/store/apps/details?id=de.tum.in.net.measrdroid.gui.stats}.

\bibitem{llamport2002interview}
\BIBentryALTinterwordspacing
D.~Milojicic, ``{A Discussion with Leslie Lamport},'' Aug. 2002. [Online].
  Available:
  \url{http://research.microsoft.com/en-us/um/people/lamport/pubs/ds-interview.pdf}
\BIBentrySTDinterwordspacing

\bibitem{gupta2014sdx}
A.~Gupta, L.~Vanbever, M.~Shahbaz, S.~P. Donovan, B.~Schlinker, N.~Feamster,
  J.~Rexford, S.~Shenker, R.~Clark, and E.~Katz-Bassett, ``{SDX: A Software
  Defined Internet Exchange},'' in \emph{ACM SIGCOMM}.\hskip 1em plus 0.5em
  minus 0.4em\relax Chicago, Illinois: ACM, Aug. 2014, pp. 551--562.

\bibitem{nelson2013balanceofpower}
T.~Nelson, A.~Guha, D.~J. Dougherty, K.~Fisler, and S.~Krishnamurthi, ``{A
  Balance of Power: Expressive, Analyzable Controller Programming},'' in
  \emph{Second ACM SIGCOMM Workshop on Hot Topics in Software Defined
  Networking}, ser. HotSDN '13.\hskip 1em plus 0.5em minus 0.4em\relax Hong
  Kong, China: ACM, Aug. 2013, pp. 79--84.

\bibitem{netkat2016state}
J.~McClurg, H.~Hojjat, N.~Foster, and P.~\v{C}ern\'{y}, ``{Event-driven Network
  Programming},'' in \emph{37th ACM SIGPLAN Conference on Programming Language
  Design and Implementation}, ser. PLDI '16.\hskip 1em plus 0.5em minus
  0.4em\relax Santa Barbara, CA: ACM, Jun. 2016, pp. 369--385.

\bibitem{nelsongithubflowlogfirewall}
\BIBentryALTinterwordspacing
T.~Nelson, ``{FlowLog},'' github, Apr. 2015, stateful firewall,
  \url{FlowLog/interpreter/examples/getting-started/sfw.flg}, commit 8be5051.
  [Online]. Available: \url{https://github.com/tnelson/FlowLog}
\BIBentrySTDinterwordspacing

\bibitem{nelson2013phdthesis}
\BIBentryALTinterwordspacing
T.~Nelson, ``{First-Order Models for Configuration Analysis},''
  etd-042513-142414, Worcester Polytechnic Institute, Apr. 2013. [Online].
  Available: \url{http://www.wpi.edu/Pubs/ETD/Available/etd-042513-142414/}
\BIBentrySTDinterwordspacing

\bibitem{alloywebsite}
\BIBentryALTinterwordspacing
D.~Jackson, A.~Milicevic, E.~Torlak, E.~Kang, J.~Near, J.~Edwards, R.~Seater,
  D.~Rayside, G.~Dennis, F.~Chang, I.~Shlyakhter, M.~Taghdiri, M.~Vaziri,
  S.~Khurshid, and M.~Sridharan, ``alloy: a language \& tool for relational
  models.'' [Online]. Available: \url{http://alloy.mit.edu/alloy/}
\BIBentrySTDinterwordspacing

\bibitem{nelson2015chimp}
T.~Nelson, A.~D. Ferguson, and S.~Krishnamurthi, \emph{{Static Differential
  Program Analysis for Software-Defined Networks}}.\hskip 1em plus 0.5em minus
  0.4em\relax Springer International Publishing, 2015, pp. 395--413.

\bibitem{guha2013formalshort}
\BIBentryALTinterwordspacing
A.~Guha, M.~Reitblatt, and N.~Foster, ``{Formal Foundations for Software
  Defined Networks},'' \emph{Open Networking Summit (ONS) Research Track},
  2013. [Online]. Available:
  \url{http://www.cs.cornell.edu/~jnfoster/papers/frenetic-verified-controllers-ons.pdf}
\BIBentrySTDinterwordspacing

\bibitem{stewart2013computationalnetcorecoq}
G.~Stewart, \emph{{Computational Verification of Network Programs in
  Coq}}.\hskip 1em plus 0.5em minus 0.4em\relax Springer International
  Publishing, Dec. 2013, pp. 33--49.

\bibitem{netocre2016severalswitchesbad}
H.~Date and N.~Yoshiura, \emph{{Computational Verification of Network Programs
  for Several OpenFlow Switches in Coq}}.\hskip 1em plus 0.5em minus
  0.4em\relax Springer International Publishing, Jul. 2016, pp. 223--238.

\bibitem{netkat2015coalgebraicdecision}
N.~Foster, D.~Kozen, M.~Milano, A.~Silva, and L.~Thompson, ``{A Coalgebraic
  Decision Procedure for NetKAT},'' in \emph{42nd Annual ACM SIGPLAN-SIGACT
  Symposium on Principles of Programming Languages}, ser. POPL '15.\hskip 1em
  plus 0.5em minus 0.4em\relax Mumbai, India: ACM, Jan. 2015, pp. 343--355.

\bibitem{githubNetKATcoalgebraic}
\BIBentryALTinterwordspacing
{frenetic-lang/netkat-automata}, ``{A Coalgebraic Decision Procedure for
  NetKAT},'' github, May 2016, revision
  d89b2ba8c20a66fda6172edaec5d4714590f7f4b. [Online]. Available:
  \url{https://github.com/frenetic-lang/netkat-automata}
\BIBentrySTDinterwordspacing

\bibitem{githubNetKATjsonparser}
\BIBentryALTinterwordspacing
{frenetic-lang/frenetic}, ``{frenetic/lib/Frenetic\_NetKAT\_Json.mli},''
  github, Jul. 2016, revision 8a8d8630cc45dd0904a538efc60b5bec8deed936.
  [Online]. Available: \url{https://github.com/frenetic-lang/frenetic/}
\BIBentrySTDinterwordspacing

\bibitem{braibant2010kaincoq}
T.~Braibant and D.~Pous, \emph{{An Efficient Coq Tactic for Deciding Kleene
  Algebras}}.\hskip 1em plus 0.5em minus 0.4em\relax Berlin, Heidelberg:
  Springer Berlin Heidelberg, 2010, pp. 163--178.

\bibitem{Pous2013katincoq}
D.~Pous, \emph{{Kleene Algebra with Tests and Coq Tools for while
  Programs}}.\hskip 1em plus 0.5em minus 0.4em\relax Berlin, Heidelberg:
  Springer Berlin Heidelberg, 2013, pp. 180--196.

\bibitem{githubnetkatrepoissue2}
\BIBentryALTinterwordspacing
{frenetic-lang/netkat}, ``{Issue \#2 -- Current State?}'' github, Sep. 2016.
  [Online]. Available:
  \url{https://web.archive.org/web/20160910140202/https://github.com/frenetic-lang/netkat/issues/2}
\BIBentrySTDinterwordspacing

\bibitem{bloom1970filters}
B.~H. Bloom, ``{Space/Time Trade-offs in Hash Coding with Allowable Errors},''
  \emph{Communications of the ACM}, vol.~13, no.~7, pp. 422--426, Jul. 1970.

\bibitem{githubmignis}
\BIBentryALTinterwordspacing
{secgroup/Mignis}, ``{Mignis is a semantic based tool for firewall
  configuration.}'' github, Mar. 2014, revision
  3448282d5b8e057ebb745bc75af994eae45ff793. [Online]. Available:
  \url{https://github.com/secgroup/Mignis}
\BIBentrySTDinterwordspacing

\bibitem{githubmignisissue2}
\BIBentryALTinterwordspacing
{secgroup/Mignis}, ``{Issue \#2 -- \_check\_filters not called on all
  user-added matching features},'' github, Sep. 2016. [Online]. Available:
  \url{https://web.archive.org/web/20160914073740/https://github.com/secgroup/Mignis/issues/2}
\BIBentrySTDinterwordspacing

\bibitem{EWD:EWD800}
\BIBentryALTinterwordspacing
E.~W. Dijkstra, ``{A bagatelle for the left hand},'' Jan. 1982, {EWD800}.
  [Online]. Available:
  \url{http://www.cs.utexas.edu/users/EWD/ewd08xx/EWD800.PDF}
\BIBentrySTDinterwordspacing

\bibitem{github2016insalata}
\BIBentryALTinterwordspacing
M.~Dorfhuber and C.~Rudolf, ``{INSALATA -- IT NetworkS AnaLysis And deploymenT
  Application},'' github. [Online]. Available:
  \url{https://github.com/tumi8/INSALATA}
\BIBentrySTDinterwordspacing

\bibitem{serverfaultpinput}
\BIBentryALTinterwordspacing
{rakemanyohneth}, ``{What does -P INPUT ACCEPT mean with regards to
  iptables?}'' {Server Fault Question}, Mar. 2016. [Online]. Available:
  \url{http://serverfault.com/questions/766198/what-does-p-input-accept-mean-with-regards-to-iptables/770971#770971}
\BIBentrySTDinterwordspacing

\bibitem{serverfaultspoofing}
\BIBentryALTinterwordspacing
{excanoe}, ``{What is the destination of the spoofed source ip packet in terms
  of the netfilter chains?}'' {Server Fault Question}, Aug. 2016. [Online].
  Available:
  \url{http://serverfault.com/questions/800229/what-is-the-destination-of-the-spoofed-source-ip-packet-in-terms-of-the-netfilte/800434#800434}
\BIBentrySTDinterwordspacing

\bibitem{serverfaultwhyblocking}
\BIBentryALTinterwordspacing
{hsivonen}, ``{Why are these ip6tables rules blocking ssh over IPv6 when the
  iptables version allows it over IPv4},'' {Server Fault Question}, Aug. 2016.
  [Online]. Available:
  \url{http://serverfault.com/questions/796630/why-are-these-ip6tables-rules-blocking-ssh-over-ipv6-when-the-iptables-version-a/796729#796729}
\BIBentrySTDinterwordspacing

\bibitem{serverfaultwhywhitelistdns}
\BIBentryALTinterwordspacing
M.~Ragoso, ``{Iptables rules which only whitelist dns and drop everything
  else},'' {Server Fault Question}, Aug. 2016. [Online]. Available:
  \url{http://serverfault.com/questions/799212/iptables-rules-which-only-whitelist-dns-and-drop-everything-else/799273#799273}
\BIBentrySTDinterwordspacing

\bibitem{serverfaultipv6fail}
\BIBentryALTinterwordspacing
R.~Koch, ``{Will using ACCEPT then DROP for a specific port/ip couple allow the
  ip but nothing else on that port?}'' {Server Fault Question}, Aug. 2016.
  [Online]. Available:
  \url{http://serverfault.com/questions/795506/will-using-accept-then-drop-for-a-specific-port-ip-couple-allow-the-ip-but-nothi/795524#795524}
\BIBentrySTDinterwordspacing

\bibitem{diekmanngithubiptablessemantics}
\BIBentryALTinterwordspacing
C.~Diekmann, ``{Iptables\_Semantics -- Verified iptables Firewall Ruleset
  Analysis},'' github. [Online]. Available:
  \url{https://github.com/diekmann/Iptables_Semantics}
\BIBentrySTDinterwordspacing

\bibitem{google2016sre}
B.~Beyer, C.~Jones, J.~Petoff, and N.~R. Murphy, \emph{{Site Reliability
  Engineering -- How Google Runs Production Systems}}.\hskip 1em plus 0.5em
  minus 0.4em\relax O'Reilly Media, Mar. 2016, {ISBN} 978-1-4919-2909-4.

\bibitem{keown2013sdnsummerschoool}
\BIBentryALTinterwordspacing
N.~McKeown, ``{Forwarding Plane Correctness},'' in \emph{Summer School on
  Formal Methods and Networks}, 2013. [Online]. Available:
  \url{http://www.cs.cornell.edu/conferences/formalnetworks/}
\BIBentrySTDinterwordspacing

\bibitem{knuth77onlyproved}
\BIBentryALTinterwordspacing
D.~E. Knuth, ``{The correspondence between Donald E.\ Knuth and Peter van Emde
  Boas on priority deques during the spring of 1977}.'' [Online]. Available:
  \url{https://staff.fnwi.uva.nl/p.vanemdeboas/knuthnote.pdf}
\BIBentrySTDinterwordspacing

\end{thebibliography}

  \cleardoublepage

\ifnetcover
\thispagestyle{empty}
\cleartoleftpage
\thispagestyle{empty}
\includepdf[pages={1},offset=-2.54cm -2.54cm]{cover_back.pdf}
\fi

\end{document}